\documentclass[aps,10pt,pra,citeautoscript,longbibliography,superscriptaddress,twocolumn]{revtex4-1}
\usepackage[T1]{fontenc}
\usepackage{graphicx}
\usepackage{amsmath,amssymb,amsfonts,bm,dsfont,units}
\usepackage{mathtools}
\usepackage{hyperref}

\usepackage{yfonts}

\newcommand{\cohosub}[1]{\scalebox{0.7}{\textswab{#1}}}
\newcommand{\coho}[1]{\textswab{#1}}

\usepackage[mathscr]{euscript}

\usepackage{braket}

\usepackage{xparse}
\usepackage{tikz}
\usetikzlibrary{decorations.pathmorphing}

\makeatletter

\newcommand{\math@param}[3]{%
  \fontdimen#3
  \ifx#1\displaystyle\textfont#2
  \else\ifx#1\textstyle\textfont#2
  \else\ifx#1\scriptstyle\scriptfont#2
  \else\scriptscriptfont#2 \fi\fi\fi
}

\newdimen\@my@yshift
\NewDocumentCommand{\zig}{m}{\mathord{\mathpalette\@myvec{#1}}}

\newcommand*{\@myvec}[2]{%
  \begin{tikzpicture}[baseline=(n.base)]
  \node (n) [inner sep=0]
    {\global\@my@yshift=0.5\math@param{#1}{2}{5}%
     $\m@th #1#2$};
    \draw[decorate, decoration={snake, amplitude=0.6pt, segment length=2pt}]
      ([yshift=\@my@yshift]n.north west) -- ([yshift=\@my@yshift]n.north east);
  \end{tikzpicture}%
}
\makeatother

\makeatletter

\newcommand{\math@paramp}[3]{%
  \fontdimen#3
  \ifx#1\displaystyle\textfont#2
  \else\ifx#1\textstyle\textfont#2
  \else\ifx#1\scriptstyle\scriptfont#2
  \else\scriptscriptfont#2 \fi\fi\fi
}

\newdimen\@my@yshift
\NewDocumentCommand{\zigp}{m}{\mathord{\mathpalette\@myvecp{#1}}}

\newcommand*{\@myvecp}[2]{%
  \begin{tikzpicture}[baseline=(n.base)]
  \node (n) [inner sep=0]
    {\global\@my@yshift=0.5\math@paramp{#1}{2}{5}%
     $\m@th #1#2$};
    \draw[decorate, decoration={snake, amplitude=0.6pt, segment length=2pt}]
      ([yshift=\@my@yshift]n.north west) -- ([yshift=\@my@yshift]n.north east);
  \end{tikzpicture}%
}
\makeatother

\usetikzlibrary{decorations.pathreplacing,calligraphy}
\makeatletter

\newcommand{\math@parampt}[3]{%
  \fontdimen#3
  \ifx#1\displaystyle\textfont#2
  \else\ifx#1\textstyle\textfont#2
  \else\ifx#1\scriptstyle\scriptfont#2
  \else\scriptscriptfont#2 \fi\fi\fi
}

\newdimen\@my@yshift
\NewDocumentCommand{\tors}{m}{\mathord{\mathpalette\@myvecpt{#1}}}

\newcommand*{\@myvecpt}[2]{%
  \begin{tikzpicture}[baseline=(n.base)]
  \node (n) [inner sep=0]
    {\global\@my@yshift=0.3\math@parampt{#1}{2}{5}%
     $\m@th #1#2$};
    \draw[decorate, decoration={brace, amplitude=1.5pt, segment length=2pt}, line width=.5pt]
      ([yshift=\@my@yshift]n.north west) -- ([yshift=\@my@yshift]n.north east);
  \end{tikzpicture}%
}
\makeatother

\newlength{\dhatheight}

\makeatletter
\newcommand{\hathat}[1]{%
\begingroup%
  \let\macc@kerna\z@%
  \let\macc@kernb\z@%
  \let\macc@nucleus\@empty%
  \hat{\raisebox{.35ex}{\vphantom{\ensuremath{#1}}}\smash{\hat{#1}}}%
\endgroup%
}
\makeatother

\makeatletter
\DeclareRobustCommand\widecheck[1]{{\mathpalette\@widecheck{#1}}}
\def\@widecheck#1#2{%
    \setbox\z@\hbox{\m@th$#1#2$}%
    \setbox\tw@\hbox{\m@th$#1%
       \widehat{%
          \vrule\@width\z@\@height\ht\z@
          \vrule\@height\z@\@width\wd\z@}$}%
    \dp\tw@-\ht\z@
    \@tempdima\ht\z@ \advance\@tempdima2\ht\tw@ \divide\@tempdima\thr@@
    \setbox\tw@\hbox{%
       \raise\@tempdima\hbox{\scalebox{1}[-1]{\lower\@tempdima\box
\tw@}}}%
    {\ooalign{\box\tw@ \cr \box\z@}}}
\makeatother

\usepackage{stackengine,scalerel}
\stackMath
\newcommand\that[1]{\ThisStyle{%
  \setbox0=\hbox{$\SavedStyle#1$}%
  \ensurestackMath{%
    \stackon[1.0\LMpt]{\copy0}{\,\rotatebox{90}{%
      $\SavedStyle\stretchto{\scaleto[3pt]{\triangleright}{\textheight}}{\dimexpr\wd0-2pt}$}}%
  }%
}}

\usepackage{stackengine,scalerel}
\stackMath
\newcommand\thattilde[1]{\ThisStyle{%
  \setbox0=\hbox{$\SavedStyle#1$}%
  \ensurestackMath{%
    \stackon[1.0\LMpt]{\copy0}{\,\rotatebox{90}{%
      $\SavedStyle\stretchto{\scaleto[3pt]{\square}{\textheight}}{\dimexpr\wd0-2pt}$}}%
  }%
}}

\usepackage{color}
\definecolor{dkgreen}{rgb}{0,0.5,0}
\definecolor{midnightblue}{rgb}{0.39,0.58,0.93}
\definecolor{ltgreen}{rgb}{0.1,.59,.43}
\definecolor{hanpurple}{rgb}{0.32, 0.09, 0.98}
\definecolor{readableyellow}{rgb}{0.55, 0.45,0}
\hypersetup{colorlinks=true,citecolor=ltgreen,linkcolor=ltgreen,urlcolor=hanpurple,pdfstartview=FitH,
pdftitle = {FSET}}

\DeclareMathAlphabet{\mathpzc}{OT1}{pzc}{m}{it}

\definecolor{purpledk}{rgb}{0.32, 0.09, 0.80}

\newcommand{\cond}[1]{\zig{#1}}
\newcommand{\tor}[1]{\that{#1}}
\newcommand{\tortilde}[1]{\thattilde{#1}}

\def\stackbelow#1#2{\underset{\displaystyle\overset{\displaystyle\shortparallel}{#2}}{#1}}

\newcommand{\vv}[1]{\mathrm{#1}}
\newcommand{\I}{\mathcal{I}}

\newcommand{\tc}{\operatorname{{\bf TC}}}
\newcommand{\ifo}{\operatorname{{\bf K}}}
\newcommand{\spt}{\operatorname{\bf SPT}}

\newcommand{\cbd}{\text{d}}
\newcommand{\res}{\mathsf{res}}
\newcommand{\im}{\operatorname{im}}

\newcommand{\mcb}{\mathcal{B}}
\newcommand{\mcg}{\mathcal{G}}

\newcommand{\bMTC}{\mathcal{C}}
\newcommand{\fMTC}{\mathcal{M}}

\newcommand{\BTC}{\mathcal{B}}
\newcommand{\base}{{\bf K}^{{(0)} \times  \text{ base}}}

\newcommand{\defectO}{\mathscr{O}}

\newcommand{\central}{\vv{w}}
\newcommand{\vviso}{\vv{u}}
\newcommand{\rr}{\vv{r}}
\newcommand{\Xdot}{\mathscr{X}}
\newcommand{\Ydot}{\mathscr{Y}}
\newcommand{\xdot}{\chi}
\newcommand{\ydot}{\lambda}
\newcommand{\bdot}{\upsilon}
\newcommand{\maj}{\pi}

\newcommand{\on}[1]{\operatorname{#1}}

\newcommand{\eff}{\operatorname{f}}
\newcommand{\empsi}{\varepsilon}
\newcommand{\Gr}{\mathsf{G}^{\mathsf{(3,2,1)}}}
\newcommand{\Grto}{\mathsf{G}^{\mathsf{(2,1)}}}
\newcommand{\Grt}{\mathsf{G}^{\mathsf{(2)}}}
\newcommand{\Gro}{\mathsf{G}^{\mathsf{(1)}}}
\newcommand{\Gtriv}{\mathsf{G}^{\mathsf{(3,2)}}}
\newcommand{\GSPT}{\mathsf{G}^{\mathsf{(3)}}}
\newcommand{\Gifo}{\mathsf{G}^{\mathsf{(3,2,1,0)}}}

\newcommand{\BZ}{\mathsf{G}^{\mathsf{(0)}}}

\newcommand{\BH}{BH}
\newcommand{\U}{U}
\newcommand{\Q}{\mathcal{Q}}
\newcommand{\V}{\mathcal{V}}
\newcommand{\Aut}[1]{\text{Aut}({#1})}
\newcommand{\Autpsi}[1]{\text{Aut}^{\psi}({#1} )}
\newcommand{\Autf}[1]{\text{Aut}^{\eff}({#1})}
\newcommand{\Autpsif}[1]{\text{Aut}^{\psi / \eff}({#1} )}

\newcommand{\ftimes}{\, \mathord{\vcenter{\hbox{\includegraphics[scale=.2]{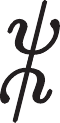}}}}}
\newcommand{\cocyleatorLold}{\mathord{\vcenter{\hbox{\includegraphics[scale=1]{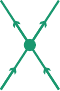}}}}}

\newcommand{\cocyleatorR}{\mathord{\vcenter{\hbox{\includegraphics[scale=1]{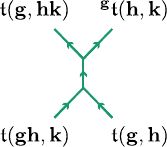}}}}}

\newcommand{\FMoveCocyR}{\mathord{\vcenter{\hbox{\includegraphics[scale=.95]{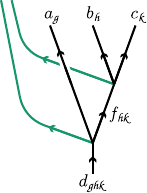}}}}}
\newcommand{\FMoveCocyL}{\mathord{\vcenter{\hbox{\includegraphics[scale=.95]{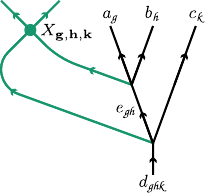}}}}}

\newcommand{\FusionSpaceL}{\mathord{\vcenter{\hbox{\includegraphics[scale=1]{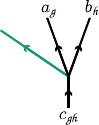}}}}}
\newcommand{\FusionSpaceR}{\mathord{\vcenter{\hbox{\includegraphics[scale=1]{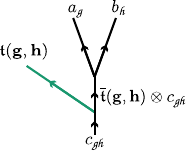}}}}}
\newcommand{\FusionSpaceHat}{\mathord{\vcenter{\hbox{\includegraphics[scale=1]{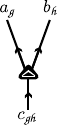}}}}}

\newcommand{\FLeftHat}{\mathord{\vcenter{\hbox{\includegraphics[scale=1]{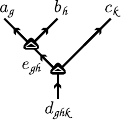}}}}}
\newcommand{\FRightHat}{\mathord{\vcenter{\hbox{\includegraphics[scale=1]{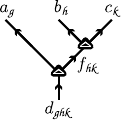}}}}}

\newcommand{\FLeft}{\mathord{\vcenter{\hbox{\includegraphics[scale=1]{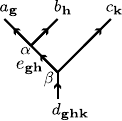}}}}}
\newcommand{\FRight}{\mathord{\vcenter{\hbox{\includegraphics[scale=1]{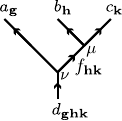}}}}}

\newcommand{\RLeft}{\mathord{\vcenter{\hbox{\includegraphics[scale=1]{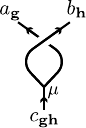}}}}}
\newcommand{\RRight}{\mathord{\vcenter{\hbox{\includegraphics[scale=1]{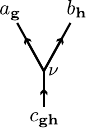}}}}}

\newcommand{\ULeft}{\mathord{\vcenter{\hbox{\includegraphics[scale=1]{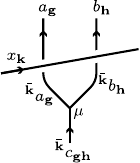}}}}}
\newcommand{\URight}{\mathord{\vcenter{\hbox{\includegraphics[scale=1]{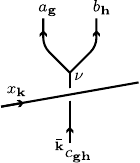}}}}}

\newcommand{\etaLeft}{\mathord{\vcenter{\hbox{\includegraphics[scale=1]{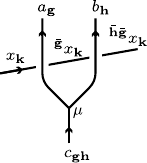}}}}}
\newcommand{\etaRight}{\mathord{\vcenter{\hbox{\includegraphics[scale=1]{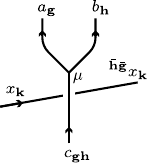}}}}}

\newcommand{\VabcmuL}{\mathord{\vcenter{\hbox{\includegraphics[scale=1]{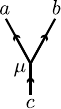}}}}}
\newcommand{\VabcmuR}{\mathord{\vcenter{\hbox{\includegraphics[scale=1]{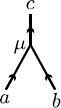}}}}}

\newcommand{\RVabc}{\mathord{\vcenter{\hbox{\includegraphics[scale=1]{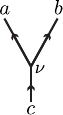}}}}}
\newcommand{\Rabc}{\mathord{\vcenter{\hbox{\includegraphics[scale=1]{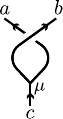}}}}}

\newcommand{\Sab}{\mathord{\vcenter{\hbox{\includegraphics[scale=1]{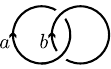}}}}}

\newcommand{\Rab}{\mathord{\vcenter{\hbox{\includegraphics[scale=1]{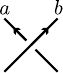}}}}}
\newcommand{\Ragbh}{\mathord{\vcenter{\hbox{\includegraphics[scale=1]{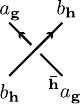}}}}}

\newcommand{\RabRef}{\mathord{\vcenter{\hbox{\includegraphics[scale=1]{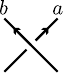}}}}}
\newcommand{\Mab}{\mathord{\vcenter{\hbox{\includegraphics[scale=1]{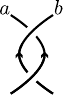}}}}}
\newcommand{\idab}{\mathord{\vcenter{\hbox{\includegraphics[scale=1]{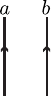}}}}}

\newcommand{\idabesolve}{\mathord{\vcenter{\hbox{\includegraphics[scale=1]{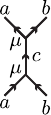}}}}}
\newcommand{\idc}{\mathord{\vcenter{\hbox{\includegraphics[scale=1]{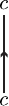}}}}}
\newcommand{\bubbleabc}{\mathord{\vcenter{\hbox{\includegraphics[scale=1]{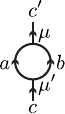}}}}}

\newcommand{\FLeftMTC}{\mathord{\vcenter{\hbox{\includegraphics[scale=1]{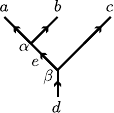}}}}}
\newcommand{\FRightMTC}{\mathord{\vcenter{\hbox{\includegraphics[scale=1]{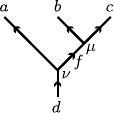}}}}}

\newcommand{\AlgL}{\mathord{\vcenter{\hbox{\includegraphics[scale=.8]{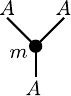}}}}}
\newcommand{\AlgRa}{\mathord{\vcenter{\hbox{\includegraphics[scale=.8]{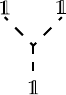}}}}}
\newcommand{\AlgRb}{\mathord{\vcenter{\hbox{\includegraphics[scale=.8]{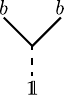}}}}}
\newcommand{\AlgRc}{\mathord{\vcenter{\hbox{\includegraphics[scale=.8]{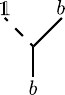}}}}}
\newcommand{\AlgRd}{\mathord{\vcenter{\hbox{\includegraphics[scale=.8]{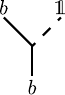}}}}}

\newcommand{\AlgAssocRight}{\mathord{\vcenter{\hbox{\includegraphics[scale=.8]{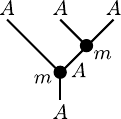}}}}}
\newcommand{\AlgAssocLeft}{\mathord{\vcenter{\hbox{\includegraphics[scale=.8]{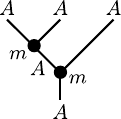}}}}}

\newcommand{\AlgCommRight}{\mathord{\vcenter{\hbox{\includegraphics[scale=.8]{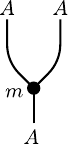}}}}}
\newcommand{\AlgCommLeft}{\mathord{\vcenter{\hbox{\includegraphics[scale=.8]{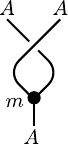}}}}}

\newcommand{\mortensa}{\mathord{\vcenter{\hbox{\includegraphics[scale=.8]{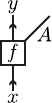}}}}}
\newcommand{\mortensb}{\mathord{\vcenter{\hbox{\includegraphics[scale=.8]{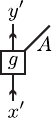}}}}}
\newcommand{\mortensc}{\mathord{\vcenter{\hbox{\includegraphics[scale=.8]{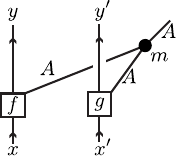}}}}}

\newcommand{\modob}{\mathord{\vcenter{\hbox{\includegraphics[scale=.8]{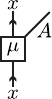}}}}}
\newcommand{\modoba}{\mathord{\vcenter{\hbox{\includegraphics[scale=.8]{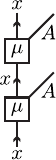}}}}}
\newcommand{\modobb}{\mathord{\vcenter{\hbox{\includegraphics[scale=.8]{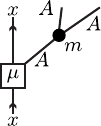}}}}}

\newcommand{\modobfusL}{\mathord{\vcenter{\hbox{\includegraphics[scale=.8]{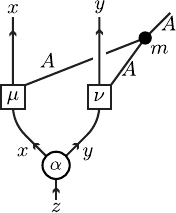}}}}}
\newcommand{\modobfusR}{\mathord{\vcenter{\hbox{\includegraphics[scale=.8]{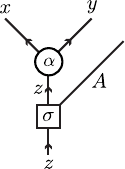}}}}}

\newcommand{\modobdual}{\mathord{\vcenter{\hbox{\includegraphics[scale=.8]{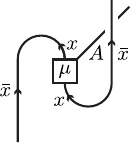}}}}}

\newcommand{\repmodob}{\mathord{\vcenter{\hbox{\includegraphics[scale=.8]{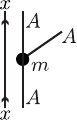}}}}}

\newcommand{\rhopp}{\mathord{\vcenter{\hbox{\includegraphics[scale=.78]{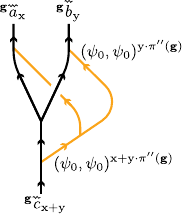}}}}}
\newcommand{\zigvabc}{\mathord{\vcenter{\hbox{\includegraphics[scale=.8]{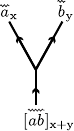}}}}}
\newcommand{\condbraid}{\mathord{\vcenter{\hbox{\includegraphics[scale=1]{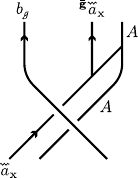}}}}}

\newcommand{\rhoppprime}{\mathord{\vcenter{\hbox{\includegraphics[scale=.8]{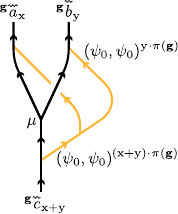}}}}}
\newcommand{\zigvabcprime}{\mathord{\vcenter{\hbox{\includegraphics[scale=.8]{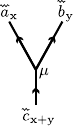}}}}}
\newcommand{\zigvabcprimenu}{\mathord{\vcenter{\hbox{\includegraphics[scale=.8]{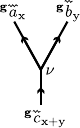}}}}}

\newcommand{\zigvabcnu}{\mathord{\vcenter{\hbox{\includegraphics[scale=1]{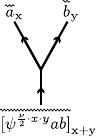}}}}}
\newcommand{\zigvabcnurhs}{\mathord{\vcenter{\hbox{\includegraphics[scale=1]{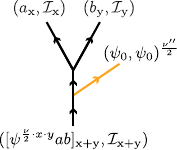}}}}}

\newcommand{\FCondL}{\mathord{\vcenter{\hbox{\includegraphics[scale=.8]{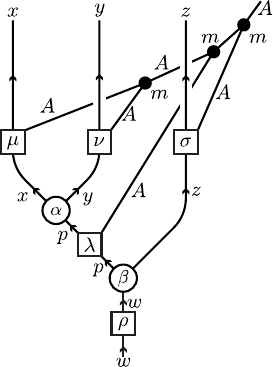}}}}}
\newcommand{\FCondR}{\mathord{\vcenter{\hbox{\includegraphics[scale=.8]{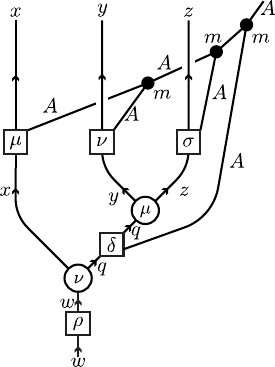}}}}}

\newcommand{\raaa}{\mathord{\vcenter{\hbox{\includegraphics[scale=.8]{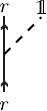}}}}}
\newcommand{\rbbb}{\mathord{\vcenter{\hbox{\includegraphics[scale=.8]{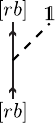}}}}}
\newcommand{\rccc}{\mathord{\vcenter{\hbox{\includegraphics[scale=.8]{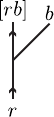}}}}}
\newcommand{\rddd}{\mathord{\vcenter{\hbox{\includegraphics[scale=.8]{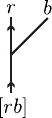}}}}}

\newcommand{\saaa}{\mathord{\vcenter{\hbox{\includegraphics[scale=.8]{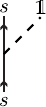}}}}}
\newcommand{\sbbb}{\mathord{\vcenter{\hbox{\includegraphics[scale=.8]{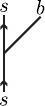}}}}}

\newcommand{\twist}{\mathord{\vcenter{\hbox{\includegraphics[scale=1]{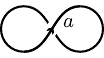}}}}}
\newcommand{\dimension}{\mathord{\vcenter{\hbox{\includegraphics[scale=1]{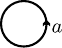}}}}}

\newcommand{\TrivialSymmetryFusionSpace}{\mathord{\vcenter{\hbox{\includegraphics[scale=1]{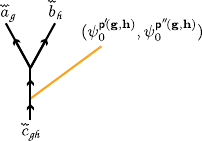}}}}}

\newcommand{\PBfa}{\mathord{\vcenter{\hbox{\includegraphics[scale=1]{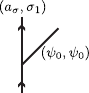}}}}}
\newcommand{\PBfb}{\mathord{\vcenter{\hbox{\includegraphics[scale=1]{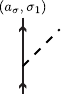}}}}}
\newcommand{\PBfc}{\mathord{\vcenter{\hbox{\includegraphics[scale=1]{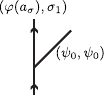}}}}}
\newcommand{\PBfd}{\mathord{\vcenter{\hbox{\includegraphics[scale=1]{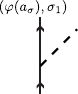}}}}}
\newcommand{\PBfe}{\mathord{\vcenter{\hbox{\includegraphics[scale=1]{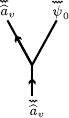}}}}}
\newcommand{\PBff}{\mathord{\vcenter{\hbox{\includegraphics[scale=1]{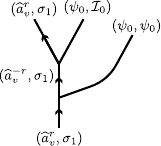}}}}}
\newcommand{\PBfg}{\mathord{\vcenter{\hbox{\includegraphics[scale=1]{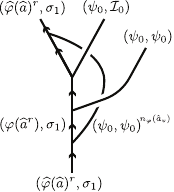}}}}}
\newcommand{\PBfh}{\mathord{\vcenter{\hbox{\includegraphics[scale=1]{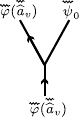}}}}}

\newcommand{\pbpba}{\mathord{\vcenter{\hbox{\includegraphics[scale=1]{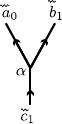}}}}}
\newcommand{\pbpbb}{\mathord{\vcenter{\hbox{\includegraphics[scale=1]{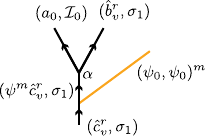}}}}}
\newcommand{\pbpbc}{\mathord{\vcenter{\hbox{\includegraphics[scale=1]{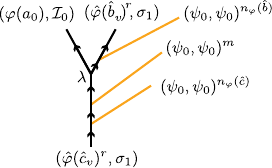}}}}}
\newcommand{\pbpbd}{\mathord{\vcenter{\hbox{\includegraphics[scale=1]{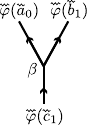}}}}}

\newcommand{\VSymmetry}{\mathord{\vcenter{\hbox{\includegraphics[scale=1]{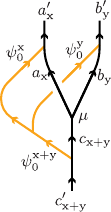}}}}}
\newcommand{\VSymmetryRHS}{\mathord{\vcenter{\hbox{\includegraphics[scale=1]{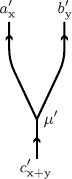}}}}}

\newcommand{\fprod}{\mathbin{\ooalign{ $\mkern12mu$ \cr\hidewidth \raisebox{0.7ex}{\hbox{$\scriptstyle\mkern8mu\bm\psi$}} \hidewidth\cr\hidewidth \raisebox{-0.4ex}{\rotatebox[origin=c]{180}{\hbox{$\scriptstyle\bm\psi\mkern5mu$}}} \hidewidth\cr }}}

\begin{document}

\title{
Characterization and Classification of Fermionic Symmetry Enriched Topological Phases
}

\author{David Aasen}
\affiliation{Microsoft Station Q, Santa Barbara, California 93106-6105 USA}
\affiliation{Kavli Institute for Theoretical Physics, University of California, Santa Barbara, California 93106, USA}

\author{Parsa Bonderson}
\affiliation{Microsoft Station Q, Santa Barbara, California 93106-6105 USA}

\author{Christina Knapp}
\affiliation{Microsoft Station Q, Santa Barbara, California 93106-6105 USA}
\affiliation{Department of Physics and Institute for Quantum Information and Matter, California Institute of Technology, Pasadena, CA 91125, USA}
\affiliation{Walter Burke Institute for Theoretical Physics, California Institute of Technology, Pasadena, CA 91125, USA}

\date{\today}

\begin{abstract}
We examine the interplay of symmetry and topological order in $2+1$ dimensional fermionic topological phases of matter.
We define fermionic topological symmetries acting on the emergent topological effective theory described using braided tensor category theory.
Connecting this to the $\mathcal{G}^{\eff}$ fermionic symmetry of the microscopic physical system, we characterize and classify symmetry fractionalization in fermionic topological phases.
We find that the physical fermion provides constraints that result in a tiered structure of obstructions and classification of fractionalization with respect to the physical fermions, the quasiparticles, and the vortices.
The fractionalization of the (bosonic) symmetry $G= \mathcal{G}^{\eff}/\mathbb{Z}_2^{\eff}$ on the physical fermions is essentially the central extension of $G$ by the $\mathbb{Z}_2^{\eff}$ fermion parity conservation that yields the fermionic symmetry $\mathcal{G}^{\eff}$.
We develop an algebraic theory of $\mathcal{G}^{\eff}$ symmetry defects for fermionic topological phases using $G$-crossed braided tensor category theory, which describes the fusion, braiding, symmetry action, and fractionalization patterns of quasiparticles, vortices, and defects.
This formalism allows us to fully characterize and classify $2+1$ dimensional fermionic symmetry enriched topological phases with on-site unitary fermionic symmetry group $\mathcal{G}^{\eff}$.
We first apply this formalism to extract the minimal data specifying a general fermionic symmetry protected topological phase, and demonstrate that such phases with fixed $\mathcal{G}^{\eff}$ form a group under fermionic stacking, i.e. pairing $\mathcal{G}^{\eff}$-symmetric phases and condensing the bound pair of their physical fermions, finding agreement with existing classifications.
Then we analyze general fermionic symmetry enriched topological phases and find their classification is given torsorially by the classification of the symmetry fractionalization of quasiparticles combined with the classification of fermionic symmetry protected topological phases.
We illustrate our results by detailing a number of examples, including all the invertible fermionic topological phases.
\end{abstract}

\maketitle
\tableofcontents

\section{Introduction}
\label{sec:intro}

Gapped quantum phases of matter may exhibit different types of topological order, which are defined independently of any symmetries and cannot be distinguished by any local observables~\cite{Wen04,Nayak08}.
The defining properties of such topological phases are encoded globally in long-range entanglement patterns, resulting in a number of remarkable nonlocal properties including topologically protected ground state degeneracies and emergent quasiparticles with exotic exchange statistics.
Imposing a symmetry on a topological phase enriches the classification by restricting, and possibly fracturing, the phase space.
Regions of phase space that were adiabatically connected in the absence of symmetry, but disconnected when symmetry is imposed, may manifest distinct phases that have the same topological order, but different symmetry characteristics.
In the case of trivial topological order, these new phases are symmetry protected topological (SPT) phases that may be distinguished by the nontrivial topological properties of their symmetry defects~\cite{Senthil15,Chen13}.
When the phase has nontrivial topological order (in the absence of symmetry), it can divide into distinct symmetry enriched topological (SET) phases in the presence of symmetry (see Ref.~\onlinecite{Bark2019}, and references therein).
Notably, nontrivial topological order allows the symmetry to act nontrivially on the emergent quasiparticles and even to fractionalize, meaning the quasiparticles can carry symmetry quantum numbers that do not correspond to linear representations of the symmetry group.
In this way, SET phases are characterized by different realizations of both symmetry fractionalization and symmetry defects.

Topological orders can be divided into two classes corresponding to whether the microscopic degrees of freedom supporting the phase are purely bosonic (e.g., spins or qubits) or whether they include fermions (e.g., electrons).
Emergent coherent excitations, such as quasiparticles, of a topological phase carry emergent quantum numbers, called topological charges, that are conserved under local operations.
Equivalence under local operations defines the notion of the topological vacuum charge $\I $, which is assigned to local bosonic operators and particles.
Fermionic topological phases additionally include a fermionic underlying particle of the system, the physical fermion, which is distinguished from the bosonic vacuum.
We denote the topological charge corresponding to the physical fermion as $\psi$.
Some key consequences of having a physical fermion in the theory include: (1) fermion parity of $\psi$ is always conserved and can, in many regards, be treated as a symmetry of the system, denoted $\mathbb{Z}_{2}^{\eff}$, (2) the $\mathbb{Z}_{2}^{\eff}$ symmetry translates into a $\mathbb{Z}_{2}$-grading on the theory distinguishing objects as quasiparticles, which have trivial braiding with $\psi$, and vortices, which have a $-1$ braiding with $\psi$, and can thus be thought of as defects of the $\mathbb{Z}_{2}^{\eff}$ symmetry, and (3) the topological charge $\psi$ couples to a spin structure, which is a reference field that tracks the phases and relative minus signs associated with transport of the physical fermion around closed paths on the spacetime manifold.

Despite the ubiquity of fermionic phases in condensed matter systems and the intense interest in understanding the interplay of symmetry and topology, a general characterization and classification of fermionic topological order has remained elusive, with efforts typically relying on exactly solvable models~\cite{Bhardwaj2017,Wang2020} or mathematical tools such as spin-cobordism~\cite{Kapustin2014,Gaiotto2016}.
On the other hand, a variety of methods have been utilized in the effort to classify bosonic SPT phases~\cite{Chen13,Lu12,Kapustin2014,Senthil15} and SET phases~\cite{Wen02,Levin12,Essin13,Lu16,Bark2019}.
The methods utilized in this paper build on the category theory description and classification of $(2+1)$D SPT and SET phases~\cite{Bark2019,turaev2000,turaev2008,ENO}.

The topological properties of a $(2+1)$D bosonic topological phase without symmetry are captured by a modular tensor category (MTC) $\bMTC$ for which the simple objects $a\in \bMTC$ correspond to the topological charges associated with different quasiparticle types of the phase.
In the presence of an on-site unitary global symmetry characterized by a group $G$, the analogous characterization of the symmetric topological phase is captured by a $G$-crossed extension $\bMTC_G^\times$ of the MTC, provided that certain obstructions vanish~\cite{Bark2019,ENO}.
The simple objects of $\bMTC_G^\times$ consist of the quasiparticles as well as symmetry defects, which are all subject to consistency conditions imposed by the interplay of the symmetry group action with fusion, associativity, and braiding~\cite{Bark2019}.
The $G$-crossed extensions of MTCs can be classified in three stages.
First, a $G$ symmetry action is specified on the topological degrees of freedom of the quasiparticles, described by the MTC $\bMTC$.
Such actions are classified by $\text{Hom}(G,\on{Aut}(\bMTC))$, the set of homomorphisms from $G$ to the topological symmetry group $\on{Aut}(\bMTC)$.
Second, for a given symmetry action $[\rho]$, the localization of the action on the physical Hilbert space allows different patterns of fractionalized quantum numbers carried by the quasiparticles of the topological phase.
Assuming a certain obstruction vanishes, the symmetry fractionalization is classified torsorially by $H^2_{[\rho]}(G,\mathcal{A})$, where $\mathcal{A}$ is the group formed by the Abelian topological charges of the MTC.
Third, for a given symmetry action and fractionalization class on the quasiparticles of the topological phase, there are different possible manifestations of symmetry defects, with different algebraic structures.
When the associated obstruction vanishes, symmetry defectification is classified by $H^3(G,\text{U}(1))$.

Any fermionic topological phase has an analogous bosonic topological phase obtained by gauging the $\mathbb{Z}_2^{\eff}$ fermion parity symmetry (promoting the physical fermion to an emergent coherent excitation of the system).
As such, one might hope to translate the $G$-crossed theory of $(2+1)$D bosonic topological phases into a theory of $(2+1)$D fermionic topological phases.
Several complications arise, however, when attempting to do so.
First, imposing conservation of fermion parity implies the full symmetry group is given by $\mathcal{G}^{\eff}$, a $\mathbb{Z}_2^{\eff}$ central extension of the global symmetry group $G$.
One might then expect the bosonic $G$-crossed classification to fracture into multiple $\mathcal{G}^{\eff}$-crossed classifications, one for each $\mathbb{Z}_2^{\eff}$ central extension of $G$.
However, not all such group extensions are necessarily valid for fermionic topological phases; for instance $\mathbb{Z}_4^{\eff}$ only exists for invertible fermionic phases with integer chiral central charge.
Additionally, not all symmetry actions on the quasiparticle sector of a fermionic topological phase can extend to the full theory including vortices (which map to quasiparticles in the bosonic theory) while preserving locality of the physical fermion.
Even when a symmetry action can both extend and fractionalize, it cannot always do so in a manner consistent with $\mathcal{G}^{\eff}$, e.g., $\mathbb{Z}_4^{\eff}$ fermionic SPT (FSPT) phases are only compatible with trivial symmetry action.
Finally, stacking and condensing paired fermions of FSPT phases is an Abelian operation that should result in a group structure on the classification of fermionic theories, analogous to (yet more complicated than) the $H^3(G,\text{U}(1))$ torsorial structure of bosonic theories under gluing bosonic SPT phases.
However, stacking FSPT phases can change both the symmetry action and fractionalization class, indicating that theories belonging to distinct bosonic classifications can map to fermionic theories within the same classification.
Even taking this into account, the number of theories in the classification set between bosonic and fermionic sides of the map do not agree, as is seen for the celebrated example of $G=\mathbb{Z}_2$~\cite{Gu14,Essin13,Lu16,Bark2019}.
Thus, the relatively simple map between related bosonic and fermionic topological orders appears to require a complicated fracturing and rearranging of the classification when symmetry is included.

In this work, we resolve these complications for $(2+1)$D fermionic symmetry enriched topological (FSET) phases, developing the theory of fermionic symmetry fractionalization and the $\mathcal{G}^{\eff}$-crossed theory for defects, which provide both characterization and classification of FSET phases.
Using the categorical description of fermionic topological order through fermionic MTCs, we examine the constraints imposed by the presence of a physical fermion.
Crucially, these restrict the various possible gauge transformations that are considered physical equivalences, resulting in physical differences compared to the analogous bosonic theory.
For instance, the fermionic topological symmetry group is not necessarily equal to the bosonic topological symmetry group and fractionalization patterns that differ by modifications of the vorticity of local operators are inequivalent.
Another important property is that the fractionalization of $G$ symmetry on the physical fermions precisely corresponds to the particular central extension of $G$ by $\mathbb{Z}_2^{\eff}$ that yields the fermionic symmetry group $\mathcal{G}^{\eff}$.
A crucial aspect of our analysis is the role played by vortices, both implicitly for classifying the patterns of symmetry fractionalization of quasiparticles, and explicitly for how symmetry can fractionalize on vortices themselves.
This all yields a new classification structure for symmetry fractionalization of fermionic topological phases, which includes finer scale structures that can be described in terms of the symmetry fractionalization of quasiparticles, together with the extension of fractionalization to the vortices.
We summarize this structure in Table~\ref{table:obstructions}, which describes the different obstructions to fermionic symmetry fractionalization and the corresponding torsorial classifications of fractionalization structures that exist when these obstructions vanish.
Nontrivial obstructions prevent theories from existing as strictly $(2+1)$D FSET phases, but such theories can occur as anomalous $(2+1)$D FSET phases that exist on the boundary of $(3+1)$D fermionic phases.
With this in mind, we perform a careful analysis of the dependence of these obstructions on the choices involved in extending the symmetric theories from the quasiparticles to the vortices, as this plays an important role in understanding the corresponding anomalous boundary theories.

After developing the theory of fractionalization for fermionic topological phases, we develop the $\mathcal{G}^{\eff}$-crossed theory, which incorporates the constraints and classification uncovered for fractionalization.
In this way, $(2+1)$D FSET phases can be classified by finding all the possible physically inequivalent $\mathcal{G}^{\eff}$-crossed extensions of a given fermionic topological order.
Brute force solving the consistency conditions does not provide a practical approach to this problem, except when the fermionic topological order is sufficiently simple, so we seek a similar classification scheme to that of bosonic SET phases.
In order to address the classification problem, we begin with FSPT phases with $G=\mathbb{Z}_2$, solving for the topological data of all such theories.
These theories are not just illustrative examples of $\mathcal{G}^{\eff}$-crossed formalism; they play a critical role in the efficient construction of a ``base theory'' for any choice of $\mathcal{G}^{\eff}$.
By taking a restricted product of a $\mathbb{Z}_2^{\eff}\times \mathbb{Z}_2$ FSPT phase with a $G$ bosonic SPT phase, we can construct a FSPT phase for each symmetry action compatible with full symmetry group $\mathbb{Z}_2^{\eff}\times G$.
This theory provides a base from which we can torsorially generate all remaining theories with the same symmetry action~\cite{Aasen21,Bark2019b}, including those with different $\mathcal{G}^{\eff}$.
We write the topological data of each such theory from that of the base theory and the cocycles relating them.
From this, we extract a minimal set of data specifying every FSPT phase.

When two phases with the same microscopic fermions are brought together, the composite of physical fermions for the two theories condenses (is identified with the vacuum charge).
Using general properties of the minimal set of data specifying each FSPT phase, we demonstrate that the set of FSPT phases with fixed $\mathcal{G}^{\eff}$ form a group under stacking.
We extract this group law to classify FSPT phases.

Combining our understanding of fermionic symmetry fractionalization and its classification, the fermionic $\mathcal{G}^{\eff}$-crossed defect theory, and the classification of FSPT phases, we are able to provide a three stage classification of all $(2+1)$D FSET phases, analogous to the classification of bosonic SET phases.
Noticing that parts of the symmetry action and fractionalization structure can be accounted for by stacking with FSPT phases, we recognize two perspectives for organizing the classification structure of FSET phases: the first views the fermionic theory of the quasiparticles together with the vortices as fundamental, and develops the symmetry action, fractionalization, and defectification as a $G$-symmetry extension; the second views the fermionic theory of just the quasiparticles as fundamental, and develops the symmetry action, fractionalization, and defectification as a $\mathcal{G}^{\eff}$-symmetry extension, including the vortices at the defectification stage.
The former view includes a torsorial action through gluing with bosonic SPT phases, while the latter includes a torsorial action through stacking with FSPT phases.
We summarize the classification structure of FSET phases from both perspectives in Table~\ref{table:classification}, as well as that of SET phases for comparison.
We demonstrate through a number of explicit examples that our findings recover previously known classification results.

Previous works have classified FSPT phases using spin-cobordism~\cite{Freed19,Kapustin15,Kapustin18}, defect decoration~\cite{Wang2016,Tarantino2016,Wang2018,Wang2020}, $G$-graded tensor categories~\cite{Cheng2018,Bhardwaj2017}, and by constructing topological invariants of band Hamiltonians~\cite{Turzillo2019}.
Spin cobordism and defect decoration classifications provide a physical picture for the ground state wavefunction of a FSPT phase, but further calculation is necessary in order to determine the algebraic properties of the topological defects.
In this work, we classify FSPT phases according to their group structure under stacking.
When the group law from these other approaches is explicitly known, we demonstrate agreement with our results.
Moreover, we find that the elements of the Pontryagin dual of the spin bordism group determine the complete set of data describing the algebraic theory of the FSPT phase, in the cases where the Pontryagin dual has been computed~\cite{Brumfiel2018a}.

Far less progress has previously been made for FSET phases.
Ref.~\onlinecite{Lan2017} proposed to classify fermionic topological phases with symmetry by classifying their fully gauged counterparts.
This relies on the one-to-one correspondence of the gauging process (equivariantization).
Since it does not explicitly construct the theory with symmetry and defects, it does not provide an explicit characterization of the symmetric theory.
In particular, the extraction of properties like symmetry fractionalization and defect fusion and braiding require a further difficult calculation associated with a $\text{Rep}(G)$ condensation (de-equivariantization).
As such, this approach lacks or obfuscates much of the organizing structure of the classification.
Moreover, this approach to classification requires a case-by-case brute force solution of the pentagon and hexagon consistency conditions for a range of possible fusion algebras to determine all possible gauged theories.
In practice, even when given the full data of the original topological order, this approach is only feasible for theories with very low rank and symmetry groups with very small order, severely limiting its usefulness and applicability.
Other works have addressed some aspects of symmetry fractionalization~\cite{Fidkowski2018} and fermionic topological order from the perspective of anomalies and bosonization~\cite{Thorngren2020,Tata2021}, though these did not produce general classifications.
The fermionic state-sum approach take in Ref.~\onlinecite{Tata2021} is complementary to the analysis here in that it is particularly useful for understanding anomalous theories, but does not yield a classification or description of defects for $(2+1)$D theories.

Our paper is organized as follows.
In Sec.~\ref{sec:fMTC}, we review the BTC description of $(2+1)$D bosonic and fermionic topological order.
We then incorporate symmetry in Sec.~\ref{sec:symmetric-fermions}, extending the notions of topological symmetry, symmetry action, and symmetry fractionalization to fermionic theories. 
The obstructions encountered in this section are collected in Table~\ref{table:obstructions} along with their associated torsors.  
Sec.~\ref{sec:fermionic-defectification} develops the $\mathcal{G}^{\eff}$-crossed formalism of FSET phases.
We then focus on FSPT phases, deriving the minimal set of data describing all such theories in Sec.~\ref{sec:fSPT}.
In Sec.~\ref{sec:classification}, we combine these results to provide the full classification of FSET phases, summarized in Table~\ref{table:classification}.
We illustrate these concepts with explicit examples in Sec.~\ref{sec:examples}, before concluding in Sec.~\ref{sec:discussion}.
Additional details and technical background material are relegated to the Appendices.

\section{Braided Tensor Category Description of \texorpdfstring{$(2+1)$}{2p1}D Topological Order}
\label{sec:fMTC}

Unitary braided tensor categories (BTCs) provide an algebraic structure that can be used to describe the universal properties of $(2+1)$D topological phases in the absence of symmetry, such as the anyonic fusion and braiding of quasiparticles.
For bosonic topological phases, the topological order is understood to be fully characterized by the chiral central charge and a unitary modular tensor category (MTC), which is a BTC with non-degenerate braiding.
The MTC can also be used to construct the low-energy effective topological quantum field theory (TQFT) describing the system on arbitrary spacetime manifolds.
For fermionic topological phases, the universal properties of quasiparticles are described by a super-modular tensor category (SMTC), which is a BTC with an object identified as a physical fermion that has trivial (pure-)braiding with all objects in the theory, and which has nondegenerate braiding aside from the 2-fold degeneracy implied by the physical fermion.
However, this does not capture the full topological order of a fermionic topological phase; to complete the picture, one again needs the chiral central charge and a fermionic modular tensor category (FMTC), which includes the SMTC.
(Actually, each SMTC has a 16-fold way of possible FMTC extensions, and specifying the chiral central charge uniquely specifies one of these 16 FMTCs.)
The extra structure encoded in the FMTC describes the universal properties of fermionic vortices, which have nontrivial braiding with the physical fermions, and their interplay with the quasiparticles.
Similarly, the FMTC can be used to construct the corresponding spin TQFT describing the system, which incorporates the spin structures necessary for defining spinors on general manifolds.
 Gauging the $\mathbb{Z}_2^{\eff}$ fermionic parity symmetry has the effect of summing over spin structures, making the order parameter fluctuating, and coupling the physical fermion to a $\mathbb{Z}_2$ gauge field to make it an emergent excitation.
As a result, the gauged theory is a MTC of a bosonic topological phase for which the fermion is no long physical, but emergent.
This section reviews the relevant material on BTCs and details needed in this paper.
Additional details can be found in Refs.~\onlinecite{Kitaev2006,Bark2019} (we primarily use the conventions of Ref.~\onlinecite{Bark2019}).

\subsection{Braided Tensor Categories}
\label{sec:BTC}

The simple objects of a BTC $\bMTC$ are topological charges associated with distinct types of quasiparticles (or other objects).
In other words, they are conserved quantum numbers that cannot be altered by operations localized within a region to which they are ascribed, such as the neighborhood of a single quasiparticle.
Physical states that can be related by local operations are considered topologically equivalent and not distinguished by the topological effective theory.
The set of topological charges (simple objects) is assumed to be finite, and we will typically use $\bMTC$ to denote the set of topological charges as well as the BTC.
The topological charges in $\bMTC$ satisfy an associative fusion algebra
\begin{align}
\label{eq:fusion}
a\otimes b &= \bigoplus_{c\in \bMTC} N_{ab}^c c
\end{align}
where the fusion multiplicities $N_{ab}^c$ are non-negative integers indicating the number of distinct ways that $a$ and $b$ can be combined to form $c$. Associativity implies $\sum_{e}N_{ab}^{e}N_{ec}^{d} = \sum_{f} N_{af}^{d}N_{bc}^{f} $.

There is a special charge $\I$ designated as the vacuum or trivial charge, which has trivial fusion and braiding.
The vacuum charge corresponds to local operators acting on the microscopic degrees of freedom, which do not affect the topological charges.
In particular, this means it is the unique charge with $N_{a \I}^{c} = N_{\I a}^{c} = \delta_{ac}$ for all $a$.
Each topological charge $a$ has a unique conjugate charge denoted $\bar{a}$ such that $a$ and $\bar{a}$ can fuse to $\I $ (in exactly one way), i.e. $N_{a b}^{\I } = \delta_{b \bar{a}}$.
It is possible, but not necessary, for charges to be self-dual, meaning $a=\bar{a}$.

When the fusion of a topological charge $a$ with all other charges has a unique outcome, i.e. $a \otimes b = c$ or $\sum_{c} N_{ab}^{c} = 1$ for all $b$, then $a$ is said to be Abelian; otherwise it is non-Abelian.
This terminology is correlated with the braiding properties of such objects, i.e. Abelian charges have strictly Abelian braiding with all objects, while quasiparticles carrying the same non-Abelian charge give rise to non-Abelian braiding transformations.
The vacuum charge $\I $ is necessarily Abelian, since $\I \otimes a = a$, thus all BTCs trivially contain at least one Abelian charge.
By associativity, two Abelian charges always fuse to an Abelian charge.
Together with each charge having a unique conjugate charge, this implies that the collection of Abelian charges $\mathcal{A}$ of any BTC $\bMTC$ forms a group under fusion. Including the full fusion and braiding structure, $\mathcal{A}$ always forms a BTC which is a subcategory of $\bMTC$.

We define the quantum dimension $d_a$ to be the largest eigenvalue of the fusion matrix ${\bf N}_{a}$ defined as $\left[ {\bf N}_a\right]_{bc} = N_{ab}^c$, with $b$ and $c$ the matrix indices.
(The Perron-Frobenius theorem guarantees $d_a \geq 1$ are positive real numbers.)
A charge $a$ is Abelian if and only if it has $d_a=1$, while non-Abelian charges have $d_a>1$.
The total quantum dimension $\mathcal{D}$ of $\bMTC$ is given by $\mathcal{D}=\sqrt{\sum_{a\in \bMTC} d_a^2}$.

The fusion of charges $a$ and $b$ to $c$ has an associated vector space $V_{ab}^c$ with $\text{dim}V_{ab}^c= N_{ab}^c$, and a dual (splitting) space $V_c^{ab}$.
The corresponding anyonic states and operators can be represented diagrammatically, which can often be useful for encapsulating complicated calculations in a simple way.
In the diagrammatic notation, topological charges label oriented line segments and trivalent vertices describe states within these vector spaces
\begin{align}
\VabcmuR &= \left(\frac{d_c}{d_a d_b}\right)^{1/4} \bra{a,b;c,\mu} \in V_{ab}^c, \label{eq:Vabc}\\
\VabcmuL &= \left(\frac{d_c}{d_a d_b}\right)^{1/4} \ket{a,b;c,\mu} \in V_c^{ab} \label{eq:Vcab}
\end{align}
where $\mu =1,\dots,N_{ab}^c$.
The normalizations are chosen to make bending a line a unitary operation.
The corresponding inner product obtained by stacking diagrams translates into a bubble removal identity
\begin{align}
\bubbleabc = \delta_{\mu \mu' }\delta_{cc'}\sqrt{\frac{d_a d_b}{d_c}}\;\idc
,
\end{align}
and a partition of identity
\begin{align}
\idab = \sum_{c,\mu} \sqrt{\frac{d_c}{d_a d_b}}\; \idabesolve
.
\end{align}
In the diagrammatic formalism, we can freely add and remove charge lines associated with $\I $.
The direction of an arrow can be reversed by changing the topological charge to its conjugate charge.

More complicated diagrams involving additional anyons are constructed by stacking trivalent vertices and connecting lines corresponding to the same topological charge.
The vector spaces for larger numbers of anyons satisfy associativity of fusion
\begin{align}
V_d^{abc} &\cong \bigoplus_e V_e^{ab} \otimes V_d^{ec} \cong \bigoplus_f V_d^{af} \otimes V_f^{bc},
\end{align}
where $\cong$ denotes an isomorphism called an $F$-move, written diagrammatically as
\begin{align}
\FLeftMTC &= \sum_{f,\mu,\nu} [F_d^{abc}]_{(e,\alpha,\beta)(f,\mu,\nu)} \FRightMTC.
\end{align}
The $F$-moves amount to changing the basis of the three-anyon splitting space; as such they correspond to unitary transformations.
Unitarity fixes
\begin{align}
\left[ (F_d^{abc})^{-1}\right]_{(f,\mu,\nu)(e,\alpha,\beta)} &= \left[(F_d^{abc})^\dagger \right]_{(f,\mu,\nu)(e,\alpha,\beta)} \nonumber
\\ &= \left[ F_d^{abc}\right]^*_{(e,\alpha,\beta)(f,\mu, \nu)}.
\end{align}

\begin{figure*}[t!]
   \centering
   \includegraphics[width=1.2\columnwidth]{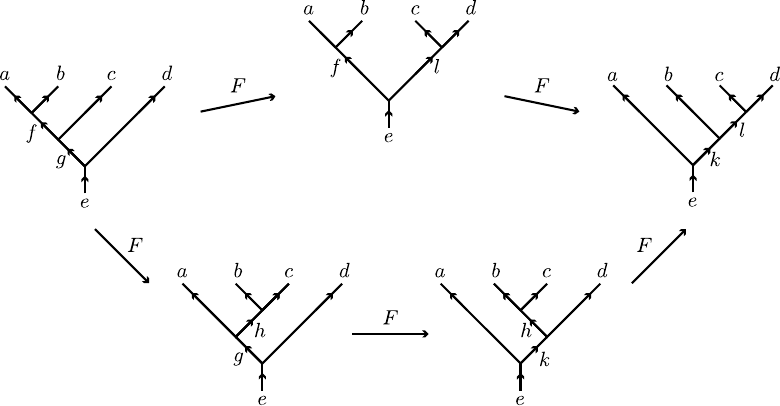}
   \caption{The pentagon equation is a consistency condition that requires equivalence of different sequences of $F$-moves that start and end at common decompositions of the fusion state space.}
   \label{fig:pentagon}
\end{figure*}

Any combination of $F$-moves that begin and end with the same diagram must be equivalent; this consistency condition results in the pentagon equation for the $F$-symbols:
\begin{widetext}
\begin{align}
\label{eq:pentagoneqn}
\sum_\delta [F_e^{fcd}]_{(g,\beta,\gamma),(l,\nu,\delta)} [F_e^{abl}]_{(f,\alpha,\delta),(k,\mu,\lambda)} = \sum_{h,\sigma,\psi,\rho} [F_g^{abc}]_{(f,\alpha,\beta),(h,\psi,\sigma)} [F_e^{ahd}]_{(g,\sigma,\gamma),(k,\rho,\lambda)} [F_k^{bcd}]_{(h,\psi,\rho),(l,\nu,\mu)}.
\end{align}
\end{widetext}
The corresponding diagrammatic equation is depicted in Fig.~\ref{fig:pentagon}.

Fusion with the vacuum charge is trivial, which means vacuum lines can be freely added and removed from diagrams.
This corresponds to canonical isomorphisms between $V^{\I a}_{a}$, $V^{a \I}_{a}$, and $\mathbb{C}$, i.e. making canonical choices of the basis state $\left| \I , a; a \right\rangle$ and $\left| a, \I ; a \right\rangle$ (including $\left| \I, \I ; \I \right\rangle$).
Physically, this can be understood as connecting the emergent effective theory's vacuum charge and state to corresponding quantities in the physical system, i.e. local operators and states of the microscopic degrees of freedom.
In terms of the $F$-symbols, the triviality of the vacuum charge corresponds to the condition $F^{abc}_{d} = \openone$ when any of $a$, $b$, or $c$ equals $\I$.

The inner product together with the $F$-moves can be used to define a pivotal structure that allows for the bending of lines and raising or lowering legs of a fusion vertex, up to unitary transformations.
Using this to with the partition of identity, we find the quantum dimensions satisfy
\begin{equation}
\label{eq:d_relation}
d_a d_b = \sum_{c\in \bMTC} N_{ab}^c d_c
.
\end{equation}

The theory described thus far is a unitary fusion tensor category (FTC), which contains no braiding structure.
In order to define a BTC, we introduce a braiding operation that allows the a pair of topological charges to exchange positions, which can be represented diagrammatically as
\begin{align}
R^{ab} = \Rab
\end{align}
for a counterclockwise braiding exchange.
This operator can be concretely defined in terms of its action on the fusion and splitting state space using $R$-moves that map between vector spaces as $R^{ab}_{c}:V_c^{ba}\to V_c^{ab}$, that is
\begin{align}
\Rabc =\sum_{\nu} \left[ R^{ab}_c \right]_{\mu \nu}\RVabc
.
\end{align}
Locality of the theory thus requires $N_{ab}^{c} = N_{ba}^{c}$, making the fusion algebra commutative, i.e. $a \otimes b = b \otimes a$.
Clockwise braiding exchanges correspond to the operator
\begin{align}
\left(R^{ab} \right)^{-1} = \RabRef
\end{align}
where unitary requires
\begin{align}
\left[ (R^{ab}_c)^{-1} \right]_{\nu \mu} = \left[ R^{ab}_c \right]^{\ast}_{\mu \nu}
.
\end{align}

\begin{figure*}[htp]
   \centering
   \includegraphics{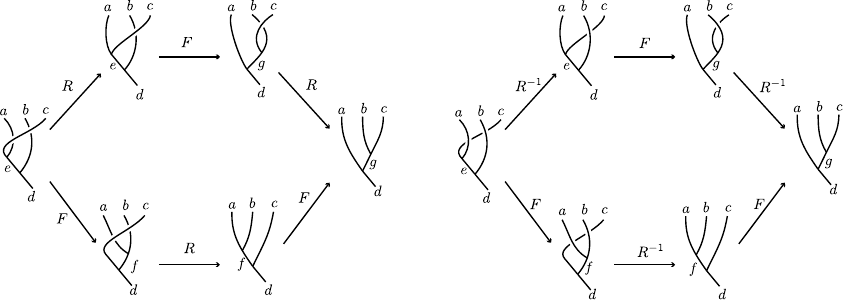}
   \caption{Consistency of braiding and fusion is given by the hexagon equations. These enforce the property that a line can freely slide over or under a vertex, which follows from locality. (This sliding property is manifested at the bottom left diagram in each of these hexagons.)
   }
   \label{fig:hexagon}
\end{figure*}

In order for braiding to be compatible with fusion, locality requires that a pair of anyons that are braided with a third and then fused together should be equivalent to first fusing the pair and braiding with the third.
This equivalence can be expressed as a pair of consistency conditions known as the hexagon equations
\begin{widetext}
\begin{align}\label{eq:hex1}
\sum_{\lambda,\gamma} [R_e^{ac}]_{\alpha,\lambda}\, [F_d^{acb}]_{(e,\lambda,\beta),(g,\gamma,\nu)} \,[R_g^{bc}]_{\gamma \mu} &= \sum_{f,\sigma,\delta,\psi} [F_d^{cab}]_{(e,\alpha,\beta),(f,\delta,\sigma)} \,[R_d^{fc}]_{\sigma\psi} \, [F_d^{abc}]_{(f,\delta,\psi),(g,\mu,\nu)}
,\\ \label{eq:hex2}
\sum_{\lambda,\gamma}\left[(R_e^{ca})^{-1}\right]_{\alpha \lambda} [F_d^{acb}]_{(e,\lambda,\beta),(g,\gamma,\nu)} \left[(R^{cb}_g)^{-1}\right]_{\gamma \mu} &=\sum_{f,\sigma,\delta,\psi} [F_d^{cab}]_{(e,\alpha,\beta),(f,\delta,\sigma)}\left[ (R_d^{cf})^{-1}\right]_{\sigma \psi} [F_d^{abc}]_{(f,\delta,\psi),(g,\mu,\nu)}
,
\end{align}
\end{widetext}
which are depicted diagrammatically in Fig.~\ref{fig:hexagon}.
The $F$- and $R$-symbols constitute the basic data characterizing a BTC.
The hexagon equations together with the condition that $F^{abc}_{d} = \openone$ when any of $a$, $b$, or $c$ equal $\I$ implies that $R^{\I a}_{a} = R^{a \I}_{a} = 1$, i.e. braiding with the vacuum charge is trivial, as it ought to be.

Unitary transformations $\Gamma_c^{ab}$ acting on the fusion/splitting spaces $V_c^{ab}$ and $V_{ab}^c$ redefine the basis states as
\begin{align}
\label{eq:vertex_basis_gauge}
\widetilde{\ket{a,b;c,\mu}} &= \sum_\nu \left[\Gamma_c^{ab}\right]_{\mu \nu} \ket{a,b;c,\nu}.
\end{align}
The unitary transformations define a gauge freedom of the theory.
The $F$- and $R$-symbols are generally not invariant under such vertex basis gauge transformations, but rather transform as
\begin{align}
\label{eq:Fgaugetransf}
& \left[\widetilde{F}_d^{abc} \right]_{(e,\alpha,\beta)(f,\mu,\nu)} = \sum_{\alpha',\beta',\mu',\nu'}
[\Gamma_e^{ab}]_{\alpha \alpha'}
[\Gamma_d^{ec}]_{\beta \beta'} \notag \\
& \quad \times \left[F^{abc}_d \right]_{(e,\alpha',\beta')(f,\mu',\nu')}
 [( \Gamma_f^{bc})^{-1} ]_{\mu' \mu}
[( \Gamma_d^{af})^{-1} ]_{\nu' \nu}
\end{align}
and
\begin{align}
\label{eq:Rgaugetransf}
\left[ \widetilde{R}_c^{ab} \right]_{\mu \nu}&= \sum_{\mu',\nu'}\left[\Gamma_c^{ba}\right]_{\mu \mu'} \left[R_c^{ab} \right]_{\mu'\nu'}\left[\left(\Gamma^{ab}_c\right)^{-1}\right]_{\nu' \nu}
.
\end{align}
Theories whose basic data are related by such gauge transformations are considered equivalent theories.

In terms of the vertex basis gauge transformations, the condition $F^{abc}_{d} = \openone$ when any of $a$, $b$, or $c$ equals $\I$ requires that the corresponding gauge transformations are fixed, that is ${\Gamma_a^{a\I } = \Gamma_b^{\I b} = \Gamma_{\I }^{\I \I }}$.
The $\Gamma_{\I }^{\I \I }$ gauge transformation does not change any of the basic data, because it always cancels out.
However, the canonical isomorphisms between $V^{\I a}_{a}$, $V^{a \I}_{a}$, and $\mathbb{C}$ indicate that we should fix $\Gamma_{\I }^{\I \I }=1$.

We now mention some important gauge invariant quantities.
We first note that the quantum dimension is an invariant that can also be expressed in terms of $F$-symbols for unitary theories
\begin{align}
d_{a} = \left|\left[F^{a \bar{a} a}_a\right]_{\I ,\I } \right|^{-1} =  \dimension
\end{align}

When a charge $a$ is self dual, i.e., $a=\bar{a}$, the Frobenius-Schur indicator of that charge is an invariant equal to
\begin{align}
\kappa_a &= d_a \left[F^{a a a}_a\right]_{\I ,\I } ,
\end{align}
and takes values $\kappa_a=\pm 1$.

The topological twist of a charge $a$ is an invariant defined by
\begin{align}
\label{eq:spin}
\theta_a &= \theta_{\bar{a}} = \sum_{c,\mu} \frac{d_c}{d_a} \left[ R_c^{aa} \right]_{\mu \mu}= \frac{1}{d_a}\twist.
\end{align}
The pure braiding of charges satisfy the ribbon property, which is expressed in terms of the twists as
\begin{align}
\label{eq:ribbon}
\sum_{\lambda} \left[R^{ab}_c \right]_{\mu \lambda} \left[R^{ba}_c \right]_{ \lambda \nu } = \frac{\theta_c}{\theta_a \theta_b} \delta_{\mu \nu}.
\end{align}

Taking the quantum trace of a pure-braid yields the topological S-matrix (up to normalization), which is the invariant defined by
\begin{align}
S_{ab} = \frac{1}{\mathcal{D}}\sum_c N_{\bar{a}b}^c \frac{\theta_c}{\theta_a \theta_b} d_c = \frac{1}{\mathcal{D}}\;\Sab
.
\end{align}

Related to these is the monodromy scalar component
\begin{align}
\label{eq:monodromyscaler}
M_{ab} = \frac{S_{ab}^{\ast} S_{\I \I}}{S_{\I a} S_{\I b}}
\end{align}
If at least one of $a$ or $b$ is Abelian, we have
\begin{align}
R^{ab} R^{ba} &= \Mab = M_{ab}\;\idab
,
\end{align}
and $M_{ab}$ is a phase.
When $M_{ae}$ is a phase for a specific choice of $a$ and $e$, it follows that
\begin{equation}
\label{eq:M_relation}
M_{ae} M_{be} = M_{ce}
\end{equation}
for any $b,c$ with $N_{ab}^{c} \neq 0$.
If Eq.~(\ref{eq:M_relation}) holds for all $a,b,c$ such that $N_{ab}^{c} \neq 0$, then $e$ is Abelian.

A useful property of BTCs is that their fusion algebra can be diagonalized.
More specifically, the associativity and commutativity of the fusion rules implies that the fusion matrices are simultaneously diagonalizable by a unitary matrix ${\bf P}$.
In this way, we can write ${\bf N}_{a} = {\bf P} {\bf \Lambda}^{(a)} {\bf P}^{-1}$, where the diagonal matrix $[{\bf \Lambda}^{(a)}]_{jk} = \lambda^{(a)}_{j} \delta_{jk}$ and the eigenvalues are $\lambda^{(a)}_{j} = {\bf P}_{aj} / {\bf P}_{\I j}$. These eigenvalues provide the characters $\lambda_{j}$ of the fusion algebra, satisfying
\begin{equation}
\label{eq:characters}
\lambda^{(a)}_{j} \lambda^{(b)}_{j} = \sum_{c\in \bMTC} N_{ab}^c \lambda^{(c)}_{j}
\end{equation}
for each $j$.
The number of characters is equal to the number of topological charge types.

A MTC is a BTC for which the $S$-matrix is unitary. This is equivalent to the condition that braiding is nondegenerate, that is for every charge $a \neq \I$, there is some charge $b$ for which $R^{ab}R^{ba} \neq \openone$.
Some important properties of MTCs include the following:
The chiral central charge $c_-$ mod 8 is given by
\begin{equation}
e^{i \frac{2 \pi}{8} c_{-}} = \frac{1}{\mathcal{D}} \sum_{a} d_a^2 \theta_a
.
\end{equation}
The $S$-matrix together with $T_{ab} = \theta_{a} \delta_{ab}$ provide a projective representation of the modular transformations for the system on a torus.
The topological $S$-matrix diagonalizes the fusion algebra, that is ${\bf P}_{ab} = S_{ab}$, and so there is a canonical labeling of the characters $\lambda^{(a)}_{b}$ by the topological charge values $b \in \bMTC$.

An important implication of this is that, for MTCs, if there are phase factors that satisfy the relation
\begin{equation}
e^{i \phi_{a}} e^{i \phi_{b}} = e^{i \phi_{c}}
\end{equation}
for all $a,b,c$ such that $N_{ab}^{c} \neq 0$, then there exists a unique Abelian topological charge $e$ such that
\begin{equation}
e^{i \phi_{a}} = M_{a e}^{\ast}
\end{equation}
for all $a$.
In other words, there is a bijection between the set of such phases and the set of Abelian topological charges $\mathcal{A}$.

\subsubsection{Example: Toric code}

One of the most well-known topological phases is the toric code, $\tc$~\cite{Kitaev2003}.
This model has four Abelian anyons: the vacuum charge $\I $, an emergent fermion $\psi$, and two Abelian bosons $e$ and $m$.
The charges satisfy $\mathbb{Z}_2\times \mathbb{Z}_2$ fusion
\begin{align}
\psi \otimes e &= m, & \psi\otimes m &= e, & e\otimes m &= \psi, & a\otimes a &= \I ,
\end{align}
The $F$-symbols are trivial, and the $R$-symbols are given by,
\begin{align}
R^{\psi \psi} = R^{e\psi} = R^{\psi m} =  R^{em} = -1.
\end{align}
The three bosons $\I $, $e$, and $m$ all have trivial topological twist, while the emergent fermion $\psi$ has $\theta_\psi = -1$.

As we will see in the next section, when $\psi$ is interpreted as a physical fermion the toric code corresponds to the trivial fermionic theory.

\subsection{Fermionic Theories}
\label{sec:SMTC_FMTC}

We now consider the algebraic theory that describes fermionic topological phases in the absence of symmetry.
When the microscopic degrees of freedom are fermions (e.g., a superconductor), there are additional physical properties that must be incorporated in the effective topological theory, such as canonical gauge fixing, fermionic vortices, and spin structures.
Two approaches to constructing the algebraic theory of the low energy excitations and spin TQFT for fermionic topological phases include using super pivotal modular tensor categories~\cite{Aasen19,WalkerSimmons} and using fermionic MTCs~\cite{Bonderson20}.
These descriptions are related by condensing a fermion~\cite{Aasen19}.
In this paper, we utilize the FMTC approach and draw various results from Ref.~\onlinecite{Bonderson20}.

We begin by identifying a topological charge $\psi$ in the theory that is identified as the \emph{physical fermion}.
The fermion has topological data
\begin{align}
\psi \otimes \psi &= \I , \\
F^{\psi \psi \psi}_{\psi} &=1, \\
R^{\psi \psi}_{\I} = \theta_\psi &=-1
.
\end{align}
This forms a BTC which we denote as $\mathbb{Z}_{2}^{\psi}$ (sometimes called SVec), which will be a subcategory of any fermionic theory.
(It is possible to define generalized fermionic theories in which there are multiple distinct types of fermions, as well as fermions with more complicated fusion, such as $\mathbb{Z}_{2N}$ fusion, but we will leave such theories to be explored in future work.)

The quasiparticles of a fermionic topological phase are described by a SMTC.
A SMTC $\mathcal{B}$ is a BTC that contains the physical fermion $\psi$, which has trivial monodromy with all quasiparticles, so that $ M_{\psi a} =1$ for all $a \in \mathcal{B}$, and only the fermion has degenerate braiding.
The last condition is equivalent to the condition that
\begin{align}
S = S_{\psi} \otimes \widehat{S}
,
\end{align}
where $\widehat{S}_{\hat{a} \hat{b}} = \sqrt{2} S_{ab}$ is unitary, and
\begin{align}
S_{\psi} = \frac{1}{\sqrt{2}} \left[
\begin{array}{rr}
1 & 1 \\
1 & 1
\end{array}
\right]
\end{align}
is the $S$-matrix of $\mathbb{Z}_{2}^{\psi}$.
Here, we have introduced ``supersectors'' of topological charge, defined to be pairs related by fusion with the physical fermion
\begin{align}
\hat{a} = \{ a , \psi \otimes a \}
.
\end{align}
We emphasize that even though the $S$-matrix of a SMTC factorizes in this way, it is not necessarily the case that the entire category $\mathcal{B}$ can be written in the factorized form $\mathbb{Z}_{2}^{\psi} \boxtimes \widehat{\mathcal{B}}$, where $\widehat{\mathcal{B}}$ is a MTC.
In fact, there are examples for which there are no BTCs, or even FTCs with the corresponding fusion rules required for such a $\widehat{\mathcal{B}}$ (see Secs.~\ref{sec:ex_SU2} and \ref{sec:ex_MooreRead}).
On the other hand, every Abelian SMTC does factorize in this way.
This follows from the more general statement that any Abelian BTC $\mathcal{A}_{\vv{0}}$ that contains a fermion $\psi$ for which $\psi \otimes \psi = \I$, $\theta_\psi =-1$, $M_{\psi a} =1$ for all $a \in \mathcal{A}_{\vv{0}}$ factorizes as $\mathcal{A}_{\vv{0}} = \mathbb{Z}_{2}^{\psi} \boxtimes \widehat{\mathcal{A}}_{\vv{0}} $ where $\widehat{\mathcal{A}}_{\vv{0}}$ is a BTC (see Appendix~\ref{app:A_factorize}).

One might expect a SMTC could be sufficient to describe a fermionic topological phase, since it describes the quasiparticles and even contains a large amount of information about the vortices, such as the fusion rules of vortex supersectors.
However, it is not sufficient to fully describe the vortices, determine the chiral central charge $c_-$ (mod 8), nor consistently assign quantum states to closed manifolds.
In order to fully describe a $(2+1)$D fermionic topological phase, one needs to specify which FMTC describes the system.
(It is, however, possible that a SMTC can be realized as an anomalous $(2+1)$D boundary theory of a $(3+1)$D theory with an appropriately matching bulk.)

The definition of a FMTC is a MTC $\mathcal{M}$ that contains the physical fermion $\mathbb{Z}_{2}^{\psi}$.
As a fermionic theory, both the vacuum $\I$ and the physical fermion $\psi$ topological charges correspond to local microscopic degrees of freedom, whereas the same MTC describing a bosonic system would treat the charge $\psi$ as an emergent fermion.
These statements have properties and implications that require explanation.

The first implication is that such a theory has a $\mathbb{Z}_{2}$ grading provided by braiding with $\psi$.
In particular, we can write
\begin{align}
\fMTC = \fMTC_{\vv{0}} \oplus \fMTC_{\vv{1}}
,
\end{align}
where each topological charge can be ascribed a label $\vv{x} \in \mathbb{Z}_{2}$ such that
\begin{align}
M_{\psi a_\vv{x} } = (-1)^{\vv{x}}
,
\end{align}
and the fusion rules respect the group structure of these labels, that is
\begin{align}
a_{\vv{x}} \otimes b_{\vv{y}} = \bigoplus_{c} N_{ab}^{c} c_{\vv{x}+\vv{y} \text{ mod }2}
.
\end{align}
That all objects have either $+1$ or $-1$ monodromy with the fermion follows from Eq.~(\ref{eq:M_relation}) and the $\mathbb{Z}_{2}$ fusion of the fermion.
This grading separates the topological charges into quasiparticles $\fMTC_{\vv{0}}$ and fermionic vortices $\fMTC_{\vv{1}}$.
Moreover, this $\mathbb{Z}_{2}$ grading is recognized as an incorporation of the $\mathbb{Z}_{2}^{\eff}$ fermionic parity conservation ``symmetry'' of the system, i.e. vortices can be thought of as defects of fermionic parity that can locally detect parity via braiding.
Clearly, $\I,\psi \in \fMTC_{\vv{0}}$, so we henceforth write these as $\I_{\vv{0}},\psi_{\vv{0}}$.
As with any $G$-graded FTC, the grading on the theory implies that the total quantum dimension of each sector is equal~\cite{Bark2019}, i.e. $\mathcal{D}_{\vv{0}} = \mathcal{D}_{\vv{1}}$.

Introducing the shorthand $[ \psi a]_{\vv{x}} = \psi_{\vv{0}} \otimes a_{\vv x}$, it follows that for a FMTC
\begin{align}
S_{a_{\vv{x}} b_{\vv{y}}} = (-1)^{\vv{x}} S_{a_{\vv{x}} [\psi b]_{\vv{y}}} = (-1)^{\vv{y}} S_{[\psi a]_{\vv{x}} b_{\vv{y}}}
\end{align}
and
\begin{align}
\theta_{[\psi a]_{\vv{x}}} = (-1)^{\vv{x} + \vv{1}} \theta_{a_{\vv{x}}}.
\end{align}

We now recognize that the quasiparticle sector $\fMTC_{\vv{0}}$ of any FMTC automatically satisfies the conditions to be a SMTC.
Thus, we can equivalently define a FMTC to be a $\mathbb{Z}_{2}^{\eff}$ modular extension of a SMTC.
It is also true that every SMTC has a (minimal) $\mathbb{Z}_{2}^{\eff}$ modular extension, which is a FMTC~\cite{JohnsonFreyd2021}.
In fact, there are always exactly 16 $\mathbb{Z}_{2}^{\eff}$ distinct modular extensions for a given SMTC~\cite{Bruillard17}.
This is a generalization of Kitaev's ``16-fold way'' of $\mathbb{Z}_{2}^{\eff}$ extensions of the trivial SMTC $\mathbb{Z}_{2}^{\psi}$~\cite{Kitaev2006} to general SMTCs $\fMTC_{\vv{0}}$.
The general 16-fold set of extensions of a SMTC can be obtained as a $\mathbb{Z}_{16}$ torsor action by fermionic stacking the extensions with the 16 extensions of the trivial fermion theory, as will be explained in Sec.~\ref{sec:f-identify}.
These 16 FMTC extensions are distinguished by their central charge $c_-$ mod 8, which is actually determined entirely by the vortices
\begin{equation}
e^{i \frac{2 \pi}{8} c_{-}} = \frac{1}{\mathcal{D}} \sum_{a_{\vv{1}} \in \fMTC_{\vv{1}}} d_{a_{\vv{1}}}^2 \theta_{a_{\vv{1}}}
.
\end{equation}
In light of this, we see that when a SMTC factorizes as $\mathcal{B} = \mathbb{Z}_{2}^{\psi} \boxtimes \widehat{\mathcal{B}}$, its $\mathbb{Z}_{2}^{\eff}$ extensions are given by the FMTCs $\ifo^{(\nu)} \boxtimes \widehat{\mathcal{B}}$, where $\ifo^{(\nu)}$ with $\nu = 0,1,\ldots,15$ are the FMTCs in Kitaev's 16-fold way.

Considering the vortices in a FMTC in more detail, since $\theta_{[\psi a]_{\vv{1}}} = \theta_{a_{\vv{1}}}$, it is possible to have $[\psi a]_{\vv{1}}$ either unequal or equal to $a_{\vv{1}}$, which we designate as ``$v$-type'' or ``$\sigma$-type'' vortices, respectively.
The supersector of a $\sigma$-type vortex is a singleton.
(Quasiparticles cannot be $\sigma$-type, since $\theta_{[\psi a]_{\vv{0}}} = -\theta_{a_{\vv{0}}}$.)
The set of vortices can then be split into subsets of these types $\fMTC_{\vv{1}} = \fMTC_{v} \sqcup \fMTC_{\sigma}$.
Applying the hexagon equations to $\sigma$-type vortices show that
\begin{align}
F^{\psi_{\vv{0}} a_{\sigma}  \psi_{\vv{0}}}_{a_{\sigma}} = R^{\psi_{\vv{0}} \psi_{\vv{0}}} =-1
.
\end{align}

When $\fMTC$ has $\sigma$-type vortices, all vortices are non-Abelian, and so have quantum dimension $d_{a_{\vv{1}}} \geq \sqrt{2}$.
In order to see this, we note that $a_{\vv{1}} \otimes b_{\sigma} \otimes \psi_{\vv{0}} = a_{\vv{1}} \otimes b_{\sigma}$, so the fusion channels of a $\sigma$-type vortex with any other vortex always come in pairs of quasiparticles related by fusion with the fermion, which is not possible if either vortex charge is Abelian.
It can also be shown that when $\fMTC$ has both $v$-type and $\sigma$-type vortices, not only are $d_{a_{\vv{1}}} \geq \sqrt{2}$, but the $\sigma$-type vortices have $d_{a_{\sigma}} \geq 2$.
This can be seen by fermionic stacking $\fMTC$ with $\ifo^{(1)}$, which transforms $v$-type vortices into $\sigma$-type vortices and vice-versa, for which their quantum dimensions are multiplied or divided by $\sqrt{2}$, respectively.

The characters of the fusion algebra of a SMTC $\fMTC_{\vv{0}}$ can be specified in terms of the fermionic modular $S$-matrix of its FMTC extensions.
In this way, there is a canonical labeling of the characters $\lambda^{(a_{\vv{0}})}_{\hat{b}}$ by the supersectors $\hat{b}$ of topological charges of $\fMTC$.
In particular, the fusion matrices of $\fMTC_{\vv{0}}$ are diagonalized by~\cite{Bonderson20}
\begin{align}
{\bf P}_{a_{\vv{0}} \hat{b}_{\vv{y}}} &= \frac{\sqrt{2}}{f_{b_{\vv{y}}}} S_{a_{\vv{0}} b_{\vv{y}}} ,\\
f_{b_{\vv{y}}} &= \left\{
\begin{array}{cll}
1 & & \text{ for } b_{\vv{y}}\in  \fMTC_{\vv{0}} \sqcup \fMTC_{v}
\\
\sqrt{2} & & \text{ for } b_{\vv{y}}\in  \fMTC_{\sigma}
\end{array}
\right.
,
\end{align}
where the $S$-matrix used here is that of the FMTC extensions.
The use of supersectors here is well-defined, since all charges in a given supersector have the same braiding with a particular quasiparticle charge.
Thus, half of these characters correspond to the quasiparticle supersectors and half correspond to the vortex supersectors (there are an equal number of such supersectors).
Since we are only considering the fusion algebra of the SMTC, this statement must be independent of which of the 16 $\mathbb{Z}_{2}^{\eff}$ extensions one chooses.
Moreover, it implies that the fusion rules of supersectors are the same for all FMTCs related by the $\mathbb{Z}_{16}$ torsor action, which are given by the superVerlinde formula~\cite{Bonderson20}
\begin{align}
\label{eq:superVerlinde}
\widehat{N}_{\hat{a}_{\vv{x}} \hat{b}_{\vv{y}}}^{\hat{c}_{\vv{x}+\vv{y}}} &=
\frac{1}{f_{a_{\vv{x}}} f_{ b_{\vv{y}}} f_{c_{\vv{x}+\vv{y}}}} \left( {N}_{a_{\vv{x}} b_{\vv{y}}}^{c_{\vv{x}+\vv{y}}} + {N}_{a_{\vv{x}} b_{\vv{y}}}^{[\psi c]_{\vv{x}+\vv{y}}} \right)
\notag\\
&= \sum_{e_{\vv{0}}} \frac{{\bf P}_{e_{\vv{0}} \hat{a}_{\vv{x}}} {\bf P}_{e_{\vv{0}} \hat{b}_{\vv{y}}} {\bf P}^{\ast}_{e_{\vv{0}} \hat{c}_{\vv{x}+\vv{y} }}}{ {\bf P}_{e_{\vv{0}} \hat{\I}_{\vv{0}}} }
.
\end{align}

Similar to the case of MTCs (and FMTCs), it follows that, for a SMTC $\fMTC_{\vv{0}}$, if there are phase factors that satisfy the relation
\begin{equation}
e^{i \phi_{a_{\vv{0}}}} e^{i \phi_{b_{\vv{0}}}} = e^{i \phi_{c_{\vv{0}}}}
\end{equation}
for all $a_{\vv{0}},b_{\vv{0}},c_{\vv{0}} \in \fMTC_{\vv{0}}$ such that $N_{a_{\vv{0}} b_{\vv{0}} }^{c_{\vv{0}}} \neq 0$, then there exists a unique superAbelian supersector of topological charge $\hat{e}_{\vv{x}}$ such that
\begin{equation}
\label{eq:supercharector}
e^{i \phi_{a_{\vv{0}} }} = \frac{ \lambda^{(a_{\vv{0}})}_{\hat{e}_{\vv{x}}} }{d_{a_{\vv{0}}} } = \frac{{\bf P}_{a_{\vv{0}} \hat{e}_{\vv{x}}} {\bf P}_{\I_{\vv{0}} \hat{\I}_{\vv{0}}} }{ {\bf P}_{\I_{\vv{0}} \hat{e}_{\vv{x}}} {\bf P}_{a_{\vv{0}} \hat{\I}_{\vv{0}}}   } = M_{a_{\vv{0}} \hat{e}_{\vv{x}}}^{\ast}
\end{equation}
for all $a_{\vv{0}}$.
(We note that $M_{a_{\vv{0}} \hat{e}_{\vv{x}}}$ is always well-defined, since $M_{a_{\vv{0}} e_{\vv{x}}} = M_{a_{\vv{0}} [\psi e]_{\vv{x}}}$.)
Here, we define a supersector $\hat{e}_{\vv{x}}$ to be superAbelian if ${\bf P}_{\I_{\vv{0}} \hat{e}_{\vv{x}}} / {\bf P}_{\I_{\vv{0}} \hat{\I}_{\vv{0}}} =1$, that is if $d_{e_{\vv{x}}} = 1$ for $e_{\vv{x}} \in \fMTC_{\vv{0}}  \sqcup \fMTC_{v}$ or $d_{e_{\vv{x}}}=\sqrt{2}$ for $e_{\vv{x}} \in \fMTC_{\sigma}$.
We denote the set of superAbelian supersectors of topological charges of $\fMTC$ as $\widehat{\bf A} = \widehat{\bf A}_{\vv{0}} \oplus \widehat{\bf A}_{\vv{1}}$, and recognize that they form a group with multiplication given by the fusion rules of supersectors.
We use different font to more clearly differentiate the superAbelian supersectors from $\mathcal{A} = \mathcal{A}_{\vv{0}} \oplus \mathcal{A}_{\vv{1}}$ the Abelian topological charges of $\fMTC$.
In particular, it is possible to have $\mathcal{A}_{\vv{1}} = \varnothing$, while $\widehat{\bf A}_{\vv{1}} \neq \varnothing$, since $\sigma$-type vortices may contribute to the latter, but not the former.
However, it is the case that $\mathcal{A}_{\vv{0}} = \mathbb{Z}_{2}^{\psi} \boxtimes \widehat{\bf A}_{\vv{0}}$, as previously noted, so we can use $\widehat{\mathcal{A}}_{\vv{0}}$ and $\widehat{\bf A}_{\vv{0}}$ interchangeably.

We emphasize that the group $\widehat{\bf A}$ is defined by the fusion rules of $\fMTC_{\vv{0}}$, with no dependence of the choice of $\mathbb{Z}_{2}^{\eff}$ extension.
More concretely, when $\widehat{\bf A}_{\vv{1}} \neq \varnothing$, we can pick an element $\hat{e}_{\vv{1}} \in \widehat{\bf A}_{\vv{1}}$, which we use to assign labels from the set $\widehat{\bf A}_{\vv{0}} \times \mathbb{Z}_{2}^{\eff}$ to all elements of $\widehat{\bf A}$ via
\begin{align}
\hat{a}_{\vv{x}} = \hat{a}_{\vv{0}} \otimes \hat{e}_{\vv{1}}^{\vv{x}} = (\hat{a}_{\vv{0}} , \vv{x})
.
\end{align}
The group multiplication is then specified by
\begin{align}
\hat{a}_{\vv{x}} \otimes \hat{b}_{\vv{y}} = \left( \hat{a}_{\vv{0}} \otimes \hat{b}_{\vv{0}} \otimes \hat{h}_{\vv{0}}^{\vv{x}\cdot \vv{y}} , \vv{x}+\vv{y} \text{ mod }2 \right)
,
\end{align}
where
\begin{align}
\hat{h}_{\vv{0}} = \hat{e}_{\vv{1}} \otimes \hat{e}_{\vv{1}}
,
\end{align}
is determined by $\delta_{\hat{c}_{\vv{0}} \hat{h}_{\vv{0}}} = \widehat{N}_{\hat{e}_{\vv{1}} \hat{e}_{\vv{1}}}^{\hat{c}_{\vv{0}}}$, using Eq.~\eqref{eq:superVerlinde}.
We see that
\begin{align}
\hat{h}(\vv{x},\vv{y}) = \hat{h}_{\vv{0}}^{\vv{x}\cdot \vv{y}}
,
\end{align}
defines a cocycle $\hat{h} \in Z^2(\mathbb{Z}_{2}^{\eff} , \widehat{\bf A}_{\vv{0}})$, and hence $\widehat{\bf A}$ is a central extension
\begin{align}
\widehat{\bf A} = \widehat{\bf A}_{\vv{0}} \times_{\hat{h}} \mathbb{Z}_{2}^{\eff}
.
\end{align}
It is straightforward to see that changing the arbitrary choice of $\hat{e}_{\vv{1}}$ to $\hat{e}_{\vv{1}}' = \hat{s}_{\vv{0}}\otimes \hat{e}_{\vv{1}}$, yields $\hat{h}'$ that differs from $\hat{h}$ by a coboundary in $B^{2}(\mathbb{Z}_{2}^{\eff} , \widehat{\bf A}_{\vv{0}})$.
Thus, $[\hat{h}]\in H^2(\mathbb{Z}_{2}^{\eff} , \widehat{\bf A}_{\vv{0}})$ is the central extension class defining $\widehat{\bf A}$.

We can similarly analyze $\mathcal{A}$, when $\mathcal{A}_{\vv{1}} \neq \varnothing$.
In this case, we pick an element $e_{\vv{1}} \in \mathcal{A}_{\vv{1}}$, which we use to assign labels from the set $\mathcal{A}_{\vv{0}} \times \mathbb{Z}_{2}^{\eff}$ to all elements of $\mathcal{A}$ via
\begin{align}
a_{\vv{x}} = a_{\vv{0}} \otimes e_{\vv{1}}^{\vv{x}} = (a_{\vv{0}} , \vv{x})
.
\end{align}
The group multiplication is then specified by
\begin{align}
a_{\vv{x}} \otimes b_{\vv{y}} = \left( a_{\vv{0}} \otimes b_{\vv{0}} \otimes h_{\vv{0}}^{\vv{x}\cdot \vv{y}} , \vv{x}+\vv{y} \text{ mod }2 \right)
,
\end{align}
where
\begin{align}
h_{\vv{0}} &= e_{\vv{1}} \otimes e_{\vv{1}} , \\
h(\vv{x},\vv{y}) &= h_{\vv{0}}^{\vv{x}\cdot \vv{y}}
,
\end{align}
defines a cocycle $h \in Z^2(\mathbb{Z}_{2}^{\eff} , \mathcal{A}_{\vv{0}})$, and hence the central extension
\begin{align}
\mathcal{A} = \mathcal{A}_{\vv{0}} \times_{h} \mathbb{Z}_{2}^{\eff}
.
\end{align}
The arbitrary choice of $e_{\vv{1}}$ corresponds to differences by coboundaries in $B^{2}(\mathbb{Z}_{2}^{\eff} , \mathcal{A}_{\vv{0}})$, so $[h]\in H^2(\mathbb{Z}_{2}^{\eff} , \mathcal{A}_{\vv{0}})$ is the central extension class defining $\mathcal{A}$.
While $\hat{h}_{\vv{0}}$ is uniquely determined by $\fMTC_{\vv{0}}$, it only determines $h_{\vv{0}}$ up to a factor of $\psi_{\vv{0}}$.
Indeed, both choices of $h_{\vv{0}} \in \hat{h}_{\vv{0}}$ are allowed, but correspond to different choices of the FMTC $\fMTC$, and these different FMTCs are related by stacking with $\ifo^{(\nu)}$ where $\nu \text{ mod }4= 2$.

A FMTC naturally couples to a spin structure through fermion condensation~\cite{Aasen19}.
Importantly the presence of the spin structure allows for two (not independent) equivalence relations on the theory.
One is that closed $\psi$-loops can be removed at the expense of a spin-structure dependant $\pm$ sign, even if the $\psi$-loop links a homologically nontrivial cycle.
If the theory lives on a manifold with trivial topology, this relation has no effect; however, when the manifold has nontrivial topology, the equivalence relation forces states to be labeled by supersectors.
The spin structure also ties vortices of the FMTC to non-bounding defects in the spin structure in a unique way~\cite{Aasen19}.
The second relation is that the spin structure allows for local encoding of all properties of the physical fermion.
The price to pay for a local encoding of the physical fermion is the theory must be enriched in super-vector spaces, and requires careful tracking of Koszul and spin-structure dependent $\pm$ signs.
This subtlety can be avoided by only considering quantum states with even global fermion parity, which can be canonically identified with states assigned by the FMTC.

\subsubsection{Canonical choices}
\label{sec:canonical}

Similar to the case of the vacuum charge, there are canonical choices that should be imposed as a result of interpreting the topological charge $\psi_\vv{0}$ as describing a physical (local) fermion, rather than an emergent fermionic quasiparticle.
Recall that, for the vacuum charge, there are canonical isomorphisms from the vector spaces $V^{\I_\vv{0} \I_\vv{0} }_{\I_\vv{0}}$, $V^{a_\vv{x}\I_\vv{0} }_{a_\vv{x}}$, and $V^{\I_\vv{0} a_\vv{x}}_{a_\vv{x}}$ to the complex numbers $\mathbb{C}$, associated with the ability to freely add and remove vacuum charges.
In terms of the basic data, this corresponds to fixing $F^{\I_\vv{0} a_\vv{x}b_\vv{y}}_{[ab]_{\vv{x}+\vv{y}}} =F^{a_\vv{x} \I_\vv{0}  b_\vv{y}}_{[ab]_{\vv{x}+\vv{y}}} = F^{a_\vv{x}b_\vv{y} \I_\vv{0} }_{[ab]_{\vv{x}+\vv{y}}} = \openone$ and $R^{a_\vv{x} \I_\vv{0}}_{a_\vv{x}} = R^{ \I_\vv{0}a_\vv{x}}_{a_\vv{x}} =1$, and restricts vertex basis gauge transformations to have $\Gamma^{\I_\vv{0} \I_\vv{0} }_{\I_\vv{0}}  = \Gamma^{\I_\vv{0}  a_\vv{x}}_{a_\vv{x}} = \Gamma^{a_\vv{x} \I_\vv{0} }_{a_\vv{x}} = 1$.
Clearly, a physical fermion will not obey exactly the same conditions as vacuum, since fermion parity conservation does not allow the creation or annihilation of a single isolated fermion, and the fusion and braiding of fermions with all objects is not completely trivial.
However, the connection to the microscopic degrees of freedom generates similar properties.

\begin{figure*}[t!]
   \centering
   \includegraphics[width=1.99\columnwidth]{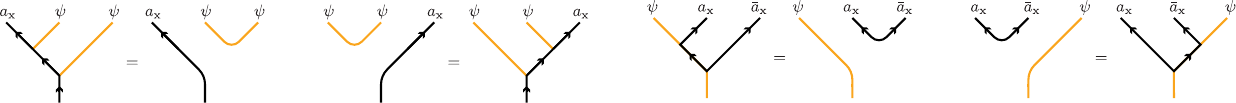}
   \caption{Canonical gauge choices are made for physical fermions that set the $F$-symbols to $1$ shown here. Diagrammatically, we can see these allow the $\psi_{\vv{0}}$ lines to be locally fused into or pulled off of other lines.}
   \label{fig:canonical}
\end{figure*}

In a fermionic topological phase, pairs of physical fermions are viewed as equivalent to the vacuum and can be locally added and removed freely.
This motivates a canonical isomorphism between $V^{\psi_\vv{0} \psi_\vv{0}}_{\I_\vv{0} }$ and $\mathbb{C}$, i.e. making a canonical choice of the basis state $\left| \psi_\vv{0} , \psi_\vv{0} ; \I_\vv{0} \right\rangle$.
This can be understood as fixing this basis state with respect to the corresponding state in the physical Hilbert space obtained by acting with the (local) creation and annihilation operators of the physical fermion.

Furthermore, we impose canonical gauge fixing of the $F$-symbols to satisfy
\begin{align}
\label{eq:Fpsipsi}
1 &= F^{a_\vv{x}\psi_\vv{0} \psi_\vv{0}}_{a_\vv{x}} = F^{\psi_\vv{0} \psi_\vv{0} a_\vv{x}}_{a_\vv{x}}
= F^{a_\vv{x} \bar{a}_\vv{x} \psi_\vv{0}}_{\psi_\vv{0}} = F^{\psi_\vv{0} a_\vv{x} \bar{a}_\vv{x}}_{\psi_\vv{0}}
.
\end{align}
These canonical choices for $F$-symbols are represented diagrammatically in Fig.~\ref{fig:canonical}.
Fixing these $F$-symbols, together with the pentagon equations implies the following $F$-symbols are also fixed
\begin{align}
\label{eq:Fpsipsi_implied}
1 &= F^{\psi_\vv{0} [\psi a]_\vv{x} \bar{a}_\vv{x}}_{\I_\vv{0}} = F^{a_\vv{x} [\psi \bar{a}]_\vv{x} \psi_\vv{0}}_{\I_\vv{0}}
.
\end{align}
It follows (also noting $F^{\psi_\vv{0} \psi_\vv{0} \psi_\vv{0}} =1$) that the physical fermion lines can freely bend up and down where they join other lines, as shown diagrammatically in Fig.~\ref{fig:bending}.
We emphasize that, in contrast to the vacuum charge, not all $F$-symbols involving $\psi$ can be made trivial.
For example, for $\sigma$-type vortices,
\begin{align}
F^{\psi a_{\sigma} \psi}_{a_{\sigma}} = \frac{ F^{a_{\sigma} \psi \bar{a}_{\sigma} }_{\psi}  }{F^{a_{\sigma} \psi \bar{a}_{\sigma} }_{\I}} = -1
\end{align}
are gauge invariant relations that follow from the consistency conditions.

\begin{figure*}[t!]
   \centering
   \includegraphics[width=1.99\columnwidth]{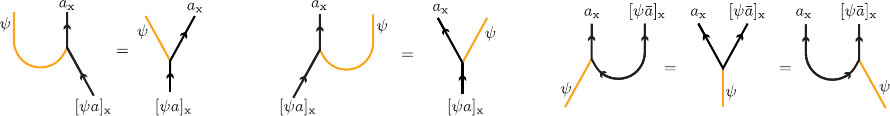}
   \caption{The canonical gauge choices for $F$-symbols involving the physical fermion imply that the $\psi_{\vv{0}}$ lines can freely bend up and down where they join other lines.}
   \label{fig:bending}
\end{figure*}

In terms of vertex basis gauge transformations, respecting the canonical isomorphisms and gauge choices reduce the gauge freedom by requiring
\begin{align}
\label{eq:gauge-1}
1&= \Gamma^{\psi_\vv{0} \psi_\vv{0}}_{\I_\vv{0}}
= \Gamma^{ a_\vv{x}\psi_\vv{0}}_{[\psi a]_\vv{x}} \Gamma^{ [\psi a]_\vv{x}\psi_\vv{0}}_{a_\vv{x}}
= \Gamma^{\psi_\vv{0} a_\vv{x}}_{[\psi a]_\vv{x}} \Gamma^{\psi_\vv{0} [\psi a]_\vv{x}}_{a_\vv{x}}
\notag \\
& = \frac{ \Gamma^{a_\vv{x} \bar{a}_\vv{x}}_{\I_\vv{0}}   }{ \Gamma^{\bar{a}_\vv{x} \psi_\vv{0} }_{[\psi \bar{a}]_\vv{x}} \Gamma^{a_\vv{x} [\psi \bar{a}]_\vv{x}}_{\psi_\vv{0}}}
=\frac{ \Gamma^{\psi_\vv{0} a_\vv{x} }_{[\psi a]_\vv{x}} \Gamma^{[\psi a]_\vv{x} \bar{a}_\vv{x}}_{\psi_\vv{0}}} { \Gamma^{a_\vv{x} \bar{a}_\vv{x}}_{\I_\vv{0}}   }
.
\end{align}
We note that for $\sigma$-type vortices, this implies
\begin{align}
\frac{ \Gamma^{a_\sigma \bar{a}_\sigma }_{\I_\vv{0}} }{ \Gamma^{a_\sigma \bar{a}_\sigma }_{\psi_\vv{0}}}
= \Gamma^{\bar{a}_\sigma \psi_\vv{0} }_{\bar{a}_\sigma } =  \Gamma^{ \psi_\vv{0} a_\sigma }_{a_\sigma } = \pm 1
.
\end{align}

We also observe that, for $\sigma$-type vortices, the canonical gauge choice combined with the pentagon equation gives
\begin{align}
\left[F^{a_{\vv{0}} b_{\sigma} \psi_{\vv{0}}}_{c_{\sigma}} \right]^2 = \left[F^{a_{\sigma} \psi_{\vv{0}} b_{\sigma} }_{c_{\vv{0}}} \right]^2 = \left[F^{\psi_{\vv{0}} a_{\sigma} b_{\vv{0}} }_{c_{\sigma}} \right]^2 = \openone
\end{align}
and combined with the hexagon equations give
\begin{align}
R^{a_{\sigma} \psi_{\vv{0}}}_{a_{\sigma}} =R^{\psi_{\vv{0}} a_{\sigma}}_{a_{\sigma}} = \pm i
.
\end{align}

\subsubsection{Example: Toric code as a trivial FMTC}
\label{sec:fToric}

We now consider the toric code as an illustrative example of a FMTC.
Recall the toric code has topological charges $\tc = \{\I, e, m, \psi\}$, whose monodromies with the fermion $\psi$ are
\begin{align}
M_{\I \psi} &= M_{\psi \psi} = 1, & M_{e\psi}&= M_{m\psi} = -1
.
\end{align}
When describing a bosonic topological phase, $\psi$ is considered an emergent fermion.
Reinterpreting this MTC as describing a fermionic topological phase, $\psi$ is considered the physical fermion $\psi_\vv{0}$, and the topological charges are divided into quasiparticles ($\I$, $\psi$) and vortices ($e$, $m$): we track their vorticity ($\mathbb{Z}_{2}^{\eff}$) labels by renaming the charges for the fermionic theory as
\begin{align}
\I &\to \I_\vv{0}, & \psi&\to \psi_\vv{0}, & e &\to \I_\vv{1}, & m &\to \psi_\vv{1}.
\end{align}
Note that the assignment of $e \to \I_{\vv{1}}$ and $m \to \psi_{\vv{1}}$ is arbitrary and these labels could be interchanged.
We denote this fermionic theory as $\ifo^{(0)}=\ifo^{(0)}_\vv{0}\oplus \ifo^{(0)}_\vv{1}$.
The topological data is inherited from the bosonic theory, and can be compactly summarized as
\begin{align}
a_\vv{x} \otimes b_\vv{y} &= [a\otimes b]_{\vv{x}+\vv{y}}
\\ F^{a_\vv{x} b_\vv{y} c_\vv{z}} &= 1
\\ R^{a_\vv{x} b_\vv{y}} &= (-1)^{(\mathsf{a}+\mathsf{x}) \cdot \mathsf{b}},
\end{align}
where we have written $a, b, c$ to denote labels $\I, \psi \in \mathbb{Z}_2^{\psi}$  with fusion (group multiplication) denoted by $\otimes$, and $\vv{x}, \vv{y}, \vv{z}$ to denote vorticity labels $\vv{0}, \vv{1} \in \mathbb{Z}_2^{\eff}$ with the group operation (addition mod 2) denoted by +.
For compactness, we have also used the associated fields  $\mathbb{F}_2^\psi$ and $\mathbb{F}_2^{\eff}$  with elements $\{0,1\}$, and with addition + and multiplication $\cdot$ given by the standard rules for addition and multiplication of integers modulo 2.
To distinguish field elements from their associated topological charges or group elements, we use the same letters in a different font, as summarized in Table~\ref{table:notation}.
We adopt this notation throughout the remainder of the paper.

Note that $\ifo^{(0)}$ corresponds to the trivial fermionic theory: it has no nontrivial topological quasiparticles, i.e. $\ifo^{(0)}_\vv{0}= \mathbb{Z}_2^{\psi}$, and is non-chiral, with $c_- = 0 \text{ mod }8$.
This theory describes an $s$-wave superconductor~\cite{Hansson04}.

\begin{table}[t]
\centering
\begin{tabular}{ c | c c }
Group & Operation~ & Elements \\
 \hline
$\mathbb{Z}_2^{\psi}$ & $\otimes$ & $a, b, \dots $\\
 $\mathbb{Z}_2^{\eff}$ & $+$ & $\vv{x}, \vv{y},\dots$
\end{tabular} \quad
\begin{tabular}{ c | c c }
Field & Operation~ & Elements \\
 \hline
$\mathbb{F}_2^{\psi}$ & $+$, $\cdot$ & $\mathsf{a}, \mathsf{b}, \dots $\\
 $\mathbb{F}_2^{\eff}$ & $+$, $\cdot$  & $\mathsf{x}, \mathsf{y},\dots$
\end{tabular}
\caption{Notation for distinguishing related labels associated with topological charge, group elements, and field elements.}
\label{table:notation}
\end{table}

\subsubsection{Example: Invertible FMTCs}\label{sec:ifo}

\begin{table*}[t]
\centering
\fbox{
\begin{tabular}{ll}
$\ifo^{(\nu=0\,\text{mod}\,4)}$: &
$a_\vv{x} \otimes b_\vv{y} = [a\otimes b]_{\vv{x} + \vv{y}} $, ~~~($\mathbb{Z}_2 \times \mathbb{Z}_2$ fusion rules)
\\ & \\
\begin{tabular}[t]{l}
 \\ \\  $F$-symbols: \\ \\ $R^{}$-symbols  \end{tabular}
& \begin{tabular}[t]{lll}
$\nu=0$ (Toric code)   & $\nu=8$ (3 Fermion) & $\nu=4,\,12$ (Semion$\times$ Semion)
\\  && \\
$F^{a_\vv{x} b_\vv{y} c_\vv{z}}= 1 $ & $F^{a_\vv{x} b_\vv{y} c_\vv{z}}=1 $ & $F^{a_\vv{x} b_\vv{y} c_\vv{z}}=(-1)^{\mathsf{x}\cdot \mathsf{y}\cdot \mathsf{z}} $
\\ &&\\
$R^{a_\vv{x} b_\vv{y}} =(-1)^{(\mathsf{a}+\mathsf{x}) \cdot \mathsf{b}} $~~~~ & $R^{a_\vv{x} b_\vv{y}} = (-1)^{\mathsf{x}\cdot \mathsf{y}+(\mathsf{a}+\mathsf{x}) \cdot \mathsf{b} } $~~~~ & $R^{a_\vv{x} b_\vv{y}} =i ^{\frac{\nu}{4} \mathsf{x }\cdot \mathsf{y} } (-1)^{(\mathsf{a}+\mathsf{x}) \cdot  \mathsf{b}} $
~
\end{tabular}
\\ & \\ \hline &\\
$\ifo^{(\nu=2\,\text{mod} \,4)}$: &$ a_\vv{x} \otimes b_\vv{y} = [\psi^{\mathsf{x}\cdot \mathsf{y}} \otimes a\otimes b]_{\vv{x} + \vv{y}} $, ~~~($\mathbb{Z}_4$ fusion rules)
\\ & \\
$F$-symbols: &
$F^{a_\vv{x} b_\vv{y} c_\vv{z}}= i^{-\frac{\nu}{2} \cdot \mathsf{x}\cdot \mathsf{y}\cdot \mathsf{z}} (-1)^{ \frac{\nu}{2} \cdot \mathsf{x} \cdot \mathsf{b}\cdot \mathsf{z}}$
\\ & \\
$R$-symbols: &
$R^{a_\vv{x} b_\vv{y}} = e^{i \frac{\pi}{8} \nu \cdot \mathsf{x }\cdot \mathsf{y} } (-1)^{(\mathsf{a}+\mathsf{b})\cdot \mathsf{x }\cdot \mathsf{y} + \mathsf{a} \cdot (\mathsf{b}+\mathsf{y}) } $
 \\ &\\ \hline & \\
$\ifo^{(\nu=1\,\text{mod}\,2)}$:&
$\psi_\vv{0} \otimes \psi_\vv{0} = \I_\vv{0},$ ~~~$ \sigma_\vv{1} \otimes \psi_\vv{0} = \sigma_\vv{1},$~~~
  $ \sigma_\vv{1}  \otimes \sigma_\vv{1}  = \I_\vv{0}\oplus \psi_\vv{0} $,~~~(Ising fusion rules)
 \\
&\\
$F$-symbols: &
\begin{tabular}[t]{l}
$F^{a_\vv{0}\sigma_\vv{1}b_\vv{0}}_{\sigma_\vv{1}} = F_{b_\vv{0}}^{\sigma_\vv{1} a_\vv{0} \sigma_\vv{1}} = (-1)^{\mathsf{a}\cdot \mathsf{b}},$~~
 $ [F_{\sigma_\vv{1}}^{\sigma_\vv{1}\sigma_\vv{1} \sigma_\vv{1}}]_{a_\vv{0} b_\vv{0}} = \kappa (-1)^{\mathsf{a}\cdot \mathsf{b}}/\sqrt{2} $, \quad $\kappa = (-1)^{(\nu^2-1)/8}$
 \end{tabular}
  \\
&\\
 $R$-symbols: &
 \begin{tabular}[t]{ll}$R^{\psi_\vv{0} \psi_\vv{0}}=-1$, &
 $ R^{\psi_\vv{0} \sigma_\vv{1}} = R^{\sigma_\vv{1}\psi_\vv{0}}=-e^{i\pi \nu/2}$,
 $R^{\sigma_\vv{1} \sigma_\vv{1}}_{\I_\vv{0}} = \kappa e^{-i \pi \nu/8} $, $R^{\sigma_\vv{1} \sigma_\vv{1}}_{\psi_\vv{0}} = \kappa e^{i 3\pi \nu/8}$
 \end{tabular}
\end{tabular}}
\caption{
The invertible FMTCs $\ifo^{(\nu)}$ of Kitaev's 16-fold way.
The notation follows the convention in Table~\ref{table:notation}.
(The somewhat unconventional gauge choices used here come from the use of the zesting functor of Sec.~\ref{sec:defect_torsor_functors}, which allow the $\nu$ even theories to be described by a single set of expressions, as given in Eqs.~\eqref{eq:Keven_fusion}-\eqref{eq:Keven_Gamma}).
}
\label{table:sixteen}
\end{table*}

The trivial fermionic theory $\ifo^{(0)}$ of the previous section is just one of 16 invertible fermionic orders $\ifo^{(\nu)}$, $\nu \in \mathbb{Z}_{16}$.
``Invertible'' refers to a theory with no nontrivial quasiparticles, that is $\ifo^{(\nu)}_\vv{0} =\mathbb{Z}_2^{\psi}$.
The vortex sectors have total quantum dimension $\mathcal{D}_{\ifo^{(\nu)}_\vv{1}}^2 = 2$, corresponding to either $\ifo^{(\nu=\text{even})}_\vv{1} = \{\I_\vv{1},\psi_\vv{1}\}$ (Abelian theories) or $\ifo^{(\nu=\text{odd})}_\vv{1}=\{\sigma_\vv{1}\}$ (non-Abelian theories).
The Abelian theories satisfy either $\mathbb{Z}_2\times \mathbb{Z}_2$ fusion ($\nu=0\,\text{mod}\,4$) or $\mathbb{Z}_4$ fusion ($\nu=2\,\text{mod}\,4$), while the non-Abelian theories all have Ising fusion rules.
The topological twists of the vortices in all of these theories are given by
\begin{align}
\theta_{a_{\vv{1}}} = e^{i \frac{\pi}{8}\nu}
,
\end{align}
which therefore distinguishes between different theories.
The chiral central charges are directly related to these and similarly given by
\begin{align}
c_- = \frac{\nu}{2} \text{ mod }8
.
\end{align}
Mathematically, the invertible fermionic orders correspond to the minimal modular extensions of $\mathbb{Z}_2^{\psi}$, colloquially known as Kitaev's 16-fold way~\cite{Kitaev2006}.
We present their basic data in Table~\ref{table:sixteen}, in a gauge chosen for convenience of presentation and which satisfies the canonical fermionic gauge choices.

The data for $\nu$ even here is obtained by applying the zesting functor described in Sec.~\ref{sec:defect_torsor_functors} to $\ifo^{(0)}$, including the vertex basis gauge transformation that returns the data into the canonical fermionic gauge.
In this way, the data for all $\nu$ even can be described by the single set of expressions
\begin{align}
\label{eq:Keven_fusion}
a_\vv{x} \otimes b_\vv{y} &= [\psi^{\frac{\nu}{2}\mathsf{x}\cdot \mathsf{y}} \otimes a\otimes b]_{\vv{x} + \vv{y}}
, \\
\label{eq:Keven_F}
F^{a_\vv{x} b_\vv{y} c_\vv{z}} &= \frac{\Gamma^{a_\vv{x}, b_\vv{y}} \Gamma^{a_\vv{x} \otimes b_\vv{y}, c_\vv{z}} }{\Gamma^{a_\vv{x} , b_\vv{y} \otimes c_\vv{z} } \Gamma^{b_\vv{y}, c_\vv{z}}}
\, i^{\frac{\nu}{2} \cdot \mathsf{x}\cdot \mathsf{y}\cdot \mathsf{z}} (-1)^{ \frac{\nu}{2} \cdot \mathsf{a} \cdot \mathsf{y}\cdot \mathsf{z}}
,\\
\label{eq:Keven_R}
R^{a_\vv{x} b_\vv{y}} &= \frac{\Gamma^{b_\vv{y} , a_\vv{x} }}{\Gamma^{a_\vv{x}, b_\vv{y}}}
\, e^{i \frac{\pi}{8} \nu \cdot \mathsf{x }\cdot \mathsf{y} } (-1)^{(\mathsf{a}+\mathsf{x}) \cdot  \mathsf{b}}
,\\
\label{eq:Keven_Gamma}
\Gamma^{a_\vv{x}, b_\vv{y}} &= (-1)^{\frac{\nu}{2} \cdot \mathsf{a} \cdot (1-\mathsf{x})\cdot \mathsf{y} }
.
\end{align}

\subsection{Stacking Fermionic Topological Phases}
\label{sec:f-identify}

Distinct fermionic topological phases can emerge from the same type of microscopic fermions, e.g. electrons.
When two such phases are brought into proximity of each other, so that the physical fermions can pass between them, a BTC description of the joint system must account for the fact that their physical fermions are the same.
Mathematically, this is done for two FMTCs $\fMTC^{(1)}$ and $\fMTC^{(2)}$ by taking their product and condensing the boson $(\psi_\vv{0}^{(1)}, \psi_\vv{0}^{(2)})$, where $a^{(j)}_{\vv{x}} \in \fMTC^{(j)}$.
Essentially, this process identifies the pair of physical fermions with the vacuum in the ``stacked'' theory such that objects related by fusion with $(\psi_\vv{0}^{(1)},\psi_\vv{0}^{(2)})$ become isomorphic.
We denote the resulting FMTC $\fMTC$ of the stacked fermionic phases as
\begin{align}
\label{eq:ftimes-first}
\fMTC =\fMTC^{(1)}\ftimes\, \fMTC^{(2)} \equiv \frac{\fMTC^{(1)}\boxtimes\, \fMTC^{(2)}}{A[\psi_\vv{0}^{(1)}, \psi_\vv{0}^{(2)}]} ,
\end{align}
where the condensate algebra object is
\begin{align}
\label{eq:alg-def}
A[\psi_\vv{0}^{(1)}, \psi_\vv{0}^{(2)}] &= (\I_\vv{0}^{(1)}, \I_\vv{0}^{(2)} ) \oplus (\psi_\vv{0}^{(1)} , \psi_\vv{0}^{(2)} )
.
\end{align}
For a review of condensation see Ref.~\onlinecite{Bais2009} or Appendix~\ref{app:condensation} for a discussion tailored to this paper.
In this way, the physical fermions from each of the original phases are identified to yield the (single) physical fermion of the resulting combined phase, that is
\begin{align}
\psi_\vv{0} \cong (\psi^{(1)}_{\vv{0}}, \I^{(2)}_{\vv{0}} ) \cong (\I^{(1)}_{\vv{0}}, \psi^{(2)}_{\vv{0}})
.
\end{align}

Condensation necessarily pairs objects of the same vorticity, i.e. quasiparticles with quasiparticles and vortices with vortices, since all opposite vorticity combinations yield confined objects (they have nontrivial braiding with the bosons in the condensate).
In light of this, the definition of fermionic stacking could have been made using a product of FMTCs that are diagonal in $\mathbb{Z}_{2}^{\eff}$, using $\boxtimes_{\mathbb{Z}_{2}^{\eff}}$ instead of $\boxtimes$.
We also note that pairing a $\sigma$-type vortex from one theory with a $v$-type vortex of the other will yield a $\sigma$-type vortex in the stacked theory, while pairing a $\sigma$-type vortex from one theory with a $\sigma$-type vortex of the other will split into two $v$-type vortices.
Specifically, $(a^{(1)}_{\sigma}, a^{(2)}_{v}) \cong a_{\sigma}$ with quantum dimension $d_{a_{\sigma}} = d_{a^{(1)}_{\sigma}}d_{a^{(2)}_{v}}$, while $(a^{(1)}_{\sigma}, a^{(2)}_{\sigma}) \cong a_{v}^{+} \oplus a_{v}^{-}$ with $d_{a_{v}^{+}} =d_{a_{v}^{-}} = \frac{1}{2} d_{a^{(1)}_{\sigma}}d_{a^{(2)}_{\sigma}}$. (For a more precise description of how objects split or combine under condensation, see Appendix~\ref{app:condensation}.)

More generally, if we have $n$ fermionic topological phases in proximity, with respective physical fermions $\psi^{(1)}_\vv{0}, \cdots, \psi^{(n)}_\vv{0}$, identifying the physical fermions reduces the symmetry $\left[ \mathbb{Z}_2^{\eff}\right]^{n}$ to the diagonal subgroup $\mathbb{Z}_2^{\eff}$.
Mathematically, identifying the physical fermions corresponds to condensing the bosons described by the algebra object
\begin{align}
\label{eq:A-def}
A[\psi^{(1)}_\vv{0}, \cdots, \psi^{(n)}_\vv{0}] = \bigoplus_{\vec{\mathsf{f}} \in \mathbb{F}_2^n: \; |\vec{\mathsf{f}}|=0} (\psi^{(1)}_\vv{0})^{\mathsf{f}_1}  (\psi^{(2)}_\vv{0})^{\mathsf{f}_2}  \cdots (\psi^{(n)}_\vv{0})^{\mathsf{f}_n}.
\end{align}
The constraint $|\vec{\mathsf{f}}|= \vec{\mathsf{f}} \cdot \vec{\mathsf{f}} =0\mod 2$ guarantees the algebra object $A[\psi^{(1)}_\vv{0},\dots,\psi^{(n)}_\vv{0}]$ is bosonic, i.e., an even number of fermions appears in each term on the right hand side.
After condensation, all the physical fermions have been identified, so that the resulting FMTC $\fMTC$ has one physical fermion $\psi_\vv{0}$.
We note that condensing the algebra object $A$ is nontrivial; condensation will result in objects being confined, and possibly even split (corresponding to multiple objects in the stacked theories).
The multiplication morphisms accompanying the algebra object $A$ are all isomorphic, see Appendix~\ref{app:associativity}.
Physically, this means there is a unique (up to isomorphism) way to identify the physical fermions.
Stacking multiple fermionic phases in this way is equivalent to successively stacking them pairwise.
This can be seen by observing that the stacking operation is associative and commutative, so the order one chooses for pairwise stacking does not matter.
We caution the reader that stacking does not necessarily preserve the fermionic canonical gauge choices of Sec.~\ref{sec:canonical}.
When it is important to do so, a vertex basis gauge transformation may be required to ensure that a FMTC obtained from stacking satisfies the fermionic canonical gauge choices.

This notion of stacking fermionic phases, i.e. taking the product and condensing physical fermion pairs, provides a natural tool for classifying fermionic topological order, even without strong interactions between the systems.
Indeed, stacking can be used to generate the 16-fold way of FMTCs with the same quasiparticle sector SMTC $\fMTC_{\vv{0}}$, and we will later use the generalization of stacking to defect theories in the classification of FSET phases.
In order to understand the general 16-fold way, we start by considering the invertible FMTCs $\ifo^{(\nu)}$, i.e. Kitaev's 16-fold way.
Stacking these FMTCs reveals a $\mathbb{Z}_{16}$ group structure
\begin{align}
\label{eq:16-group}
\ifo^{(\nu)} \ftimes \, \ifo^{(\nu')} = \ifo^{([\nu+\nu']\text{ mod }16)}
.
\end{align}
In this case, $\ifo^{(0)}$ is the identity element and any $\ifo^{(\nu)}$ with $\nu$ odd is a generator of the group.

For the general 16-fold way, we consider stacking a FMTC $\fMTC$ with $\ifo^{(\nu)}$ to obtain a new FMTC
\begin{equation}
\fMTC' = \fMTC \ftimes \, \ifo^{(\nu)}
.
\end{equation}
We note that: (1) $\fMTC'_{\vv{0}} = \fMTC_{\vv{0}}$, because $\ifo^{(\nu)}_{\vv{0}}$ is trivial, (2) $\fMTC' = \fMTC$ when $\nu \text{ mod }16 =0$, (3) $c'_{-} = [c_{-} + \frac{\nu}{2} ] \text{ mod }8$, so distinct $\nu \text{ mod }16$ yield distinct $\fMTC'$, (4) $\fMTC \ftimes \, \ifo^{(\nu)} \ftimes \, \ifo^{(\nu')} = \fMTC \ftimes \, \ifo^{(\nu + \nu ')}$, and (5) except when $\fMTC$ is an invertible FMTC, there is no canonical ``0'' element in the set of 16 FMTCs generated this way.
Thus, we see that the set of $\mathbb{Z}_2^{\eff}$ extensions of a given SMTC $\fMTC_{\vv{0}}$ forms a $\mathbb{Z}_{16}$ torsor under the action of stacking with the invertible FMTCs.

\section{Symmetry and Fractionalization in Fermionic Topological Phases}
\label{sec:symmetric-fermions}

\begin{table*}
\begin{center}
 \begin{tabular}{l c p{0.45\textwidth}}
 Obstruction & Torsor & Interpretation \\ [0.5ex]
 \hline
 &&\\[-.5ex]
$[O^{\rho}] \in H^{2}(G , \mathbb{Z}_2^{\V})$ & $H^{1}(G, \mathbb{Z}_2^{\V})$ &
\begin{minipage}[c]{.99\linewidth}\raggedright
Extension of symmetry action $\rho^{(\vv{0})}$ on $\fMTC_{\vv{0}}$ to $\rho$ on $\fMTC$. Prerequisite: $[ \rho_{\bf g}^{(\vv{0})}] \in \im (\res_{\fMTC_{\vv{0}}})$. Torsor action changes symmetry action. See Eq.~\eqref{eq:Orho} and \eqref{eq:O2obs}.
\end{minipage}
\\[4ex]
$[\coho{O}] \in H^3_{[\rho]}(G,{\mathcal{A}})$ & $\frac{Z_{[\rho]}^{2}(G,\mathcal{A})}{B_{[\rho]}^{2}(G,\mathcal{A}_{\vv{0}})}$ &
\begin{minipage}[c]{.99\linewidth}\raggedright
Symmetry fractionalization on $\fMTC$ with $\rho$. Torsor action potentially changes $\mathcal{G}^{\eff}$. See Eq.~\eqref{eq:H3obsruction}.
\end{minipage}
\\[4ex]
$[\coho{O}^{\central}] \in \left\{
\begin{array}{lll}
 H^3_{[\rho^{(\vv{0})}]}(G, \mathcal{A}_{\vv{0}}) \times Z^2(G,\mathbb{Z}_2^{\eff}) &  \mathcal{A}_{\vv{1}} = \varnothing \\
 H^3_{[\rho^{(\vv{0})}]}(G, \mathcal{A}_{\vv{0}}) &\mathcal{A}_{\vv{1}} \neq \varnothing
\end{array}
\right. $& $H^2_{[\rho^{(\vv{0})}]}(G,\mathcal{A}_{\vv{0}})$ &
\begin{minipage}[c]{.99\linewidth}\raggedright
Symmetry fractionalization manifesting $\mathcal{G}^{\eff} = \mathbb{Z}_2^{\eff} \times_{\central} G$ on $\fMTC$ with $\rho$. Torsor action preserves $\mathcal{G}^{\eff}$. See Eqs.~\eqref{eq:Gf_obstruction_general_1}, \eqref{eq:O_central_general}, and \eqref{eq:Gf_obstruction_general_2}.
\end{minipage}
\\[4ex]
$[\coho{O}^{(\vv{0})}] \in H^3_{[\widehat{\rho}]}(G,\widehat{\bf A})$ & $\frac{Z_{[\widehat{\rho}]}^{2}(G,\widehat{{\bf A}})}{B_{[\widehat{\rho}]}^{2}(G,\widehat{{\bf A}}_{\vv{0}})}$ &
\begin{minipage}[c]{.99\linewidth}\raggedright
Symmetry fractionalization on $\fMTC_{\vv{0}}$ with $\rho^{(\vv{0})}$. Torsor action can potentially change $\mathcal{G}^{\eff}$. See Eq.~\eqref{eq:H3SMTC}.
\end{minipage}
\\[4ex]
 $[{\coho{O}}^{(\vv{0})\central}] \in \left\{
\begin{array}{lll}
 H^3_{[\widehat{\rho}^{(\vv{0})}]}(G, \widehat{\mathcal{A}}_{\vv{0}}) \times Z^2(G,\mathbb{Z}_2^{\eff}) & \widehat{{\bf A}}_{\vv{1}} = \varnothing \\
 H^3_{[\widehat{\rho}^{(\vv{0})}]}(G, \widehat{\mathcal{A}}_{\vv{0}}) &  \widehat{{\bf A}}_{\vv{1}} \neq \varnothing
\end{array}
\right.$&$H_{[\widehat{\rho}^{(\vv{0})}]}^{2}(G,\widehat{\mathcal{A}}_{\vv{0}} )$&
\begin{minipage}[c]{.99\linewidth}\raggedright
Symmetry fractionalization manifesting $\mathcal{G}^{\eff} = \mathbb{Z}_2^{\eff} \times_{\central} G$ on $\fMTC_{\vv{0}}$ with $\rho^{(\vv{0})}$. Torsor action preserves $\mathcal{G}^{\eff}$. See Eq.~\eqref{eq:Gf_obstruction_general_M0_1} and \eqref{eq:Gf_obstruction_general_M0_2}.
\end{minipage}
\\[4ex]
$[{\coho{O}}^{\eta} ] \in H^3(G,\mathbb{Z}_{2}^{\psi})$ & $H^2 (G,\mathbb{Z}_{2}^{\psi})$ &
\begin{minipage}[c]{.99\linewidth}\raggedright
Extension of symmetry fractionalization $\eta^{(\vv{0})}$ on $\fMTC_{\vv{0}}$ with $\rho^{(\vv{0})}$ to $\eta$ on $\fMTC$ with $\rho$. Prerequisite: $\coho{w}^{(\vv{0})} \in C^2(G,\widehat{\mathcal{A}})$. See Eq.~\eqref{eq:O_extension} and \eqref{eq:obstructioneta}.
\end{minipage}
\end{tabular}
\caption{Obstructions and torsorial classifications (when unobstructed) appearing in Sec.~\ref{sec:symmetric-fermions} for fermionic topological phases.
$\fMTC_{\vv{0}}$ is a SMTC describing the quasiparticles of a fermionic topological phases and $\fMTC$ is one of its 16 possible $\mathbb{Z}_{2}^{\eff}$ modular extensions, i.e. $\fMTC$ is a FMTC which describes the quasiparticles and vortices.
$\mathcal{G}^{\eff} = \mathbb{Z}_{2}^{\eff} \times_\central G$ is the fermionic symmetry group, which is a $\mathbb{Z}_{2}^{\eff}$ central extension of $G$.
$\rho$ and $\rho^{(\vv{0})}$ are symmetry actions on $\fMTC$ and $\fMTC_{\vv{0}}$, respectively.
$\widehat{\rho}$ and $\widehat{\rho}^{(\vv{0})}$ are their corresponding restrictions to acting on supersectors of topological charge, which are both fully determined by $\rho^{(\vv{0})}$.
$\mathbb{Z}_{2}^{\V}$ is the subgroup of the fermionic topological symmetry group generated by the vortex permuting symmetry of Eq.~\eqref{eq:Vsymm}.
The map $\res_{\fMTC_{\vv{0}}}$ is the restriction of the topological symmetries on $\fMTC$ to $\fMTC_{\vv{0}}$, as defined in Eq.~\eqref{eq:resdef}.
The Abelian groups $\mathcal{A} = \mathcal{A}_{\vv{0}} \oplus \mathcal{A}_{\vv{1}}$ and $\mathcal{A}_{\vv{0}}$ are formed from Abelian topological charges and Abelian quasiparticles, respectively.
The groups $\widehat{\bf A} = \widehat{\bf A}_{\vv{0}} \oplus \widehat{\bf A}_{\vv{1}}$ and $ \widehat{\bf A}_{\vv{0}}$ are formed from superAbelian supersectors of topological charge and Abelian supersectors of quasiparticles, respectively.
While $\widehat{\bf A}_{\vv{0}} = \widehat{\mathcal{A}}_{\vv{0}}$, it is not necessarily the case that the set of superAbelian supersectors of vortices $\widehat{\bf A}_{\vv{1}}$ is equal to the set of supersectors of Abelian vortices $\widehat{\mathcal{A}}_{\vv{1}}$ (see paragraph following Eq.~\eqref{eq:supercharector});
in general, $\widehat{\mathcal{A}} \lhd \widehat{\bf A}$.
(The cohomology groups listed for extension of the symmetry action assume $\ker(\res_{\fMTC_{\vv{0}}}) = \mathbb{Z}_{2}^{\V}$.
Though we believe this assumption is generally true, a more general $\ker(\res_{\fMTC_{\vv{0}}})$ has not been ruled out for FMTCs with $\widehat{\bf A}_{\vv{1}} = \varnothing$.
As discussed in Sec.~\ref{sec:symmetry_action} and Appendix~\ref{app:rho_extension_general}, the more general case would require replacing $\mathbb{Z}_{2}^{\V}$ with $\ker(\res_{\fMTC_{\vv{0}}})$ in the cohomology groups for symmetry action extensions.)
}
\label{table:obstructions}
\end{center}
\end{table*}

In this section, we develop the notions of topological symmetry, symmetry action, and symmetry fractionalization for fermionic topological phases, which apply to both SMTCs and FMTCs.
Much of this theory is similar to the same notions for bosonic topological phases developed in Ref.~\onlinecite{Bark2019}, which we follow closely for background and notation.
However, there are important distinctions that arise due to the inclusion of physical fermions in the theory, which results in a different and more intricate structure that we detail.

Topological symmetries, which are invertible maps from the emergent topological theory to itself, are constrained to leave the physical fermion fixed and preserve the canonical isomorphisms of the topological state space associated with it.
Moreover, equivalences of such maps under local transformations of objects, i.e. natural isomorphism, are subject to the condition that they act completely trivially on the fermion.

Closed fermionic systems have an inherent $\mathbb{Z}_2^{\eff}$ fermion parity conservation, which is similar to an ordinary (bosonic) symmetry in some ways, but behaves differently in others.
We have already seen that $\mathbb{Z}_2^{\eff}$ manifests as a grading on FMTCs that distinguishes quasiparticles from vortices.
For a symmetric fermionic topological phase, there is an additional ordinary global symmetry of the system, described by a group $G$, which is sometimes called the ``bosonic symmetry group.''
The full fermionic symmetry group $\mathcal{G}^{\eff}$ corresponds to a $\mathbb{Z}_2^{\eff}$ central extension of $G$, which can be defined as the one for which the physical fermions have linear representations.
The global symmetry action on the topological theory is given by an action of $G$ via fermionic topological symmetries.

Implementing the symmetry action on the physical Hilbert space leads to the notions of localization of the symmetry action on quasiparticles and vortices, as well as fractionalization of the symmetry.
The fractionalization of the $G$ symmetry action on the physical fermion $\psi_{\vv 0}$ is encoded in projective phase factors
\begin{equation}
\eta_{\psi_{\vv 0}}({\bf g,h})=(-1)^{\central ({\bf g,h})}
,
\end{equation}
where $\central \in Z^2 (G,\mathbb{Z}_2^{\eff})$ is a 2-cocycle (for a review of group cohomology, see Appendix~\ref{app:cohomology}).
As this corresponds to the projective representation of $G$ on the physical fermions, it is directly correlated with the choice of central extension $\mathcal{G}^{\eff} = \mathbb{Z}_2^{\eff} \times_{\central} G$ for which the representation can be made linear (non-projective) for the physical fermions.

The equivalences of symmetry fractionalization classes is subject to the condition that action of the localized operators on the physical fermions is fixed, i.e. the vorticity of the localized operators cannot change.
These constraints associated with the physical fermion modify the analysis of obstruction and classification of symmetry fractionalization, as compared to bosonic systems.
For a fixed $\mathcal{G}^{\eff}$, when unobstructed, the classification of fractionalization on the FMTC $\fMTC$ is given torsorially by
$H^2_{[\rho^{(\vv{0})}]}(G,\mathcal{A}_\vv{0})$.
Similarly, the classification for fixed $\mathcal{G}^{\eff}$ of fractionalization on the quasiparticles described by the SMTC $\fMTC_{\vv{0}}$ is given torsorially by
$H^2_{[\rho^{(\vv{0})}]}(G,\widehat{\mathcal{A}}_\vv{0})$.

When viewed from the perspective of extending the $G$ symmetry fractionalization patterns, from the physical fermions $\mathbb{Z}_2^{\psi}$ to the SMTC $\fMTC_{\vv{0}}$ describing the quasiparticles and then to the FMTC $\fMTC$ describing quasiparticles and vortices, this results in tiers of obstructions and classifications that we summarize in Table~\ref{table:obstructions}.
When unobstructed, the extensions of fractionalization on the quasiparticles $\fMTC_{\vv{0}}$ to fractionalization on the full FMTC $\fMTC$ including vortices is torsorially classified by $H^2(G,\mathbb{Z}_2^{\psi})$.
In terms of the classification of quasiparticle fractionalization and extensions to vortices, the classification of fractionalization on $\fMTC$ takes the form
\begin{align}
\label{eq:frac-factor}
H^2_{[\rho^{(\vv{0})}]}(G,\mathcal{A}_\vv{0})
& \cong i_{\ast}\left( H^2 (G,\mathbb{Z}_{2}^{\psi}) \right) \times_{\varepsilon} p_{\ast} \left( H^2_{[\rho^{(\vv{0})}]}(G,\mathcal{A}_\vv{0}) \right)
.
\end{align}
Here, $i_\ast : H^2(G,\mathbb{Z}_2^{\psi}) \to H^2_{[\rho^{(\vv{0})}]}(G,\mathcal{A}_\vv{0})$ is the homomorphism induced by the inclusion map $i: \mathbb{Z}_2^{\psi} \to \mathcal{A}_\vv{0}$; $p_{\ast}: H^2_{[\rho^{(\vv{0})}]}(G,\mathcal{A}_\vv{0}) \to H^2_{[\widehat{\rho}^{(\vv{0})}]}(G,\widehat{\mathcal{A}}_\vv{0})$ is the induced homomorphism from the map $p: \mathcal{A}_\vv{0} \to \widehat{\mathcal{A}}_\vv{0}$ from topological charges to their supersectors; and $\varepsilon \in Z^2(\im(p_\ast) , \im(i_{\ast}))$ indicates some central extension.

$p_{\ast} ( H^2_{[\rho^{(\vv{0})}]}(G,\mathcal{A}_\vv{0}) )$ is the subgroup of $H^2_{[\widehat{\rho}^{(\vv{0})}]}(G,\widehat{\mathcal{A}}_\vv{0})$ corresponding to vanishing $[{\coho{O}}^{\eta}]$ obstruction.
In other words, it corresponds to the subset of quasiparticle fractionalization classes that can be extended to the vortices to provide fractionalization classes on the full FMTC $\fMTC$.

$i_{\ast}( H^2 (G,\mathbb{Z}_{2}^{\psi}))$ is not necessarily isomorphic to $H^2 (G,\mathbb{Z}_{2}^{\psi})$, as $i_{\ast}$ may map a nontrivial element of $H^2 (G,\mathbb{Z}_{2}^{\psi})$ to a trivial element of $H^2_{[\rho^{(\vv{0})}]}(G,\mathcal{A}_\vv{0})$.
This means different extensions of the quasiparticle fractionalization to vortices may actually yield the same fractionalization on the FMTC $\fMTC$.

The symmetry action $[\rho^{(\vv{0})}]$ plays an important role in determining the structure in Eq.~\eqref{eq:frac-factor}, such as $i_{\ast}$.
In many instances, this takes the simple form $H^2_{[\rho^{(\vv{0})}]}(G,\mathcal{A}_\vv{0}) \cong H^2(G,\mathbb{Z}_2^{\psi}) \times p_{\ast} ( H^2_{[\rho^{(\vv{0})}]}(G,\mathcal{A}_\vv{0}) )$, but this need not always be the case.

When we develop the theory of symmetry defects for FSET phases, we will see in Sec.~\ref{sec:classification} that this structure is accounted for in a natural way in the classification of FSET phases.
In particular, the $H^2(G,\mathbb{Z}_2^{\psi})$ factor corresponding to extending quasiparticle fractionalization to vortices is generated through fermionic stacking with $\mathcal{G}^{\eff}$ FSPT phases and the $p_{\ast} ( H^2_{[\rho^{(\vv{0})}]}(G,\mathcal{A}_\vv{0}) ) \lhd H^2_{[\widehat{\rho}^{(\vv{0})}]}(G,\widehat{\mathcal{A}}_\vv{0})$ factor is the quasiparticle fractionalization that is unchanged by the FSPT stacking torsor action.

Ref.~\onlinecite{Galindo2017} provided closely related, but distinct definitions for categorical fermionic actions, and analysis of the obstruction and classification of fractionalization manifesting a supergroup $(G, [\central])$.
In our terminology, their definitions and analysis do not constrain equivalences of symmetry actions and fractionalization classes to be completely trivial for the physical fermion.
That is, they do not require $\gamma_{\psi_{\vv 0}}({\bf g})=1$ on symmetry action gauge transformations nor $\zeta_{\psi_{\vv 0}}({\bf g,h})=1$ on the equivalences of fractionalizaion classes.
These correspond to physical differences stemming from the physical nature of the fermion, and lead to important differences in the resulting classification of fractionalization, as we will explain in Sec.~\ref{sec:symmetry_fractionalization}.

While the theory of symmetry and fractionalization for fermionic topological phases that we develop applies to both SMTCs and FMTCs, a strictly $(2+1)$D fermionic topological phase is necessarily described by a FMTC.
This is because the vortices are an essential part of the fermionic theory, as previously discussed.
It is nonetheless useful to consider the theory for SMTCs, because when a symmetry and fractionalization can be defined on a SMTC, but cannot be extended to the corresponding FMTC, it may be viewed as anomalous.
Such theories can potentially be realized as surface terminations of a $(3+1)$D topological phase, where the obstruction invariants of the $(2+1)$D boundary theory match invariants of the $(3+1)$D bulk~\cite{Vishwanath2013,Fidkowski2013,Fidkowski2018}.
More specifically, we can identify the obstructions $[O^{\rho}] \in H^{2}(G , \mathbb{Z}_2^{\V})$ and $[{\coho{O}}^{\eta}] \in H^3(G,\mathbb{Z}_{2}^{\psi})$ as anomalies of a surface termination that respectively match the bulk $(3+1)$D FSPT classifying elements valued in $H^{2}(G , \mathbb{Z}_2)$, describing Majorana chain decorations of tri-junctions of $G$-foams, and $H^3(G,\mathbb{Z}_{2})$, describing fermionic parity decorations of quad-junctions of $G$-foams.
We also note that the obstruction to extending a quasiparticle fractionalization class due to $\coho{w}^{(\vv{0})} \notin C^2(G,\widehat{\mathcal{A}})$, i.e. a quasiparticle fractionalization class for which $\coho{w}^{(\vv{0})}({\bf g,h}) \in \widehat{\bf A}_{\vv{1}} \neq \widehat{\mathcal{A}}_{\vv{1}}$ for some ${\bf g,h}$, may correspond to an anomalous $(2+1)$D theory in which the tri-junction of ${\bf g}$-, ${\bf h}$-, and ${\bf gh}$-defect branch lines carries a Majorana zero mode.
This would also necessitate a matching $(3+1)$D bulk with Majorana chain decorations of tri-junctions of $G$-foams.

\subsection{Fermionic Topological Symmetries}
\label{sec:Topo_symmetry}

The topological symmetries of a BTC $\BTC$ are invertible braided auto-equivalence maps $\varphi: \BTC \to \BTC$ which leave all topological data invariant.
We primarily focus on unitary, on-site symmetries that are space-time parity preserving, but will discuss more general types of symmetries in Sec.~\ref{sec:loc_pres_symmetry}.
At the level of topological charges, an auto-equivalence map $\varphi$ is simply a permutation, which we will often write using the shorthand
\begin{align}
\varphi (a) = a'
.
\end{align}
However, arbitrary permutations are not allowed.
Importantly, the vacuum $\I$ topological charge must be left fixed,
\begin{align}
\varphi(\I) = \I
,
\end{align}
as must all gauge invariant quantities, most notably
\begin{align}
N_{a' b'}^{c'} &= N_{a b}^{c},\\
d_{a'} &= d_{a} ,\\
\theta_{a'} &=\theta_{a},\\
S_{a' b'} &= S_{a b}
.
\end{align}

Quantities that are not gauge invariant must be left unchanged by the permutation of objects, up to a vertex basis gauge transformation.
As such, the action of an auto-equivalence map on vertex basis states can be written in the form
\begin{align}
\label{eq:udef}
\varphi(\ket{a,b; c, \mu}) & = \widetilde{\ket{a',b'; c', \mu}}\nonumber \\
&= \sum_{\mu'} \left[ u_{c'}^{a'b'} \right]_{\mu \mu'}\ket{a',b'; c', \mu'}
\end{align}
where $\left[ u_{c'}^{a'b'} \right]_{\mu \mu'}$ is a unitary vertex basis transformation that is required to leave all of the topological data invariant.
Specifically, this means the transformations must satisfy
\begin{align}
\varphi \left(\left[F^{abc}_{d} \right]_{(e, \alpha, \beta)(f, \mu, \nu)} \right)& = \left[\widetilde{F}^{a' b' c' }_{d'} \right]_{(e', \alpha, \beta)(f', \mu, \nu)}
\notag \\
&= \left[F^{a b c }_{d} \right]_{(e, \alpha, \beta)(f, \mu, \nu)}
,
\label{eq:uFcond} \\
\varphi \left( \left[R^{a b}_{c}\right]_{\mu \nu} \right)  &=
\left[\widetilde{R}^{a' b' }_{c' }\right]_{\mu \nu}  = \left[R^{a b}_{c}\right]_{\mu \nu}
.
\label{eq:uRcond}
\end{align}

The canonical isomorphisms between $V^{a \I}_{a}$, $V^{ \I a}_{a}$, and $\mathbb{C}$ for general BTCs discussed in Sec.~\ref{sec:BTC} require that we fix
\begin{align}
\label{eq:upsicondvac}
u^{a \I}_{a} = u^{ \I a}_{a} = u^{ \I \I}_{\I}=1
,
\end{align}
for all $a$.
In the case of $a \neq \I$, these are automatically implied by $\varphi$ preserving the canonical gauge choices of $F$-symbols ($F^{abc}_{d} = \openone$ when any of  $a$, $b$, or $c$ equals $\I$), together with imposing the condition $u^{ \I \I}_{\I} = 1$ reflecting the canonical identification of vacuum.

Autoequivalence maps of the form
\begin{align}
\label{eq:nat_iso}
\Upsilon(a) &= a ,\\
\Upsilon \left(\ket{a,b; c, \mu} \right) &= \frac{\gamma_{a} \gamma_{b}}{\gamma_{c}} \ket{a,b; c, \mu}
,
\end{align}
where $\gamma_{a}$ are phases, are called natural isomorphisms.
Eq.~(\ref{eq:upsicondvac}) implies that we must fix $\gamma_{\I} = 1$.

Natural isomorphisms always leave the topological data invariant, and so are considered trivial auto-equivalence maps.
Defining an equivalence of auto-equivalence maps related by natural isomorphism, the equivalence classes $[\varphi]$ form elements of a group denoted $\text{Aut}(\BTC)$ that we call the topological symmetry group.
Multiplication in this group is given by composition: $[\varphi_{3}] = [\varphi_{1}] \cdot [\varphi_{2}] = [\varphi_{1} \circ \varphi_{2}]$, and can be expressed in terms of representatives of the equivalence classes by $\varphi_{3} = \kappa \circ \varphi_{1} \circ \varphi_{2}$, where $\kappa$ is a natural isomorphism.

There is also a redundancy of natural isomorphisms in terms of how one decomposes the phases $\gamma_{a}$.
In particular, if we instead use the phases $\breve{\gamma}_{a} = \lambda_a \gamma_a$, such that the phases $\lambda_a$ satisfy $\lambda_a \lambda_b = \lambda_c$ when $N_{ab}^{c} \neq 0$, this defines exactly the same natural isomorphism autoequivalence map $\Upsilon$.
As shown in Sec.~\ref{sec:BTC}, when $\BTC$ is a MTC, there is a bijection between phases that obey this property and the set of Abelian topological charges $\mathcal{A} \subset \BTC$, given by
\begin{align}
\label{eq:lambda}
\lambda_{a} = M_{a e}^*
\end{align}
for $e \in \mathcal{A}$.

The topological symmetries of a fermionic theory must incorporate additional conditions reflecting the physical nature of the fermion.
(From here on, we only consider fermionic auto-equivalence maps.)
Since the topological charge associated with the physical fermion is treated as canonical, a fermionic auto-equivalence map must leave this topological charge fixed, that is
\begin{align}
\label{eq:psinvariance}
\varphi(\psi_\vv{0}) = \psi_\vv{0}
.
\end{align}
As a direct consequence, a fermionic topological symmetry cannot change the vorticity of objects in $\fMTC$, since
\begin{align}
M_{\psi_\vv{0},\varphi(a_{\vv{x}})} = M_{\varphi(\psi_\vv{0}),\varphi(a_{\vv{x}})} = M_{\psi_\vv{0},a_{\vv{x}}}
.
\end{align}
Thus, at the level of topological charges, we can write
\begin{align}
\varphi(a_{\vv{x}}) = a'_{\vv{x}}
\end{align}
for some $a'_{\vv{x} } \in \fMTC_{\vv{x}}$.
Eq.~(\ref{eq:psinvariance}) also implies supersectors of topological charge are permuted together, since we have $\varphi( \psi_{\vv{0}} \otimes a_{\vv{x}}) = \psi_{\vv{0}} \otimes \varphi (a_{\vv{x}})$.

The canonical isomorphisms for fermionic theories discussed in Sec.~\ref{sec:canonical} additionally require fermionic auto-equivalence maps to satisfy
\begin{align}
\label{eq:upsicond}
u_{\I_{\vv{0}}}^{\psi_{\vv{0}}\psi_{\vv{0}}}  =1
,
\end{align}
to preserve the canonical isomorphism between $V_{\I_{\vv{0}}}^{\psi_{\vv{0}}\psi_{\vv{0}}}$ and $\mathbb{C}$.
Enforcing the canonical gauge choices for the $F$-symbols will further require
\begin{align}
\label{eq:ugaugeconst}
1 &= u^{ a_\vv{x}\psi_\vv{0}}_{[\psi a]_\vv{x}} u^{ [\psi a]_\vv{x}\psi_\vv{0}}_{a_\vv{x}}
= u^{\psi_\vv{0} a_\vv{x}}_{[\psi a]_\vv{x}} u^{\psi_\vv{0} [\psi a]_\vv{x}}_{a_\vv{x}}
\notag \\
& = \frac{ u^{a_\vv{x} \bar{a}_\vv{x}}_{\I_\vv{0}}   }{ u^{\bar{a}_\vv{x} \psi_\vv{0} }_{[\psi \bar{a}]_\vv{x}} u^{a_\vv{x} [\psi \bar{a}]_\vv{x}}_{\psi_\vv{0}}}
=\frac{ u^{\psi_\vv{0} a_\vv{x} }_{[\psi a]_\vv{x}} u^{[\psi a]_\vv{x} \bar{a}_\vv{x}}_{\psi_\vv{0}}} { u^{a_\vv{x} \bar{a}_\vv{x}}_{\I_\vv{0}}   }
.
\end{align}

We define the ``$\psi$-fixed topological symmetry group'' $\Autpsi{\fMTC}$ by forming equivalence classes of fermionic auto-equivalence maps (subject to the above conditions) under natural isomorphisms.
Natural isomorphisms that satisfy Eq.~(\ref{eq:upsicond}) are constrained to have
\begin{align}
\label{eq:ferm_nat_cond}
\gamma_{\psi_{\vv{0}}} =\pm 1
,
\end{align}
so this condition must be respected for the natural isomorphisms that relate fermionic auto-equivalence maps.
It is clear that this forms a subgroup $\Autpsi{\fMTC} < \Aut{\fMTC}$ of the ordinary topological symmetry group, since Eq.~(\ref{eq:psinvariance}) potentially reduces the number of distinct topological symmetries.

In fact, the local nature of the physical fermions requires the stricter constraint
\begin{align}
\label{eq:ferm-nat-iso}
\gamma_{\psi_{\vv{0}}} =1
\end{align}
on the natural isomorphisms that are allowed to equate different auto-equivalence maps for the physical fermion.~\footnote{If generalizing to a system with multiple physical fermions $\psi_\vv{0}^{(j)}$, all must satisfy $\gamma_{\psi_\vv{0}^{(j)}}=1$.}
We will call such natural isomorphisms that satisfy this condition ``fermionic natural isomorphisms.''
This constraint can be understood from the perspective that the action of operators on the physical fermions in the emergent theory are determined by the underlying microscopic details of the physical system.
The natural isomorphisms with nontrivial $\gamma_{\psi_{\vv{0}}}$ correspond to physically nontrivial operations on the physical fermions.
More specifically, a $\gamma_{\psi_{\vv{0}}} = -1$ transformation would change the vorticity of the localized global symmetry action on the physical fermions, meaning it would create differences that are physically measurable using the fermions, and hence should not be viewed as a trivial gauge freedom.
When we develop the symmetry defect theory for fermionic topological phases in Sec.~\ref{sec:fermionic-defectification}, we will also see that such natural isomorphism change the vorticity of the symmetry defects.

Restricting fermionic natural isomorphisms to be those with $\gamma_{\psi_{\vv{0}}} =1$ also restricts the redundancy of natural isomorphisms in terms of how they are decomposed.
In particular, for the transformations of fermionic natural isomorphisms as $\breve{\gamma}_{a_{\vv{x}}} = \lambda_{a_{\vv{x}}} \gamma_{a_{\vv{x}}}$, fixing $\breve{\gamma}_{\psi_{\vv{0}}} =1$ requires that $\lambda_{\psi_{\vv{0}}} = 1$.
Through the relation in Eq.~(\ref{eq:lambda}), this implies
$\lambda_{a_{\vv{x}} } = M_{a_{\vv{x}} {e}_{\vv{0}}}^*$
for some ${e}_{\vv{0}} \in \mathcal{A}_{\vv{0}}$.

We similarly define the ``fermionic topological symmetry group'' $\Autf{\fMTC}$ by forming equivalence classes of fermionic auto-equivalence maps related by fermionic natural isomorphisms.
In order to understand the difference between $\text{Aut}(\fMTC)$, $\Autpsi{\fMTC}$, and $\Autf{\fMTC}$ in more detail, it is useful to give special attention to the auto-equivalence map
\begin{align}
\label{eq:Qiso}
\Q&(a_\vv{x}) = a_\vv{x} ,\\
\Q& \left(\ket{a_\vv{x},b_\vv{y}; c_{\vv{x}+\vv{y}}, \mu} \right) \notag
\\ &= (-1)^{\delta_{a_\vv{x} \psi_\vv{0}} + \delta_{b_\vv{y} \psi_\vv{0}} +\delta_{c_{\vv{x}+\vv{y}} \psi_\vv{0}}}  \ket{a_\vv{x},b_\vv{y}; c_{\vv{x}+\vv{y}}, \mu}
,
\label{eq:Q2}
\end{align}
which corresponds to the natural isomorphism with $\gamma_{\psi_\vv{0}}=-1$ and $\gamma_{a_\vv{x}}=1$ for $a_\vv{x} \neq \psi_\vv{0}$.
Clearly, $[\Q]$ is always trivial in $\text{Aut}(\fMTC)$ and $\Autpsi{\fMTC}$, since it is a natural isomorphism.
Whether $\Q$ is a fermionic natural isomorphism, i.e. whether it can be written as a natural isomorphism that has $\gamma_{\psi_\vv{0}}=1$ and $\gamma_{a_\vv{x}} \neq 1$ for certain other $a_\vv{x} \neq \psi_\vv{0}$, depends on $\fMTC$.
For example, $[\Q]$ is clearly nontrivial whenever $\fMTC$ has $\sigma$-type vortices.

More generally, $[\Q]$ is trivial in $\Autf{\fMTC}$ if and only if $\mathcal{A}_{\vv{1}} \neq \varnothing$.
This is because all partitions of $\Q$ can be written as $\gamma_{a_{\vv{x}}}=(-1)^{\delta_{a_{\vv{x}} \psi_{\vv{0}}}} M_{a_{\vv{x}} e_{\vv{y}}}$ for $e_{\vv{y}} \in \mathcal{A}$.
We note that if there exist $a_{\vv{x}}, b_{\vv{y}}, c_{\vv{x+y}} \in \fMTC$ such that $N_{a_{\vv{x}} b_{\vv{y}} }^{c_{\vv{x+y}}} N_{a_{\vv{x}} b_{\vv{y}} }^{[\psi c]_{\vv{x+y}}} \neq 0$, then $[\Q]$ is nontrivial.
When $[\Q]$ is a nontrivial element of $\Autf{\fMTC}$, i.e. when $\mathcal{A}_{\vv{1}} = \varnothing$, it defines a central subgroup $\mathbb{Z}_{2}^{\Q}$ of $\Autf{\fMTC}$, since $\Q ^2 = \openone$ and $\Q$ commutes with all auto-equivalence maps.
It follows that $\Autpsi{\fMTC} \cong \Autf{\fMTC} / \mathbb{Z}_{2}^{\Q}$ when $[\Q]$ is nontrivial, and $\Autpsi{\fMTC} \cong \Autf{\fMTC}$ when $[\Q]$ is trivial.
We emphasize that these differences allow for $\text{Aut}(\fMTC)$, $\Autpsi{\fMTC}$, and $\Autf{\fMTC}$ to be inequivalent for the same FMTC $\fMTC$.

It is useful to compute some examples to gain intuition for $\Autf{\fMTC}$.
For this, we consider the invertible fermionic phases $\ifo^{(\nu)}$ described in Sec.~\ref{sec:ifo}.
When $\nu$ is even (the Abelian theories), the auto-equivalence maps are specified by whether they leave all topological charges fixed or permute $\I_{\vv{1}} \leftrightarrow \psi_{\vv{1}}$.
It is straightforward to see that we can write $\Q$ as a fermionic natural isomorphism, e.g. with $\gamma_{\I_{\vv{1}}} =-1$ and $\gamma_{a_{\vv{x}}}=1$ for $a_{\vv{x}} \neq \I_{\vv{1}}$.
When $\nu$ is odd (the non-Abelian theories), the topological charges cannot be permuted by a symmetry.
However, these theories all have $\sigma$-type vortices, so $\Q$ is a nontrivial fermionic topological symmetry.
Thus, we have shown
\begin{align}
\label{eq:autKnu}
\Autf{\ifo^{(\nu)} } = \mathbb{Z}_2
\end{align}
for all $\nu$.
In contrast, $\Autpsi{\ifo^{(\nu)}}=\mathbb{Z}_1$ for $\nu$ odd, in agreement with the analogous bosonic topological symmetry group $\text{Aut}(\text{Ising}^{(\nu)}) = \mathbb{Z}_1$.
For an example where the $\psi$-fixed topological symmetry group differs from the bosonic topological symmetry group, we consider $\nu =8$.
Here, the $\Autpsi{\ifo^{(8)}} \cong \Autf{\ifo^{(8)}}=\mathbb{Z}_2$, while the analogous bosonic topological symmetry group is distinct, $\text{Aut}(\text{SO}(8)_1)=S_3$.

The vortex permuting symmetry described for $\ifo^{(\nu)}$ with $\nu$ even can be generalized to a nontrivial fermionic auto-equivalence for all FMTCs.
Denoting this special vortex permuting symmetry by $\V$, we define it to act trivially on quasiparticles and interchange vortices within a supersector, that is
\begin{align}
\label{eq:Viso}
\V(a_{\vv{x}}) = \psi^\vv{x}_\vv{0} \otimes a_{\vv{x}}
.
\end{align}
It is easy to check that this permutation preserves gauge invariant quantities.
For example, the topological spin satisfies
$\theta_{\V(a_{\vv{x}})} = \theta_{[\psi^{\vv{x}} a]_{\vv{x}}} = \theta_{a_{\vv{x}}}$.
However, the corresponding $u$-symbols for $\V$ are nontrivial, and we write these as
\begin{align}
&\V \left(\ket{a_{\vv{x}},b_{\vv{y}}; c_{\vv{x}+\vv{y}}, \mu} \right) \notag \\
& \qquad = \sum_{\mu '} \left[\V(a'_{\vv{x}},b'_{\vv{y}}; c'_{\vv{x}+\vv{y}})\right]_{\mu \mu'}   \ket{a'_{\vv{x}},b'_{\vv{y}}; c'_{\vv{x}+\vv{y}}, \mu'}
\end{align}
and define them diagrammatically to be
\begin{align}
\label{eq:Vsymm}
\VSymmetry &= \sum_{\mu '}\left[\V(a'_{\vv{x}},b'_{\vv{y}}; c'_{\vv{x}+\vv{y}})\right]_{\mu \mu'} \VSymmetryRHS
\end{align}
where $a_{\vv{x}}' \equiv \V(a_{\vv{x}})$.
One can directly see that the $\V$ defined this way will satisfy Eqs.~\eqref{eq:uFcond} and \eqref{eq:uRcond}, using the commutative diagrams displayed in Fig.~\ref{fig:vsymmery_fig}.
 \begin{figure*}[htp]
   \centering
   \includegraphics{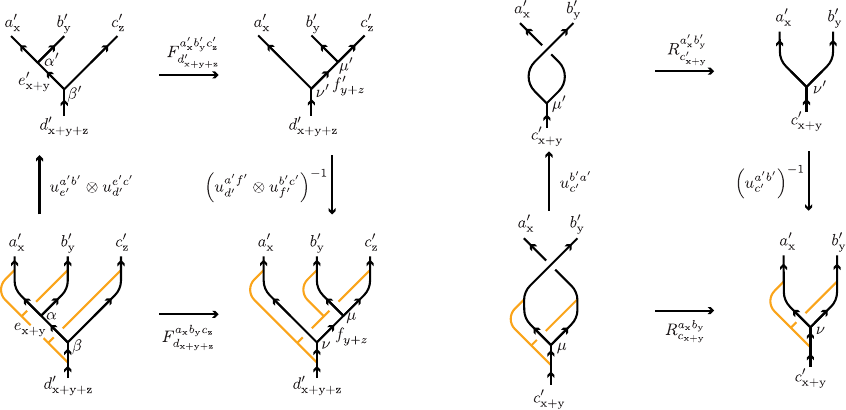}
   \caption{We show that the map $\V$ with $\V(a_{\vv{x}}) = [\psi^{\vv{x}} a]_{\vv{x}} = a'_{\vv{x}}$ and $ u_{c'_{\vv{x}+\vv{y}}}^{a'_{\vv{x}}b'_{\vv{y}}}  \equiv \V(a'_{\vv{x}},b'_{\vv{y}}; c'_{\vv{x}+\vv{y}})$ defined in Eq.~\eqref{eq:Vsymm} is a fermionic topological symmetry.
   (The value $\I$ or $\psi$ of the orange lines here can be inferred from the definition.)
   In the left panel, we demonstrate that Eq.~\eqref{eq:uFcond} is satisfied.
   This uses the fact that $\psi_{\vv{0}}$-strands can freely pass under trivalent vertices, together with the fermionic canonical gauge choice described in Sec.~\ref{sec:canonical}.
   In the right panel, we demonstrate that Eq.~\eqref{eq:uRcond} is satisfied.
   This also uses the property $M_{\psi_{\vv{0}}^{\vv{y}} a_{\vv{x}}'}R^{\psi_{\vv{0}}^{\vv{x}}\psi_{\vv{0}}^{\vv{y}}} = 1$.
   }
   \label{fig:vsymmery_fig}
\end{figure*}

Evaluating the diagrams in Eq.~\eqref{eq:Vsymm} and using the canonical gauge choices of Eq.~\eqref{eq:Fpsipsi}, we have
\begin{align}
\label{eq:VsymmetryU}
&\left[\V(a'_{\vv{x}}, b'_{\vv{y}} ; c'_{\vv{x} + \vv{y}}) \right]_{\mu \mu'} = \sum_{\nu}
\left[ \left( F^{a_{\vv{x}} \psi_{\vv{0}}^{\vv{y}} b'_{\vv{y}}}_{c_{\vv{x} + \vv{y}} } \right)^{-1}\right]_{(b_{\vv{y}} ,\mu)([\psi^{\vv{y}} a]_{\vv{x}}, \nu)}
\notag \\
& \qquad \qquad \times R^{\psi_{\vv{0}}^{\vv{y}}a_{\vv{x}}}
\left[ \left( F^{ \psi_{\vv{0}}^{\vv{x}+\vv{y}} [\psi^{\vv{y}} a]_{\vv{x}} b'_{\vv{y}}}_{c'_{\vv{x}+ \vv{y}}} \right)^{-1} \right]_{(c_{\vv{x}+ \vv{y}}, \nu) (a'_{\vv{x}} , \mu')}
.
\end{align}

We note that $[\V]^2 = [\V^2] =[\openone]$, since
\begin{align}
\label{eq:kappaV}
\kappa(a_{\vv{x}},b_{\vv{y}}; c_{\vv{x}+\vv{y}})\delta_{\mu \nu} &= \left[\V(a'_{\vv{x}},b'_{\vv{y}}; c'_{\vv{x}+\vv{y}}) \V(a_{\vv{x}},b_{\vv{y}}; c_{\vv{x}+\vv{y}}) \right]_{\mu \nu} \notag \\
&= R^{\psi_{\vv{0}}^{\vv{x}} \psi_{\vv{0}}^{\vv{y}}} \delta_{\mu\nu} = (-1)^{\vv{x} \cdot \vv{y}} \delta_{\mu\nu}
\end{align}
is a fermionic natural isomorphism given by $\gamma_{a_{\vv{x}}} = i^{\vv{x}}$.
The evaluation in going to the second line is most easily performed diagrammatically, particularly in light of the canonical gauge choices.

We can also see that $[\V]$ commutes with all elements of $\Autpsif{\fMTC}$.
To verify this, we consider another fermionic auto-equivalence map $\varphi$ and check that $\V \circ \varphi$ and $\varphi \circ \V$ differ by a fermionic natural isomorphism.
First, we observe that
\begin{align}
\V \circ \varphi(a_{\vv{x}}) =  \psi_{\vv{0}}^{\vv{x}} \otimes \varphi(a_{\vv{x}}) =  \varphi(\psi_{\vv{0}}^{\vv{x}} \otimes a_{\vv{x}})  = \varphi \circ \V (a_{\vv{x}}),
\end{align}
since $\varphi(\psi_{\vv{0}}) = \psi_{\vv{0}}$.
Examining the action on fusion vertices, the diagrammatic definition of $\V$ makes it easy to see that
\begin{align}
\label{eq:Vcommute}
\Upsilon \circ \V \circ \varphi \left(\ket{a_{\vv{x}},b_{\vv{y}}; c_{\vv{x}+\vv{y}}, \mu} \right) &= \varphi \circ \V  \left(\ket{a_{\vv{x}},b_{\vv{y}}; c_{\vv{x}+\vv{y}}, \mu} \right)
,
\end{align}
where
\begin{align}
\gamma_{a_{\vv{x}}} = U_{\varphi} (\psi_{\vv{0}}^{\vv{x}} , a_{\vv{x}} ; [\psi^{\vv{x}} a]_{\vv{x}})
.
\end{align}
Thus, we have the claimed result
\begin{align}
[\V] \cdot [\varphi] = [\varphi] \cdot [\V]
.
\end{align}
This shows $[\V]$ always generates a central subgroup $\mathbb{Z}_2^{\V}$ of $\Autpsif{\fMTC}$ for any FMTC $\fMTC$.

One can similarly consider the fermionic auto-equivalence maps of a SMTC $\fMTC_{\vv{0}}$.
All of the discussion above applies directly, except with vorticity labels always set to $\vv{x} = \vv{0}$, and the redundancy of natural isomorphisms is given by superAbelian supersectors of charge $\hat{z} \in \widehat{{\bf A}}$, which is further restricted to $\widehat{{\bf A}}_{\vv{0}}$ if we constrain $\gamma_{\psi_{\vv{0}}} = 1$.
We note that $\V = \openone$ when restricted to $\fMTC_{\vv{0}}$.
On the other hand, $[\Q]$ can be either a trivial or nontrivial element of $\Autf{\fMTC_{\vv{0}}}$, depending on $\fMTC_{\vv{0}}$.
In particular, $[\Q]$ is trivial in $\Autf{\fMTC_{\vv{0}}}$ if and only if $\widehat{{\bf A}}_{\vv{1}} \neq \varnothing$.
We note that if there exist $a_{\vv{0}}, b_{\vv{0}}, c_{\vv{0}} \in \fMTC_{\vv{0}}$ such that $N_{a_{\vv{0}} b_{\vv{0}} }^{c_{\vv{0}}} N_{a_{\vv{0}} b_{\vv{0}} }^{[\psi c]_{\vv{0}}} \neq 0$, then $[\Q]$ is nontrivial on $\fMTC_{\vv{0}}$.
The invertible fermionic topological phases $\ifo^{(\nu)}$ with $\nu$ odd are examples where $[\Q]$ is nontrivial in $\Autf{\fMTC}$, but trivial in $\Autf{\fMTC_{\vv{0}}}$.
The $\text{SO}(3)_{4n+2}$ SMTCs described in Sec.~\ref{sec:ex_SU2} provide examples where $[\Q]$ is nontrivial in $\Autf{\fMTC_{\vv{0}}}$.

We can now make a precise statement regarding when $[\V ]$ and $[\Q]$ are equal:
$[\V ] = [\Q]$ on $\fMTC$ if and only if $\mathcal{A}_{\vv{1}} = \varnothing$ and $\widehat{\bf A}_{\vv{1}} \neq \varnothing$.
Recall that this is equivalent to the condition that $\fMTC$ has only $\sigma$-type vortices with at least one being superAbelian, i.e. has quantum dimension $d_{a_{\sigma}} =\sqrt{2}$.
That this is necessary for $[\V ] = [\Q]$ is clear, since when it is not the case, either $\V$ is nontrivial when $\Q$ is trivial on $\fMTC$, or $\Q$ is nontrivial on $\fMTC_{\vv{0}}$.
That this is sufficient for $[\V ] = [\Q]$ can be see by writing an explicit fermionic natural isomorphism that relates $\V$ and $\Q$.
For this, we first choose a superAbelian $\sigma$-type vortex $e_{\vv{1}} \in \widehat{\bf A}_{\vv{1}} $, and then define $a_{\vv{0}}^{e}$ such that $N_{a_{\vv{0}}^{e} e_{\vv{1}}}^{a_{\vv{1}}} = 1$ and $q_{\vv{0}}$ such that $N_{q_{\vv{0}} e_{\vv{1}}}^{\overline{e_{\vv{1}}}} = 1$.
(There are exactly two choices for each $a_{\vv{0}}^{e}$ and $q_{\vv{0}}$, related by fusion with $\psi_{\vv{0}}$, and the choices of which to use will not matter.)
We then define the fermionic natural isomorphism $\Upsilon^{\eff}$ by
\begin{align}
\label{eq:VQequiv0}
\gamma_{a_{\vv{0}}} &= M_{a_{\vv{0}} e_{\vv{1}}} (-1)^{\delta_{a_{\vv{0}} \psi_{\vv{0}}} } ,
\\
\label{eq:VQequiv1}
\gamma_{a_{\vv{1}}} &= \frac{M_{a_{\vv{0}}^{e} e_{\vv{1}}} }{ F^{\psi_{\vv{0}} e_{\vv{1}} a_{\vv{0}}^{e} }_{a_{\vv{1}} } }
\left( \frac{
R^{\psi_{\vv{0}} e_{\vv{1}} } F^{e_{\vv{1}} \psi_{\vv{0}} \overline{e_{\vv{1}}} }_{\I_{\vv{0}}}
F^{\psi_{\vv{0}} e_{\vv{1}} q_{\vv{0}}}_{ \overline{e_{\vv{1}}} }
}
{ M_{q_{\vv{0}} e_{\vv{1}} } } \right)^{\frac{1}{2}}
,
\end{align}
which gives $\V = \Upsilon^{\eff} \Q$.
This follows from the respective definitions, together with applications of the pentagon and hexagon consistency equations.
We can now see that $[\V ] = [\Q]$ for $\ifo^{(\nu)}$ with $\nu$ odd, so $\Autf{\ifo^{(\nu)}} = \mathbb{Z}_2^{\V}$ for all $\nu$.

We note that specifying how a fermionic auto-equivalence map ${\varphi}$ permutes the topological charges of $\fMTC_{\vv{0}}$ is equivalent to specifying how it permutes the supersectors of quasiparticle topological charges.
This is because the quasiparticle charges in a supersector have distinct topological twist factors $\theta_{[\psi a]_{\vv{0}}} =-\theta_{a_{\vv{0}}}$, so there is a unique lift from the permutation of supersectors to the permutation of charges.
We denote such permutations of supersectors as $\widehat{\varphi} (\hat{a}_{\vv{x}})$ or $\widehat{\varphi}^{(\vv{0})} (\hat{a}_{\vv{0}})$, even when $\fMTC_{\vv{0}}$ does not factorize as $\mathbb{Z}_{2}^{\psi} \boxtimes \widehat{\fMTC_{\vv{0}}}$.
These can be defined by $\widehat{\varphi} (\hat{a}_{\vv{x}}) = \widehat{\varphi (a_{\vv{x}})}$ and $\widehat{\varphi}^{(\vv{0})} (\hat{a}_{\vv{0}}) = \widehat{\varphi^{(\vv{0})} (a_{\vv{0}})}$.

Moreover, specifying how ${\varphi}$ permutes the topological charges of $\fMTC_{\vv{0}}$ uniquely determines how it permutes the supersectors of vortex topological charges.
In order to demonstrate this, we use the fact, discussed in Sec.~\ref{sec:fMTC}, that the unitary matrix ${\bf P}$ that diagonalizes the fusion algebra of a SMTC $\fMTC_{\vv{0}}$ can be expressed in terms of the $S$-matrix of its $\mathbb{Z}_{2}^{\eff}$ extensions (independent of which extension).
Recall that matrix indices of ${\bf P}_{a_{\vv{0}} \hat{b}_{\vv{y}}}$ are $a_{\vv{0}} \in \fMTC_{\vv{0}}$ and the supersectors of charges in $\fMTC$, which we will denote as $\hat{b}_{\vv{y}} \in \widehat{\fMTC}$.
Since the $S$-matrix is invariant under the symmetry action, this implies ${\bf P}_{\varphi^{(\vv{0})}(a_{\vv{0}}) \widehat{\varphi}(\hat{b}_{\vv{y}})} = {\bf P}_{a_{\vv{0}} \hat{b}_{\vv{y}}}$.
From this we see that the permutation of supersectors of vortices by a fermionic auto-equivalence map $\varphi$ is completely determined by how it permutes the quasiparticles, that is
\begin{align}
\delta_{\widehat{\varphi}(\hat{b}_{\vv{1}}) , \hat{b}_{\vv{1}}'  } &= \sum_{a_{\vv{0}} \in \fMTC_{\vv{0}}} {\bf P}_{a_{\vv{0}} \widehat{\varphi}(\hat{b}_{\vv{1}})} {\bf P}_{a_{\vv{0}} \hat{b}'_{\vv{1}}}^{\ast} \notag \\
&= \sum_{a_{\vv{0}} \in \fMTC_{\vv{0}}} {\bf P}_{ \varphi^{(\vv{0})-1}(a_{\vv{0}}) \hat{b}_{\vv{1}}} {\bf P}_{a_{\vv{0}} \hat{b}'_{\vv{1}}}^{\ast}
\end{align}
shows that
\begin{align}
\label{eq:vortex_lifted_permutation}
\widehat{\varphi}(\hat{b}_{\vv{1}}) = \hat{b}_{\vv{1}}'
\end{align}
is determined entirely by $\varphi^{(\vv{0})}$ and the fusion algebra of $\fMTC_{\vv{0}}$.

It is useful to consider extensions of a fermionic topological symmetry of a SMTC $\fMTC_{\vv{0}}$, i.e. $[\varphi^{(\vv{0})}] \in  \Autpsif{\fMTC_{\vv{0}}}$, to a fermionic topological symmetry of one of its $\mathbb{Z}_{2}^{\eff}$ extensions $\fMTC$, $[\varphi] \in \Autpsif{\fMTC}$.
For this, we introduce the restriction map $\res_{\fMTC_{\vv{0}}} $ defined by
\begin{align}
\label{eq:resdef}
\res_{\fMTC_{\vv{0}}} \left( \varphi \right) = \left. \varphi \right|_{\fMTC_{\vv{0}}}
,
\end{align}
and similarly for equivalence classes of auto-equivalence maps.
In fact, $\res_{\fMTC_{\vv{0}}}$ is a homomorphism, so its kernel, i.e. all $\psi$-fixed or fermionic topological symmetries of $\fMTC$ which act trivially on $\fMTC_{\vv{0}}$, forms a normal subgroup $ \ker(\res_{\fMTC_{\vv{0}}}) \lhd \Autpsif{\fMTC}$.
This can also be seen from the fact that $\varphi^{-1} \Upsilon \varphi = \Upsilon'$ with $\gamma'_{a} = \gamma_{\varphi(a)}$.
Additionally, its image is isomorphic to the quotient
\begin{align}
\im (\res_{\fMTC_{\vv{0}}}) = \res_{\fMTC_{\vv{0}}}( \Autpsif{\fMTC} ) \cong \frac{\Autpsif{\fMTC}}{\ker(\res_{\fMTC_{\vv{0}}}) }
.
\end{align}
For $[\varphi^{(\vv{0})}] \in \Autpsif{\fMTC_{\vv{0}}}$, we ask: does it even lift?
We see that, if $\res_{\fMTC_{\vv{0}}}( \Autpsif{\fMTC}) \neq \Autpsif{\fMTC_{\vv{0}}}$, there are $[\varphi^{(\vv{0})}] \notin  \res_{\fMTC_{\vv{0}}}( \Autpsif{\fMTC} )$, which cannot be extended to $\Autpsif{\fMTC}$.~\footnote{We do not know an example where $\res_{\fMTC_{\vv{0}}}( \Autpsif{\fMTC} ) \neq \Autpsif{\fMTC_{\vv{0}}}$ when topological symmetries are not space or time reflecting. However, they are ubiquitous when considering symmetries that reflect space or time. For example, when $\fMTC$ has a space or time reflecting topological symmetry, there are no nontrivial space or time reflecting topological symmetries of $\fMTC \boxtimes \ifo^{(\nu)}/A[\psi^{(1)}, \psi^{(2)}]$ for $\nu \neq 0$ or $8$, since such topological symmetries imply $c_{-} = 0 \text{ mod } 4$ for a MTC.}

Assuming $[\varphi^{(\vv{0})}] \in  \res_{\fMTC_{\vv{0}}}( \Autpsif{\fMTC})$, we can extend it to a $[\varphi] \in \Autpsif{\fMTC}$, and this extension is unique up to an element of $\ker(\res_{\fMTC_{\vv{0}}})$.
It is clear that $\mathbb{Z}_{2}^{\V} \lhd \ker(\res_{\fMTC_{\vv{0}}})$.
When $\widehat{{\bf A}}_{\vv{1}} \neq \varnothing$, we can show that $\ker(\res_{\fMTC_{\vv{0}}}) = \mathbb{Z}_{2}^{\V}$.
In fact, $\ker(\res_{\fMTC_{\vv{0}}}) = \mathbb{Z}_{2}^{\V}$ for all the examples that we know, so one might conjecture that it is generally true.
However, the possibility that $\ker(\res_{\fMTC_{\vv{0}}}) \neq \mathbb{Z}_{2}^{\V}$ when $\widehat{{\bf A}}_{\vv{1}} = \varnothing$ remains an open question.
In Appendix~\ref{app:kerres}, we prove that $\ker(\res_{\fMTC_{\vv{0}}}) = \mathbb{Z}_{2}^{\V}$ when $\widehat{{\bf A}}_{\vv{1}} \neq \varnothing$, and discuss the general properties of $\ker(\res_{\fMTC_{\vv{0}}})$ when $\widehat{{\bf A}}_{\vv{1}} = \varnothing$.

\subsection{Fermionic Global Symmetry Action}
\label{sec:symmetry_action}

We now consider a global fermionic symmetry of the physical system, i.e. the microscopic Hamiltonian, described by the group $\mathcal{G}^{\eff}$.
For any (closed) fermionic system, fermion parity conservation is treated as a symmetry, since the fermion parity operator $(-1)^{\bf F}$ always commutes with the Hamiltonian.
Moreover, it commutes with all other symmetries, since they cannot change the fermionic parity of the system.
In other words, fermionic parity generates a central subgroup $\mathbb{Z}_2^{\eff}$ of $\mathcal{G}^{\eff}$.
As such, $\mathcal{G}^{\eff}$ is a $\mathbb{Z}_2^{\eff}$ central extension of $G = \mathcal{G}^{\eff}/\mathbb{Z}_2^{\eff}$, which is sometimes referred to as the ``bosonic symmetry group'' and considered the standard symmetry of the system.
We denote the identity element of $G$ as ${\bf 0}$.

The possible central group extensions $\mathcal{G}$ of the group $G$ by $\mathbb{Z}_2$ are classified by the 2nd cohomology group $H^2(G,\mathbb{Z}_2)$.
In particular, a 2-cocycle $\central \in Z^2(G,\mathbb{Z}_2)$ may be used to specify the extension $\mathcal{G} = \mathbb{Z}_2 \times_{\central} G$, where the group elements can be written as $\mathpzc{g} = (\vv{x} , {\bf g})$, for $\mathpzc{g} \in \mathcal{G}$, $\vv{x} \in \mathbb{Z}_2$, and ${\bf g}\in G$, with group multiplication given by
\begin{align}
\label{eq:gfmultiplicationprime}
\mathpzc{gh}=
(\vv{x},{\bf g} ) (\vv{y}, {\bf h})  = (\vv{x} + \vv{y} + \central({\bf g,h}), {\bf gh} ).
\end{align}
Associativity of group multiplication imposes the 2-cocycle condition
\begin{align}
0 &= \cbd \central({\bf g,h,k}) \notag \\
&= \central({\bf h,k}) - \central({\bf gh,k}) + \central({\bf g,hk}) - \central({\bf g,h})
.
\end{align}
Taking the quotient by 2-coboundaries, i.e. cocycles of the form $\central({\bf g,h}) = \rm{d} \vviso({\bf g,h}) = \vviso({\bf h}) - \vviso({\bf gh}) +\vviso({\bf g})$, is recognizing the equivalence of group extensions that are related by relabeling the elements as $\mathpzc{g}' = (\vv{x} + \vviso({\bf g}) , {\bf g})$.

While these relabeling by coboundaries relate cocycles corresponding to equivalent group extensions, we will see that there is a finer level of distinction for symmetry fractionalization and defects, for which cocycles differing by coboundaries should be considered physically distinct.
We will see that this distinction corresponds to imposing the stricter $\gamma_{\psi_{\vv{0}}}({\bf g}) = +1$ condition, which maps to $\vviso({\bf g}) = 0$ in the context of group extensions.
In this case, when we refer to $\mathcal{G}^{\eff}$ and a $\mathbb{Z}_2^{\eff}$ central extension of $G$, it will mean with respect to a particular 2-cocycle $\central \in Z^2(G,\mathbb{Z}_2)$, not the corresponding 2-cohomology class $[\central] \in H^2(G,\mathbb{Z}_2)$.

In the case of bosonic topological phases, the action of the global symmetry group $G$ on the emergent topological degrees of freedom $\mathcal{B}$ is represented by a homomorphism
\begin{align}
[\rho]: G \to \text{Aut}{(\mathcal{B})}
\end{align}
from the global symmetry group to the topological symmetry group.
This means the group multiplication of $G$ is respected, i.e. $[\rho_{\bf g}] \cdot [\rho_{\bf h}] = [\rho_{\bf gh}]$, with equivalences by natural isomorphisms.

In the case of fermionic topological phases, we similarly define the action of the global symmetries on the emergent topological degrees of freedom encoded in the FMTC $\fMTC$ to be given by an action
\begin{align}
[\rho]: G \to \Autf{\fMTC}
,
\end{align}
with equivalences by fermionic natural isomorphisms.
While it is often suitable to require this fermionic symmetry action to be a homomorphism, a slightly less strict definition is generally permissible.
We define the fermionic symmetry action $[\rho]$ to be a ``$\Q$-projective action,'' by which we mean the group multiplication of $G$ is respected up to $[\Q]$, that is
\begin{equation}
[\Q]^{\phi({\bf g,h})} \cdot [\rho_{\bf g}] \cdot [\rho_{\bf h}] = [\rho_{\bf gh}]
,
\end{equation}
where associativity of group multiplication requires $\phi \in Z^{2}(G,\mathbb{Z}_2)$.
Of course, this is only different from an ordinary group action when $[\Q] \neq [\openone]$, i.e. when $\mathcal{A}_{1} = \varnothing$.
When $\Q$ is a fermionic natural isomorphism, the $\Q$ factors are simply absorbed into the fermionic natural isomorphisms of the classes $[\rho_{\bf g}]$ and this definition reduces to an ordinary homomorphism $[\rho]: G \to \Autf{\fMTC}$.

The $\Q$-projective generalization can be understood in the context of the fermionic symmetry group $\mathcal{G}^{\eff}$ being a $\mathbb{Z}_2^{\eff}$ central extension of $G$ in a manner nearly identical to projective representations.~\footnote{
The notion of a ``projective homomorphisms,'' generalizing the notion of projective representations, can be defined to be a map $p: G \to K$ between two groups such that $p ({\bf g}_1 \cdot {\bf g}_2) = q({\bf g}_1,{\bf g}_2) p({\bf g}_1) p({\bf g}_2)$, where $q({\bf g}_1,{\bf g}_2) \in C$ is in a central subgroup $C \lhd Z(K)$.
In order to be well-defined, we are required to have $q \in Z^2(G,C)$.
Introducing an equivalence relation of projective homomorphisms $h' \sim k \cdot h$ translates into an equivalence $q' = \cbd k \cdot q$.
This allows us to take the quotient by $B^2(G,C)$ to classify projective homomorphisms by cohomology classes $[q] \in H^2(G,C)$.
Similar to projective representations, one can lift a projective homomorphism $p$ to an ordinary homomorphism $h: \mathcal{G} \to K$ from a central extension $\mathcal{G} = C \times_{\varepsilon} G$ with $\varepsilon \in Z^2(G,C)$ when there is a homomorphism $r$ such that $r \circ \varepsilon =q$.
}
In particular, when $[\Q]$ is nontrivial, it must be viewed as acting nontrivially on the physical fermion.
As such, the projective factors of $\Q$ are only allowed if the projective action can be lifted to an ordinary group action (homomorphism) from $\mathcal{G}^{\eff}$ to $\Autf{\fMTC}$.
Consequently, if $\phi({\bf g,h}) \neq 0$ for any ${\bf g,h}\in G$, then we must have $[\phi] = [\central]$.
In other words, nontrivial $\Q$-projective actions and the fermionic symmetry group $\mathcal{G}^{\eff}$ are correlated.
If $\phi = 0$, then the fermionic symmetry action provides no direct constraint on $\central$.
(When we consider symmetry fractionalization in Sec.~\ref{sec:symmetry_fractionalization}, we will find the stronger constraint that $\phi = \central$ for any $\phi$, when $[\Q]$ is nontrivial.)

One might wonder why the symmetry action is not instead defined to be a homomorphism $[\rho]: G \to \Autpsi{\fMTC}$, as the projective $\Q$ factors would be absorbed into the equivalences by natural isomorphisms (which are allowed to have nontrivial $\gamma_{\psi_{\vv{0}}}$).
Indeed, this is the definition of fermionic action used in Ref.~\onlinecite{Galindo2017}.
Such a definition would miss important physical considerations stemming from the locality of the physical fermion and lead to different obstruction and classification results, as we will explain.
If one used this definition with the extra ad hoc constraint that only natural isomorphisms with $\gamma_{\psi_{\vv{0}}}=1$ are used for equivalences of the $\rho_{\bf g}$, i.e. the equivalence classes $[\rho_{\bf g}]$ are under fermionic natural isomorphism, but their multiplication is under all natural isomorphisms, it would be equivalent to the $\Q$-projective action defined above.

In more detail, we denote a specific auto-equivalence map as $\rho_{\bf g} \in [\rho_{\bf g}]$.
Equivalent representatives of $[\rho_{\bf g}]$ are given by $\check{\rho}_{\bf g} = \Upsilon_{\bf g} \circ \rho_{\bf g}$, where $\Upsilon_{\bf g}$ is a fermionic natural isomorphism.
In terms of representatives, $\rho_{\bf g}$ and $\rho_{\bf h}$ compose to $\rho_{\bf gh}$ up to a natural isomorphism $\kappa_{\bf g,h}$ or, equivalently, they compose $\Q$-projectively up to a fermionic natural isomorphism $\kappa^{\eff}_{\bf g,h} = \Q^{\phi({\bf g,h})} \kappa_{\bf g,h}$
\begin{align}
\label{eq:kappa}
\kappa_{{\bf g},{\bf h}} \circ \rho_{\bf g} \circ \rho_{\bf h} = \Q^{\phi({\bf g,h})} \circ \kappa^{\eff}_{{\bf g},{\bf h}} \circ \rho_{\bf g} \circ \rho_{\bf h}  = \rho_{\bf gh}.
\end{align}
(When $[\Q]$ is trivial, we require $\kappa_{{\bf g},{\bf h}}$ to be a fermionic natural isomorphism.)
Associativity of composing  $\rho_{\bf g} $, $\rho_{\bf h}$, and $\rho_{\bf k}$ implies
\begin{align}
\label{eq:natiso}
\kappa_{\bf g, hk}\circ \rho_{\bf g} \circ \kappa_{\bf h,k} \circ \rho_{\bf g}^{-1} &= \kappa_{\bf gh,k}\circ \kappa_{\bf g, h},
\\
\kappa^{\eff}_{\bf g, hk}\circ \rho_{\bf g} \circ \kappa^{\eff}_{\bf h,k} \circ \rho_{\bf g}^{-1} &= \kappa^{\eff}_{\bf gh,k}\circ \kappa^{\eff}_{\bf g, h}.
\end{align}
This equation will play a special role in defining the obstruction to symmetry fractionalization.
In this way, we can see that the property $\phi \in Z^2(G, \mathbb{Z}_{2})$ follows from the associativity of group multiplication in $\Autpsi{\fMTC}$ and the fact that $[\Q]$ is nontrivial in $\Autf{\fMTC}$.

We write the global symmetry action on topological charges as
\begin{align}
\label{eq:rho_charge}
\rho_{\bf g}(a_\vv{x}) &= \,^{\bf g} a_\vv{x},
\end{align}
and will also use the shorthand $\bar{\bf g} = {\bf g}^{-1}$.
The symmetry action must leave the fusion rules invariant
\begin{align}
\rho_{\bf g} \left( N_{a_{\vv{x}} b_{\vv{y}}}^{c_{\vv{x+y}}} \right) = N_{\,^{\bf g}a_{\vv{x}} \,^{\bf g}b_{\vv{y}}}^{\,^{\bf g}c_{\vv{x+y}}} = N_{a_{\vv{x}} b_{\vv{y}}}^{c_{\vv{x+y}}} .
\label{eq:rho_g_N}
\end{align}

The action on fusion vertex basis states takes the form
\begin{widetext}
\begin{align}
\label{eq:rho_vertex}
\rho_{\bf g} \left| a_\vv{x},b_\vv{y};c_{\vv{x}+\vv{y}} , \mu \right\rangle &= \sum_{\mu'}  \left[U_{\bf g}(\,^{\bf g}a_\vv{x},\,^{\bf g}b_\vv{y};\,^{\bf g}c_{\vv{x}+\vv{y}})\right]_{\mu \mu'} \left| \,^{\bf g}a_\vv{x},\,^{\bf g}b_\vv{y};\,^{\bf g}c_{\vv{x}+\vv{y}} , \mu' \right\rangle
,
\end{align}
where the $U_{\bf g}$-symbols are required to leave the basic data of $\fMTC$ invariant
\begin{eqnarray}
&& \rho_{\bf g} \left( \left[F_{d}^{abc}\right]_{(e,\alpha,\beta)(f,\mu,\nu)} \right) = \sum_{\alpha',\beta',\mu'\nu'} \left[U_{\bf g}(\,^{\bf g}a,\,^{\bf g}b ;\,^{\bf g}e )\right]_{\alpha \alpha'} \left[U_{\bf g}(\,^{\bf g}e, \,^{\bf g}c ;\,^{\bf g}d)\right]_{\beta \beta'} \left[F_{\,^{\bf g}d}^{ \,^{\bf g}a \,^{\bf g}b \,^{\bf g}c}\right]_{( ^{\bf g}e,\alpha',\beta')(^{\bf g}f,\mu',\nu')} \notag \\
&& \qquad \qquad \qquad \qquad \qquad \qquad  \times \left[U_{\bf g}(\,^{\bf g}b,\,^{\bf g}c ;\,^{\bf g}f)^{-1}\right]_{\mu' \mu} \left[U_{\bf g}(\,^{\bf g}a,\,^{\bf g}f ;\,^{\bf g}d)^{-1}\right]_{\nu' \nu} = \left[F_{d}^{abc}\right]_{(e,\alpha,\beta)(f,\mu,\nu)}   \qquad
\label{eq:rho_g_F} \\
&& \rho_{\bf g} \left( \left[{R}_{c}^{ab} \right]_{\mu \nu} \right)
= \sum_{\mu',\nu'} \left[U_{\bf g}(\,^{\bf g}b,\,^{\bf g}a ;\,^{\bf g}c)\right]_{\mu \mu'}  \left[R_{\,^{\bf g}c}^{\,^{\bf g}a \,^{\bf g}b} \right]_{\mu' \nu'} \left[U_{\bf g}(\,^{\bf g}a,\,^{\bf g}b ;\,^{\bf g}c)^{-1}\right]_{\nu' \nu}
=  \left[ R_{c}^{ab} \right]_{\mu \nu} ,
\label{eq:rho_g_R}
.
\end{eqnarray}
In the above expressions, brackets with subscripts denote matrix elements, and vorticity labels are left implicit.
These also give the corresponding
\begin{eqnarray}
&& \kappa_{\bf g,h} \left( \left| a,b;c,\mu \right\rangle \right) = \sum_{\nu} \left[ \kappa_{\bf g,h} (a,b;c) \right]_{\mu \nu} \left| a,b;c,\nu \right\rangle ,
\label{eq:kappa_vertex}
\\
&& \left[ \kappa_{\bf g,h} (a,b;c) \right]_{\mu \nu} = \sum_{\alpha , \beta} \left[U_{\bf g}(a,b;c)^{-1}\right]_{\mu \alpha}  \left[U_{\bf h}(\,^{\bf \bar g}a,\,^{\bf \bar g}b ;\,^{\bf \bar g}c)^{-1}\right]_{\alpha \beta}  \left[U_{\bf gh}(a,b;c)\right]_{\beta \nu}
\label{eq:kappa_U}
.
\end{eqnarray}
\end{widetext}

The canonical isomorphisms for the vacuum fixes
\begin{align}
\label{eq:CanU}
U_{\bf g}(\I_\vv{0},\I_\vv{0};\I_\vv{0}) = U_{\bf g}(a_\vv{x},\I_\vv{0};a_\vv{x}) = U_{\bf g}(\I_\vv{0},a_\vv{x};a_\vv{x}) =1.
\end{align}
and similarly for the physical fermion [Eq.~\eqref{eq:upsicond}] fixes
\begin{align}
U_{\bf g}(\psi_\vv{0},\psi_\vv{0};\I_\vv{0}) =1.
\end{align}
The constraints on $U_{\bf g}$-symbols from the other canonical gauge choices associated with gauge fixing $F$-symbols can also be imposed or allowed to be imposed automatically by the symmetry invariance of the basic data.
An important one for $\sigma$-type vortices is
\begin{align}
U_{\bf g}(\psi_{\vv{0}},a_{\sigma}; a_{\sigma}) =U_{\bf g}(a_{\sigma},\psi_{\vv{0}};a_{\sigma}) = \pm 1
.
\end{align}
Requiring $\rho_{\bf 0} = \mathds{1}$ fixes $[U_{\bf 0}(a_\vv{x},b_\vv{y};c_{\vv{x}+\vv{y}})]_{\mu \nu}  = \delta_{\mu \nu}$.

The fact that $\kappa_{{\bf g},{\bf h}}$ is a natural isomorphism implies it can be written as,
\begin{align}
\label{eq:kappaiso}
\left[ \kappa_{\bf g,h}(a_\vv{x},b_\vv{y};c_{\vv{x}+\vv{y}}) \right]_{\mu \nu} = \frac{\beta_{a_\vv{x}}({\bf g},{\bf h}) \beta_{b_\vv{y}}({\bf g},{\bf h})}{\beta_{c_{\vv{x}+\vv{y}}}({\bf g},{\bf h}) }\delta_{\mu \nu}
\end{align}
where the $\beta_{a_\vv{x}}({\bf g},{\bf h}) $ are complex phases satisfying the constraints
\begin{align}
\label{eq:canonical_beta}
\beta_{\I_{\vv{0}}}({\bf g},{\bf h}) = \beta_{\psi_{\vv{0}}}({\bf g},{\bf h})^{2} = 1
.
\end{align}

When $[\Q] = [\openone]$, $\kappa_{{\bf g},{\bf h}}$ can always be written as a fermionic natural isomorphism, and we can further require
\begin{align}
\label{eq:fermionic_beta}
\beta_{\psi_{\vv{0}}}({\bf g},{\bf h}) = 1
.
\end{align}
When $[\Q] \neq [\openone]$, $\beta_{\psi_{\vv{0}}}({\bf g},{\bf h})= (-1)^{\phi({\bf g},{\bf h})}$ cannot be removed by a gauge transformation.
We factor out $\Q$, when necessary, to write the fermionic natural isomorphisms $\kappa^{\eff}_{\bf g,h} = \Q^{\phi({\bf g,h})} \kappa_{\bf g,h}$, which can be decomposed into $\beta^{\eff}_{a_\vv{x}}({\bf g},{\bf h}) $ that satisfy
\begin{align}
\label{eq:fermionic_betaf}
\beta^{\eff}_{\I_{\vv{0}}}({\bf g},{\bf h}) = \beta^{\eff}_{\psi_{\vv{0}}}({\bf g},{\bf h}) = 1
.
\end{align}

Identical natural isomorphisms are found by shifting
\begin{align}
\label{eq:betafreedom}
\breve{\beta}_{a_\vv{x}}({\bf g,h}) = \nu_{a_\vv{x}}({\bf g},{\bf h}) \beta_{a_\vv{x}}({\bf g,h})
\end{align}
as long as $\nu_{a_\vv{x}}({\bf g},{\bf h}) \nu_{b_\vv{y}}({\bf g},{\bf h}) = \nu_{c_{\vv{x}+\vv{y}}}({\bf g},{\bf h})$ whenever $N_{a_\vv{x}b_\vv{y}}^{c_{\vv{x}+\vv{y}}} \neq 0 $. This is equivalent to the condition that
\begin{align}
\label{eq:nu_relation}
\nu_{a_\vv{x}}({\bf g},{\bf h}) = M^{\ast}_{a_\vv{x} \cohosub{v}({\bf g},{\bf h})}
\end{align}
for some $\coho{v}({\bf g},{\bf h}) \in C^2(G,\mathcal{A})$.
For fermionic natural isomorphisms, one can always choose a factorization with $\beta_{\psi_\vv{0}}({\bf g,h}) =1$.
Then, in order to leave $\beta_{\psi_\vv{0}}({\bf g,h}) =1$ fixed, the equivalences between factorizations of the fermionic natural isomorphisms must be further constrained to correspond to $\coho{v}({\bf g},{\bf h}) \in C^2(G,\mathcal{A}_{\vv{0}})$.
Since $\kappa_{\bf g,0} = \kappa_{\bf 0,g} = \openone$, we can use this gauge freedom to always choose
\begin{align}
\beta_{a_\vv{x}}({\bf 0,0}) = \beta_{a_\vv{x}}({\bf g,0}) = \beta_{a_\vv{x}}({\bf 0,g} ) = 1.
\end{align}

We recall that $[\Q]$ is nontrivial in $\Autf{\fMTC}$ if and only if $\mathcal{A}_{\vv{1}} = \varnothing$.
Thus, for $\Q$-projective actions, equivalent factorizations of $\kappa_{\bf g,h}$ into $\beta_{a_\vv{x}}({\bf g,h})$ are automatically constrained to have $\coho{v}({\bf g},{\bf h}) \in C^2(G,\mathcal{A}_{\vv{0}})$.
This means $\beta_{\psi_\vv{0}}({\bf g,h})$ is fixed, though possibly nontrivial, when $[\Q]$ is nontrivial.
Of course, this is necessary in order for the factorizations of $\kappa^{\eff}_{\bf g,h}$ into $\beta^{\eff}_{a_\vv{x}}({\bf g,h})$ to leave $\beta^{\eff}_{\psi_\vv{0}}({\bf g,h})=1$ fixed.
Moreover, we see that $\beta_{\psi_\vv{0}}({\bf g,h}) = (-1)^{\phi ({\bf g,h})} \in Z^2(G,\mathbb{Z}_{2})$ is a 2-cocycle when $[\Q]$ is nontrivial.

Since fermionic auto-equivalence maps related by fermionic natural isomorphisms are considered equivalent, the transformation of the symmetry action as
\begin{align}
\label{eq:symmetry_action_gauge}
\check{\rho}_{\bf g} = \Upsilon_{\bf g} \circ \rho_{\bf g}
\end{align}
is considered a gauge transformation.
This transforms the $U_{\bf g}$-symbols and $\beta_{a_\vv{x}}$ (up to their redundancy) as
\begin{align}
[\check{U}_{\bf g}(a_\vv{x},b_\vv{y};c_{\vv{x}+\vv{y}})]_{\mu \nu} &= \frac{\gamma_{a_\vv{x}}({\bf g}) \gamma_{b_\vv{y}}({\bf g}) }{\gamma_{c_{\vv{x}+\vv{y}}}({\bf g})} [U_{\bf g}(a_\vv{x},b_\vv{y};c_{\vv{x}+\vv{y}})]_{\mu \nu} ,\\
\label{eq:gauge2}
\check{\beta}_{a_\vv{x}}({\bf g},{\bf h} ) &= \frac{\gamma_{a_\vv{x}}({\bf gh} )}{\gamma_{{}^{\bf \bar{g}}a_\vv{x}}({\bf h}) \gamma_{a_\vv{x}}({\bf g}) }  \beta_{a_\vv{x}}({\bf g},{\bf h} ) ,
\end{align}
with the constraints
\begin{align}
\gamma_{\I_{\vv{0}}}({\bf g})  = \gamma_{\psi_{\vv{0}}}({\bf g})  = 1
.
\end{align}

The discussion for fermionic symmetry action on FMTCs above applies directly for SMTCs by restricting the vorticity labels to $\vv{x}=\vv{0}$.
We denote a fermionic symmetry action on a SMTC $\fMTC_{\vv{0}}$ as
\begin{align}
[\rho^{(\vv{0})}] : G \to \Autf{\fMTC_{\vv{0}}}
.
\end{align}
When $[\Q]$ is a nontrivial element of $\Autf{\fMTC_{\vv{0}}}$, i.e. when $\widehat{\bf A}_{\vv{1}} = \varnothing$, the action may be $\Q$-projective
\begin{align}
[\Q]^{\phi^{(\vv{0})}({\bf g,h})} \cdot [\rho^{(\vv{0})}_{\bf g}] \cdot [\rho^{(\vv{0})}_{\bf h}] = [\rho^{(\vv{0})}_{\bf gh}]
.
\end{align}
For SMTCs, the redundancy of factoring $\kappa^{(\vv{0})}_{{\bf g},{\bf h}}$ into $\beta^{(\vv{0})}_{a_\vv{0}}({\bf g},{\bf h})$ is given by $\nu_{a_\vv{0}}({\bf g},{\bf h}) = M^{\ast}_{a_\vv{0} \cohosub{v}({\bf g},{\bf h})}$ where $\coho{v}({\bf g},{\bf h}) \in C^2(G,\widehat{\bf A} )$ and $\widehat{\bf A}$ is the group formed by the superAbelian supersectors of topological charges of $\fMTC$ (recall that $\widehat{\bf A}$ is not necessarily equal to $\widehat{\mathcal{A}}$).
This is further restricted to $\coho{v}({\bf g},{\bf h}) \in C^2(G,\widehat{\bf A}_{\vv{0}} )$ to leave $\beta^{(\vv{0})}_{\psi_\vv{0}}({\bf g},{\bf h})$ fixed.

It is useful to analyze the extensions of a global symmetry action $[\rho^{(\vv{0})}]$ on $\fMTC_{\vv{0}}$ to global symmetry actions $[\rho]$ on $\fMTC$.
From the discussion in Sec.~\ref{sec:Topo_symmetry} on extensions of topological symmetries of $\fMTC_{\vv{0}}$ to those of $\fMTC$, we see that such an extension is impossible if $[\rho^{(\vv{0})}_{\bf g}] \notin \im (\res_{\fMTC_{\vv{0}}})$ for any ${\bf g}$, because it does not even lift.
However, even when $[\rho^{(\vv{0})}_{\bf g}] \in \res_{\fMTC_{\vv{0}}}( \Autf{\fMTC} )$ for all ${\bf g}$, there is still another possible obstruction to extending the symmetry action from $\fMTC_{\vv{0}}$ to $\fMTC$.
The obstruction and classification of extensions of the symmetry action will significantly involve $\ker(\res_{\fMTC_{\vv{0}}})$.
For all examples we know, $\ker(\res_{\fMTC_{\vv{0}}}) = \mathbb{Z}_{2}^{\V}$, and we have proven it is always true when $\widehat{{\bf A}}_{\vv{1}} \neq \varnothing$ in Appendix~\ref{app:kerres}.
Thus, we will assume this condition here and leave the analysis for more general $\ker(\res_{\fMTC_{\vv{0}}})$ in Appendix~\ref{app:rho_extension_general}.

An important aspect of the extended fermionic symmetry action $[\rho]$ is its $\Q$-projective structure, as determined by the exponents $\phi({\bf g,h})$, and we will analyze extensions with respect to a fixed $\Q$-projective structure, i.e. a fixed $\phi$.
When $[\V] \neq [\Q]$, the $\Q$-projective structure is determined by $[\rho^{(\vv{0})}]$, and the extension must match, requiring that $\phi = \phi^{(\vv{0})}$.
This is because $[\V] \neq [\Q]$ implies that either $[\Q] = [\openone]= [\openone^{(\vv{0})}]$, for which we simply have $\phi=\phi^{(\vv{0})} = 0$; or $[\Q] \neq [\openone^{(\vv{0})}]$.
On the other hand, $[\V] = [\Q]$ if and only if $\mathcal{A}_{\vv{1}} =\varnothing$ and $\widehat{\bf A}_{\vv{1}} \neq \varnothing$, i.e. there are only $\sigma$-type vortices, at least one of which is superAbelian.
In this case, $[\Q]$ is trivial in $\Autf{\fMTC_{\vv{0}}} $, but nontrivial in $\Autf{\fMTC}$, and $\ker (\res_{\fMTC_{\vv{0}}} )= \mathbb{Z}_{2}^{\Q}$.
Thus, when $[\V] = [\Q]$, we have $\phi^{(\vv{0})}=0$, but can potentially have nontrivial $\Q$-projective factors $\phi$ for the extension.

To characterize the obstruction, we first define an arbitrary section on the image of the restriction, that is $s: \res_{\fMTC_{\vv{0}}}( \Autf{\fMTC} ) \to \Autf{\fMTC}$, such that $\res_{\fMTC_{\vv{0}}} \circ s ([\varphi^{(\vv{0})}]) = [\varphi^{(\vv{0})}]$.
Then we define
\begin{align}
\label{eq:Orho}
O^{\rho}({\bf g,h}) &= [\Q]^{\phi({\bf g,h}) } s[ \rho^{(\vv{0})}_{\bf gh} ] \cdot s[ \rho^{(\vv{0})}_{\bf h} ]^{-1} \cdot s[ \rho^{(\vv{0})}_{\bf g} ]^{-1}
.
\end{align}
Here, $\phi$ is fixed and will constrain the $\Q$-projective structure of the possible extensions.

Since $[\rho^{(\vv{0})}]$ is a fermionic symmetry action and the restriction map is a homomorphism, it follows that $\res_{\fMTC_{\vv{0}}}( O^{\rho}({\bf g,h}) )= [\openone^{(\vv{0})}]$, and hence $O^{\rho}({\bf g,h}) \in \ker (\res_{\fMTC_{\vv{0}}} )= \mathbb{Z}_{2}^{\V}$.
We can also see that
\begin{align}
\cbd O^{\rho}({\bf g,h,k}) &= O^{\rho}({\bf h,k})  \cdot O^{\rho}({\bf gh,k})^{-1} \notag \\
 & \qquad \cdot O^{\rho}({\bf g,hk}) \cdot O^{\rho}({\bf g,h})^{-1}
\notag \\
&= [\openone]
,
\end{align}
where we used the fact that $\mathbb{Z}_{2}^{\V}$ is central and $\cbd \phi =0$.

If we had made a different arbitrary choice of section, $\grave{s}: \res_{\fMTC_{\vv{0}}}( \Autf{\fMTC} ) \to \Autf{\fMTC}$, we would have $\grave{s}[ \rho^{(\vv{0})}_{\bf g} ] = \varsigma({\bf g})^{-1} \cdot s[ \rho^{(\vv{0})}_{\bf g} ]$ for some $\varsigma({\bf g}) \in C^{1}(G , \mathbb{Z}_{2}^{\V} )$.
This different choice of section would yield
\begin{align}
\grave{O}^{\rho}({\bf g,h}) &= [\Q]^{\phi({\bf g,h}) } \grave{s}[ \rho^{(\vv{0})}_{\bf gh} ] \cdot \grave{s}[ \rho^{(\vv{0})}_{\bf h} ]^{-1} \cdot \grave{s}[ \rho^{(\vv{0})}_{\bf g} ]^{-1}
\notag \\
&= [\Q]^{\phi({\bf g,h}) } s[ \rho^{(\vv{0})}_{\bf gh} ] \cdot s[ \rho^{(\vv{0})}_{\bf h} ]^{-1} \cdot s[ \rho^{(\vv{0})}_{\bf g} ]^{-1} \notag \\
& \qquad \cdot  \varsigma({\bf h}) \cdot \varsigma({\bf gh})^{-1} \cdot \varsigma({\bf g})
\notag \\
&= {O}^{\rho}({\bf g,h}) \cdot \cbd \varsigma({\bf g,h})
.
\end{align}
Since the arbitrary choice of section should not matter in the definition of an obstruction, we treat these as equivalent definitions, and define the obstruction to be the equivalence class under multiplication with 2-coboundaries $B^{2}(G , \mathbb{Z}_{2}^{\V} )$.
Thus, we have
\begin{align}
\label{eq:O2obs}
[O^{\rho}] \in H^{2}(G , \mathbb{Z}_{2}^{\V} )
,
\end{align}
which is independent of the choice of section.

In order to see that $[O^{\rho}]$ represents an obstruction to extending the fermionic symmetry action for a fixed $\phi$, we note that if $[\rho_{\bf g}]$ is an extension of $[\rho^{(\vv{0})}_{\bf g}]$, then it can be written as $[\rho_{\bf g}] = \xi({\bf g}) \cdot  s[ \rho^{(\vv{0})}_{\bf g} ]$ for some $\xi({\bf g}) \in C^1(G,\mathbb{Z}_{2}^{\V} )$.
Then we see the condition that $[\rho]:G \to \Autf{\fMTC}$ is a $\Q$-projective homomorphism with $\phi$ translates into the condition
\begin{align}
O^{\rho}({\bf g,h}) &= \xi({\bf h}) \cdot  \xi^{-1}({\bf gh}) \cdot \xi({\bf g}) = \cbd \xi ({\bf g, h})
.
\end{align}
This shows $[O^{\rho}]$ is indeed an obstruction to extending the symmetry action $[\rho^{(\vv{0})}]$ for a fixed $\phi$, as it is impossible to satisfy this equation unless $O^{\rho} \in B^{2}(G , \mathbb{Z}_{2}^{\V})$ is a 2-coboundary, that is unless $[O^{\rho}] = [\openone]$.

It also follows from this analysis that, for any $[\rho]$ that is an extension of $[\rho^{(\vv{0})}]$ with $\Q$-projective structure $\phi$, we can define another valid extension by $[\rho'_{\bf g}] = [\varphi _{\bf g}] \cdot [\rho_{\bf g}]$, where $[\varphi _{\bf g}] \in H^{1}(G, \mathbb{Z}_{2}^{\V})$.
Thus, the extensions of $[\rho^{(\vv{0})}] $ to fermionic topological symmetries with $\Q$-projective structure $\phi$ on $\fMTC$ are torsorially classified by
\begin{align}
\label{eq:H1rhotorsor}
H^{1}(G, \mathbb{Z}_{2}^{\V})
.
\end{align}

In Ref.~\onlinecite{Bark2021cascade}, it was conjectured that $[O^{\rho}]$ (which they define for the case $\phi=\vv{0}$) takes the same value for all 16 modular extensions of $\fMTC_{\vv{0}}$.
Assuming that $\ker(\res_{\fMTC_{\vv{0}}}) = \mathbb{Z}_{2}^{\V}$, we can use stacking to prove that, for modular extensions $\fMTC$ and $\fMTC' = \fMTC \fprod \ifo^{(\nu)}$, we have
\begin{align}
\label{eq:Orho_nu_relation}
[O^{\rho}]' &= [\V]^{\, \nu \cdot [\phi]} \cdot [O^{\rho}]
.
\end{align}
This shows that $[O^{\rho}]$ is the same for modular extensions that differ by stacking with $\ifo^{(\nu)}$ for $\nu$ odd if and only if $\phi \in B^2(G,\mathbb{Z}_{2})$.
It also shows that $[O^{\rho}]$ is always the same for modular extensions that differ by stacking with $\ifo^{(\nu)}$ for $\nu$ even.

In order to prove Eq.~\eqref{eq:Orho_nu_relation}, we only need to analyze the case of stacking $\fMTC$ with $\ifo^{(1)}$.
(The case of theories related by $\nu$ even can be shown much more easily and in full detail, as we will show in Sec.~\ref{sec:symmetry_frac_extension}.
It is also straightforward to remove the assumption that $\ker(\res_{\fMTC_{\vv{0}}}) = \mathbb{Z}_{2}^{\V}$ for $\nu$ even, since stacking with $\ifo^{(2)}$ establishes a simple isomorphism between $\Autf{\fMTC}$ and $\Autf{\fMTC'}$.)
For this, we consider a general topological symmetry $\varphi$ on $\fMTC$ and trivial symmetry $\openone$ on $\ifo^{(1)}$, and compute the resulting topological symmetry $\zig{\varphi}$ on $\zig{\fMTC} = \fMTC \fprod \ifo^{(1)}$ induced by stacking, using the condensation formalism reviewed in Appendix~\ref{app:condensation}.

Since we assume $\ker(\res_{\fMTC_{\vv{0}}}) = \mathbb{Z}_{2}^{\V}$, we can restrict our attention to $\varphi(a_{v})$ for $v$-type vortices and $U_{\varphi}( a_{\sigma}, \psi_{\vv{0}}; a_{\sigma} )$ for $\sigma$-type vortices, as these will fully specify whether $O^{\rho}({\bf g,h}) = [\openone]$ or $[\V]$.
It will be convenient for this analysis to introduce the notation $\hat{a}_{v}^{r}$ with $r=\pm$ for the two charges $a_{v}$ and $[\psi a]_{v}$ in a $v$-type supersector, with an arbitrary choice of which is labeled $+$ or $-$.
It will also be useful to define the quantity $n_{\varphi}$ by
\begin{align}
\varphi \left( \hat{a}_{v}^{r}  \right) = \psi_{\vv{0}}^{n_{\varphi} (\hat{a}_{v})} \otimes \widehat{\varphi} \left( \hat{a}_{v} \right)^{r}
,
\end{align}
which tracks the action of $\varphi$ within $v$-type vortex supersectors with respect to this arbitrary choice.
($n_{\varphi} (\hat{a}_{v})$ is independent of $r$, but depends on $\hat{a}_{v}$.)

Under stacking with $\ifo^{(1)}$, $v$-type vortices of $\fMTC$ become $\sigma$-type vortices of $\fMTC'$, and vice-versa, while the quasiparticle charges are left unchanged.
More specifically, we can write the vortex charges of $\zig{\fMTC}$ in terms of those of $\fMTC$ and $\ifo^{(1)}$, using representative choices.
The post-condensation $\sigma$-type vortices are labeled by
\begin{align}
\zig{\hat{a}_{v}} = (\hat{a}_{v}^{r} , \sigma_{\vv{1}})
.
\end{align}
For the post-condensation $v$-type vortices, we represent these diagrammatically by
\begin{align}
\zig{a_{\sigma}}^{r} &= \PBfb \,\, +r \PBfa \,\,
.
\end{align}
We caution the reader that we are using the charge labels of $\fMTC$ to represent the charges of $\zig{\fMTC}$, so the $v$-type charges of $\zig{\fMTC}$ are labeled using $\sigma$-type charges of $\fMTC$ with $\zig{}$ over them, and $\sigma$-type charges of $\zig{\fMTC}$ are labeled using $v$-type charge of $\fMTC$ with $\zig{}$ over them.

In this way, we can compute the symmetry action on post-condensation $v$-type vortices to be
\begin{align}
\zig{\varphi}\left( \zig{a_{\sigma}}^{r} \right) &= \PBfd \,\, +r U_{\varphi}( \varphi( a_{\sigma}), \psi_{\vv{0}}; \varphi( a_{\sigma}) ) \PBfc \,\,
\notag \\
&= \psi_{\vv{0}}^{\zig{n}_{\zig{\varphi}}(\widehat{\zig{a_{\sigma}}})} \otimes \zig{ \varphi\left( a_{\sigma} \right)}^{r}
,
\end{align}
where we find
\begin{align}
\zig{n}_{\zig{\varphi}} (\widehat{\zig{a_{\sigma}}}) &= \frac{ 1 - U_{\varphi}(\varphi( a_{\sigma}), \psi_{\vv{0}}; \varphi( a_{\sigma}) ) }{2}
.
\end{align}

\begin{widetext}
For the post-condensation $\sigma$-type vortices, we have $\zig{\varphi}\left( \zig{\hat{a}_{v}} \right) = \zig{ \widehat{\varphi}\left( \hat{a}_{v}  \right)}$.
Inspecting the action on the vertex states, we find
\begin{align}
\zig{\varphi}\left( \PBfe \right) &= \varphi \left( \PBff \right)
=\PBfg
= (-1)^{n_{\varphi} (\hat{a}_{v}) } \PBfh
,
\label{eq:rhozig_a}
\end{align}
which gives the relation
\begin{align}
\zig{U}_{\zig{\varphi}}\left( \zig{\varphi}\left( \zig{\hat{a}_{v}} \right) , \psi_{\vv{0}}; \zig{\varphi}\left( \zig{\hat{a}_{v}} \right) \right) &= (-1)^{n_{\varphi} (\hat{a}_{v})}
.
\end{align}

Thus, we find that $n_{\varphi} (\hat{a}_{v})$ and $U_{\varphi}(\varphi( a_{\sigma}), \psi_{\vv{0}}; \varphi( a_{\sigma}) )$ of $\fMTC$ with $\varphi$ are respectively correlated with $\zig{U}_{\zig{\varphi}}\left( \zig{\varphi}\left( \zig{\hat{a}_{v}} \right) , \psi_{\vv{0}}; \zig{\varphi} \left( \zig{\hat{a}_{v}} \right) \right)$ and $\zig{n}_{\zig{\varphi}} (\widehat{\zig{a_{\sigma}}})$ of $\zig{\fMTC}$ with $\zig{\varphi}$.

We now use this construction to define a choice of section $\zig{s}$ on $\zig{\fMTC}$ for every $[\varphi^{(0)}]$ by
\begin{align}
\zig{s}[\varphi^{(0)}] &= \zig{ s[\varphi^{(0)}] }
.
\end{align}
With this choice, we have
\begin{align}
O^{\rho} ({\bf g,h}) &= [\V]^{ N_{O^{\rho} ({\bf g,h})} }
,\\
N_{O^{\rho} ({\bf g,h})} &= n_{ s[ \rho^{(\vv{0})}_{\bf gh} ]} (\hat{a}_{v}) + n_{ s[ \rho^{(\vv{0})}_{\bf h} ]} (\hat{a}_{v}) + n_{ s[ \rho^{(\vv{0})}_{\bf g} ]} ( \,^{\bf h} \hat{a}_{v})  \\
&= \phi({\bf g,h}) + \frac{ 1 - U_{s[ \rho^{(\vv{0})}_{\bf gh} ]}( s[ \rho^{(\vv{0})}_{\bf gh} ](a_{\sigma}), \psi_{\vv{0}};  s[ \rho^{(\vv{0})}_{\bf gh} ](a_{\sigma}) ) }{2}
+\frac{ 1 - U_{s[ \rho^{(\vv{0})}_{\bf h} ]}( s[ \rho^{(\vv{0})}_{\bf h} ](a_{\sigma}), \psi_{\vv{0}};  s[ \rho^{(\vv{0})}_{\bf h} ](a_{\sigma}) ) }{2}
\notag \\
&
\qquad \qquad \qquad + \frac{ 1 - U_{s[ \rho^{(\vv{0})}_{\bf g} ]}( s[ \rho^{(\vv{0})}_{\bf g} ] (  \,^{\bf h} a_{\sigma}), \psi_{\vv{0}};  s[ \rho^{(\vv{0})}_{\bf g} ] ( \,^{\bf h} a_{\sigma}) ) }{2}
.
\end{align}
Similarly defining $\zig{N}_{\zig{O}^{\zig{\rho}}}$ for $\zig{O}^{\zig{\rho}}$, the above correlation of $\varphi$ and $\zig{\varphi}$ quantities yields the relations
\begin{align}
\zig{N}_{\zig{O}^{\zig{\rho}}({\bf g,h})} & = \phi({\bf g,h}) + N_{O^{\rho} ({\bf g,h})}
,
\\
\zig{O}^{\zig{\rho}} &= [\V]^{\phi} \cdot O^{\rho}
.
\end{align}
This proves Eq.~\eqref{eq:Orho_nu_relation}.
\end{widetext}

\subsection{\texorpdfstring{$H^3_{[\rho]}(G,\mathcal{A})$}{H3GA} Invariance Class of the Fermionic Symmetry Action}
\label{sec:O3invclass}

The fermionic symmetry action $[\rho]$ on $\fMTC$ has an associated invariant valued in $H^3_{[\rho]}(G,\mathcal{A})$.
In the next section, we explain why this invariant is an obstruction to realizing symmetry fractionalization.
Physically, the symmetry fractionalization obstruction indicates that the corresponding symmetry action on the emergent topological degrees of freedom is incompatible with the global symmetry action on the physical Hilbert space in a locality preserving manner.
In mathematical parlance, the obstruction signifies a failure for the fermionic symmetry action $[\rho]: G \to \Autf{\fMTC}$ to lift to a categorical action $\underline{\rho}: \underline{G} \to \underline{\Autf{\fMTC}}$.
The fermionic setting mirrors the bosonic one, except that we incorporate the restricted class of fermionic topological symmetries and fermionic natural isomorphisms.
The discussion follows a streamlined version of that presented in Ref.~\onlinecite{Bark2019}.

We begin by defining
\begin{align}
\label{eq:obstructiontool}
\Omega_{a_\vv{x}}({\bf g},{\bf h},{\bf k}) = \frac{\beta_{\rho_{\bf g}^{-1}(a_\vv{x})}({\bf h},{\bf k})\beta_{a_\vv{x}}({\bf g},{\bf hk})}{\beta_{a_\vv{x}}({\bf gh},{\bf k })\beta_{a_\vv{x}}({\bf g},{\bf h})}.
\end{align}
Eq.~\eqref{eq:natiso} implies
\begin{align}
\label{eq:omegachar}
\Omega_{a_\vv{x}}({\bf g,h,k} ) \Omega_{b_\vv{y}}({\bf g,h,k}) = \Omega_{c_{\vv{x}+\vv{y}}}({\bf g,h,k})
\end{align}
whenever $N_{a_\vv{x}b_\vv{y}}^{c_{\vv{x}+\vv{y}}} \neq 0 $.

Using the relation through the characters of the fusion algebra (reviewed in Sec.~\ref{sec:BTC}), this indicates we can write
\begin{align}
\label{eq:cohoO}
\Omega_{a_\vv{x}}({\bf g,h,k} ) = M_{a_\vv{x} \cohosub{O}({\bf g,h,k})}^*
\end{align}
for some $\coho{O} \in C^3(G,\mathcal{A})$.

One can directly show from the definition that $\Omega_{a_\vv{x}}({\bf g},{\bf h},{\bf k})$ satisfies the cocycle-like condition
\begin{align}
\label{eq:cocylelike}
\frac{
\Omega_{\rho_{\bf g}^{-1}(a_\vv{x})}({\bf h},{\bf k},{\bf l}) \Omega_{a_\vv{x}}({\bf g},{\bf hk},{\bf l} ) \Omega_{a_\vv{x}}({\bf g},{\bf h},{\bf k})}{\Omega_{a_\vv{x}}({\bf gh},{\bf k},{\bf l})\Omega_{a_\vv{x}}({\bf g},{\bf h},{\bf kl} )} = 1.
\end{align}
Translating through Eq.~(\ref{eq:cohoO}), this can equivalently be written as
\begin{align}
\cbd \coho{O} = \I_\vv{0},
\end{align}
where the boundary operator here includes the action on $\mathcal{A}$ induced by the symmetry action $[\rho]$ on topological charges.
Thus, $\coho{O} \in Z^3_{\rho}(G,\mathcal{A})$.

Using Eq.~\eqref{eq:gauge2}, it is straightforward to see that the symmetry action gauge transformation gives $\check{\coho{O}} = \coho{O}$.
In other words, $\coho{O}$ is independent of the choice of representative $\rho \in [\rho]$ used for the symmetry action.

Finally, as described by Eqs.~\eqref{eq:betafreedom} and \eqref{eq:nu_relation}, we have the freedom to redefine $\beta_{a_\vv{x}}({\bf g},{\bf h})$ to $\breve{\beta}_{a_\vv{x}}({\bf g,h} ) =  \nu_{a_\vv{x}}({\bf g,h}) \beta_{a_\vv{x}}({\bf g,h})$ where
\begin{align}
\nu_{a_\vv{x}}({\bf g},{\bf h}) = M_{a_\vv{x},\cohosub{v}({\bf g},{\bf h} )}^*
\end{align}
for some $\coho{v} \in C^2(G,\mathcal{A})$.
Under such redefinitions, one finds $\breve{\coho{O}} = \coho{O} \otimes \cbd \coho{v}$.
Thus, one should treat $\coho{O}$ as equivalent when related by 3-coboundaries $\cbd \coho{v} \in B^3_{[\rho]}(G,\mathcal{A})$, and define the equivalence class
\begin{align}
\label{eq:obstruction3}
[\coho{O}] \in H^3_{[\rho]}(G,\mathcal{A}).
\end{align}

Recall that $[\Q]$ is nontrivial if and only if $\mathcal{A}_{\vv{1}} = \varnothing$.
In this case, $[\coho{O}] \in H^3_{[\rho]}(G,\mathcal{A}_{\vv{0}})$.
When we have a $\Q$-projective symmetry action (which is only the case when $[\Q]$ is nontrivial), since $\kappa^{\eff}_{\bf g,h} = \Q^{\phi({\bf g,h})} \kappa_{\bf g,h}$, we can let $\beta^{\eff}_{a_\vv{x}}({\bf g},{\bf h}) = (-1)^{\phi({\bf g,h}) \delta_{a_{\vv{x}} \psi_{\vv{0}}}} \beta_{a_\vv{x}}({\bf g},{\bf h})$.
Then plugging this into Eq.~(\ref{eq:obstructiontool}), we find that $\Omega^{\eff}_{a_\vv{x}}({\bf g},{\bf h},{\bf k}) = \Omega_{a_\vv{x}}({\bf g},{\bf h},{\bf k})$ and $\coho{O}^{\eff} = \coho{O}$, since $\phi$ is a 2-cocycle.
The fact that $\beta^{\eff}_{\psi_\vv{0}}({\bf g},{\bf h}) =1$ implies that $\Omega^{\eff}_{\psi_\vv{0}}({\bf g},{\bf h},{\bf k}) = \Omega_{\psi_\vv{0}}({\bf g},{\bf h},{\bf k}) = 1$.

Repeating the analysis for a fermionic symmetry action $[\rho^{(\vv{0})}] : G \to \Autf{\fMTC_{\vv{0}}}$, we replace Abelian topological charges with superAbelian supersectors of topological charge and recall that $[\rho^{(\vv{0})}]$ uniquely determines $[\widehat{\rho}]$ to similarly define the class
\begin{align}
\label{eq:obstruction3_SMTC}
[\coho{O}^{(\vv{0})}] \in H^3_{[\widehat{\rho}]}(G,\widehat{\bf A}).
\end{align}
When $[\rho^{(\vv{0})}]$ extends to a symmetry action $[\rho]$ on $\fMTC$, we see that $[\coho{O}^{(\vv{0})}] = [\widehat{\coho{O}}]$ for the corresponding $[\coho{O}]$ projected onto supersectors.
Similarly, when $[\Q]$ is a nontrivial element of $\Autf{\fMTC_{\vv{0}}}$, $\widehat{\bf A}_{\vv{1}} = \varnothing$, and $[\coho{O}^{(\vv{0})}] \in H^3_{[\widehat{\rho}^{(\vv{0})}]}(G,\widehat{\bf A}_{\vv{0}})$.

We note that modifying the symmetry action by $\Q$ does not change $[\coho{O}]$ or $[\coho{O}^{(\vv{0})}]$.
To see this in detail, we consider modifying a symmetry action $\rho_{\bf g}$ to
\begin{align}
\rho'_{\bf g} = \Q^{\rr({\bf g})} \rho_{\bf g}
,
\end{align}
where $\rr \in C^{1}(G ,\mathbb{Z}_{2})$ is not required to be a homomorphism.
Since $\Q^2 = \openone$, this yields the modified
\begin{align}
\kappa'_{\bf g,h} &= Q^{\cbd \rr({\bf g,h})} \kappa_{\bf g,h} ,
\end{align}
which can be factorized into
\begin{align}
\beta'_{a_{\vv{x}}}({\bf g,h}) &= (-1)^{\cbd \rr({\bf g,h}) \delta_{a_{\vv{x}} \psi_{\vv{0}}}}  \beta_{a_{\vv{x}}}({\bf g,h})
.
\end{align}
(If $r$ were a homomorphism, we would have $\kappa'_{\bf g,h} = \kappa_{\bf g,h}$ and $\beta'_{a_{\vv{x}}}({\bf g,h}) = \beta_{a_{\vv{x}}}({\bf g,h})$.)
Since $\cbd \rr$ is a coboundary, it follows that $\Omega'_{a_\vv{x}}({\bf g,h,k} ) =\Omega_{a_\vv{x}}({\bf g,h,k} )$ and $\coho{O}'=\coho{O}$.
We note that $\beta'_{\psi_{\vv{0}}}({\bf g,h}) =  (-1)^{\cbd \rr({\bf g,h})} \beta_{\psi_{\vv{0}}}({\bf g,h})$.
However, when $\mathcal{A}_{\vv{1}} \neq \varnothing$, we can alternatively use $\breve{\beta}'_{a_{\vv{x}}}({\bf g,h}) = M_{a_{\vv{x}} z_{\vv{1}} }^{\cbd \rr({\bf g,h})} (-1)^{\cbd \rr({\bf g,h}) \delta_{a_{\vv{x}} \psi_{\vv{0}}}}  \beta_{a_{\vv{x}}}({\bf g,h})$ for some $z_{\vv{1}} \in \mathcal{A}_{\vv{1}}$, which will have $\breve{\beta}'_{\psi_{\vv{0}}}({\bf g,h}) =  \beta_{\psi_{\vv{0}}}({\bf g,h})$.

Modifying the symmetry action by $\V$ clearly does not affect $\coho{O}^{(\vv{0})}$, but can change $\coho{O}$ in a more complicated way.
We will show that it leaves $[\coho{O}]$ unchanged for $\widehat{\bf A}_{\vv{1}} \neq \varnothing$.
To see this in more detail, we consider modifying a symmetry action $\rho_{\bf g}$ to
\begin{align}
\rho'_{\bf g} = \V^{\pi({\bf g})} \rho_{\bf g}
,
\end{align}
where $\pi : G \to \mathbb{Z}_{2}$ is a homomorphism.
From Eqs.~(\ref{eq:kappaV}) and \eqref{eq:Vcommute}, we see that this yields the modified
\begin{align}
\kappa'_{\bf g,h} &= \Lambda^{\pi({\bf gh})}  \circ Y_{\bf g}^{-\pi({\bf h}) (1-2 \pi({\bf g})) } \circ \V^{-2 \pi({\bf g})\pi({\bf h})} \circ 
\kappa_{\bf g,h}
,
\end{align}
where $\Lambda$ and $Y_{\bf g}$ are fermionic natural isomorphisms specified by the respective factors
\begin{align}
\lambda_{a_{\vv{x}}} &= \frac{\beta_{[\psi^{\vv{x}} a]_{\vv{x}}}({\bf g,h})}{\beta_{\psi_{\vv{0}}^{\vv{x}}}({\bf g,h}) \beta_{a_{\vv{x}}}({\bf g,h})}
,
\\
y_{a_{\vv{x}}} &= U_{\bf g}(\psi_{\vv{0}}^{\vv{x}} ,  a_{\vv{x}} ; [\psi^{\vv{x}} a]_{\vv{x}} ) 
.
\end{align} 
Note that $y_{[\psi^{\vv{x}}a]_{\vv{x}}} = y_{a_{\vv{x}}}^{-1}$ due to the fermionic canonical gauge choice fixing $F^{\psi_{\vv{0}} \psi_{\vv{0}} a_{\vv{x}}} =1$.
With this, the natural isomorphism $\kappa'_{\bf g,h}$ can be factorized as
\begin{widetext}
\begin{align}
\beta'_{a_{\vv{x}}}({\bf g,h}) &= 
\left( \frac{\beta_{[\psi^{\vv{x}} a]_{\vv{x}}}({\bf g,h})}{\beta_{\psi_{\vv{0}}^{\vv{x}}}({\bf g,h}) \beta_{a_{\vv{x}}}({\bf g,h})} \right)^{\pi({\bf gh})} 
U_{\bf g}(\psi_{\vv{0}}^{\vv{x}} ,  a_{\vv{x}} ; [\psi^{\vv{x}} a]_{\vv{x}} )^{-\pi({\bf h}) (1-2 \pi({\bf g})) } 
\, i^{\vv{x}\cdot \pi({\bf g}) \cdot \pi({\bf h}) } 
\, \beta_{a_{\vv{x}}}({\bf g,h})
.
\end{align}
\end{widetext}
Clearly, we have $\beta'_{a_{\vv{0}}}({\bf g,h}) =\beta_{a_{\vv{0}}}({\bf g,h})$.
For vortices, it is useful to consider the two types separately.

For $v$-type vortices, it is always possible to make a symmetry action gauge choice such that $\widecheck{U}_{\bf g}(\psi_{\vv{0}} ,  a_{v} ; [\psi a]_{v} ) =1$, using a gauge transformation with $\gamma_{[\psi a]_{v}} = U_{\bf g}(\psi_{\vv{0}},  a_{v} ; [\psi a]_{v} ) \gamma_{a_{v}}$.
This gauge choice also give $\check{\beta}_{\psi_{\vv{0}}}({\bf g,h}) \check{\beta}_{a_{v}}({\bf g,h}) = \check{\beta}_{[\psi a]_{v}}({\bf g,h})$.
This convenient gauge choice yields
\begin{align}
\check{\beta}'_{a_{v}}({\bf g,h}) &= i^{\pi({\bf g}) \cdot \pi({\bf h}) } \, \check{\beta}_{a_{v}}({\bf g,h})
.
\end{align}
Using this, and noting that $\widecheck{\Omega}_{a_{\vv{x}}} = \Omega_{a_{\vv{x}}}$ is independent of symmetry action gauge choices, we find
\begin{align}
\Omega'_{a_{v}}({\bf g,h,k}) &= \beta_{\psi_{\vv{0}}}({\bf h,k})^{\pi({\bf g})} \, \Omega_{a_{v}}({\bf g,h,k})
.
\end{align}
Here, we used the fact that $\pi$ is a homomorphism.
(We emphasize that $\Omega'$ is defined using $\beta'$ and $\rho'$.)

For $\sigma$-type vortices, $U_{\bf g}(\psi_{\vv{0}} ,  a_{\sigma} ; a_{\sigma} ) = \pm 1$ is invariant under fermionic symmetry action gauge transformations (and is invariant under vertex basis gauge transformations when $^{\bf g} a_{\sigma} =a_{\sigma} $), so we cannot make a similar gauge choice.
However, there is some simplification in this case to
\begin{align}
\beta'_{a_{\sigma}}({\bf g,h}) =& 
\beta_{\psi_{\vv{0}}}({\bf g,h})^{\pi({\bf gh})} 
U_{\bf g}(\psi_{\vv{0}} ,  a_{\sigma} ; a_{\sigma} )^{\pi({\bf h})} 
\notag \\
& \, \times i^{\pi({\bf g}) \cdot \pi({\bf h}) } 
\, \beta_{a_{\sigma}}({\bf g,h})
.
\end{align}
When there are $\sigma$-type vortices, $\mathcal{A}_1 = \varnothing$ and $[\Q]\neq [\openone]$, so $\beta_{\psi_{\vv{0}}}({\bf g,h}) = (-1)^{\phi({\bf g,h})}$ is a 2-cocycle describing the $\Q$-projective nature of the symmetry action.
Using these, together with Eqs.~\eqref{eq:kappa_U} and \eqref{eq:kappaiso}, we again find
\begin{align}
\Omega'_{a_{\sigma}}({\bf g,h,k}) &= \beta_{\psi_{\vv{0}}}({\bf h,k})^{\pi({\bf g})} \, \Omega_{a_{\sigma}}({\bf g,h,k})
.
\end{align}
Remarkably, the same modification arises in very different manners for the two types of vortices.

Writing $\beta_{\psi_{\vv{0}}}({\bf g,h}) = (-1)^{\mathsf{b}({\bf g,h})}$ for $\mathsf{b} \in C^2(G,\mathbb{Z}_{2})$, these results combine to give
\begin{align}
\Omega'_{a_{\vv{x}}}({\bf g,h,k}) &= \beta_{\psi_{\vv{0}}}({\bf h,k})^{\vv{x} \cdot \pi({\bf g})} \, \Omega_{a_{\vv{x}}}({\bf g,h,k})
\notag \\
&= (-1)^{\vv{x} \cdot  \pi({\bf g}) \cdot \mathsf{b}({\bf h,k})} \, \Omega_{a_{\vv{x}}}({\bf g,h,k})
,
\end{align}
which yields
\begin{align}
\label{eq:O'}
\coho{O}' &= \psi^{\pi \cup \mathsf{b}} \otimes \coho{O}.
\end{align}

We now examine this change for the three types of FMTCs.
In the case where $\mathcal{A}_{\vv{1}} \neq \varnothing$, we can always choose a gauge in which $\breve{\beta}_{\psi_{\vv{0}}}({\bf g,h}) = 1$, which gives $\breve{\coho{O}}' =\breve{\coho{O}}$, and hence $[\coho{O}'] = [\coho{O}]$.
In the case where $\mathcal{A}_{\vv{1}} = \varnothing$, this is not possible as $\beta_{\psi_{\vv{0}}}({\bf g,h})$ is invariant.
In fact, in this case, we have seen that $\mathsf{b} = \phi$, and we will see that $\mathsf{b} = \central$ in the next subsection.
Thus, for $\mathcal{A}_{\vv{1}} = \varnothing$, we have $\coho{O}' = \psi^{\pi \cup \central} \otimes \coho{O}$.
However, when $\mathcal{A}_{\vv{1}} = \varnothing$ and $\widehat{\bf A}_{\vv{1}} \neq \varnothing$, we can trivialize $\psi^{\pi \cup \central}$ in $H^3_{[\rho]}(G,\mathcal{A}_{\vv{0}})$.
In order to see this, we can note that $[\V] = [\Q]$ in this case, and hence the above discussion for modifying the symmetry action by $\Q$ implies that $[\coho{O}'] = [\coho{O}]$.
A significant implication of this is that, for this case, we can find $\coho{v} \in C^{2}(G,\mathcal{A}_{\vv{0}})$ such that $\cbd \coho{v} = \psi^{\pi \cup \central}$.
In particular, we can obtain such a $\coho{v}$ through the relation $\nu_{a_{\vv{x}}}({\bf g,h}) = M^{\ast}_{a_{\vv{x}} \cohosub{v}({\bf g,h})}$ with
\begin{align}
\label{eq:special_nu}
\nu_{a_{\vv{x}}}({\bf g,h}) = \frac{ \beta_{\psi_{\vv{0}}^{\vv{x}}}({\bf g,h})^{\pi({\bf gh})} 
U_{\bf g}(\psi_{\vv{0}}^{\vv{x}} ,  a_{\vv{x}} ; a_{\vv{x}} )^{\pi({\bf h})} 
\, i^{\vv{x} \cdot \pi({\bf g}) \cdot \pi({\bf h}) }  }
{ \gamma_{a_{\vv{x}}}^{\pi({\bf gh})} \gamma_{\,^{\bf \bar{g}}a_{\vv{x}}}^{-\pi({\bf h})} \gamma_{a_{\vv{x}}}^{-\pi({\bf g})}}
,
\end{align}
where $\gamma_{a_{\vv{x}}}$ are defined as in Eqs.~\eqref{eq:VQequiv0} and \eqref{eq:VQequiv1}.
It is straightforward to check that this $\nu_{a_{\vv{x}}}$ has all the claimed properties.
When $\widehat{\bf A}_{\vv{1}} = \varnothing$, $[\V] \neq [\Q]$, so we cannot generally find a similar trivialization of $\psi^{\pi \cup \central}$, and hence we have $[\coho{O}'] = [\psi^{\pi \cup \central} \otimes \coho{O}]$.

\subsection{Fermionic Symmetry Fractionalization: Obstruction and Classification}
\label{sec:symmetry_fractionalization}

Following the discussion in Ref.~\onlinecite{Bark2019}, symmetry fractionalization may occur when the action of the global symmetry on the physical Hilbert space exhibits localization on the low energy subspace describing the topological phase, which is the case when the symmetry acts in a locality preserving manner.
In particular, let $| \Psi_{a_1 , \ldots , a_n } \rangle$ be a ground state of the system localizing $n$ quasiparticles, with the $j$th quasiparticle carrying topological charge $a_{j}$. (To be concrete, we consider the system to be defined on a genus zero surface.)
When a global symmetry acts in a unitary on-site manner (which is locality preserving), the symmetry action $R_{\bf g}$ on the low energy Hilbert space acts as
\begin{align}
\label{eq:localized_symmetry}
R_{\bf g}| \Psi_{a_1 , \ldots , a_n } \rangle &=  \prod_{j=1}^{n} U^{(j)}_{\bf g} \rho_{\bf g}| \Psi_{a_1 , \ldots , a_n } \rangle
,\end{align}
where $U^{(j)}_{\bf g}$ is a unitary operator whose support is localized in a neighborhood of the $j$th quasiparticle.
Here, $\rho_{\bf g}$ is an operator on the physical Hilbert space that acts on the topological degrees of freedom ascribed to the states $| \Psi_{a_1 , \ldots , a_n } \rangle$ in exactly the same way that $\rho_{\bf g}$ acts on the MTC $\fMTC$.
(The $U^{(j)}_{\bf g}$ operators should not be confused with the $[U_{\bf g}(a,b;c)]_{\mu \nu}$ symbols, which enter Eq.~(\ref{eq:localized_symmetry}) through $\rho_{\bf g}$.)

It can then be shown that the localized symmetry operators $U^{(j)}_{\bf g}$ will obey a projective-like multiplication
\begin{align}
U^{(j)}_{\bf g} \rho_{\bf g} U^{(j)}_{\bf h} \rho_{\bf g}^{-1}| \Psi_{a_1 , \ldots , a_n } \rangle &= \eta_{a_j}({\bf g,h}) U^{(j)}_{\bf gh}| \Psi_{a_1 , \ldots , a_n } \rangle
,
\end{align}
for projective phases $\eta_{a}({\bf g,h})$.
Together with Eq.~(\ref{eq:localized_symmetry}), this implies
\begin{align}
\kappa_{\bf g,h}(a, b; c)  &= \frac{\eta_{a }({\bf g,h})\eta_{b }({\bf g,h})}{\eta_{c }({\bf g,h})}
.
\label{eq:kapp-eta}
\end{align}
Associativity of the global symmetry operators $R_{\bf g}$ require these phases to satisfy
\begin{align}
\label{eq:eta_relation}
\frac{\eta_{\rho_{\bf g}^{-1}(a)}({\bf h,k}) \eta_{a}({\bf g,hk})}{ \eta_{a}({\bf gh,k}) \eta_{a}({\bf g,h})} = 1
.
\end{align}
The projective phases $\eta_{a}({\bf g,h})$, together with the symmetry action $\rho$, encode the fractionalization of the global symmetry, in that they can be interpreted as assignments of fractional quantum numbers of the symmetry group $G$ to the quasiparticles (topological charges) of the topological phase.
However, there may be no or multiple distinct solutions to Eqs.~(\ref{eq:kapp-eta}) and (\ref{eq:eta_relation}).

It is worth mentioning that the equivalence of the topological symmetry action under natural isomorphism $\check{\rho}_{\bf g} = \Upsilon_{\bf g} \rho_{\bf g}$, enter as a gauge freedom involved in writing the global symmetry operators, which physically justifies the equivalence.
In particular, such a modification of the topological symmetry action can be canceled in the global symmetry action $R_{\bf g}$ by modifying the local operators as $\check{U}^{(j)}_{\bf g} = {U}^{(j)}_{\bf g}  Y^{(j) \, -1}_{\bf g}$, where $Y^{(j)}_{\bf g} | \Psi_{a_1 , \ldots , a_n } \rangle = \gamma_{a_j}({\bf g}) | \Psi_{a_1 , \ldots , a_n } \rangle$.
The effect of this symmetry action gauge transformation on the projective phases is
\begin{align}
\label{eq:eta_check}
\check{\eta}_{a}({\bf g,h}) = \frac{\gamma_{a}({\bf gh})}{\gamma_{\,^{\bf \bar{g}}a}({\bf h}) \gamma_{a}({\bf g})} \eta_{a}({\bf g,h})
.
\end{align}

In order to understand when Eqs.~(\ref{eq:kapp-eta}) and (\ref{eq:eta_relation}) have solutions and how to classify them, it is useful to define the phases
\begin{align}
\label{eq:omega_beta_etaprime}
\omega_{a}({\bf g,h}) = \frac{\beta_{a}({\bf g,h})}{\eta_{a}({\bf g,h})}
,
\end{align}
which can be seen to obey $\omega_{a}({\bf g,h}) \omega_{b}({\bf g,h}) = \omega_{c}({\bf g,h})$ when $N_{ab}^{c} \neq 0$.
(The gauge freedom associated with decomposing the $\kappa_{\bf g,h}$ into $\beta_{a}({\bf g,h})$ can be assumed to modify the $\omega_{a}$ phases in a correlated manner that leaves the projective phases $\eta_{a}$ unchanged.)
It follows that
\begin{align}
\label{eq:omega_beta_eta}
\omega_{a}({\bf g,h}) = M^{\ast}_{a \cohosub{w}({\bf g,h})}
\end{align}
for some $\coho{w}({\bf g,h}) \in C^2(G,\mathcal{A})$.
In this way, Eq.~(\ref{eq:eta_relation}) translates into the condition
\begin{align}
\label{eq:obstrvansh}
\coho{O}({\bf g,h,k}) &= \,^{\bf g}\coho{w}({\bf h,k}) \otimes \overline{\coho{w}({\bf gh,k})} \otimes \coho{w}({\bf g,hk}) \otimes \overline{\coho{w}({\bf g,h})}
\notag \\
&= \cbd \coho{w}({\bf g,h,k})
.
\end{align}
This condition must be satisfied for the symmetry action $\rho$ on the emergent topological theory $\fMTC$ to be compatible with the localization of the symmetry action on the physical Hilbert space.
Since the right hand side of this equation is a 3-coboundary, it can only be satisfied when $\coho{O}$ is also a 3-coboundary.
This reveals a potential obstruction to symmetry fractionalization, as there can be no solutions to Eq.~(\ref{eq:obstrvansh}) when $\coho{O}$ is not a coboundary.

When $\coho{O}$ is a 3-coboundary, there are solutions to Eq.~(\ref{eq:obstrvansh}), and hence Eq.~(\ref{eq:eta_relation}).
Moreover, when $\coho{w} \in C^{2}(G,\mathcal{A})$ provides a solution, it is clear that
\begin{align}
\coho{w}'({\bf g,h}) = \coho{t}({\bf g,h}) \otimes \coho{w}({\bf g,h})
\end{align}
will also provide a solution for any $\coho{t}({\bf g,h}) \in Z_{[\rho]}^{2}(G,\mathcal{A})$, and that all solutions are obtained this way.

It remains to determine whether such solutions should be considered equivalent or distinct.
For this, we recognize that, for fixed $\rho_{\bf g}$, if we redefine the local operators $U^{(j)}_{\bf g}$ by
\begin{align}
\grave{U}^{(j)}_{\bf g} = {U}^{(j)}_{\bf g}  Z^{(j) \, -1}_{\bf g}
,
\end{align}
where $Z^{(j)}_{\bf g}$ are similarly localized unitary operators, then the action of $R_{\bf g}$ on the low energy Hilbert space will not change as long as
\begin{align}
\prod_{j =1}^{n} Z^{(j)}_{\bf g} | \Psi_{a_1 , \ldots , a_n } \rangle &=  | \Psi_{a_1 , \ldots , a_n } \rangle
.
\end{align}
Since such operators will act on the low-energy Hilbert space as
\begin{align}
Z^{(j)}_{\bf g} | \Psi_{a_1 , \ldots , a_n } \rangle &= \zeta_{a_j}({\bf g}) | \Psi_{a_1 , \ldots , a_n } \rangle
,
\end{align}
where $\zeta_{a_j}({\bf g})$ are phases, this constraint becomes
\begin{align}
\prod_{j =1}^{n} \zeta_{a_j}({\bf g}) &=  1
.
\end{align}
This is equivalent to the condition that these phases respect the fusion rules, i.e. $\zeta_{a}({\bf g}) \zeta_{b}({\bf g}) = \zeta_{c}({\bf g})$ whenever $N_{ab}^{c} \neq 0$, which establishes the bijection
\begin{align}
\zeta_{a}({\bf g}) = M^{\ast}_{a \cohosub{z}({\bf g})}
\end{align}
for some $\coho{z} \in C^{1}(G,\mathcal{A})$.
This translates into the redefinitions
\begin{align}
\label{eq:eta_grave}
\grave{\eta}_{a}({\bf g,h}) &= \frac{\zeta_{a}({\bf gh})}{\zeta_{\,^{\bf \bar{g}}a}({\bf h}) \zeta_{a}({\bf g})} \eta_{a}({\bf g,h}),
\end{align}
and
\begin{align}
\grave{\coho{w}}({\bf g,h}) &= \,^{\bf g}\coho{z}({\bf h}) \otimes \overline{\coho{z}({\bf gh})} \otimes \coho{z}({\bf h}) \otimes  \coho{w}({\bf g,h})
\notag \\
&= \cbd \coho{z}({\bf g,h}) \otimes  \coho{w}({\bf g,h})
.
\label{eq:cohow_grave}
\end{align}

In the case of bosonic topological phases, any such redefinition of $U^{(j)}_{\bf g}$ by $Z^{(j)}_{\bf g}$ yields a physically equivalent description of the symmetry fractionalization, as the distinction is physically undetectable.
As such, the solutions of Eq.~(\ref{eq:obstrvansh}) are considered equivalent when related by a 2-coboundary in $B_{[\rho]}^{2}(G,\mathcal{A})$, so the classification of symmetry fractionalization is given torsorially by $H_{[\rho]}^{2}(G,\mathcal{A})$, through the action of $\coho{t} \in Z_{[\rho]}^{2}(G,\mathcal{A})$.
In this way, the notion of a symmetry fractionalization class on $\fMTC$ is specified by an equivalence class $[\rho , \eta]$, where the equivalence relations on these pairs are
\begin{align}
\label{eq:frac_equivalence}
(\rho , \eta) \sim (\check{\rho} , \check{\eta}) \sim (\rho , \grave{\eta})
.
\end{align}
One can also include equivalences under vertex basis gauge transformations of the FMTC, i.e. $\fMTC \sim \widetilde{\fMTC}$ as in Eqs.~\eqref{eq:vertex_basis_gauge}, \eqref{eq:Fgaugetransf}, and \eqref{eq:Rgaugetransf}, which yield the equivalence relation $(\fMTC, \rho , \eta) \sim (\widetilde{\fMTC}, \tilde{\rho} , {\eta})$.
Here, $\tilde{\rho}$ is defined by $\tilde{\rho}(a) = \rho(a)$ and
\begin{align}
[\widetilde{U}_{\bf g} (a,b;c)]_{\mu \nu} &= \sum_{\mu',\nu'} [\Gamma^{\,^{\bf \bar{g}}a \,^{\bf \bar{g}}b}_{\,^{\bf \bar{g}}c}]_{\mu \mu'}
\notag \\
& \qquad \times [{U}_{\bf g} (a,b;c)]_{\mu' \nu'} [(\Gamma^{a b}_{c})^{-1}]_{\nu' \nu}
.
\end{align}

In order to apply this analysis to fermionic topological phases, we include the fermionic vorticity labels on topological charges and the additional constraints that result from the symmetry action applying to the physical fermions of the system.
We previously argued that the natural transformations were required to act trivially on the physical fermion, that is $\gamma_{\psi_{\vv{0}}} = 1$.
Now we see this is additionally justified, as nontrivial $\gamma_{\psi_{\vv{0}}}$ would modify how the symmetry acts on the physical fermions in the physical Hilbert space, and would constitute a physically observable difference.
The localization of symmetry action applies similarly, except the equivalence of identifications of the local operators $U^{(j)}_{\bf g}$ under redefinition by $Z^{(j)}_{\bf g}$ are further constrained.
In particular, we additionally require that $Z^{(j)}_{\bf g}$ have eigenvalues $\zeta_{\psi_{\vv{0}}}({\bf g}) = 1$ when acting on the physical fermion, as nontrivial $\zeta_{\psi_{\vv{0}}}({\bf g})$ would change the local operators in a manner that is physically observable using the physical fermions, as opposed to an unmeasurable gauge transformation on the emergent degrees of freedom.
This translates into the condition that $\coho{z} \in C^{1}(G,\mathcal{A}_{\vv{0}})$ and that the equivalence of fractionalization classes corresponds to a quotient by $B_{[\rho]}^{2}(G,\mathcal{A}_{\vv{0}})$.
On the other hand, the solutions of Eq.~(\ref{eq:obstrvansh}) are still given by torsorial action of $\coho{t} \in Z_{[\rho]}^{2}(G,\mathcal{A}_{\vv{0}}\oplus \mathcal{A}_{\vv{1}})$.
Thus, we find that symmetry fractionalization of fermionic topological phases has the obstruction similarly given by
\begin{align}
\label{eq:H3obsruction}
[\coho{O}] \in H^3_{[\rho]}(G,\mathcal{A})
.
\end{align}
When $[\coho{O}] = [\I]$, i.e. $\coho{O} \in B^3_{[\rho]}(G,\mathcal{A})$, the obstruction vanishes and the classification of symmetry fractionalization is given torsorially by
\begin{align}
\frac{Z_{[\rho]}^{2}(G,\mathcal{A}_{\vv{0}}\oplus \mathcal{A}_{\vv{1}})}{B_{[\rho]}^{2}(G,\mathcal{A}_{\vv{0}})}
.
\end{align}
Here, we write $\mathcal{A} = \mathcal{A}_{\vv{0}}\oplus \mathcal{A}_{\vv{1}}$ to emphasize the difference with the term in the quotient.
The notion of fractionalization class is defined as in Eq.~\eqref{eq:frac_equivalence}, but with the constraints $\gamma_{\psi_{\vv{0}}} ({\bf g})= 1$ and $\zeta_{\psi_{\vv{0}}}({\bf g}) = 1$ imposed on the transformations involved in the equivalence relations.

This result for obstruction and classification of symmetry fractionalization in fermionic topological phases includes finer detailed structure that is important to unpack.
In this regard, we first focus on the meaning of symmetry fractionalization of the physical fermion of the theory.
Recall that $\kappa_{\bf g,h}(\psi_{\vv{0}},\psi_{\vv{0}}; \I_{\vv{0}})  =1$.
Together with Eq.~(\ref{eq:kapp-eta}), this implies $\eta_{\psi_{\vv{0}} }({\bf g,h})^{2} = 1$.
Thus, we have
\begin{align}
\label{eq:eta_psi_central}
\eta_{\psi_{\vv{0}} }({\bf g,h}) = \beta_{\psi_{\vv{0}} }({\bf g,h}) M_{\psi_{\vv{0}} \cohosub{w}({\bf g,h})} = (-1)^{\central( {\bf g,h} )}
,
\end{align}
where the only aspect of $\coho{w}({\bf g,h})$ that enters this expression is its vorticity.
We emphasize that, for the case $\mathcal{A}_{1} = \varnothing$, $[\Q]$ is nontrivial and $\beta_{\psi_{\vv{0}} }({\bf g,h}) = (-1)^{\phi ({\bf g,h})}$, giving the relation $\central = \phi$.

In a fermionic theory, which is constrained to have $\gamma_{\psi_\vv{0}}({\bf g}) =1$, this is a gauge invariant quantity.
Since the symmetry action leaves $\psi_{\vv{0}}$ fixed, Eq.~(\ref{eq:eta_relation}) implies the 2-cocycle condition
\begin{align}
\label{eq:psi2cocycle}
\frac{\eta_{\psi_{\vv{0}}}({\bf h,k})\eta_{\psi_{\vv{0}}}({\bf g,hk} )}{\eta_{\psi_{\vv{0}}}({\bf gh,k})\eta_{\psi_{\vv{0}}}({\bf g,h})} =1
.
\end{align}
Together, this shows
\begin{align}
\central \in Z^2(G,\mathbb{Z}_2^{\eff})
.
\end{align}

The use of the symbol $\central$ here was intentionally the same as that of the 2-cocycle that determines the fermionic symmetry group $\mathcal{G}^{\eff} = \mathbb{Z}_2^{\eff} \times_{\central} G$ describing the system.
Indeed, the $G$ symmetry fractionalization of the physical fermion determines the fermionic symmetry group $\mathcal{G}^{\eff}$.
We can understand this by heuristically writing the local operators $U^{(j)}_{\bf g}$ acting on a physical fermion $\psi_{\vv{0}}$ as $U^{(\psi)}_{\bf g}$.
Then we see $U^{(\psi)}_{\bf g}$ forms a projective representation of $G$, that is
\begin{align}
U^{(\psi)}_{\bf g} U^{(\psi)}_{\bf h} = \eta_{\psi_{\vv{0}} }({\bf g,h}) U^{(\psi)}_{\bf gh}
.
\end{align}
Projective representations of a group can generally be understood in terms of linear representations of the central extensions of that group.
Specifically, this lifting is constructed by using the projective phase factors of the projective representation to define the central extension's group multiplication.
The fermionic symmetry group $\mathcal{G}^{\eff}$ is the group including fermion parity conservation, for which the physical fermions carry linear representations, as expected for local microscopic objects.
In other words, the group multiplication of $\mathcal{G}^{\eff}$ given by
\begin{align}
\label{eq:gfmultiplication}
\mathpzc{gh}=
(\vv{x},{\bf g} ) (\vv{y}, {\bf h})  = (\vv{x} + \vv{y} + \central({\bf g,h}), {\bf gh} )
,
\end{align}
makes $(-1)^{\central( {\bf g,h} )} \in B^2(\mathcal{G}^{\eff},\mathbb{Z}_2^{\eff})$ a 2-coboundary.

It is worth emphasizing an important distinction between taking equivalences of the fermionic symmetry action by fermionic natural isomorphisms requiring $\gamma_{\psi_\vv{0}}({\bf g})=1$ and equivalences of fermionic fractionalization classes by $B_{[\rho]}^{2}(G,\mathcal{A}_{\vv{0}})$ (i.e. $\zeta_{\psi_{\vv{0}} }({\bf g,h}) = 1$), as opposed to equivalences under natural isomorphisms which allow $\gamma_{\psi_\vv{0}}({\bf g})=\pm 1$ and fractionalization classes related by $B_{[\rho]}^{2}(G,\mathcal{A})$.
Were we to allow equivalence under natural isomorphisms with $\gamma_{\psi_\vv{0}}({\bf g})= \pm 1$, then $\check{\eta}_{\psi_{\vv{0}} }({\bf g,h}) = \frac{\gamma_{\psi_\vv{0}}({\bf gh})}{\gamma_{\psi_\vv{0}}({\bf h}) \gamma_{\psi_\vv{0}}({\bf g})} \eta_{\psi_{\vv{0}} }({\bf g,h})$ would equate $\central$ and $\check{\central}$ that differ by 2-coboundaries $B^2(G,\mathbb{Z}_2^{\eff})$.
Similarly, equivalence of fractionalization classes related by $B_{[\rho]}^{2}(G,\mathcal{A})$, which correspond to $\zeta_{\psi_{\vv{0}} }({\bf g,h}) = \pm 1$, would have $\grave{\eta}_{\psi_{\vv{0}} }({\bf g,h}) = \frac{\zeta_{\psi_{\vv{0}}}({\bf gh})}{\zeta_{\psi_{\vv{0}}}({\bf h}) \zeta_{\psi_{\vv{0}}}({\bf g})} \eta_{\psi_{\vv{0}}}({\bf g,h})$, also equating $\central$ and $\grave{\central}$ that differ by $B^2(G,\mathbb{Z}_2^{\eff})$.
This would yield $[\central] \in H^2(G,\mathbb{Z}_2^{\eff})$ as the corresponding invariant, rather than $\central \in Z^2(G,\mathbb{Z}_2^{\eff})$.
While the possible group extensions $\mathcal{G}^{\eff}$ are classified by $[\central] \in H^2(G,\mathbb{Z}_2^{\eff})$, the fractionalization class contains a finer level of specificity with $\central \in Z^2(G,\mathbb{Z}_2^{\eff})$.
The distinction will become even more pronounced when we include symmetry defects, as $\gamma_{\psi}({\bf g})=-1$ would change the vorticity of ${\bf g}$-defects, and $\zeta_{\psi_{\vv{0}}}({\bf g}) = -1$ would correspond to redefining the topological charges of ${\bf g}$-defects by fusion with a vortex-valued charge.
Physically, the global symmetry group labels correspond to extrinsically measurable quantities, as they are fixed with respect to some physical reference.
(This is why $G$-crossed MTCs related by $\text{Aut}(G)$ are not considered physically equivalent.)
Thus, different realizations $\central \in [\central]$ of the same symmetry group $\mathcal{G}^{\eff}$ correspond to isomorphic, but physically distinguishable realizations of the symmetry.
Henceforth, when we refer to $\mathcal{G}^{\eff}$, we will mean the extension associated with a particular cocycle $\central$, not the cohomology class $[\central]$, unless explicitly stated.

We can now re-examine the classification of symmetry fractionalization with respect to a specific $\mathcal{G}^{\eff} = \mathbb{Z}_2^{\eff} \times_{\central} G$.
The first observation is that not all possible central extensions of the group $G$ by $\mathbb{Z}_2^{\eff}$ will necessarily be manifested as possible symmetry fractionalization classes.
In other words, the set of possible fermionic symmetry fractionalization classes for given $\fMTC$, $G$, and $\rho$ encodes a possible obstruction to realizing a particular fermionic symmetry group $\mathcal{G}^{\eff}$.
The second observation is that the torsorial action of 2-cocycles in $Z_{[\rho]}^{2}(G,\mathcal{A}_{\vv{0}}\oplus \mathcal{A}_{\vv{1}})$ that have $\mathcal{A}_{\vv{1}}$-valued elements change which $\mathcal{G}^{\eff}$ is manifested.
Given a symmetry fractionalization class that manifests one particular $\mathbb{Z}_2^{\eff}$ central extension of $G$, the 2-cocycles in $Z_{[\rho]}^{2}(G,\mathcal{A}_{\vv{0}}\oplus \mathcal{A}_{\vv{1}})$ informs us which other $\mathbb{Z}_2^{\eff}$ central extensions of $G$ can also arise.
When no vortex-valued 2-cocycles exist, only one possible $\mathcal{G}^{\eff}$ can be manifested.
For example, when $\mathcal{A}_{\vv{1}} = \varnothing$ (as is always the case when $\fMTC$ has $\sigma$-type vortices), $\fMTC$ can only manifest one $\mathcal{G}^{\eff}$ for a given symmetry action $\rho$.
We note that this condition always applies for nontrivial $\Q$-projective symmetry actions.
The third observation is that for a fixed fermionic symmetry group $\mathcal{G}^{\eff}$, the classification of symmetry fractionalization is given by restricting to the $\mathcal{A}_{\vv{0}}$-valued 2-cocycles, so it is torsorially classified by
\begin{align}
H_{[\rho^{(\vv{0})}]}^{2}(G,\mathcal{A}_{\vv{0}} ) = \frac{Z_{[\rho^{(\vv{0})}]}^{2}(G,\mathcal{A}_{\vv{0}} )}{B_{[\rho^{(\vv{0})}]}^{2}(G,\mathcal{A}_{\vv{0}})}
.
\end{align}
The fourth observation is that since the group of Abelian quasiparticles factorizes as $\mathcal{A}_\vv{0} = \mathbb{Z}_2^\psi \times \widehat{\mathcal{A}}_\vv{0}$ (see Appendix~\ref{app:A_factorize}), this classifying cohomology group can be expressed in terms of a factor associated with the symmetry fractionalization of the quasiparticles and a factor associated with the symmetry fractionalization of vorticity.

These observations can be formalized from the perspective of extensions of symmetry fractionalization.
For this, we begin by considering the physical fermion $\mathbb{Z}_2^{\psi}$ by itself.
We first note that $\widehat{\bf A} = \mathbb{Z}_{2}^{\eff}$ in this case (as can be seen from the $\ifo^{(\nu)}$ theories).
Clearly the symmetry action on $\mathbb{Z}_2^{\psi}$ can only be trivial, with all $U_{\bf g}$-symbols trivial and $\kappa_{\bf g,h} =1$.
We can choose
\begin{align}
\label{eq:betapsib}
\beta_{\psi_{\vv{0}}} ({\bf g,h})  &= (-1)^{\mathsf{b}({\bf g,h})}
,
\end{align}
for any $\mathsf{b} \in C^2(G,\mathbb{Z}_2)$.
(While it may seem natural to choose $\beta_{\psi_{\vv{0}}} ({\bf g,h}) = 1$ to make this a fermionic natural isomorphism, we leave this general, so that we can match it to any SMTC or FMTC that contains it.)
Symmetry fractionalization of the physical fermion is then classified by $\eta_{\psi_{\vv{0}}}({\bf g,h}) = (-1)^{\central({\bf g,h})}$, where $\central \in Z^{2}(G , \mathbb{Z}_{2}^{\eff})$.
This precisely corresponds to the classification of the possible fermionic symmetry groups $\mathcal{G}^{\eff} = \mathbb{Z}_2^{\eff} \times_{\central} G$ through Eq.~(\ref{eq:eta_psi_central}).

We next consider a FMTC $\fMTC$ with a particular fermionic symmetry action $[\rho] : G \to \Autf{\fMTC}$.
While $\eta_{\psi_{\vv{0}}}= (-1)^{\central}$ with any $\central \in Z^{2}(G , \mathbb{Z}_{2}^{\eff})$ may occur on $\mathbb{Z}_{2}^{\psi}$, there may not always be a fractionalization class on $\fMTC$ with $[\rho]$ that has a given $\central$.
We can now define an obstruction to realizing the fermionic symmetry group $\mathcal{G}^{\eff}$ with $\central$ through a symmetry fractionalization class of $\fMTC$ and $[\rho]$ in the following way.

In the case where $\mathcal{A}_{\vv{1}} = \varnothing$, we have seen that $\central = \mathsf{b} = \phi$ is the only possibility, where $\phi$ is the $\Q$-projective structure of the symmetry action discussed in Sec.~\ref{sec:symmetry_action}.
Of course, fractionalization classes on $\fMTC$ with $[\rho]$ will exist if and only if $[\coho{O}]$ is nontrivial.
Thus, for $\mathcal{A}_{\vv{1}} = \varnothing$, we define the obstruction class to manifesting $\mathcal{G}^{\eff} = \mathbb{Z}_2^{\eff} \times_{\central} G$ on $\fMTC$ with $[\rho]$ to be the pair
\begin{align}
\label{eq:Gf_obstruction_general_1}
[\coho{O}^{\central}] =\big( [\coho{O}] , (\central - \phi) \big) \in  H^3_{[\rho^{(\vv{0})}]}(G, \mathcal{A}_{\vv{0}}) \times Z^2(G,\mathbb{Z}_2^{\eff})
.
\end{align}

In the case where $\mathcal{A}_{\vv{1}} \neq \varnothing$, different values of $\central$ may be possible.
For this case, let us pick an arbitrary Abelian vortex, which we denote $e_{\vv{1}} \in \mathcal{A}_{\vv{1}}$.
In order for a fractionalization class specified by $\rho$ and $\eta_{a_{\vv{x}}} = \beta_{a_{\vv{x}}} M_{a_{\vv{x}} \cohosub{w}}$ to have a given $\central$, it must correspond to a 2-cochain of the form $\coho{w} = \coho{y} \otimes e_{\vv{1}}^{\central} \otimes \bar{e}_{\vv{1}}^{\mathsf{b}}$, where $\coho{y} \in C^2(G, \mathcal{A}_{\vv{0}})$.
To express this as an obstruction class, we define
\begin{align}
&\coho{O}^{\central} ({\bf g,h,k})= \coho{O}({\bf g,h,k}) \otimes \cbd ( {e}_{\vv{1}}^{\mathsf{b}} \otimes \bar{e}_{\vv{1}}^{\central}  )({\bf g,h,k})
\notag \\
&\quad =  \coho{O}({\bf g,h,k}) \otimes \cbd {e}_{\vv{1}}^{\mathsf{b}}({\bf g,h,k}) \notag \\
& \quad \quad \otimes \,^{\bf g} \bar{e}_{\vv{1}}^{\central({\bf h,k})} \otimes {e}_{\vv{1}}^{{\central}({\bf gh,k})} \otimes \bar{e}_{\vv{1}}^{\central({\bf g,hk})} \otimes {e}_{\vv{1}}^{\central({\bf g,h})}
.
\end{align}
More compactly, we can write this as
\begin{align}
\label{eq:O_central_general}
\coho{O}^{\central} &= \coho{O} \otimes \cbd ( {e}_{\vv{1}}^{\mathsf{b}} \otimes \bar{e}_{\vv{1}}^{\central}  )
.
\end{align}
We can also reexpress Eq.~\eqref{eq:O_central_general} in the convenient form
\begin{align}
\label{eq:O_central_cups}
\coho{O}^{\central} & = \coho{O} \otimes \cbd {e}_{\vv{1}}^{\mathsf{b}} \otimes \coho{H}_{\rho}^{\tilde{\central}} \otimes h_{\vv{0}}^{\tilde{\central} \cup_{1} \tilde{\central}}
.
\end{align}
(Recall that $e_{\vv{1}} \otimes e_{\vv{1}} = h_{\vv{0}} \in \mathcal{A}_{\vv{1}}$.)
Here, we have defined the 1-cochain
\begin{align}
\coho{H}_{\rho}({\bf g}) &= e_{\vv{1}} \otimes \rho_{\bf g}(\bar{e}_{\vv{1}} ) \in C^1(G,\mathcal{A}_{\vv{0}})
,
\end{align}
(noting the fact that symmetry action preserves vorticity,) and we write $\coho{H}_{\rho}^{\tilde{\central}}({\bf g,h,k}) =\coho{H}_{\rho_{\bf g}}^{\tilde{\central}({\bf h,k})}$.
We have also defined $\tilde{\central} \in \{ 0,1\}$ to take the value of $\central$ with integer addition, rather than modulo 2 addition, and used the property 
\begin{align}
\cbd \tilde{\central}({\bf g,h,k}) & = \tilde{\central}({\bf h,k})-\tilde{\central}({\bf gh,k})+\tilde{\central}({\bf g,hk})-\tilde{\central}({\bf g,h})
\notag \\
&= 2 \left( \tilde{\central}({\bf g,hk})\cdot\tilde{\central}({\bf h,k}) -\tilde{\central}({\bf gh,k})\cdot \tilde{\central}({\bf g,h}) \right)
\notag \\
&= -2 \left( \tilde{\central} \cup_{1} \tilde{\central} \right) ({\bf g,h,k})
,
\label{eq:dtildew}
\end{align}
which holds for any $\central \in Z^2(G,\mathbb{Z}_2^{\eff})$.
(See Appendix~\ref{app:cup} for a general definition of $\cup_1$.)

We now verify that $\coho{O}^{\central}$ in Eq.~\eqref{eq:O_central_general} defines an obstruction class
\begin{align}
\label{eq:Gf_obstruction_general_2}
[\coho{O}^{\central}] \in  H^3_{[\rho^{(\vv{0})}]}(G, \mathcal{A}_{\vv{0}})
,
\end{align}
associated with manifesting $\mathcal{G}^{\eff} = \mathbb{Z}_2^{\eff} \times_{\central} G$ on $\fMTC$ with $[\rho]$, in the case $\mathcal{A}_{\vv{1}} \neq \varnothing$.
First, we note that $M_{\psi_{\vv{0}}, \cohosub{O}} = (-1)^{\cbd \mathsf{b}} = M_{\psi_{\vv{0}}, \cbd {e}_{\vv{1}}^{\mathsf{b}}}$, and hence $\coho{O}^{\central} ({\bf g,h,k}) \in \mathcal{A}_{\vv{0}}$.
It is also straightforward to see that $\cbd \coho{O}^{\central} ({\bf g,h,k}) = \I_{\vv{0}}$, hence $\coho{O}^{\central} \in Z^3_{\rho^{(\vv{0})}}(G, \mathcal{A}_{\vv{0}})$.

Next, we check that the arbitrary (gauge) choices in the definition of $\coho{O}^{\central}$ will leave the cohomology class invariant.
It is easy to see that $\coho{O}^{\central}$ is unchanged by the choice of representative fermionic symmetry action $\rho \in [\rho]$, since $\check{\coho{O}} = \coho{O}$.

In order to see the invariance under redefinitions of $\beta_{a_{\vv{x}}}$ to $\breve{\beta}_{a_{\vv{x}}} = \nu_{a_{\vv{x}}} \beta_{a_{\vv{x}}}$, we first define $\nu_{\psi_{\vv{0}}} = (-1)^{\mathsf{v}}$, which characterizes the vorticity of $\coho{v}$. 
From this, we find $\breve{\mathsf{b}} = \mathsf{b} + \mathsf{v}$, and thus
\begin{align}
\breve{\coho{O}}^{\central} &= \breve{\coho{O}} \otimes \cbd ( {e}_{\vv{1}}^{\breve{\mathsf{b}}} \otimes \bar{e}_{\vv{1}}^{\central}  ) 
\notag \\
&= \coho{O}^{\central} \otimes \cbd ( \coho{v} \otimes {e}_{\vv{1}}^{\mathsf{v}}  )
,
\end{align}
where $\coho{v} \otimes {e}_{\vv{1}}^{\mathsf{v}} \in C^2 (G, \mathcal{A}_{\vv{0}})$, by definition.
We note that we can always use this freedom, in the case where $\mathcal{A}_{\vv{1}} \neq \varnothing$, to choose a convenient gauge in which $\mathsf{b} = \vv{0}$.

Additionally, if we used a different Abelian vortex $\acute{e}_{\vv{1}} \in \mathcal{A}_{\vv{1}}$, then we would have
\begin{align}
\acute{\coho{O}}^{\central} &= \coho{O} \otimes \cbd ( \acute{e}_{\vv{1}}^{\mathsf{b}} \otimes \bar{\acute{e}}_{\vv{1}}^{\central}  )
\notag \\
&= \coho{O}^{\central} \otimes \cbd ( c_{\vv{0}}^{\mathsf{b}} \otimes \bar{c}_{\vv{0}}^{\central}  )
,
\end{align}
where $c_{\vv{0}} = \acute{e}_{\vv{1}} \otimes \bar{e}_{\vv{1}} \in \mathcal{A}_{\vv{0}}$.
Thus, the arbitrary choices in defining $\coho{O}^{\central}$ can only change it by an element of $B^3_{[\rho^{(\vv{0})}]}(G, \mathcal{A}_{\vv{0}})$, and it follows that $[\coho{O}^{\central}]\in  H^3_{[\rho^{(\vv{0})}]}(G, \mathcal{A}_{\vv{0}})$ is a well-defined invariant.

Finally, we observe that a symmetry fractionalization class with $\central$ exists if and only if
\begin{align}
\cbd \coho{y} = \coho{O}^{\central}
,
\end{align}
where $\coho{y} \in C^2(G, \mathcal{A}_{\vv{0}})$, i.e. when $\coho{O}^{\central} \in B^3_{[\rho^{(\vv{0})}]}(G, \mathcal{A}_{\vv{0}})$.
Thus, $[\coho{O}^{\central}]$ is indeed an obstruction for realizing $\central$.
Notice that, for the $\mathcal{A}_{\vv{1}} \neq \varnothing$ case, $[\coho{O}^{\central}] = [\I_{\vv{0}}]$ implies $[\coho{O}] = [\I_{\vv{0}}]$, so the vanishing of the fractionalization obstruction need not be separately imposed.

When the obstruction $[\coho{O}^{\central}]$ is trivial (for either $\mathcal{A}_{\vv{1}}$ empty or nonempty), there is a symmetry fractionalization class with $\rho$ and $\eta_{a_{\vv{x}}}({\bf g,h})$ that manifests $\mathcal{G}^{\eff} = \mathbb{Z}_2^{\eff} \times_{\central} G$.
In this case, ${\eta}'_{a_{\vv{x}}}({\bf g,h}) = M_{a_{\vv{x}} \cohosub{t}_{\vv{0}}({\bf g,h})} \eta_{a_{\vv{x}}}({\bf g,h})$ with $\coho{t}_{\vv{0}} \in Z^2_{[\rho^{(\vv{0})}]}(G, \mathcal{A}_{\vv{0}})$ also represents a symmetry fractionalization class corresponding to $\central$, and all fractionalization classes corresponding to $\central$ are obtained this way.
As previously mentioned, for fermionic topological phases, symmetry fractionalization classes are formed by equivalence under product with 2-coboundaries in $B_{[\rho^{(\vv{0})}]}^{2}(G,\mathcal{A}_{\vv{0}})$.
Thus, the fermionic symmetry fractionalization classes for an unobstructed fermionic $\mathcal{G}^{\eff}$ and action $[\rho]$ are again seen to be torsorially classified by \begin{align}
H_{[\rho^{(\vv{0})}]}^{2}(G,\mathcal{A}_{\vv{0}} ).
\end{align}

We can also define a relative obstruction $[\coho{O}^{\central - \central'}_{r}]$ to manifesting fermionic symmetry group with $\central$ relative to $\central'$ by
\begin{align}
[\coho{O}^{\central - \central'}_{r}] = (\central - \central') \in Z^2(G,\mathbb{Z}_2^{\eff})
,
\end{align}
for $\mathcal{A}_{\vv{1}} = \varnothing$, and
\begin{align}
[\coho{O}^{\central - \central'}_{r}] = \left[ \cbd \left( e_{\vv{1}}^{\central'} \otimes \bar{e}_{\vv{1}}^{\central} \right) \right] \in H^3_{[\rho^{(\vv{0})}]}(G, \mathcal{A}_{\vv{0}})
,
\end{align}
for $\mathcal{A}_{\vv{1}} \neq \varnothing$.
In this way, we have
\begin{align}
\label{eq:Gf_obstruction_relative}
\left[\coho{O}^{\central - \central'}_{r}\right] = \left\{
\begin{array}{ll}
\left[\coho{O}^{\central} \right] -  [\coho{O}^{\central'}]  & \text{ if } \mathcal{A}_{\vv{1}} = \varnothing \\
\left[\coho{O}^{\central} \right] \otimes [\overline{\coho{O}}^{\central'}] & \text{ if } \mathcal{A}_{\vv{1}} \neq \varnothing
\end{array}
\right.
,
\end{align}
justifying the description as a relative obstruction.

When the fractionalization obstruction vanishes, $[\coho{O}] = [\I_{\vv{0}}]$, we can pick an arbitrary fractionalization class, which we specify through $\rho_{\bf g}$ and $\dot{\eta}_{a_{\vv{x}}}({\bf g,h})$, with corresponding $\dot{\coho{w}}$, such that $\cbd \dot{\coho{w}} = \coho{O}$.
In this case, we can alternatively define the obstruction to manifesting $\mathcal{G}^{\eff} = \mathbb{Z}_2^{\eff} \times_{\central} G$ on $\fMTC$ with $[\rho]$ to be
\begin{align}
\label{eq:Gf_obstruction_dot}
[\coho{O}^{\central}] = \left\{
\begin{array}{ll}
(\central - \dot{\central}) \in Z^2(G,\mathbb{Z}_2^{\eff}) & \text{ if } \mathcal{A}_{\vv{1}} = \varnothing \\
\left[\cbd \left( e_{\vv{1}}^{\dot{\central}} \otimes \bar{e}_{\vv{1}}^{\central} \right) \right] \in H^3_{[\rho^{(\vv{0})}]}(G, \mathcal{A}_{\vv{0}}) & \text{ if } \mathcal{A}_{\vv{1}} \neq \varnothing
\end{array}
\right.
.
\end{align}
Clearly, we can view this to be a relative obstruction, $[\coho{O}^{\central}] = [\coho{O}^{\central - \dot{\central}}_{r}]$, with respect to a known unobstructed fractionalization class with $\dot{\central}$, using Eq.~\eqref{eq:Gf_obstruction_relative}.
If we use a different unobstructed fractionalization class given by $\ddot{\eta}_{a_{\vv{x}}}({\bf g,h})$ with respective $\ddot{\central}( {\bf g,h} )$, then we would instead have
\begin{align}
\acute{\coho{O}}^{\central} &= \cbd \left( e_{\vv{1}}^{\ddot{\central}} \otimes \bar{e}_{\vv{1}}^{\central}  \right) =  \coho{O}^{\central} \otimes \coho{O}^{\dot{\central} - \ddot{\central}}_{r}
.
\end{align}
Since fractionalization classes exist (by assumption) for both $\dot{\central}$ and $\ddot{\central}$, i.e. they are unobstructed, it follows that $\coho{O}^{\dot{\central} - \ddot{\central}}_{r} \in B^3_{[\rho^{(\vv{0})}]}(G, \mathcal{A}_{\vv{0}})$.
Thus, the obstruction class defined this way does not depend on the choice of known fractionalization class.

The obstruction defined in Eq.~(\ref{eq:Gf_obstruction_dot}) is very similar to an obstruction defined in Ref.~\onlinecite{Galindo2017} for manifesting the group extensions corresponding to $[\central]$.
Our definitions use equivalences with the constraints $\gamma_{\psi_{\vv{0}}} = 1$ and $\zeta_{\psi_{\vv{0}}}({\bf g,h}) =1$, as explained above, whereas the definitions in Ref.~\onlinecite{Galindo2017} use equivalences that permit $\gamma_{\psi_{\vv{0}}} = \pm 1$ and $\zeta_{\psi_{\vv{0}}}({\bf g,h}) =\pm 1$, which consequently characterize the fermionic symmetry group by cohomology classes $[\central]\in H^2(G,\mathbb{Z}_2^{\eff})$.
A similar analysis using these less constrained equivalences yields the obstruction to manifesting $[\central]$ given by
\begin{align}
\label{eq:Gf_obstruction_CC}
[\coho{O}^{[\central]}] = \left\{
\begin{array}{ll}
[\central - \dot{\central}] \in H^2(G,\mathbb{Z}_2^{\eff}) & \text{ if } \mathcal{A}_{\vv{1}} = \varnothing \\
\left[\cbd \left( e_{\vv{1}}^{\dot{\central} - \central} \right) \right] \in H^3_{[\rho^{(\vv{0})}]}(G, \mathcal{A}_{\vv{0}}) & \text{ if } \mathcal{A}_{\vv{1}} \neq \varnothing
\end{array}
\right.
.
\end{align}
The corresponding classification of fractionalization classes manifesting $[\central]$ is given torsorially by
\begin{align}
\label{eq:ker_classification}
\ker \left( r_{\ast} : H_{[\rho]}^{2}(G,\mathcal{A}) \to H^2(G,\mathbb{Z}_2^{\eff}) \right),
\end{align}
where $r_{\ast}$ is the map induced from the restriction $r: \mathcal{A} \to \mathbb{Z}_2^{\eff}$ given by $a_{\vv{x}} \mapsto \vv{x}$.
More specifically, this is the set of cohomology classes $[\coho{t}] \in H_{[\rho]}^{2}(G,\mathcal{A})$ such that $M_{\psi_{\vv{0}} \cohosub{t}({\bf g,h})} = (-1)^{\cbd \vviso ({\bf g,h})}$ for some $\vviso \in C^1(G,\mathbb{Z}_2^{\eff})$.
The group defined in Eq.~(\ref{eq:ker_classification}) is generally not equivalent to $H_{[\rho^{(\vv{0})}]}^{2}(G,\mathcal{A}_{\vv{0}} )$.
An important example of such a difference occurs for the $G=\mathbb{Z}_{2}$ FSPT phases, in particular for $\fMTC = \ifo^{(0)}$ with $\rho_{\bf 1} = \V$.
For this example, $\ker ( r_{\ast} ) = \mathbb{Z}_{1}$, whereas $H_{[\rho^{(\vv{0})}]}^{2}(G,\mathcal{A}_{\vv{0}} ) = \mathbb{Z}_{2}$.
In other words, using this classification would mistakenly identify distinct $\mathbb{Z}_{2}$ FSPT phases, while ours properly distinguishes them.

\subsection{Fermionic Symmetry Fractionalization of Quasiparticles and Extensions to Vortices}
\label{sec:symmetry_frac_extension}

The symmetry fractionalization obstruction and classification discussion for FMTCs can be directly applied to SMTCs by using the fact that the characters of the fusion algebra of $\fMTC_{\vv{0}}$ are mapped to supersectors of topological charges of (any of) its $\mathbb{Z}_2^{\eff}$ extensions.
In this way, we can write the results for $\fMTC_{\vv{0}}$ by making the following modifications.
Symmetry fractionalization for $\fMTC_{\vv{0}}$ has the obstruction
\begin{align}
\label{eq:H3SMTC}
[{\coho{O}}^{(\vv{0})}] \in H^3_{[\widehat{\rho}]}(G,\widehat{{\bf A}})
,
\end{align}
and when the obstruction is trivial, i.e. $[{\coho{O}}^{(\vv{0})}] = [\widehat{\I}_{\vv{0}}]$, the classification of symmetry fractionalization is given torsorially by
\begin{align}
\frac{Z_{[\widehat{\rho}]}^{2}(G,\widehat{{\bf A}})}{B_{[\widehat{\rho}]}^{2}(G,\widehat{{\bf A}}_{\vv{0}})}
.
\end{align}
We re-emphasize that $\widehat{{\bf A}}$ is not necessarily equal to $\widehat{\mathcal{A}}$, due to the possibility of superAbelian $\sigma$-type vortices, but $\widehat{{\bf A}}_{\vv{0}} = \widehat{\mathcal{A}}_{\vv{0}}$.

Again, the fractionalization class determines the fermionic symmetry group $\mathcal{G}^{\eff} = \mathbb{Z}_2^{\eff} \times_{\central} G$ through $\eta^{(\vv{0})}_{\psi_{\vv{0}} }({\bf g,h}) = (-1)^{\central( {\bf g,h} )}$.
As with the full FMTC, we can define an obstruction $[\coho{O}^{(\vv{0})\central}]$ to manifesting $\central$ on $\fMTC_{\vv{0}}$ with $[\rho^{(\vv{0})}]$.

In the case where $\widehat{\bf A}_{\vv{1}} = \varnothing$, this is given by the pair
\begin{align}
\label{eq:Gf_obstruction_general_M0_1}
[\coho{O}^{(\vv{0})\central}] =\big( [\coho{O}^{(\vv{0})}] , (\central - \phi^{(\vv{0})}) \big)
,
\end{align}
where $(\central - \phi^{(\vv{0})}) \in   Z^2(G,\mathbb{Z}_2^{\eff})$, and $\beta_{\psi_{\vv{0}}}^{(0)} = (-1)^{\phi^{(\vv{0})}}$ is the $\Q$-projective structure of the symmetry action.

In the case where $\widehat{\bf A}_{\vv{1}} \neq \varnothing$, the obstruction to manifesting $\central$ is defined as
\begin{align}
\label{eq:Gf_obstruction_general_M0_2}
[\coho{O}^{(\vv{0})\central}] = [\coho{O}^{(\vv{0})} \otimes \cbd ( \hat{e}_{\vv{1}}^{\mathsf{b}^{(\vv{0})}} \otimes \bar{\hat{e}}_{\vv{1}}^{\central}  )] \in H^3_{[\widehat{\rho}^{(\vv{0})}]}(G, \widehat{\mathcal{A}}_{\vv{0}})
,
\end{align}
where $\beta_{\psi_{\vv{0}}}^{(0)} = (-1)^{\mathsf{b}^{(\vv{0})}}$, and $\hat{e}_{\vv{1}} \in \widehat{{\bf A}}_{\vv{1}}$ is an arbitrary choice of superAbelian vortex supersector.
We can verify that this definition has the appropriate properties using the same arguments used for FMTCs.

When the $[\coho{O}^{(\vv{0})\central}]$ obstruction is trivial, the fermionic symmetry fractionalization classes for $\mathcal{G}^{\eff}$ and $[\rho^{(\vv{0})}]$ are torsorially classified by 
\begin{align}
H_{[\widehat{\rho}^{(\vv{0})}]}^{2}(G,\widehat{\mathcal{A}}_{\vv{0}} )
.
\end{align}
Clearly this describes the symmetry fractionalization of the quasiparticles of the fermionic topological phase.

It is worth considering the case $\widehat{{\bf A}}_{\vv{1}} \neq \varnothing$ in more detail.
We first recall that the symmetry action $\widehat{\rho}$ on supersectors of vortices, and hence $\widehat{{\bf A}}_{\vv{1}}$, is determined entirely by the symmetry action $\rho^{(0)}$ on $\fMTC_{\vv{0}}$.
Using $\hat{e}_{\vv{1}} \otimes \hat{e}_{\vv{1}} = \hat{h}_{\vv{0}}$ and the 1-cochain
\begin{align}
\widehat{\coho{H}}_{\widehat{\rho}}({\bf g}) &= \hat{e}_{\vv{1}} \otimes \widehat{\rho}_{\bf g}(\bar{\hat{e}}_{\vv{1}} ) \in C^1(G,\widehat{\mathcal{A}}_{\vv{0}})
,
\end{align}
we can similarly write the obstruction in the convenient form
\begin{align}
\label{eq:O_central_cups_M0}
\coho{O}^{(\vv{0})\central} &=  \coho{O}^{(\vv{0})} \otimes \cbd \hat{e}_{\vv{1}}^{\mathsf{b}^{(\vv{0})}} \otimes \widehat{\coho{H}}_{\widehat{\rho}}^{\tilde{\central}} \otimes  \hat{h}_{\vv{0}}^{\tilde{\central} \cup_{1} \tilde{\central} }
.
\end{align}
In this case, for a fractionalization class specified by $\rho$ and $\eta^{(\vv{0})}_{a_{\vv{0}}}({\bf g,h}) = \beta^{(\vv{0})}_{a_{\vv{0}}}({\bf g,h}) M_{a_{\vv{0}} \cohosub{w}^{(\vv{0})}}$ to have a given $\central$, it must correspond to a 2-cochain of the form $\coho{w}^{(\vv{0})} = \coho{y}^{(\vv{0})} \otimes \hat{e}_{\vv{1}}^{\central} \otimes \bar{\hat{e}}_{\vv{1}}^{\mathsf{b}^{(\vv{0})}}$, where $\coho{y}^{(\vv{0})} \in C^2(G, \widehat{\mathcal{A}}_{\vv{0}})$ satisfies the obstruction condition
\begin{align}
\cbd\coho{y}^{(\vv{0})} &= \coho{O}^{(\vv{0})\central}
.
\end{align}

When $\mathcal{A}_{\vv{1}} \neq \varnothing$, we can choose $\hat{e}_{\vv{1}}$ to be the supersector containing the $e_{\vv{1}}$ chosen in the definition of ${\coho{O}}^{\central}$.
With this choice, it is clear that
\begin{align}
\coho{O}^{(\vv{0})\central} &= \widehat{\coho{O}^{\central}}
.
\end{align}

We can also define a relative obstruction $[\coho{O}^{(\vv{0}) \central - \central'}_{r}]$ to manifesting $\central$ relative to $\central'$.
In the case where $\widehat{{\bf A}}_{\vv{1}} = \varnothing$, this is defined to be
\begin{align}
\label{eq:Gf_obstruction_relative_M0_1}
\left[\coho{O}^{(\vv{0}) \central - \central'}_{r}\right] = (\central - \central') \in Z^2(G,\mathbb{Z}_2^{\eff})
,
\end{align}
while in the case where $\widehat{{\bf A}}_{\vv{1}} \neq \varnothing$, it is
\begin{align}
\label{eq:Gf_obstruction_relative_M0_2}
\left[\coho{O}^{(\vv{0}) \central - \central'}_{r}\right] = \left[\cbd \left( \hat{e}_{\vv{1}}^{\central'} \otimes \bar{\hat{e}}_{\vv{1}}^{\central} \right) \right] \in H^3_{[\widehat{\rho}^{(\vv{0})}]}(G, \widehat{\mathcal{A}}_{\vv{0}})
.
\end{align}
When the fractionalization obstruction vanishes, $[{\coho{O}}^{(\vv{0})}] = [\widehat{\I}_{\vv{0}}]$, we can pick an arbitrary unobstructed fractionalization class with corresponding $\dot{\coho{w}}$, and alternatively define the obstruction to manifesting $\central$ on $\fMTC_{\vv{0}}$ with $[\rho^{(\vv{0})}]$ to be $[{\coho{O}}^{(\vv{0})\central}] = \left[\coho{O}^{(\vv{0}) \central - \dot{\central}}_{r}\right]$.

Finally, we consider the extension of fermionic symmetry fractionalization of a SMTC $\fMTC_{\vv{0}}$ to one of its $\mathbb{Z}_2^{\eff}$ extensions $\fMTC$.
The obstruction for such an extension under simplifying conditions was discussed in Ref.~\onlinecite{Fidkowski2018}; we consider the general case and obtain the classification in addition to the obstruction.
For this, we assume that we have a fermionic symmetry action $[\rho^{(\vv{0})}]$ on $\fMTC_{\vv{0}}$ that extends to a fermionic symmetry action $[\rho]$ on $\fMTC$ (see Sec.~\ref{sec:symmetry_action} for the corresponding obstruction and classification of this symmetry action extension).
We also assume $[{\coho{O}}^{(\vv{0})}]$ is trivial and pick a symmetry fractionalization class on $\fMTC_{\vv{0}}$ specified by $\rho^{(\vv{0})}$ and $\eta^{(\vv{0})}_{a_{\vv{0}}}({\bf g,h})$.
This choice also fixes the fermionic symmetry group $\mathcal{G}^{\eff}$.
We pick the representative action $\rho$ such that $\res_{\fMTC_{\vv{0}}} (\rho_{\bf g}) = \rho^{(\vv{0})}_{\bf g}$ and the corresponding $\beta^{(\vv{0})}_{a_{\vv{0}}}$ such that $\beta_{a_{\vv{0}}}({\bf g,h}) = \beta^{(\vv{0})}_{a_{\vv{0}}}({\bf g,h})$.
(We note that it is possible to have valid choices of $\beta^{(\vv{0})}_{a_{\vv{0}}}$ that are not restrictions of valid choices of $\beta_{a_{\vv{x}}}$.)
This implies $\Omega_{a_{\vv{0}}} = \Omega^{(\vv{0})}_{a_{\vv{0}}}$ and hence $\widehat{\coho{O}} ={\coho{O}}^{(\vv{0})}$, i.e. $M_{a_{\vv{0}} \cohosub{O}} = M_{a_{\vv{0}} \cohosub{O}^{(\vv{0})}}$.
We write $\coho{w}^{(\vv{0})}({\bf g,h})$ given by $M_{a_{\vv{0}} \cohosub{w}^{(\vv{0})}({\bf g,h})} = \eta^{(\vv{0})}_{a_{\vv{0}}}({\bf g,h}) / \beta^{(\vv{0})}_{a_{\vv{0}}}({\bf g,h})$, which satisfies $\coho{O}^{(\vv{0})} = \cbd \coho{w}^{(\vv{0})}$.
With these choices, the symmetry fractionalization extension problem can be expressed as determining whether there exist $\coho{w} \in C^{2}(G,\mathcal{A})$ such that $\cbd \coho{w} = \coho{O}$ and $\widehat{\coho{w}}({\bf g,h}) = \coho{w}^{(\vv{0})}({\bf g,h})$.

We proceed by defining an arbitrary section on the set of supersectors of topological charge $\ell : \widehat{\fMTC} \to \fMTC$, such that $\widehat{\ell(\hat{a}_{\vv{x}})} = \hat{a}_{\vv{x}}$.
We recall that $\widehat{{\bf A}}_{\vv{1}} \neq \widehat{\mathcal{A}}_{\vv{1}}$ if and only if there are $\sigma$-type superAbelian vortices, in which case $\mathcal{A}_{\vv{1}} = \varnothing$.
This allows for the possibility that $\coho{w}^{(\vv{0})} \in C^{2}(G,\widehat{{\bf A}})$, but  $\coho{w}^{(\vv{0})} \notin C^{2}(G,\widehat{\mathcal{A}})$, in which case the corresponding symmetry fractionalization class on $\fMTC_{\vv{0}}$ cannot be extended to a symmetry fractionalization class of $\fMTC$.
Thus, a necessary condition for extending the fractionalization class to $\fMTC$ is that $\coho{w}^{(\vv{0})} \in C^{2}(G,\widehat{\mathcal{A}})$, so that $\ell(\coho{w}^{(\vv{0})}) \in C^{2}(G,\mathcal{A})$.

When $\coho{w}^{(\vv{0})} \in C^{2}(G,\widehat{\mathcal{A}})$, we define
\begin{align}
& {\coho{O}}^{\eta}({\bf g,h,k}) = \coho{O}({\bf g,h,k}) \otimes \cbd \overline{\ell(\coho{w}^{(\vv{0})})}({\bf g,h,k}) \notag \\
& \qquad = \coho{O}({\bf g,h,k}) \otimes \,^{\bf g}\overline{\ell(\coho{w}^{(\vv{0})}({\bf h,k}))} \otimes {\ell(\coho{w}^{(\vv{0})}({\bf gh,k}))}
\notag \\
& \qquad \qquad \qquad \otimes \overline{\ell(\coho{w}^{(\vv{0})}({\bf g,hk}))} \otimes {\ell(\coho{w}^{(\vv{0})}({\bf g,h}))}
,
\end{align}
which we write more compactly as
\begin{align}
{\coho{O}}^{\eta} = \coho{O}\otimes \cbd \overline{\ell(\coho{w}^{(\vv{0})})}
.
\label{eq:O_extension}
\end{align}

We can see that ${\coho{O}}^{\eta}({\bf g,h,k}) \in \mathbb{Z}_{2}^{\psi}$, since
\begin{align}
M_{a_{\vv{0}} {\cohosub{O}}^{\eta}({\bf g,h,k})} &= M_{a_{\vv{0}} {\cohosub{O}}({\bf g,h,k})} M_{a_{\vv{0}} \cbd \cohosub{w}^{(\vv{0})}({\bf g,h,k}) }^{\ast} =1
\end{align}
for all $a_{\vv{0}} \in \fMTC_{\vv{0}}$.
It is clear that $\cbd {\coho{O}}^{\eta}({\bf g,h,k}) = \I_{\vv{0}}$, and hence ${\coho{O}}^{\eta} \in Z^3(G,\mathbb{Z}_{2}^{\psi})$.

We now show that the cohomology class
\begin{align}
\label{eq:obstructioneta}
[{\coho{O}}^{\eta}] \in H^3(G,\mathbb{Z}_{2}^{\psi})
\end{align}
is an invariant of the quasiparticle fractionalization class $[\rho^{(\vv{0})} , \eta^{(\vv{0})}]$ on $\fMTC_{\vv{0}}$ and the extended symmetry action $[\rho]$ on $\fMTC$, corresponding to an obstruction to extending this quasiparticle fractionalization class to a fractionalization on $\fMTC$ with $[\rho]$.

We first verify that the various gauge choices in the definition of ${\coho{O}}^{\eta}$ only change it by coboundaries in $B^3(G,\mathbb{Z}_{2}^{\psi})$.
We observe that a different arbitrary choice of section $\acute{\ell}(\hat{a}_{\vv{x}}) = \psi^{f(\hat{a}_{\vv{x}})} \otimes {\ell}(\hat{a}_{\vv{x}})$, where $f: \widehat{\fMTC} \to \mathbb{Z}_{2}$, only changes the definition by a 3-coboundary
\begin{align}
\acute{\coho{O}}^{\eta} &= \coho{O} \otimes \cbd \overline{\acute{\ell}(\coho{w}^{(\vv{0})})} = {\coho{O}}^{\eta} \otimes  \cbd \psi_{\vv{0}}^{ f(\cohosub{w}^{(\vv{0})})}
,
\end{align}
since $\psi_{\vv{0}}^{ f(\cohosub{w}^{(\vv{0})}({\bf g,h}) ) )} \in C^{2}(G,\mathbb{Z}_{2}^{\psi})$.

It is easy to see that symmetry action gauge transformations leave $\check{\coho{O}}^{\eta} = {\coho{O}}^{\eta}$ unchanged, because $\check{\coho{O}} = \coho{O}$ and $\check{\coho{w}}^{(\vv{0})} = {\coho{w}}^{(\vv{0})}$.
We emphasize that applying a symmetry action gauge transformation to $\rho$ requires the that same transformation be applied to $\eta^{(\vv{0})}$ in order to represent the same the quasiparticle fractionalization class.
That is, when we change $\rho$ to $\check{\rho}_{\bf g}$, we should also change $\eta^{(\vv{0})}$ to $\check{\eta}^{(\vv{0})}$, as in Eq.~\eqref{eq:eta_check}.

The redefinitions of $\beta_{a_{\vv{x}}}$ to $\breve{\beta}_{a_{\vv{x}}} = \nu_{a_{\vv{x}}} \beta_{a_{\vv{x}}}$ gives $\breve{\coho{O}} = \coho{O} \otimes \cbd \coho{v}$ and $\breve{\coho{w}}^{(\vv{0})} = \hat{\coho{v}} \otimes {\coho{w}}^{(\vv{0})}$.
From this, we find
\begin{align}
\breve{\coho{O}}^{\eta} &= \coho{O} \otimes \cbd \coho{v}  \otimes \cbd \overline{\ell(\hat{\coho{v}} \otimes \coho{w}^{(\vv{0})})}
\notag \\
&= {\coho{O}}^{\eta} \otimes \cbd \left( \coho{v} \otimes \overline{\ell(\hat{\coho{v}})} \otimes \psi_{\vv{0}}^{n_{\ell}(\hat{\cohosub{v}}, \cohosub{w}^{(\vv{0})} ) } \right)
,
\end{align}
where $\coho{v} \otimes \overline{\ell(\hat{\coho{v}})} \otimes \psi_{\vv{0}}^{n_{\ell}(\hat{\cohosub{v}}, \cohosub{w}^{(\vv{0})} )} \in C^{2}(G,\mathbb{Z}_{2}^{\psi})$.
Here, we defined the function $n_{\ell}: \widehat{\mathcal{A}} \times \widehat{\mathcal{A}} \to \mathbb{Z}_{2}$ by
\begin{align}
\label{eq:ell_relation}
\ell(\hat{a}_{\vv{x}} \otimes \hat{b}_{\vv{y}}) = \psi_{\vv{0}}^{n_{\ell}(\hat{a}_{\vv{x}} , \hat{b}_{\vv{y}})} \otimes \ell(\hat{a}_{\vv{x}} ) \otimes \ell(\hat{b}_{\vv{y}})
,
\end{align}
which depends on $\fMTC$ and the particular choice of $\ell$.

Additionally, we consider redefinitions of the symmetry localization, which correspond to changing $\eta^{(\vv{0})}$ to $\grave{\eta}^{(\vv{0})}$ as in Eq.~\eqref{eq:eta_grave} with $\zeta^{(\vv{0})}_{a_{\vv{0}}} = M^{\ast}_{a_{\vv{0}} , \cohosub{z}^{(\vv{0})}}$, and $\coho{w}^{(\vv{0})}$ to $\grave{\coho{w}}^{(\vv{0})} = \coho{w}^{(\vv{0})} \otimes \cbd \coho{z}^{(\vv{0})}$ for $\coho{z}^{(\vv{0})} \in C^{1}(G,\widehat{\mathcal{A}}_{\vv{0}})$.
These yield the transformation
\begin{align}
\grave{\coho{O}}^{\eta} &= \coho{O} \otimes \cbd \overline{\ell(\coho{w}^{(\vv{0})} \otimes \cbd \coho{z}^{(\vv{0})} )}
\notag \\
&= {\coho{O}}^{\eta} \otimes \cbd \left( \coho{T}_{\rho^{(0)}}(\coho{z}^{(\vv{0})}) \otimes \psi_{\vv{0}}^{m_{\ell}( \cohosub{w}^{(\vv{0})}, \cohosub{z}^{(\vv{0})})} \right)
,
\end{align}
where $\coho{T}_{\rho^{(0)}}(\coho{z}^{(\vv{0})}) \otimes \psi_{\vv{0}}^{m_{\ell}(\cohosub{w}^{(\vv{0})}, \cohosub{z}^{(\vv{0})})} \in C^{2}(G,\mathbb{Z}_{2}^{\psi})$ is obtained by multiple applications of Eq.~\eqref{eq:ell_relation} that separate $\coho{w}^{(\vv{0})}$ from $\cbd \coho{z}^{(\vv{0})}$ and then convert $\ell(\cbd \coho{z}^{(\vv{0})} )$ into $\coho{T}_{\rho^{(0)}}(\coho{z}^{(\vv{0})}) \otimes \cbd \overline{\ell( \coho{z}^{(\vv{0})} )}$.
Here, we have defined the important quantity $\coho{T}_{\rho^{(0)}_{\bf g}}(\hat{a}_{\vv{0}} ) \in \mathbb{Z}_{2}^{\psi}$ by
\begin{align}
\coho{T}_{\rho^{(0)}_{\bf g}}(\hat{a}_{\vv{0}} ) &= \overline{ \rho_{\bf g}^{(\vv{0})} \left(\ell( \hat{a}_{\vv{0}}  ) \right)} \otimes \ell\left(  \widehat{\rho}_{\bf g}^{(\vv{0})} ( \hat{a}_{\vv{0}}  ) \right)
,
\end{align}
and we generally write $\coho{T}_{\rho^{(0)}}(\coho{b}^{(0)})({\bf g}_{1},\ldots ,{\bf g}_{n+1}) = \coho{T}_{\rho^{(0)}_{{\bf g}_{1}}}( \coho{b}^{(0)}({\bf g}_{2},\ldots ,{\bf g}_{n+1}) )$ for any $\coho{b}^{(0)} \in C^{n}(G,\widehat{\mathcal{A}}_{\vv{0}})$.

In order to see that $[{\coho{O}}^{\eta}]$ defines an obstruction to extending the quasiparticle symmetry fractionalization to $\fMTC$ with $[\rho]$, we observe that the condition $\widehat{\coho{w}}({\bf g,h}) = \coho{w}^{(\vv{0})}({\bf g,h})$ is equivalent to writing ${\coho{w}}({\bf g,h}) = \coho{p}({\bf g,h}) \otimes \ell(\coho{w}^{(\vv{0})}({\bf g,h}) )$ for some $\coho{p}({\bf g,h}) \in C^2(G,\mathbb{Z}_{2}^{\psi})$.
With this, the condition $\coho{O}=  \cbd \coho{w}$ becomes the obstruction condition
\begin{align}
{\coho{O}}^{\eta}&= \cbd \coho{p}
.
\end{align}
This condition can be satisfied if and only if ${\coho{O}}^{\eta} \in B^3(G,\mathbb{Z}_{2}^{\psi})$, i.e. when $[{\coho{O}}^{\eta}] = [\I_{\vv{0}}]$ is trivial.

When the obstruction $[{\coho{O}}^{\eta}]$ is trivial, we find that the symmetry fractionalization class on $\fMTC_{\vv{0}}$ specified by $[\rho^{(\vv{0})}, \eta^{(\vv{0})}]$ can be extended to a symmetry fractionalization on $\fMTC$ with $[\rho]$.
From this, it is clear that if $[{\coho{O}}^{\eta}]$ is trivial, then $[\coho{O}]$ is trivial, since it establishes the existence of fractionalization on $\fMTC$ with $[\rho]$.
Moreover, when $\eta_{a_{\vv{x}}}({\bf g,h})$ corresponds to such an extension of the quasiparticles' fractionalization class, we see that ${\eta}'_{a_{\vv{x}}}({\bf g,h}) = M_{a_{\vv{x}} \cohosub{q}({\bf g,h})} \eta_{a_{\vv{x}}}({\bf g,h})$ with $\coho{q} \in Z^2(G, \mathbb{Z}_{2}^{\psi})$ corresponds to another such extension.
The equivalences of fractionalization classes on $\fMTC$ that leave the fractionalization class on $\fMTC_{\vv{0}}$ fixed are given by $\coho{q} \in B^2(G, \mathbb{Z}_{2}^{\psi})$.
Thus, the classification of extensions of symmetry fractionalization from $\fMTC_{\vv{0}}$ to $\fMTC$ is given torsorially by
\begin{align}
H^2 (G,\mathbb{Z}_{2}^{\psi})
.
\end{align}
Since these are extensions of the symmetry fractionalization from the quasiparticle theory to the full theory including vortices, we think of the extensions as encoding the symmetry fractionalization of the vorticity.

For the purposes of evaluating and comparing the obstructions, it is useful to rewrite $[{\coho{O}}^{\eta}]$ in terms of previously defined obstruction objects.
It will often be convenient to use a particular choice of lift for which $\ell(\hat{a}_{\vv{0}}) = (\I_{\vv{0}} ,\hat{a}_{\vv{0}}) \in \mathbb{Z}_{2}^{\psi} \times \widehat{\mathcal{A}}_{\vv{0}} = \mathcal{A}_{\vv{0}}$.
When $\mathcal{A}_{\vv{1}} \neq \varnothing$, we further choose $\ell$ such that $\ell(\hat{e}_{\vv{1}}) = e_{\vv{1}}$, so that $\ell(\hat{a}_{\vv{x}}) = (\I_{\vv{0}} ,\hat{a}_{\vv{0}}) \otimes e_{1}^{\vv{x}}$.
We consider the different classes of FMTCs separately, to rewrite them in convenient forms.

In the case where $\mathcal{A}_{\vv{1}} \neq \varnothing$, we had
\begin{align}
\coho{w}^{(\vv{0})} =\coho{y}^{(\vv{0})} \otimes \hat{e}_{\vv{1}}^{\central} \otimes \bar{\hat{e}}_{\vv{1}}^{\mathsf{b}^{(\vv{0})}}
,
\end{align}
where $\coho{y}^{(\vv{0})} \in C^2(G,\widehat{\mathcal{A}}_{\vv{0}})$ such that $\cbd \coho{y}^{(\vv{0})} = \coho{O}^{(0)\central} $.
It is often convenient to choose a gauge in which $\mathsf{b}^{(\vv{0})} = 0$
With this, we find
\begin{align}
\label{eq:Oeta_case1}
[\coho{O}^{\eta}] &= [\coho{O}^{\central} \otimes \overline{\ell(\coho{O}^{(0)\central} )} \otimes \coho{T}_{\rho^{(0)}}( \coho{y}^{(0)} ) ]
,
\end{align}
where $\coho{T}_{\rho^{(0)}}( \coho{y}^{(0)} )({\bf g,h,k}) = \coho{T}_{\rho^{(0)}_{\bf g}}( \coho{y}^{(0)}({\bf h,k}) )$.
We note that when $[\central]=[\vv{0}]$, this becomes
\begin{align}
\label{eq:Oeta_case1_central0}
[\coho{O}^{\eta}] &= [\coho{O} \otimes \overline{\ell(\coho{O}^{(0)})} \otimes \coho{T}_{\rho^{(0)}}( \coho{y}^{(0)} ) ]
.
\end{align}

In the case where $\mathcal{A}_{\vv{1}} = \varnothing$ and $\widehat{\bf A}_{\vv{1}} \neq \varnothing$, we also had 
\begin{align}
\coho{w}^{(\vv{0})} =\coho{y}^{(\vv{0})} \otimes \hat{e}_{\vv{1}}^{\central} \otimes \bar{\hat{e}}_{\vv{1}}^{\mathsf{b}^{(\vv{0})}}
.
\end{align}
One can also gauge fix $\mathsf{b}^{(\vv{0})}$ in this case, but we may not want to set it to zero when extending to $\fMTC$.
The obstruction $[\coho{O}^{\central}] = ([\coho{O}], (\central - \phi))$ requires fractionalization classes on $\fMTC$ to have $\central = \mathsf{b}= \phi$.
In other words, we can only extend $\coho{w}^{(0)}$ to a fractionalization class on vortices when $\mathsf{b}^{(\vv{0})} = \central$.
In this case, ${\coho{w}}^{(\vv{0})} = \coho{y}^{(0)} \in C^{2}(G,\widehat{\mathcal{A}}_{\vv{0}})$, and we find
\begin{align}
\label{eq:Oeta_case2}
[\coho{O}^{\eta}] &= [\coho{O} \otimes \overline{\ell(\coho{O}^{(0)} )} \otimes \coho{T}_{\rho^{(0)}}( \coho{w}^{(0)} ) ]
.
\end{align}

In the case where $\widehat{\bf A}_{\vv{1}} = \varnothing$, we had $[\coho{O}^{(0) \central}] = ([\coho{O}^{(0)}],(\central - \phi^{(\vv{0})})$, so ${\coho{w}}^{(\vv{0})}$ automatically has $\central = \phi^{(\vv{0})}$ and $\coho{w}^{(0)} \in C^{2}(G,\widehat{\mathcal{A}}_{\vv{0}})$.
For this, we again find
\begin{align}
\label{eq:Oeta_case3}
[\coho{O}^{\eta}] &= [\coho{O} \otimes \overline{\ell(\coho{O}^{(0)} )} \otimes \coho{T}_{\rho^{(0)}}( \coho{w}^{(0)} ) ]
.
\end{align}

When the fractionalization obstruction $[\coho{O}]$ is trivial, we can provide an equivalent definition of $[\coho{O}^{\eta}]$.
In this case, there exists a $\dot{\coho{w}} \in C^{2}(G,\mathcal{A})$ such that $\cbd \dot{\coho{w}} = \coho{O}$.
We use this to define a quasiparticle fractionalization corresponding to $\dot{\coho{w}}^{(\vv{0})} = \widehat{\dot{\coho{w}}}$, which is guaranteed to be extendible to a fractionalization on $\fMTC$.
Then we use
\begin{align}
\label{eq:obstructioneta_alt}
{\coho{O}}^{\eta}({\bf g,h,k}) &= \cbd \ell( \dot{\coho{w}}^{(\vv{0})} \otimes \overline{\coho{w}}^{(\vv{0})})({\bf g,h,k})
,
\end{align}
which is equivalent to Eq.~\eqref{eq:O_extension}, up to an element of $B^3(G,\mathbb{Z}_{2}^{\psi})$.
It is straightforward to check that this definition is independent of the arbitrary choice of $\dot{\coho{w}}$, since $\cbd \ell( \dot{\coho{w}}^{(\vv{0})} \otimes \overline{\ddot{\coho{w}}}^{(\vv{0})})$ is a coboundary for another choice $\ddot{\coho{w}}$.
We note that generally $\dot{\coho{w}}^{(\vv{0})} \otimes \overline{\coho{w}}^{(\vv{0})} \in Z_{[\widehat{\rho}]}^{2}(G,\widehat{\mathcal{A}})$.
However, if we restrict to a particular fixed value of $\central$, then $\dot{\coho{w}}^{(\vv{0})} \otimes \overline{\coho{w}}^{(\vv{0})} \in Z_{[\widehat{\rho}^{(\vv{0})}]}^{2}(G,\widehat{\mathcal{A}}_{\vv{0}})$, and we find that
\begin{align}
\label{eq:Oeta_fixedw}
[{\coho{O}}^{\eta}] &= [\coho{T}_{\rho^{(0)}}(\dot{\coho{w}}^{(\vv{0})} \otimes \overline{\coho{w}}^{(\vv{0})}) ]
.
\end{align}
This can be understood in terms of the connecting homomorphism of a long exact sequence.

We note that there is a short exact sequence
\begin{align}
\label{eq:short_exact_A}
0 \to \mathbb{Z}_{2}^{\psi} \xrightarrow{i} \mathcal{A}_{\vv{0}} \xrightarrow{p} \widehat{\mathcal{A}}_{\vv{0}} \to 0
,
\end{align}
where $i: \psi \mapsto \psi_{\vv{0}}$ is the inclusion homomorphism and $p : a_{\vv{x}} \mapsto \hat{a}_{\vv{x}}$ is the homomorphism corresponding to mapping topological charges to their supersectors.
This induces a long exact sequence of group cohomology
\begin{align}
\cdots &\xrightarrow{} H_{[\widehat{\rho}^{(\vv{0})}]}^{n-1}(G,\widehat{\mathcal{A}}_{\vv{0}} ) \xrightarrow{\delta_{n-1}} H^n (G,\mathbb{Z}_{2}^{\psi}) \xrightarrow{i_{\ast}} H_{[{\rho}^{(\vv{0})}]}^{n}(G,{\mathcal{A}}_{\vv{0}} )
\notag \\
&\quad \xrightarrow{p_{\ast}} H_{[\widehat{\rho}^{(\vv{0})}]}^{n}(G,\widehat{\mathcal{A}}_{\vv{0}} ) \xrightarrow{\delta_n} H^{n+1}(G,\mathbb{Z}_{2}^{\psi}) \xrightarrow{} \cdots
.
\label{eq:long_exact_sequence}
\end{align}
Here, $i_{\ast}: [\coho{q}] \mapsto [i(\coho{q})]$ and $p_{\ast}: [\coho{t}] \mapsto [p(\coho{t})]$ are the corresponding induced homomorphisms, and the connecting homomorphism $\delta_n$ can be defined using $\coho{T}_{\rho^{(0)}}$ as
\begin{align}
\delta_{n}([\coho{b}^{(0)}]) &= [\cbd \ell(\coho{b}^{(0)})] = [\coho{T}_{\rho^{(0)}}(\coho{b}^{(0)})]
,
\end{align}
where $\coho{b}^{(0)} \in Z^{n}_{[\widehat{\rho}^{(0)}]}(G,\widehat{\mathcal{A}}_{\vv{0}})$.
With this, we see that, for a fixed $\central$, Eq.~\eqref{eq:Oeta_fixedw} is just $[\coho{O}^{\eta}]  = \delta_{2}( [\dot{\coho{w}}^{(\vv{0})} \otimes \overline{\coho{w}}^{(\vv{0})}] )$, when there is an unobstructed $\dot{\coho{w}}^{(\vv{0})}$.
In this case, the unobstructed ${\coho{w}}^{(\vv{0})}$ are those that differ from $\dot{\coho{w}}^{(\vv{0})}$ by an element of $\ker (\delta_{2}) = \im (p_\ast)$.

Now that we have the classifications of fractionalization on FMTCs and SMTCs for a given $\mathcal{G}^{\eff} = \mathbb{Z}_2^{\eff} \times_{\central} G$, as well as the classification of extensions of a fractionalization class from SMTCs to FMTCs, we can examine the relation between all of these in detail.
For this, we consider the long exact sequence of Eq.~\eqref{eq:long_exact_sequence}.
The properties of long exact sequences allows us to infer various relations.
We also use the fact that cohomology groups are Abelian, so all their subgroups are central.

Since $\ker(p_\ast) = \im(i_\ast)$, we know
\begin{align}
\frac{H_{[{\rho}^{(\vv{0})}]}^{2}(G,{\mathcal{A}}_{\vv{0}} ) }{ \im(i_{\ast}) } \cong \im(p_\ast)
.
\end{align}
Since $\ker(p_\ast)$ is a central subgroup of $H_{[{\rho}^{(\vv{0})}]}^{2}(G,{\mathcal{A}}_{\vv{0}} )$, it follows that $H^2_{[\rho^{(\vv{0})}]}(G,\mathcal{A}_\vv{0})$ can be written as a central extension
\begin{align}
H^2_{[\rho^{(\vv{0})}]}(G,\mathcal{A}_\vv{0})
& \cong i_{\ast}\left( H^2 (G,\mathbb{Z}_{2}^{\psi}) \right) \times_{\varepsilon} p_{\ast} \left( H^2_{[\rho^{(\vv{0})}]}(G,\mathcal{A}_\vv{0}) \right)
,
\end{align}
for some $\varepsilon \in Z^2(\im(p_\ast) , \im(i_{\ast}))$.

We see that $p_{\ast} \left( H^2_{[\rho^{(\vv{0})}]}(G,\mathcal{A}_\vv{0}) \right)$ is a central subgroup of $H_{[\widehat{\rho}^{(\vv{0})}]}^{2}(G,\widehat{\mathcal{A}}_{\vv{0}} )$, since $\im(p_\ast) = \ker(\delta_2)$.
As the image of $p_\ast$, we can interpret this subgroup as corresponding to the quasiparticle fractionalization classes which can be extended to fractionalization classes on the full FMTC.
As the kernel of $\delta_2$, we can equivalently interpret it as the subgroup corresponding to vanishing fractionalization extension obstruction $[\coho{O}^{\eta}]$, by recognizing $\delta_2$ as the map that yields the obstruction class $[\coho{O}^{\eta}]$ for a given quasiparticle fractionalization class, as in Eq.~\eqref{eq:obstructioneta_alt}.

We emphasize that $i_{\ast}$ need not be injective, so its image is not necessarily isomorphic to $H^2(G,\mathbb{Z}_2^{\psi})$.
In particular, $i_{\ast}$ could map a nontrivial element of $H^2(G,\mathbb{Z}_2^{\psi})$ to a trivial element of $H^2_{[\rho^{(\vv{0})}]}(G,\mathcal{A}_\vv{0})$.
This means distinct extensions of fractionalization from quasiparticles to vortices may result in equivalent fractionalization classes on the full FMTC $\fMTC$.

The fermionic symmetry action on quasiparticles $[\rho^{(\vv{0})}]$ determines much of the structure in this decomposition of $H^2_{[\rho^{(\vv{0})}]}(G,\mathcal{A}_\vv{0})$.
Recall that $\mathcal{A}_{\vv{0}} = \mathbb{Z}_2^{\psi} \boxtimes \widehat{\mathcal{A}}_{\vv{0}}$ factorizes as a BTC, as does the corresponding group.
However, the corresponding $G$-modules may not factorize, because of nontrivial symmetry action.

In the case where $[\rho^{(\vv{0})}]$ factorizes so that it only acts nontrivially on the $\widehat{\mathcal{A}}_{\vv{0}}$ sector, i.e. $\rho^{(\vv{0})}(a^{(\psi)}, \hat{a}_{\vv{0}}) = (a^{(\psi)}, \widehat{\rho}^{(\vv{0})} (\hat{a}_{\vv{0}}))$, the corresponding $G$-module also factorizes in this manner, and we find that
\begin{align}
H^2_{[\rho^{(\vv{0})}]}(G,\mathcal{A}_\vv{0}) \cong
H^2(G,\mathbb{Z}_2^{\psi}) \times H^2_{[\widehat{\rho}^{(\vv{0})}]}(G,\widehat{\mathcal{A}}_\vv{0})
.
\end{align}

For a more general action $[\rho^{(\vv{0})}]$, we can analyze $\im(i_\ast) \cong H^2(G,\mathbb{Z}_2^{\psi}) / \ker(i_\ast)$ by examining the kernel $\ker(i_\ast) = \im (\delta_1)$.
From the definition of the connecting homomorphism, we see that the elements of $\ker(i_\ast)$ take the form
\begin{align}
[\coho{T}_{\rho^{(0)}}(\coho{b}^{(\vv{0})})({\bf g,h})] &= [\psi_{\vv{0}}^{n_{\bf g}( \cohosub{b}^{(\vv{0})} ({\bf h}) )}]
,
\end{align}
with $\coho{b}^{(\vv{0})} \in Z_{[\widehat{\rho}^{(\vv{0})}]}^{1}(G,\widehat{\mathcal{A}}_{\vv{0}} )$ and $n_{\bf g}$ as function valued in $\mathbb{Z}_{2}$.

When $G=\mathbb{Z}_{2}$, $\ker(i_\ast)$ is always trivial, and hence $\im(i_{\ast}) \cong H^2(G,\mathbb{Z}_2^{\psi})$.
This can be seen through the following properties.
$\coho{b}^{(\vv{0})} \in Z_{[\widehat{\rho}^{(\vv{0})}]}^{1}(\mathbb{Z}_{2},\widehat{\mathcal{A}}_{\vv{0}} )$ gives the constraint that $\widehat{\rho}_{\bf 1}^{(\vv{0})}(\coho{b}^{(\vv{0})} ({\bf 1}) ) = \overline{\coho{b}}^{(\vv{0})} ({\bf 1})$.
$[\rho^{(\vv{0})}]$ being a topological symmetry requires $\theta_{\ell( \cohosub{b}^{(\vv{0})} ({\bf 1}) )} = \theta_{\rho_{\bf 1}^{(\vv{0})} \left(\ell( \cohosub{b}^{(\vv{0})} ({\bf 1}) )\right)} = (-1)^{n_{\bf 1}( \cohosub{b}^{(\vv{0})} ({\bf 1}) )} \theta_{\ell\left(  \widehat{\rho}_{\bf 1}^{(\vv{0})} ( \cohosub{b}^{(\vv{0})} ({\bf 1})  ) \right)}$.
Using the choice of lift $\ell(\hat{a}_{\vv{0}}) = (I_{\vv{0}} ,\hat{a}_{\vv{0}})$ together with the fact that $\theta_{\bar{a}} = \theta_{a}$, this implies $n_{\bf 1}( \coho{b}^{(\vv{0})} ({\bf 1}) ) =0$, and hence $\ker(i_\ast)=\{ \I_{\vv{0}} \}$, as claimed.

We end this subsection by comparing the value of the obstruction $[\coho{O}^{\eta}]$ for different extensions $\fMTC$ with $[\rho]$ and $\fMTC'$ with $[\rho']$ of the same SMTC $\fMTC_{\vv{0}}$ with $[\rho^{(\vv{0})}]$.
We allow for different quasiparticle fractionalization classes $[\rho^{(\vv{0})}, \eta^{(\vv{0})}]$ $[\rho^{(\vv{0})}, \eta^{(\vv{0})\prime}]$ with the same $\central$ for such comparisons.
These comparisons of obstructions $[\coho{O}^{\eta}]$ will later be understood in terms of the obstruction structure of invertible FSET phases, and will be seen to play a role in understanding the classification structure of general FSET phases obtained by stacking with invertible FSET phases.
This also has relevance for understanding anomalous $(2+1)$D FSET phases that can occur at the boundary of a $(3+1)$D FSET phase.~\footnote{In v2 of Ref.~\onlinecite{Bark2021cascade}, it was conjectured that the obstruction class $[\coho{O}^{\eta}]$ is the same for all modular extensions of $\fMTC_{\vv{0}}$ for which there is a valid extension of the symmetry action $[\rho^{(0)}]$.
In v2 of our paper, we computed the dependence of $[\coho{O}^{\eta}]$ on different extensions and showed that this conjecture is generally false.
In light of this dependence, the authors of Ref.~\onlinecite{Bark2021cascade} have revised their statements regarding the independence of the obstructions on modular extensions to involve quotients by terms corresponding to the boundary anomalous FSPT states associated with $(3+1)$D FSPT phases described in Ref.~\onlinecite{Wang2020}.
These terms that they quotient by precisely match the $\nu$ and $\pi$ dependence presented in v4 of our paper, which proves their statements.
}

In order to make precise comparisons for different $[\fMTC, \rho, \eta^{(\vv{0})}]$ and $[\fMTC', \rho', \eta^{(\vv{0})\prime}]$, we need to be precise about the relation between these different theories.
Leaving the representative choices of $\rho^{(\vv{0})}$ and $\beta^{(\vv{0})}$ fixed, the different quasiparticle fractionalization classes described by $\eta^{(\vv{0})}$ and $\eta^{(\vv{0}) \prime}$ correspond to ${\coho{w}}^{(\vv{0})}$ and ${\coho{w}}^{(\vv{0}) \prime} = {\coho{t}}^{(\vv{0})} \otimes {\coho{w}}^{(\vv{0})}$, where ${\coho{t}}^{(\vv{0})} \in Z^2_{[\rho^{(\vv{0})}]}(G,\widehat{\mathcal{A}}_\vv{0})$.

For a fixed $\fMTC$, we can always change the extension $[\rho]$ of the symmetry action $[\rho^{(0)}]$ to $[\rho'] = [\V]^{\pi} \cdot [\rho]$ for any $\pi \in H^{1}(G,\mathbb{Z}_{2}^{\V})$, and these are the only ways we can change the action while preserving $[\rho^{(0)}]$ and $\central$.
In order to leave the representative choice of $\rho^{(0)}$ fixed, we choose to use the representative $\rho' = \V^{\pi} \circ \rho$, since $\res_{\fMTC_{\vv{0}}}(\V) = \openone^{(\vv{0})}$.
This allows us to continue to compare $\eta^{(\vv{0})}$ and $\eta^{(\vv{0})\prime}$ as before, without tracking any modifications due to symmetry action gauge transformations.

We can change the $\mathbb{Z}_{2}^{\eff}$ extension $\fMTC$ by stacking with an invertible fermionic topological phase to obtain a different extension $\fMTC' = \fMTC \fprod \ifo^{(\nu)}$.
However, when we do this, we must also specify how the symmetry action changes accordingly, which requires specifying the symmetry action on the invertible phase $\ifo^{(\nu)}$ with which we stack.
The na\"ive choice (and simplest to calculate) would be to assume the trivial symmetry action $\openone$ on $\ifo^{(\nu)}$.
For a given topological symmetry $\varphi$ on $\fMTC$, stacking in this way yields $(\fMTC' , \zig{\varphi}) = (\fMTC , \varphi) \fprod (\ifo^{(\nu)} , \openone)$.
While the resulting $\zig{\varphi}$ obtained in this way is well-defined, there is no guarantee that $\zig{\rho_{\bf g}}$ obtained from $\rho_{\bf g}$ in this way forms a valid fermionic symmetry action.
Indeed, Eq.~\eqref{eq:Orho_nu_relation} showed it is not always possible to have a fermionic symmetry action $[\rho']$ on $\fMTC'$ that extends the same $[\rho^{(0)}]$ on $\fMTC_{\vv{0}}$ as the fermionic symmetry action $[\rho]$ on $\fMTC$.
When $\nu$ is even, $\fMTC$ and $\fMTC'$ have the same obstruction $[O^{\rho}]' = [O^{\rho}]$ for extending $[\rho^{(0)}]$, meaning the existence of $[\rho]$ implies the existence of $[\rho']$.
In fact, $[\rho']=[\zig{\rho}]$ is a valid fermionic symmetry action on $\fMTC'$ for $\nu$ even, assuming $[\rho]$ is a fermionic symmetry action on $\fMTC$.
When $\nu$ is odd, $[O^{\rho}]' = [\V]^{[\central]}\cdot[O^{\rho}]$, since at least one of the two FMTCs will have $\mathcal{A}_{\vv{1}}=\varnothing$, which implies $\phi = \central$.
Consequently, for $\nu$ odd, there can only be symmetry action extensions $[\rho]$ and $[\rho']$ of the same $[\rho^{(0)}]$ when $[\central]=[\vv{0}]$.
When $[\central] \neq [\vv{0}]$, either none of the modular extensions permit an extension of $[\rho^{(0)}]$, or exactly eight of them (related by $\nu$ even stacking) permit extensions of $[\rho^{(0)}]$.
In this case, we only compare $[\coho{O}^{\eta}]$ for different modular extensions related by $\nu$ even stacking.
For $\nu$ odd with $\central= \cbd \vviso$, $[\rho']=[\V^{\vviso} \circ \zig{\rho}]$ is a valid fermionic symmetry action on $\fMTC'$.
In this case, there is no canonical choice for $\vviso$, since we cannot have $\vviso = \vv{0}$, so an arbitrary choice must be made to make the comparison; any other $\vviso'$ such that $\cbd \vviso' = \cbd \vviso$ could alternatively be used.
(We will later understand the properties described in this paragraph as a reflection of the fact that we must have matching $\central$ in FSET phases that we stack together, and $\ifo^{(\nu)}$ with $\nu$ odd can only have $[\central]=[\vv{0}]$.)

We note that one could generally include the change of the symmetry action by $\V^{\pi}$ in the stacking step, by using $(\ifo^{(\nu)} , \V^{\pi})$ instead of $(\ifo^{(\nu)} , \openone)$.
While this yields the same result (up to gauge equivalences), it makes the condensation calculation significantly more difficult, so we prefer to change the symmetry action as a separate step from the change of the modular extension.

Comparing the obstruction $[\coho{O}^{\eta}]$ defined for $[\fMTC, \rho, \eta^{(\vv{0})}]$ and $[\coho{O}^{\eta}]'$ defined for $[\fMTC', \rho' , \eta^{(\vv{0})\prime}]$, where $\fMTC' = \fMTC \fprod \ifo^{(\nu)}$, $\rho' = \V^{\pi}\circ \zig{\rho}$, and $\eta_{a_{\vv{0}}}^{(\vv{0})\prime} = M_{{a_{\vv{0}}} {\cohosub{t}}^{(\vv{0})}} \, \eta_{a_{\vv{0}}}^{(\vv{0})}$, when $\nu$ is even, we can prove that
\begin{align}
\label{eq:Oeta_relation}
[\coho{O}^{\eta}]' = [ \psi^{\pi \cup \central + \frac{\nu}{2} \cdot \central \cup_1 \central }\otimes \coho{T}_{\rho^{(0)}}( \coho{t}^{(0)} )] \otimes [\coho{O}^{\eta}]
.
\end{align}
When $\nu$ is odd and $\central = \cbd \vviso$, comparing the same quantities, but with $\rho' = \V^{\pi + \vviso} \circ \zig{\rho}$, we can prove that
\begin{align}
\label{eq:Oeta_relation_w0}
[\coho{O}^{\eta}]' = [ \coho{T}_{\rho^{(0)}}( \coho{t}^{(0)} )] \otimes [\coho{O}^{\eta}]
,
\end{align}
for FMTCs with $\widehat{\bf A}_{\vv{1}} \neq \varnothing$.
We conjecture that this same relation is also true for FMTCs with $\widehat{\bf A}_{\vv{1}} = \varnothing$.
We note that $\pi \cup \cbd \vviso = \cbd ( \pi \cup \vviso )$ and $\cbd \vviso \cup_{1} \cbd \vviso = \cbd (\vviso \cup_1 \cbd \vviso + \vviso \cup \vviso)$, so the expressions for all the cases are compatible.
In light of this, one can simply use Eq.~\eqref{eq:Oeta_relation} to describe all cases, with the understanding that $\nu$ odd comparisons only exist when $[\central] = [\vv{0}]$.

Determining the equivalences of the $[\coho{O}^{\eta}]$ obstruction for different $[\fMTC, \rho, \eta^{(\vv{0})}]$ is thus reduced to the problem of determining when the cohomology class
\begin{align}
[ \psi_{\vv{0}}^{\pi \cup \central + \frac{\nu}{2} \cdot \central \cup_1 \central } \otimes \coho{T}_{\rho^{(0)}}( \coho{t}^{(0)} )] \in H^3(G,\mathbb{Z}_{2}^{\psi})
\end{align}
is trivial for some ${\coho{t}}^{(\vv{0})} \in Z^2_{[\rho^{(\vv{0})}]}(G,\widehat{\mathcal{A}}_\vv{0})$.
When $[ \psi^{\pi \cup \central + \frac{\nu}{2} \cdot \central \cup_1 \central }]$ is nontrivial, this indicates that the different theories will not both have extensions of the same quasiparticle fractionalization class; however, it may be possible to have extensions of different fractionalization classes, if there is a trivialization.
This is a matter of determining when $[ \psi^{\pi \cup \central + \frac{\nu}{2} \cdot \central \cup_1 \central }] \in \ker (i_\ast) = \im (\delta_2)$.
We note that it is possible to have individual terms in this expression be nontrivial, yet the combination conspires to yield a trivial total product, leading to equivalences of the obstruction under nontrivial transformations between $\fMTC$ and $\fMTC'$.
It is also possible to have a particular $\central$ for which this term is always nontrivial when $\nu = 2 \text{ mod }4$, allowing for no equivalences of the obstruction across such FMTCs.
In Sec.~\ref{sec:ifo}, we provide simple examples of both of these possibilities using $\fMTC = \ifo^{(\nu)}$ with $G =  \mathbb{Z}_{2} \times \mathbb{Z}_{2}$.

In order to prove the relations in Eqs.~\eqref{eq:Oeta_relation} and \eqref{eq:Oeta_relation_w0}, we can utilize Eqs.~\eqref{eq:O'}, \eqref{eq:O_central_cups}, \eqref{eq:O_central_cups_M0}, \eqref{eq:Oeta_case1}, \eqref{eq:Oeta_case2}, and \eqref{eq:Oeta_case3}.
The only information we still require for this is the dependence of $\coho{O}$ on the choice of $\mathbb{Z}_{2}^{\eff}$ modular extension of $\fMTC_{\vv{0}}$.
We now compute this dependence by performing the condensation calculations associated with the relevant stacking operations.

We first compute $\zig{\varphi}$ from the stacking $(\fMTC' , \zig{\varphi}) = (\fMTC , \varphi) \fprod (\ifo^{(\nu)} , \openone)$ for $\nu$ even.
(One can alternatively use the zesting functor described in Sec.~\ref{sec:defect_torsor_functors}, restricted to the FMTC and symmetry action, to obtain gauge equivalent results.)
We see that the topological charges of $\fMTC$ and $\fMTC'$ are all the same, i.e. $\zig{a_{\vv{x}}} = a_{\vv{x}}$, but their fusion rules are related by
\begin{align}
\label{eq:Mnueven_fusion}
a_{\vv{x}} \otimes' b_{\vv{y}} = \psi_{\vv{0}}^{\frac{\nu}{2} \cdot \mathsf{x} \cdot \mathsf{y}} \otimes a_{\vv{x}} \otimes b_{\vv{y}}
.
\end{align}
Performing a relatively simple condensation calculation, we find the action of $\zig{\varphi}$ on $\fMTC'$ is specified by
\begin{align}
\zig{\varphi}(a_{\vv{x}}) &= \varphi(a_{\vv{x}}) , \\
\left[\zig{U}_{\zig{\varphi}}(a_{\vv{x}} , b_{\vv{y}} ; c_{\vv{x+y}})\right]_{\alpha \beta} & = U_{\varphi}([\psi^{\frac{\nu}{2} \cdot \mathsf{x}\cdot \mathsf{y}} c]_{\vv{x+y}} , \psi_{\vv{0}}^{\frac{\nu}{2} \cdot \mathsf{x}\cdot \mathsf{y}}; c_{\vv{x+y}}) \notag \\
& \times [U_{\varphi}(a_{\vv{x}} , b_{\vv{y}} ; [\psi^{\frac{\nu}{2} \cdot \mathsf{x}\cdot \mathsf{y}}c]_{\vv{x+y}})]_{\alpha \beta}
.
\end{align}
From this, we find
\begin{align}
\zig{\kappa}_{\bf g,h}(a_{\vv{x}} , b_{\vv{y}} ; c_{\vv{x+y}}) & = {\kappa}_{\bf g,h}([\psi^{\frac{\nu}{2} \cdot \mathsf{x}\cdot \mathsf{y}} c]_{\vv{x+y}} , \psi_{\vv{0}}^{\frac{\nu}{2} \cdot \mathsf{x}\cdot \mathsf{y}}; c_{\vv{x+y}}) \notag \\
& \qquad \times {\kappa}_{\bf g,h}(a_{\vv{x}} , b_{\vv{y}} ; [\psi^{\frac{\nu}{2} \cdot \mathsf{x}\cdot \mathsf{y}}c]_{\vv{x+y}})
\notag\\
&= \frac{\beta_{a_{\vv{x}}} ({\bf g,h}) \beta_{b_{\vv{y}}} ({\bf g,h}) }{ \beta_{c_{\vv{x+y}}} ({\bf g,h}) } \beta_{\psi_{\vv{0}}} ({\bf g,h})^{\frac{\nu}{2} \cdot \mathsf{x}\cdot \mathsf{y}} 
.
\label{eq:zig_kappa}
\end{align}

For FMTCs with $\mathcal{A}_{\vv{1}} \neq \varnothing$, we can make a convenient gauge choice that sets $\beta_{\psi_{\vv{0}}} =1$.
In this case, Eq.~\eqref{eq:zig_kappa} yields $\zig{\beta}_{a_{\vv{x}}} = \beta_{a_{\vv{x}}}$, and hence
\begin{align}
\zig{\coho{O}} =\coho{O}
\end{align}
for $\zig{\rho}$ on $\fMTC'$.
Eq.~\eqref{eq:Mnueven_fusion} shows that $h'_{\vv{0}} = \psi_{\vv{0}}^{\frac{\nu}{2}} \otimes h_{\vv{0}}$ for the FMTCs with $\mathcal{A}_{\vv{1}} \neq \varnothing$ that are related by $\nu$ even stacking.
Combining these results with Eqs.~\eqref{eq:O'}, \eqref{eq:O_central_cups}, and \eqref{eq:Oeta_case1} yields
\begin{align}
\label{eq:Ocentral_relation}
[\coho{O}^{\central}]' = [ \psi^{\pi \cup \central + \frac{\nu}{2} \cdot \central \cup_1 \central }] \otimes [\coho{O}^{\central}]
,
\end{align}
as well as Eq.~\eqref{eq:Oeta_relation}.
In particular, $\coho{O}'=\zig{\coho{O}} =\coho{O}$, and we find that the $\psi^{\pi \cup \central}$ and $\psi^{\frac{\nu}{2} \cdot \central \cup_1 \central}$ terms respectively arise from the $\coho{H}_{\rho}^{\tilde{\central}}$ and $h_{\vv{0}}^{\tilde{\central} \cup_{1} \tilde{\central}}$ terms in Eq.~\eqref{eq:O_central_cups}.

For FMTCs with $\mathcal{A}_{\vv{1}} = \varnothing$, we necessarily have $\beta_{\psi_{\vv{0}}} =(-1)^{\central}$.
In this case, Eq.~\eqref{eq:zig_kappa} yields
\begin{align}
\zig{\beta}_{a_{\vv{x}}} = i^{\frac{\nu}{2} \cdot \mathsf{x} \cdot \central} \beta_{a_{\vv{x}}}
.
\end{align}
It follows that
\begin{align}
\zig{\Omega}_{a_{\vv{x}}} &= (-1)^{\frac{\nu}{2} \cdot \mathsf{x} \cdot \central \cup_1 \central} {\Omega}_{a_{\vv{x}}},
\\
\zig{\coho{O}} &= \psi^{\frac{\nu}{2} \cdot \central \cup_1 \central} \otimes \coho{O}
\end{align}
for $\zig{\rho}$ on $\fMTC'$.
Combining these results with Eqs.~\eqref{eq:O'}, \eqref{eq:Oeta_case2}, and \eqref{eq:Oeta_case3} yields Eq.~\eqref{eq:Oeta_relation}.
In this case, the $\psi^{\pi \cup \central}$ and $\psi^{\frac{\nu}{2} \cdot \central \cup_1 \central}$ terms arise from
\begin{align}
\coho{O}' &= \psi^{\pi \cup \central} \otimes \zig{\coho{O}} \notag \\
&= \psi^{\pi \cup \central} \otimes \psi^{\frac{\nu}{2} \cdot \central \cup_1 \central} \otimes \coho{O}
.
\end{align}
Remarkably, the same relative factors arise for the $\mathcal{A}_{\vv{1}} \neq \varnothing$ and $\mathcal{A}_{\vv{1}} = \varnothing$ cases in rather different ways.

Next, we compute $\zig{\varphi}$ from the stacking $(\fMTC' , \zig{\varphi}) = (\fMTC , \varphi) \fprod (\ifo^{(\nu)} , \openone)$ for $\nu$ odd when $\central = \vv{0}$.
This case is more complicated and we have only been able to complete the stacking calculation for FMTCs with $\widehat{\bf A}_{\vv{1}} \neq \varnothing$.
In this case, the FMTC either has all $v$-type vortices ($\mathcal{A}_{\vv{1}} \neq \varnothing$) or all $\sigma$-type vortices ($\mathcal{A}_{\vv{1}} = \varnothing$).
We will assume $\fMTC$ to have all $v$-type vortices, which implies $\fMTC'$ has all $\sigma$-type vortices.
Since stacking with an invertible topological phase is an invertible operation, this is sufficient to establish the result for all $\nu$.

We use the same notation used in the proof of Eq.~\eqref{eq:Orho_nu_relation}, but in this case, $\fMTC$ has only $v$-type vortices.
We also introduce the quantity $m(a,b;c) \in \mathbb{Z}_{2}$, which is defined to make the fusion of representative choices of $\zig{a_{\vv{x}}}$, $\zig{b_{\vv{y}}}$, and $\zig{c_{\vv{x+y}}}$ permissible.
For example, we write a fusion vertex of $\fMTC'$ as
\begin{align}
\pbpba = \pbpbb
\,\,
.
\end{align}
For the post-condensation symmetry action, the action on quasiparticles and ($\sigma$-type) vortices of $\fMTC'$ is again
\begin{align}
\zig{\varphi}\left( \zig{a_{\vv{0}}} \right) &= \zig{\varphi(a_{\vv{0}}) }
,
\\
\zig{\varphi}\left( \zig{\hat{a}_{v}} \right) &= \zig{ \widehat{\varphi}(\hat{a}_{v})}
.
\end{align}
For the post-condensation symmetry action on vertex states, we have
\begin{widetext}
\begin{align}
\zig{\varphi}\left( \pbpba \right)&= \sum_{\lambda} U_{\varphi} (\psi^{m} \varphi(\hat{c}^{r}_{v}) , \psi_{\vv{0}}^{m}; \varphi(\hat{c}^{r}_{v}) ) [U_{\varphi}( \varphi(a_{\vv{0}}) , \varphi(\hat{b}^{r}_{v}) ; \psi^{m(a,b;c)} \varphi(\hat{c}^{r}_{v})]_{\alpha \lambda} \pbpbc
\notag \\
&=\sum_{\beta} \left[\zig{U}_{\zig{\varphi}}\left( \zig{\varphi}(\zig{a_{\vv{0}}}) , \zig{\varphi}(\zig{b_{\vv{1}}}); \zig{\varphi}(\zig{c_{\vv{1}}}) \right)\right]_{\alpha \beta} \pbpbd
,
\end{align}
which yields
\begin{align}
\left[\zig{U}_{\zig{\varphi}}\left( \zig{\varphi}(\zig{a_{\vv{0}}}) , \zig{\varphi}(\zig{b_{\vv{1}}}); \zig{\varphi}(\zig{c_{\vv{1}}}) \right)\right]_{\alpha \beta} =& \sum_{\lambda} [U_{\varphi}( \varphi(a_{\vv{0}}) , \varphi(\hat{b}^{r}_{v}) ; \psi^{m(a,b;c)} \varphi(\hat{c}^{r}_{v})]_{\alpha \lambda}
\notag \\
& \qquad  \times U_{\varphi} (\psi^{m(a,b;c)} \varphi(\hat{c}^{r}_{v}) , \psi_{\vv{0}}^{m(a,b;c)}; \varphi(\hat{c}^{r}_{v}) )
\left[F^{\varphi(a_{\vv{0}}) , \hat{\varphi}(\hat{b}_{v})^{r} , \psi_{\vv{0}}^{n_{\varphi}(b)}}_{\psi^{m(a,b;c) } \varphi(\hat{c}^{r}_{v})  } \right]^{-1}_{\lambda \beta}
.
\end{align}
It will be convenient to make a vertex basis gauge choice for $\fMTC$ such that $[F^{a_{\vv{x}} b_{\vv{y}} \psi_{\vv{0}}}_{ [\psi c]_{\vv{x+y}} }] = \openone$, and a symmetry action gauge choice such that $U_{\varphi} (a_{\vv{x}}  , \psi_{\vv{0}}; [\psi a]_{\vv{x}} ) =1$;
this is always possible when $\mathcal{A}_{\vv{1}} \neq \varnothing$.
With this convenient choice of gauge, we find
\begin{align}
\left[\zig{U}_{\zig{\varphi}}\left( \zig{\varphi}(\zig{a_{\vv{0}}}) , \zig{\varphi}(\zig{b_{\vv{1}}}); \zig{\varphi}(\zig{c_{\vv{1}}}) \right)\right] 
&= \left[U_{\varphi}( \varphi(a_{\vv{0}}) , \varphi(\hat{b}^{r}_{v}) ; \psi^{m(a,b;c)} \varphi(\hat{c}^{r}_{v}))\right]
,
\\
\left[\zig{U}_{\zig{\varphi}}\left( \zig{\varphi}(\zig{a_{\vv{1}}}) , \zig{\varphi}(\zig{b_{\vv{0}}}); \zig{\varphi}(\zig{c_{\vv{1}}}) \right)\right] 
&= \left[U_{\varphi}( \varphi(\hat{a}^{r}_{v} ) , \varphi(b_{\vv{0}} ) ; \psi^{m(a,b;c)} \varphi(\hat{c}^{r}_{v})) \right] 
[R^{\varphi(\hat{a}^{r}_{v} ) \varphi(b_{\vv{0}} ) }_{\psi^{m(a,b;c)} \varphi(\hat{c}^{r}_{v})} ]^{-1} 
[R^{\hat{\varphi}(\hat{a}_{v} )^{r} \varphi(b_{\vv{0}} ) }_{\psi^{m(a,b;c)+ n_{\varphi}(a) } \varphi(\hat{c}^{r}_{v})} ]
,
\\
\left[\zig{U}_{\zig{\varphi}}\left( \zig{\varphi}(\zig{a_{\vv{1}}}) , \zig{\varphi}(\zig{b_{\vv{1}}}); \zig{\varphi}(\zig{c_{\vv{0}}}) \right)\right] 
&= \left[U_{\varphi}( \varphi(\hat{a}^{r}_{v}) , \varphi(\hat{b}^{r}_{v}) ; \psi^{m(a,b;c)} \varphi(c_{\vv{0}}) ) \right]
[R^{\varphi(\hat{a}^{r}_{v} ) \hat{\varphi}(\hat{b}_{v} )^{r} }_{\psi^{m(a,b;c)+n_{\varphi}(b)} \varphi(c_{\vv{0}})} ]^{-1} 
\notag \\
& \qquad \times [R^{\hat{\varphi}(\hat{a}_{v} )^{r} \hat{\varphi}(\hat{b}_{v} )^{r} }_{\psi^{m(a,b;c)+n_{\varphi}(a)+n_{\varphi}(b)} \varphi(c_{\vv{0}})} ]
(-1)^{n_{\varphi}(a)(m(a,b;c)+n_{\varphi}(a)+n_{\varphi}(b))} i^{-\nu n_{\varphi}(a)}
.
\end{align}
\end{widetext}

Using these expressions to relate $\zig{\rho_{\bf g}}$ and ${\rho_{\bf g}}$ for each ${\bf g} \in G$, it follows that
\begin{align}
\label{eq:zig_kappa_nu_odd}
\zig{\kappa}_{\bf g,h}(\zig{a_{\vv{x}}} , \zig{b_{\vv{y}}}; \zig{c_{\vv{x+y}}}) & = \kappa_{\bf g,h}(a_{\vv{x}} , b_{\vv{y}} ; c_{\vv{x+y}})
,
\end{align}
where the charges values on the right hand side must be chosen to be allowed by the fusion rules.
Obtaining Eq.~\eqref{eq:zig_kappa_nu_odd} from the expressions for $\zig{U}$-symbols requires the application of various consistency relations.
For example, our convenient gauge choice combined with Eq.~\eqref{eq:rho_g_F} gives the relation $U_{\varphi}( a_{\vv{x}} , [\psi b]_{\vv{y}} ; [\psi c]_{\vv{x+y}} ) =U_{\varphi}( a_{\vv{x}} , b_{\vv{y}} ; c_{\vv{x+y}} )$, which is all that is required to compute the $\vv{x}=\vv{0}$ terms.

Thus, we can write
\begin{align}
\zig{\beta}_{\zig{a_{\vv{x}}}} = \beta_{a_{\vv{x}}}
.
\end{align}
Note that this is only consistent when $\zig{\beta_{\psi_{\vv{0}}}} =\beta_{\psi_{\vv{0}}} =1$.
From this, we obtain
\begin{align}
\zig{\Omega}_{a_{\vv{x}}} &= {\Omega}_{a_{\vv{x}}},
\\
\zig{\coho{O}} &= \coho{O}
\end{align}
Combining this with Eqs.~\eqref{eq:O'}, \eqref{eq:O_central_cups}, \eqref{eq:O_central_cups_M0}, \eqref{eq:Oeta_case1}, \eqref{eq:Oeta_case2}, and \eqref{eq:Oeta_case3} yields Eq.~\eqref{eq:Oeta_relation_w0} for $\nu$ odd when $\central = \vv{0}$.

When $\central = \cbd \vviso \neq \vv{0}$, rather than compute the obstruction using $\rho' = \V^{\pi + \vviso} \circ \zig{\rho}$ directly, we will use an easier method of comparing these theories.
Specifically, the operation we use to relate the different theories will be to apply an isomorphism that maps the $\central= \cbd \vviso$ theory to a $\central = \vv{0}$ theory, apply the stacking and symmetry action changes, and then apply another isomorphism that maps the post-stacking $\central = \vv{0}$ theory back to a $\central= \cbd \vviso$ theory.
Such isomorphisms that change $\central$ by $\cbd \vviso$ can be constructed using $\Q$, as detailed in Sec.~\ref{sec:centralisomorphism}.
The result of this ends up being to use $\rho' = \V^{\pi} \circ \Q^{\vviso} \circ \zig{\Q^{\vviso} \circ \rho}$ rather than $\V^{\pi +\vviso} \circ \zig{\rho}$.
It is straightforward to see that $\Q^{\vviso} \circ \zig{\Q^{\vviso} \circ \rho}$ is equivalent to $\V^{\vviso} \circ \zig{\rho}$, when $\ker(\res_{\fMTC{\vv{0}}}) = \mathbb{Z}_{2}^{\V}$ (which we have proven to be true when $\widehat{\bf A}_{\vv{1}} \neq \varnothing$).
In particular, it is clear that $\res_{\fMTC_{\vv{0}}} (\Q^{\vviso} \circ \zig{\Q^{\vviso} \circ \rho}) = \res_{\fMTC_{\vv{0}}} (\V^{\vviso} \circ \zig{\rho}) = \rho^{(\vv{0})}$, since the effects of the two $\Q^{\vviso}$ factors cancel on the $\fMTC_{\vv{0}}$ sector, and the action on vortices is seen to be identical by comparing the permutation of $v$-type vortices and the action on vertex states for $\sigma$-type vortices, similar to the argument used in the proof of Eq.~\eqref{eq:Orho_nu_relation}.
Moreover, the effects of the two $\Q^{\vviso}$ factors cancel on the quasiparticle fractionalization phases $\eta^{(\vv{0})}$.
The isomorphisms using $\Q$ are (non-fermionic) natural isomorphisms, so they do not change the fractionalization obstructions.
Thus, we can use the $\central =\vv{0}$ results, with no further calculation, to obtain the same result in Eq.~\eqref{eq:Oeta_relation_w0} for the case when $\central = \cbd \vviso \neq \vv{0}$.

\subsection{Locality Preserving Symmetries}
\label{sec:loc_pres_symmetry}

The discussion and analysis of symmetry and fractionalization in this section so far has focused on unitary on-site symmetries.
However, the notions of symmetry localization and fractionalization are similar for ``locality preserving'' symmetries, which preserve the locality of operators in the sense that operators that act nontrivially on disjoint regions of the system remain disjoint after they are acted on by the symmetry~\cite{Bark2019}.
As such, the analysis can be adapted to apply to more general locality preserving symmetries, including space and time reflecting symmetries~\cite{Bonderson13,Bark2019,Barkeshli2019b} and translational symmetries~\cite{Cheng2016}.
For our purposes, a symmetry is time-reversing if and only if it is anti-unitary.

For a global symmetry group $G$, we can designate space and time reflecting symmetries via a $\mathbb{Z}_2 \times \mathbb{Z}_2$ grading on $G$.
We denote the corresponding homomorphisms as
\begin{align}
\mathsf{p}\left( {\bf g} \right) &= \left\{
\begin{array}{lll}
0  & & \text{ if ${\bf g}$ is spatial parity preserving} \\
1  & & \text{ if ${\bf g}$ is spatial parity reflecting}
\end{array}
\right.
,
\\
\mathsf{q}\left( {\bf g} \right) &= \left\{
\begin{array}{lll}
0  & & \text{ if ${\bf g}$ is not time-reversing} \\
1  & & \text{ if ${\bf g}$ is time-reversing}
\end{array}
\right.
.
\end{align}

In order to apply this to a topological phase, we require a generalization of the topological symmetry group that also includes space and time reflecting auto-equivalences on a BTC $\BTC$.
We write the full group of quantum symmetries of the topological theory as
\begin{equation}
{\bf Aut}(\BTC) = \bigsqcup_{\mathsf{q},\mathsf{p} \in \mathbb{Z}_{2}} \text{Aut}_{\mathsf{q},\mathsf{p}}(\BTC)
,
\end{equation}
where $\text{Aut}_{\mathsf{0},\mathsf{0}}(\BTC)= \text{Aut}(\BTC)$ is the usual on-site, unitary topological symmetry group.
We can make similar definitions of ${\bf Aut}^{\psi}$ and ${\bf Aut}^{\eff}$ for $\psi$-fixed and fermionic topological symmetries, respectively.

It is simple to write the transformation of certain gauge invariant quantities under such generalized topological symmetries.
For this, we define
\begin{align}
s &= \left\{
\begin{array}{lll}
1  & & \text{ if $(\mathsf{p}+\mathsf{q}) \text{ mod }2 =0$} \\
\ast & & \text{ if $(\mathsf{p}+\mathsf{q}) \text{ mod }2 =1$}
\end{array}
\right.
,
\end{align}
which gives complex conjugation when the space-time reflection parity of a symmetry is odd.
Then for a generalized auto-equivalence map $\varphi$, we have
\begin{align}
\varphi(a) &= a' \\
N_{a' b'}^{c'} &= N_{a b}^{c},\\
d_{a'} &= d_{a} ,\\
\theta_{a'} &=\theta_{a}^{s(\varphi)},\\
S_{a' b'} &= S_{a b}^{s(\varphi)}
.
\end{align}

We will not duplicate a detailed description of the transformations on the full topological data in this paper, but such details can be found in Refs.~\onlinecite{Bonderson13,Bark2019,Barkeshli2019b}.
We simply comment that the modifications for locality preserving symmetries all directly apply to the case of fermionic topological phases with the incorporation of the fermionic constraints and conditions discussed earlier in this section.
We highlight some properties of the space-time reflecting symmetries that will be useful.

Applying the symmetry fractionalization analysis produces the same results with the modification of Eq.~\eqref{eq:eta_relation} to
\begin{equation}
\frac{ K^{\mathsf{q}({\bf g})} \eta_{\rho_{\bf g}^{-1}(a)}({\bf h,k}) K^{\mathsf{q}({\bf g})} \eta_{a}({\bf g,hk})}
{\eta_{a}({\bf gh,k}) \eta_{a}({\bf g,h}) }= 1
,
\label{eq:eta_cocycle_anti}
\end{equation}
where $K$ is the complex conjugation operator.
Since $S_{a' b'} = S_{a b}^{s(\varphi)}$, and hence
\begin{align}
K^{\mathsf{q}({\bf g})} M_{\rho_{\bf g}^{-1}(a) \cohosub{b}} K^{\mathsf{q}({\bf g})} = \left\{
\begin{array}{lll}
M_{a \rho_{\bf g}(\cohosub{b})} & & \text{ for } \mathsf{p}({\bf g})=0 \\
M_{a \overline{\rho_{\bf g}(\cohosub{b})}} & & \text{ for } \mathsf{p}({\bf g})=1
\end{array}
\right.
,
\label{eq:LPMrelation}
\end{align}
the map into the cohomological classifying structure through the characters of the fusion algebra yields the same coboundary operator when $\mathsf{p}({\bf g})=0$ (even if $\mathsf{q}({\bf g})=1$), and incorporates the symmetry action of ${\bf g}$ on $\mathcal{A}$-valued cochains with an additional topological charge conjugation, i.e. $^{\bf g} \coho{b} = \overline{\rho_{\bf g}(\coho{b})}$ for $\coho{b} \in C^{n}(G,\mathcal{A})$, when $\mathsf{p}({\bf g})=1$.
For example, the coboundary operation on $\coho{w}$ when $\mathsf{p}({\bf g})=1$ becomes
\begin{align}
\cbd \coho{w}({\bf g,h,k}) &= \rho_{\bf g}\left( \overline{\coho{w}}({\bf h,k}) \right) \otimes \overline{\coho{w}}({\bf gh,k})
\notag \\
&\qquad \otimes \coho{w}({\bf g,hk}) \otimes \overline{\coho{w}}({\bf g,h})
.
\end{align}
This yields obstructions and classifying cohomology groups that are defined exactly as before for $\mathsf{p}({\bf g})=0$ and with this modified coboundary operator for $\mathsf{p}({\bf g})=1$.

A couple important modifications to the form of gauge transformations include the modification to the transformation of $U$-symbols under vertex basis gauge transformations, which becomes
\begin{align}
\left[\widetilde{U}_{{\bf k}}(a,b; c) \right]_{\mu \nu} &= \sum_{\mu'\nu'}
K^{\mathsf{q}({\bf k})}\left[\Gamma^{{}^{\bar{{\bf k}}}a {}^{\bar{{\bf k}}}b}_{{}^{\bar{{\bf k}}} c} \right]_{\mu \mu'}K^{\mathsf{q}({\bf k})}
\notag \\
& \times \left[U_{{\bf k}}(a,b; c) \right]_{\mu' \nu'}
\left[ \left( \Gamma^{a b}_{c} \right)^{-1} \right]_{\nu' \nu}
,
\end{align}
and $\eta$ under symmetry action gauge transformations, which becomes
\begin{align}
\check{\eta}_{c}({\bf g},{\bf h}) &= \frac{\gamma_{c}({\bf gh})}
{ K^{\mathsf{q}({\bf g})} \gamma_{\,^{\bf \bar{g}}c}({\bf h}) K^{\mathsf{q}({\bf g})} \gamma_{c}({\bf g})}
\eta_{c}({\bf g},{\bf h})
.
\end{align}

In the case where the fermionic symmetry group includes anti-unitary, but not spatial reflecting  symmetries (i.e. restrict to $\mathsf{p}=0$), we find some modifications to the comparisons between obstructions for different extensions.
We first note that anti-unitary topological symmetries can only exist for FMTCs with $c_{-} = 0 \text{ mod } 4$.
Thus, for anti-unitary symmetries, it is only meaningful to compare obstructions for different $\mathbb{Z}_{2}^{\eff}$ modular extensions of a given SMTC $\fMTC_{\vv{0}}$ when they are related by $\nu=8$ stacking, that is $\fMTC' = \fMTC' \fprod \ifo^{(8)}$.
The relation in Eq.~\eqref{eq:Orho_nu_relation} comparing obstructions of extending a symmetry action $[\rho^{(\vv{0})}]$ on $\fMTC_{\vv{0}}$ to a symmetry action on $\fMTC$ and $\fMTC'$ is unchanged, and since $\nu=8$, it is given by
\begin{align}
\label{eq:Orho'_antiunitary}
[O^\rho]' = [O^{\rho}]
.
\end{align}
Ref.~\onlinecite{Bark2021cascade} proved Eq.~\eqref{eq:Orho'_antiunitary} for anti-unitary symmetries when the FMTCs have $v$-type vortices and $\ker(\res_{\fMTC_{\vv{0}}}) = \mathbb{Z}_{2}^{\V}$; our proof removes these conditions on the FMTCs.
We find a modification of Eq.~\eqref{eq:O'}, comparing obstructions of fractionalization for different extensions $[\rho]$ and $[\rho']$ of the same symmetry action $[\rho^{(\vv{0})}]$, on the same $\fMTC$, which becomes
\begin{align}
\label{eq:O'_antiunitary}
\coho{O}' &= \psi^{\pi \cup \mathsf{b} + \mathsf{q} \cup \pi \cup \pi} \otimes \coho{O}
.
\end{align}
Similarly, we find a modification of Eq.~\eqref{eq:Oeta_relation}, comparing obstructions of extending fractionalization from quasiparticles to vortices with different $[\fMTC, \rho, \eta^{(\vv{0})}]$ and $[\fMTC, \rho', \eta^{(\vv{0}) \prime}]$, which becomes~\footnote{We thank the authors of Ref.~\onlinecite{Bark2021cascade} for sharing an early draft of their v3 revision, where they predicted, from anomaly matching considerations combined with our result for the unitary case, that the anti-unitary case should have a $q \cup \pi \cup \pi$ term when it has a $\pi \cup \central$ term.}
\begin{align}
\label{eq:Oeta'_antiunitary}
[\coho{O}^{\eta}]' &= [\psi^{\pi \cup \central + \mathsf{q} \cup \pi \cup \pi} \otimes  \coho{T}_{\rho^{(0)}}( \coho{t}^{(0)} )] \otimes [\coho{O}].
\end{align}
Again, we emphasize that $\fMTC$ and $\fMTC'$ are necessarily related by stacking with $\nu=0$ or $8$ for anti-unitary symmetries.

In order to derive these results, we first note that the when a topological symmetry $\varphi$ on $\fMTC$ is anti-unitary, we obtain a related $\zig{\varphi}$ on $\fMTC'$ by the stacking $(\fMTC' , \zig{\varphi}) = (\fMTC , \varphi) \fprod (\ifo^{(8)} , K)$.
The topological charges and fusion rules of $\fMTC'$ are identical to those of $\fMTC$.
With conventional gauge choices, all the topological data of $\ifo^{(8)}$ are real-valued, so $K$ do not permute topological charges of $\ifo^{(8)}$ and its $U$-symbols are all trivial.
It follows that the resulting $\zig{\varphi}$ has the same permutation of topological charges and $U$-symbols as $\varphi$.
Applying this to produce related lifts of $[\rho^{(\vv{0})}_{\bf g}]$ to $\fMTC$ and $\fMTC'$ for each ${\bf g}$ [as in the proof of Eq.~\eqref{eq:Orho_nu_relation}], it is clear that we obtain Eq.~\eqref{eq:Orho'_antiunitary}.
Next, we observe that the relation in Eq.~\eqref{eq:Vcommute} holds, with the same natural isomorphism $\Upsilon$, for anti-unitary $\varphi$.
This allows us to follow the same derivation as before.
The only difference arises from the $i^{\vv{x} \cdot \pi ({\bf g}) \cdot \pi({\bf h})}$ term in $\beta'_{a_{\vv{x}}}({\bf g,h})$, since all of its other terms are real-valued (at least when using the convenient gauge choice for $v$-type vortices).
Using
\begin{align}
\frac{K^{\mathsf{q}({\bf g})}i^{ \pi({\bf h}) \cdot \pi({\bf k}) } K^{\mathsf{q}({\bf g})} \cdot i^{ \pi({\bf g}) \cdot \pi({\bf hk}) }}{ i^{ \pi({\bf gh}) \cdot \pi({\bf k}) } i^{ \pi({\bf g }) \cdot \pi({\bf h}) } } = (-1)^{\mathsf{q}({\bf g}) \cdot \pi({\bf h}) \cdot \pi({\bf k})}
,
\end{align}
we find the modified result
\begin{align}
\Omega'_{a_{\vv{x}}}&= (-1)^{\vv{x} \cdot \left( \pi \cup \mathsf{b} + \mathsf{q} \cup \pi \cup \pi \right)} \, \Omega_{a_{\vv{x}}}
,
\end{align}
which proves Eq.~\eqref{eq:O'_antiunitary}.
For Eq.~\eqref{eq:Oeta'_antiunitary}, the analysis is the same as that leading to Eq.~\eqref{eq:Oeta_relation}, but with the above modifications included.

We now turn to the two important examples of $G=\mathbb{Z}_{2}$ time-reversal and space-reflection symmetries, which have important invariants that can occur for fermionic topological phases.

\subsubsection{Time-reversal symmetry and fermionic Kramers degeneracy}

In Ref.~\onlinecite{Bark2019}, $\mathbb{Z}_{2}^{\bf T}$ time-reversal symmetry in bosonic topological phases was analyzed using this symmetry action and fractionalization formalism.
For topological charges with the property $a = \,^{\bf T}a $, the quantity
\begin{align}
\eta_{a}^{\bf T} \equiv \eta_{a}({\bf T,T}) = \pm 1
\end{align}
was shown to be an invariant that could be identified as the ``local ${\bf T}^2$'' value ascribed to $a$.
This invariant is associated with local Kramers degeneracy~\cite{Levin12}.
Among other properties for $a = \,^{\bf T}a $, it was shown in Ref.~\onlinecite{Bark2019} that $\theta_{a}=\pm 1$ and if there is a topological charge $b$ such that $N_{b \,^{\bf T}b}^{a}$ is odd, then $\eta_{a}^{\bf T} = \theta_{a}$.

For fermionic systems with $G=\mathbb{Z}_{2}^{\bf T}$, we can continue to use the above invariant for $a_{\vv{x}} = \,^{\bf T}a_{\vv{x}}$.
In particular, $\eta_{\psi_{\vv{0}}}^{\bf T} = \pm 1$ since $\psi_{\vv{0}} = \,^{\bf T}\psi_{\vv{0}}$, which corresponds to the physical fermions having ${\bf T}^{2} = (\pm 1)^{\bf F}$, and the corresponding fermionic symmetry groups $\mathcal{G}^{\eff} = \mathbb{Z}_{2}^{\eff} \times \mathbb{Z}_{2}^{\bf T}$ and $\mathcal{G}^{\eff} = \mathbb{Z}_{4}^{{\bf T},\eff}$, respectively.

There is a similar notion of local fermionic Kramers degeneracy for topological charges with the property $a_{\vv{x}} = \psi_{\vv{0}} \otimes \,^{\bf T}a_{\vv{x}} = \,^{\bf T}[\psi a]_{\vv{x}}$~\cite{Fidkowski2013,Metlitski2014,Wang2017}.
For charges $a_{\vv{x}} = \,^{\bf T}[\psi a]_{\vv{x}}$, we consider an invariant defined in Ref.~\onlinecite{Tata2021} by
\begin{align}
\label{eq:etaTdef}
\eta_{a_{\vv{x}}}^{\bf T} \equiv F^{a_{\vv{x}}\psi_{\vv{0}} \psi_{\vv{0}} }_{a_{\vv{x}}} U_{\bf T}(a_{\vv{x}} , \psi_{\vv{0}} ; [\psi a]_{\vv{x}} ) \eta_{a_{\vv{x}}}({\bf T,T})
,
\end{align}
which we re-derive here, showing its invariance under vertex basis and symmetry action gauge transformations.
For fermionic theories, we can use the canonical gauge choice $F^{a_{\vv{x}}\psi_{\vv{0}} \psi_{\vv{0}} }_{a_{\vv{x}}}=1$, but we will leave this term unfixed to make the role of the canonical isomorphism explicit.
We also show that
\begin{align}
\label{eq:etaT_property}
\eta_{a_{\vv{x}}}^{\bf T} = ( \eta_{[\psi a]_{\vv{x}}}^{\bf T} )^{-1} = \pm \sqrt{\eta_{\psi_{\vv{0}}}^{\bf T}}
,
\end{align}
where $\eta_{\psi_{\vv{0}}}^{\bf T} = \pm 1$, since $\psi_{\vv{0}} = \,^{\bf T}\psi_{\vv{0}}$.
When the physical fermion have ${\bf T}^2 = (-1)^{\bf F}$, that is $\eta_{\psi_{\vv{0}}}^{\bf T} = - 1$, this indicates $\eta_{a_{\vv{x}}}^{\bf T} = \pm i$.

For charges $a_{\vv{x}} = \,^{\bf T}[\psi a]_{\vv{x}}$, we also have the property
\begin{align}
\theta_{a_{\vv{x}} }^{2} = (-1)^{\vv{x} +1}
,
\end{align}
so this requires $\theta_{a_{\vv{0}} }=-\theta_{[\psi a]_{\vv{0}} }= \pm i$ for quasiparticles and  $\theta_{a_{\vv{1}} }=\theta_{[\psi a]_{\vv{1}} }= \pm 1$ for vortices.
This follows from the transformation of topological twists under time-reversing symmetry $\theta_{\,^{\bf T} a_{\vv{x}}} = \theta_{a_{\vv{x}}}^{\ast}$ and the ribbon property of braiding [Eq.~\eqref{eq:ribbon}].

We now check the gauge invariance of $\eta_{a_{\vv{x}}}^{\bf T}$ under vertex basis gauge transformations, which is given by
\begin{align}
\widetilde{\eta_{a_{\vv{x}}}^{\bf T}} &= \frac{\Gamma^{a_{\vv{x}}\psi_{\vv{0}}} \Gamma^{[\psi a]_{\vv{x}} \psi_{\vv{0}}}}{\Gamma^{\psi_{\vv{0}} \psi_{\vv{0}}}} \frac{( \Gamma^{\,^{\bf T} a_{\vv{x}}\psi_{\vv{0}}} )^{\ast} }{ \Gamma^{a_{\vv{x}}\psi_{\vv{0}}} } {\eta}_{a_{\vv{x}}}^{\bf T}
= \frac{1}{\Gamma^{\psi_{\vv{0}} \psi_{\vv{0}}}}  {\eta}_{a_{\vv{x}}}^{\bf T}
.
\end{align}
Here, we have written $\Gamma^{\psi_{\vv{0}} \psi_{\vv{0}}}$ explicitly to highlight the fact that the canonical fermionic gauge constraint $\Gamma^{\psi_{\vv{0}} \psi_{\vv{0}}} =1$ is necessary for this to be gauge invariant.
Similarly, we check the symmetry action gauge transformations, which are given by
\begin{align}
\check{\eta}_{a_{\vv{x}}}^{\bf T} &= \frac{\gamma_{a_{\vv{x}}}({\bf T}) \gamma_{\psi_{\vv{0}}}({\bf T})  }{\gamma_{[\psi a]_{\vv{x}}}({\bf T})} \frac{1}{ \gamma_{\,^{\bf T}a_{\vv{x}}}({\bf T})^{\ast}  \gamma_{a_{\vv{x}}}({\bf T})} {\eta}_{a_{\vv{x}}}^{\bf T}
= \gamma_{\psi_{\vv{0}}}({\bf T}) {\eta}_{a_{\vv{x}}}^{\bf T}
.
\end{align}
Again, we have written $\gamma_{\psi_{\vv{0}}}({\bf T})$ explicitly to highlight the fact that the canonical fermionic gauge constraint $\gamma_{\psi_{\vv{0}}}({\bf g}) =1$ is necessary for this to be gauge invariant.
It also shows that if we were to allow $\gamma_{\psi_{\vv{0}}}({\bf g}) = \pm 1$, then $\eta_{a_{\vv{x}}}^{\bf T}$ would only be well-defined up to an overall sign.

In order to show Eq.~\eqref{eq:etaT_property}, we need the following properties.
From the pentagon equation, we have
\begin{align}
\label{eq:etaT_F}
F^{a_{\vv{x}}\psi_{\vv{0}} \psi_{\vv{0}} }_{a_{\vv{x}}} = F^{[\psi a]_{\vv{x}}\psi_{\vv{0}} \psi_{\vv{0}} }_{[\psi a]_{\vv{x}}}
.
\end{align}
The transformation of $F$-symbols under a time-reversing (anti-unitary) symmetry yields
\begin{align}
\label{eq:etaT_FT}
&\rho_{\bf T} \left(  F^{a_{\vv{x}}\psi_{\vv{0}} \psi_{\vv{0}} }_{a_{\vv{x}}} \right) =\left(  F^{a_{\vv{x}}\psi_{\vv{0}} \psi_{\vv{0}} }_{a_{\vv{x}}} \right)^{\ast}
\\
&= U_{\bf T}(\,^{\bf T}a_{\vv{x}} , \psi_{\vv{0}} ; \,^{\bf T}[\psi a]_{\vv{x}} ) U_{\bf T}(\,^{\bf T}[\psi a]_{\vv{x}} , \psi_{\vv{0}} ; \,^{\bf T}a_{\vv{x}} ) F^{\,^{\bf T} a_{\vv{x}}\psi_{\vv{0}} \psi_{\vv{0}} }_{\,^{\bf T} a_{\vv{x}}}
.
\notag
\end{align}
Here, we used the canonically fixed $U_{\bf T}(\psi_{\vv{0}} , \psi_{\vv{0}} ; I_{\vv{0}} ) =1$.
The natural isomorphism $\kappa_{\bf T,T}$ requires
\begin{align}
& \frac{1}{\kappa_{\bf T,T} (a_{\vv{x}} , \psi_{\vv{0}} ; [\psi a]_{\vv{x}} )} = \frac{\eta_{[\psi a]_{\vv{x}}}({\bf T,T})}{\eta_{a_{\vv{x}}}({\bf T,T}) \eta_{\psi_{\vv{0}}}({\bf T,T})}
\notag \\
&\qquad = U_{\bf T}(a_{\vv{x}} , \psi_{\vv{0}} ; [\psi a]_{\vv{x}} )^{\ast} U_{\bf T}(\,^{\bf T}a_{\vv{x}} , \psi_{\vv{0}} ; \,^{\bf T}[\psi a]_{\vv{x}} )
.
\label{eq:etaT_kappa}
\end{align}
Eq.~\eqref{eq:eta_cocycle_anti} for a time-reversing symmetry becomes
\begin{align}
\label{eq:etaT_cocycle}
\eta_{\,^{\bf T}a_{\vv{x}}}({\bf T,T})^{\ast} = \eta_{a_{\vv{x}}}({\bf T,T})
.
\end{align}
Using $a_{\vv{x}} = \,^{\bf T}[\psi a]_{\vv{x}}$, Eqs.~\eqref{eq:etaT_F} and \eqref{eq:etaT_kappa} yield $\eta_{a_{\vv{x}}}^{\bf T} \eta_{\psi_{\vv{0}}}^{\bf T} = \eta_{[\psi a]_{\vv{x}}}^{\bf T} $, while Eqs.~\eqref{eq:etaT_FT} and \eqref{eq:etaT_cocycle} yield $\eta_{a_{\vv{x}}}^{\bf T}  = (\eta_{[\psi a]_{\vv{x}}}^{\bf T})^{-1}$.
These combine to give Eq.~\eqref{eq:etaT_property}.

\subsubsection{Space-reflection fermionic fractionalization invariant}

In Ref.~\onlinecite{Barkeshli2019b}, $\mathbb{Z}_{2}^{\bf r}$ space-reflection symmetry in bosonic topological phases was analyzed using this symmetry action and fractionalization formalism.
Analyzing the configuration with two quasiparticles of charge $a$ and $\bar{a}$ pair-produced from vacuum that are located opposite each other across the reflection axis, it was shown that for $a = \,^{\bf r} \bar{a}$, the quantity
\begin{align}
\eta_{a}^{\bf r} \equiv U_{\bf r}(a , \bar{a}; \I ) \eta_{a}({\bf r,r}) = \pm 1
\end{align}
is an invariant, analogous to $\eta_{a}^{\bf T}$ for time-reversal.
Such charges must also have $\theta_{a}=\pm 1$.

For fermionic systems with $G=\mathbb{Z}_{2}^{\bf r}$, we can similar use this invariant for $a_{\vv{x}} = \,^{\bf T}a_{\vv{x}}$.
In particular, $\eta_{\psi_{\vv{0}}}^{\bf r} = \pm 1$ since $\psi_{\vv{0}} = \,^{\bf r}\psi_{\vv{0}}$ and $U_{\bf g}(\psi_{\vv{0}} , \psi_{\vv{0}} ; \I_{\vv{0}} )=1$, which corresponds to the physical fermions having ${\bf r}^{2} = (\pm 1)^{\bf F}$, and the corresponding fermionic symmetry groups $\mathcal{G}^{\eff} = \mathbb{Z}_{2}^{\eff} \times \mathbb{Z}_{2}^{\bf r}$ and $\mathcal{G}^{\eff} = \mathbb{Z}_{4}^{{\bf r},\eff}$, respectively.

We can also define a quantity analogous to the fermionic Kramers invariant when $a_{\vv{x}} = \psi_{\vv{0}} \otimes \,^{\bf r}\bar{a}_{\vv{x}} = \,^{\bf r}[\psi \bar{a}]_{\vv{x}}$ by
\begin{align}
\label{eq:etaRdef}
\eta_{a_{\vv{x}}}^{\bf r} \equiv U_{\bf r}(a_{\vv{x}} , [\psi \bar{a}]_{\vv{x}} ; \psi_{\vv{0}} ) \eta_{a_{\vv{x}}}({\bf r,r})
,
\end{align}
where we are considering the pair of topological charge $a_{\vv{x}}$ and $[\psi \bar{a}]_{\vv{x}}$ located opposite each other across the reflection axis.
We show that it is, in fact invariant under vertex basis and symmetry action gauge transformations, and that
\begin{align}
\label{eq:etaR_property}
\eta_{a_{\vv{x}}}^{\bf r} = - \eta_{[\psi \bar{a}]_{\vv{x}}}^{\bf r} = \pm \sqrt{\eta_{\psi_{\vv{0}}}^{\bf r}}
,
\end{align}
where $\eta_{\psi_{\vv{0}}}^{\bf r} = \pm 1$, since $\psi_{\vv{0}} = \,^{\bf r} \bar{\psi}_{\vv{0}}$.
When the physical fermions have ${\bf r}^{2} = (-1)^{\bf F}$, that is $\eta_{\psi_{\vv{0}}}^{\bf r} = - 1$, this indicates $\eta_{a_{\vv{x}}}^{\bf r} = \pm i$.

For charges $a_{\vv{x}} = \,^{\bf r}[\psi \bar{a}]_{\vv{x}}$, we also have the property
\begin{align}
\theta_{a_{\vv{x}} }^{2} = (-1)^{\vv{x} +1}
,
\end{align}
so this requires $\theta_{a_{\vv{0}} }=-\theta_{[\psi a]_{\vv{0}} }= \pm i$ for quasiparticles and  $\theta_{a_{\vv{1}} }=\theta_{[\psi a]_{\vv{1}} }= \pm 1$ for vortices.
This follows from the transformation of topological twists under space-reflecting symmetry $\theta_{\,^{\bf r} a_{\vv{x}}} = \theta_{a_{\vv{x}}}^{\ast} = \theta_{\bar{a}_{\vv{x}}}^{\ast}$ and the ribbon property of braiding [Eq.~\eqref{eq:ribbon}].

We now check the gauge invariance of $\eta_{a_{\vv{x}}}^{\bf r}$ under vertex basis gauge transformations, which is given by
\begin{align}
\widetilde{\eta_{a_{\vv{x}}}^{\bf r}} &= \frac{\Gamma^{\,^{\bf r}[\psi \bar{a}]_{\vv{x}} \,^{\bf r}a_{\vv{x}}} }{\Gamma^{a_{\vv{x}} [\psi \bar{a}]_{\vv{x}}} } \eta_{a_{\vv{x}}}^{\bf r} = \eta_{a_{\vv{x}}}^{\bf r}
.
\end{align}
Similarly, we check the symmetry action gauge transformations, which are given by
\begin{align}
\check{\eta}_{a_{\vv{x}}}^{\bf r} &= \frac{\gamma_{a_{\vv{x}}}({\bf r}) \gamma_{[\psi a]_{\vv{x}}}({\bf r})}{\gamma_{\psi_{\vv{0}}}({\bf r})}
\frac{1}{\gamma_{\,^{\bf r}a_{\vv{x}}}({\bf r}) \gamma_{a_{\vv{x}}}({\bf r})} = \frac{1}{\gamma_{\psi_{\vv{0}}}({\bf r})} {\eta}_{a_{\vv{x}}}^{\bf r}
.
\end{align}
We have written $\gamma_{\psi_{\vv{0}}}({\bf r})$ explicitly here to highlight the fact that the canonical fermionic gauge constraint $\gamma_{\psi_{\vv{0}}}({\bf g}) =1$ is necessary for this to be gauge invariant; if we were to allow $\gamma_{\psi_{\vv{0}}}({\bf g}) = \pm 1$, then $\eta_{a_{\vv{x}}}^{\bf r}$ would only be well-defined up to an overall sign.

In order to show Eq.~\eqref{eq:etaR_property}, we need the following properties.
The transformation of $R$-symbols under a spatial-reflection symmetry yields
\begin{align}
\rho_{\bf r} \left(  R^{a_{\vv{x}} [\psi \bar{a}]_{\vv{x}}}_{\psi_{\vv{0}}} \right) &= \frac{U_{\bf r}(\,^{\bf r}[\psi \bar{a}]_{\vv{x}} , \,^{\bf r}a_{\vv{x}} ; \psi_{\vv{0}} ) } {U_{\bf r}(\,^{\bf r}a_{\vv{x}} , \,^{\bf r}[\psi \bar{a}]_{\vv{x}} ; \psi_{\vv{0}} ) } R^{\,^{\bf r}a_{\vv{x}} \,^{\bf r}[\psi \bar{a}]_{\vv{x}}}_{\psi_{\vv{0}}}
\notag \\
&= \left(  R^{a_{\vv{x}} [\psi \bar{a}]_{\vv{x}}}_{\psi_{\vv{0}}} \right)^{-1}
.
\label{eq:etaR_R}
\end{align}
The natural isomorphism $\kappa_{\bf r,r}$ requires
\begin{align}
& \frac{1}{\kappa_{\bf r,r} (a_{\vv{x}} , [\psi \bar{a}]_{\vv{x}} ; \psi_{\vv{0}} )} = \frac{\eta_{\psi_{\vv{0}}}({\bf r,r})}{\eta_{a_{\vv{x}}}({\bf r,r}) \eta_{[\psi \bar{a}]_{\vv{x}}}({\bf r,r})}
\notag \\
&\qquad = U_{\bf r}(\,^{\bf r}[\psi \bar{a}]_{\vv{x}} , \,^{\bf r}a_{\vv{x}}; \psi_{\vv{0}} ) U_{\bf r}(a_{\vv{x}} , [\psi \bar{a}]_{\vv{x}} ; \psi_{\vv{0}} )
.
\label{eq:etaR_kappa}
\end{align}
Eq.~\eqref{eq:eta_cocycle_anti} for a space-reflection symmetry becomes
\begin{align}
\label{eq:etaR_cocycle}
\eta_{\,^{\bf r}a_{\vv{x}}}({\bf r,r}) = \eta_{a_{\vv{x}}}({\bf r,r})
.
\end{align}
Using $a_{\vv{x}} = \,^{\bf r}[\psi \bar{a}]_{\vv{x}}$, Eqs.~\eqref{eq:etaR_R} and \eqref{eq:etaR_cocycle}, together with the ribbon equation and transformation of topological twists, yields $\eta_{a_{\vv{x}}}^{\bf r} = - \eta_{[\psi a]_{\vv{x}}}^{\bf r}$, while Eqs.~\eqref{eq:etaR_kappa} and \eqref{eq:etaR_cocycle} yield $(\eta_{a_{\vv{x}}}^{\bf r})^{2} = \eta_{\psi_{\vv{0}}}^{\bf r}$.
These combine to give Eq.~\eqref{eq:etaR_property}.

\subsection{Examples}

\subsubsection{Invertible FMTCs}
\label{sec:Ex_IFMTCs}

In the case of invertible FMTCs $\ifo^{(\nu)}$, the quasiparticle sector is simply $\ifo^{(\nu)}_{\vv{0}} = \mathbb{Z}_{2}^{\psi}$ and $\Autf{\ifo^{(\nu)}} = \ker(\res_{\fMTC_{\vv{0}}}) =  \mathbb{Z}_{2}^{\V}$.
The symmetry action on $\mathbb{Z}_{2}^{\psi}$ is trivial and can always be extended (i.e $[O^{\rho}]=[\openone]$) to a symmetry action on $\ifo^{(\nu)}$.
For this example, the symmetry fractionalization extension problem from $\mathbb{Z}_{2}^{\psi}$ to $\fMTC$ and from $\fMTC_{\vv{0}}$ to $\fMTC$ are the same.
It is useful to consider $\nu$ even and odd separately.

In the case of $\nu$ even, the symmetry action on $\ifo^{(\nu)}$ takes the form $[\rho_{\bf g}] = [\V]^{\pi({\bf g})}$, where $\pi: G \to \mathbb{Z}_{2}$ is a homomorphism.
We analyze this with the representative auto-equivalence maps $\rho_{\bf g} = \V^{\pi({\bf g})}$ for $\V$ defined in Eq.~(\ref{eq:Vsymm}).
For $\nu$ even in the choice of gauge used in Table~\ref{table:sixteen}, the coefficients of $\V$ acting on basis states are given by
\begin{align}
\label{eq:Knu_even_V}
\V (a_{\vv{x}} , b_{\vv{y}} ; [\psi^{\frac{\nu}{2}}ab]_{\vv{x}+\vv{y}}) & = (-1)^{(\mathsf{a}+\mathsf{x})\cdot \mathsf{b}}
.
\end{align}
For this, we can choose $\beta^{(\vv{0})}_{\psi_{\vv{0}}}({\bf g,h}) = 1$ and $\beta_{a_{\vv{x}}}({\bf g,h}) = i^{\vv{x} \cdot \pi({\bf g}) \cdot \pi({\bf h})}$.
As discussed, the symmetry fractionalization classes of $\mathbb{Z}_{2}^{\psi}$ encoded in $\eta^{(\vv{0})}_{\psi_{\vv{0}}}({\bf g,h}) = (-1)^{\central({\bf g,h})}$ correspond to the $\central \in Z^2 (G,\mathbb{Z}_{2}^{\eff})$ characterizing the fermionic symmetry group $\mathcal{G}^{\eff} = \mathbb{Z}_2^{\eff} \times_{\central} G$.
The symmetry fractionalization obstruction $\coho{O}=\I_{\vv{0}}$ is trivial.
It is straightforward to check that $\eta_{a_{\vv{x}}}({\bf g,h}) = i^{\vv{x}\cdot\pi({\bf g}) \cdot \pi({\bf h})}$ describes a fractionalization class and has $\central({\bf g,h}) = 0$.
Using $\mathsf{b}=\mathsf{0}$, $e_{\vv{1}} = \I_{\vv{1}}$, and $\ell(\hat{\I}_{\vv{1}}) = \I_{\vv{1}}$ in Eqs.~\eqref{eq:O_central_cups} and \eqref{eq:O_extension} [or Eq.~\eqref{eq:Gf_obstruction_dot} with $\dot{\central}=\vv{0}$], we find the obstruction is
\begin{align}
&\coho{O}^{\central}({\bf g,h,k}) = {\coho{O}}^{\eta}({\bf g,h,k}) = \cbd \overline{\I}_{\vv{1}}^{\central}({\bf g,h,k}) \notag \\
&\quad = (\,^{\bf g}\overline{\I}_{\vv{1}} \otimes {\I}_{\vv{1}})^{\central({\bf h,k})} \otimes \overline{\I}_{\vv{1}}^{\cbd \tilde{\central}({\bf g,h,k})}
\notag \\
&\quad = \psi_{\vv{0}}^{\pi({\bf g})\cdot \central({\bf h,k}) + \frac{\nu}{2} \left(\tilde{\central}({\bf gh,k})\cdot\tilde{\central}({\bf g,h}) - \tilde{\central}({\bf g,hk})\cdot \tilde{\central}({\bf h,k}) \right)}
.
\end{align}
Here, we have used the fact that $\I_{\vv{1}} \otimes \I_{\vv{1}}= \psi_{\vv{0}}^{\nu /2}$.
We note that this can be written in the more compact form
\begin{align}
\label{eq:Knu_even_extension_obstruction}
\coho{O}^{\central} & = \coho{O}^{\eta} = \psi_{\vv{0}}^{\pi \cup \central + \frac{\nu}{2} \cdot \central \cup_{1} \central}
.
\end{align}
Thus, fractionalization for $\ifo^{(\nu)}$ with $\nu$ even and $\rho$ only exists for a given $\mathcal{G}^{\eff} = \mathbb{Z}_2^{\eff} \times_{\central} G$ when $\psi_{\vv{0}}^{\pi \cup \central + \frac{\nu}{2} \central \cup_{1} \central} \in B^3(G,\mathbb{Z}_{2}^{\psi})$.
When this is the case, fractionalization is torsorially classified by $H^2 (G,\mathbb{Z}_{2}^{\psi})$, with corresponding
\begin{align}
\eta_{a_{\vv{x}}}({\bf g,h}) &=i^{\vv{x} \cdot \pi({\bf g})\cdot \pi({\bf h})}  M_{a_{\vv{x}} \I_{\vv{1}}}^{\central({\bf g,h})} (-1)^{\vv{x} \cdot \mathsf{p}({\bf g,h})}
,
\end{align}
where $\coho{p}({\bf g,h}) = \psi^{ \mathsf{p}({\bf g,h})} \in C^2(G,\mathbb{Z}_{2}^{\psi})$ such that
\begin{align}
\cbd \coho{p} & = \psi_{\vv{0}}^{\pi \cup \central + \frac{\nu}{2} \cdot \central \cup_{1} \central}
.
\end{align}
This is equivalent to the condition $\cbd \coho{w} = \I_{\vv{0}}$ with $\coho{w}({\bf g,h}) = \coho{p}({\bf g,h}) \otimes \I_{\vv{1}}^{\central({\bf g,h})}$.
Thus, we see the symmetry action $[\rho]$ is classified by $H^1(G,\mathbb{Z}_{2}^{\V}) = Z^1(G,\mathbb{Z}_{2}^{\V})$ (the set of homomorphisms from $G$ to $\mathbb{Z}_{2}^{\V}$), the fermionic symmetry group $\mathcal{G}^{\eff}$ (or equivalently the fractionalization of quasiparticles $\ifo^{(\nu)}_{\vv{0}} = \mathbb{Z}_{2}^{\psi}$) is classified by $Z^2(G,\mathbb{Z}_{2}^{\eff})$, and the fractionalization of vortices (represented by $\coho{p}$) is classified torsorially by $H^2 (G,\mathbb{Z}_{2}^{\psi})$.

In the case of $\nu$ odd, $\mathcal{A}_{\vv{1}} = \varnothing$ and $[\Q] = [\V]$ is nontrivial.
The symmetry action now takes the form $[\rho_{\bf g}] = [\Q]^{\rr({\bf g})}$, where $\rr({\bf g}) \in C^1 (G, \mathbb{Z}_{2})$ is not required to be a homomorphism.
We analyze this with the representative auto-equivalence maps $\rho_{\bf g} = \Q^{\rr({\bf g})}$, which have $\kappa_{\bf g,h} = \Q^{\cbd \rr({\bf g,h})}$.
When $\rr \in H^1 (G, \mathbb{Z}_{2})$, i.e. the action is a homomorphism, it follows that $\kappa_{\bf g,h} = \openone$ and we could choose $\beta_{a_{\vv{x}}}({\bf g,h}) = 1$.
More generally, $\beta_{a_{\vv{x}}}({\bf g,h}) = (-1)^{\cbd \rr ({\bf g,h}) \delta_{a_{\vv{x}} \psi_{\vv{0}}} }$.
The symmetry fractionalization obstruction $\coho{O}=\I_{\vv{0}}$ is trivial.
Since we are required to have $\beta_{\psi_{\vv{0}}}({\bf g,h}) = (-1)^{\cbd \rr ({\bf g,h}) }$ in this case, we have a $\Q$-projective action with $\phi = \cbd \rr$.
Even though $[\rho_{\bf g}^{(0)}] = [\openone]$, in order to match $\beta_{\psi_{\vv{0}}}$ of the extension, we must have $\beta^{(\vv{0})}_{\psi_{\vv{0}}}({\bf g,h}) = (-1)^{\cbd \rr ({\bf g,h}) }$ and $\central = \cbd \rr$.
Since $\mathcal{A}_{\vv{1}} = \varnothing$, $\coho{O}^{\central} = \central - \cbd \rr$, which also demonstrates the requirement that $\central = \cbd \rr$.
Thus, fractionalization for $\ifo^{(\nu)}$ and $\rho$ only exists for $\mathcal{G}^{\eff} = \mathbb{Z}_2^{\eff} \times_{\central} G$ with $\central = \cbd \rr$.
The obstruction for extending fractionalization is ${\coho{O}}^{\eta}({\bf g,h,k}) = \I_{\vv{0}}$, so fractionalization is torsorially classified by $H^2 (G,\mathbb{Z}_{2}^{\psi})$, with corresponding
\begin{align}
\eta_{a_{\vv{x}}}({\bf g,h}) &= (-1)^{\central({\bf g,h}) \delta_{a_{\vv{x}} \psi_{\vv{0}}} } (-1)^{\vv{x} \cdot \mathsf{p}({\bf g,h})}
,
\end{align}
where $\central = \cbd \rr$ and $\coho{p}({\bf g,h}) = \psi^{ \mathsf{p}({\bf g,h})} \in Z^2(G,\mathbb{Z}_{2}^{\psi})$.
Thus, we see the symmetry action $[\rho]$ are classified by $C^1(G,\mathbb{Z}_{2}^{\Q})$, the fermionic symmetry group $\mathcal{G}^{\eff}$ is fixed by the symmetry action to be $\central = \cbd \rr$, and the fractionalization of vortices (represented by $\coho{p}$) is classified torsorially by $H^2 (G,\mathbb{Z}_{2}^{\psi})$.
We note that $C^1(G,\mathbb{Z}_{2}^{\Q}) \cong H^1(G,\mathbb{Z}_{2}^{\Q}) \times B^2(G,\mathbb{Z}_{2}^{\Q})$, where $H^1(G,\mathbb{Z}_{2}^{\Q})$ is the classification for fixed $\phi$ and the $B^2(G,\mathbb{Z}_{2}^{\Q})$ factor is the classification of $\phi = \cbd \rr$.
Since $\central = \phi$, one can also think of $\mathcal{G}^{\eff}$ as classified by $B^2(G,\mathbb{Z}_{2}^{\Q})$ in this way.

The case where $G = \mathbb{Z}_{2} \times \mathbb{Z}_{2}$ provides interesting examples of how the fractionalization obstructions can behave, in particular for the $\nu$ even case given in Eq.~\eqref{eq:Knu_even_extension_obstruction}. We write the group elements as ${\bf g} = (g_1, g_2)$. We note that $H^1(G,\mathbb{Z}_{2}^{\V}) = \mathbb{Z}_{2}^{2}$, $H^2(G,\mathbb{Z}_{2}^{\V}) = \mathbb{Z}_{2}^{3}$, and $H^3(G,\mathbb{Z}_{2}^{\V}) = \mathbb{Z}_{2}^{4}$.

The first example is a choice of $\pi$ and $\central$ for which $[\pi \cup \central]$ and $[\central \cup_1 \central]$ are nontrivial, but $[\pi \cup \central + \central \cup_1 \central]=[0]$.
This gives a $\central$ for which $[\rho] = [\openone]$ is obstructed, but $[\rho] \neq [\openone]$ is unobstructed.
It also allows for $[\coho{O}^{\central}]=[\coho{O}^{\eta}]$ to be nontrivial for $\nu = 0 \text{ mod } 4$, but trivial for $\nu = 2 \text{ mod } 4$, for the same $\pi$ and $\central$.
We construct such an example by letting
\begin{align}
\pi( {\bf g}) &= [g_1 + g_2] \text{ mod }2 , \\
\central ({\bf g,h}) &= g_1 \cdot h_2
.
\end{align}
Noting that these can be written as $\pi = \pi' + \pi''$ and $\central = \pi' \cup \pi''$ for the homomorphisms $\pi'( {\bf g}) = g_1$ and $\pi''( {\bf g}) = g_2$, it is straightforward to show that
\begin{align}
\pi \cup \central &= \pi \cup \pi' \cup \pi''
,\\
\central \cup_1 \central &= \pi' \cup \pi \cup \pi''
,\\
\cbd ( (\pi \cup_1 \pi') \cup \pi'' ) &= \pi \cup \pi' \cup \pi'' + \pi' \cup \pi \cup \pi''
.
\end{align}

The second example is a choice of $\central$ for which $[\central \cup_1 \central]$ is nontrivial, and $[\pi \cup \central] \neq [\central \cup_1 \central]$ for any $\pi \in H^1(G,\mathbb{Z}_{2}^{\V})$.
For such $\central$, $[\coho{O}^{\central}]=[\coho{O}^{\eta}]$ is nontrivial for all $\pi$ when $\nu = 2 \text{ mod } 4$, but is trivial (at least) for $\pi =0$ when $\nu = 0 \text{ mod } 4$.
Such an example is given by choosing $\central$ to have
\begin{align}
\vv{1}&=\central ((1,0),(1,0)) = \central ((0,1),(0,1)) = \central ((1,1),(1,1)) \notag\\
&=\central ((1,0),(1,1)) = \central ((0,1),(1,0)) = \central ((1,1),(0,1))
\end{align}
and all other $\central({\bf g,h}) =\vv{0}$.
For this example, the eight distinct combinations of $\nu$ and $\pi$ yields eight distinct values of $[\coho{O}^{\central}]=[\coho{O}^{\eta}]$, so none of the $\nu = 0 \text{ mod } 4$ values equal any of the $\nu = 2 \text{ mod } 4$ values.

\subsubsection{FMTCs with trivial symmetry action on the quasiparticles}
\label{sec:Ex_no_perm}

In the case of a general FMTC $\fMTC$ with symmetry that acts trivially on the quasiparticles, the fractionalization analysis is very similar to that of the invertible fermionic topological phases.
The trivial symmetry action on $\fMTC_{\vv{0}}$ will be represented by $\rho_{\bf g}^{(0)} = \openone$, which can always be extended to a symmetry action on $\fMTC$, since $[O^{\rho}]=[\openone]$.
We will assume that $\ker(\res_{\fMTC_{\vv{0}}}) =  \mathbb{Z}_{2}^{\V}$, but if more general examples exist, it is a straightforward generalization.
It is convenient to divide the FMTCs into three classes based on whether $\mathcal{A}_{\vv{1}}$ and $\widehat{\bf A}_{\vv{1}}$ are empty.

We first consider FMTCs with $\mathcal{A}_{\vv{1}} \neq \varnothing$.
In this case, $\widehat{{\bf A}} = \widehat{\mathcal{A}}$.
The extended symmetry action on $\fMTC$ takes the form $[\rho_{\bf g}] = [\V]^{\pi({\bf g})}$, where $\pi \in H^1(G , \mathbb{Z}_{2}^{\V})$ is a homomorphism.
We analyze this with the representative $\rho_{\bf g} = \V^{\pi({\bf g})}$, in which case we can choose $\beta^{(\vv{0})}_{\psi_{\vv{0}}}({\bf g,h}) = 1$ and $\beta_{a_{\vv{x}}}({\bf g,h}) = i^{\vv{x} \cdot \pi({\bf g}) \cdot \pi({\bf h})}$.
It follows that $\coho{O}^{(0)}=\hat{\I}_{\vv{0}}$ and $\coho{O}=\I_{\vv{0}}$.

One possible fractionalization class on $\fMTC_{\vv{0}}$ is given by $\eta^{(0)}_{a_{\vv{0}}}({\bf g,h}) = 1$, corresponding to the trivially extended fermionic symmetry group.
Using $\mathsf{b}^{(\vv{0})}=\mathsf{0}$ and an arbitrary $\hat{e}_{\vv{1}} \in \widehat{\mathcal{A}}_{\vv{1}}$ in Eq.~\eqref{eq:O_central_cups_M0}, we find the obstruction
\begin{align}
\coho{O}^{(\vv{0})\central}({\bf g,h,k}) & = \cbd \bar{\hat{e}}_{\vv{1}}^{\central}({\bf g,h,k}) = \bar{\hat{e}}_{\vv{1}}^{\cbd \tilde{\central}({\bf g,h,k})} \notag \\
& = \hat{h}_{\vv{0}}^{ \tilde{\central}({\bf gh,k})\cdot\tilde{\central}({\bf g,h}) - \tilde{\central}({\bf g,hk})\cdot \tilde{\central}({\bf h,k})}
,
\end{align}
where $\hat{h}_{\vv{0}} = {\hat{e}}_{\vv{1}} \otimes {\hat{e}}_{\vv{1}} \in \widehat{\mathcal{A}}_{\vv{0}}$.
This can be written in the more compact form
\begin{align}
\label{eq:trivialactionO(0)w}
\coho{O}^{(\vv{0})\central} & = \hat{h}_{\vv{0}}^{\tilde{\central} \cup_{1} \tilde{\central}}
.
\end{align}
Thus, $\fMTC_{\vv{0}}$ with trivial symmetry action can only manifest a fermionic symmetry group $\mathcal{G}^{\eff} = \mathbb{Z}_{2}^{\eff} \times_{\central} G$ with $\central \in Z^2(G, \mathbb{Z}_{2}^{\eff})$ when $\hat{h}_{\vv{0}}^{\tilde{\central} \cup_{1} \tilde{\central}} \in B^3(G,\widehat{\mathcal{A}}_{\vv{0}})$.
When $\widehat{{\bf A}} = \widehat{\mathcal{A}}_{\vv{0}} \times \mathbb{Z}_{2}^{\eff}$, we can choose $\hat{e}_{\vv{1}}$ such that $\hat{h} = \hat{\I}_{\vv{0}}$, and it follows that all $\central \in Z^2(G, \mathbb{Z}_{2}^{\eff})$  can be manifested.

When the obstruction $[\coho{O}^{(\vv{0})\central}]$ vanishes, the fractionalization classes on $\fMTC_{\vv{0}}$ with fixed $\central$ are classified by $H^2(G,\widehat{\mathcal{A}}_{\vv{0}})$, with
\begin{align}
\eta^{(0)}_{a_{\vv{0}}}({\bf g,h}) &=  M_{a_{\vv{0}} \cohosub{w}^{(0)}({\bf g,h})}
,
\\
\coho{w}^{(0)}({\bf g,h}) &= \coho{y}^{(0)}({\bf g,h}) \otimes \hat{e}_{\vv{1}}^{\central({\bf g,h})}
,
\end{align}
where $\coho{y}^{(0)} \in C^2(G,\widehat{\mathcal{A}}_{\vv{0}})$ such that
\begin{align}
\cbd \coho{y}^{(0)} &= \hat{h}_{\vv{0}}^{\tilde{\central} \cup_{1} \tilde{\central}}
.
\end{align}

Considering the full FMTC $\fMTC$, we can either analyze the fractionalization directly or as the extension from $\fMTC_{\vv{0}}$.
Using $\mathsf{b}^{(\vv{0})}=\mathsf{0}$ and an arbitrary $e_{\vv{1}} \in \mathcal{A}_{\vv{1}}$ in Eq.~\eqref{eq:O_central_cups}, we find the obstruction is
\begin{align}
&\coho{O}^{\central}({\bf g,h,k})  = \cbd \bar{e}_{\vv{1}}^{\central}({\bf g,h,k}) \notag \\
& \quad = (\,^{\bf g} \bar{e}_{\vv{1}} \otimes {e}_{\vv{1}})^{\central({\bf h,k})} \otimes \bar{e}_{\vv{1}}^{\cbd \tilde{\central}({\bf g,h,k})} \notag \\
& \quad = \psi_{\vv{0}}^{\pi({\bf g})\cdot \central({\bf h,k})} \otimes h_{\vv{0}}^{\tilde{\central}({\bf gh,k})\cdot\tilde{\central}({\bf g,h}) - \tilde{\central}({\bf g,hk})\cdot \tilde{\central}({\bf h,k})}
,
\end{align}
where $h_{\vv{0}} = e_{\vv{1}} \otimes e_{\vv{1}} \in \mathcal{A}_{\vv{0}}$.
This can be written in the more compact form
\begin{align}
\label{eq:trivialactionOw}
\coho{O}^{\central} & = \psi_{\vv{0}}^{\pi \cup \central} \otimes h_{\vv{0}}^{\tilde{\central} \cup_{1} \tilde{\central}}
.
\end{align}
Thus, fractionalization on $\fMTC$ with $\rho$ and fixed $\central$ only exists when $\psi_{\vv{0}}^{\pi \cup \central} \otimes h_{\vv{0}}^{\tilde{\central} \cup_{1} \tilde{\central}} \in B^3_{\rho^{(0)}}(G,\mathcal{A}_{\vv{0}})$.

From the perspective of extension from $\fMTC_{\vv{0}}$, we first note that we can use $e_{\vv{1}}$ to write the vortices' topological charges as $a_{\vv{1}} = a_{\vv{0}} \otimes e_{\vv{1}}$.
As shown in Sec.~\ref{sec:fMTC}, we can use this to write $\mathcal{A} = \mathcal{A}_{\vv{0}} \times_{h} \mathbb{Z}_{2}^{\eff}$.
Also noting that $\mathcal{A}_{\vv{0}} = \mathbb{Z}_{2}^{\psi} \times \widehat{\mathcal{A}}_{\vv{0}}$, we can thus write the elements of $\mathcal{A}$ as $a_{\vv{x}} =  (a_{\vv{0}}^{(\psi)}, \hat{a}_{\vv{0}})\otimes e_{1}^{\vv{x}}$, where $a_{\vv{0}}^{(\psi)}$ is the $\mathbb{Z}_{2}^{\psi}$ component of $a_{\vv{0}}$.
We can now make a convenient choice of lift from $\widehat{\mathcal{A}}$ to $\mathcal{A}$ to be $\ell( \hat{a}_{\vv{x}} ) = (\I_{\vv{0}} , \hat{a}_{\vv{0}} )\otimes e_{1}^{\vv{x}}$.
From this, we can compute
\begin{align}
&\coho{O}^{\eta}({\bf g,h,k})  = \cbd \overline{\ell (\coho{w}^{(0)})({\bf g,h,k}) } \notag \\
&\quad = \cbd \overline{(\I_{\vv{0}} , \coho{y}^{(0)} , {\central})({\bf g,h,k}) } \notag \\
&\quad = \overline{(\I_{\vv{0}} , \cbd \coho{y}^{(0)}({\bf g,h,k}) , \vv{0} )} \otimes \cbd (\bar{e}_{\vv{1}}^{\central})({\bf g,h,k}) \notag \\
&\quad = \psi_{\vv{0}}^{\pi({\bf g}) \cdot \central({\bf h,k}) + \mathsf{h}_{\psi} \left( \tilde{\central}({\bf gh,k})\cdot\tilde{\central}({\bf g,h}) - \tilde{\central}({\bf g,hk})\cdot \tilde{\central}({\bf h,k}) \right)}
,
\end{align}
where
\begin{align}
\psi_{\vv{0}}^{\mathsf{h}_{\psi}} = h_{\vv{0}} \otimes \overline{\ell( \hat{h}_{\vv{0}} ) }
\end{align}
is the $\mathbb{Z}_{2}^{\psi}$ component of $h_{\vv{0}}$ (for this choice of lift).
This can be written in the more compact form
\begin{align}
\label{eq:trivialactionOeta}
\coho{O}^{\eta} &= \overline{ \ell( \coho{O}^{(\vv{0})\central}  )} \otimes \coho{O}^{\central} = \psi_{\vv{0}}^{\pi \cup \central + \mathsf{h}_{\psi} \cdot \central \cup_{1} \central}
.
\end{align}
From this it is clear that, in this case, the obstruction $[\coho{O}^{\central}]$ vanishes if and only if the obstructions $[\coho{O}^{(\vv{0})\central}]$ and $[\coho{O}^{\eta}]$ independently vanish.
Thus, extensions of fractionalization from $\fMTC_{\vv{0}}$ to $\fMTC$ for fixed $\central$ only exist when $\psi_{\vv{0}}^{\pi \cup \central + \mathsf{h}_{\psi} \cdot \central \cup_{1} \central} \in B^3_{\rho^{(0)}}(G,\mathbb{Z}_{2}^{\psi})$.

When the obstructions vanishes, the extension of fractionalization from $\fMTC_{\vv{0}}$ to $\fMTC$ for fixed $\central$ is torsorially classified by $H^{2}(G,\mathbb{Z}_{2}^{\psi})$, and the full fractionalization by $H^2(G,\mathcal{A}_{\vv{0}}) = H^{2}(G,\mathbb{Z}_{2}^{\psi}) \times H^2(G,\widehat{\mathcal{A}}_{\vv{0}})$.
The corresponding projective phases are
\begin{align}
\eta_{a_{\vv{x}}}({\bf g,h}) &=i^{\vv{x} \cdot \pi({\bf g})\cdot \pi({\bf h})}  M_{a_{\vv{x}} \cohosub{w}({\bf g,h})} ,
\\
\coho{w}({\bf g,h}) &= \coho{y}({\bf g,h}) \otimes e_{\vv{1}}^{\central({\bf g,h})}
\\
\coho{y} &= (\coho{p}, \coho{y}^{(0)} , \vv{0} ) ,
\end{align}
where $\coho{y}\in C^2(G,\mathcal{A}_{\vv{0}})$ and $\coho{p} \in C^2(G,\mathbb{Z}_{2}^{\psi})$ such that
\begin{align}
\cbd \coho{y} &= \psi_{\vv{0}}^{\pi \cup \central} \otimes h_{\vv{0}}^{\tilde{\central} \cup_{1} \tilde{\central}} ,
\\
\cbd \coho{p} &= \psi_{\vv{0}}^{\pi \cup \central + \mathsf{h}_{\psi} \cdot \central \cup_{1} \central}
.
\end{align}

Next, we consider FMTCs with $\mathcal{A}_{\vv{1}}  = \varnothing$ and ${\bf A}_{\vv{1}} \neq \varnothing$.
In this case, $[\Q] = [\V]$, so the extended symmetry action on $\fMTC$ takes the form $[\rho_{\bf g}] = [\Q]^{\rr({\bf g})}$, where $\rr \in C^1(G , \mathbb{Z}_{2}^{\Q})$ is not required to be a homomorphism.
We analyze this with the representative $\rho_{\bf g} = \Q^{\rr({\bf g})}$, in which case we can choose $\beta_{a_{\vv{x}}}({\bf g,h}) = (-1)^{\cbd \rr ({\bf g,h}) \delta_{a_{\vv{x}} \psi_{\vv{0}}} }$.
Even though $\rho_{\bf g}^{(\vv{0})} = \openone^{(\vv{0})}$, we must choose $\beta^{(\vv{0})}$ to match the extension, so we choose $\beta^{(0)}_{a_{\vv{0}}}({\bf g,h}) = (-1)^{\cbd \rr ({\bf g,h}) \delta_{a_{\vv{0}} \psi_{\vv{0}}} }$.
It follows that $\coho{O}^{(0)}=\hat{\I}_{\vv{0}}$ and $\coho{O}=\I_{\vv{0}}$.

One possible fractionalization class on $\fMTC_{\vv{0}}$ is given by $\eta^{(0)}_{a_{\vv{0}}}({\bf g,h}) = (-1)^{\cbd \rr ({\bf g,h}) \delta_{a_{\vv{0}} \psi_{\vv{0}}} }$.
Consequently, we can choose $\dot{\central}= \mathsf{b}^{(\vv{0})}=\cbd \rr$ and an arbitrary $\hat{e}_{\vv{1}} \in \widehat{{\bf A}}_{\vv{1}}$ to find
\begin{align}
\coho{O}^{(\vv{0})\central} & = \cbd ({\hat{e}}_{\vv{1}}^{\cbd \rr} \bar{\hat{e}}_{\vv{1}}^{\central}) = \cbd( \bar{\hat{e}}_{\vv{1}}^{{\central}}) \otimes \cbd (\hat{h}_{\vv{0}}^{\tilde{\vv{R}}} )
\notag \\
& = \hat{h}_{\vv{0}}^{\tilde{\central} \cup_{1} \tilde{\central} + \cbd \tilde{\vv{R}} }
,
\end{align}
where we have defined
\begin{align}
\tilde{\vv{R}}({\bf g,h}) =& 2 \tilde{r}({\bf g})  \tilde{r}({\bf h}) \tilde{r}({\bf gh})- \tilde{r}({\bf g}) \tilde{r}({\bf h})
\notag \\
& -\tilde{r}({\bf g}) \tilde{r}({\bf gh})- \tilde{r}({\bf h}) \tilde{r}({\bf gh}) + \tilde{r}({\bf gh})
,
\end{align}
and used the property
\begin{align}
\widetilde{\cbd r} = \cbd \tilde{r}  + 2 \tilde{\vv{R}}
.
\end{align}
Thus, $\fMTC_{\vv{0}}$ with trivial symmetry action can only manifest a fermionic symmetry group $\mathcal{G}^{\eff} = \mathbb{Z}_{2}^{\eff} \times_{\central} G$ with $\central \in Z^2(G, \mathbb{Z}_{2}^{\eff})$ when $\hat{h}_{\vv{0}}^{\tilde{\central} \cup_{1} \tilde{\central}} \in B^3(G,\widehat{\mathcal{A}}_{\vv{0}})$.
When $\widehat{{\bf A}} = \widehat{\mathcal{A}}_{\vv{0}} \times \mathbb{Z}_{2}^{\eff}$, we can choose $\hat{e}_{\vv{1}}$ such that $\hat{h} = \hat{\I}_{\vv{0}}$, and it follows that all $\central \in Z^2(G, \mathbb{Z}_{2}^{\eff})$  can be manifested.

For fixed $\central$, the fractionalization classes on $\fMTC_{\vv{0}}$ are classified by $H^2(G,\widehat{\mathcal{A}}_{\vv{0}})$, with
\begin{align}
\eta^{(0)}_{a_{\vv{0}}}({\bf g,h}) &=  (-1)^{\cbd \rr ({\bf g,h}) \delta_{a_{\vv{0}} \psi_{\vv{0}}} }  M_{a_{\vv{0}} \cohosub{w}^{(0)}({\bf g,h})}
,
\\
\coho{w}^{(0)}({\bf g,h}) &= \coho{y}^{(0)}({\bf g,h}) \otimes \hat{e}_{\vv{1}}^{\central({\bf g,h})} \otimes \bar{\hat{e}}_{\vv{1}}^{\cbd \rr ({\bf g,h})}
,
\end{align}
where $\coho{y}^{(0)} \in C^2(G,\widehat{\mathcal{A}}_{\vv{0}})$ such that
\begin{align}
\cbd \coho{y}^{(0)}&= \hat{h}_{\vv{0}}^{\tilde{\central} \cup_{1} \tilde{\central} + \cbd \tilde{\vv{R}} }
.
\end{align}

For the full FMTC $\fMTC$, we similarly have one of the possible fractionalization classes given by $\eta_{a_{\vv{x}}}({\bf g,h}) = (-1)^{\cbd \rr ({\bf g,h}) \delta_{a_{\vv{x}} \psi_{\vv{0}}} }$, and hence $\dot{\central}=\mathsf{b} =\cbd \rr$.
However, since $\mathcal{A}_{\vv{1}}  = \varnothing$, the obstruction is given by
\begin{align}
\label{eq:trivialactionOcentraldr}
\coho{O}^{\central} & = \central - \cbd \rr
.
\end{align}
In other words, $[\rho_{\bf g}] = [\Q]^{\rr({\bf g})}$ requires $\central = \cbd \rr$.

From the perspective of extension from $\fMTC_{\vv{0}}$, since $\mathcal{A}_{\vv{1}}  = \varnothing$, we must impose the restriction $\coho{w}^{(0)} \in C^2(G,\widehat{\mathcal{A}}_{\vv{0}})$ in order to be able to extend the fractionalization to $\fMTC$.
This similarly requires $\central = \cbd \rr$.
Writing elements of $\mathcal{A}_{\vv{0}} = \mathbb{Z}_{2}^{\psi} \times \widehat{\mathcal{A}}_{\vv{0}}$ as $a_{\vv{0}} = (a_{\vv{0}}^{(\psi)}, \hat{a}_{\vv{0}})$, we choose the lift from $\widehat{\mathcal{A}}_{\vv{0}}$ to $\mathcal{A}_{\vv{0}}$ to be $\ell( \hat{a}_{\vv{0}} ) = (\I_{\vv{0}} , \hat{a}_{\vv{0}} )$.
From this, we can compute the obstruction
\begin{align}
\label{eq:OetaA1nothing}
\coho{O}^{\eta}({\bf g,h,k}) & = \cbd \overline{\ell (\coho{w}^{(0)})({\bf g,h,k}) } = \I_{\vv{0}}.
\end{align}

Thus, we find that fractionalization on $\fMTC$ only exists when $\central = \cbd \rr$.
The extension of fractionalization from $\fMTC_{\vv{0}}$ to $\fMTC$ for $\central = \cbd \rr$ is torsorially classified by $H^{2}(G,\mathbb{Z}_{2}^{\psi})$, and the full fractionalization by $H^2(G,\mathcal{A}_{\vv{0}}) = H^{2}(G,\mathbb{Z}_{2}^{\psi}) \times H^2(G,\widehat{\mathcal{A}}_{\vv{0}})$.
The corresponding projective phases are
\begin{align}
\eta_{a_{\vv{x}}}({\bf g,h}) &=  (-1)^{\cbd \rr ({\bf g,h}) \delta_{a_{\vv{x}} \psi_{\vv{0}}} }   M_{a_{\vv{x}} \cohosub{w}({\bf g,h})} ,
\\
\coho{w} &= (\coho{p}, \coho{w}^{(0)} ) ,
\end{align}
where $\coho{w}\in Z^2(G,\mathcal{A}_{\vv{0}})$, $\coho{p} \in Z^2(G,\mathbb{Z}_{2}^{\psi})$, and $\coho{w}^{(0)}\in Z^2(G,\widehat{\mathcal{A}}_{\vv{0}})$.

Finally, we consider FMTCs with $\mathcal{A}_{\vv{1}}  = \varnothing$ and ${\bf A}_{\vv{1}} = \varnothing$.
In this case, $[\Q] \neq [\V]$, so the extended symmetry action on $\fMTC$ takes the form $[\rho_{\bf g}] = [\V]^{\pi(\bf g)}$, where $\pi \in H^1(G , \mathbb{Z}_{2}^{\V})$ is a homomorphism.
We analyze this with the representative $\rho_{\bf g} = \V^{\pi({\bf g})}$, in which case we can choose $\beta^{(\vv{0})}_{\psi_{\vv{0}}}({\bf g,h}) = 1$ and $\beta_{a_{\vv{x}}}({\bf g,h}) = i^{\vv{x} \cdot \pi({\bf g}) \cdot \pi({\bf h})}$.
It follows that $\coho{O}^{(0)}=\hat{\I}_{\vv{0}}$ and $\coho{O}=\I_{\vv{0}}$.

One possible fractionalization class on $\fMTC_{\vv{0}}$ is given by $\eta^{(0)}_{a_{\vv{0}}}({\bf g,h}) = 1$, so we have $\dot{\central}= \mathsf{b}^{(\vv{0})}= \vv{0}$.
However, since $\widehat{{\bf A}}_{\vv{1}}  = \varnothing$, the obstruction is given by
\begin{align}
\coho{O}^{(\vv{0})\central} & = \central
.
\end{align}
In other words, this case requires $\central = \vv{0}$.
For this, the fractionalization classes on $\fMTC_{\vv{0}}$ are classified by $H^2(G,\widehat{\mathcal{A}}_{\vv{0}})$, with
\begin{align}
\eta^{(0)}_{a_{\vv{0}}}({\bf g,h}) &=  M_{a_{\vv{0}} \cohosub{w}^{(0)}({\bf g,h})}
,
\end{align}
where $\coho{w}^{(0)} \in Z^2(G,\widehat{\mathcal{A}}_{\vv{0}})$.

For the full FMTC $\fMTC$, we have one of the possible fractionalization classes given by $\eta_{a_{\vv{x}}}({\bf g,h}) = i^{\vv{x} \cdot \pi({\bf g}) \cdot \pi({\bf h})}$, and so we choose $\dot{\central}= \mathsf{b}=\vv{0}$.
Since $\mathcal{A}_{\vv{1}}  = \varnothing$, the obstruction is also given by
\begin{align}
\label{eq:trivialactionOcentral}
\coho{O}^{\central} & = \central
,
\end{align}
so $\central = \vv{0}$ is again required.

From the perspective of extension from $\fMTC_{\vv{0}}$, since $\widehat{{\bf A}}_{\vv{1}}  = \varnothing$, this similarly requires $\central = \vv{0}$.
Writing elements of $\mathcal{A}_{\vv{0}} = \mathbb{Z}_{2}^{\psi} \times \widehat{\mathcal{A}}_{\vv{0}}$ as $a_{\vv{0}} = (a_{\vv{0}}^{(\psi)}, \hat{a}_{\vv{0}})$, we choose the lift from $\widehat{\mathcal{A}}_{\vv{0}}$ to $\mathcal{A}_{\vv{0}}$ to be $\ell( \hat{a}_{\vv{0}} ) = (\I_{\vv{0}} , \hat{a}_{\vv{0}} )$.
From this, we can compute the obstruction
\begin{align}
\coho{O}^{\eta}({\bf g,h,k}) & = \cbd \overline{\ell (\coho{w}^{(0)})({\bf g,h,k}) } = \I_{\vv{0}}.
\end{align}

Thus, we find that fractionalization on $\fMTC_{\vv{0}}$ and $\fMTC$ with $[\rho_{\bf g}^{(0)}] = [\openone]$ only exists when $\central = \vv{0}$.
The extension of fractionalization from $\fMTC_{\vv{0}}$ to $\fMTC$ is torsorially classified by $H^{2}(G,\mathbb{Z}_{2}^{\psi})$, and the full fractionalization by $H^2(G,\mathcal{A}_{\vv{0}}) = H^{2}(G,\mathbb{Z}_{2}^{\psi}) \times H^2(G,\widehat{\mathcal{A}}_{\vv{0}})$.
The corresponding projective phases are
\begin{align}
\eta_{a_{\vv{x}}}({\bf g,h}) &=  i^{\vv{x} \cdot \pi({\bf g}) \cdot \pi({\bf h})} M_{a_{\vv{x}} \cohosub{w}({\bf g,h})} ,
\\
\coho{w} &= (\coho{p}, \coho{w}^{(0)} ) ,
\end{align}
where $\coho{w}\in Z^2(G,\mathcal{A}_{\vv{0}})$, $\coho{p} \in Z^2(G,\mathbb{Z}_{2}^{\psi})$, and $\coho{w}^{(0)}\in Z^2(G,\widehat{\mathcal{A}}_{\vv{0}})$.

For this case, it is easy to repeat the analysis for the slightly nontrivial symmetry action $\rho^{(0)}_{\bf g} = \Q^{\rr({\bf g})}$ with extension $\rho_{\bf g} = \Q^{\rr({\bf g})} \V^{\pi(\bf g)}$.
The result is fractionalization on $\fMTC_{\vv{0}}$ and $\fMTC$ only exists when $\central = \cbd \rr$, and is classified by $H^2(G,\mathcal{A}_{\vv{0}}) = H^2(G,\widehat{\mathcal{A}}_{\vv{0}}) \times H^{2}(G,\mathbb{Z}_{2}^{\psi})$, with the corresponding projective phases
\begin{align}
\eta^{(0)}_{a_{\vv{0}}}({\bf g,h}) &=  (-1)^{\cbd \rr ({\bf g,h}) \delta_{a_{\vv{0}} \psi_{\vv{0}}} }  M_{a_{\vv{0}} \cohosub{w}^{(0)}({\bf g,h})}
,
\\
\eta_{a_{\vv{x}}}({\bf g,h}) &= (-1)^{\cbd \rr ({\bf g,h}) \delta_{a_{\vv{0}} \psi_{\vv{0}}} } i^{\vv{x} \cdot \pi({\bf g}) \cdot \pi({\bf h})} M_{a_{\vv{x}} \cohosub{w}({\bf g,h})} ,
\\
\coho{w} &= (\coho{p}, \coho{w}^{(0)} ) ,
\end{align}
where $\coho{w}^{(0)} \in Z^2(G,\widehat{\mathcal{A}}_{\vv{0}})$, $\coho{w}\in Z^2(G,\mathcal{A}_{\vv{0}})$, and $\coho{p} \in Z^2(G,\mathbb{Z}_{2}^{\psi})$.

\subsubsection{\texorpdfstring{$\text{SO}(3)_{6}$}{SO36} with Time-Reversal Symmetry}
\label{sec:Ex_SO3_6}

In this section, we consider the SMTC $\text{SO}(3)_{6}$, which can be viewed as the restriction of the FMTC $\text{SU}(2)_{6}$ to topological charges labeled by integers
\begin{align}
\fMTC_{\vv{0}} &= \{ 0, 1, 2, 3 \}
.
\end{align}
As a fermionic theory, the physical fermion is given by $\psi_{\vv{0}} = 3$.
The fusion rules are given by
\begin{align}
1 \otimes 1 &= 2 \otimes 2 = 0 \oplus 1 \oplus 2 ,\\
1 \otimes 2 &=  1 \oplus 2 \oplus 3 ,\\
1 \otimes 3 &=  2 ,\\
3 \otimes 3 &= 0.
\end{align}
The quantum dimensions are $d_0 = d_3 =1$ and $d_1 = d_2 = 1+ \sqrt{2}$, and the topological twists are $\theta_{0}=1$, $\theta_{1}=i$, $\theta_{2}=-i$, and $\theta_{3}=-1$.
The full topological data of can be found in Ref.~\onlinecite{Bond2007}.
This example has $\widehat{\bf A}_{\vv{1}} = \varnothing$, so $\Q$ acts nontrivially.
In fact, this is the only nontrivial topological symmetry, so
\begin{align}
\text{Aut}^{\eff}(\fMTC_{\vv{0}}) &= \mathbb{Z}_{2}^{\Q}
.
\end{align}
The case of unitary on-site global symmetries will be analyzed in Sec.~\ref{sec:ex_SU2} along with all $\text{SO}(3)_{4n+2}$, so we will only focus on time-reversing symmetries here.
The theory of $\text{SO}(3)_{6}$ with $G = \mathbb{Z}_{2}^{\bf T}$ time-reversing global symmetry was previously studied in Ref.~\onlinecite{Fidkowski2013} as an anomalous surface termination theory, where it was shown to necessarily have local ${\bf T}^{2} = (-1)^{\bf F}$ behavior on the physical fermions.
However, the fermionic constraints, $U_{\bf g}(\psi_{\vv{0}},\psi_{\vv{0}}; \I_{\vv{0}}) =1$ and $\gamma_{\psi_{\vv{0}}}({\bf g})=1$, and their implications were not previously understood, so it is worth re-examining this example in light of these developments.
(In fact, this example motivated our development of the notion of $\Q$-projective actions, since it is required in this example, as we will show.)

A time-reversing (anti-unitary) topological symmetry $\tau$ acting on $\fMTC_{\vv{0}} = \text{SO}(3)_{6}$ is required to permute the quasiparticles $\tau: 1 \leftrightarrow 2$, while leaving $0$ and $3$ fixed.
The full details of the $U$-symbols of such a topological symmetry were computed in Ref.~\onlinecite{Tata2021}.
It is obvious that $\Q \tau$ is also a time-reversing topological symmetry.
On the other hand, from the explicit $U$-symbols, we find that
\begin{align}
\tau^2 &= \Q
.
\end{align}
Thus, we find
\begin{align}
{\bf Aut}^{\eff}_{\tau}(\fMTC_{\vv{0}}) &= \mathbb{Z}_{4}^{\tau,\Q}
,
\end{align}
where we use ${\bf Aut}^{\eff}_{\tau}$ to denote the $p=0$ subgroup of ${\bf Aut}^{\eff}$ that includes time-reversing, but not spatial parity reversing symmetries.

Considering fermionic symmetry actions $[\rho]$ for $G = \mathbb{Z}_{2}^{\bf T}$ time-reversing global symmetry, we can now see that it is impossible to have a strict homomorphism from $G$ to ${\bf Aut}^{\eff}_{\tau}(\fMTC_{\vv{0}})$, since the only homomorphisms from $\mathbb{Z}_2$ to $\mathbb{Z}_4$ would map the nontrivial group element ${\bf T}$ to one of the unitary elements $[\openone]$ or $[\Q]$.
In order to respect the anti-unitarity of ${\bf T}$, we must choose $\rho_{\bf T} = \tau$ or $\tau^{3}$.
This leads to $\rho_{\bf T}^{2} = \Q$, which is a $\Q$-projective action with nontrivial $\Q$-projective factor $\phi({\bf T,T}) =1$.
This nontrivial $\Q$-projective behavior immediately shows us that the fermionic global symmetry group must be
\begin{align}
\mathcal{G}^{\eff} &= \mathbb{Z}_{2}^{\eff} \times_\central  \mathbb{Z}_{2}^{\bf T}= \mathbb{Z}_{4}^{{\bf T},\eff}
.
\end{align}
In other words, trivial $\central$ is obstructed and only $\central({\bf T,T}) =1$ is allowed to occur for $\text{SO}(3)_{6}$.
More generally, a symmetry group $G$ with time-reversing symmetries will necessarily have nontrivial $\Q$-projective action if any of the time-reversing group elements have order $2 \text{ mod }4$; if the time-reversing group elements all have order $0 \text{ mod }4$, then $\phi$ can be trivial.

Turning to fermionic symmetry fractionalization for $G = \mathbb{Z}_{2}^{\bf T}$, we see that $\kappa_{\bf T,T} = \Q$, and thus
\begin{align}
\beta_{a_{\vv{0}}}({\bf T,T}) &= (-1)^{\delta_{a_{\vv{0}} \psi_{\vv{0}} }}
.
\end{align}
This yields $\coho{O}^{(0)} = \widehat{\I}_{\vv{0}}$, so fractionalization is not obstructed.
Since $\widehat{A} = \{ \widehat{\I}_{\vv{0}} \}$, we always have $H^2_{[\rho^{(0)}]}(G,\widehat{\mathcal{A}}_{\vv{0}}) = \mathbb{Z}_{1}$, i.e. there is only ever at most one fractionalization class on $\text{SO}(3)_{6}$.
In this case, we see that the fermionic symmetry fractionalization class is characterized by
\begin{align}
\eta_{\psi_{\vv{0}}}({\bf T,T}) &= -1
.
\end{align}

The $\mathbb{Z}_{2}^{\eff}$ modular extensions of $\text{SO}(3)_{6}$, which include $\fMTC =\text{SU}(2)_{6}$, all have two $v$-type vortices and one $\sigma$-type vortex.
They also have chiral central charge $c_{-} = [\frac{9}{4} + \frac{\nu}{2}]\text{ mod }8$, where $\nu$ is an integer specifying the particular modular extension obtained from $\text{SU}(2)_{6}$ by stacking as $\fMTC ' = \fMTC \fprod \ifo^{(\nu)}$.
This makes it clear that the time-reversing topological symmetries of $\text{SO}(3)_{6}$ cannot be extended to topological symmetries of any of the modular extensions, as such symmetries are only possible for FMTCs with $c_{-} = 0 \text{ mod }4$, i.e. the extension of $\tau$ to the modular extensions is obstructed.
However, one can na\"ively attempt to determine the manner in which such an extension would act on the topological charge supersectors of vortices by applying Eq.~\eqref{eq:vortex_lifted_permutation}.
Doing so reveals that a na\"ive extension of a time-reversing topological symmetry would interchange the two vortex supersectors.
Of course, this is impossible for a fixed modular extension, since a $v$-type supersector pair of vortices cannot be interchanged with a $\sigma$-type vortex.
However, it is compatible with the generalized notion of extensions of time-reversing topological symmetries discussed in Ref.~\onlinecite{Bark2021cascade}, where the generalized extensions are allowed to be maps between different modular extensions of the same SMTC.
In this case, interchanging vortex supersectors together with mapping to a time-reversal partner theory ($c_-$ to $[-c_{-}]\text{ mod }8$) will map $v$-type vortices of the original theory to $v$-type vortices of the partner theory, and similarly the $\sigma$-type will map to $\sigma$-type.
(Recall that stacking with $\ifo^{(\nu)}$ for $\nu$ odd changes $v$-type vortices to $\sigma$-type, and vice-versa.)

\section{Symmetry Defects in Fermionic Topological Phases}
\label{sec:fermionic-defectification}

In this section, we describe the algebraic theory of symmetry defects in a fermionic topological phase whose quasiparticles are described by the SMTC $\fMTC_{\vv{0}}$ and vortices by $\fMTC_{\vv{1}}$; collectively described by the FMTC $\fMTC$.

Thus far we have described how the global symmetries act on the topological states, captured by $\fMTC$, $\rho$, $U$ and $\eta$.
We now incorporate symmetry defects into our theory.
These correspond to topological charges which carry labels $\mathpzc{g}\in \mathcal{G}^{\eff}$ indicating that they are $\mathpzc{g}$-defects.
Correspondingly, we will need to extend all of the topological data ($F$, $R$, $U$ and $\eta$) to these point defects.
This data will be subject to several consistency conditions.
In the bosonic case, these are the extensions of the pentagon equation [Eq.~\eqref{eq:pentagoneqn}] and generalizations of the hexagon equations [Eqs.~(\ref{eq:hex1}-\ref{eq:hex2})] for BTCs.
It is not always possible to solve these $G$-crossed consistency conditions;
each symmetry fractionalization class has a ``defectification obstruction'' $[\defectO] \in H^4(G,\text{U}(1))$ that must vanish in order to extend to a full defect theory~\cite{ENO}.
Fermionic defectification has additional constraints associated with the physical fermion.
We first review the bosonic case, before introducing the necessary modifications for the fermionic setting.

\subsection{ \texorpdfstring{${G}$}{G}-crossed BTCs}

We begin by reviewing the definition of a ${G}$-crossed BTC, without requiring that it correspond to either a bosonic or fermionic topological phase.
In general, one can view a $G$-crossed BTC as the extension of a BTC with specified symmetry action and fractionalization to a theory of symmetry defects.
We first pick a BTC $\mcb$ and a symmetry group $G$, and then incorporate the symmetry defects.

The fusion structure of the symmetry defects is given by a $G$-graded FTC
\begin{align}
\mcb_{G}=\bigoplus_{{\bf g} \in G} \mcb_{{\bf g}}
,
\end{align}
such that $\mcb_{{\bf 0}}=\mcb$.
For a group element ${\bf g}\in G$, there may be topologically distinct ${{\bf g}}$-defects, which are given by the simple objects $a \in \mcb_{{\bf g}}$.
It is useful to explicitly denote the symmetry group label of defects by writing the topological charges as $a_{{\bf g}}$.
The $G$-grading does not alter any of the defining properties of a FTC, but simply introduces the additional constraint that the fusion rules are $G$-graded, meaning
\begin{align}
a_{{\bf g}} \otimes b_{{\bf h}} = \bigoplus_{c_{{\bf gh}} \in \mcb_{{\bf gh}} }  N_{a_{{\bf g}} b_{{\bf h}}}^{c_{{\bf gh}}} c_{{\bf gh}}
\end{align}
Similar to quasiparticles, defect fusion is associative, with associativity isomorphism given by the $F$-symbols,
\begin{align}
\FLeft \!\!\!\!\!\!= \sum_{f_{{\bf hk}}, \mu,\nu}
\left[ F^{a_{{\bf g}}b_{{\bf h}} c_{{\bf k}}}_{d_{{\bf ghk}} }  \right]_{(e_{{\bf gh}},\alpha,\beta)(f_{{\bf hk}},\mu,\nu)}\!\!\!\!\!\!\!\!\!\!\!
\FRight.
\end{align}
The $F$-symbols must again satisfy the pentagon equation (Eq.~\ref{eq:pentagoneqn}).
The fusion rules of defects generally do not need to be commutative, though the fusion rules of the $\mcb_{{\bf 0}}$ are necessarily commutative, since it is a BTC.

In principle, the $G$-graded extension does not need to be faithful, meaning $\mcb_{{\bf g}}$ is not required to be non-empty for all ${\bf g}$, as long as the fusion rules of defects are nonetheless consistent.
When the grading is not faithful, the non-empty sectors correspond to a subgroup $H <G$, which defines a faithfully $H$-extended subcategory.

Next, we incorporate the symmetry action, which enters through the $G^{}$-crossed braiding transformations of symmetry defects.
These defect braiding transformations generalize the ordinary notion of braiding.
For this, we introduce the shorthand notations
\begin{align}
{ \bar{{\bf g}} }&= {{\bf g}}^{-1}, \\
{}^{{\bf g}} {{\bf h}} &={\bf gh\bar{g}}, \\
{}^{{\bf g}} b_{{\bf h}} &= \rho_{{\bf g}}(b_{{\bf h}}) \in \mathcal{B}_{{\bf gh\bar{g}}}
,
\end{align}
where $\rho$ acting on $\mcb_{{\bf 0}}$ is a representative of the symmetry action $[\rho] : G \to \text{Aut}(\mcb)$, and the possible extensions of the action to the symmetry defects will be determined by imposing consistency conditions that we will discuss.

On the spatial manifold, a ${\bf g}$-defect is the endpoint of a ${\bf g}$-symmetry branch line.
We can picture the corresponding worldline of a ${\bf g}$-defect as the termination of a ${\bf g}$-branch worldsheet.
This is represented for a fusion vertex in Fig.~\ref{fig:branchsheets}.

\begin{figure}[t!]
   \centering
   \includegraphics[width=\columnwidth]{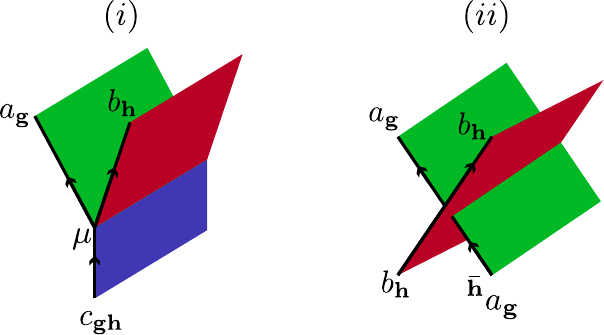}
   \caption{(i) A fusion vertex of symmetry defects, showing the symmetry branch worldsheets that terminate on the corresponding defect worldlines.
   (ii) A braiding exchange of symmetry defects, showing the $a_{\bf g}$ defect passing through the ${\bf h}$-branch line.}
   \label{fig:branchsheets}
\end{figure}

In the diagrammatic formalism, we leave the defect branch sheets implicit and assume they emanate from the lines labeled by defect topological charges back into the page.
With this picture in mind, we see that when a ${\bf h}$-defect line passes in front of an object, it acts on that object with the ${\bf h}$-symmetry action (see Fig.~\ref{fig:branchsheets}).
Thus, we define the $G$-crossed braiding operator diagrammatically as
\begin{align}
R^{a_{{\bf g}} b_{{\bf h}}} = \Ragbh.
\end{align}
The operator $R^{a_{{\bf g}} b_{{\bf h}}}$ provides an isomorphism between the vector space $V_{c_{{\bf gh}}}^{b_{{\bf h}} {}^{{\bf \bar{h}}}a_{{\bf g}}}$ and $V_{c_{{\bf gh}}}^{a_{{\bf g}} b_{{\bf h}}}$.
We note that this implies $^{{\bf g}}a_{{\bf g}} = a_{{\bf g}}$.

Resolving $R^{a_{{\bf g}} b_{{\bf h}}}$ in the basis of the fusion spaces, we find the diagrammatic definition of the $R$-symbol given by
\begin{align}
\RLeft = \sum_{\nu} \left[ R^{a_{{\bf g}} b_{{\bf h}}}_{c_{{\bf gh}}} \right]_{\mu \nu}\RRight.
\end{align}
The inverse $G$-crossed braiding operator is defined in the obvious manner.

The symmetry action on vertex basis states, i.e. the $U_{{\bf k}}$-symbols, is generated by sliding defect lines over the vertices
\begin{align}
\label{eq:U-2}
\ULeft = \sum_{\nu} \left[ U_{{\bf k}}(a_{{\bf g}}, b_{{\bf h}}; c_{{\bf gh}})\right]_{\mu \nu}\; \URight
.
\end{align}

On the other hand, the symmetry fractionalization $\eta$-symbols are generated by sliding defect lines under vertices
\begin{align}
\label{eq:eta-2}
\etaLeft = \eta_{x_{{\bf k}}}({{\bf g}},{{\bf h}})\;\etaRight.
\end{align}
Note that these sliding moves are trivial in a BTC (where all objects are ${{\bf 0}}$-defects).
It should be clear that $U_{{\bf k}}(a_{{\bf 0}}, b_{{\bf 0}}; c_{{\bf 0}})$ corresponds to the symmetry action on $\mcb$, while $\eta_{x_{{\bf 0}}}({{\bf g}},{{\bf h}})$ corresponds to the symmetry fractionalization projective phases of $\mcb$.

The ability to freely remove ${{\bf 0}}$-branch sheets and $\I_{{\bf 0}}$-lines allows us to fix
\begin{align}
\label{Uetatrivial}
\left[U_{{\bf 0}}(a_{{\bf g}},b_{{\bf h}}; c_{{\bf gh}})\right]_{\mu\nu}&=\delta_{\mu \nu}
\\
\left[U_{{\bf k}}(a_{{\bf g}},\I_{{\bf 0}}; a_{{\bf g}})\right]_{\mu \nu} &= \left[U_{{\bf k}}(\I_{{\bf 0}},b_{{\bf h}}; b_{{\bf h}})\right]_{\mu \nu}= \delta_{\mu \nu}
\\ \eta_{\I_{{\bf 0}}}({{\bf g}},{{\bf h}})&= \eta_{x_{{\bf k}}}({{\bf 0}},{{\bf h}}) = \eta_{x_{{\bf k}}}({{\bf g}},{{\bf 0}}) = 1.
\end{align}

The symmetry fractionalization pattern must respect associativity of the symmetry action, and be compatible with the $U$-symbols.
This imposes the following consistency conditions:
\begin{align}\label{eq:eta1}
{\eta}_{a_{{\bf g}}}({{\bf h}}, {{\bf k}}) \eta_{a_{{\bf g}}}({{\bf hk}},{{\bf l}}) &= \eta_{a_{{\bf g}}}({{\bf h}},{{\bf kl}}) \eta_{{}^{{\bf \bar{ h}}}{a_{{\bf g}}}}({{\bf k}},{{\bf l}}),
\end{align}
and
\begin{widetext}
\begin{align}
 \frac{\eta_{a_{{\bf g}}}({{\bf k}},{{\bf l}}) \eta_{b_{{\bf h}}}( {{\bf k}},{{\bf l}})  }{\eta_{c_{{\bf gh}}}({{\bf k}},{{\bf l}})} =
 \sum_{\alpha, \beta}
 \left[ U_{{\bf k}}(a_{{\bf g}}, b_{{\bf h}}; c_{{\bf gh}})^{-1} \right]_{\mu \alpha}
 \left[U_{{\bf l}}\left({}^{{\bf \bar{k}}}a_{{\bf g}}, {}^{{\bf \bar{k}}}b_{{\bf h}}; {}^{{\bf \bar{k}}}c_{{\bf gh}} \right)^{-1} \right]_{\alpha \beta}
 \left[ U_{{\bf k}l}(a_{{\bf g}}, b_{{\bf h}}; c_{{\bf gh}}) \right]_{\beta \nu}
=\kappa_{{{\bf k}},{{\bf l}}}(a,b;c) \delta_{\mu \nu},
\label{eq:sliding}
\end{align}
where $\kappa_{{{\bf g}},{{\bf h}}}$ is the natural isomorphism that matches the representative symmetry actions $\kappa_{{{\bf g}},{{\bf h}}} \circ \rho_{{\bf g}} \circ \rho_{{\bf h}} = \rho_{{\bf gh}}$.
Eqs.~\eqref{eq:eta1} and \eqref{eq:sliding} are the natural generalizations of Eqs.~\eqref{eq:kappa_U}, \eqref{eq:kapp-eta}, and \eqref{eq:eta_relation}, extended to the symmetry defects.

\begin{figure*}[t!]
   \centering
   \includegraphics{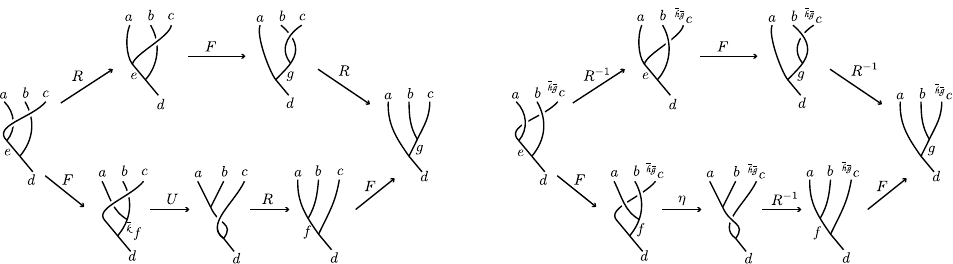}
   \caption{The heptagon equations that enforce consistency of $G^{}$-crossed braiding with fusion.
   For convenience we have left the group labels implicit in figure.
   In the left panel, they are given by $a_{{\bf g}}$, $b_{{\bf h}}$, $c_{{\bf k}}$, $d_{{\bf ghk}}$, $e_{{\bf gk}}$, $f_{{\bf gh}}$, and $g_{{\bf hk}}$.
   In the right panel, they are $a_{{\bf g}}$, $b_{{\bf h}}$, $c_{{\bf k}}$, $d_{{\bf kgh}}$, $e_{{\bf kg}}$, $f_{{\bf gh}}$, and $g_{{\bf \bar{g}kgh}}$.
}
   \label{fig:heptagon}
\end{figure*}

Consistency of the $R$-symbols with $F$-, $U$-, and $\eta$-symbols results in the heptagon equations for counterclockwise braids
\begin{align}
&
\sum_{\lambda,\gamma}
\left[ R^{a_{{\bf g}}c_{{\bf k}}}_{e_{{\bf gk}}} \right]_{\alpha \lambda}
\left[F_{d_{{\bf ghk}}}^{a_{{\bf g}}c_{{\bf k}} \,^{{\bf \bar{k}}}b_{{\bf h}}}\right] _{(e_{{\bf gk}},\lambda,\beta),(g_{{\bf hk}},\gamma,\nu)}
\left[R_{g_{{\bf hk}}}^{b_{{\bf h}}c_{{\bf k}}} \right]_{\gamma \mu}\notag \\
& \quad \quad \quad  =\sum_{f_{{\bf gh}},\sigma,\delta, \eta, \psi}
\left[ F_{d_{{\bf ghk}}}^{c_{{\bf k}} \,^{{\bf \bar{k}}}a_{{\bf g}} \,^{{\bf \bar{k}}}b_{{\bf h}}}\right] _{( e_{{\bf gk}},\alpha,\beta),(\,^{{\bf \bar{k}}}f_{{\bf gh}},\delta,\sigma)}
\left[ U_{{\bf k}}\left( a_{{\bf g}} ,b_{{\bf h}} ;f_{{\bf gh}} \right)\right]_{\delta\eta}
\left[R_{d_{{\bf ghk}}}^{f_{{\bf gh}} c_{{\bf k}}}\right]_{\sigma \psi}
\left[F_{d_{{\bf ghk}}}^{a_{{\bf g}}b_{{\bf h}}c_{{\bf k}}}\right] _{(f_{{\bf gh}},\eta,\psi),(g_{{\bf hk}},\mu,\nu)}
,
\label{eq:heptagon+}
\end{align}
and for clockwise braids
\begin{align}
&
\sum_{\lambda,\gamma}\left[ \left( R_{e_{{\bf kg}}}^{c_{{\bf k}}a_{{\bf g}}}\right) ^{-1}\right]_{\alpha \lambda}
\left[ F_{d_{{\bf kgh}}}^{a_{{\bf g}} \,^{{\bf \bar{g}}}c_{{\bf k}}b_{{\bf h}}}\right] _{(e_{{\bf kg}},\lambda,\beta),(g_{{\bf \bar{g}kgh}},\gamma,\nu)}
\left[ \left( R_{g_{{\bf \bar{g}kgh}}}^{\,^{{\bf \bar{g}}}c_{{\bf k}} b_{{\bf h}}}\right) ^{-1}\right]_{\gamma \mu} \notag \\
&\quad  \quad  \quad =\sum_{f_{{\bf gh}},\sigma,\delta,\psi}
\left[ F_{d_{{\bf kgh}}}^{c_{{\bf k}}a_{{\bf g}}b_{{\bf h}}}\right] _{(e_{{\bf kg}},\alpha,\beta),(f_{{\bf gh}},\delta,\sigma)}
\eta_{c_{{\bf k}}}\left({{\bf g}},{{\bf h}}\right)
\left[\left(R_{d_{{\bf kgh}}}^{c_{{\bf k}}f_{{\bf gh}}}\right) ^{-1}\right]_{\sigma,\psi}
\left[ F_{d_{{\bf kgh}}}^{a_{{\bf g}}b_{{\bf h}}\,^{{\bf \bar{h}\bar{g}}}c_{{\bf k}}}\right]_{(f_{{\bf gh}},\delta,\psi),(g_{{\bf \bar{g}kgh}},\mu,\nu)}.
\label{eq:heptagon-}
\end{align}
\end{widetext}
The heptagon equations are depicted diagrammatically in Fig.~\ref{fig:heptagon}.
Clearly, the heptagon equations are generalizations of the hexagon equations of BTCs, now including the nontrivial step of sliding a charge line over or under a vertex.

The $N$, $F$, $R$, $G$, $\rho$, $U$, and $\eta$, subject to the consistency conditions, constitute the basic data of a $G$-crossed BTC, which we denote as $\mcb_{G}^{\times}$.
Similar to BTCs, different sets of basic data can represent the same $G^{}$-crossed BTC if they are related by gauge transformations.
The vertex basis gauge transformations reviewed for BTCs generalize in the natural way to defect theories.
In particular, if we define
\begin{align}
\widetilde{\ket{a_{{\bf g}},b_{{\bf h}}; c_{{\bf gh}}, \mu}} = \sum_{\mu '} \left[ \Gamma^{a_{{\bf g}},b_{{\bf h}}}_{c_{{\bf gh}}} \right]_{\mu \mu'}
\ket{a_{{\bf g}},b_{{\bf h}}; c_{{\bf gh}}, \mu'}
,
\end{align}
then the $F$-, $R$-, $U$-, and $\eta$-symbols transform as
\begin{widetext}
\begin{align}
\left[\widetilde{F}_{d_{{\bf ghk}}}^{a_{{\bf g}} b_{{\bf h}} c_{{\bf k}}} \right]_{(e_{{\bf gh}},\alpha,\beta)(f_{{\bf hk}},\mu,\nu)} &= \sum_{\alpha',\beta',\mu',\nu'}
[\Gamma_{e_{{\bf gh}}}^{a_{{\bf g}} b_{{\bf h}}}]_{\alpha \alpha'}
[\Gamma_{d_{{\bf ghk}}}^{e_{{\bf gh}} c_{{\bf k}}}]_{\beta \beta'}
\left[F^{a_{{\bf g}} b_{{\bf h}} c_{{\bf k}} }_{d_{{\bf ghk}}} \right]_{(e_{{\bf gh}},\alpha',\beta')(f_{{\bf hk}},\mu',\nu')}
[( \Gamma_{f_{{\bf hk}}}^{b_{{\bf h}} c_{{\bf k}}})^{-1} ]_{\mu' \mu}
[( \Gamma_{d_{{\bf ghk}}}^{a_{{\bf g}} f_{{\bf hk}}})^{-1} ]_{\nu' \nu}
,\\
\left[ \widetilde{R}^{a_{{\bf g}} b_{{\bf h}}}_{c_{{\bf gh}}} \right]_{\mu \nu} &= \sum_{\mu', \nu'} \left[\Gamma^{b_{{\bf h}} {}^{\bar{{\bf h}}}a_{{\bf g}}}_{c_{{\bf gh}}} \right]_{\mu \mu'} \left[R^{a_{{\bf g}} b_{{\bf h}}}_{c_{{\bf gh}}} \right]_{\mu' \nu'} \left[ \left( \Gamma^{a_{{\bf g}}b_{{\bf h}}}_{c_{{\bf gh}}} \right)^{-1} \right]_{\nu' \nu}
,
\\
\left[\widetilde{U}_{{\bf k}}(a_{{\bf g}},b_{{\bf h}}; c_{{\bf gh}}) \right]_{\mu \nu}  &= \sum_{\mu'\nu'}
\left[\Gamma^{{}^{\bar{{\bf k}}}a_{{\bf g}}{}^{\bar{{\bf k}}}b_{{\bf h}}}_{{}^{\bar{{\bf k}}} c_{{\bf gh}}} \right]_{\mu \mu'}
\left[U_{{\bf k}}(a_{{\bf g}},b_{{\bf h}}; c_{{\bf gh}}) \right]_{\mu' \nu'}
\left[ \left( \Gamma^{a_{{\bf g}}b_{{\bf h}}}_{c_{{\bf gh}}} \right)^{-1} \right]_{\nu' \nu}
,
\\
\widetilde{\eta}_{c_{{\bf k}}}({{\bf g}},{{\bf h}}) &= \eta_{c_{{\bf k}}}({{\bf g}},{{\bf h}})
.
\end{align}
\end{widetext}
The gauge transformations must also respect the canonical isomorphisms associated with the vacuum charge, which fixes
\begin{align}
\Gamma_{a_{\bf g}}^{a_{\bf g}\I_{\bf 0}}=\Gamma_{b_{\bf h}}^{\I_{\bf 0} b_{\bf h}}=\Gamma_{\I_{\bf 0}}^{\I_{\bf 0}\I_{\bf 0}}
.
\end{align}

There is another type of gauge transformation associated with the symmetry action being modified by a natural isomorphism $\Upsilon$, i.e. $\check{\rho}_{{\bf g}} = \Upsilon \circ \rho_{{\bf g}}$, extending the notion of equivalence defined in Eq.~\eqref{eq:nat_iso}.
The symmetry action gauge transformation acts on the basic data according to
\begin{align}
\check{F}^{a_{{\bf g}} b_{{\bf h}} c_{{\bf k}}}_{d_{{\bf ghk}}}
&=
F^{a_{{\bf g}} b_{{\bf h}} c_{{\bf k}}}_{d_{{\bf ghk}}}
,
\\
\check{R}_{c_{\bf gh}}^{a_{\bf g} b_{\bf h}}&= \gamma_{a_{{\bf g}}}({{\bf h}})
R_{c_{{\bf gh}}}^{a_{{\bf g}} b_{{\bf h}}}
,
\\
\check{U}_{{\bf k}}(a_{{\bf g}},b_{{\bf h}};c_{{\bf gh}}) &= \frac{\gamma_{a_{{\bf g}}}({{\bf k}})\gamma_{b_{{\bf h}}}({{\bf k}})}{\gamma_{c_{{\bf gh}}}({{\bf k}})}
U_{{\bf k}}(a_{{\bf g}},b_{{\bf h}};c_{{\bf gh}})
,
\\
\check{\eta}_{c_{{\bf k}}}({{\bf g}},{{\bf h}}) &= \frac{\gamma_{c_{{\bf k}}}({{\bf gh}})}{\gamma_{{}^{{\bf \bar{g}}} c_{{\bf k}}}({{\bf h}}) \gamma_{c_{{\bf k}}}({{\bf g}})} {\eta}_{c_{{\bf k}}}({{\bf g}},{{\bf h}})
.
\end{align}
Preserving the canonical isomorphisms associated with the vacuum charge and ensuring that the trivial symmetry ${{\bf 0}}$ acts completely trivially fixes
\begin{align}
\gamma_{\I_{\bf 0}}({{\bf h}}) = \gamma_{a_{\bf g}}({{\bf 0}}) =1
.
\end{align}
When the basic data of two $G^{}$-crossed BTCs are related through the application of vertex basis and symmetry action gauge transformations, they are considered equivalent theories.

For a given BTC $\BTC$, symmetry action, and fractionalization class, the pentagon and heptagon equations only admit solutions when the defectification obstruction $[\defectO] \in H^4(G,\text{U}(1))$ vanishes~\cite{ENO}.
The explicit form of this obstruction is only known in special cases, for instance when the symmetry action does not permute topological charges~\cite{Bark2019}.
When this obstruction vanishes, the defectification classes for $G$-crossed extensions of the BTC with specified symmetry action and fractionalization class are torsorially classified by $H^3(G,\text{U}(1))$.
When $\BTC$ is a MTC, the classification of all $G$-crossed extensions for a specified symmetry action is given torsorially by $H^2_{[\rho]}(G,\mathcal{A})$ (assuming that the fractionalization obstruction also vanishes) and $H^3(G,\text{U}(1))$.
This classification is with respect to equivalences of theories under the vertex basis and symmetry action gauge transformations described above.

These torsorial classifications are associated with torsor functors that map the different $G$-crossed extensions to each other.
Explicit torsor functors were introduced in Ref.~\onlinecite{Bark2019b} for the $F$-symbols and fully developed in Ref.~\onlinecite{Aasen21} for the complete $G$-crossed data.
These map the basic data of a $G$-crossed theory $\BTC_G^\times$ to a theory $\tor{\BTC}_G^\times$ related by a $2$-cocycle ${\coho{t}\in Z^2_{[\rho]}(G,\mathcal{A})}$ and $3$-cochain $\Xdot \in C^3(G,\text{U}(1))$ that satisfies $\cbd \Xdot = \defectO_{r}(\coho{t},\rho)^{-1}$ for the relative obstruction given in Eq.~\eqref{eq:pwformula}.
This condition on $\Xdot$ ensures that the $F$-symbols of the theory obtained from applying the torsor functor satisfy the pentagon equation.
The theories $\BTC_G^\times$ and $\tor{\BTC}_G^\times$ have the same symmetry action on $\BTC_{\bf 0}$, but distinct fractionalization and defectification classes related by $\coho{t}$ and $\Xdot$, respectively.
Denoting these functor maps by
\begin{align}
\mathcal{F}_{\cohosub{t},\Xdot}: \BTC_G^\times \mapsto \tor{\BTC}_G^\times,
\end{align}
we refer to $\BTC_G^\times$ as the ``pre-torsor functor theory'' and $\tor{\BTC}_G^\times$ as the ``post-torsor functor theory.''
We denote quantities in $\tor{\BTC}_G^\times$ as $\tor{Q}=\mathcal{F}_{\cohosub{t},\Xdot}(Q)$, with the exception of the topological charge labels which are unchanged by the torsor action.
When $\BTC$ is an MTC, if we have the complete data of one unobstructed $G$-crossed extension of $\BTC$ with a given symmetry action $\rho$, we can use it as a ``base'' theory $\BTC_{G,\rho}^{\times \, \text{base}}$ from which we can obtain all other unobstructed $G$-crossed extension of $\BTC$ with $\rho$, by applying the torsor functor for all unobstructed $\coho{t}$ and $\Xdot$.~\footnote{Anomalous theories that do not fully satisfy the $G$-crossed consistency conditions can also be generated by applying the torsor functor for $\coho{t}$ and $\Xdot$ that have nontrivial obstructions.
These may occur, for example, as surface termination states at the boundaries of $(3+1)$D SET phases.}

Another type of equivalence that can be considered is when two $G$-crossed BTCs can be related by relabeling the topological charges.
Mathematically, any relabeling of objects constitutes an isomorphism of $G$-crossed BTCs.
However, relabelings that change the symmetry group labels are not a physical equivalence.
This is because the symmetry group elements are extrinsically measurable quantities, which can by viewed as fixed with respect to some physical reference.
Thus, two $G$-crossed BTCs that are related by relabelings of the form $a_{{\bf g}} \mapsto a'_{{\bf g}}$ are considered physically equivalent.
Such relabeling equivalences can break the $H^2_{[\rho]}(G,\mathcal{A})$ and $H^3(G,\text{U}(1))$ classification, in the sense that na\"ively distinct theories in this classification may be equivalent under these relabelings.

We note for later use that, for a $G$-crossed BTC with topological charge $e_{{\bf 0}}$ that satisfies $e_{{\bf 0}} \otimes e_{{\bf 0}} = \I_{{\bf 0}}$ and ${}^{{\bf g}}e_{{\bf 0}}=e_{{\bf 0}}$ for all ${\bf g}\in G$, it follows that monodromy with all other topological charges obeys
\begin{align}
\label{eq:defectmono}
\left(M_{a_{{\bf g}} e_{{\bf 0}}} \right)^2=
U_{\bf g}(e_{{\bf 0}},e_{{\bf 0}};\I_{{\bf 0}}).
\end{align}
The monodromy is generally invariant under vertex basis gauge transformations, that is $\widetilde{M}_{a_{{\bf g}}e_{{\bf 0}}}={M}_{a_{{\bf g}} e_{{\bf 0}}}$.
On the other hand, it is generally not invariant under symmetry action gauge transformations, as
\begin{align}
\label{eq:Mcheck}
\widecheck{M}_{a_{{\bf g}} e_{{\bf 0}}} = \gamma_{e_{{\bf 0}}}({\bf g}) {M}_{a_{{\bf g}} e_{{\bf 0}}} .
\end{align}
We can also see that
\begin{align}
\label{eq:Mfusion}
M_{a_{{\bf g}} e_{{\bf 0}}} M_{b_{{\bf h}} e_{{\bf 0}}} = \eta_{e_{{\bf 0}} } ({\bf g},{\bf h}) M_{c_{{\bf gh}} e_{{\bf 0}}},
\end{align}
when $N_{ab}^{c} \neq 0$.

\subsection{\texorpdfstring{$\mathcal{G}^{\eff}$}{Gf}-crossed SMTCs}
\label{sec:GfSMTC}

In order for a $G$-crossed BTC $\mcb_{G}^{\times}$ to characterize a bosonic SET phase with symmetry group $G$, the only additional condition required is that $\mcb_{\bf 0}$ is the MTC describing the quasiparticles of the topological phase.
The modularity of $\mcb_{\bf 0}$ implies that its $G$-crossed extensions have $G$-crossed modularity, which allows the full defect theory to be consistently defined on arbitrary manifolds.

A fermionic SET (FSET) phase with fermionic symmetry group $\mathcal{G}^{\eff} = \mathbb{Z}_{2}^{\eff} \times_{\central} G$ can similarly be characterized by a $\mathcal{G}^{}$-crossed BTC $\fMTC_{\mathcal{G}}^{\times}$ by imposing additional conditions that follow from the incorporation of a physical fermion.
As these conditions do not modify the pentagon and heptagon equations, FSET phases have the same defectification obstruction $[\defectO] \in H^4(G,\text{U}(1))$ as bosonic phases.

As before, we denote group elements of $\mcg^{\eff}$ by $\mathpzc{g}=(\vv{x},{\bf g})$, where $\vv{x} \in \mathbb{Z}_{2}^{\eff}$ and ${\bf g} \in G$, with group multiplication
\begin{align}
\mathpzc{g}\mathpzc{h} = (\vv{x},{\bf g}) (\vv{y},{\bf h}) = (\vv{x}+\vv{y}+\central({\bf g,h}),{\bf gh})
.
\end{align}
We will denote a $\mathcal{G}^{\eff}$-crossed SMTC as
\begin{align}
\fMTC_{\mathcal{G}^{\eff}}^{\times} = \bigoplus_{\mathpzc{g} \in \mathcal{G}^{\eff}} \fMTC_{\mathpzc{g}}
.
\end{align}
We require that $\fMTC_{\mathpzc{0}} = \fMTC_{(\vv{0},{\bf 0})}$ is equal to the SMTC describing the quasiparticles of the fermionic topological phase, which includes the physical fermion  $\psi_{\mathpzc{0}} \in \fMTC_{\mathpzc{0}}$.
This should satisfy the previously described properties, such as
\begin{align}
\label{eq:rhopsi}
\rho_{\mathpzc{g}}(\psi_{\mathpzc{0}}) &= \psi_{\mathpzc{0}} ,
\\
\label{eq:Upzcpsipsi}
U_{\mathpzc{g}}(\psi_{\mathpzc{0}},\psi_{\mathpzc{0}};\I_{\mathpzc{0}}) &= 1
,
\end{align}
preserving the canonical isomorphism between $V^{\psi_{\mathpzc{0}} \psi_{\mathpzc{0}}}_{\I_{\mathpzc{0}}}$ and $\mathbb{C}$ for physical fermions.

We can allow for the possibility that a $\mathcal{G}^{\eff}$-crossed SMTC is not faithfully graded, i.e. some of the $\fMTC_{\mathpzc{g}}$ sectors may be empty.
It is particularly significant whether $\fMTC_{(\vv{1},{\bf 0})} $ is empty, as this sector comprises the fermionic vortices.

When $\fMTC_{(\vv{1},{\bf 0})} = \varnothing$, the non-empty sectors form a faithfully graded $\mathcal{H}$-crossed extension of $\fMTC_{\mathpzc{0}}$, where $\mathcal{H} \lhd \mathcal{G}^{\eff}$ is isomorphic to a normal subgroup of $G$, i.e. $\mathcal{H} \cong H \lhd G$.
This is because $\fMTC_{(\vv{1},{\bf 0})} = \varnothing$ requires the subgroup $\mathcal{H}$ of nonempty sectors to have $\central |_{\mathcal{H}} \in B^{2}(\mathcal{H} , \mathbb{Z}_{2})$.
An important consequence of this is that a $\mathcal{G}^{\eff}$-crossed SMTC without vortices can only be faithfully $G$-graded when $[\central] = [\vv{0}]$.

When $\fMTC_{(\vv{1},{\bf 0})} \neq \varnothing$, we require the subcategory $\fMTC_{(\vv{0},{\bf 0})}\bigoplus\fMTC_{(\vv{1},{\bf 0})}$ to be a FMTC, which is a $\mathbb{Z}_2^{\eff}$ \emph{braided} extension of the SMTC $\fMTC_{(\vv{0},{\bf 0})}$.
Since this $\mathbb{Z}_2^{\eff}$ subcategory of a $\mcg^{\eff}$-crossed SMTC is not crossed-braided, the symmetry action and fractionalization must be constrained accordingly.
The extension of the symmetry action and fractionalization from the SMTC $\fMTC_{\vv{0},{\bf 0}} = \fMTC_{\vv{0}}$ to the FMTC $\fMTC_{\vv{0},{\bf 0}}\bigoplus\fMTC_{\vv{1},{\bf 0}}=\fMTC$ is subject to the obstructions and classification detailed in Sec.~\ref{sec:symmetric-fermions}.

While it may provide a locally consistent algebraic theory of symmetry defects of a fermionic system, the theory described thus far may have pathologies that prevent it from providing a consistent theory for a strictly $(2+1)$D FSET phase.
For example, it may not provide a consistent description of the theory on arbitrary surfaces and $G$-symmetry may need to be broken in order to gauge the $\mathbb{Z}_2^{\eff}$.
These pathologies may be viewed as additional types of obstructions for strictly $(2+1)$D FSET phases.
Theories that can be defined with nontrivial obstructions are viewed as anomalous, and we expect that at least some of the pathologies can be remedied via anomaly inflow by treating the anomalous theory as a description of a $(2+1)$D boundary of a $(3+1)$D fermionic topological phase with matching quantum invariants, as can be done with other obstructions.
We will leave a detailed exploration of anomalous theories for future work, and focus here on defining the non-anomalous theory.

As discussed in Sec.~\ref{sec:SMTC_FMTC}, the vortices must necessarily be included to describe a strictly $(2+1)$D fermionic topological phase.
Moreover, it must be possible to gauge the $\mathbb{Z}_2^{\eff}$ fermion parity conservation.
Extending these properties to a $\mcg^{\eff}$-crossed SMTC $\fMTC_{\mathcal{G}^{\eff}}^{\times}$ is equivalent to requiring that it be a $G$-crossed extension of a FMTC $\fMTC$, subject to the previously discussed fermionic constraints.
In this case, there is no nontrivial dependence on $\mathbb{Z}_2^{\eff}$ for the symmetry action, fractionalization, or defectification; the $\mathbb{Z}_2^{\eff}$ only enters as an extension of the $G$-grading to $\mathcal{G}^{\eff}$-grading on the $G$-crossed theory, given by braiding with the physical fermion, as we will explain.
As we will focus on the non-anomalous $(2+1)$D theories, we will define a $\mathcal{G}^{\eff}$-crossed SMTC to be a $G$-crossed FMTC, with the understanding that a more general definition applies for anomalous theories.

Viewing $\mathcal{G}^{\eff}$-crossed theories as $G$-crossed FMTCs, we only keep the vorticity labels on topological charges, and drop the $\mathbb{Z}_2^{\eff}$ dependence of $\rho$, $U$, $\eta$, and $\gamma$, letting
\begin{align}
\rho_{(\vv{x},{\bf g})}  &\to \rho_{\bf g},
\\
U_{(\vv{x},{\bf g})}(a_{\mathpzc{g}},b_{\mathpzc{h}}; c_{\mathpzc{gh}}) &\to
U_{\bf g}(a_{\mathpzc{g}},b_{\mathpzc{h}}; c_{\mathpzc{gh}}),
\\
\eta_{c_{\mathpzc{k}}}((\vv{x},{\bf g}),(\vv{y},{\bf h})) & \to \eta_{c_{\mathpzc{k}}}({\bf g},{\bf h}),
\\
\gamma_{b_{\mathpzc{h}}}((\vv{x},{\bf g})) &\to \gamma_{b_{\mathpzc{h}}}({\bf g}).
\end{align}

The equivalences of fermionic theories under vertex basis and symmetry action gauge transformations are again constrained to reflect the physical nature of the fermion, as in Eqs.~\eqref{eq:gauge-1} and \eqref{eq:ferm-nat-iso}, which imposes the conditions
\begin{align}
\label{eq:GammaPsiPsi}
\Gamma^{\psi_{\mathpzc{0}} \psi_{\mathpzc{0}}}_{\I_{\mathpzc{0}}}  &= 1,\\
\label{eq:gammapsi}
\gamma_{\psi_{\mathpzc{0}}}({\bf g})  &=1.
\end{align}

The properties discussed in Sec.~\ref{sec:symmetric-fermions} are incorporated for the fermionic symmetry action and fractionalization applying to the FMTC sector $\fMTC_{(\vv{0},{\bf 0})}\bigoplus\fMTC_{(\vv{1},{\bf 0})}$.
In particular, $\mcg^{\eff}$ and $\eta_{\psi_{\mathpzc{0}}}({\bf g,h})$ are correlated through the relation
\begin{align}
\label{eq:etapsicob}
\eta_{\psi_{\vv{0}} } ({\bf g,h})  =(-1)^{\central({\bf g},{\bf h})}.
\end{align}
We note that even though $\central \in Z^2(G, \mathbb{Z}_{2}^{\eff})$, the fermionic symmetry group $\mcg^{\eff}$ is defined such that $\eta_{\psi_{\mathpzc{0}}}(\mathpzc{g},\mathpzc{h}) \in B^2(\mcg^{\eff},\mathbb{Z}_{2})$.
This reflects the fact that the physical fermion is a local particle, which must carry a linear representation of $\mcg^{\eff}$.
(However, we do not take the extra step of trivializing the projective phases for these linear representations by making $\eta_{\psi_{\vv{0}} }(\mathpzc{g},\mathpzc{h})$ equal to 1, as this would require using symmetry action gauge transformations with nontrivial $\gamma_{\psi_{\mathpzc{0}}}(\mathpzc{g})$, which are not permitted.)

The notion of vorticity of the symmetry defects is directly related to the conditions stemming from the physical fermion.
More specifically, we now see that Eqs.~\eqref{eq:defectmono} and \eqref{eq:Upzcpsipsi} imply ${M}_{a_{\mathpzc{g}} \psi_{\mathpzc{0}} } = \pm 1$, while Eqs.~\eqref{eq:Mcheck} and \eqref{eq:gammapsi} imply that ${M}_{a_{\mathpzc{g}} \psi_{\mathpzc{0}} }$ is invariant under fermionic vertex basis and symmetry action gauge transformations.
Physically, this provides an invariant notion of the fermionic vorticity of symmetry defects, which defines the $\mathbb{Z}_{2}^{\eff}$ label of a defect via
\begin{align}
\label{eq:psimon}
M_{ a_{(\vv{x},{\bf g})} \psi_{\mathpzc{0}}} = (-1)^{\vv{x}}.
\end{align}
Finally, Eqs.~\eqref{eq:Mfusion} and \eqref{eq:etapsicob} show that this notion of vorticity of symmetry defects matches with the group multiplication of $\mathcal{G}^{\eff}$, extending the $G$-grading of defects to a $\mathcal{G}^{\eff}$-grading.

We can now see that were we to allow nontrivial $\gamma_{\psi_{\mathpzc{0}}}(\mathpzc{g})$, the vorticity of defects would not be well-defined.
For example, $\gamma_{\psi_{\mathpzc{0}}}((\vv{0},{\bf g})) = \gamma_{\psi_{\mathpzc{0}}}((\vv{1},{\bf g})) =-1$ would flip the vorticity of ${\bf g}$-defects, while $\gamma_{\psi_{\mathpzc{0}}}((\vv{x},{\bf g})) = (-1)^{\vv{x}}$ would make their vorticity all trivial.
Indeed, in order for a symmetry action gauge transformation to make $\eta_{\psi_{\vv{0}} }(\mathpzc{g},\mathpzc{h})$ equal to 1, it would necessarily be of this disallowed form.

In this vein, we also have restrictions on the equivalences of fermionic defect theories obtained by relabeling the topological charges.
We already mentioned that relabelings that change the symmetry group labels provide a mathematical isomorphism of theories, but should not be viewed as a physical equivalence.
For fermionic theories, we emphasize that this applies for the fermionic symmetry group $\mathcal{G}^{\eff}$, not just $G$.
In particular, relabelings that change the vorticity label of objects are not considered a physical equivalence.
We have, in fact, already encountered this restriction in Sec.~\ref{sec:symmetry_fractionalization} in a slightly obfuscated form when we discussed the equivalence of fractionalization classes that differ by modifications by $\zeta_{c}({\bf g})=M^{\ast}_{c \cohosub{z}({\bf g}) }$.
The restriction of these equivalences to having $\zeta_{\psi_{\mathpzc{0}}}({\bf g})=1$ is equivalent to requiring $\cohosub{z}({\bf g}) \in \mathcal{A}_{\vv{0}}$ for FMTCs.
When translated into defect theories, these equivalences of fractionalization classes can be interpreted as relabeling equivalences, where the defect topological charges are relabeled as $a'_{(\vv{x},{\bf g})} = \coho{z}({\bf g}) \otimes a_{(\vv{x},{\bf g})}$.
A consequence of this is there can be multiple distinct $\mathcal{G}^{\eff}$-crossed theories that correspond to the same ${G}$-crossed theory after gauging fermion parity.

\subsection{Torsor Functors and Defect Zesting or \texorpdfstring{$\mathcal{G}^{\eff}$}{Gf}-Crossed Torsor Functors}
\label{sec:defect_torsor_functors}

Since a fermionic defect theory $\fMTC_{\mathcal{G}^{\eff}}^{\times}$ can be viewed as $\mathcal{G}$-crossed BTC with additional conditions imposed on it, we can generate new theories by applying $\mathcal{G}$-crossed torsor functors $\mathcal{F}_{\cohosub{t},\Xdot}$ to it, where ${\coho{t}\in Z^2_{[\rho]}(\mathcal{G},\mathcal{A})}$ and $\Xdot \in C^3(\mathcal{G},\text{U}(1))$ such that $\cbd \Xdot = \defectO_{r}(\coho{t},\rho)^{-1}$.
The complete data of the theory generated by applying these torsor functors can be found in Ref.~\onlinecite{Aasen21} by replacing ${\bf g}\in G$ with $\mathpzc{g}\in \mathcal{G}$.
While the action of such a functor is guaranteed to leave the SMTC $\fMTC_{\mathpzc{0}}$ fixed, it generally need not map $\fMTC =\fMTC_{\vv{0},{\bf 0}} \oplus \fMTC_{\vv{1},{\bf 0}}$ back to itself, or even to a FMTC, as it only has to map to a $\mathbb{Z}_2$-crossed BTC extension of $\fMTC_{\vv{0},{\bf 0}}$.

One way to ensure that the torsor functor leaves $\fMTC$ fixed is to choose ${\coho{t}\in Z^2_{[\rho]}(G,\mathcal{A})}$ and $\Xdot \in C^3(G,\text{U}(1))$.
In other words, choosing $\mathcal{F}_{\cohosub{t},\Xdot}$ to be a $G$-crossed torsor functor maps $\fMTC$ back to $\fMTC$.
This is a particularly natural subset of functors to consider for the non-anomalous $(2+1)$D fermionic theories, which are required to be $G$-crossed FMTCs.
Indeed, from the theory of (bosonic) $G$-crossed MTCs, we can see that, given the data of a single unobstructed $G$-crossed extension (base theory) for a given $\fMTC$ and symmetry action $\rho$, we can apply these $G$-crossed torsor functors to generate all non-anomalous $(2+1)$D fermionic theories for a given $\fMTC$ and symmetry action $\rho$, including those with different $\central$.
As such, these torsor functors can be used in the classification of $(2+1)$D FSET phases; since there is additional structure to develop for a complete understanding of the FSET classification, we defer a more detailed discussion of this to Sec.~\ref{sec:FSET_classification}.

One can also produce functors that map $\fMTC$ back to a different FMTC.
In particular, the ``zesting'' procedure of Ref.~\onlinecite{delaney2020} can be applied to $\fMTC$ to generate $\fMTC' =\fMTC \ftimes \ifo^{(\nu)} $ for $\nu$ even.
We now introduce a hybrid of this zesting procedure and the torsor method of Ref.~\onlinecite{Aasen21} that can be applied to a $\mathbb{Z}_2^{\eff} \times G$-crossed SMTC $\fMTC_{\mathbb{Z}_2^{\eff} \times G}^{\times}$ to generate the complete set of data for theories $\fMTC_{\mathbb{Z}_2^{\eff} \times G}^{\prime \, \times}$.
For this, we define the ``defect zesting fuctor'' as
\begin{align}
\label{eq:zesty}
\mathcal{Z}^{(\nu)}
= \Upsilon^{(\nu)} \circ \mathcal{F}_{\cohosub{p}^{(\nu)},\Xdot^{(\nu)}}
,
\end{align}
where
\begin{align}
\label{eq:zestp}
\coho{p}^{(\nu)}(\mathpzc{g},\mathpzc{h})  &= \psi_{\mathpzc{0}}^{\frac{\nu}{2} \cdot \vv{x} \cdot \vv{y}}
,\\
\label{eq:zestX}
\Xdot^{(\nu)}(\mathpzc{g},\mathpzc{h},\mathpzc{k}) &= e^{i \frac{\pi}{4} \nu \cdot \vv{x} \cdot \vv{y} \cdot \vv{z}}
,\\
\label{eq:zestgamma}
\gamma^{(\nu)}_{a_{\mathpzc{g}}}(\mathpzc{h} )  &= e^{i \frac{\pi}{8} \nu \cdot \vv{x} \cdot \vv{y}}
,
\end{align}
where $\vv{x}$, $\vv{y}$, and $\vv{z}$ are the vorticity labels of $\mathpzc{g}$, $\mathpzc{h}$, and $\mathpzc{k}$, respectively.
We note that the relative defectification obstruction for $\coho{p}^{(\nu)}$ is found [using Eq.~\eqref{eq:pwformula}] to be
\begin{align}
\defectO_r(\coho{p}^{(\nu)})({\mathpzc{g}},{\mathpzc{h}}, {\mathpzc{k}},{\mathpzc{l}}) = (-1)^{\frac{\nu}{2} \cdot \vv{x} \cdot \vv{y} \cdot \vv{z} \cdot \vv{l}}
,
\end{align}
where $\vv{l}$ is the vorticity label of $\mathpzc{l}$, and that $\widetilde{\Xdot}$ in Eq.~\eqref{eq:zestX} satisfies $\cbd \widetilde{\Xdot}  =\defectO_r(\coho{p}^{(\nu)})^{-1}$, which explicitly shows that the relative obstruction is always trivial.

We emphasize that $\nu (\text{ mod }16)$ completely parameterizes this defect zesting functor, thus Eq.~\eqref{eq:zesty} is a $\mathbb{Z}_{8}$ torsor connecting eight different theories.
From this definition, we see that $\mathcal{F}_{\cohosub{p}^{(\nu)},\Xdot^{(\nu)}}$ is actually a $\mathbb{Z}_{2}$-crossed torsor functor and $\Upsilon^{(\nu)}$ is a $\mathbb{Z}_{2}$ symmetry action gauge transformation.
However, their combination $\mathcal{Z}^{(\nu)}$ is designed to be a $\mathbb{Z}_2^{\eff}$-crossed torsor functor, in the sense that it maps $\fMTC$ to another FMTC with the same SMTC, which we will show to be $\fMTC \ftimes \ifo^{(\nu)} $.
We will verify this claim after we produce the complete data of the post-zesting theory, which we denote as
\begin{align}
\tortilde{\fMTC}_{\mathbb{Z}_2^{\eff} \times G}^{\times} =
\mathcal{Z}^{(\nu)} ( \fMTC_{\mathbb{Z}_2^{\eff} \times G}^{\times} )
.
\end{align}

We now use Ref.~\onlinecite{Aasen21} to explicitly write the complete data of $\tortilde{\fMTC}_{\mathbb{Z}_2^{\eff} \times G}^{\times}$ in terms of that of $\fMTC_{\mathbb{Z}_2^{\eff} \times G}^{\times}$.
First, the fusion rules of the post-zesting theory are given by
\begin{align}
a_{\mathpzc{g}} \tortilde{\otimes }b_{\mathpzc{h}}
=\psi_{\mathpzc{0}}^{\frac{\nu}{2} \cdot \vv{x} \cdot \vv{y}} \otimes a_{\mathpzc{g}} \otimes b_{\mathpzc{h}}
.
\end{align}
In terms of fusion coefficients, this is
\begin{align}
\label{eq:zest_N_explicit}
\tortilde{N}_{a_{\mathpzc{g}} b_{\mathpzc{h}}}^{c_{\mathpzc{gh}}}
={N}_{a_{\mathpzc{g}} b_{\mathpzc{h}}}^{c'_{\mathpzc{gh}}}
,
\end{align}
where $c'_{\mathpzc{gh}} = \psi_{\mathpzc{0}}^{\frac{\nu}{2} \cdot \vv{x} \cdot \vv{y}} \otimes c_{\mathpzc{gh}}$.
While defect zesting (or torsor) functors do not change the topological charge labels, since they can change the fusion rules, the conjugate of a topological charge may not be the same after applying the functor.

The $F$-symbols of the theory after applying the zesting functor are given by
\begin{widetext}
\begin{align}
\label{eq:zest_F_explicit}
\left[\tortilde{F}^{a_{\mathpzc{g}} b_{\mathpzc{h}} c_{\mathpzc{k}}} _{d_{\mathpzc{ghk}}}\right]_{(e_{\mathpzc{gh}},\alpha, \beta)(f_{\mathpzc{hk}},\mu,\upsilon )}& =
e^{i \frac{\pi}{4} \nu \cdot \vv{x} \cdot \vv{y} \cdot \vv{z}} \sum_{\beta',\upsilon',\upsilon''}
\left[{F}^{\psi_{\mathpzc{0}}^{\frac{\nu}{2} \cdot \vv{x} \cdot \vv{y}} e'_{\mathpzc{gh}} c_{\mathpzc{k}}}_{d'_{\mathpzc{ghk}}}\right]_{(e_{\mathpzc{gh}} , \beta)( d''_{\mathpzc{ghk}},\beta' )}
\left[{F}^{a_{\mathpzc{g}} b_{\mathpzc{h}} c_{\mathpzc{k}}} _{d''_{\mathpzc{ghk}}}\right]_{(e'_{\mathpzc{gh}},\alpha,\beta')(f'_{\mathpzc{hk}},\mu,\upsilon'' )}
\notag \\
& \qquad \times
\left[\left( F^{\psi_{\mathpzc{0}}^{\frac{\nu}{2} \cdot \vv{y} \cdot \vv{z}} a_{\mathpzc{g}} f'_{\mathpzc{hk}}}_{d'''_{\mathpzc{ghk}}}  \right)^{-1}\right]_{(d''_{\mathpzc{ghk}},\upsilon'')(a'_{\mathpzc{g}},\upsilon')}
\left( R^{\psi_{\mathpzc{0}}^{\frac{\nu}{2} \cdot \vv{y} \cdot \vv{z}} a_{\mathpzc{g}}}_{ a'_{\mathpzc{g}} }\right)^{-1}
\left[ F^{a_{\mathpzc{g}} \psi_{\mathpzc{0}}^{\frac{\nu}{2} \cdot \vv{y} \cdot \vv{z}} f'_{\mathpzc{hk}}}_{d'''_{\mathpzc{ghk}}} \right]_{(a'_{\mathpzc{g}},\upsilon')(f_{\mathpzc{hk}},\upsilon)}
,
\end{align}
\end{widetext}
where
$e'_{\mathpzc{gh}} = \psi_{\mathpzc{0}}^{\frac{\nu}{2} \cdot \vv{x} \cdot \vv{y}}  \otimes e_{\mathpzc{gh}}$,
$f'_{\mathpzc{hk}} = \psi_{\mathpzc{0}}^{\frac{\nu}{2} \cdot \vv{y} \cdot \vv{z}}  \otimes f_{\mathpzc{hk}}$,
$a_{\mathpzc{g}}' = \psi_{\mathpzc{0}}^{\frac{\nu}{2} \cdot \vv{y} \cdot \vv{z}}  \otimes a_{\mathpzc{g}}$,
$d'_{\mathpzc{ghk}} = \psi_{\mathpzc{0}}^{\frac{\nu}{2} \cdot (\vv{x}+\vv{y}) \cdot \vv{z}}  \otimes d_{\mathpzc{ghk}}$,
$d''_{\mathpzc{ghk}}= \psi_{\mathpzc{0}}^{\frac{\nu}{2} \cdot (\vv{x} \cdot \vv{y} + \vv{x} \cdot \vv{z} + \vv{y} \cdot \vv{z})}  \otimes d_{\mathpzc{ghk}} $, and
$d'''_{\mathpzc{ghk}}= \psi_{\mathpzc{0}}^{\frac{\nu}{2} \cdot \vv{x} \cdot (\vv{y+z})}\otimes d_{\mathpzc{ghk}}$.

For the defect zesting functor, the symmetry action on charges does not change, that is
\begin{align}
\label{eq:zest_rho_explicit}
\tortilde{\rho}_{\mathpzc{k}} (a_{\mathpzc{g}})&= {\rho}_{\mathpzc{k}} (a_{\mathpzc{g}})= {\rho}_{{\bf k}} (a_{\mathpzc{g}}).
\end{align}
The remaining topological data is
\begin{align}
\label{eq:zest_R_explicit}
\left[ \tortilde{R}^{a_{\mathpzc{g}}b_{\mathpzc{h}}}_{c_{\mathpzc{gh}}} \right]_{\alpha \beta} &= e^{i \frac{\pi}{8} \nu \cdot \vv{x} \cdot \vv{y}}  \left[R^{a_{\mathpzc{g}}b_{\mathpzc{h}}}_{c'_{\mathpzc{gh}}}\right]_{\alpha \beta}
,
\\
\label{eq:zest_U_explicit}
\left[ \tortilde{U}_{\mathpzc{k}}(a_{\mathpzc{g}},b_{\mathpzc{h}}; c_{\mathpzc{gh}}) \right]_{\alpha \beta} &=
{U}_{{\bf k}}\left( \psi_{\mathpzc{0}}^{\frac{\nu}{2} \cdot \vv{x} \cdot \vv{y}}  , c'_{\mathpzc{gh}} \right)
\notag \\
& \quad \times \left[{U}_{{\bf k}}\left( a_{\mathpzc{g}}, b_{\mathpzc{h}}; c'_{\mathpzc{gh}} \right) \right]_{\alpha \beta}
,
\\
\label{eq:zest_eta_explicit}
\tortilde{\; \eta}_{c_{\mathpzc{k}} }(\mathpzc{g},\mathpzc{h}) &={\eta}_{c_{\mathpzc{k}} }({\bf g},{\bf h}).
\end{align}

It is now straightforward to see that the resulting data is actually that of a $G$-crossed FMTC.
In particular, $\rho_{\mathpzc{k}}$ and $\tortilde{U}_{\mathpzc{k}}$ do not depend on $\vv{z}$, and $\tortilde{\; \eta}_{c_{\mathpzc{k}} }(\mathpzc{g},\mathpzc{h})$ does not depend on $\vv{x}$ or $\vv{y}$, which indicates that the ${\bf g}={\bf 0}$ sector is braided, not $\mathbb{Z}_2$-crossed braided.
The data of the $\mathpzc{g} = \mathpzc{0}$ sector, i.e. $\fMTC_{\mathpzc{0}}$, is left unchanged, so it corresponds to the same SMTC.
We can identify the resulting FMTC by merely examining the topological twist factors of the vortices, i.e. the $\mathpzc{g} =(\vv{1},{\bf 0})$ sector.
From Eq.~\eqref{eq:zest_R_explicit}, we find
\begin{align}
\tortilde{R}^{a_{\vv{x},{\bf 0}}b_{\vv{y},{\bf 0}}}_{c_{\vv{x+y},{\bf 0}}} &= e^{i \frac{\pi}{8} \nu \cdot \vv{x} \cdot \vv{y}}  {R}^{a_{\vv{x},{\bf 0}}b_{\vv{y},{\bf 0}}}_{[\psi^{\frac{\nu}{2}}c]_{\vv{x+y},{\bf 0}}}
,\\
\tortilde{\; \theta}_{a_{\vv{x},{\bf 0}}} &= e^{i \frac{\pi}{8} \nu \cdot \vv{x}} \, {\theta}_{a_{\vv{x},{\bf 0}}}
,\\
\tortilde{\; c}_{-} &= c_{-} + \frac{\nu}{2} \text{ mod }8
.
\end{align}
Thus, we see that $\tortilde{\fMTC} =\fMTC \ftimes \ifo^{(\nu)} $.

We note that the $\tortilde{F}$-symbols in Eq.~\ref{eq:zest_F_explicit} are not fully in canonical gauge for $\nu = 2 \text{ mod }4$, in particular because $\tortilde{F}^{\psi_{\mathpzc{0}} b_{\mathpzc{h}} \overline{b_{\mathpzc{h}}}} _{\psi_{\mathpzc{0}}} = (-1)^{\frac{\nu}{2} \cdot \vv{y}}$.
This can be corrected by applying a vertex basis gauge transformation given by
\begin{align}
\label{eq:zest_gauge_trans}
\Gamma^{a_{\mathpzc{g}} b_{\mathpzc{h}}}_{c_{\mathpzc{gh}}} &= (-1)^{\frac{\nu}{2} \cdot \delta_{a_{\mathpzc{g}} \psi_{\mathpzc{0}} } \cdot \vv{y}}
\end{align}
after the defect zesting functor.
This gauge transformation changes the $F$- and $R$-symbols, but leaves the $U$- and $\eta$-symbols unchanged.

\subsection{Isomorphic but Inequivalent \texorpdfstring{$\mathcal{G}^{\eff}$}{Gf}-crossed SMTCs}
\label{sec:centralisomorphism}

We previously discussed the possibility of having $G$-crossed BTCs that are mathematically isomorphic, but physically inequivalent, in the context of relabeling topological charges.
In particular, relabeling topological charges always gives an isomorphism, but it is only considered a physical equivalence when the symmetry group labels are left fixed by the relabeling.
Thus, a simple class of relabelings that are not physical equivalences is obtained by applying the nontrivial elements of $\text{Aut}(G)$ to the symmetry group labels.
The notion of relabeling isomorphisms and equivalences extends to $\mathcal{G}^{\eff}$-crossed SMTCs, even though the $\mathbb{Z}_2^{\eff}$ is not treated on the same footing as $G$.
Since $\mathcal{G}^{\eff}$ is the fermionic symmetry group, relabelings that change the $\mathbb{Z}_2^{\eff}$ labels are not physical equivalences.
When the $\mathbb{Z}_2^{\eff}$ labels are changed by a relabeling isomorphism, one must also apply a transformation that changes the corresponding vorticity (monodromy with the physical fermion) of the defects to match their new labels.

More concretely, at the level of groups, $\mathcal{G}$ and $\mathcal{G}^{'}$ are considered equivalent $\mathbb{Z}_2$ extensions of $G$, with corresponding $\central$ and $\central'$, when $\central  = \central' + \cbd \vviso$ for some $\vviso \in C^1(G,\mathbb{Z}_2)$.
These equivalences correspond to the group isomorphisms given by $\mathpzc{g} \mapsto \mathpzc{g}'=(\vv{x}+\vviso({\bf g}),{\bf g})$.
We note that it is also possible to have inequivalent central extensions that are nonetheless isomorphic groups, but the corresponding isomorphisms involve nontrivial permutations of the $G$ labels.

We can directly translate the central extension group equivalences into isomorphisms between $\mathcal{G}^{\eff}$-crossed SMTCs and $\mathcal{G}^{\eff '}$-crossed SMTCs that are not physical equivalences.
For this, we replace all $\mathpzc{g}$-labels on the topological charges according to the isomorphism between the groups
\begin{align}
\label{eq:charge-iso}
a_{\mathpzc{g}}=a_{(\vv{x},{\bf g})} \mapsto a_{\mathpzc{g}'} = a_{(\vv{x}+\vviso({\bf g}),{\bf g})}.
\end{align}
The change in vorticity implies that the monodromy with the physical fermion must change by $(-1)^{\vviso({\bf g})}$ to satisfy Eq.~\eqref{eq:charge-iso}, i.e.
\begin{align}
M_{\psi_\vv{0}a_{(\vv{x},{\bf g})}} = (-1)^{\vv{x}} \mapsto M_{\psi_\vv{0}a_{(\vv{x}+\vviso({\bf g}),{\bf g})}} = (-1)^{\vv{x}+\vviso({\bf g})}.
\end{align}
This can only be implemented by a symmetry action transformation with
\begin{align}
\label{eq:gammapsiillegal}
\gamma_{\psi_{\mathpzc{0}}}((\vv{x},{\bf g}))  = (-1)^{\vviso({\bf g})}.
\end{align}
We emphasize that such a transformation changes the theory in an observable way, in particular the gauge invariant quantity $M_{\psi_\vv{0}a_{(\vv{x},{\bf g})}}$ changes, so it is not permitted as a fermionic symmetry action gauge transformation for the purposes of physical equivalences of theories.

Combining Eqs.~\eqref{eq:charge-iso} and \eqref{eq:gammapsiillegal}, we have an isomorphism that modifies the basic data as
\begin{widetext}
\begin{align}
{N'}_{a_{\mathpzc{g}'}b_{\mathpzc{h}'}}^{c_{\mathpzc{g}'\mathpzc{h}'}} &= N_{a_{\mathpzc{g}}b_{\mathpzc{h}}}^{c_{\mathpzc{gh}}}
,
\\
\left[{F'}^{a_{\mathpzc{g}'}b_{\mathpzc{h}'} c_{\mathpzc{k}'}}_{d_{\mathpzc{g}'\mathpzc{h}'\mathpzc{k}'}} \right]_{(e_{\mathpzc{g}'\mathpzc{h}'},\alpha,\beta)(f_{\mathpzc{h}'\mathpzc{k}'},\mu,\nu)} &=
\left[{F}^{a_{\mathpzc{g}}b_{\mathpzc{h}} c_{\mathpzc{k}}}_{d_{\mathpzc{ghk}}} \right]_{(e_{\mathpzc{gh}},\alpha,\beta)(f_{\mathpzc{hk}},\mu,\nu)}
,
\\
\left[ {R'}^{a_{\mathpzc{g}'} b_{\mathpzc{h}'}}_{c_{\mathpzc{g}'\mathpzc{h}'}} \right]_{\mu,\nu} &= (-1)^{\vviso({\bf h}) \cdot \delta_{a_{\mathpzc{g}},\psi_{\mathpzc{0}}}}
\left[ R^{a_{\mathpzc{g}} b_{\mathpzc{h}}}_{c_{\mathpzc{gh}}} \right]_{\mu,\nu},
\\
\left[U'_{\mathpzc{k}}(a_{\mathpzc{g}'}, b_{\mathpzc{h}'};c_{\mathpzc{g}'\mathpzc{h}'})\right]_{\mu,\nu} &=
(-1)^{\vviso({\bf k}) \cdot(
\delta_{a_{\mathpzc{g}},\psi_{\mathpzc{0}}}+
\delta_{b_{\mathpzc{h}},\psi_{\mathpzc{0}}}+
\delta_{c_{\mathpzc{g}\mathpzc{h}},\psi_{\mathpzc{0}}})}
\left[U_{\mathpzc{k}}(a_{\mathpzc{g}}, b_{\mathpzc{h}};c_{\mathpzc{gh}})\right]_{\mu \nu}
,
\\
\eta'_{c_{\mathpzc{k}'}}(\mathpzc{g}',\mathpzc{h}') &= (-1)^{\cbd \vviso({\bf g,h}) \cdot \delta_{c_{\mathpzc{k}},\psi_{\mathpzc{0}}}} \eta_{c_{\mathpzc{k}}}(\mathpzc{g},\mathpzc{h}).
\end{align}
\end{widetext}
It is easy to see this yields $\eta'_{\psi_{\mathpzc{0}}}(\mathpzc{g}',\mathpzc{h}') = (-1)^{\central({\bf g,h}) + \cbd \vviso({\bf g,h})}$ and, hence, $\central '= \central + \cbd \vviso$, as intended.
Thus, for every $\mathcal{G}^{\eff}$-crossed SMTC for a given $\central$, we have an isomorphic $\mathcal{G}^{\eff '}$-crossed SMTC with $\central' = \central + \cbd \vviso$ and ${\mathcal{G}^{\eff}}'$.

Finally, we note that vertex basis and symmetry action gauge transformations pull through the $\vviso$-isomorphism so that the gauge equivalence of $\mathcal{G}^{\eff}$-crossed theories translates into gauge equivalence of $\mathcal{G}^{\eff '}$-crossed after applying the $\vviso$-isomorphism.
Moreover, there is special set of $\vviso$-isomorphisms which leave $\central$ unchanged, given by $\vviso \in Z^1(G,\mathbb{Z}_{2})$, so that $\cbd \vviso = 0$, i.e. $\vviso$ is a homomorphism.
When $[\V] = [\Q]$ for $\fMTC$, such a $\vviso$-isomorphism on ${\fMTC}_{\mathcal{G}^{\eff}}^{\times}$ can be realized by stacking with a FSPT phase that has symmetry action $\rho_{\bf g} = \V^{\vviso({\bf g}) }$.
This isomorphism cannot be realized by stacking with a FSPT phase when $[\V] \neq [\Q]$.

\subsection{Combining, Restricting, Gluing, and Stacking Defect Theories}
\label{sec:stacking}

There are several simple ways of generating new defect theories from two known defect theories.
The product of a $G_{1}$-crossed BTC ${\BTC}_{G_1}^{(1)\times}$ and a $G_{2}$-crossed BTC ${\BTC}_{G_2}^{(2)\times}$ is a $G$-crossed BTC
\begin{align}
{\BTC}_{G}^{\times}  = {\BTC}_{G_1}^{(1)\times} \boxtimes {\BTC}_{G_2}^{(2)\times}
,
\end{align}
with $G=G_1 \times G_2$ and $\BTC = \BTC^{(1)} \boxtimes \BTC^{(2)}$.
This describes the combination of two systems that have their own symmetries and essentially do not interact.
In particular, it will yield $\rho_{\bf g} = \rho_{{\bf g}^{(1)}} \boxtimes \rho_{{\bf g}^{(2)}}$ and all the topological data can be written as a product of that of the two theories.
Either or both of the original theories can be fermionic, and there will be two distinct types of fermions if both theories are fermionic.

Taking a restriction to a subcategory that is closed under fusion can be used to produce a new theory.
This can be used to generate an $H$-crossed BTC where $H < G$ is a subgroup of $G$, by simply restricting to objects with group labels in $H$, that is
\begin{align}
{\BTC}_{H}^{\times} = \left. {\BTC}_{G}^{\times} \right|_{\{a_{\bf g} | {\bf g}\in H \}}
.
\end{align}

One way of applying this to produce a new $G$-crossed BTC from two $G$-crossed BTCs is by ``gluing'' them so the topological charges of the two theories are constrained to carry the same symmetry group label, which is to say taking a $G$-diagonal product of the theories
\begin{align}
{\BTC}_{G}^{\times} = {\BTC}_{G}^{(1)\times} \underset{G}{\boxtimes} {\BTC}_{G}^{(2)\times} = \left. {\BTC}_{G}^{(1)\times} \boxtimes
{\BTC}_{G}^{(2)\times} \right|_{\{(a^{(1)}_{\bf g} , a^{(2)}_{\bf g} )\}}
,
\end{align}
which has $\BTC = \BTC^{(1)} \boxtimes \BTC^{(2)}$.
This describes the combination of two systems that have a shared symmetry, but are otherwise not interacting.
If ${\BTC}_{G}^{(1)\times}$ is a $\mathcal{G}^{\eff}$-crossed SMTC and ${\BTC}_{G}^{(2)\times}$ is a $G$-crossed MTC, their glued theory will be a $\mathcal{G}^{\eff}$-crossed SMTC.
In principle, one can glue together a $\mathcal{G}^{\eff}_{1}$-crossed SMTC and a $\mathcal{G}^{\eff}_{2}$-crossed SMTC along $G$ if $G_1 = G_2 = G$, or along $\mathcal{G}^{\eff}$ if $\mathcal{G}^{\eff}_{1}=\mathcal{G}^{\eff}_{2} = \mathcal{G}^{\eff}$ to obtain a theory with two distinct types of fermions.

Another way of using restricted products to produce $G$-crossed BTCs with general symmetry action $\rho : G \to \text{Aut}(\BTC)$, is to first produce an $H$-crossed BTC $\BTC_{H, \iota}^{\times}$, where $H = \im (\rho) < \text{Aut}(\BTC)$ with symmetry action given by the inclusion homomorphism $\iota : H \to \text{Aut}(\BTC)$ (i.e. $\iota([\varphi]) = [\varphi]$), and then take the restricted product with a bosonic SPT phase $\spt^{[\alpha]}_G$ (see Sec.~\ref{sec:bSPT} for an explanation of the notation)
\begin{align}
{\BTC}_{G}^{\times} &=  \left. \BTC_{H,\iota}^{\times} \boxtimes \spt_{G}^{[\alpha]} \right|_{\mathcal{S}_{\rho}}
,
\\
\mathcal{S}_{\rho} & = \{ (a_{[\rho_{\bf g}]} , \I_{\bf g})  \}
.
\end{align}
When $\BTC$ is an MTC, if we have one such $G$-crossed BTC ${\BTC}_{G}^{\times}$ for a given symmetry action $\rho$, it serves as a base theory from which we can obtain all other $G$-crossed extensions of $\BTC$ with the same $\rho$ by application of the $H^2_{\rho}(G,\mathcal{A})$ torsor functor.
When there is an unobstructed $\BTC_{H, \iota}^{\times}$ for $H =\text{Aut}(\BTC)$, it can be used to produce a $G$-crossed base theory for any $G$ and $\rho$, and hence torsorially generate all $G$-crossed extensions of $\BTC$.

In order to combine two $\mathcal{G}^{\eff}$-crossed SMTCs into a $\mathcal{G}^{\eff}$-crossed SMTC with a single type of physical fermion, we use the fermionic stacking procedure described in Sec.~\ref{sec:f-identify} for FMTCs, and extend it to the $\mathcal{G}^{\eff}$-crossed theories.
For this, we take a $\mathcal{G}^{\eff}$-diagonal product of two $\mathcal{G}^{\eff}$-crossed SMTCs and condense the bound pair of physical fermions of the two theories.
This produces a $\mathcal{G}^{\eff}$-crossed SMTC
\begin{align}
{\fMTC}_{\mathcal{G}^{\eff}}^{\times}  ={\fMTC^{(1)}}_{\mathcal{G}^{\eff}}^{\times}
 \underset{\mathcal{G}^{\eff}}{\ftimes}
 {\fMTC^{(2)}}_{\mathcal{G}^{\eff}}^{\times}
\, = \frac{{\fMTC^{(1)}}_{\mathcal{G}^{\eff}}^{\times}
 \underset{\mathcal{G}^{\eff}}{\boxtimes}
 {\fMTC^{(2)}}_{\mathcal{G}^{\eff}}^{\times}
}
{A[\psi^{(1)}_\vv{0} , \psi^{(2)}_\vv{0}]}
,
\end{align}
which has $\fMTC = \fMTC^{(1)} \ftimes \fMTC^{(2)}$.
The condensation enforces the diagonal pairing of vorticity labels, we nonetheless include it in the diagonal product for emphasis.
In this way, the stacked theories are required to have the same fermionic symmetry group $\mathcal{G}^{\eff}$, i.e. they have the same $\eta_{\psi_{\vv{0}}}$.

We note that if $\fMTC^{(2)} = \ifo^{(0)}$, this generates a new $\mathcal{G}^{\eff}$-crossed SMTC for the same FMTC $\fMTC = \fMTC^{(1)}$.
If $\fMTC^{(2)} = \ifo^{(\nu)}$, this can be used to generate $\mathcal{G}^{\eff}$-crossed extensions of the same SMTC $\fMTC_{\vv{0}}$, but different FMTC $\fMTC' = \fMTC \ftimes \ifo^{(\nu)}$ related by the 16-fold way.
Typically, the complete data is difficult to compute for such fermionic stacking because of the condensation step, but in the case of $\nu$ even and $\mathcal{G}^{\eff} = \mathbb{Z}_{2}^{\eff} \times G$, we can obtain the complete data using the defect zesting functor of Sec.~\ref{sec:defect_torsor_functors}.

\section{\texorpdfstring{$\mathcal{G}^{\operatorname{f}}$}{Gf}-crossed FSPT Phases}
\label{sec:fSPT}

Beyond being the simplest examples of topological phases in the presence of symmetry, SPT phases play an important role in the classification of SET phases.
In this section, we generate the minimal set of data specifying a FSPT phase, which will be utilized in the next section to classify all FSPT phases.
Additionally, we present explicit formulas for the full set of topological data.
The results of this section have overlap with previous works~\cite{Bhardwaj2017,Cheng2018,Kapustin15,Wang2018}.
Notably, we extend the results of Refs.~\onlinecite{Cheng2018,Bhardwaj2017} to include nontrivial $\mathcal{G}^{\eff}$, and also provide the complete topological data describing vortices and defects.

Recall that the trivial fermionic theory $\ifo^{(0)}$ has four topological charges, with quasiparticle sector
\begin{align}
\ifo^{(0)}_\vv{0} = \{ \I_\vv{0},\psi_\vv{0} \},
\end{align}
corresponding to the two local particles common to all fermionic theories,
and vortex sector
\begin{align}
\ifo^{(0)}_\vv{1} = \{\I_\vv{1},\psi_\vv{1}\}.
\end{align}
In contrast, the trivial bosonic MTC has a single topological charge $\bMTC=\{\I\}$.
As a result, FSPT phases exhibit a richer structure than their bosonic counterparts.
While the bosonic SPT phases for fixed symmetry group $G$ are only distinguished by their $H^3({G},\text{U}(1))$ defectification classes, FSPT phases can differ in their fermionic symmetry group $\mathcal{G}^{\eff}$ (i.e. the fractionalization class of the physical fermion), their symmetry action on vortices, and their vortex fractionalization class.

We begin by reviewing bosonic SPT phases, the ${G}$-crossed extensions of $\bMTC=\{\I\}$.
We then turn to FSPT phases, the $\mathcal{G}^{\eff}$-crossed extensions of the trivial fermionic theory $\ifo^{(0)}$.
Before considering theories with general symmetry, we first focus on the simplest case of $G=\mathbb{Z}_2$, corresponding to $\mathcal{G}^{\eff} = \mathbb{Z}_2^{\eff}\times \mathbb{Z}_2$ and $\mathcal{G}^{\eff}=\mathbb{Z}_4^{\eff}$.
We present the basic data of these theories in Tables~\ref{table:Z2xZ2-Ab}-\ref{table:Z4}, along with their group structure.
These theories illustrate key differences in comparison to the analogous bosonic classification of $\mathbb{Z}_2$-crossed extensions of the toric code.

To extend to general $G$, we combine the topological data of the $\mathbb{Z}_2$ theories and a bosonic SPT phase to construct a ``base theory'' for each choice of symmetry action $\maj$ and full symmetry group $\mathcal{G}^{\eff}$.
Our approach extends the Ising pullback developed in Ref.~\onlinecite{Cheng2018,Bhardwaj2017} to construct bona fide $\mathcal{G}^{\eff}$-crossed theories for $\ifo^{(0)}$.
These base theories are not the most general $\mathcal{G}^{\eff}$-crossed FSPT phases, but they provide a base from which we can generate all other theories with the same symmetry action.
By exploiting the torsorial nature of the $G$-crossed classification, we can write the complete topological data of each such theory in terms of the base theory and the cocycles relating them~\cite{Aasen21}.
This result is surprising, as it collapses the highly nontrivial consistency conditions of general $\mathcal{G}^{\eff}$-crossed FSPT phases onto the much simpler set of consistency conditions for: (1) a bosonic $G$-crossed SPT phase, (2) a $\mathbb{Z}_2^{\eff}\times \mathbb{Z}_2$-crossed  FSPT phase, and (3) an $H^4(G,\text{U}(1))$-valued obstruction.
Finally, we use this torsor method to define equivalence relations on FSPT phases, thereby systematizing the relabeling redundancy generally present in the $G$-crossed formalism.

\subsection{Bosonic SPT Phases}
\label{sec:bSPT}

Bosonic SPT phases can be described by $G$-crossed extensions of the trivial theory $\bMTC=\{\I\}$, whose topological data is fully specified by a  cohomology class $[\alpha] \in H^3(G ,\text{U}(1))$.
We will denote the bosonic SPT as $\spt_{G}^{[\alpha]}$.
Each $\bMTC_{{\bf g}}$-sector has one simple object, denoted
\begin{align}
\bMTC_{{\bf g}}= \{\I_{{\bf g}} \}.
\end{align}
and fusion rules given by group multiplication,
\begin{align}
\I_{{\bf g}} \otimes \I_{{\bf h}} = \I_{{\bf gh}}.
\end{align}
The topological data may be specified using a representative $\alpha$ of the cohomology class $[\alpha]$, that is,  a normalized $3$-cocycle $\alpha \in Z^3(G,\text{U}(1))$.
In a particular gauge choice, the data may be written as~\cite{Bark2019}
\begin{align}
F^{\I_{{\bf g}}\I_{{\bf h}}\I_{{\bf k}}} &= \alpha({\bf g},{\bf h},{\bf k}) ,
\label{eq:F-alpha}\\
R^{\I_{{\bf g}}\I_{{\bf h}}} &= 1 ,
\label{eq:R-alpha}\\
U_{{\bf k}}(\I_{\bf g}, \I_{{\bf h}}; \I_{{\bf gh}}) &=
\frac{\alpha({\bf g},{\bf k}, {}^{\bar{{\bf k}}}{\bf h})}{\alpha({\bf g},{\bf h},{\bf k}) \alpha({\bf k},{}^{\bar {\bf k}}{\bf g},{}^{\bar {\bf k}}{\bf h})} ,
\label{eq:U-alpha} \\
\eta_{\I_{{\bf k}}} ({\bf g},{\bf h})&= \frac{\alpha({\bf g},{}^{\bar {{\bf g}}}{\bf k},{\bf  h})}{\alpha({\bf g},{\bf h},{}^{\bar{{\bf  h}}\bar{{\bf g}}}{\bf k})\alpha({\bf  k},{\bf g},{\bf  h})} .
\label{eq:eta-alpha}
\end{align}

Bosonic SPT phases torsorially classify the defectification classes of $G$-crossed theories.
We can obtain a theory $\bMTC_{G }^{\times \prime}$ from another theory ${\bMTC}_{G}^\times$ with the same symmetry action and fractionalization class by gluing in $\spt_{G}^{[\alpha]}$, i.e. taking the $G$-diagonal product, such that
\begin{align}
\label{bSPTgrouplaw}
\bMTC_{G }^{\times \prime} &=
\spt_{G}^{[\alpha]} \underset{G}{\boxtimes} \, \bMTC_{G}^\times
.
\end{align}
The basic data of $\bMTC_{G }^{\times \prime}$ is given by a product of the data of $\bMTC_{{G}}^\times$ and $\spt_{{G}}^{[\alpha]}$.
As each theory independently satisfies the pentagon and heptagon equations, the new theory automatically satisfies the ${G}$-crossed consistency conditions.

Lastly, we note that bosonic SPT phases satisfy a group structure under gluing
\begin{align}
\spt_{{G}}^{[\alpha_1]} \underset{{G}}{\boxtimes} \spt_{{G}}^{[\alpha_2]} &= \spt_{{G}}^{[\alpha_1 \alpha_2]}
\end{align}
where we have used the multiplicative group law in $H^3({G},\text{U}(1))$.

We therefore see that bosonic SPT phases provide the $H^3({G},\text{U}(1))$ part of the torsorial ${G}$-crossed classification.
Gluing in a nontrivial bosonic SPT phase to a ${G}$-crossed BTC generates a ${G}$-crossed BTC with the same symmetry action and fractionalization class and different defectification class, up to the relabeling redundancy.

\subsection{FSPT Phases with \texorpdfstring{${G}=\mathbb{Z}_2$}{GeZ2}}
\label{sec:FSPTphasesZ2}

We now consider FSPT phases with on-site unitary $\mathbb{Z}_2$ symmetry.
There are two $\mathbb{Z}_2^{\eff}$ extensions of $\mathbb{Z}_2$:  the trivial extension given by $\mathcal{G}^{\eff} =\mathbb{Z}_2^{\eff} \times \mathbb{Z}_2$, and the nontrivial $\mathcal{G}^{\eff}  = \mathbb{Z}_4^{\eff}$ extension.

\subsubsection{\texorpdfstring{$\mathcal{G}^{\operatorname{f}}=\mathbb{Z}_2^{\operatorname{f}} \times \mathbb{Z}_2$}{GZ2}}
\label{sec:Z8}

FSPT phases with $\mathcal{G}^{\eff}=\mathbb{Z}_2^{\eff}\times \mathbb{Z}_2$ have two possible symmetry actions: trivial (no topological charge permutation) or nontrivial (vortex permutation $\rho_{\bf 1} = \V$).
As discussed in Sec.~\ref{sec:O3invclass}, the symmetry fractionalization obstruction $\coho{O}$ vanishes for both the trivial and vortex-permuting symmetry action.

In order to specify the topological data of these theories, we first choose a symmetry action given by a group homomorphism $[\rho]:\mathbb{Z}_2 \to \on{Aut}^{\eff}(\ifo^{(0)}) = \mathbb{Z}_{2}^{\V}$.
The possible choices are classified by $H^1(\mathbb{Z}_2, \mathbb{Z}_2^{\V}) = \mathbb{Z}_2$.
The next piece of data is to choose a pattern of symmetry fractionalization.
Recall from Sec.~\ref{sec:Ex_IFMTCs} that for a fixed $\mathcal{G}^{\eff}$, the vortex symmetry fractionalization obstruction ${\coho{O}}^{\eta}$ is given by $\psi^{\pi \cup \central}$.
In this case, $\coho{O}^{\eta}$ is trivial, implying the vortex symmetry fractionalization is classified torsorially by $[\coho{q}]\in H^2(\mathbb{Z}_2, \mathbb{Z}_2^{\psi}) = \mathbb{Z}_2$.
There are two vortex symmetry fractionalization classes corresponding to $\coho{q}({\bf 1},{\bf 1}) = \I_{\vv{0}} $ or $\psi_{\vv{0}} $.
Given a symmetry action and vortex symmetry fractionalization class, one needs to solve the pentagon and heptagon equations, which is only possible when the $H^4(G,\text{U}(1))$ defectification obstruction vanishes.
If unobstructed (which will be the case here), the defectification will be classified torsorially by $H^3(\mathbb{Z}_2,\text{U}(1)) = \mathbb{Z}_2$.
Thus, we expect eight distinct theories, for which we now provide the explicit detail.

\begin{table}[t!]
\centering
\fbox{
\begin{tabular}{ll}
$\ifo^{(0)}_{\mathbb{Z}_2^{\eff}\times \mathbb{Z}_2}$: & trivial symmetry action $\rho_{\bf 1} = \openone$
\\[1ex] &  $\fMTC_{\vv{x},{\bf g}} = \{\I_{\vv{x},{\bf g}}, \psi_{\vv{x},{\bf g}}\}$
\\[1ex]
& $a_{\vv{x},{\bf g}}\otimes b_{\vv{y},{\bf h}} =[ \coho{q}({\bf g,h}) a  b]_{\vv{x}+\vv{y},{\bf gh}}$,
\\[1ex]&
$\coho{q}({\bf 1,1}) = \psi^{\mathsf{q}({\bf 1,1})}$, ~~$\Xdot({\bf 1},{\bf 1},{\bf 1}) = (-1)^{\mathsf{s} } i^{\mathsf{q}({\bf 1,1})}$
\\ [.1ex] \\
  $F$: & \begin{tabular}[t]{l}
$F^{a_{\vv{x},{\bf g}} b_{\vv{y},{\bf h}} c_{\vv{z},{\bf k}} } = \Xdot({\bf g},{\bf h},{\bf k})/R^{\cohosub{q}({\bf h,k}) a_{\vv{0},{\bf 0}}}$
\end{tabular}
 \\ &\\
  $R$: & \begin{tabular}[t]{l}
 $ R^{a_{\vv{x},{\bf g}} b_{\vv{y},{\bf h}}} =  (-1)^{(\mathsf{a}+\mathsf{x})\cdot \mathsf{b}}$
 \end{tabular}
  \\ &\\
 $U$: &
 \begin{tabular}[t]{l}
 $U_{\bf k}(a_{\vv{x},{\bf g}},b_{\vv{y},{\bf h}}) =\Xdot({\bf g},{\bf h},{\bf k})^{-1}$
 \end{tabular}
  \\ &\\
 $\eta$: &
\begin{tabular}[t]{l}
 $\eta_{c_{\vv{z},{\bf k}}}({\bf g},{\bf h}) =M_{\cohosub{q}({\bf g,h}) c_{\vv{z},{\bf 0}}} \Xdot({\bf g},{\bf h},{\bf k} )^{-1}$
 \end{tabular}
\end{tabular}
}
\caption{FSPT phases with $\mathcal{G}^{\eff} =\mathbb{Z}_2^{\eff}\times\mathbb{Z}_2$ and trivial symmetry action.
There are four such theories, parameterized by $\mathsf{q}({\bf 1,1}) , \mathsf{s}  \in \mathbb{F}_2$, corresponding to the vortex fractionalization class and the defectification class, respectively.
}\label{table:Z2xZ2-Ab}
\end{table}

\begin{table}[t!]
\centering
\begin{tabular}{c| cccccccc}
$n$~ & 0 & 2 & 4 & 6
\\ $(\mathsf{s},\mathsf{q}({\bf 1,1}))$~ & ~$(\mathsf{0},\mathsf{0} )$ & $(\mathsf{1},\mathsf{1})$ & $(\mathsf{1}, \mathsf{0})$ & $(\mathsf{0}, \mathsf{1})$
\end{tabular}
\caption{An assignment of even integers $n \in \{0,2,4,6\}$ to the four theories given in Table~\ref{table:Z2xZ2-Ab}.
This assignment corresponds to the invariant $T^2(a_{\vv{0},{\bf 1}}) = (-1)^{\mathsf{s}}(- i)^{\mathsf{q}({\bf 1,1})} = e^{i \frac{\pi}{4} n}$.
Stacking these theories labeled this way is additive modulo 8 in this label, forming a $\mathbb{Z}_4$ subgroup of the $\mathbb{Z}_8$ classification of FSPT phases with $\mathbb{Z}_2$ symmetry.
}
\label{table:Z8}
\end{table}

\begin{table*}[t]
\centering
\fbox{
\begin{tabular}{ll}
  $\ifo^{(0)}_{\mathbb{Z}_2^{\eff}\times\mathbb{Z}_2}$: &  nontrivial symmetry action $\rho_{\bf 1} = \V$, ~~ $n \in \{ 1,3,5, 7 \}$
\\[1ex]
 &$\fMTC_{\vv{x},{\bf 0}} = \{\I_{\vv{x},{\bf 0}}, \psi_{\vv{x},{\bf 0}}\}$,  $\fMTC_{\vv{x},{\bf 1}} = \{\sigma_{\vv{x},{\bf 1}}\}$, \quad $a_{\vv{x},{\bf g}} \otimes  b_{\vv{y},{\bf h}} = \left[ a\otimes b\right]_{\vv{x}+\vv{y},{\bf gh}}$, $a,b\in \textbf{Ising}$
  \\ [3ex]
 $F$: & \begin{tabular}[t]{l}
 $\left[ F^{a_{\vv{x},{\bf g}} b_{\vv{y},{\bf h}}c_{\vv{z},{\bf k}}}_{d_{\vv{x}+\vv{y}+\vv{z},{\bf gh k}}} \right]_{e_{\vv{x} + \vv{y},{\bf gh}} f_{\vv{y}+\vv{z},{\bf hk}} }\! = \left[F^{abc}_d \right]_{ef}$~~(Ising $F$-symbol)
 \end{tabular}
\\ &\\
 $R$: & \begin{tabular}[t]{l}
$ R^{a_{\vv{x},{\bf 0}} b_{\vv{y},{\bf 0}}}  =  (-1)^{(\mathsf{a}+\mathsf{x})\cdot \mathsf{b}}$,~~~~$R^{\sigma_{\vv{0},{\bf 1}} \sigma_{\vv{x},{\bf 1}}}_{\I_{\vv{x},{\bf 0}}} =
\left(R^{\sigma_{\vv{0},{\bf 1}} \sigma_{\vv{x},{\bf 1}}}_{\psi_{\vv{x},{\bf 0}}} \right)^*
= R^{\sigma_{\vv{1},{\bf 1}} \sigma_{\vv{1+x},{\bf 1}}}_{\I_{\vv{x},{\bf 0}}}
= \left(-R^{\sigma_{\vv{1},{\bf 1}} \sigma_{\vv{1+x},{\bf 1}}}_{\psi_{\vv{x},{\bf 0}}}\right)^* = e^{- i n \pi/4}$
\\
\\
\footnotesize{$\begin{pmatrix}
R^{\sigma_{\vv{0},{\bf 1}} \I_{\vv{x},{\bf 0}}} &
R^{\sigma_{\vv{1},{\bf 1}} \I_{\vv{x},{\bf 0}}} \\
R^{\sigma_{\vv{0},{\bf 1}} \psi_{\vv{x},{\bf 0}}} &
R^{\sigma_{\vv{1},{\bf 1}} \psi_{\vv{x},{\bf 0}}} \end{pmatrix} = \begin{pmatrix}1&1\\-i^n&i^n\end{pmatrix}$
~~~~
$\begin{pmatrix}
R^{ \I_{\vv{x},{\bf 0}}\sigma_{\vv{0},{\bf 1}}}
&R^{ \I_{\vv{x},{\bf 0}}\sigma_{\vv{1},{\bf 1}}}\\
R^{ \psi_{\vv{x},{\bf 0}}\sigma_{\vv{0},{\bf 1}}}&
R^{ \psi_{\vv{x},{\bf 0}}\sigma_{\vv{1},{\bf 1}}}\end{pmatrix} = i^{n\vv{x}}\begin{pmatrix}1&1\\i^n&i^n
\end{pmatrix}$
}
 \end{tabular}
 \\ &\\
 $U$:  &
 $ U_{\bf 1}({a_{\vv{x},{\bf 0}}, b_{\vv{y},{\bf 0}}}; [ab]_{\vv{x+y}}) = (-1)^{(\mathsf{a} + \mathsf{x})\cdot \mathsf{y}}$~~~~
\\[2ex]
&
\footnotesize{$\begin{pmatrix}
 U_{\bf 1}\left(\sigma_{\vv{0},{\bf 1}},\I_{\vv{0},{\bf 0}};\sigma_{\vv{0},{\bf 1}}\right) & U_{\bf 1}\left(\I_{\vv{0},{\bf 0}},\sigma_{\vv{0},{\bf 1}};\sigma_{\vv{1},{\bf 1}}\right) \\
 U_{\bf 1}\left(\sigma_{\vv{0},{\bf 1}},\psi_{\vv{0},{\bf 0}} ;\sigma_{\vv{0},{\bf 1}} \right) &  U_{\bf 1}\left(\psi_{\vv{0},{\bf 0}},\sigma _{\vv{0},{\bf 1}};\sigma_{\vv{1},{\bf 1}}\right) \\
  U_{\bf 1}\left(\sigma_{\vv{0},{\bf 1}},\I_{\vv{1},{\bf 0}};\sigma_{\vv{1},{\bf 1}}\right) & U_{\bf 1}\left(\I_{\vv{1},{\bf 0}},\sigma_{\vv{0},{\bf 1}};\sigma_{\vv{0},{\bf 1}}\right) \\
 U_{\bf 1}\left(\sigma_{\vv{0},{\bf 1}},\psi_{\vv{1},{\bf 0}} ;\sigma_{\vv{1},{\bf 1}} \right) & U_{\bf 1}\left(\psi_{\vv{1},{\bf 0}},\sigma _{\vv{0},{\bf 1}};\sigma_{\vv{0},{\bf 1}}\right) \\
\end{pmatrix}
 =
\begin{pmatrix}
 1 & 1 \\
 -1 & -1 \\
-1 & i^n\\
-1 & -i^n\\
\end{pmatrix}
$}
~~~~
\footnotesize{$\begin{pmatrix}
 U_{\bf 1}\left(\sigma_{\vv{1},{\bf 1}},\I_{\vv{0},{\bf 0}};\sigma_{\vv{0},{\bf 1}}\right) & U_{\bf 1}\left(\I_{\vv{0},{\bf 0}},\sigma_{\vv{1},{\bf 1}};\sigma_{\vv{1},{\bf 1}}\right) \\
 U_{\bf 1}\left(\sigma_{\vv{1},{\bf 1}} ,\psi_{\vv{0},{\bf 0}};\sigma_{\vv{0},{\bf 1}} \right) & U_{\bf 1}\left(\psi_{\vv{0},{\bf 0}},\sigma _{\vv{1},{\bf 1}};\sigma_{\vv{1},{\bf 1}}\right) \\
  U_{\bf 1}\left(\sigma_{\vv{1},{\bf 1}},\I_{\vv{1},{\bf 0}};\sigma_{\vv{1},{\bf 1}}\right) & U_{\bf 1}\left(\I_{\vv{1},{\bf 0}},\sigma_{\vv{1},{\bf 1}};\sigma_{\vv{0},{\bf 1}}\right) \\
 U_{\bf 1}\left(\sigma_{\vv{1},{\bf 1}},\psi_{\vv{1},{\bf 0}} ;\sigma_{\vv{1},{\bf 1}} \right) & U_{\bf 1}\left(\psi_{\vv{1},{\bf 0}},\sigma _{\vv{1},{\bf 1}};\sigma_{\vv{0},{\bf 1}}\right) \\
\end{pmatrix}
 =
\begin{pmatrix}
 1 & 1 \\
 1 & 1 \\
1 & i^n\\
-1 & i^n\\
\end{pmatrix}
$}\\
&\\
&
\!\!\!\!\!\!\!\!\!\!\!\!\!\!\! \footnotesize{$\begin{pmatrix}
U_{\bf 1}(\sigma_{\vv{0},{\bf 1}},\sigma_{\vv{0},{\bf 1}}; \I_{\vv{0},{\bf 0}}) &
U_{\bf 1}(\sigma_{\vv{1},{\bf 1}},\sigma_{\vv{1},{\bf 1}}; \I_{\vv{0},{\bf 0}})\\
U_{\bf 1}(\sigma_{\vv{0},{\bf 1}},\sigma_{\vv{0},{\bf 1}}; \psi_{\vv{0},{\bf 0}})&
U_{\bf 1}(\sigma_{\vv{1},{\bf 1}},\sigma_{\vv{1},{\bf 1}}; \psi_{\vv{0},{\bf 0}})\\
\end{pmatrix}
=e^{i \frac{\pi n}{4}}
\begin{pmatrix}
-i^n & -1\\
i^n&-1\\
\end{pmatrix}
$}
~~~~\footnotesize{$\begin{pmatrix}
U_{\bf 1}(\sigma_{\vv{0},{\bf 1}},\sigma_{\vv{1},{\bf 1}}; \I_{\vv{1},{\bf 0}}) &
U_{\bf 1}(\sigma_{\vv{1},{\bf 1}},\sigma_{\vv{0},{\bf 1}}; \I_{\vv{1},{\bf 0}})\\
U_{\bf 1}(\sigma_{\vv{0},{\bf 1}},\sigma_{\vv{1},{\bf 1}}; \psi_{\vv{1},{\bf 0}})&
U_{\bf 1}(\sigma_{\vv{1},{\bf 1}},\sigma_{\vv{0},{\bf 1}}; \psi_{\vv{1},{\bf 0}})\\
\end{pmatrix}
=e^{i \frac{\pi n}{4}}
\begin{pmatrix}
i^n & -1\\
i^n&1\\
\end{pmatrix}
$}
 \\ &\\
 $\eta$: &
 \begin{tabular}[t]{l}
 $\eta_{\I_{\vv{1},{\bf 0}}}({\bf 1},{\bf 1}) = \eta_{\psi_{\vv{1},{\bf 0}}}({\bf 1},{\bf 1}) = i^n$,~~~~$\eta_{\sigma_{\vv{0},{\bf 1}}}({\bf 1},{\bf 1}) = e^{i \pi n/4}$,~~~~$\eta_{\sigma_{\vv{0},{\bf 1}}}({\bf 1},{\bf 1}) =- e^{-i \pi n/4}$
 \end{tabular}
\end{tabular}}
\caption{FSPT phases with $\mathcal{G}^{\eff} =\mathbb{Z}_2^{\eff}\times\mathbb{Z}_2$ and nontrivial symmetry action.
There are four theories, parameterized by an odd integer $n\in \{ 1,3,5, 7 \}$.
 }\label{table:Z2xZ2-nonAb}
\end{table*}

First consider theories with trivial symmetry action.
We can parameterize the fractionalization classes by
\begin{align}
\coho{q}({\bf 1},{\bf 1}) &= \psi^{\mathsf{q}({\bf 1,1})}
,
\end{align}
where $\mathsf{q} \in Z^2(\mathbb{Z}_2, \mathbb{F}_2)$.
We then need to solve the pentagon and heptagon equations.
When the symmetry action does not permute topological charges, Ref.~\onlinecite{Bark2019} gave explicit expressions for the defectification obstruction and the complete topological data of the $G$-crossed extensions.
Evaluating the expression for the defectification obstruction in the present case yields the only nontrivial element given by
\begin{align}
\defectO({\bf 1},{\bf 1},{\bf 1},{\bf 1}) = R^{\cohosub{q}({\bf 1,1}) \cohosub{q}({\bf 1,1})}= (-1)^{\mathsf{q}({\bf 1,1}) \cdot \mathsf{q}({\bf 1,1})}
,
\end{align}
so we can write the obstruction as
\begin{align}
\defectO = (-1)^{\mathsf{q} \cup \mathsf{q}}
.
\end{align}
This $\defectO$ is a coboundary, for which a solution to $\cbd \Xdot = \defectO$ is given by
\begin{align}
\Xdot({\bf 1},{\bf 1},{\bf 1}) =(-1)^{\mathsf{s}}i^{\mathsf{q}({\bf 1,1}) }
,
\end{align}
where $\mathsf{s} \in \mathbb{F}_2$ parametrizes the two distinct defectification classes coming from $H^3(\mathbb{Z}_2,\text{U}(1))  = \mathbb{Z}_2$.
We produce a complete set of solutions to the pentagon and heptagon equations in Table~\ref{table:Z2xZ2-Ab}.
We show below that all solutions are physically distinct $\mathcal{G}^{\eff}$-crossed SMTCs and distinguished by a particular gauge invariant quantity.
With foresight, we will label these four theories by even integers $n \in \{ 0,2,4,6\}$ as indicated in Table~\ref{table:Z8}.

When the symmetry action permutes topological charges, we solve the pentagon and heptagon equations explicitly, rather than solving for the obstruction objects.
We find that for $\mathcal{G}^{\eff}=\mathbb{Z}_2^{\eff}\times \mathbb{Z}_2$ there are four solutions to the $G$-crossed consistency conditions.
This agrees with the expectation from $H^2(\mathbb{Z}_2, \mathbb{Z}_2^{\psi}) = \mathbb{Z}_2$ and $H^3(\mathbb{Z}_2,\text{U}(1))  = \mathbb{Z}_2$.
We show below that they can be distinguished by a particular gauge invariant quantity.
Their topological data is presented in Table~\ref{table:Z2xZ2-nonAb}, and parameterized by an odd integer $n\in \{1, 3, 5, 7\}$.

The eight FSPT phases with $\mathcal{G}^{\eff}=\mathbb{Z}_2^{\eff}\times\mathbb{Z}_2$ have a $\mathbb{Z}_8$ structure under stacking~\cite{Gu14,Lu16,Bark2019}.
This group structure can be seen by computing the quantity
\begin{align}
\label{eq:Tsquared}
T^2(a_{\vv{x},{\bf g}}) &= \theta^2_{a_{\vv{x},{\bf g}}}\eta_{a_{\vv{x},{\bf g}}}({\bf g},{\bf g})
,
\end{align}
which is gauge-invariant for $G=\mathbb{Z}_{2}$.
Under stacking, $T^2$ of a given defect sector of the condensed theory corresponds to the product of $T^2$ from the same sector for the two stacked theories.
This is a consequence of the fact that the condensed theory inherits the braiding of the parent (product) theory, since the theory acts trivially on $\psi_{\vv{0},{\bf 0}}$.
The $T^2$ eigenvalues of the eight theories take the form
\begin{align}
T^2(a_{\vv{x},{\bf 1}}) = i^{-n \mathsf{x}} e^{i \frac{\pi}{4} n}
,
\end{align}
where $n$ are the integers we have assigned to the various theories.
The gauge invariant quantity $T^2(a_{\vv{x},{\bf g}})$ distinguishes all eight $\mathbb{Z}_2$ FSPT phases.
One can check that the gauge invariant quantities are additive under stacking and, hence, we see the $\mathbb{Z}_8$ group structure emerge,
\begin{align}
\label{eq:Z8classification}
\left[\ifo^{(0)\times}_{\mathbb{Z}_2^{\eff}\times \mathbb{Z}_2}\right]_n  \underset{\mathbb{Z}_2^{\eff}\times \mathbb{Z}_2}{\ftimes}\, \left[\ifo^{(0)\times}_{\mathbb{Z}_2^{\eff}\times \mathbb{Z}_2}\right]_{n'} &= \left[\ifo^{(0)\times}_{\mathbb{Z}_2^{\eff}\times \mathbb{Z}_2}\right]_{[n+n']\text{mod}8}.
\end{align}

In summary, for $\mathcal{G}^{\eff} = \mathbb{Z}_2^{\eff} \times \mathbb{Z}_2$, there are four distinct FSPT phases with trivial symmetry action, and four more with nontrivial symmetry action.
We have also shown that there is a $\mathbb{Z}_8$ classification (consistent with previous works~\cite{Gu14,Lu16,Bark2019}) and have parameterized the theories according to this integer.
We remark that sending $n \to n+4$ glues in the nontrivial bosonic SPT phase in $H^3(\mathbb{Z}_2, \text{U}(1))$, and applying the $\vviso$-isomorphism of Sec.~\ref{sec:centralisomorphism} with $\vviso({\bf 1)} = 1$ leaves $\central =0$ unchanged, but sends $n \to -n$.

\subsubsection{\texorpdfstring{$\mathcal{G}^{\operatorname{f}}=\mathbb{Z}_4^{\operatorname{f}}$}{Z4f}}
\label{sec:z4fclassification}

We now classify FSPT phases with $\mathcal{G}^{\eff}  = \mathbb{Z}_4^{\eff}$.
In this case, we use a nontrivial $\mathbb{Z}_2^{\eff}$ extension of the on-site symmetry group $\mathbb{Z}_2$.
The extension is given by the nontrivial $\central \in Z^2(\mathbb{Z}_2 , \mathbb{Z}_2^{\eff})$, whose only non-zero element is
\begin{align}
\central({\bf 1,1}) = \vv{1}.
\end{align}
In this language, one of the $\mathbb{Z}_4^{\eff}$ generators is given by $\mathpzc{g} = (\vv{0},{\bf 1})$, and the rest of the elements are given by
\begin{align*}
\mathpzc{g}^2 &= (\vv{0},{\bf 1})^2 =(\vv{0} + \vv{0} + \central({\bf 1},{\bf 1}), {\bf 1} + {\bf 1}) = (\vv{1},{\bf 0}),
\\
\mathpzc{g}^3 &= (\vv{0},{\bf 1})\cdot (\vv{1},{\bf 0}) =(\vv{0} + \vv{1} + \central({\bf 1},{\bf 0}), {\bf 1} + {\bf 0}) = (\vv{1},{\bf 1}),
\\
\mathpzc{g}^4 &= (\vv{1},{\bf 0})^2 = (\vv{1}+\vv{1}+\central({\bf 0},{\bf 0}), {\bf 0} + {\bf 0}) = (\vv{0},{\bf 0}).
\end{align*}
Below we present the data for theories with trivial symmetry action.
For nontrivial symmetry action, the vortex symmetry fractionalization obstruction ${\coho{O}}^{\eta} = \psi^{\pi \cup \central}$ is a nontrivial 3-cocycle (not a coboundary), so there are no solutions to the pentagon and heptagon equations, as can also be confirmed directly.

For trivial symmetry action, the vortex symmetry fractionalization classes are torsorially classified by $H^2(\mathbb{Z}_2,\mathbb{Z}_2^\psi)$.
This can be parameterized as
\begin{align}
\coho{q}({\bf 1},{\bf 1}) &= \psi^{\mathsf{q}({\bf 1,1})}
,
\end{align}
where $\mathsf{q} \in \mathbb{F}_2$ denotes whether the 2-cocycle is trivial ($\coho{q}({\bf 1,1}) = \I$) or nontrivial ($\coho{q}({\bf 1,1}) = \psi$).
The defectification obstruction vanishes, $\defectO({\bf g},{\bf h},{\bf k},{\bf l}) = 1$, hence the remaining data can be parameterized by the $3$-cocycle $\Xdot$, whose only nontrivial element is
\begin{align}
\Xdot({\bf 1,1,1}) = (-1)^{\mathsf{m}}.
\end{align}
Thus, we have four candidate theories parameterized by $(\mathsf{m}, \mathsf{q}({\bf 1,1}))\in \mathbb{F}_2\times\mathbb{F}_2$, specifying the vortex fractionalization and defectification classes, respectively.

\begin{table}[t!]
\centering
\fbox{
\begin{tabular}{ll}
  $\ifo^{(0)}_{\mathbb{Z}_4^{\eff}}$: & trivial symmetry action $\rho_{\bf 1} = \openone$
\\[1ex] & $\fMTC_{\vv{x},{\bf g}} = \{\I_{\vv{x},{\bf g}}, \psi_{\vv{x},{\bf g}}\}$
  \\ & $a_{\vv{x},{\bf g}} \otimes  b_{\vv{y},{\bf h}} = \begin{cases}
  [a \otimes b]_{\vv{x+y+1},{\bf 0}} &\text{if ${\bf g} = {\bf h} = {\bf 1}$}\\
  [a \otimes b]_{\vv{x+y},{\bf gh}} &\text{otherwise}
  \end{cases}$
  \\ [3ex]
  $F$: & \begin{tabular}[t]{l}
$F^{a_{\vv{x},{\bf 1}} b_{\vv{y},{\bf 1}} c_{\vv{z},{\bf k}}} =  (-1)^{\mathsf{c}}$
 \end{tabular}
 \\ &\\
  $R$: & \begin{tabular}[t]{l}
 $ R^{a_{\vv{x},{\bf g}} b_{\vv{y},{\bf h}}} =  (-1)^{(\mathsf{a}+\mathsf{x})\cdot \mathsf{b}}$
 \end{tabular}
  \\ &\\
 $U$: &
 \begin{tabular}[t]{l}
 $U_{\bf 1}(a_{\vv{x},{\bf 1}},b_{\vv{y},{\bf 1}}) = 1$
 \end{tabular}
  \\ &\\
 $\eta$: &
\begin{tabular}[t]{l}
 $\eta_{c_{\vv{z},{\bf k}}}({\bf 1},{\bf 1}) =(-1)^{\mathsf{c}}$
 \end{tabular}
\end{tabular}}
\caption{FSPT phase with $\mathcal{G}^{\eff}=\mathbb{Z}_4^{\eff}$.
While the formulas for the topological data of theories with trivial symmetry action~\cite{Bark2019} na\"ively suggest there are four theories, they can all be identified by the relabelings discussed in the text.
Here, we present the data for the representative with $\mathsf{q}({\bf 1,1}) =\mathsf{m} =0$.
}\label{table:Z4}
\end{table}

Noting that $\coho{w} = \coho{q}_{\vv{w}}$, the topological data of these theories are given by~\cite{Bark2019}
\begin{align}
\label{eq:Z4ffusionring}
a_{\vv{x},{\bf g}} \otimes b_{\vv{y},{\bf h}}  &= [\coho{q}({\bf g,h}) ab]_{\vv{x}+ \vv{y} + \central({\bf g,h}) ,{\bf gh}},
\\
F^{a_{\vv{x},{\bf g}} b_{\vv{y},{\bf h} }c_{\vv{z},{\bf k}} } &= (-1)^{(\mathsf{q}({\bf g,h}) + \central({\bf g,h}))\cdot \mathsf{c} }\Xdot({\bf g,h,k}) ,
\\
R^{a_{\vv{x},{\bf g}} b_{\vv{y},{\bf h} }} & = (-1)^{(\mathsf{a}+\mathsf{x}) \cdot \mathsf{b}} ,
\\
U_{\bf k}(a_{\vv{x},{\bf g}}, b_{\vv{y},{\bf h}})  &= \Xdot({\bf g,h,k})^{-1} ,
\\
\eta_{c_{\vv{z},{\bf k}} }({\bf g,h}) & = (-1)^{\vv{z} \cdot \mathsf{q}({\bf g,h}) + \mathsf{c} \cdot \central({\bf g,h})}  \Xdot({\bf g,h,k})^{-1} ,
\end{align}
corresponding to four solutions of the pentagon and heptagon equations.
However, these na\"ively distinct theories can be identified by relabeling the topological charges, as we now explain.

Relabeling the objects as
\begin{align}
{a}_{\vv{x},{\bf g}} \to a'_{\vv{x},{\bf g}} ={[\psi^{\vv{x}} a]}_{\vv{x},{\bf g}}
\end{align}
identifies the theories $\coho{q}({\bf 1,1})\to \coho{q}({\bf 1,1})'=\psi \otimes \coho{q}({\bf 1,1})$.
We note that this relabeling is permitted because it changes charges by a $\psi_{\vv{0}}$ value (not a vortex value), and has trivial $G$ dependence, i.e. it is just a relabeling of the vortices of $\ifo^{(0)}$.
To see this makes the claimed identification of theories, note that the relabeling modifies the fusion rules as
\begin{align}
a'_{\vv{x},{\bf g}} \otimes b'_{\vv{y},{\bf h}} = [ \psi^{\central({\bf g,h})} \coho{q}({\bf g,h}) ab]'_{\vv{x+y}+\central({\bf g,h}), {\bf gh}}
,
\end{align}
indicating that
\begin{align}
\coho{q}({\bf g,h}) \to \coho{q}'({\bf g,h})  = \coho{q}({\bf g,h})\psi^{\central({\bf g,h})}_{\vv{0}}.
\end{align}
The rest of the topological data can be made to match using a vertex basis gauge transformation.
Thus, the two na\"ively different vortex symmetry fractionalization classes are actually identical.

Similarly, relabeling the objects as
\begin{align}
{a}_{\vv{x},{\bf g}} \to a'_{\vv{x},{\bf g}} ={[\coho{z}({\bf g}) a]}_{\vv{x},{\bf g}},
\end{align}
with $\coho{z}({\bf 0}) = \I$ and $\coho{z}({\bf 1}) = \psi$, identifies $\mathsf{m} \to \mathsf{m}'=\mathsf{m+1}$.
This is an allowed relabeling because it changes charges by a $\psi_{\vv{0}}$ value (not a vortex value), and has trivial $\mathbb{Z}_{2}^{\eff}$ dependence, i.e. it leaves $\mathcal{G}^{\eff}$-sectors fixed.
The equivalence of theories can be seen using Sec.~IV of Ref.~\onlinecite{Aasen21}; the relabeled theory has an $\spt_{G}^{[\alpha]}$ glued into it with
\begin{align}
\alpha({\bf g,h,k})  = M_{[\cohosub{q}({\bf g,h})]_{\central({\bf g,h})}\cohosub{z}({\bf k})} =
\begin{cases}
-1 &\text{if ${\bf g}={\bf h} = {\bf k} ={\bf 1}$} \\
1 &\text{otherwise}.
\end{cases}
\end{align}
Thus $\alpha$ is the nontrivial element of $H^3(\mathbb{Z}_2,\text{U}(1))$, indicating the defectification class has shifted.
The rest of the topological data can be made to match using vertex basis and symmetry action gauge transformations.

The relabelings above equate the four candidate theories to represent a single $\mathbb{Z}_4^{\eff}$ FSPT phase, whose data is given in Table~\ref{table:Z4} in a convenient gauge.
These results agree with those obtained using defect decoration~\cite{Wang2020} and spin-cobordism classification~\cite{Kapustin15,Garcia2019}.

\subsubsection{Comparison to \texorpdfstring{${G}=\mathbb{Z}_2$}{GeZ2bosonic} extensions of the toric code}

Recall that gauging fermion parity promotes the trivial fermionic theory $\ifo^{(0)}$ to the toric code bosonic MTC $\tc$, mapping
\begin{align}
\I_\vv{0} \to \I, && \psi_\vv{0}\to \psi, && \I_\vv{1}\to e, && \psi_\vv{1}\to m.
\end{align}
It is instructive to compare the classification of FSPT phases with $G=\mathbb{Z}_2$ to the analogous bosonic classification of $\mathbb{Z}_2$-extensions of the toric code, which was computed as an example in Refs.~\onlinecite{Lu16} and \onlinecite{Bark2019} (though the latter reference did not identify the relabelings that equate the $\coho{w}({\bf 1,1})=e$ and $m$ fractionalization classes, as well as their corresponding defectification classes).

The first difference between the bosonic and fermionic classifications is that bosonic theories are classified with respect to a fixed symmetry action, while FSPT phases reference a fixed fermionic symmetry group $\mathcal{G}^{\eff}$.
The next significant departure is the number of distinct theories.
The relevant cohomology groups appearing in the bosonic classification of toric code with $G = \mathbb{Z}_2$ are
\begin{align}
H^2_{[\rho]}(\mathbb{Z}_2,\mathbb{Z}_2\times\mathbb{Z}_2)
= \begin{cases}
\mathbb{Z}_2 \times \mathbb{Z}_2 & \text{$\rho_{\bf 1}$ trivial},\\
\mathbb{Z}_1 & \text{$ \rho_{\bf 1}$ nontrivial},\\
\end{cases}
\end{align}
and $H^3(\mathbb{Z}_2,\text{U}(1)) = \mathbb{Z}_2$.
Thus, there are na\"ively 10 different $G$-crossed theories, eight with trivial symmetry action and two with nontrivial symmetry action.
However, there are relabelings that equate some of the theories with trivial symmetry action.
For the $\coho{w}({\bf 1,1})=\psi$ fractionalization class, the relabelings $a'_{\bf 1} = e \otimes a_{\bf 1}$ or $a'_{\bf 1} = m \otimes a_{\bf 1}$ (together with gauge transformations) equate its two defectification classes.
For the $\coho{w}({\bf 1,1})=e$ and $m$ fractionalization classes, relabeling the quasiparticles $e \leftrightarrow m$ equates the two fractionalization classes and a relabeling $a'_{\bf 1} = \psi \otimes a_{\bf 1}$ can be used to equate their corresponding defectification classes, so all four theories are equivalent.
Thus, there are actually four physically distinct SET phases with trivial symmetry action and two with nontrivial action for $\tc$ with $\mathbb{Z}_{2}$ symmetry.
In contrast, there were eight distinct $\mathbb{Z}_2^{\eff}\times \mathbb{Z}_2$ FSPT phases, four with trivial symmetry action and four with nontrivial action; and there was one $\mathbb{Z}_4^{\eff}$ FSPT phase, which had trivial symmetry action.

Comparing to the corresponding $G=\mathbb{Z}_{2}$ SET and FSPT phases, we see that the $\mathbb{Z}_4^{\eff}$ FSPT phase corresponds to the bosonic $\mathbb{Z}_{2}$-crossed $\tc$ with trivial symmetry action and $\coho{w}({\bf 1,1})=e$ and $m$ symmetry fractionalization.
The identification of these four bosonic $G$-crossed theories as one SET phase is precisely the same identification of the four na\"ive $\mathbb{Z}_4^{\eff}$-crossed theories as one FSPT phase, so these match up exactly.

On the other hand, the $\mathbb{Z}_2^{\eff}\times \mathbb{Z}_2$ FSPT phases correspond to the remaining bosonic $\mathbb{Z}_{2}$-crossed $\tc$ theories, with important differences.
The first is that the $\mathbb{Z}_2^{\eff}\times \mathbb{Z}_2$-crossed theories with trivial symmetry action and $\coho{w}({\bf 1,1})=\psi$ fractionalization class do not have an allowed relabeling that equates the two defectification classes.
The relabeling that allowed such a relation for the bosonic $\mathbb{Z}_{2}$-crossed $\tc$ theories would map to a vortex-valued relabeling for the corresponding fermionic theory.
In other words, a relabeling equating the FSPT defectifiation classes would actually involve changing the $\mathcal{G}^{\eff}$ sectors $(\vv{0},{\bf 1}) \leftrightarrow (\vv{1},{\bf 1})$, which is not a permissible equivalence of fermionic theories.
The second difference is for the theories with nontrivial symmetry action.
For the bosonic theories, there is only one distinct fractionalization class, whereas the FSPT phases have fractionalization classified by $H^2(\mathbb{Z}_2,\mathbb{Z}_2^{\psi}) = \mathbb{Z}_2$.
This difference is due to the fact that $\coho{w}({\bf 1,1})=\psi$ is related to $\coho{w}({\bf 1,1})=\I$ for the bosonic theories by a coboundary $\cbd \coho{z}$, where $\coho{z}({\bf 1}) = e$ or $m$.
This coboundary maps to a vortex-valued coboundary for the fermionic theory, which is not a permissible equivalence, as explained in Sec.~\ref{sec:symmetry_fractionalization}.

Lastly, we note that stacking $\mathbb{Z}_2$-crossed extensions of toric code and condensing $(\psi,\psi)$ does not admit a consistent group structure for the bosonic theories, as is clear by comparison with the fermionic results.
This reflects the fact that $(\psi,\psi)$ has no additional physical significance compared to the many other condensible bosons in the stacked theory.
The physical fermion $\psi_\vv{0}$ is a local particle, thus condensing a bound pair does not require strong interactions between the stacked theories.
In contrast, the fermion $\psi$ of the toric code is an emergent quasiparticle and generally will require strong interactions for its bound pairs to form a condensate.
As such, while stacking provides a natural tool for classifying fermionic topological order, it plays no special role for understanding the structure of the analogous bosonic theories.

\subsection{General \texorpdfstring{$G$}{genG} FSPT Phases}

We now turn to general symmetry group $G$ to generate representatives of all FSPT phases.
Our results rely on three key developments: (1) the efficient construction of the topological data for a particular $\mathbb{Z}_2^{\eff}\times G$ FSPT phase given the data of a bosonic SPT phase and a $\mathbb{Z}_2^{\eff}\times \mathbb{Z}_2$ FSPT phase; (2) a method of torsorially generating the data of all theories with the same symmetry action (and possibly distinct $\mathcal{G}^{\eff}$) given the data of any one theory in that classification; and (3) a systematization of the relabeling redundancy for FSPT phases.
For (2), the data of the theory obtained by applying the torsor functor is subject to a relative defectification obstruction, which must vanish in order for the new $F$-symbols to satisfy the pentagon equation.
Using these developments, we are able to produce the basic data of all $\mathcal{G}^{\eff}$-crossed FSPT phases.
The next section will describe the classification of $\mathcal{G}^{\eff}$-crossed FSPT phases into a group structure under stacking.

\subsubsection{All Your Base Theories}
\label{sec:AYBT}

The $\mathbb{Z}_2$-symmetric invertible fermionic theories are not just illustrative examples of the $\mathcal{G}^{\eff}$-crossed formalism; they play a critical role in the efficient construction of a $\mathbb{Z}_2^{\eff}\times G$ FSPT ``base theory.''
This base theory is formed through a restricted product of a $\mathbb{Z}_2^{\eff}\times \mathbb{Z}_2$ FSPT phase with a $G$-symmetric bosonic SPT phase.
Our approach extends the Ising pullbacks developed in Ref.~\onlinecite{Cheng2018,Bhardwaj2017} to construct bona fide $\mathcal{G}^{\eff}$-crossed SMTCs.
While we focus here on FSPT phases, this method can also be applied to all invertible fermionic orders $\ifo^{(\nu)}$, as we do in Secs.~\ref{sec:ex-A} and~\ref{sec:invfSET}.

We begin by recalling that the trivial fermionic theory has topological symmetry group $\operatorname{Aut}^{\eff}(\ifo^{(0)}) = \mathbb{Z}_2$, so that the global symmetry action for a $\mathcal{G}^{\eff}$ FSPT phase is specified by a group homomorphism $\maj: G\to \mathbb{Z}_2$, given by
\begin{align}
\rho_{\bf g} = \V^{\maj({\bf g})},
\end{align}
where $\V$ is the vortex-permuting symmetry action described in Sec.~\ref{sec:symmetry_action}.
To construct a base theory with $\mathcal{G}^{\eff} = \mathbb{Z}_2^{\eff}\times G$ and symmetry action $\rho$, we use a restricted product
\begin{align}
\label{eq:allyourbasearebelongtous}
\base_{\mathbb{Z}_2^{\eff}\times G,\rho} &= \left[\ifo^{(0)}_{\mathbb{Z}_2^{\eff}\times \mathbb{Z}_2}\right]_1 \boxtimes \spt_G^{[1]} \bigg|_{\mathcal{S}_\rho}.
\end{align}
where $\left[\ifo^{(0)}_{\mathbb{Z}_2^{\eff}\times \mathbb{Z}_2}\right]_{1}$ is the generator of the $\mathbb{Z}_{8}$ classification of $\mathbb{Z}_2^{\eff}\times \mathbb{Z}_2$ FSPT phases [see Eq.~\eqref{eq:Z8classification}], $\spt_G^{[1]}$ is trivial bosonic SPT phase, and the restriction is to the topological charges specified by
\begin{align}
\mathcal{S}_\rho= \{(a_{\vv{x},\maj({\bf g})},{\I_{\bf g}} )\}.
\end{align}
We have fixed these choices of initial fermionic and bosonic SPT phases for specificity, but the method will apply equally well for more general choices, e.g. $1\to n$ and $[1]\to [\alpha]$.
The restriction to $\mathcal{S}_\rho$ defines a subcategory of the full $\mathbb{Z}_2^{\eff}\times \mathbb{Z}_2\times G$-crossed product theory $\left[\ifo^{(0)}_{\mathbb{Z}_2^{\eff}\times\mathbb{Z}_2}\right]_1 \boxtimes \spt_G^{[1]}$.
This subcategory is seen to be a $\mathbb{Z}_2^{\eff} \times G$-crossed SMTC.

The topological data of $\base_{\mathbb{Z}_2^{\eff}\times G,\rho}$ can be inferred from the product structure by multiplying the $F$-, $R$-, $U$-, and $\eta$-symbols given in Table~\ref{table:Z2xZ2-nonAb} with the corresponding bosonic SPT phase data given in Sec.~\ref{sec:bSPT}.
For convenience, we use the shorthand
\begin{align}
a_{\mathpzc{g}} \equiv a_{\vv{x},{\bf g}} \equiv (a_{\vv{x},\maj({\bf g})},{\I_{\bf g}} ).
\end{align}
With this notation, we see that the defect sectors are given by
\begin{align}
\left[ \base_{\mathbb{Z}_2^{\eff}\times G,\rho}\right]_{\mathpzc{g}} =
\begin{cases}
\{\I_{\mathpzc{g}} ,\psi_{\mathpzc{g}} \} &\text{if $\maj({\bf g}) ={\bf 0}$}\\
\{\sigma_{\mathpzc{g}} \}  & \text{if $\maj({\bf g}) ={\bf 1}$}\\
\end{cases}
,
\end{align}
i.e., $\maj({\bf g})$ specifies whether we retain the Abelian or non-Abelian defects of the full product theory.
The symmetry action on the topological charges can be succinctly written as
\begin{align}
\rho_{\bf h}(a_{\mathpzc{g}}) = \psi_{\mathpzc{0}}^{\vv{x} \cdot \maj({\bf h})} \otimes a_{\mathpzc{hg \bar{h} }}
.
\end{align}
Here, we note that conjugating by $\mathpzc{h}$ is the same as conjugating by ${\bf h}$ and that $\psi_{\mathpzc{0}} \otimes \sigma_{\mathpzc{g}} = \sigma_{\mathpzc{g}}$.
The fusion rules are given by
\begin{align}
a_{\mathpzc{g}} \otimes b_{\mathpzc{h}} =  [a \otimes b]_{\mathpzc{gh}}
,
\end{align}
with $a\otimes b$ inherited from the Ising fusion category.
From this, we see that these fusion rules imply that the resulting theory has ${\mathcal{G}^{\eff} = \mathbb{Z}_2^{\eff} \times G}$ grading.
The $F$-, $R$-, $U$- and $\eta$-symbols of $\base_{\mathbb{Z}_2^{\eff}\times G,\rho}$ are
\begin{align}
\left[F^{a_{\mathpzc{g}}b_{\mathpzc{h}} c_{\mathpzc{k}}}_{d_{\mathpzc{ghk}}}\right]_{e_\mathpzc{gh} f_\mathpzc{hk}}
&=\left[F^{a_{\maj(\mathpzc{g})}b_{\maj(\mathpzc{h})} c_{\maj(\mathpzc{k})}}_{d_{\maj(\mathpzc{ghk})}}\right]_{e_{\maj(\mathpzc{gh})} f_{\maj(\mathpzc{hk})}} \\
R^{a_{\mathpzc{g}}b_{\mathpzc{h}}}_{c_{\mathpzc{gh}}}
& = R^{a_{\maj(\mathpzc{g})}b_{\maj(\mathpzc{h})}}_{c_{\maj(\mathpzc{gh})}}\\
 U_{\bf k}(a_{\mathpzc{g}},b_{\mathpzc{h}}; c_{\mathpzc{gh}})  &=
U_{\maj({\bf k})}(a_{\maj(\mathpzc{g})},b_{\maj(\mathpzc{h})}; c_{\maj(\mathpzc{gh})}) \\
 \eta_{c_{\mathpzc{k}}}({\bf g},{\bf h})
 &=\eta_{c_{\maj(\mathpzc{k})}}({\bf g},{\bf h}),
\end{align}
where the $F$-, $R$-, $U$- and $\eta$-symbols on the right hand sides of these equations are those of $\left[\ifo^{(0)}_{\mathbb{Z}_2^{\eff}\times \mathbb{Z}_2}\right]_{1}$ and we use the shorthand ${\maj(\mathpzc{g}) \equiv (\vv{x},\maj({\bf g} ) )}$.
Note that each base theory is fully specified by the external symmetry group $G$ and the symmetry action $\rho$.

\subsubsection{Torsorial generation of all FSPT phases}
\label{sec:torsor-method}

The base theories described above are not the most general $\mathcal{G}^{\eff}$-crossed FSPT phases.
However, they provide a base from which we can generate all remaining FSPT phases with the same symmetry action.
By exploiting the torsorial nature of the $G$-crossed classification, we can write the topological data of each such theory from that of the base theory and the cocycles relating them.
This result is surprising, as it collapses the highly nontrivial consistency conditions of all symmetric invertible FSPT phases onto a much simpler set of consistency conditions.

In order to generate the $\mathcal{G}^{\eff}$-crossed theories of all FSPT phases, we use the results of Ref.~\onlinecite{Aasen21}, which generate the complete data of the post-torsor functor theory
\begin{align}
\tor{\bMTC}_G^\times = \mathcal{F}_{\cohosub{t},\Xdot}( \bMTC_G^\times),
\end{align}
in terms of the pre-torsor functor theory $\bMTC_G^\times$.
We denote quantities in $\tor{\bMTC}_G^\times$ as $\tor{Q}=\mathcal{F}_{\cohosub{t},\Xdot}(Q)$, with the exception of the topological charge labels which are unchanged by the torsor action.

In the context of FSPT phases, $\bMTC = \ifo^{(0)}$, and we can write $\coho{t}({\bf g,h})  = \coho{p}_{\epsilon}({\bf g,h}) \in Z^2_{[\rho]}(G,\mathcal{A})$, where $\coho{p}({\bf g,h}) \in C^2(G,\mathbb{Z}_{2}^{\psi})$ with the vorticity label $\epsilon({\bf g,h}) \in Z^2(G,\mathbb{Z}_{2}^{\eff})$, so that
\begin{align}
\label{eq:p-to-p}
\coho{p}_{\epsilon} =\coho{p} \otimes \I_{\epsilon} = \psi_{\vv{0}}^{\mathsf{p}} \otimes \I_{\vv{1}}^{\epsilon}
.
\end{align}
(In this way, we can think of $\mathcal{A} = \mathbb{Z}_{2}^{\psi} \times \mathbb{Z}_{2}^{\eff}$.)
Applying a torsor functor with $2$-cocycle $\coho{p}_{\epsilon}$ to a $\mathcal{G}^{\eff}$-crossed theory $\fMTC_{\mathcal{G}^{\eff}}^\times$ modifies the defect fusion rules as
\begin{align}
\label{eq:torsor-fusion}
a_{\mathpzc{g}}\tor{\otimes}b_{\mathpzc{h}} &= \coho{p}_{\epsilon}({\bf g},{\bf h}) \otimes \left( a_{\mathpzc{g}}\otimes b_{\mathpzc{h}}\right)
.
\end{align}
Crucially, this implies the torsor action modifies the fermionic symmetry group from $\mathcal{G}^{\eff} = \mathbb{Z}_{2}^{\eff} \times_\central G$ to
\begin{align}
\tor{\mathcal{G}}^{\eff} &= \mathbb{Z}_{2}^{\eff} \times_{\tor{\central}} G
,\\
\tor{\central} &= \central + \epsilon
.
\end{align}
We argue below that a general $\mathcal{G}^{\eff}$ FSPT phase can be obtained by applying the torsor functor to one of the $\mathbb{Z}_{2}^{\eff} \times G$-crossed base theories $\base_{\mathbb{Z}_2^{\eff}\times G,\rho}$ of Eq.~\eqref{eq:allyourbasearebelongtous}, which is to say it takes the form
\begin{align}
\mathcal{F}_{\cohosub{p}_{\central}, \Xdot}(\base_{\mathbb{Z}_2^{\eff}\times G,\rho})
.
\end{align}
In this way, a general $\mathcal{G}^{\eff}$ FSPT phase can be represented by the triple
\begin{align}
\label{eq:triple}
(\xdot,\mathsf{p},\maj)_{\central},
\end{align}
where we have defined $\xdot \in C^3(G,\mathbb{R}/\mathbb{Z})$ through
\begin{align}
\Xdot = e^{i 2\pi \xdot}
,
\end{align}
and used $\mathsf{p} \in C^2(G,\mathbb{Z}_{2})$ as defined in Eq.~\eqref{eq:p-to-p}, in order to better match existing conventions in the literature.
We have written $\central$ separately, because classification will be given with respect to a fixed $\mathcal{G}^{\eff}$.
The remainder of this section summarizes the results of Ref.~\onlinecite{Aasen21} in the context of FSPT phases, to justify representing a general FSPT phase by Eq.~\eqref{eq:triple}.

Tracking the modification of the defect fusion rules [Eq.~\eqref{eq:torsor-fusion}] fixes the relation between fusion multiplicities of the post-torsor functor theory to those of the pre-torsor functor theory,
\begin{align}
\tor{N}^{c_{\mathpzc{gh}}}_{a_\mathpzc{g}b_\mathpzc{h}} &= N^{\bar{\cohosub{t}}({\bf g},{\bf h}) \otimes c_\mathpzc{gh}}_{a_\mathpzc{g} b_\mathpzc{h}}.
\end{align}
The $2$-cocycle condition
\begin{align}
\label{eq:t2cocycle}
{}^{\bf g}\coho{t}({\bf h}, {\bf k}) \otimes \bar{\coho{t}}({\bf gh}, {\bf k})\otimes \coho{t}({\bf g}, {\bf hk})\otimes \bar{\coho{t}}({\bf g}, {\bf h}) = \I_\vv{0}
\end{align}
 implies that the fusion rules are associative.
The fusion/splitting spaces of $\tor{\bMTC}_G^\times$ are similarly modified as
\begin{align}
\tor{V}_{c_\mathpzc{gh}}^{a_\mathpzc{g}b_\mathpzc{h}} &\cong V_{c_\mathpzc{gh}}^{\cohosub{t}({\bf g},{\bf h}) \, a_\mathpzc{g}\, b_\mathpzc{h}} \cong V_{\bar{\cohosub{t}}({\bf g},{\bf h}) \otimes c_\mathpzc{gh}}^{a_\mathpzc{g}\,b_\mathpzc{h}}
.
\end{align}
Other valid choices for the isomorphisms between fusion spaces of $\tor{\fMTC}_G^\times$ and $\fMTC_G^\times$ can be related by a braid.

The data of the post-torsor functor theory can be calculated explicitly in terms of the data of the pre-torsor functor theory by introducing an explicit isormorphism between $\tor{V}^{a_\mathpzc{g} b_\mathpzc{h}}_{c_\mathpzc{gh}}$  and ${V}^{a_\mathpzc{g} b_\mathpzc{h}}_{c_\mathpzc{gh}}$.
Diagrammatically, we write~\cite{Bark2019b}
 \begin{align}
\label{torsorfusionspace}
\FusionSpaceHat \cong \FusionSpaceR \equiv \FusionSpaceL
,
\end{align}
where the right-most diagram introduces a shorthand to simplify the diagrams.
Diagrams on the right of $\cong$ are defined according to the pre-torsor functor theory $\bMTC_G^\times$, and we indicate the ``phantom'' lines that disappear when passing to the left-hand side of the expression using green lines.
We note that every FSPT is fusion multiplicity-free and so will not include the fusion multiplicity labels in this subsection.

Using this notation, the $F$-moves of $\tor{\fMTC}_G^\times$ are given by
\begin{align}
&\FLeftHat
=\sum_{f_{\mathpzc{hk}}} \left[ \tor{F}^{a_{\mathpzc{g}} b_{\mathpzc{h}} c_{\mathpzc{k}} }_{d_{\mathpzc{ghk}}} \right]_{e_{\mathpzc{gh}}f_{\mathpzc{hk}}}
\FRightHat
,
\end{align}
where the sum runs over all charges with the specified group label and the nontrivial elements are those obeying the new fusion rules.
We determine the $F$-symbols from the ${\fMTC}_G^\times$ data using the isomorphism of Eq.~\eqref{torsorfusionspace}.
Matching the phantom $\coho{t}$ lines on both sides of the equation requires an additional operation which we write as
\begin{align}
\label{eq:F-hat}
&\!\!\!\!\FMoveCocyL\!\!\!\!\!\!\!\!\!=\sum_{f_{\mathpzc{hk}}} \left[ \tor{F}^{a_{\mathpzc{g}} b_{\mathpzc{h}} c_{\mathpzc{k}} }_{d_{\mathpzc{ghk}}} \right]_{e_{\mathpzc{gh}},f_{\mathpzc{hk}}}\!\!\!\!\!\!
\FMoveCocyR
.
\end{align}
We call the extra operator the ``cocycleator'' -- it is an isomorphism
\begin{align}
X_{{\bf g},{\bf h}, {\bf k}} : V^{\cohosub{t}({\bf gh}, {\bf k}) , \cohosub{t}({\bf g}, {\bf h})}_{\cohosub{t}({\bf gh}, {\bf k}) \otimes \cohosub{t}({\bf g}, {\bf h})} \rightarrow V^{\cohosub{t}({\bf g}, {\bf hk}) , {}^{\bf g}\cohosub{t}({\bf h}, {\bf k})}_{\cohosub{t}({\bf g}, {\bf hk}) \otimes {}^{\bf g}\cohosub{t}({\bf h}, {\bf k})},
\end{align}
diagramatically defined by
\begin{equation}
\label{cocycleator}
\cocyleatorLold\!\!\!\!X_{{\bf g},{\bf h},{\bf k}}~~ \equiv \Xdot ( {\bf g},{\bf h},{\bf k} ) \cocyleatorR .
\end{equation}
Note that we can leave the topological charge labels on the left side implicit as they are determined by the group labels on $X$.

The complex coefficients $\Xdot ( {\bf g},{\bf h},{\bf k} ) \in C^{3}(G,\text{U}(1))$ are chosen so that $F$-symbols of the post-torsor functor theory satisfy the pentagon equation.
This defines a relative defectification obstruction to generating the theory $\tor{\fMTC}_{\tor{\mathcal{G}}^{\eff}}^\times$ from the unobstructed theory $\fMTC_{\mathcal{G}^{\eff}}^\times$.
Here, the post-torsor functor theory is unobstructed when $\Xdot$ is a solution to
\begin{align}
\label{eq:X-pw}
\cbd \Xdot =\defectO_{r}(\coho{t})^{-1}
\end{align}
where $\text{d}$ is the usual coboundary operator, i.e.,
\begin{align}
\cbd \Xdot({\bf g},{\bf h},{\bf k},{\bf l}) &= \frac{\Xdot({\bf h},{\bf k},{\bf l})\Xdot({\bf g},{\bf hk},{\bf l})\Xdot({\bf g},{\bf h},{\bf k})} {\Xdot({\bf gh},{\bf k},{\bf l}) \Xdot({\bf g},{\bf h},{\bf kl})}.
\end{align}
$\defectO_{r}(\coho{t})$ is the relative obstruction~\cite{Cui2016,Bark2019b,Aasen21}, which for FSPT phases is given by
\begin{align}
\label{eq:defectO}
&\defectO_{r}(\coho{t})({\bf g},{\bf h},{\bf k},{\bf l})
\notag \\
&\,\, = \eta_{{}^{\bf gh}\cohosub{t}({\bf k},{\bf l})}({\bf g},{\bf h})
\frac{U_{\bf g}({}^{\bf g} \coho{t}({\bf h},{\bf kl}),{}^{\bf gh}\coho{t}({\bf k},{\bf l})) }
{U_{\bf g}({}^{\bf g}\coho{t}({\bf hk},{\bf l}),{}^{\bf g}\coho{t}({\bf h},{\bf k}))}
R^{{}^{\bf gh} \cohosub{t}({\bf k},{\bf l}) \cohosub{t}({\bf g},{\bf h})}
.
\end{align}

The $\tor{F}$-symbols of the post-torsor functor theory can be written in terms of the data of the pre-torsor functor theory by evaluating the diagrams in Eq.~\eqref{eq:F-hat}.
The general expression can be found in Ref.~\onlinecite{Aasen21}; for FSPT phases it is given by
\begin{widetext}
\begin{align}
\label{eq:F-hatprime}
 \left[\tor{F}^{a_\mathpzc{g} b_\mathpzc{h} c_\mathpzc{k}} _{d_\mathpzc{ghk}}\right]_{e_\mathpzc{gh},f_\mathpzc{hk}} =&\Xdot({\bf g},{\bf h}, {\bf k})
\left[{F}^{\cohosub{t}({\bf g},{\bf h}) e'_\mathpzc{gh} c_\mathpzc{k}}_{d'_\mathpzc{ghk}}\right]_{e_\mathpzc{gh},d''_\mathpzc{ghk}}
\left[{F}^{a_\mathpzc{g} b_\mathpzc{h} c_\mathpzc{k}} _{d''_\mathpzc{ghk}}\right]_{e'_\mathpzc{gh},f'_\mathpzc{hk}}
\left( F^{\cohosub{t}({\bf gh},{\bf k} )\cohosub{t}({\bf g},{\bf h})d''_\mathpzc{ghk}}_{d_\mathpzc{ghk}} \right)^{-1}
F^{\cohosub{t}({\bf g},{\bf hk}) {}^{\bf g}\cohosub{t}({\bf h},{\bf k} )d''_\mathpzc{ghk}}_{d_\mathpzc{ghk}} \notag
\\ &\times
\left[\left( F^{{}^{\bf g}\cohosub{t}({\bf h},{\bf k})a_\mathpzc{g} f'_\mathpzc{hk}}_{d'''_\mathpzc{ghk}}  \right)^{-1}\right]_{d''_\mathpzc{ghk}, a'_\mathpzc{g}}
\left( R^{{}^{\bf g} \cohosub{t}({\bf h},{\bf k}) a_\mathpzc{g}}_{ a'_\mathpzc{g} }\right)^{-1}
\left[ F^{a_\mathpzc{g} \cohosub{t}({\bf h},{\bf k})f'_\mathpzc{hk}}_{d'''_\mathpzc{ghk}} \right]_{a'_\mathpzc{g}, f_\mathpzc{hk}}
\end{align}
where  $e'_\mathpzc{gh} = \bar{\coho{t}}({\bf g},{\bf h})  \otimes e_\mathpzc{gh}$, $f'_\mathpzc{hk} = \bar{\coho{t}}({\bf h},{\bf k}) \otimes f_\mathpzc{hk}$, $a_\mathpzc{g}' = \bar{\coho{t}}({\bf h},{\bf k}) \otimes a_\mathpzc{g} = {}^{\bf g} \bar{\coho{t}}({\bf h},{\bf k}) \otimes a_\mathpzc{g}$, $d'_\mathpzc{ghk} = \bar{\coho{t}}({\bf gh},{\bf k}) \otimes d_\mathpzc{ghk}$, $d''_\mathpzc{ghk}= \bar{\coho{t}}({\bf g},{\bf h}) \otimes \bar{\coho{t}}({\bf gh},{\bf k}) \otimes d_\mathpzc{ghk} = {}^{\bf g}\bar{\coho{t}}({\bf h},{\bf k}) \otimes \bar{\coho{t}}({\bf g},{\bf hk}) \otimes d_\mathpzc{ghk}$, and $d'''_\mathpzc{ghk}= \bar{\coho{t}}({\bf g},{\bf hk}) \otimes d_\mathpzc{ghk}$.

In order to determine the remaining topological data of the post-torsor functor theory, we need to relate the symmetry action on the topological charges of the post-torsor functor theory to that of the pre-torsor functor theory.
This is given by
\begin{align}
{}^{{\tor{\,\bf k}}}a_\mathpzc{g} &= \tor{\rho}_{\bf k} (a_\mathpzc{g}) = {}^{\bf k} \bar{\mathfrak{q}}({\bf g},\bar{\bf k}) \otimes \rho_{\bf k} (a_\mathpzc{g})
\end{align}
where
\begin{align}
\mathfrak{q}({\bf g},{\bf h}) &= \bar{\coho{t}}({\bf g},{\bf h}) \otimes \coho{t}({\bf h},\bar{\bf h}{\bf gh})
\end{align}
and $\tor{\rho}_{\bf k}(a_{\vv{x},{\bf 0}}) = \rho_{\bf k}(a_{\vv{x},{\bf 0}})$, as expected, since the torsor method changes the fractionalization class while leaving the symmetry action on the ${\bf g}={\bf 0}$ sector unchanged.
Note that $\mathfrak{q}$ can be vortex-valued when $\central({\bf g},{\bf h})\neq \central({\bf h},{\bf g})$.

Ref.~\onlinecite{Aasen21} presents diagrammatic expressions for the $R$-, $U$-, and $\eta$-symbols of the post-torsor functor theory analogous to Eq.~\eqref{eq:F-hat} for the $F$-symbols.
Here we summarize the result of evaluating these expressions for FSPT phases.
The $\tor{R}$-symbols can be written as
\begin{align}\label{eq:Rhatp}
\tor{R}^{a_\mathpzc{g} b_\mathpzc{h}}_{c_\mathpzc{gh}}
&=R^{a'_\mathpzc{g} b_\mathpzc{h}}_{c'_\mathpzc{gh}}\,
F^{\cohosub{t}({\bf g },{\bf h})\mathfrak{q}({\bf g},{\bf h}) c'_\mathpzc{gh}}_{c_\mathpzc{gh}}
\left[ \left(F^{\mathfrak{q}({\bf g},{\bf h}) a'_\mathpzc{g} b_\mathpzc{ h}}_{c''_\mathpzc{gh}}\right)^{-1} \right]_{c'_\mathpzc{gh},a_\mathpzc{g}}
\end{align}
where $a'_\mathpzc{g} = \bar{\coho{t}}({\bf g} , {\bf h}) \otimes a_\mathpzc{g}$, $c'_\mathpzc{gh} = \bar{\coho{t}}({\bf h},{\bf \bar{h}gh}) \otimes c_\mathpzc{gh}$, and $c''_\mathpzc{gh} = \bar{\coho{t}}({\bf g},{\bf h}) \otimes c_\mathpzc{gh}$.

The $\tor{U}$-symbols can be written explicitly as
\begin{align}\label{eq:Uhat}
\tor{U}_{\bf k}\left( a_\mathpzc{g}, b_\mathpzc{h}; c_\mathpzc{gh} \right)&=
\frac{\Xdot({\bf g},{\bf k}, {\bf \bar{k}hk}) }{\Xdot({\bf g},{\bf h}, {\bf k}) \Xdot({\bf k}, {\bf \bar{k}gk}, {\bf \bar{k}hk})}
{U}_{\bf k}\left( a'_\mathpzc{g}, b'_\mathpzc{h}; c'''_\mathpzc{gh} \right)
{U}_{\bf k}\left( {}^{\bf k}\coho{t}({\bf \bar{k}gk},{\bf \bar{k}hk}) , c'''_\mathpzc{gh} ; c'_\mathpzc{gh} \right)
{U}_{\bf g}\left( {}^{\bf g}\coho{t}({\bf h},{\bf k}) , {}^{\bf g}\mathfrak{q}({\bf h},{\bf k}) \right)
\notag \\
&  \times
\left( R^{{}^{\bf g} \mathfrak{q}({\bf h},{\bf k}) a_\mathpzc{g}} \right)^{-1}
\left[ \left(F^{\mathfrak{q}({\bf g},{\bf k}) a'_\mathpzc{g} b'_\mathpzc{h}}_{c''''_\mathpzc{gh}}\right)^{-1} \right]_{c'''_\mathpzc{gh}, a_\mathpzc{g}}
\left[ \left(F^{{}^{\bf g}\mathfrak{q}({\bf h},{\bf k}) a_\mathpzc{g} b'_\mathpzc{h}}_{c''_\mathpzc{gh}}\right)^{-1} \right]_{c''''_\mathpzc{gh}, [\mathfrak{q}({\bf h},{\bf k})a_\mathpzc{g}] }
\left[ F^{a_\mathpzc{g} \mathfrak{q}({\bf h},{\bf k}) b'_\mathpzc{h}}_{c''_\mathpzc{gh}} \right]_{[\mathfrak{q}({\bf h},{\bf k})a_\mathpzc{g}],b_\mathpzc{h}}
\notag \\
&  \times
\frac{F^{[\cohosub{t}({\bf gk},{\bf \bar{k}hk})  \cohosub{t}({\bf g},{\bf k})] \mathfrak{q}({\bf g},{\bf k})  c'''_\mathpzc{gh}  }
F^{[{}^{\bf g}\cohosub{t}({\bf h},{\bf k}) \cohosub{t}({\bf g},{\bf hk}) ] {}^{\bf g}\mathfrak{q}({\bf h},{\bf k})   c''''_\mathpzc{gh}  }
F^{\cohosub{t}({\bf gh},{\bf k}) \cohosub{t}({\bf g},{\bf h})  c''_\mathpzc{gh}  }
}
{F^{\cohosub{t}({\bf k},{\bf \bar{k}ghk}) {}^{\bf k}\cohosub{t}({\bf \bar{k}gk},{\bf \bar{k}hk})  c'''_\mathpzc{gh}  }
F^{\cohosub{t}({\bf gh},{\bf k}) \mathfrak{q}({\bf gh},{\bf k})  c'_\mathpzc{gh}  }
}
\end{align}
where $a'_\mathpzc{g} = \bar{\mathfrak{q}}({\bf g},{\bf k}) \otimes a_\mathpzc{g}$, $b'_\mathpzc{h} = \bar{\mathfrak{q}}({\bf h},{\bf k}) \otimes b_\mathpzc{h}$, $c'_\mathpzc{gh} = \bar{\mathfrak{q}}({\bf gh},{\bf k}) \otimes c_\mathpzc{gh}$, $c''_\mathpzc{gh} = \bar{\coho{t}}({\bf g},{\bf h}) \otimes c_\mathpzc{gh}$, $c'''_\mathpzc{gh} = {}^{\bf k}\bar{\coho{t}}({\bf \bar{k}gk},{\bf \bar{k}hk}) \otimes c'_\mathpzc{gh}$, and $c''''_\mathpzc{gh} = \mathfrak{q}({\bf g},{\bf k}) \otimes c'''_\mathpzc{gh}$.

The explicit expression for the $\tor{\eta}$-symbols is
\begin{align}
\label{eq:etahat}
 \tor{\eta}_{a_\mathpzc{k}} \left( {\bf g}, {\bf h} \right) =&
\frac{\Xdot({\bf g},{\bf \bar{g}kg}, {\bf h}) }{\Xdot({\bf g},{\bf h}, {\bf \bar{h}\bar{g}kgh}) \Xdot({\bf k}, {\bf g}, {\bf h})}
 \eta_{a'_\mathpzc{k}} \left( {\bf g}, {\bf h} \right)
\frac{{U}_{\bf g}\left( {}^{\bf g}\mathfrak{q}({\bf \bar{g}kg},{\bf h}) , a'_\mathpzc{k} \right)}{{U}_{\bf g}\left( {}^{\bf g}\coho{t}({\bf \bar{g}kg},{\bf h}) , {}^{\bf g}\mathfrak{q}({\bf \bar{g}kg},{\bf h}) \right)}
R^{{}^{\bf k}\cohosub{t}({\bf g},{\bf h}) a_\mathpzc{k}}R^{ a'_\mathpzc{k} \cohosub{t}({\bf g},{\bf h})}
\notag \\
& \qquad \times
\frac{
F^{\cohosub{t}({\bf gh},{\bf \bar{h}\bar{g}kgh}) \cohosub{t}({\bf g},{\bf h}) a'_\mathpzc{k} }
F^{\cohosub{t}({\bf k},{\bf gh}) \mathfrak{q}({\bf k},{\bf gh})  a'''_\mathpzc{k}  }
}{F^{\cohosub{t}({\bf k},{\bf gh}) {}^{\bf k}\cohosub{t}({\bf g},{\bf h}) a_\mathpzc{k} }
F^{[\cohosub{t}({\bf kg},{\bf h})  \cohosub{t}({\bf k},{\bf g})] \mathfrak{q}({\bf k},{\bf g})  a''_\mathpzc{k}  }
F^{[\cohosub{t}({\bf g},{\bf \bar{g}kgh}) {}^{\bf g}\cohosub{t}({\bf \bar{g}kg},{\bf h}) ] {}^{\bf g}\mathfrak{q}({\bf \bar{g}kg},{\bf h})   a'_\mathpzc{k}  }
F^{\mathfrak{q}({\bf k},{\bf gh}) a'_\mathpzc{k} \cohosub{t}({\bf g},{\bf h}) }
}
\end{align}
\end{widetext}
where $a'_\mathpzc{k} = \bar{\mathfrak{q}}({\bf k},{\bf gh}) \otimes a_\mathpzc{k}$, $a''_\mathpzc{k} = \bar{\mathfrak{q}}({\bf k},{\bf g}) \otimes a_\mathpzc{k}$, and $a'''_\mathpzc{k} = {\coho{t}}({\bf g},{\bf h}) \otimes a'_\mathpzc{k}$.
We point out that,
\begin{align}
\label{eq:etatrivialsector}
\tor{\eta}_{a_{\vv{x},{\bf 0}}}({\bf g,h})  = \eta_{a_{\vv{x},{\bf 0}}}({\bf g,h})M_{a_{\vv{x},{\bf 0}} \cohosub{t}({\bf g,h})}.
\end{align}

Note that $\Xdot$ appears in Eqs.~(\ref{eq:F-hatprime})-(\ref{eq:etahat}) in precisely the same way as the $3$-cocycle $\alpha \in Z^3(G,\text{U}(1))$ in the topological data of a bosonic SPT phase $\spt_G^{[\alpha]}$ in Eqs.~(\ref{eq:F-alpha})-(\ref{eq:eta-alpha}).
Therefore, it follows that shifting $\Xdot \to \alpha \Xdot $ is equivalent to gluing in a bosonic SPT phase
\begin{align}
\label{eq:bSPT-torsor}
\mathcal{F}_{\cohosub{t}, \alpha \Xdot} = \spt_G^{[\alpha]} \underset{G}{\boxtimes}\mathcal{F}_{\cohosub{t}, \Xdot},
\end{align}
where $\alpha \Xdot$ denotes multiplication in $C^{3}(G,\text{U}(1))$
\begin{align}\label{eq:X-alpha}
\alpha \Xdot ({\bf g},{\bf h},{\bf k}) = \alpha({\bf g},{\bf h},{\bf k}) \Xdot({\bf g},{\bf h},{\bf k}).
\end{align}

Finally, we extract a minimal set of data needed to specify a FSPT phase.
We first pick a symmetry action $\rho: G \to \on{Aut}^{\eff}(\ifo^{(0)}) = \mathbb{Z}_2^{\V}$.
The symmetry action is determined by a group homomorphism
\begin{align}
\pi \in H^1(G,\mathbb{Z}_2^{\V}).
\end{align}
The symmetry action on objects in $\ifo^{(0)}$ is given by $\rho_{\bf g}(a_{\vv{x}}) = \psi^{\vv{x} \cdot \pi({\bf g})}\otimes a_{\vv{x}}$.
Thus, the choice of symmetry action uniquely fixes the base theory $\base_{\mathbb{Z}_2^{\eff}\times G,\rho}$ to which we apply the torsor functor.
Next, we specify the vortex symmetry fractionalization class relative to the base theory through the 2-cocycle $\coho{t} \in Z^2_{[\rho]}(G,\mathcal{A})$.
Writing $\coho{t}= \coho{p}_{\central}$ the cocycle condition for $\coho{t}$ presented in Eq.~\eqref{eq:t2cocycle} becomes
\begin{align}
\cbd \coho{p} = \psi^{\pi \cup \central},
\end{align}
with $\coho{p} \in C^2(G,\mathbb{Z}_2^{\psi})$.
This is the simplest example of a $[\coho{O}^{\central}]$ obstruction presented in Eq.~\eqref{eq:O_central_general} and computed for the present case in Eq.~\eqref{eq:Knu_even_extension_obstruction}.
Lastly, we specify the defectification class with
\begin{align}
\cbd \Xdot = \defectO_r(\coho{p}_{\central})^{-1}.
\end{align}
The classification of $G$-crossed MTCs guarantees the torsor method generates the topological data of every FSPT phase.
It follows that the most general $\mathcal{G}^{\eff}$ FSPT phase $\mathcal{F}_{\cohosub{p}_{\central},\Xdot}(\base_{\mathbb{Z}_2^{\eff}\times G,\rho})$ can be parameterized by the triple
\begin{align}
(\xdot, \mathsf{p}, \maj)_{\central}
.
\end{align}
In the next section, we will generate the group law of $\mathcal{G}^{\eff}$ FSPT phases for fixed $\central$ using general properties of the torsor method.
Before doing so, we account for the relabeling redundancy in the triples mentioned above.

\subsubsection{Equivalence relations}
\label{sec:equivalence_relations}

Na\"ively distinct $\mathcal{G}^{\eff}$-crossed FSPT theories may be equivalent under certain conditions.
For example, we saw that any two bosonic SPT phases whose topological data are related by a 3-coboundary are equivalent.
This produces the equivalence relation
\begin{align}
\label{eq:3coboundaryFSPequiv}
(\xdot,\mathsf{p},\maj)_{\central} \sim(\xdot + \cbd \bdot ,\mathsf{p},\maj)_{\central}
\end{align}
where $\bdot \in C^2(G,\mathbb{R}/\mathbb{Z})$.
More generally, the cohomological classification of $G$-crossed theories can be reduced due to relabelings, as discussed in Sec.~\ref{sec:fermionic-defectification}.
As the torsor method presented above utilizes this classification, not all of the FSPT theories generated by the formulas of the previous sections are necessarily distinct.
Rather, some of these theories can be identified using generalized versions of the relabeling equivalences encountered for $\mathcal{G}^{\eff}=\mathbb{Z}_4^{\eff}$ FSPT phases in Sec.~\ref{sec:z4fclassification}.

Relabeling topological charges within a fixed $\mathpzc{g}$-sector will always result in a physically equivalent theory representing the same FSPT phase, possibly with a distinct cataloging of topological data.
The goal here is to understand how the triple of cochains $(\xdot, \mathsf{p},\maj)_{\central}$, which label a representative of every FSPT phase, are identified under such relabelings.
Consequently, we only need to consider the relabelings which preserve the set of admissible cochain triples.
Within each $\mathpzc{g}$-sector the relabeling can either be trivial (sending $a_{\mathpzc{g}} \to a_{\mathpzc{g}}$), or nontrivial (sending $a_{\mathpzc{g}} \to \psi_{\mathpzc{0}} \otimes a_{\mathpzc{g}}$).
Thus, we recognize a relabeling as a $\psi_{\mathpzc{0}}$-valued 1-cochain $\coho{z}\in C^1(\mathcal{G}^{\eff}, \mathbb{Z}_2^{\psi})$, sending $a_{\mathpzc{g}} \to [ \coho{z}(\mathpzc{g}) a]_{\mathpzc{g}}$.
We require that the relabeling does not interchange the vacuum and physical fermion, implying the 1-cochain is normalized.
Moreover, we require the relabeling to preserve the set of cochain triples, and so we only consider the subset
\begin{align}
\label{eq:relabeling}
\coho{z} \in C^1(G,\mathbb{Z}_2^{\psi}) \quad \text{and} \quad
\coho{z}' \in C^1(\mathbb{Z}_2^{\eff}, \mathbb{Z}_2^{\psi}).
\end{align}
Such normalized 1-cochains can be lifted to normalized $C^1(\mathcal{G}^{\eff}, \mathbb{Z}_2^{\psi})$ through the maps ${\bf g} \mapsto (\vv{x},{\bf g})$ and $\vv{x} \mapsto (\vv{x},{\bf g})$, respectively.
Both types of relabelings can be physically motivated.
The relabeling $\coho{z}$ is the standard relabeling we expect from the vortex symmetry fractionalization classification discussed in Sec.~\ref{sec:symmetric-fermions}.
The relabeling $\coho{z}'$ can be understood as saying that it should not matter which vortex in $\ifo^{(0)}$ we label as $\I_{\vv{1}}$.
The goal now is to develop the equivalence relations which appear in the (generalized) cohomology theory that classify FSPT phases, which we discuss in Sec.~\ref{sec:fSPTClassification}.
For a fixed $\mathcal{G}^{\eff}$, since the FSPT phases are labeled by triples $(\xdot,\mathsf{p},\maj)_{\central}$, this amounts to saying when two triples are related by a relabeling of the corresponding theory, up to vertex basis and fermionic symmetry action gauge transformations.

The first type of relabeling sends
\begin{align}
a_{\vv{x},{\bf g}} \to [\coho{z}({\bf g}) a]_{\vv{x},{\bf g}}
\end{align}
and can modify the symmetry fractionalization and defectification classes by a coboundary.
In Ref.~\onlinecite{Aasen21}, it was shown that such a relabeling is equivalent to applying a torsor functor $\mathcal{F}_{\cbd \cohosub{z},\mathscr{Z}_{\cohosub{z}}}$ with
\begin{align}
\mathscr{Z}_{\cohosub{z}}({\bf g,h,k}) =
 \eta_{{\cohosub{z}}({\bf k})}({\bf g},{\bf h})
R^{ \cbd \cohosub{z}({\bf h},{\bf k}) , {\cohosub{z}}({\bf g})}
\end{align}
where we have used that $\coho{z}$ is valued in $\mathbb{Z}_2^{\psi}$.
This can be thought of as either a $G$-crossed or $\mathcal{G}$-crossed torsor functor, in this context, as it always maps a $G$-crossed BTC to a $G$-crossed BTC.
Writing $\coho{z}({\bf g})  = \psi_{\mathpzc{0}}^{\mathsf{z}({\bf g})} $, we find
\begin{align}
\mathscr{Z}_{\cohosub{z}} = (-1)^{\central \cup \mathsf{z} + \mathsf{z} \cup \cbd \mathsf{z} }
.
\end{align}
We can also pull the relabeling equivalence into the functor, yielding an equivalence of functors under $\coho{z}({\bf g})$ relabelings given by
\begin{align}
\mathcal{F}_{\cbd \cohosub{z},\mathscr{Z}_{\cohosub{z}}} \circ \mathcal{F}_{\cohosub{p}_{\central},\Xdot}  =
\mathcal{F}_{\cohosub{p}_{\central} \cbd \cohosub{z},(-1)^{\mathsf{p} \cup_1 \cbd \mathsf{z} +\pi \cup(\central \cup_2 \cbd \mathsf{z})}\mathscr{Z}_{\cohosub{z}} \Xdot}
.
\end{align}
We emphasize that the fractionalization class stays the same (only changes by a coboundary) under this relabeling, and only the defectification class has potential to change.
We can now use this relation to indicate when two theories are equivalent, that is
\begin{widetext}
\begin{align}
& (\xdot,\mathsf{p},\maj)_{\central} \label{eq:first-relab}
\sim (\xdot + \frac{1}{2}( \mathsf{p} \cup_1 \cbd \mathsf{z}  + \central \cup \mathsf{z} + \pi \cup(\central \cup_2 \cbd \mathsf{z})+ \mathsf{z} \cup \cbd \mathsf{z} )  ,\mathsf{p}_{}+ \cbd \mathsf{z},\maj)_{\central}.
\end{align}
\end{widetext}
We remark that even if $\cbd \coho{z} = \I_{\vv{0}}$, the relabeling can result in a nontrivial modification of the defectification class, given by the 3-cocycle $\frac{1}{2} \central \cup \mathsf{z}$.
Notice we can account for the asymmetry by using that $[\central \cup \mathsf{z}] = [ \mathsf{z} \cup \central]$ since $\central \cup \mathsf{z} = \mathsf{z} \cup \central + \cbd(\central \cup_1 \mathsf{z} )$.
In particular, such a relabeling is only nontrivial when $\mathcal{G}^{\eff}$ is a nontrivial $\mathbb{Z}_2^{\eff}$ extension of $G$.

The second type of relabeling sends
\begin{align}
a_{\vv{x},{\bf g}} \to  [\coho{z}'(\vv{x}) a]_{\vv{x},{\bf g}}.
\end{align}
There are only two options for $\coho{z}'$, either it is trivial, or $\coho{z}'(\vv{1}) = \psi_{\mathpzc{0}}$.
This relabeling is related to the fact that every FMTC has the vortex permuting symmetry $a_{{\vv{x}}} \to \psi^{\vv{x}}_{\vv{0}} a_{\vv{x}}$.
Again, such a relabeling is equivalent to applying a torsor functor $\mathcal{F}_{\cbd \cohosub{z}',\mathscr{Z}_{\cohosub{z}'}}$ with $\coho{z}'(\mathpzc{g}) = \psi_{\mathpzc{0}}^{\vv{x}}$.
In this case, it is necessarily a $\mathcal{G}$-crossed torsor functor and it does not map a $G$-crossed FMTC back to a $G$-crossed FMTC (the vortices no longer exhibit ordinary braiding).
However, we can follow $\mathcal{F}_{\cbd \cohosub{z}',\mathscr{Z}_{\cohosub{z}'}}$ with a vertex basis gauge transformation given by $\Gamma^{a_{\vv{x},{\bf g}}, b_{\vv{y},{\bf h}}}_{c_{\vv{x+y}+\central({\bf g,h}),{\bf gh}}}  = (-1)^{\vv{x} \cdot \vv{y}}$, and the combination returns a $G$-crossed FMTC, and acts as a $G$-crossed functor.
In more detail, we have
\begin{align}
\cbd \coho{z}' (\mathpzc{g},\mathpzc{h}) &= \central ({\bf g},{\bf h}) , \\
\mathscr{Z}_{\cohosub{z}'}(\mathpzc{g},\mathpzc{h},\mathpzc{k}) &= (-1)^{\central({\bf g},{\bf h}) \cdot \vv{z} + \vv{x} \cdot \central({\bf h},{\bf k})}
,
\end{align}
and find the equivalence of functors
\begin{align}
\Gamma \circ \mathcal{F}_{\cbd \cohosub{z}',\mathscr{Z}_{\cohosub{z}'}} \circ \mathcal{F}_{\cohosub{p}_{\central},\Xdot}  &=
\mathcal{F}_{[\cohosub{p}+ \central]_{\central},(-1)^{\mathsf{p} \cup_1 \central + \pi \cup \central} \, \Xdot}
.
\end{align}
In this case, the relabeling changes the fractionalization class for nontrivial $[\central]$.
The resulting equivalence of theories under this second type of relabeling is therefore given by
\begin{align}
\label{eq:Gfequiv}
(\xdot,\mathsf{p}_{}, \maj)_{\central} \sim (\xdot + \frac{1}{2}(\mathsf{p} \cup_1 \central + \maj \cup \central) ,\mathsf{p}_{}+{\central}, \maj )_{\central} .
\end{align}
Notice that this equivalence relation can be obtained from the first type of relabeling equivalence relation [together with Eq.~\eqref{eq:3coboundaryFSPequiv}] when $[\central]=[0]$, and is totally trivial when $\central=0$.

Gathering these results, we can denote equivalence classes of $\mathcal{G}^{\eff} =\mathbb{Z}_2^{\eff}\times_\central G$ FSPT phases
\begin{align}\label{eq:equiv-triple}
[\xdot,\mathsf{p},\maj]_{\central}
\end{align}
defined by the relations in Eqs.~\eqref{eq:3coboundaryFSPequiv}, \eqref{eq:first-relab}, and \eqref{eq:Gfequiv}.
In the following section, we classify $[\xdot,\mathsf{p},\maj]_{\central}$ under fermionic stacking.
This classification can be interpreted as defining a generalized cohomology~\cite{Gaiotto2019}.

\section{Classification of Fermionic Symmetry Enriched Topological Phases}
\label{sec:classification}

We present the $\mathcal{G}^{\eff}$-crossed classification of FSET phases.
We first derive the group $\Gr$ formed by FSPT phases under fermionic stacking, making contact with previous results obtained using alternate formalisms.
We then explain how this group fits into the multi-stage torsorial classification of FSET phases.
We emphasize similarities and differences to the bosonic case, elucidating how the additional obstructions and associated torsors appearing for symmetric fermionic topological phases (Table~\ref{table:obstructions}) can be absorbed into the richer structure of FSPT phases compared to their bosonic counterparts.

\subsection{Classification of FSPT Phases}
\label{sec:fSPTClassification}

\begin{table}
\begin{center}
 \begin{tabular}{l l}
 Group  &  Interpretation \\ [0.5ex]
 \hline
 &\\[-0.5ex]
 $
\Gr$ &
\begin{minipage}[c]{.8\linewidth}\raggedright
FSPT phases. See Eq.~\eqref{eq:general-group} and Eq.~\eqref{eq:GFSPTgrouplaw}.
\end{minipage}
\\[4ex]
$\GSPT$ &
\begin{minipage}[c]{.8\linewidth}\raggedright
Bosonic SPT phases not trivialized by a FSPT. See Eq.~\eqref{eq:GSPT}.\end{minipage}
\\[4ex]
$\Gtriv$ &
\begin{minipage}[c]{.8\linewidth}\raggedright
FSPT phases with trivial symmetry action. See Eq.~\eqref{eq:afsptprod}.\end{minipage}
\\[4ex]
$\Grt$ &
\begin{minipage}[c]{.8\linewidth}\raggedright
Symmetry fractionalization with trivial symmetry action. Equal to $\Gtriv/\GSPT$. See Eq.~\eqref{eq:grt}.\end{minipage}
\\[4ex]
$\Gro$ &
\begin{minipage}[c]{.8\linewidth}\raggedright
Symmetry actions which admit symmetry fractionalization and defectifcation. Equal to ${\Gr}/{\Gtriv}= \Grto/\Grt$. See Eq.~\eqref{eq:gro}.
\end{minipage}
\\[6ex]
$\Grto $ &
\begin{minipage}[c]{.8\linewidth}\raggedright
Symmetry fractionalizations which admit defectifications. Equal to ${\Gr}/{\GSPT}$. See Eq.~\eqref{eq:grta}.
\end{minipage}
\end{tabular}
\caption{Classifying groups appearing in Sec.~\ref{sec:fSPTClassification} for $\mathcal{G}^{\eff} = \mathbb{Z}_{2}^{\eff} \times_{\central} G$ FSPT phases.
The superscript indices on $\mathsf{G}$ indicate which cochains are used to describe the corresponding group elements.
Each group is a function of the fermionic symmetry group $\mathcal{G}^{\eff}$, so depends on the specified $G$ and $\central$.
}
\label{table:fsptgroups}
\end{center}
\end{table}

In this section, we show that fermionic stacking of FSPT phases is an Abelian operation that results in a group structure on the classification of fermionic theories analogous to (yet more complicated than) the $H^3(G,\text{U}(1))$ torsorial structure of bosonic theories under stacking bosonic SPT phases.
Using the parameterization of $\mathcal{G}^{\eff}=\mathbb{Z}_2^{\eff}\times_{\central}G$ FSPT phases by equivalence classes of the triples $[\xdot,\mathsf{p},\maj]_{\central}$ defined in Eq.~\eqref{eq:equiv-triple}, we derive the fermionic stacking relation
\begin{align}
\label{eq:general-group}
&[{\xdot},{\mathsf{p}}, {\maj}]_{\central} = [\xdot',\mathsf{p}_{}',\maj']_{\central}  \underset{\mathcal{G}^{\eff}}{\ftimes} [\xdot'',\mathsf{p}_{}'',\maj'']_{\central} =
\\
&\, [\xdot'+\xdot''+\ydot(\mathsf{p}',\maj';\mathsf{p}'',\maj''), \mathsf{p}'+\mathsf{p}''+{\maj' \cup \maj''}, \maj' +\maj'']_{\central}
\notag
\end{align}
for some $3$-cochain $\ydot \in C^3(G,\mathbb{R}/\mathbb{Z})$ that is subject to a consistency condition reported in Eq.~\eqref{eq:general-Y}.

The value of $\ydot$ can be determined uniquely, up to equivalences, by performing the condensation calculation in the stacking operation.
We compute $\ydot$ explicitly this way in several cases.
However, we leave the general calculation for future work (or ambitious readers), as it is laborious, and finding the gauge transformation to relate the stacked theory back to the image of the torsor functor is particularly challenging.
We use the condensation calculation to show that $\ydot$ generally only depends on $\mathsf{p}',\maj', \mathsf{p}''$ and $\maj''$, and is subject to the consistency condition of Eq.~\eqref{eq:general-Y}.
In Ref.~\onlinecite{Brumfiel2018a}, a 3-cochain $\ydot$ was reported that satisfies Eq.~\eqref{eq:general-Y} when $\central = 0$.
In Ref.~\onlinecite{Bark2021inv}, a 3-cochain $\ydot$ satisfying Eq.~\eqref{eq:general-Y} was reported for nontrivial $\central$.
In Appendix~\ref{app:Y-details}, we reproduce these calculations in the notations and conventions used in our paper.
Restricting the resulting expression to FSPTs (i.e. setting $\nu' = \nu'' = \nu = 0$) yields
\begin{align}
\label{eq:lambdaeqmaiantext}
\lambda &= \frac{1}{4}\left(\pi ' \cup \pi '' \cup \pi'' + (\pi ' \cup_1 \pi'' )\cup \central \right)
\notag \\
&+\frac{1}{2}\left(
 \pi' \cup(\pi' \cup_1 \pi'')\cup \pi''
+ (\pi' \cup \pi'')  \cup_1 (\mathsf{p}' + \mathsf{p}'') \right.
\notag \\
&\qquad \qquad \left. + \mathsf{p}' \cup_1 \mathsf{p}''
+(\cbd \mathsf{p}' )\cup_2 \mathsf{p}'' \right)
\notag \\
&+ \frac{1}{2}\left( f(\mathsf{p}',\pi') +f(\mathsf{p}'',\pi'') - f(\mathsf{p},\pi) \right)
,
\end{align}
where we have defined
\begin{align}
\label{eq:eqnforf}
f(\mathsf{p},\pi)  &= \pi \cup (\mathsf{p}\cup_2 \central) + \mathsf{p} \cup_2 \cbd \mathsf{p}
.
\end{align}
We emphasize that satisfying Eq.~\eqref{eq:general-Y} only determines $\ydot$ up to a 3-cocycle, which should be some universal function of $(\mathsf{p}', \maj')$, $(\mathsf{p}'', \maj'')$ and $\central$ that is symmetric in $(\mathsf{p}', \maj')$ and $(\mathsf{p}'', \maj'')$ up to 3-coboundaries.
Ref.~\onlinecite{Bark2021inv} conjectured that this expression for $\ydot$ has no 3-cocycle correction.

We show the triples $[{\xdot},{\mathsf{p}}, {\maj}]_{\central}$ form the group of FSPT phases $\Gr$, with group operation Eq.~\eqref{eq:general-group}.
We show the group $\Gr$ takes a universal form given by,
\begin{align}
\label{eq:GFSPTgrouplaw}
\Gr \cong \GSPT \times_{\varepsilon_{\mathsf{(3,21)}}} \left( \Grt  \times_{\varepsilon_{\mathsf{(2,1)}}} \Gro  \right)
\end{align}
where $\GSPT$, $\Grt$, and $\Gro$ are various classifying groups which can be used to understand $\Gr$, see Table~\ref{table:fsptgroups} for their interpretations.
We determine the central extension $\varepsilon_{\mathsf{(3,21)}}$ in several cases and $\varepsilon_{\mathsf{(2,1)}}$ in all cases using condensation.

We first discuss general properties of $[\xdot,\mathsf{p},\maj]_{\central}$.
In Sec.~\ref{sec:group-law}, we assume Eq.~\eqref{eq:general-group} and prove that it defines a group multiplication for the group $\Gr$ of $\mathcal{G}^{\eff}$ FSPT phases.
The following Secs.~\ref{sec:subgroup-3}-\ref{sec:generalFSPTphases} build up to Eq.~\eqref{eq:general-group} by considering a nested structure of subgroups.
We summarize the classifying groups appearing in this section in Table~\ref{table:fsptgroups} in the order in which they are encountered.
The notation we use is meant to convey the relevant $n$-cochains participating in the specified group.
For example, elements of $\Gr$ are labeled by $3$-, $2$-, and $1$-cochains, indicated by the superscript $\mathsf{(3,2,1)}$.
We will use the same convention throughout this section, so that, for example if only $2$- and $1$-cochains participate in a particular group, we will denote the group by $\Grto$.
The simplest subgroup $\GSPT$ corresponds to the equivalence classes $[\xdot,0,0]_{\central}$ of the FSPT phases that differ from the identity $[0,0,0]_{\central}$ by gluing in a bosonic SPT phase.
These theories form a subgroup of $\Gtriv$, the group of FSPT phases $[\xdot,\mathsf{p},0]_{\central}$ with trivial symmetry action, which in turn forms a subgroup of $\Gr$.
As fermionic stacking is commutative by definition, all subgroups are Abelian and
\begin{align}
\GSPT \lhd \Gtriv \lhd \Gr.
\end{align}
At each step, we define the group under consideration by specifying its elements and group multiplication, and relate it to the group from the previous step of the calculation.

When deriving the group multiplication of $\mathcal{G}^{\eff}$ FSPT phases, we need to fix the $2$-cocycle $\central \in Z^2(G,\mathbb{Z}_2^{\eff})$ defining $\mathcal{G}^{\eff}$, rather than just the cohomology class $[\central] \in H^2(G,\mathbb{Z}_2^{\eff})$.
As noted in Sec.~\ref{sec:centralisomorphism}, while $2$-cocycles $\central$ and $\central'$ related by a coboundary define isomorphic groups $\mathcal{G}^{\eff}\cong \mathcal{G}^{\eff'}$, extending the isomorphism to the $\mathcal{G}^{\eff}$-crossed SMTCs can result in physically inequivalent theories.
It is therefore important to work with fixed $\central$, for example, when demonstrating the existence of an identity element in the classifying group.

Next, we collect several properties of the triple of cochains specifying a $\mathcal{G}^{\eff}$ FSPT phase.
The previous section demonstrated that every $\mathcal{G}^{\eff}$ FSPT phase can be parameterized by a triple of cochains
\begin{align}
(\xdot, \mathsf{p},\maj)_{\central} \in C^3(G,\mathbb{R}/\mathbb{Z}) \times C_{}^2(G,\mathbb{Z}_{2}^{\psi}) \times Z^1(G,\mathbb{Z}_{2}^{\V}),
\end{align}
where
\begin{align}
\rho = \V^{\maj} &\in Z^1(G,\mathbb{Z}_2^{\V}),
\\
\coho{p}_{\central} = \psi^{\mathsf{p}}_{\vv{0}}\otimes \I_{\vv{1}}^{\central} &\in Z^2_{[\rho]}(G , \mathbb{Z}_{2}^{\psi}\times \mathbb{Z}_{2}^{\eff}),
\\
\Xdot = e^{i 2\pi \xdot} &\in C^3(G,\text{U}(1))
.
\end{align}
Written in this way, the group structure of the triple of cochains is additive.
The cochains are subject to the conditions
\begin{align}
\label{constraint_pi}
\cbd\maj &= 0,
\\
\label{constraint_p}
\cbd\mathsf{p} &= {\maj\cup\central},
\\
\label{constraint_X}
\cbd\Xdot &= \defectO_r(\psi^{\mathsf{p}}_{\vv{0}}\otimes \I_{\vv{1}}^{\central},\pi)^{-1}.
\end{align}
Here, we have explicitly included the symmetry action dependence in $\defectO_r$, as we will use combinations of $\defectO_r$ that involve distinct symmetry actions.
Recall that this data specifies the $\mathcal{G}^{\eff}$-crossed SMTC relative to a base theory $\base_{\mathbb{Z}_2^{\eff}\times G,\rho} = (0,0,\pi)_{\vv{0}}$ through an application of the torsor functor
\begin{align}
(\xdot, \mathsf{p},\maj)_{\central} = \mathcal{F}_{\cohosub{p}_{\central}, \Xdot}(\base_{\mathbb{Z}_2^{\eff}\times G,\rho})
,
\end{align}
using the torsor method of Ref.~\onlinecite{Aasen21} (reviewed in Sec.~\ref{sec:torsor-method}) with the relative vortex symmetry fractionalization class $\coho{p}_{\central}$ and relative defectficiation class $\Xdot$.
Explicitly evaluating $\defectO_r(\psi^{\mathsf{p}}_{\vv{0}} \otimes \I_{\vv{1}}^\central,\pi)$, we find
\begin{align}
\label{eq:generaldefectoFSPT}
&\defectO_r(\psi^{\mathsf{p}}_{\vv{0}} \otimes \I_{\vv{1}}^\central,\pi)
\notag \\
& \qquad  = i^{\pi \cup \pi \cup \central} (-1)^{
\mathsf{p} \cup \mathsf{p}
+\mathsf{p} \cup \central
+\pi \cup(\mathsf{p} \cup_1 \central)
+ (\pi \cup_1 \mathsf{p}) \cup \central
}
.
\end{align}
In the additive notation Eq.~\eqref{constraint_X} becomes,
\begin{align}
\cbd \xdot &= -\frac{1}{4}\widetilde{\pi} \cup \widetilde{\pi} \cup \widetilde{\central} + \frac12(\mathsf{p} \cup \mathsf{p}+\mathsf{p} \cup \central) \notag \\
&\quad+\frac12(\pi \cup(\mathsf{p} \cup_1 \central)
+ (\pi \cup_1 \mathsf{p}) \cup \central) .
\end{align}

These triples of cochains labeling FSPT theories can be collected into equivalence classes
\begin{align}
[\xdot,\mathsf{p},\maj]_{\central}
\end{align}
defined by the equivalence relations of  Sec.~\ref{sec:equivalence_relations}:
\begin{widetext}
\begin{align}\label{eq:equiv1}
(\xdot,\mathsf{p}_{},\maj)_{\central} &\sim
(\xdot+\cbd \bdot ,\mathsf{p}_{},\maj)_{\central}
\\
&  \sim(\xdot + \frac{1}{2}( \mathsf{p} \cup_1 \cbd \mathsf{z}  + \central \cup \mathsf{z} + \pi \cup(\central \cup_2 \cbd \mathsf{z})+ \mathsf{z} \cup \cbd \mathsf{z} )  ,\mathsf{p}_{}+ \cbd \mathsf{z},\maj)_{\central}
\label{eq:equiv2}
\\
& \sim (\xdot + \frac{1}{2}(\mathsf{p} \cup_1 \central + \maj \cup \central) ,\mathsf{p}_{}+{\central}, \maj )_{\central}
\label{eq:equiv3}
\end{align}
\end{widetext}
for $\mathsf{z} \in C^1(G,\mathbb{Z}_{2}^{\psi})$ and $\bdot \in C^2(G,\mathbb{R}/\mathbb{Z})$.
Note that there are no equivalence relations on $\maj$, as changing the symmetry action would be physically detectable.

\subsubsection{FSPT phases form a group via fermionic stacking}
\label{sec:group-law}

We first assume that Eq.~\eqref{eq:general-group} is true and use general properties of condensation (reviewed in Appendix~\ref{app:condensation}) to show that fermionic stacking is a group operation on the equivalence classes of FSPT theories with fixed $\central$.
For this, we need to show that the operation maps FSPT theories back to FSPT theories, that the operation is associative, that there is an identity element, and that each element has an inverse.
Associativity of fermionic stacking follows from the fact that there is a unique way to identify physical fermions when multiple theories are stacked, as discussed in Sec.~\ref{sec:stacking}; we relegate the more technical details of this to Appendix~\ref{app:associativity}.
By definition, stacking is also commutative.
Since $\ifo^{(0)} \underset{}{\ftimes} \ifo^{(0)}  = \ifo^{(0)}  $, we are guaranteed that $[\xdot', \mathsf{p}', \maj']_{\central}\underset{\mathcal{G}^{\eff}}{\ftimes}[\xdot'', \mathsf{p}'', \maj'']_{\central}$ is also a FSPT, and therefore the stacked theory can be written as a torsor functor applied to one of the FSPT base theories.
Before explicitly computing the torsor functor in several cases in the subsequent subsections, we show that Eq.~\eqref{eq:general-group} satisfies all the necessary conditions for the stacked theory.

First, we show that the triples of cochains on the right hand side of Eq.~\eqref{eq:general-group} satisfy the conditions of FSPT triples when the triples of cochains on the left hand side do.
Eq.~\eqref{constraint_pi} is immediate, since $\cbd (\pi' + \pi'') = \cbd \pi' + \cbd \pi'' =0$.
Eq.~\eqref{constraint_p} is shown by
\begin{align}
\cbd (\mathsf{p}'+\mathsf{p}''+{\pi' \cup \pi''})= \cbd \mathsf{p}'+ \cbd \mathsf{p}'' = (\pi' +\pi'')\cup \central
,
\end{align}
which uses the fact that $\cbd (\pi' \cup \pi'') =0$.
Eq.~\eqref{constraint_X}, imposes a consistency condition on the $3$-cochain $\ydot$.
As with $\Xdot$, we define both an additive and multiplicative version of $\ydot$, related by
\begin{align}
\Ydot = e^{i 2 \pi \ydot} \in C^3(G,\text{U}(1)).
\end{align}
Then, Eq.~\eqref{constraint_X} implies that $\cbd \ydot = \cbd \xdot - \cbd \xdot' - \cbd \xdot''$, which is equivalent to requiring that $\Ydot$ satisfy
\begin{align}
\label{eq:general-Y}
\cbd \Ydot &= \frac{\defectO_r(\psi^{\mathsf{p}'}_{\vv{0}}\otimes \I_{\vv{1}}^{\central},\pi')\defectO_r(\psi^{\mathsf{p}''}_{\vv{0}} \otimes \I_{\vv{1}}^{\central},\pi'')}{\defectO_r(\psi^{ \mathsf{p}'+\mathsf{p}''+{\pi' \cup \pi''}}_{\vv{0}}\otimes \I_{\vv{1}}^{\central},\pi' +\pi'')}
.
\end{align}
This determines $\Ydot$, and hence $\ydot$, up to a 3-cocycle.
Ref.~\onlinecite{Brumfiel2018a} addressed the $\central = 0$ case, while Ref.~\onlinecite{Bark2021inv} extended the result to $\central \neq 0$ with the conjecture that the expression had no 3-cocycle correction.
As in Refs.~\onlinecite{Brumfiel2018a} and ~\onlinecite{Bark2021inv}, we provide an explicit parametrization of $\ydot$, given in Eq.~\eqref{eq:lambdaeqmaiantext}, which is proven to be correct up to a 3-cocycle, but conjectured to have no 3-cocycle correction.
In particular, we show $\ydot$ satisfies Eq.~\eqref{eq:general-Y} in Appendix~\ref{app:Y-details}.
Thus, the stacked theory is indeed a FSPT phase if the two theories being stacked are FSPT phases.
In Secs.~\ref{sec:subgroup-3} and \ref{sec:Gtriv}, we compute $\Ydot$ explicitly through condensation for the subgroups $\GSPT$ and $\Gtriv$, verifying that Eq.~\eqref{eq:lambdaeqmaiantext} has no 3-cocycle correction in these cases.
In Sec.~\ref{sec:generalFSPTphases}, we use condensation to show that $\Ydot$ can only depend on $\mathsf{p}', \maj', \mathsf{p}'', \maj''$ and $\central$.

Next, we check that $[0,0,0]_{\central}$ acts as the identity element under stacking.
It is easy to see that for this to be true, Eq.~\eqref{eq:general-group} would require $\ydot(\mathsf{p}',\maj';0,0)$ to be a 3-coboundary.
On the other hand Eq.~\eqref{eq:general-Y} only implies that $\ydot(\mathsf{p}',\maj';0,0)$ is a 3-cocycle, so a stronger constraint is needed.
We derive such a constraint explicitly by computing the condensation in the fermionic stacking
\begin{align}
\label{eq:identity}
(\xdot,\mathsf{p},\maj)_{\central}   \underset{\mathcal{G}^{\eff}}{\ftimes} (0,0,0)_{\central}
,
\end{align}
using the techniques reviewed in Appendix~\ref{app:condensation}.
The explicit topological data for the $(0,0,0)_{\central}$ FSPT theory can be obtained from the torsor method, and is found to be
\begin{align}
a_{\mathpzc{g}} \otimes b_{\mathpzc{h}} &= [ab]_{\mathpzc{gh}} ,
\\
F^{a_{\mathpzc{g}} b_{\mathpzc{h}} c_{\mathpzc{k}} } &= (-1)^{\mathsf{a} \cdot \central({\bf h,k})},
\\
R^{a_{\mathpzc{g}} b_{\mathpzc{h}}} &= (-1)^{[\mathsf{a}+\vv{x} + \central({\bf g,h})]\cdot \mathsf{b}} ,
\\
U_{\bf k}(a_{\mathpzc{g}} , b_{\mathpzc{h}}) &= (-1)^{\mathsf{a} \cdot [\central({\bf h,k}) + \central({\bf k,\bar{k}hk})]} ,
\\
\eta_{c_{\mathpzc{k}}}({\bf g,h}) &= (-1)^{\mathsf{c} \cdot \central({\bf g,h})}
.
\end{align}
We pick a representative set of simple objects of the condensed theory given by $\cond{a}_{\mathpzc{g}}  = (a_{\mathpzc{g}}, \I_{\mathpzc{g}})$.
One can check that each such object braids trivially with the condensate, and that the representative set $\{\cond{a}_{\mathpzc{g}}\}$ is in one-to-one correspondence with the set of simple objects of the $(\xdot,\mathsf{p},\maj)_{\central}$ FSPT theory.
One can also check that the $\zig{a}_{\mathpzc{g}}$ have identical fusion rules to the $a_{\mathpzc{g}}$.
From the above data, we see that by restricting to the representative charges $\I_{\mathpzc{g}}$, the $F$-, $R$-, $U$-, and $\eta$-symbols from the $(0,0,0)_{\central}$ theory are all equal to 1.
As such, the topological data of the condensed theory are identical to that of the $(\xdot,\mathsf{p},\maj)_{\central}$ theory, and thus
\begin{align}
[\xdot,\mathsf{p},\maj]_{\central} \underset{\mathcal{G}^{\eff}}{\ftimes}[0,0,0]_{\central} = (\xdot,\mathsf{p},\maj)_{\central}
.
\end{align}

Finally, we show that every FSPT theory $[\xdot,\mathsf{p},\maj]_{\central}$ has an inverse theory $\overline{[\xdot,\mathsf{p}, \maj]}_{\central}$, which we defined by the representative
\begin{align}
\label{eq:inverse}
\overline{(\xdot,\mathsf{p},\maj)}_{\central} &=(-\xdot- \ydot(\mathsf{p},\maj;\mathsf{p}+{\maj \cup \maj},\maj) ,\mathsf{p}+{\maj \cup \maj},\maj)_{\central}
,
\end{align}
which also satisfies the conditions in Eqs.~\eqref{constraint_pi}, \eqref{constraint_p}, and \eqref{constraint_X} necessary to represent a valid FSPT phase.
The first condition, $\cbd\maj=0$ is immediately satisfied.
The second condition, $\cbd (\mathsf{p} + ({\maj \cup \maj}))= {\maj\cup \central}$ is satisfied by using the fact that $\cbd ({\maj \cup \maj})= 0$.
The third condition,
\begin{align}
& \cbd (\Xdot^{-1}\Ydot(\mathsf{p},\maj;\mathsf{p}+{\maj \cup \maj},\maj)^{-1})
\notag \\
& \quad = (\cbd \Xdot)^{-1} (\cbd \Ydot(\mathsf{p},\maj;\mathsf{p}+{\maj \cup \maj},\maj))^{-1}
\notag \\
&\quad  = \defectO_r(\psi^{\mathsf{p}}_{\vv{0}}\otimes \I_{\vv{1}}^{\central},\pi) \frac{\defectO_r(\I_{\vv{1}}^{\central},0)}{\defectO_r(\psi^{\mathsf{p}}_{\vv{0}} \otimes \I_{\vv{1}}^{\central},\pi) \defectO_r(\psi^{\mathsf{p} + \maj \cup \maj}_{\vv{0}}\otimes \I_{\vv{1}}^{\central},\pi)}
\notag \\
&\quad = \defectO_r(\psi^{\mathsf{p}+\maj \cup \maj}_{\vv{0}} \otimes \I_{\vv{1}}^{\central},\pi)^{-1}.
\end{align}
follows from Eq.~\eqref{eq:general-Y} and the fact that $\defectO_r(\I_{\vv{1}}^{\central},0)=1$.
Then, using Eq.~\eqref{eq:general-group}, we have
\begin{align}
&[\xdot,\mathsf{p},\maj]_{\central} \underset{\mathcal{G}^{\eff}}{\ftimes} \overline{[\xdot,\mathsf{p},\maj]}_{\central} = [0,0,0]_{\central}.
\end{align}
Hence, every theory $(\xdot,\mathsf{p},\maj)_{\central}$ has an inverse given by Eq.~\eqref{eq:inverse}.

It follows that fermionic stacking of FSPT theories forms an Abelian group $\Gr$.
We now establish Eq.~\eqref{eq:general-group} is indeed the structure provided by fermionic stacking FSPT theories.

\subsubsection{FSPT phases with trivial symmetry action and relative fractionalization class}
\label{sec:subgroup-3}

Consider the FSPT phases
\begin{align}\label{eq:00alpha}
[\xdot,0,0]_{\central} &=\spt_G^{[\Xdot]}\underset{G}{\boxtimes} [0,0,0]_{\central}
,
\end{align}
that differ from the identity $[0,0,0]_{\central}$ by a bosonic SPT phase $\spt_G^{[\Xdot]}$ where $[\Xdot]\in H^3(G,U(1))$ (recall $\Xdot = e^{i 2\pi \xdot}$).
It is straightforward to check that this triple satisfies Eqs.~\eqref{constraint_pi}, \eqref{constraint_p}, and \eqref{constraint_X}.
The topological data of $(\xdot,0,0)_{\central}$ is given by that of $(0,0,0)_{\central}$ multiplied by the appropriate $\Xdot$ factors of $\spt_G^{[\Xdot]}$.
Stacking such a FSPT theory with a general FSPT theory thus yields
\begin{align}
[\xdot + \xdot',\mathsf{p},\maj]_{\central}  &= [\xdot,\mathsf{p},\maj]_{\central} \underset{\mathcal{G}^{\eff}}{\ftimes} [\xdot',0,0]_{\central}
\\
&= \spt_G^{[\Xdot']}\underset{G}{\boxtimes} [\xdot,\mathsf{p},\maj]_{\central}
.
\label{eq:FSPT_stack}
\end{align}
We can verify this following a nearly identical line of arguments to the paragraph following Eq.~\eqref{eq:identity}, except the topological data of the $(\xdot',0,0)_{\central}$ theory introduces the corresponding $\Xdot'$ factors of $\spt_G^{[\Xdot']}$.

In particular, this implies
\begin{align}
[\xdot',0,0]_{\central}  \underset{\mathcal{G}^{\eff}}{\ftimes} [\xdot'',0,0]_{\central}  = [\xdot'+ \xdot'',0,0]_{\central}.
\end{align}
Therefore, the set of equivalence classes of the $\{[\xdot,0,0]_{\central}\}$ FSPT theories form a subgroup $\GSPT$ of $\Gr$.

In the absence of the equivalence relation Eq.~\eqref{eq:equiv2}, the set $\{(\xdot,0,0)_{\central}\}$ is in one-to-one correspondence with bosonic SPT phases, and thus forms the group $H^3(G,\mathbb{R}/\mathbb{Z})$.
However, fermionic degrees of freedom impose additional equivalences that change this result.
In particular, applying Eq.~\eqref{eq:equiv2} with $\mathsf{z} \in Z^1(G,\mathbb{Z}_2^{\psi})$ equates
\begin{align}
[0,0,0]_{\central} = [\frac{1}{2}{\central \cup \mathsf{z}},0,0]_{\central}.
\end{align}
Thus, even though $\spt_{G}^{[(-1)^{\central \cup \mathsf{z}}]}$ may be a nontrivial bosonic SPT phase, it can be trivialized by the fermionic degrees of freedom.
The group $\GSPT$ is given by the quotient
\begin{align}
\label{eq:GSPT}
\GSPT = \frac{H^3(G,\mathbb{R}/\mathbb{Z})}{\im( \frac{1}{2}\central \cup)}.
\end{align}
Here, we define the map
\begin{align}
\frac{1}{2} \central \cup: H^1(G,\mathbb{Z}_2^{\psi}) &\to H^3(G,\mathbb{R}/\mathbb{Z}),\\
\mathsf{z} &\mapsto \left[ \frac{1}{2} \central \cup \mathsf{z} \right].
\end{align}
The group $\GSPT$ has multiplication inherited from $H^3(G,\mathbb{R}/\mathbb{Z}) $.

Lastly, as $\GSPT \lhd \Gr$, we can place $\Gr$ into a short exact sequence,
\begin{align}
\label{eq:fullgroup}
0\to \GSPT \to\Gr \to \Grto \to 0
,
\end{align}
where the quotient group
\begin{align}
\Grto \equiv \frac{\Gr}{\GSPT}
,
\end{align}
is given by FSPT phases modulo bosonic SPT phases.
The notation of $\Grto$ indicates that its elements consist of equivalence classes of $1$- and $2$-cochains $[\maj,\mathsf{p}]_{\central}$, where each representative $(\maj,\mathsf{p})_{\central}$ admits an $\Xdot$ satisfying Eq.~\eqref{constraint_X}.
We determine the group operation of $\Grto$ in Sec.~\ref{sec:generalFSPTphases}.
The central extension in Eq.~\eqref{eq:fullgroup} is specified by an element of $Z^2(\Grto , \GSPT)$ that can be determined by $\ydot$.

\subsubsection{FSPT phases with trivial symmetry action}
\label{sec:Gtriv}

The next class of FSPT phases we consider are those with trivial symmetry action, but possibly nontrivial vortex fractionalization class, that is $\{[\xdot,\mathsf{p},0]_{\central}\}$.
With fixed $\mathcal{G}^{\eff}$ determined by $\central \in Z^2(G,\mathbb{Z}_2^{\eff}) $, these theories are determined by the 2-cocycle $\mathsf{p}\in Z^2(G,\mathbb{Z}_2^{\psi})$, and a 3-cochain $\Xdot \in C^3(G,\text{U}(1))$ satisfying~\footnote{The defectification obstruction of FSPT phases with trivial symmetry action is equal to the relative defectification obstruction, i.e., $[\defectO_r(\coho{p}_{\central},0)] = [\defectO(\coho{p}_{\central} ,0)]$. }
\begin{align}
\label{H4obs}
\cbd \Xdot = \defectO_r(\psi^{\mathsf{p}}_{\vv{0}} \otimes \I_{\vv{1}}^{\central},0)^{-1} = (-1)^{\mathsf{p} \cup \mathsf{p} + \mathsf{p} \cup \central}.
\end{align}
Equivalently, $\cbd \xdot = \frac{1}{2}(\mathsf{p} \cup \mathsf{p} + \mathsf{p} \cup \central)$.

Stacking two FSPT phases with trivial symmetry action necessarily results in a theory with trivial symmetry action
\begin{align}
[\xdot, \mathsf{p}_{},0]_{\central}&= [\xdot', \mathsf{p}'_{},0]_{\central} \underset{\mathcal{G}^{\eff}}{\ftimes} [\xdot'', \mathsf{p}_{}'',0]_{\central},
\end{align}
as the resulting symmetry action of the condensed theory is inherited from the product of the symmetry actions of the parent theories.
Therefore, these theories form a subgroup $\Gtriv$ of $\Gr$.
We specify $\Gtriv$ by solving for $\mathsf{p}$ and $\Xdot$ in terms of $\mathsf{p}', \mathsf{p}'', \Xdot', \Xdot''$.
The condensation calculation proceeds as follows.
(1) We pick a representative set of simple objects from which we can compute the fusion rules, thereby determining $\mathsf{p}$.
(2) We specify the fusion spaces for our representative objects, from which we can determine $\Xdot$ by computing the $F$-symbol of the stacked theory.
Our calculation extends the condensation calculation of Ref.~\onlinecite{Bhardwaj2017} to include all possible $\mathcal{G}^{\eff}$.

For (1), we take our representative set of simple objects to be
\begin{align}
\label{eq:rep}
\cond{a}_\mathpzc{g}=\cond{a}_{\vv{x},{\bf g} } \equiv ( a_{\vv{x},{\bf g}},\I_{\vv{x},{\bf g} }) ,
\end{align}
where the right side is written in terms of the objects of the stacked theories.
The fusion rules follow from
\begin{align}
\label{eq:fusion-cond}
&\cond{a}_{\mathpzc{g}}\otimes \cond{b}_{\mathpzc{h}} =
(a_{\mathpzc{g}} \otimes b_{\mathpzc{h}}, \I_{\mathpzc{ g}} \otimes \I_{\mathpzc{h}})
\\ &= ([\psi^{\mathsf{p}'({\bf g},{\bf h}) }a  b]_{\mathpzc{gh}}, [\psi^{\mathsf{p}''({\bf g},{\bf h})}]_{\mathpzc{ gh}})
\\  &\cong[\cond{\psi^{\mathsf{p}'({\bf g},{\bf h})+\mathsf{p}''({\bf g},{\bf h})} ab}]_{\mathpzc{gh}},
\end{align}
where the last step uses $(\psi_{\vv{0}}^{\mathsf{p}'({\bf g,h}) },\psi_{\vv{0}}^{\mathsf{p}''({\bf g,h}) }) \cong(\psi_{\vv{0}}^{\mathsf{p}'({\bf g,h})+\mathsf{p}''({\bf g,h})},\I_{\vv{0}} )$ to transform the second line into a representative simple object, i.e. into the form of Eq.~\eqref{eq:rep}.
Therefore, we see that
\begin{align}\label{eq:ppp}
\mathsf{p} = \mathsf{p}'+\mathsf{p}''.
\end{align}

For (2), we need to pick a representative of the fusion space, and then compute the $F$-symbol, from which we can extract $\Xdot$.
We take
\begin{align}
\TrivialSymmetryFusionSpace \in {V}^{\cond{a}_{\mathpzc{g}}\cond{b}_{\mathpzc{h}}}_{\cond{c}_{\mathpzc{gh}}}
,
\end{align}
where again we have used that $(\psi_{\vv{0}}^{\mathsf{p}'({\bf g,h}) },\psi_{\vv{0}}^{\mathsf{p}''({\bf g,h}) }) \cong(\psi_{\vv{0}}^{\mathsf{p}'({\bf g,h})+\mathsf{p}''({\bf g,h})},\I_{\vv{0}} )$.
Now we can compute the $F$-symbol of the condensed theory using the techniques described in Appendix~\ref{app:condensation}.
By direct computation, one finds
\begin{align}
\label{eq:Fpp}
{F^{}}^{\cond{a}_{\mathpzc{g}} \cond{b}_{\mathpzc{h}} \cond{c}_{\mathpzc{k}}}
=&\Xdot'({\bf g,h,k}) \Xdot''({\bf g,h,k}) \notag\\
&\times(-1)^{\cond{\mathsf{a}} \cdot (\mathsf{p}'({\bf h,k}) +\central({\bf h,k})) }(-1)^{\mathsf{p}''({\bf g,h}) \cdot \cond{\mathsf{c}}}
.
\end{align}

In order to extract $\Xdot$ from Eq.~\eqref{eq:Fpp}, we need to write the condensed theory in terms of a torsor functor applied to a base theory, recalling that $(\xdot,\mathsf{p},\maj)_{\central} = \mathcal{F}_{\cohosub{p}_{\central},\Xdot}(\base_{\mathbb{Z}_2^{\eff}\times G,\rho})$.
To recognize the $F$-symbol as one resulting from the torsor method, we apply the vertex basis gauge transformation~\cite{Bhardwaj2017}
\begin{align}
\Gamma^{\cond{a}_{\mathpzc{g}}\cond{b}_{\mathpzc{h}}} = (-1)^{(\cond{\mathsf{a}} +\cond{\mathsf{b}})\cdot \mathsf{p}''({\bf g,h}) } i^{\mathsf{p}''({\bf g,h})}
\end{align}
and use the relation
\begin{align}
\cbd(i^{ \mathsf{p}''}) = (-1)^{\mathsf{p}'' \cup_1 \mathsf{p}''}.
\end{align}
Together, these transform the $F$-symbol to
\begin{align}
{\widetilde{F}}^{\cond{a}_{\mathpzc{g}} \cond{b}_{\mathpzc{h}} \cond{c}_{\mathpzc{k}}}
=& \Xdot'({\bf g,h,k}) \Xdot''({\bf g,h,k}) (-1)^{[\mathsf{p}' \cup_1 \mathsf{p}'']({\bf g,h,k}) } \notag \\
&\times (-1)^{\cond{\mathsf{a}} \cdot (\mathsf{p}'({\bf h,k})+\mathsf{p}''({\bf h,k}) + \central({\bf h,k}))},
\end{align}
corresponding to the post-torsor functor $F$-symbol for $\mathcal{F}_{\psi^{\mathsf{p}'+ \mathsf{p}''}_{\vv{0}} \otimes \I_{\vv{1}}^{\central}, \Xdot' \Xdot'' (-1)^{\mathsf{p}' \cup_1 \mathsf{p}''}}(\base_{\mathbb{Z}_2^{\eff}\times G,\openone})$.
The classification of $G$-crossed MTCs by their symmetry fractionalization class and defectification class implies that gauge transformations exist to translate the $R$-, $U$-, and $\eta$-symbols into the form given by the torsor method.

Thus, we conclude that
\begin{align}
&[\xdot',\mathsf{p}' , 0]_{\central} \underset{\mathcal{G}^{\eff}}{\ftimes}
[\xdot'',\mathsf{p}'' , 0]_{\central}
\notag \\
&\quad =[\xdot' +\xdot'' + \frac{1}{2}{\mathsf{p}' \cup_1 \mathsf{p}''},\mathsf{p}'+ \mathsf{p}'', 0]_{\central}
.
\label{eq:afsptprod}
\end{align}
This implies
\begin{align}
\label{eq:ydotpizero}
\ydot(\mathsf{p}',0;\mathsf{p}'',0) \sim \frac{1}{2}{\mathsf{p}' \cup_1\mathsf{p}''},
\end{align}
where the equivalence is up to a 3-coboundary.
We note that the right hand side of Eq.~\eqref{eq:afsptprod} is not strictly symmetric in $\mathsf{p}'$ and $\mathsf{p}''$, but it is up to a coboundary, i.e.
\begin{align}
\mathsf{p}' \cup_1 \mathsf{p}'' = \mathsf{p}'' \cup_1 \mathsf{p}' + \cbd( \mathsf{p}' \cup_2 \mathsf{p}'')
,
\end{align}
so it is consistent with commutativity of stacking.
It is straightforward to check that this $\ydot$ satisfies Eq.~\eqref{eq:general-Y} for this case, given Eq.~\eqref{H4obs}.

Finally, while Eq.~\eqref{eq:afsptprod} fully specifies $\Gtriv$, we can further understand the group structure by noting that $\mathsf{p}$ and ${\central}$ fix $\Xdot$ up to an element of $\GSPT$.
Therefore, we can consider the subgroup
\begin{align}
\label{eq:grt}
\frac{\Gtriv}{\GSPT}  \equiv \Grt \lhd H^2(G,\mathbb{Z}_2^{\psi}),
\end{align}
consisting of the vortex symmetry fractionalization classes with trivial symmetry action, for which the defectification obstruction vanishes.
Mathematically, $\Grt$ is the set of 2-cocycles for which $\defectO_r(\psi^{\mathsf{p}}_{\vv{0}} \otimes \I_{\vv{1}}^\central,0)$ is a coboundary.
In Ref.~\onlinecite{Gu2014}, $\Grt$ was denoted $\BH^2(G,\mathbb{Z}_2)$.
We can place the group $\Gtriv$ into a short exact sequence given by,
\begin{align}
0\to \GSPT \to \Gtriv \to \Grt \to 0.
\end{align}
The central extension here is specified by an element of $Z^2(\Grt , \GSPT)$ that can be determined by $\ydot$.

\subsubsection{General FSPT phases}
\label{sec:generalFSPTphases}

We now consider stacking two general FSPT phases
\begin{align}
[\xdot,\mathsf{p},\maj]_{\central} = [\xdot',\mathsf{p}',\maj']_{\central} \underset{\mathcal{G}^{\eff}}{\ftimes} [\xdot'',\mathsf{p}'',\maj'']_{\central} .
\end{align}
To derive $[\xdot,\mathsf{p},\maj]_{\central}$, it is helpful to break the group law into simpler pieces.
First we compute the quotient groups $\Gro\equiv \Gr/\Gtriv$ and $\Grto \equiv\Gr/\GSPT  $, which determine part of the structure of $\Gr$.
To completely specify the group, we need a general formula for $\Ydot$.
Ref.~\onlinecite{Brumfiel2018a} provided such a solution when $\central=0$, which we quote below.
However, we lack an explicit closed form expression for $\Ydot$ in the case of general $\central$.

The first quotient $\Gr/\Gtriv$ is the group of FSPT phases modulo FSPT phases with trivial symmetry action.
This is the set of unobstructed symmetry actions, i.e. $\rho_{\bf g} = \V^{\maj({\bf g})}$ with $\maj \in H^1(G, \mathbb{Z}_2^{\V})$, for which there exist $\mathsf{p}$ and $\Xdot$ such that $\cbd \mathsf{p} = {\maj \cup \central}$ and $\cbd \Xdot = \defectO_r(\psi_{\vv{0}}^{\mathsf{p}}\otimes \I_{\vv{1}}^\central,\pi)$.
We denote the set of unobstructed symmetry actions by $\Gro$.
Now consider two unobstructed symmetry actions $\maj'$ and $\maj''$.
To determine $\maj$, we simply look at how the symmetry acts on the quasiparticle and vortex sectors of the condensed theory.
Again working with the representatives $\cond{a}_{\vv{x},{\bf 0} }= (a_{\vv{x},{\bf 0}},\I_{\vv{x},{\bf 0}})$ of the post-condensation charges, the symmetry action $\rho_{\bf g}$ is given by $\rho'_{\bf g} \boxtimes \rho''_{\bf g}$, up to a condensate isomorphism that ensures the action maps the set of representative topological charges by to itself.
In particular, we require
\begin{align}
\,^{\bf g}\zig{a}_{\vv{x}} &= \rho_{\bf g} \left( \zig{a}_{\vv{x}} \right) = ([\psi^{\vv{x} \cdot \pi({\bf g})} a]_{\vv{x}} , \I_{\vv{x}})
,
\end{align}
and have
\begin{align}
\rho'_{\bf g} \boxtimes \rho''_{\bf g} \left( \zig{a}_{\vv{x}} \right) &= ([\psi^{\vv{x} \cdot \pi'({\bf g})} a]_{\vv{x}} , \psi^{\vv{x} \cdot \pi''({\bf g})}_{\vv{x}})
\notag \\
& \cong ([\psi^{\vv{x} \cdot (\pi'({\bf g})+\pi''({\bf g}) )} a]_{\vv{x}} , \I_{\vv{x}})
,
\end{align}
where the isomorphism involved fusion with the condensate bosons $(\psi_{\vv{0}}, \psi_{\vv{0}})$.
Equating these, we see that
\begin{align}
\label{eq:maj''}
\maj = \maj' +\maj''.
\end{align}
Clearly $ \cbd \maj = \cbd \maj' + \cbd \maj'' = 0$.
Hence,
\begin{align}
\label{eq:gro}
\frac{\Gr}{\Gtriv} \equiv \Gro \lhd H^1(G,\mathbb{Z}_2^{\V})
.
\end{align}
We also note that $\Gr/\Gtriv  = \Grto/\Grt$.

Determining the 2-cochain $\mathsf{p}$ requires a bit more work.
In Sec.~\ref{sec:Ex_IFMTCs} we observed that symmetry fractionalization classes of $\ifo^{(0)}$ are completely determined by
\begin{align}
\label{eq:TCsymmetryaction}
 \rho_{\bf g} &= \V^{\pi({\bf g})},\\
 \label{eq:TCU}
U_{\bf g}(a_{\vv{x}},b_{\vv{y}}) &=(-1)^{(\mathsf{a}+\vv{x})\cdot \vv{y} \cdot \pi({\bf g})},\\
\label{eq:etaTC}
\eta_{a_{\vv{x}}}({\bf g,h}) &= i^{\vv{x}\cdot \pi({\bf g})\cdot \pi({\bf h}) } (-1)^{\vv{x} \cdot \mathsf{p}({\bf g,h}) + \mathsf{a} \cdot \central({\bf g,h})}
.
\end{align}
These three pieces of data determine the symmetry fractionalization class of $\ifo^{(0)}$; moreover, this standardizes the presentation of the data so that we can readily extract differences of symmetry fractionalization classes.

We now determine $\mathsf{p}$ by computing the vortex symmetry fractionalization class of the condensed theory.
The initial topological data coming from condensation will not be of the form Eq.~\eqref{eq:TCU} and \eqref{eq:etaTC}, but rather it will differ by a suitable symmetry action gauge transformation.
We again use the representative simple objects $\zig{a}_{\vv{x}} = (a_{\vv{x}} , \I_{\vv{x}})$.
The fusion space of the condensed theory is represented by ${V}^{\zig{a}_{\vv{x}} \zig{b}_{\vv{y}}}_{[\zig{ab}]_{\vv{x+y}}} \cong V^{\prime a_{\vv{x}} b_{\vv{y}}}_{[ab]_{\vv{x+y}} } \otimes V^{\prime \prime \I_{\vv{x}} \I_{\vv{y}}}_{\I_{x+y}}$.
With the representative fusion spaces, we can compute the symmetry action $\rho_{\bf g}$, which, as previously stated, is given by $\rho'_{\bf g} \boxtimes \rho''_{\bf g}$ up to a condensate isomorphism that ensures the set of representative topological charges is mapped to itself.
We define $\rho_{\bf g}$ on the trivalent fusion spaces ${V}^{\zig{a}_{\vv{x}} \zig{b}_{\vv{y}}}_{[\zig{ab}]_{\vv{x+y}}}$ by the diagrammatic equation
\begin{align}
\label{eq:Uaction}
\rho_{\bf g} \left( \zigvabc \right) =
\text{\footnotesize{$U'_{\bf g}({}^{\bf g}a_{\vv{x}},{}^{\bf g}b_{\vv{y}})
 U''_{\bf g}({}^{\bf g}\I_{\vv{x}},{}^{\bf g}\I_{\vv{y}})$}}
 \!\!\!\!\!\! \rhopp.
\end{align}
where the additional $(\psi_{\vv{0}},\psi_{\vv{0}})$ correspond to the condensate isomorphisms needed in order to remain within the set of representative objects of the condensed theory.
We take the $(\psi_{\vv{0}},\psi_{\vv{0}})$ strands to the right in order to match our condensation conventions in Appendix~\ref{app:condensation}.
The prefactors appear due to the parent symmetry actions.
We can compute $U_{\bf g}$ through the relation Eq.~\eqref{eq:rho_vertex} by evaluating the diagram on the right hand side of Eq.~\eqref{eq:Uaction} to obtain
\begin{align}
\label{eq:stackUeval}
U_{\bf g}(\zig{a}_{\vv{x}},\zig{b}_{\vv{y}}) &= (-1)^{(\mathsf{a} + \mathsf{x})\cdot \mathsf{y} \cdot\maj({\bf g})}
\notag \\
&\times (-1)^{(\mathsf{a \cdot y} + \mathsf{b \cdot x})\cdot \maj''({\bf g}) + \mathsf{x \cdot y}\cdot \maj({\bf g})\cdot \maj''({\bf g}) }
.
\end{align}
We can return this to the standard form of Eq.~\eqref{eq:TCU} by applying a symmetry action gauge transformation given by
\begin{align}
\label{eq:sym_ac_g}
\gamma_{\zig{a}_{\vv{x}}}({\bf g})  = (-1)^{\mathsf{a} \cdot \mathsf{x} \cdot \pi''({\bf g} ) }i^{\mathsf{x} \cdot \pi''({\bf g})}i^{\mathsf{x} \cdot {\pi}'({\bf g}) \cdot \pi''({\bf g})}
,
\end{align}
which yields
\begin{align}
\widecheck{{U}}_{\bf g}(a_{\vv{x}}, b_{\vv{y}}) = (-1)^{(\mathsf{a} + \mathsf{x})\cdot \mathsf{y} \cdot \maj({\bf g})}
.
\end{align}

The $\eta$-symbols of the condensed theory can be computed once we define the braiding
\begin{align}
\label{eq:condbraid}
\condbraid.
\end{align}
We have used the simple module object as defined in Eq.~\eqref{eq:representative_simple} of Appendix~\ref{app:condensation}.
We also introduced an additional isomorphism given by the unlabeled line after the braid, so that the braid takes representative simple objects to representative simple objects.
The unlabeled line is given by $(\psi_{\vv{0}}^{\pi''({\bf g}) \cdot \vv{x}},\psi_{\vv{0}}^{\pi''({\bf g}) \cdot \vv{x}})$ .
Using this definition of the braid, one can explicitly compute $\eta_{a_{\vv{x}}}({\bf g,h} )$ using Eq.~\eqref{eq:eta-2} to find
\begin{align}
\label{eq:etaprimeprime}
{\eta}_{\zig{a}_{\vv{x}}}({\bf g,h}) = \eta'_{a_{\vv{x}}}({\bf g,h})\eta_{\I_{\vv{x}}}''({\bf g,h}).
\end{align}
Note that the simple product of $\eta$-symbols appearing on the right hand side is due to the gauge we have chosen in Eq.~\eqref{eq:TCU} for the $U$-symbols of the parent theory, in particular $U_{\bf{g}}(a_{\vv{x}}, \psi_{\vv{0}})  = 1$ in our choice of gauge for $\ifo^{(0)}$.
We can now use Eq.~\eqref{eq:etaTC} in the right hand side of Eq.~\eqref{eq:etaprimeprime}, and apply the symmetry action gauge transformation of Eq.~\eqref{eq:sym_ac_g} to obtain
\begin{align}
\label{eq:etacheck}
\widecheck{{\eta}}_{\zig{a}_{\vv{x}}}({\bf g,h}) &= i^{\vv{x}\cdot \pi({\bf g})\cdot \pi({\bf h}) } (-1)^{\vv{x} \cdot \mathsf{p}({\bf g,h}) + \mathsf{a} \cdot \central({\bf g,h})}
,
\end{align}
where
\begin{align}
\label{eq:p''}
\mathsf{p} =\mathsf{p}'+ \mathsf{p}'' +{\pi' \cup \pi''}
.
\end{align}
In deriving this, we have used the property
\begin{align}
\frac{ i^{ \pi({\bf g}) \cdot \pi({\bf h}) }}
{ i^{ \pi'({\bf g})\cdot \pi'({\bf h}) } i^{ \pi''({\bf g}) \cdot \pi''({\bf h}) } }
&=
\frac{i^{{\pi}'({\bf gh}) \cdot \pi''({\bf gh})} }
{i^{{\pi}'({\bf g}) \cdot \pi''({\bf g})} i^{{\pi}'({\bf h}) \cdot \pi''({\bf h})}}
.
\end{align}
Thus, we see that Eq.~\eqref{eq:p''} provides the value of $\mathsf{p}$ under stacking.
Note that this symmetry fractionalization addition rule for stacking FSPT phases is independent of $\central$, and it agrees with Eq.~\eqref{eq:ppp} when $\maj' = \maj''  = 0$.
The asymmetry in Eq.~\eqref{eq:p''} can be accounted for by the fact that $\mathsf{p}$ is identified under $\mathbb{Z}_2^{\psi}$-valued coboundaries, and that
\begin{align}
\maj' \cup \maj'' =   \maj'' \cup \maj' +\cbd (\maj' \cup_1 \maj'')
.
\end{align}
Our result agrees with the independent calculation done in Ref.~\onlinecite{Bhardwaj2017} for $\central=0$, but our calculation shows it applies to the more general scenario of nontrivial $\central$.

The set of unobstructed symmetry actions and vortex fractionalization classes $\{(\mathsf{p},\maj)_{\central}\}$ form the group
\begin{align}
\label{eq:grta}
\frac{\Gr}{\GSPT}  \equiv \Grto ,
\end{align}
with group multiplication given by
\begin{align}
\label{eq:grto}
(\mathsf{p}',\pi') \times (\mathsf{p}'',\pi'') = ( \mathsf{p}'+ \mathsf{p}''  +{\pi' \cup \pi''} ,\pi' +\pi'').
\end{align}
We can describe $\Grto$ as a central extension in the short exact sequence
\begin{align}
0 \to \stackbelow{\frac{\Gtriv}{\GSPT}}{\Grt}  \to \stackbelow{\frac{\Gr}{\GSPT}}{\Grto} \to \stackbelow{\frac{\Gr}{\Gtriv}}{\Gro} \to 0,
\end{align}
with the extension class determined by $\cup$.
Equivalently, $\Grto = \Grt  \times_{\varepsilon_{\mathsf{(2,1)}}} \Gro$ with $\varepsilon_{\mathsf{(2,1)}} = \cup$.
The extension class is universal, in the sense that it makes no reference to $G$.

Lastly, we discuss the addition of the 3-cochains.
The 3-cochains $\Xdot$ and $\Ydot$ can, in principle, be computed from the associator of the theory after condensation of the paired fermions.
However, finding a gauge in which the resulting FSPT phase can be recognized as the image of a torsor functor is difficult in the general case.
Nevertheless, by looking at the definition of the $F$-symbol of the condensed theory, given in Eq.~\eqref{eq:fcondmove}, we can make two observations: (1) we see that $\Xdot'$ and $\Xdot''$ appear multiplicatively in $F$, and therefore $\Xdot$, since they appear multiplicatively in $F'$ and $F''$ [see Eq.~\eqref{eq:F-hatprime}], and (2) the remaining data determining $F$ from condensation only involves $F$- and $R$-symbols from the parent theories that involve at least one $\psi$ strand.
Therefore, $\Ydot$ is independent of $\Xdot'$ and $\Xdot''$, justifying why in Eq.~\eqref{eq:general-group} we write $\ydot(\mathsf{p}',\maj';\mathsf{p}'',\maj'')$.
We also comment that this is consistent with the generalized cohomology viewpoint of SPT phases (see, for example, Ref.~\onlinecite{Freyd2019}, and references therein).

Recall that under stacking the set $\{(\mathsf{p},\maj)_{\central}\}$ forms the quotient group $\Grto = \Gr/\GSPT$.
Therefore, the group of FSPT phases $\Gr$ can be written as a central extension in the short exact sequence
\begin{align}
0 \to \GSPT \to\Gr\to \Grto \to 0
.
\end{align}
The 2-cocycle $\varepsilon_{\mathsf{(3,21)}}$ in $Z^2(\Grto, \GSPT )$ determining the central extension is defined from $\Ydot$ through the relation
\begin{align}
\label{eq:X''}
\Xdot &\sim \Xdot'\Xdot''\Ydot(\mathsf{p}',\maj';\mathsf{p}'',\maj'')
,
\end{align}
where equivalence is up to coboundaries.
Recall that this relation combined with Eq.~\eqref{constraint_X}, imposes Eq.~\eqref{eq:general-Y}.

The condensation calculation required to move beyond Eq.~\eqref{eq:X''} to find an explicit expression for $\Ydot$ when $\pi$ is nontrivial is challenging and we do not pursue it here.
Following Ref.~\onlinecite{Bark2021inv}, we show in Appendix~\ref{app:Y-details} that Eq.~\eqref{eq:lambdaeqmaiantext} indeed satisfies Eq.~\eqref{eq:general-Y}.
It is easy to check that when $\maj' = \maj'' = 0$, we recover Eq.~\eqref{eq:ydotpizero}, and when $\central = 0$, we recover the result of Ref.~\onlinecite{Brumfiel2018a}.
In Appendix~\ref{app:fSPTclassificationExamples}, we show that Eq.~\eqref{eq:lambdaeqmaiantext} recovers the results of Sec.~\ref{sec:FSPTphasesZ2}.
It is straightforward to extract $\varepsilon_{\mathsf{(3,21)}}$ from $\ydot$, and verify it takes a universal form which does not reference $G$.

\subsection{Classification of FSET Phases}
\label{sec:FSET_classification}

\begin{table*}[t]
\centering
\fbox{
\begin{tabular}{l|l|l|l}
 & Bosonic  & Fermionic FMTC & Fermionic SMTC
\\ \hline
Initial data & MTC $\bMTC$, $G$ &  FMTC $\fMTC$, $\mathcal{G}^{\eff} = \mathbb{Z}_2^{\eff}\times_{\central}G$  & SMTC $\fMTC_{\vv{0}}$, $\mathcal{G}^{\eff} = \mathbb{Z}_2^{\eff}\times_{\central}G$
\\ \hline
Symmetry action
& \begin{tabular}{l} $[\rho]$ \\ $\text{Hom}(G,\on{Aut}(\bMTC))$  \end{tabular}
& \begin{tabular}{l} $[\rho]$ \\   $\text{Hom}_{\phi}(G , \on{Aut}^{\eff}(\fMTC))$  \end{tabular}
&  \begin{tabular}{l} $[\rho^{(0)}]$ \\  $\text{Hom}_{\phi^{(\vv{0})}}(G , \on{Aut}^{\eff}(\fMTC_{\vv{0}}))$  \end{tabular}
\\
\hline
Fractionalization
& \begin{tabular}{l} $[\eta ,\rho]$ \\ Obstruction: $[\coho{O}]$  \\  Torsor: $H^2_{[\rho]}(G,\mathcal{A})$  \end{tabular}
& \begin{tabular}{l} $[\eta ,\rho]$ \\ Obstruction: $[\coho{O}^{\central}]$  \\  Torsor: $H^2_{[\rho^{(0)}]}(G,\mathcal{A}_\vv{0})$  \end{tabular}
& \begin{tabular}{l} $[\eta^{(0)} ,\rho^{(0)}]$ \\ Obstruction: $[\coho{O}^{(0)\central}]$  \\  Torsor: $H^2_{[\widehat{\rho}^{(0)}]}(G,\widehat{\mathcal{A}}_\vv{0})$  \end{tabular}
\\ \hline
Defectification
& \begin{tabular}{l} $\bMTC_{G}^{\times}$ \\ Obstruction: $[\defectO]$  \\  Torsor: $H^3(G,\text{U}(1))$  \end{tabular}
& \begin{tabular}{l} $\fMTC_{\mathcal{G}^{\eff}}^{\times}$ \\ Obstruction: $[\defectO]$   \\  Torsor: $H^3(G,\text{U}(1))$ (SPT gluing)   \end{tabular}
&
\begin{tabular}{l} $\fMTC_{\mathcal{G}^{\eff}}^{\times}$ \\ Obstructions: $[O^\rho]$, $[\coho{O}^{\eta}]$, $[\defectO]$ \\ Torsor: $\mathbb{Z}_{16}$, $\Gr$ (FSPT stacking) \end{tabular}
\end{tabular}
}
\caption{Summary of the multi-stage classifications of $(2+1)$D SET phases described in Sec.~\ref{sec:FSET_classification}.
The first column reviews the classification of bosonic SET phases, while the second and third columns describe two complementary perspectives on classifying FSET phases: (a) viewing the FMTC as fundamental, including the vortices from the beginning; and (b) viewing the SMTC as fundamental and vortices as $\mathbb{Z}_2^{\eff}$ symmetry defects that are introduced during defectification.
The three stages of building up the SET structure are (1) defining the symmetry action on the initial data, (2) define symmetry fractionalization, and (3) extend these to a full theory of symmetry defects.
At each stage of the classification, one first calculates the obstructions, which are defined in terms of the previously specified data; when the obstructions vanish, the step can be achieved and the resulting equivalence classes of theories are torsorially classified by the corresponding groups; the classes of data with non-vanishing obstructions are not present in the subsequent classification stage, altering the prior steps' classification structure.
For example, in the defectification entry for Fermionic SMTC (bottom right), only a subset of the $\mathbb{Z}_{16}$ modular extension torsor (16-fold way) survives the full defectification extension when $[\central] \neq [\vv{0}]$; we will see in Sec.~\ref{sec:inv-class} that $\mathbb{Z}_{16}$ and $\Gr$ in this entry combine into a $\Gifo$ torsor corresponding to the stacking with invertible FSET phases.
There are potentially additional relabeling equivalences that may also alter the classification structure.
}
\label{table:classification}
\end{table*}

We now classify $(2+1)$D FSET phases with on-site unitary symmetry using the fermionic $\mathcal{G}^{\eff}$-crossed defect theory developed in Sec.~\ref{sec:fermionic-defectification}.
The classification of FSET phases is analogous to that of bosonic SET phases, but with a richer structure deriving from the physical nature of the fermion.
We have already seen manifestations of this richer structure when analyzing fermionic symmetry fractionalization and FSPT phases.
For FSET phases, we synthesize the obstruction and classification structures that we developed for fermionic symmetry and fractionalization in Sec.~\ref{sec:symmetric-fermions} with the FSPT obstruction and classification structure that we analyzed in Sec.~\ref{sec:fSPTClassification} in the context of the $\mathcal{G}^{\eff}$-crossed SMTC formalism.
These classification structures can be tiered and partitioned in various ways, but we will focus on two natural presentations: the first in terms of quasiparticle and vortex fractionalization and bosonic defectification with torsor action by gluing SPT phases, and the second in terms of quasiparticle fractionalization and fermionic defectification with torsor action by stacking FSPT phases.
Before describing the FSET classification is more detail, we review the classification of $(2+1)$D bosonic SET phases with on-site unitary symmetry.
All three classifications are summarized in Table~\ref{table:classification}.

The classification of bosonic SET phases~\cite{Bark2019} begins by specifying a MTC $\bMTC$ that describes the quasiparticles of the topological phase (without symmetry), a global symmetry group $G$ of the system, and a symmetry action $[\rho]:G\to \on{Aut}(\bMTC)$.
Though this is often the starting point for classification, we can also include the choice of symmetry action in the classification.
As these symmetry actions are homomorphisms, they are classified by $\text{Hom}(G,\on{Aut}(\bMTC))$, the set of homomorphism from $G$ to $\on{Aut}(\bMTC)$.
Each $[\rho]$ defines an invariant $[\coho{O}]\in H^3_{[\rho]}(G,\mathcal{A})$ corresponding to a possible obstruction to localizing the symmetry action [see Eq.~\eqref{eq:H3obsruction}].
When the fractionalization obstruction $[\coho{O}]$ vanishes, symmetry fractionalization can occur and is torsorially classified by $H^2_{[\rho]}(G,\mathcal{A} ) $.
Each fractionalization class, specified by $\rho$ (including the $U$-symbols) and the $\eta$-symbols, defines an invariant $[\defectO]\in H^4(G,\text{U}(1))$ corresponding to a possible obstruction to extending the theory to include symmetry defects, as described by a $G$-crossed MTC.
When the defectification obstruction $[\defectO]$ vanishes, defect theories can be obtained, i.e. solutions to the pentagon and heptagon consistency equations exist.
When unobstructed, the extensions of the MTC with a given symmetry action and fractionalization class to $G$-crossed MTCs are then torsorially classified by $H^3(G,\text{U}(1))$.
This classification accounts for equivalences of $G$-crossed MTCs under vertex basis and symmetry action gauge transformations.
The $H^3(G,\text{U}(1))$-torsor can be generated by gluing $G$-SPT phases, which are classified by $[\alpha] \in H^3(G,\text{U}(1))$, to the $G$-crossed MTC.

Ref.~\onlinecite{Aasen21} constructed an explicit torsor functor of $H^2_{[\rho]}(G,\mathcal{A})$ and $H^3(G,\text{U}(1))$ acting on $G$-crossed BTCs that yield the transformation of the complete topological data.
These torsor functors allows us to explicitly see how the equivalences by coboundaries in the fractionalization and defectification classification enter the $G$-crossed theory through relabeling of the defects' topological charges by $1$-cochain relabelings, together with vertex basis and symmetry action gauge transformations.

The classification by $\text{Hom}(G,{\text{Aut}}(\bMTC))$, $H^2_{[\rho]}(G,\mathcal{A})$, and $H^3(G,\text{U}(1))$ generally does not have a group structure simply obtained from the product of these objects, as there can be complicated transformations that feed up the structure sequence, and $\text{Hom}(G,{\text{Aut}}(\bMTC))$ is not generally a group.
Moreover, nontrivial obstructions will break the $\text{Hom}(G,{\text{Aut}}(\bMTC))$, $H^2_{[\rho]}(G,\mathcal{A})$, and $H^3(G,\text{U}(1))$ classification structure in the sense that only a subset of possible symmetry actions may permit fractionalization and only a subset of possible fractionalization classes may permit defectification.
In fact, for a fixed symmetry action, the unobstructed subsets are not guaranteed to form subgroups of the $H^2_{[\rho]}(G,\mathcal{A})$ and $H^3(G,\text{U}(1))$ classifying groups.

Another way the cohomological classification structure may be broken is through the equivalence of theories via relabeling topological charges.
Na\"ively different $G$-crossed MTCs related torsorially by nontrivial $H^2_{[\rho]}(G,\mathcal{A})$ and $H^3(G,\text{U}(1))$ elements, may actually be equivalent (up to gauge transformations) through a relabeling of the topological charges of quasiparticles and/or defects that leave the $G$ labels fixed.
As mentioned above, some defect relabelings are already built into the $H^2_{[\rho]}(G,\mathcal{A})$ classification structure as $1$-cochain relabelings of defects correspond to the coboundaries $B^2_{[\rho]}(G,\mathcal{A})$ in the quotient of the fractionalization classification.
However, there can be relabeling equivalences that are not captured by these; for example, nontrivial $1$-cocycles, which correspond to trivial $2$-coboundaries, may produce a nontrivial change in the defectification class; additionally, relabelings of the quasiparticles' topological charges may change the fractionalization class.

The classification of $(2+1)$D FSET phases begins by specifying a FMTC $\fMTC$ that describes the quasiparticles and vortices of the fermionic topological phase and a global fermionic symmetry group $\mathcal{G}^{\eff} = \mathbb{Z}_2^{\eff}\times_{\central}G$ of the system, where $\central \in Z^2(G,\mathbb{Z}_2^{\eff})$.
Next, we specify a fermionic symmetry action $[\rho]: G \to \on{Aut}^{\eff}(\fMTC)$.
Recall that the fermionic action can be $\Q$-projective when $[\Q]$ is nontrivial, i.e. when $\mathcal{A}_{\vv{1}} = \varnothing$, in which case we must have $\phi = \central$.
Thus, we fix the $\Q$-projective structure (since we have already chosen $\mathcal{G}^{\eff}$) and find that the corresponding possible fermionic symmetry actions are classified by $\text{Hom}_{\phi}(G , \on{Aut}^{\eff}(\fMTC))$, the set of $\Q$-projective homomorphisms with $\Q$-projective structure $\phi$.

Identical to the bosonic case, each $[\rho]$ defines an invariant $[\coho{O}]\in H^3_{[\rho]}(G,\mathcal{A})$ corresponding to a possible obstruction to localizing the symmetry action.
When $[\coho{O}]$ vanishes, the $G$ symmetry can be fractionalized.
However, the fractionalization classes are correlated with $\central$ through $\eta_{\psi_{\vv{0}}}= (-1)^{\central}$ [see Eq.~\eqref{eq:eta_psi_central}], and there is another invariant $[\coho{O}^{\central}]$ corresponding to a possible obstruction of fractionalization manifesting with the particular fermionic symmetry group $\mathcal{G}^{\eff} = \mathbb{Z}_2^{\eff}\times_{\central}G$ [see Eqs.~\eqref{eq:Gf_obstruction_general_1}, \eqref{eq:O_central_general}, and \eqref{eq:Gf_obstruction_general_2}].
When $[\coho{O}^{\central}]$ vanishes, the $G$ symmetry can fractionalize in a manner consistent with the fermionic symmetry group $\mathcal{G}^{\eff}$ specified by $\central$.
In this case, symmetry fractionalization for fixed $\mathcal{G}^{\eff}$ and $[\rho]$ is torsorially classified by $H^2_{[\rho^{(0)}]}(G,\mathcal{A}_\vv{0})$.
We note that $[\coho{O}^{\central}]$ vanishing implies that $[\coho{O}]$ also vanishes, so we need not consider it separately.

The final step is to extend to a $\mathcal{G}^{\eff}$-crossed SMTC $\fMTC_{\mathcal{G}^{\eff}}^{\times}$ that describes the symmetry defects.
The fermionic symmetry action and fractionalization structure on $\fMTC$ is all directly incorporated in $\fMTC_{\mathcal{G}^{\eff}}^{\times}$ through the subset of $\rho$, $U$-symbols, and $\eta$-symbols of $\fMTC_{\mathcal{G}^{\eff}}^{\times}$ that act on $\fMTC$.
The remaining structure comes from completing the defect theory, which means satisfying the pentagon and heptagon consistency conditions.
Since non-anomalous $\mathcal{G}^{\eff}$-crossed SMTCs are $G$-crossed FMTCs, the consistency conditions are exactly the same for bosonic and fermionic $G$-crossed theories.
The important distinctions between bosonic versus fermionic enters through constraints and equivalences on the theories.
This allows us to translate our knowledge about obstruction and classification of $G$-crossed MTCs (bosonic SET phases) to provide an understanding of obstruction and classification of $\mathcal{G}^{\eff}$-crossed SMTCs.
In particular, we see that the differences have already been incorporated in the symmetry action and fractionalization structure.
Identical to the bosonic case, the fractionalization class, specified by $\rho$, $U$-symbols, and $\eta$-symbols acting on $\fMTC$, defines the defectification obstruction $[\defectO] \in H^4(G,\text{U}(1))$.
When $[\defectO]$ vanishes, defect theories can be obtained, i.e. solutions to the pentagon and heptagon consistency equations exist.
When unobstructed, the extensions of the FMTC $\fMTC$ with $\mathcal{G}^{\eff}$ symmetry and a given fermionic symmetry action and fractionalization class to $\mathcal{G}^{\eff}$-crossed SMTCs are then torsorially classified by $H^3(G,\text{U}(1))$.
This classification includes the equivalences of $\mathcal{G}^{\eff}$-crossed SMTCs under fermionic vertex basis and symmetry action gauge transformations that respect the canonical isomorphism associated with the physical fermions, i.e. the equivalences are restricted as compared to the bosonic SET phases.
As with the bosonic case, this $H^3(G,\text{U}(1))$-torsor is generated by gluing (bosonic) $G$-SPT phases to the $\mathcal{G}^{\eff}$-crossed SMTCs.
The torsor functor of Ref.~\onlinecite{Aasen21} generates the theories related by $H^2_{[\rho^{(0)}]}(G,\mathcal{A}_\vv{0})$ and $H^3(G,\text{U}(1))$ torsor actions.

Thus, for FSET phases with fermionic symmetry group $\mathcal{G}^{\eff} = \mathbb{Z}_2^{\eff}\times_{\central}G$, there is a $\text{Hom}_{\phi}(G , \on{Aut}^{\eff}(\fMTC))$, $H^2_{[\rho^{(0)}]}(G,\mathcal{A}_\vv{0})$, and $H^3(G,\text{U}(1))$ classification structure.
As in the case of bosonic SET phases, nontrivial obstructions can (partially) break this classifying structure, as can equivalences given by relabeling topological charges, both of which can result in non-group subsets of this structure.
In the FSET case, we recall that the allowed relabelings must leave the $\mathcal{G}^{\eff}$ labels fixed; in particular, relabelings that change the vorticity ($\mathbb{Z}_2^{\eff}$ label) are not physical equivalences.
For FSPT phases, where $\on{Aut}^{\eff}(\fMTC)= \mathbb{Z}_{2}^{\V}$, $\mathcal{A}_\vv{0} = \mathbb{Z}_{2}^{\psi}$, and $\text{Hom}_{\phi}(G , \on{Aut}^{\eff}(\fMTC)) = H^1(G , \mathbb{Z}_{2}^{\V})$, we saw that the obstructions and equivalences conspired to combine this tiered cohomology group classification structure into a single classifying group $\Gr$ under the fermionic stacking operation.

The above perspective on classifying FSET phases treated the FMTC describing quasiparticles and vortices as fundamental, with the torsor action by the classifying cohomology groups relating different FSET phases.
Another perspective is to treat the SMTC describing quasiparticles as more fundamental, and structure the classification using extensions from the SMTC to the FMTC together with fermionic stacking with FSPT phases.
This generates a picture in which defectification is viewed as the extension of the SMTC to a $\mathcal{G}^{\eff}$-crossed defect theory, i.e. $\mathbb{Z}_{2}^{\eff}$ vortices are thought of as fermionic symmetry defects.
In this approach, one starts by specifying a SMTC $\fMTC_{\vv{0}}$ describing the quasiparticles of the fermionic topological phase with a fermionic symmetry group $\mathcal{G}^{\eff} = \mathbb{Z}_2^{\eff}\times_{\central}G$.
Then the first step is to choose a $\mathbb{Z}_2^{\eff}$ modular extension to a FMTC $\fMTC = \fMTC_\vv{0}\oplus \fMTC_\vv{1}$, where such extensions are classified torsorially by $\mathbb{Z}_{16}$.
Next, we specify a fermionic symmetry action on the quasiparticle sector $[\rho^{(0)}]: G \to \on{Aut}^{\eff}(\fMTC_\vv{0})$.
Again, this can potentially be $\Q$-projective when $[\Q]$ is nontrivial, i.e. when $\widehat{\bf A}_{\vv{1}} = \varnothing$, in which case we must have $\phi^{(\vv{0})} = \central$.
Fixing $\phi^{(\vv{0})}$, the possible fermionic symmetry actions on quasiparticles are classified by $\text{Hom}_{\phi^{(\vv{0})}}(G , \on{Aut}^{\eff}(\fMTC_{\vv{0}}))$.
Each symmetry action $[\rho^{(0)}]$ defines an invariant $[O^\rho]\in H^2(G,\mathbb{Z}_{2}^{\V})$ corresponding to a possible obstruction to extending $[\rho^{(0)}]$ to a fermionic symmetry action $[\rho]$ on $\fMTC$ [see Eq.~\eqref{eq:O2obs}].
(For $\widehat{\bf A}_{\vv{1}} \neq \varnothing$ and $\mathcal{A}_{\vv{1}} = \varnothing$, $[\rho^{(0)}]$ will not have $\Q$-projective structure, but $[\rho]$ can, so we fix the possibly nontrivial $\phi = \central$ of $[\rho]$ in this case as well.)
When $[O^\rho]$ vanishes, the symmetry action can be extended to $\fMTC$ and such extensions are classified torsorially by $H^1(G,\mathbb{Z}_{2}^{\V})$.

The next stage of the classification is symmetry fractionalization.
As with the full FMTC, each symmetry action $[\rho^{(0)}]$ defines an invariant $[\coho{O}^{(0)}]\in H^3_{[\widehat{\rho}]}(G,\widehat{\bf A})$ corresponding to a possible obstruction of localizing the symmetry action on quasiparticles.
Similarly, another invariant  $[\coho{O}^{(0)\central}]$ is a possible obstruction of fractionalization manifesting with the particular fermionic symmetry group $\mathcal{G}^{\eff} = \mathbb{Z}_2^{\eff}\times_{\central}G$ [see Eqs.~\eqref{eq:Gf_obstruction_general_M0_1} and \eqref{eq:Gf_obstruction_general_M0_2}.]
When $[\coho{O}^{(0)\central}]$ vanishes, the $G$ symmetry can fractionalize on quasiparticles in a manner consistent with the fermionic symmetry group $\mathcal{G}^{\eff}$ specified by $\central$.
In this case, the symmetry fractionalization on $\fMTC_\vv{0}$ for fixed $\mathcal{G}^{\eff}$ and $[\rho^{(0)}]$ is torsorially classified by $H^2_{[\widehat{\rho}^{(0)}]}(G,\widehat{\mathcal{A}}_\vv{0})$.
We note that $[\coho{O}^{(0)\central}]$ vanishing implies that $[\coho{O}^{(0)}]$ also vanishes.

Assuming a choice $[\rho]$ that extends $[\rho^{(0)}]$ to a fermionic action on $\fMTC$, we can consider the extension of symmetry fractionalization from $\fMTC_\vv{0}$ to $\fMTC$.
As before, $[\rho]$ defines an invariant $[\coho{O}]\in H^3_{[\rho]}(G,\mathcal{A})$ corresponding to a possible obstruction to localizing the symmetry action on $\fMTC$, and when $[\coho{O}]$ vanishes, the $G$ symmetry can be fractionalized.
However, we can instead consider another invariant $[\coho{O}^{\eta}]$ defined by $[\rho]$ and a particular fractionalization class on $\fMTC_\vv{0}$, specified by $[\rho^{(0)}, \eta^{(0)}]$, which corresponds to a possible obstruction to extending the fractionalization class from the quasiparticles to the vortices [see Eq.~\eqref{eq:obstructioneta}].
($[\coho{O}^{(0)\central}]$ is assumed to vanish and the quasiparticle fractionalization class to be extended is required to have $\eta^{(0)}_{\psi_{\vv{0}}} = (-1)^\central$.)
When $[\coho{O}^{\eta}]$ vanishes, the fractionalization class specified by $[\rho^{(0)}, \eta^{(0)}]$ on $\fMTC_\vv{0}$ can be extended to a fractionalization class $[\rho, \eta]$ on $\fMTC$.
Such unobstructed extensions for a fixed $[\rho^{(0)}, \eta^{(0)}]$ are classified torsorially by $H^2(G,\mathbb{Z}_2^{\psi})$.
We note that $[\coho{O}^{\eta}]$ vanishing implies that $[\coho{O}]$ also vanishes, so we need not consider it separately.

The relation between $H^2_{[\rho^{(0)}]}(G,\mathcal{A}_\vv{0})$, $H^2_{[\widehat{\rho}^{(0)}]}(G,\widehat{\mathcal{A}}_\vv{0})$, and $H^2(G,\mathbb{Z}_2^{\psi})$ was shown to be
\begin{align}
\label{eq:frac-fac-2}
H^2_{[\rho^{(\vv{0})}]}(G,\mathcal{A}_\vv{0}) \cong
i_{\ast}\left( H^2(G,\mathbb{Z}_2^{\psi}) \right) \times_{\varepsilon} p_{\ast} \left( H^2_{[\rho^{(\vv{0})}]}(G,\mathcal{A}_\vv{0}) \right)
.
\end{align}
Here, $p_{\ast} \left( H^2_{[\rho^{(\vv{0})}]}(G,\mathcal{A}_\vv{0}) \right)$ is the subgroup of $H^2_{[\widehat{\rho}^{(0)}]}(G,\widehat{\mathcal{A}}_\vv{0})$ corresponding to the quasiparticle fractionalization classes that have vanishing $[\coho{O}^{\eta}]$, i.e. the ones that can extend to $\fMTC$.
The homomorphism $i_{\ast}$ is the is the induced inclusion of the classifying group of extensions of fractionalization from quasiparticles to vortices into the classifying group of fractionalization on $\fMTC$.
We note that $i_{\ast}\left( H^2(G,\mathbb{Z}_2^{\psi}) \right)$ is not necessarily isomorphic to $H^2(G,\mathbb{Z}_2^{\psi})$, since the inclusion allows some of the extension classes to be related to each other by coboundaries in $B^2_{[\rho^{(\vv{0})}]}(G,\mathcal{A}_\vv{0})$.
It is, however, a central subgroup of $H^2_{[\rho^{(\vv{0})}]}(G,\mathcal{A}_\vv{0})$, where the central extension is indicated by some $\varepsilon$.

The final step of extending to the full defect theory described by $\mathcal{G}^{\eff}$-crossed SMTCs is identical to before, with the possible defectification obstruction $[\defectO]\in H^4(G,\text{U}(1))$ that must vanish for a defect theory to exist.
However, we now see that the fermionic symmetry action and fractionalization has additional structure coming from the extension from the SMTC $\fMTC_{\vv{0}}$ to the FMTC $\fMTC$, and that the structure associated with extensions can be introduced through fermionic stacking with $\mathcal{G}^{\eff}$ FSPT phases.
More specifically, we notice that for FSPT phases the classifying cohomology groups are $H^1(G , \mathbb{Z}_{2}^{\V})$, $H^2(G,\mathbb{Z}_{2}^{\psi})$, and $H^3(G,\text{U}(1))$.
As mentioned, this structure actually combines for FSPT phases into a single classifying group
\begin{align}
\Gr \cong \GSPT \times_{\varepsilon_{\mathsf{(3,21)}}} \left( \Grt  \times_{\varepsilon_{\mathsf{(2,1)}}} \Gro  \right)
,
\end{align}
where $\Gro \lhd H^1(G , \mathbb{Z}_{2}^{\V})$ is the subgroup corresponding to symmetry actions that are unobstructed (for fractionalization and defectification); $\Grt \lhd H^2(G,\mathbb{Z}_{2}^{\psi})$ is the subgroup corresponding to fractionalization classes for which the defectification obstruction vanishes (sometimes denoted $\BH^2(G,\mathbb{Z}_2)$); and $\GSPT \cong H^3(G,\text{U}(1)) / \sim$ corresponds to gluing bosonic SPT phases, up to fermionic equivalences.
These match up with the classification and torsors associated with extending the fermionic symmetry action from quasiparticles to vortices, extending the symmetry fractionalization from quasiparticles to vortices, and defectification.

Combing these observations, we can restructure the classification of $\mathcal{G}^{\eff}$ FSET phases to be given in terms of $\text{Hom}_{\phi}(G , \on{Aut}^{\eff}(\fMTC_{\vv{0}}))$, $p_{\ast} \left( H^2_{[\rho^{(\vv{0})}]}(G,\mathcal{A}_\vv{0}) \right) \lhd H^2_{[\widehat{\rho}^{(0)}]}(G,\widehat{\mathcal{A}}_\vv{0})$, and $\Gr$.
In this approach, which is summarized in the Fermionic SMTC column of Table~\ref{table:classification}, we first introduce and classify the fermionic symmetry action and fractionalization on the SMTC $\fMTC_{\vv{0}}$.
Then the defectification step involves not just the introduction of $G$ defects, but of all the $\mathcal{G}^{\eff}$ defects, including vortices.
In this manner, the $\mathbb{Z}_{16}$ torsor of $\mathbb{Z}_{2}^{\eff}$ modular extensions of the SMTC enters at the defectification step, along with the $\Gr$ torsor, which includes the classifications of extensions of the symmetry action and fractionalization from quasiparticles to vortices.
The torsorial action of $\Gr$ is introduced by stacking the $\mathcal{G}^{\eff}$ FSET phases with $\mathcal{G}^{\eff}$ FSPT phases.
The $\mathbb{Z}_{16}$ and $\Gr$ torsors may actually be combined into a single $\Gifo$ torsor, corresponding to stacking with invertible FSET phases, as we will explain in Sec.~\ref{sec:inv-class}.

This classification structure is again subject to the effects of obstructions and equivalences, though some of these are already incorporated in the FSPT structure.
For example, we saw that only the subset of quasiparticle fractionalization classes corresponding to $p_{\ast} \left( H^2_{[\rho^{(\vv{0})}]}(G,\mathcal{A}_\vv{0}) \right) \lhd H^2_{[\widehat{\rho}^{(0)}]}(G,\widehat{\mathcal{A}}_\vv{0})$ can actually be extended to fractionalization classes on the full FMTC.
As another example, we can see that the subgroup $\Grt \lhd H^2(G,\mathbb{Z}_{2}^{\psi})$ is the same when the FSPT gluing torsor action is applied to a general $\mathcal{G}^{\eff}$ FSET.
To see this, we can consider the torsor action of $\coho{p} \in Z^2(G,\mathbb{Z}_2^{\psi})$ on a general (unobstructed) $\mathcal{G}^{\eff}$-crossed SMTC.
The relative defectification obstruction is found to be
\begin{align}
\defectO_r(\coho{p} )  = (-1)^{ \mathsf{p} \cup \mathsf{p} + \central \cup \mathsf{p}}.
\end{align}
From this, we see that
\begin{align}
\defectO_r(\coho{p}\coho{p}') &= \defectO_r(\coho{p})\defectO_r(\coho{p}')(-1)^{\mathsf{p} \cup \mathsf{p}' + \mathsf{p}' \cup \mathsf{p}} \\
& = \defectO_r(\coho{p})\defectO_r(\coho{p}') \cbd (-1)^{\mathsf{p} \cup_1 \mathsf{p}'}
.
\end{align}
Additionally, for $\coho{z} \in C^1(G,\mathbb{Z}_2^{\psi}) $, we have
\begin{align}
\defectO_r(\cbd \coho{z})  &= (-1)^{\cbd \mathsf{z} \cup \cbd \mathsf{z} + \central \cup \cbd \mathsf{z}} \\
&= \cbd (-1)^{\mathsf{z} \cup \cbd \mathsf{z} + \central \cup \mathsf{z}}
.
\end{align}
In terms of equivalence classes related by coboundaries, we thus see that $\defectO_r$ defines a map $[ \coho{p} ] \mapsto [\defectO_r(\coho{p}) ]$, which is a homomorphism $[\defectO_r]: H^2(G,\mathbb{Z}_2^{\psi}) \to H^4(G,U(1))$.
As such, the subset of fractionalization extensions for which the defectification obstruction $[\defectO]$ vanishes corresponds to the subgroup $\ker{[\defectO_r]} = \Grt   \lhd H^2(G,\mathbb{Z}_2^{\psi})$.

Moreover, it is possible to have distinct FSPT phases that are trivialized when stacked with a particular FSET phase.
We saw this, for example, in the potential for $i_{\ast}$ to map a nontrivial element of $H^2(G,\mathbb{Z}_2^{\psi})$ to a trivial element of $H^2_{[\rho^{(\vv{0})}]}(G,\mathcal{A}_\vv{0})$.

\subsection{Stacking Invertible FSET Phases and the 16-Fold Way in General FSET Phases}
\label{sec:inv-class}

In this section, we discuss the group law for stacking invertible $\mathcal{G}^{\eff}$ FSET phases, and use it to explain how the 16-fold way, i.e. the $\mathbb{Z}_{16}$ torsor that cycles through the modular extensions of a given SMTC, can be incorporated into the FSET classification.
Our discussion of the $[\central]\neq[\vv{0}]$ case benefits from the explicit addition laws of $2$- and $3$-cochain presented in Ref.~\onlinecite{Bark2021inv} for stacking invertible FSET phases.

An invertible $\mathcal{G}^{\eff}$ FSET phase is fully specified by its chiral central charge, fermionic symmetry group, action, fractionalization class, and defectification class.
As such, we label each invertible $\mathcal{G}^{\eff}$ FSET phase using the quantities $\nu$, $G$, $\central$, $\maj$, $\mathsf{p}$, and $\xdot$, up to equivalences of theories, which we write as
\begin{align}
\ifo^{(\nu) \times}_{\mathcal{G}^{\eff}, [\xdot,\mathsf{p},\maj]_{\central}}.
\end{align}

We first consider the case $\mathcal{G}^{\eff} = \mathbb{Z}_2^{\eff}\times G$, i.e. $\central = \vv{0}$.
In this case, we can simplify the analysis of the stacking group structure by observing that all $\nu$ have a $[\rho] = [\openone]$ base theory given by
\begin{align}
\ifo^{(\nu) \times}_{\mathcal{G}^{\eff}, [0,0,0]_{\vv{0}}} &= \ifo^{(\nu)} \boxtimes \spt_G^{[1]}
,
\end{align}
and that all invertible FSET phases with $\central =\vv{0}$ can be represented as
\begin{align}
\label{eq:w0invfSET}
\ifo^{(\nu) \times}_{\mathcal{G}^{\eff}, [\xdot,\mathsf{p},\maj]_{\vv{0}}} &= \ifo^{(0) \times}_{\mathcal{G}^{\eff}, [\xdot,\mathsf{p},\maj]_{\vv{0}}} \underset{\mathcal{G}^{\eff}}{\ftimes}  \ifo^{(\nu) \times}_{\mathcal{G}^{\eff}, [0,0,0]_{\vv{0}}}
.
\end{align}
Fermionic stacking two such phases results in
\begin{widetext}
\begin{align}
\ifo^{(\nu')\times }_{\mathcal{G}^{\eff}, [\xdot',\mathsf{p}',\maj']_{\vv{0}}} \underset{\mathcal{G}^{\eff}}{\ftimes} \ifo^{(\nu'')\times}_{\mathcal{G}^{\eff}, [\xdot'',\mathsf{p}'',\maj'']_{\vv{0}}}
&=
\left( \ifo^{(0) \times}_{\mathcal{G}^{\eff}, [\xdot',\mathsf{p}',\maj']_{\vv{0}}}
\underset{\mathcal{G}^{\eff}}{\ftimes}
\ifo^{(0) \times}_{\mathcal{G}^{\eff}, [\xdot'',\mathsf{p}'',\maj'']_{\vv{0}}} \right)
\underset{\mathcal{G}^{\eff}}{\ftimes}
\left( \ifo^{(\nu') \times}_{\mathcal{G}^{\eff}, [0,0,0]_{\vv{0}}}
\underset{\mathcal{G}^{\eff}}{\ftimes}
\ifo^{(\nu'') \times}_{\mathcal{G}^{\eff}, [0,0,0]_{\vv{0}}}
\right)
\notag \\
&= \ifo^{(0) \times}_{\mathcal{G}^{\eff}, [\xdot,\mathsf{p},\maj]_{\vv{0}}}
\underset{\mathcal{G}^{\eff}}{\ftimes}
\ifo^{(\nu ) \times}_{\mathcal{G}^{\eff}, [0,0,0]_{\vv{0}}}
= \ifo^{( \nu ) \times}_{\mathcal{G}^{\eff}, [\xdot,\mathsf{p},\maj]_{\vv{0}}}
,
\end{align}
\end{widetext}
where $\nu = [\nu' + \nu''] \text{ mod } 16$ and $[\xdot,\mathsf{p},\maj]_{\vv{0}}$ is given by the FSPT stacking rule of Eq.~\eqref{eq:general-group}.
This follows from associativity and commutativity of fermionic stacking.
The equivalence relations of the theories with $\central =\vv{0}$ are inherited from the corresponding FSPT phases, meaning the label $[\xdot,\mathsf{p},\maj]_{\central}$ is viewed as an equivalence class of labels subject to the equivalence relations found in Eqs.~\eqref{eq:3coboundaryFSPequiv}, \eqref{eq:first-relab}, and \eqref{eq:Gfequiv}.
Thus, the group structure for stacking invertible FSET phases with $\central = \vv{0}$ is simply
\begin{align}
\label{eq:GIFO-simple}
\Gifo &= \Gr \times \mathbb{Z}_{16}
,
\end{align}
as the first term is set by the group law of stacking FSPT phases, while the latter is fixed by stacking invertible topological orders.
When $\central = \cbd \vv{u}$ is not strictly equal to $\vv{0}$, more care is needed in the analysis, since the $\nu$ odd theories have $\phi = \cbd \vv{u}$ $\Q$-projective symmetry action, and the $\nu$ even theories have their fractionalization classes modified with respect to their $\central = \vv{0}$ counterparts by  vortex-valued coboundaries.
However, the resulting stacking group structure is isomorphic to that of $\central = \vv{0}$.

Next, we consider the case where $[\central] \neq [\vv{0}]$.
In this case, the $\nu$ odd theories are all obstructed (see Sec.~\ref{sec:Ex_IFMTCs}), so they drop out of the analysis.
For the $\nu$ even theories, the analysis is complicated by the fact that $[\rho] = [\openone]$ is not always unobstructed, and hence we cannot represent all of the theories as we did for $\central = \vv{0}$ in Eq.~\eqref{eq:w0invfSET}.
In this case, we first check which $[\rho] = [\V]^{\maj}$ have trivial $[\coho{O}^{\central}]$.
Then for the unobstructed $[\rho]$, we can define the unobstructed invertible FSET phase by
\begin{align}
\ifo^{(\nu) \times}_{\mathcal{G}^{\eff}, [\xdot,\mathsf{p},\maj]_{\central}} &= \mathcal{F}_{\psi_{\vv{0}}^{\mathsf{p}} \otimes \I_{\vv{1}}^{\central},{\Xdot}} \left( \ifo^{(\nu) \times}_{\mathcal{G}^{\eff}, [0,0,\pi]_{\central}} \right)
,
\end{align}
where $\mathcal{F}$ is the torsor functor with $\mathsf{p}$ and $\Xdot$ that satisfy the conditions
\begin{align}
\cbd \mathsf{p} &= \maj \cup \central + \frac{\nu}{2} \central \cup_1 \central ,\\
\cbd\Xdot &= \left(\defectO_r^{(\nu)}(\psi^{\mathsf{p}}_{\vv{0}} \otimes \I_\vv{1}^\central, \maj) \right)^{-1}
,
\end{align}
as explained in Sec.~\ref{sec:invfSET}.

Again, the label $[\xdot,\mathsf{p},\maj]_{\central}$ corresponds to an equivalence class of characterizing data.
We compute the equivalence relations in Sec.~\ref{sec:ifoEquivalences}, which are found to be slightly modified from the FSPT equivalences for $\nu = 2 \text{ mod }4$, and are given by
\begin{widetext}
\begin{align}
(\xdot,\mathsf{p},\maj ; \nu)_{\central} &\sim (\xdot + \cbd \bdot ,\mathsf{p},\maj ; \nu)_{\central}
\\
&\sim
\left( \xdot + \frac{1}{2}( \mathsf{p} \cup_1 \cbd \mathsf{z}  + \central \cup \mathsf{z} + \pi \cup(\central \cup_2 \cbd \mathsf{z})+ \mathsf{z} \cup \cbd \mathsf{z} + \frac{\nu}{2} (\central \cup_1 \central) \cup_{2} \cbd \mathsf{z} )  ,\mathsf{p}_{}+ \cbd \mathsf{z},\maj ;\nu \right)_{\central}
\\
&\sim (\xdot + \frac{1}{2}(\mathsf{p} \cup_1 \central + \maj \cup \central) ,\mathsf{p}_{}+{\central}, \maj ;\nu)_{\central}
.
\end{align}
\end{widetext}

We solve the addition rules that provide the group law for stacking these theories in Sec.~\ref{sec:ifoAddition}, resulting in
\begin{align}
\label{eq:group-law-even-1}
\ifo^{(\nu) \times }_{\mathcal{G}^{\eff},[\xdot,\mathsf{p}, \maj]_{\central} }=
\ifo^{(\nu') \times }_{\mathcal{G}^{\eff},[\xdot',\mathsf{p}', \maj']_{\central} }
\underset{\mathcal{G}^{\eff}}{\ftimes}
\ifo^{(\nu'') \times }_{\mathcal{G}^{\eff},[\xdot'',\mathsf{p}'', \maj'']_{\central} }
,
\end{align}
where
\begin{align}
\nu &= [\nu' + \nu''] \text{ mod }16
,\\
\maj &= [\maj' +\maj''] \text{ mod } 2
,
\label{eq:majadd}
\\
\mathsf{p} &= [\mathsf{p}' + \mathsf{p}'' + \maj'\cup\maj''] \text{ mod } 2
,\\
\xdot &= [\xdot' + \xdot'' + \lambda] \text{ mod } 1,
\label{eq:xdotadd}
\end{align}
and $\lambda$ is the 3-cochain given in Eq.~\eqref{eq:3cochnaddeven}, which depends on $\central$, $\nu'$, $\nu''$, $\maj'$, $\maj''$, $\mathsf{p}'$, and $\mathsf{p}''$.

The resulting group structure for stacking invertible FSET phases with $[\central ]\neq [ \vv{0}]$ is given by a central extension
\begin{align}
\Gifo \cong \Gr \times_{\varepsilon_{\mathsf{(321,0)}}} \BZ ,
\end{align}
where $\BZ$ is the subgroup of $\mathbb{Z}_{16}$ corresponding to invertible FMTCs $\ifo^{(\nu)}$ that admit $\mathcal{G}^{\eff}$ extensions, i.e. have at least one  unobstructed $\ifo^{(\nu) \times}_{\mathcal{G}^{\eff}, [\xdot,\mathsf{p},\maj]_{\central}}$, and $\varepsilon_{\mathsf{(321,0)}} \in Z^2 (\BZ ,\Gr)$ is the 2-cocycle specifying the extension class.
When $[\central]  = [\vv{0}]$, we see that $\varepsilon_{\mathsf{(321,0)}}$ is trivial, as indicated by Eq.~\eqref{eq:GIFO-simple}.
When $[\central] \neq [\vv{0}]$, the group $\Gifo$ can be a nontrivial $\Gr$ extension of $\BZ$.
The extension class specified by $\varepsilon_{\mathsf{(321,0)}}$ is completely determined by Eqs.~\eqref{eq:majadd}-\eqref{eq:xdotadd}.

We can now define a torsorial action of $\Gifo$ on general $\mathcal{G}^{\eff}$ FSET phases through stacking with invertible FSET phases.
This combines the torsorial actions of $\mathbb{Z}_{16}$ (16-fold way) and $\Gr$ (FSPT stacking) into a single $\Gifo$ action in the lower right entry of Table~\ref{table:classification}.
Given an unobstructed $\mathcal{G}^{\eff}$ FSET phase described by $\fMTC_{\mathcal{G}^{\eff}}^{\times}$, stacking with an unobstructed invertible FSET phase yields another unobstructed $\mathcal{G}^{\eff}$ FSET phase
\begin{align}
\label{eq:IFO_stacking_torsor}
\fMTC_{\mathcal{G}^{\eff}}^{\prime \times} = \ifo^{(\nu) \times }_{\mathcal{G}^{\eff},[\xdot,\mathsf{p}, \maj]_{\central} }
\underset{\mathcal{G}^{\eff}}{\ftimes} \fMTC_{\mathcal{G}^{\eff}}^{\times}
.
\end{align}
The resulting $\fMTC_{\mathcal{G}^{\eff}}^{\prime \times}$ and $\fMTC_{\mathcal{G}^{\eff}}^{\times}$ share all the same structure on the quasiparticle sector, i.e. they have the same SMTC $\fMTC_{\vv{0}}$, fermionic symmetry action and fractionalization $[\rho^{(0)}, \eta^{(0)}]$, and, of course, the same $\central$ defining the fermionic symmetry group.
However, this torsor action cycles through different extensions of these quantities to the vortex sector, potentially changing the FMTC $\fMTC$, the extended action and fractionalization classes $[\rho, \eta]$, as well as the defectification class of the $\mathcal{G}^{\eff}$-crossed FSET.

As seen in Eq.~\eqref{eq:GIFO-simple}, this amounts to a trivial product structure for trivial $\central$, where one can simply cycle through the 16-fold way of FMTCs ($\mathbb{Z}_{2}^{\eff}$ modular extensions of a SMTC $\fMTC_{\vv{0}}$) independently of the FSPT stacking torsor.
Thus, every FMTC $\fMTC$ extending $\fMTC_{\vv{0}}$ has the same $\mathbb{Z}_{2}^{\eff} \times G$ FSET classification of $\mathcal{G}^{\eff}$-crossed extensions.

For nontrivial $\central$, the situation is more complicated, but aspects of it can be understood using relations found in Sec.~\ref{sec:symmetric-fermions}.
The most clear disparity arises from the fact that there are no invertible FSET phases with $\nu$ odd, so stacking with invertible FSETs can only relate phases differing by $\nu$ even.
This can be understood via the relation of Eq.~\eqref{eq:Orho_nu_relation}.
Stacking with $\nu$ odd relates two FMTCs for which either or both have $\mathcal{A}_{\vv{1}} =\varnothing$, and hence the $\Q$-projective factor involved will be $\phi = \central$.
This shows that, for nontrivial $\central$, when there is an unobstructed $[\rho]$ extending $[\rho^{(0)}]$ for $\fMTC$, the extension of $[\rho^{(0)}]$ for $\fMTC' = \ifo^{(\nu)} \fprod \fMTC$ will be obstructed for $\nu$ odd and unobstructed for $\nu$ even.
The obstruction $[O^{\rho}]$ needs to be computed to determine which half, if either, of the FMTCs are unobstructed.

The interplay of the fractionalization obstruction structure of $\ifo^{(\nu)}$ for $\nu$ even and FMTCs related by stacking with such phases can be understood via Eq.~\eqref{eq:Oeta_relation}.
In particular, when the $\ifo^{(\nu)}$ obstruction $[\psi_{\vv{0}}^{\pi \cup \central + \frac{\nu}{2} \cdot \central \cup_{1} \central}]$ is nontrivial for all $\pi$, there is no unobstructed $\fMTC_{\mathcal{G}^{\eff}}^{\prime \times}$ with the same $[\rho^{(0)}, \eta^{(0)}]$ as $\fMTC_{\mathcal{G}^{\eff}}^{\times}$.
However, it is possible to have an unobstructed $\fMTC_{\mathcal{G}^{\eff}}^{\prime \times}$ with $[\rho^{(0)}, \eta^{\prime (0)}]$, if and only if there is a $[\coho{k}^{(0)}] \in H_{[\widehat{\rho}^{(\vv{0})}]}^{2}(G,\widehat{\mathcal{A}}_{\vv{0}})$ such that $[\coho{T}_{\rho^{(0)}}( \coho{k}^{(0)} )] = [ \psi^{\pi \cup \central + \frac{\nu}{2} \cdot \central \cup_1 \central }]$.
When this is the case, $\fMTC$ and $\fMTC'$ can have disjoint classification structures that are not connected by stacking with invertible FSET phases.
As such, one should be careful not to rule out the existence of FSET phases for $\fMTC'$, simply because they cannot be obtained from stacking FSET phases for $\fMTC$ with invertible FSET phases.
A similar situation may occur for the defectification obstruction structure, but we will not examine it in detail here.
Finally, we remark that distinct fractionalization and defectification classes of unobstructed $\ifo^{(\nu) \times }_{\mathcal{G}^{\eff},[\xdot,\mathsf{p}, \maj]_{\central} }$ may be trivialized by $\fMTC_{\mathcal{G}^{\eff}}^{\times}$, so stacking with two distinct invertible FSET phases will not necessarily yield two distinct FSET phases $\fMTC_{\mathcal{G}^{\eff}}^{\prime \times}$.
This was already seen for the fractionalization classification by the fact that $i_\ast (H^{2}(G,\mathbb{Z}_{2}^{\psi}))$ is not necessarily isomorphic to $H^{2}(G,\mathbb{Z}_{2}^{\psi})$.

\section{Examples}\label{sec:examples}

We now examine the defect theories and classification of several examples of $\mathcal{G}^{\eff}$ FSET phases in detail.
We begin with the invertible FMTCs $\ifo^{(\nu)}$.
As these phases have trivial quasiparticle sector, they are completely classified by their stacking with FSPT phases.
We first illustrate this by focusing on $\nu$ odd theories with $\mathbb{Z}_2$ symmetry in Sec.~\ref{sec:ex-IsingZ2}, for which we recover the $\mathbb{Z}_8$ torsorial classification described in Sec.~\ref{sec:Z8}.
We then treat the general case with $G$ symmetry for $\nu$ odd in Sec.~\ref{sec:ex-A} and for $\nu$ even in Sec.~\ref{sec:invfSET}.

We next consider FMTCs with trivial symmetry action on the quasiparticles.
Building upon the fractionalization analysis of Sec.~\ref{sec:Ex_no_perm}, we split these theories into three cases corresponding to whether or not $\mathcal{A}_\vv{1}$ and ${\mathbf{A}}_{\vv{1}}$ are empty.

In Sec.~\ref{sec:ex_KM0}, we examine FMTCs that take a factorized form $\fMTC=\ifo^{(\nu)}\boxtimes \widehat{\fMTC}_\vv{0}$, with symmetry action that similarly factorizes.
This product structure allows us to analyze the sectors individually, combining the results of the classification of invertible fermionic phases and bosonic MTCs.
While many FMTCs take the form $\fMTC=\ifo^{(\nu)}\boxtimes \widehat{\fMTC}_\vv{0}$, including all those containing a $\ifo^{(\nu)}$ subcategory, there are several notable exceptions.
One class of FMTCs that do not take this product structure are the Pfaffian states, which we analyze in Sec.~\ref{sec:ex_MooreRead}.
These are of significant physical interest due to their description of a number of non-Abelian fractional quantum Hall phases.
Another class of FMTCs that do not take this product structure are given by the modular extensions of the SMTCs $\operatorname{\text{SO}(3)}_{4n+2}$, which include $\fMTC = \operatorname{\text{SU}(2)}_{4n+2}$, which we examine in Sec.~\ref{sec:ex_SU2}.
These theories are notable in that they have $\widehat{\bf A}_{\vv{1}} = \varnothing$.

\subsection{Invertible Fermionic Topological Phases \texorpdfstring{$\ifo^{(\nu)}$}{Knuodd} for \texorpdfstring{$\nu$}{nu} odd with \texorpdfstring{$G = \mathbb{Z}_2$}{Z2Ising} Symmetry}
\label{sec:ex-IsingZ2}

In this section, we consider the FMTCs $\ifo^{(\nu)}$ for $\nu$ odd with $G = \mathbb{Z}_2$ symmetry.
These are the invertible fermionic topological phases with trivial quasiparticle sector $\ifo^{(\nu)}_\vv{0} = \{\I_\vv{0},\psi_\vv{0}\}$ and vortex sector $\ifo^{(\nu)}_\vv{1}=\{ \sigma_\vv{1}\}$.
Their full data is given in the bottom panel of Table~\ref{table:sixteen}.
We compare and contrast the fermionic classification with the analogous bosonic classification of $\mathbb{Z}_2$-crossed extensions of $\operatorname{Ising}^{(\nu)}$.

In this case, it is convenient to specify the fermionic symmetry action by $\rho_{\bf g} = \Q^{\maj({\bf g})} $ where $\maj \in H^1(\mathbb{Z}_2,\mathbb{Z}_2^{\Q})  = \mathbb{Z}_2$.
Note that $C^1(\mathbb{Z}_2,\mathbb{Z}_2) = H^1(\mathbb{Z}_2,\mathbb{Z}_2)$, so the fermionic symmetry action is automatically an ordinary group homomorphism $[\rho]: \mathbb{Z}_2 \to \Autf{\ifo^{(\nu)}} = \mathbb{Z}_2^{\Q}$, i.e. $\Q$-projectiveness is automatically trivial for $G = \mathbb{Z}_2$.
In Sec.~\ref{sec:Ex_IFMTCs}, we saw that the symmetry fractionalization obstruction vanishes, and for $G = \mathbb{Z}_2$, we must have $\central = \vv{0}$, that is $\mathcal{G}^{\eff} = \mathbb{Z}_2^{\eff} \times \mathbb{Z}_2$.
The symmetry fractionalization is torsorially classified by $\coho{p} \in H^2(\mathbb{Z}_2,\mathbb{Z}_2^{\psi})= \mathbb{Z}_2$, with the generator given by $\coho{p}({\bf 1,1}) = \psi_{\vv{0}}$.
Writing $\coho{p} = \psi_{\vv{0}}^{\mathsf{p}}$ allows us to write the defectification obstruction as $\defectO(\coho{p}) = (-1)^{\mathsf{p} \cup \mathsf{p}}$.
This obstruction vanishes, as we can write $\cbd \Xdot = \defectO$ for the $3$-cochain whose only nontrivial element is given by
\begin{align}
\Xdot({\bf 1},{\bf 1},{\bf 1}) &= \begin{cases}
\pm 1 &\text{ if } \mathsf{p}({\bf 1},{\bf 1}) = 0\\
\pm i &\text{ if } \mathsf{p}({\bf 1},{\bf 1}) =1\\
\end{cases}
.
\end{align}
Thus, we see that there are eight $\mathbb{Z}_2^{\eff} \times \mathbb{Z}_2$-crossed extensions of $\ifo^{(\nu )}$ for $\nu$ odd, corresponding to the possible combinations of the symmetry actions $\maj$, fractionalization classes $\coho{p}$, and defectification classes $\Xdot$.

The topological data for four of these theories can be found using the torsor method.
We begin with  $\ifo^{(\nu)} \boxtimes \spt_{\mathbb{Z}_2}^{[1]}$, which has $\pi =\vv{0}$, and then use the four possible torsor functors specified by the four combinations of $\coho{p}$ and $\Xdot$.
The remaining theories with $\pi({\bf 1}) =\vv{1}$ are found using the $\vviso$-isomorphism described in Sec.~\ref{sec:centralisomorphism}, with $\vviso({\bf 0}) = \vv{0}$ and $\vviso({\bf 1}) = \vv{1}$ (note $\cbd \vviso = 0$).
Rather than presenting the full data of these theories, in Table~\ref{table:Z2Isingtheories} we give a set of gauge invariant quantities which distinguishes them.
\begin{table}
\begin{center}
\bgroup
\def\arraystretch{1}
\setlength\tabcolsep{1.5ex}
 \begin{tabular}{c c c c c c}
    $n$
& $\Xdot({\bf 1,1,1})$
& $\coho{p}({\bf 1,1})$
& $ \rho_{\bf 1} $
& $T^2(\widehat{\I}_{\vv{0},{\bf 1}})$
& $T^2(\widehat{\sigma}_{\vv{0},{\bf 1}})$
\\ [0.5ex]
 \hline
 &&&&&\\[-1.5ex]
$0$ & $1$ & $\I$ & $\openone$ & $1$ &\\[1ex]
$1$ & $e^{-i \pi \frac{(\nu -1)}{4}}$ & $\psi^{\nu(\nu-1)/2}$ & $\Q$& &$e^{i \pi /4}$\\[1ex]
$2$ & $-i$ & $\psi$ & $\openone$ &$i$&\\[1ex]
$3$ & $e^{-i \pi \frac{(\nu -3)}{4}}$ & $ \psi^{\nu(\nu+1)/2}$ & $\Q$& &$e^{3i \pi /4}$\\[1ex]
$4$ & $-1$ & $\I$ & $\openone$ &$-1$&\\[1ex]
$5$ & $e^{-i \pi \frac{(\nu -5)}{4}}$ & $\psi^{\nu(\nu-1)/2}$ & $\Q$& &$e^{5i \pi /4}$\\[1ex]
$6$ & $i$ & $\psi$ & $\openone$ & $-i$ &\\[1ex]
$7$ & $e^{-i \pi\frac{(\nu -7)}{4}}$ & $\psi^{\nu(\nu+1)/2}$ & $\Q$& &$e^{7i \pi /4}$\\
\end{tabular}
\egroup
\caption{The $\mathbb{Z}_8$ structure assigned to $\mathbb{Z}_2^{\eff} \times \mathbb{Z}_2$-crossed extensions of $\ifo^{(\nu)}$ for $\nu$ odd.
The row $n=0$ is $\ifo^{(\nu)} \boxtimes \spt_{\mathbb{Z}_2}^{[1]}$.
The remaining theories are found from stacking with $n$ copies of the generating FSPT phase defined in Sec.~\ref{sec:FSPTphasesZ2}.
The columns labeled $\Xdot$, $\coho{p}$, and $\rho_{\bf 1}$ specify the torsor functor data and $u$-isomorphism that take the $n=0$ theory in the first row to the theory labeled by row $n$.
Each theory can be uniquely distinguished by the topological invariant $T^2(\widehat{a}_{\vv{0},{\bf 1}})$ listed in the last two columns [see Eq.~\eqref{eq:Tsquared}].
The nontrivial $\vviso$-isomorphism applied to theory $n$ results in theory $\nu - n \mod 8$.
}
\label{table:Z2Isingtheories}
\end{center}
\end{table}

We now compare these results to that of the bosonic counterparts, $\mathbb{Z}_2$-crossed extensions of $\operatorname{Ising}^{(\nu)}$, which were completely classified in Ref.~\onlinecite{Bark2019}.
The main difference is that $\operatorname{Aut}(\operatorname{Ising}^{(\nu)}) = \mathbb{Z}_1$, and so the symmetry action is always trivial.
This stems from the fact that both $U_{\bf 1}(\psi,\psi; \I)$ and $\gamma_{\psi}({\bf 1})$ are unconstrained in the bosonic theory.
Consequently, the classification is given only by the two symmetry fractionalization classes, $\coho{p}({\bf 1,1})=\I$ and $\psi$, and the two defectifcation classes associated with $H^3(\mathbb{Z}_2,\text{U}(1)) = \mathbb{Z}_2$.
Thus, the $n=\operatorname{even}$ and $n = \operatorname{odd}$ rows of the eight theories appearing in Table~\ref{table:Z2Isingtheories} collapse onto one another, resulting in only four distinct bosonic theories.

\subsection{Invertible Fermionic Topological Phases \texorpdfstring{$\ifo^{(\nu)}$}{Knuodd} for \texorpdfstring{$\nu$}{nu} odd }
\label{sec:ex-A}

In this section, we consider the general $\mathcal{G}^{\eff}$-crossed extensions of $\ifo^{(\nu)}$ for $\nu$ odd.
The fermionic topological symmetry group and symmetry action were found in  Sec.~\ref{sec:Ex_IFMTCs} to be
\begin{align}
\operatorname{Aut}^{\eff}(\ifo^{(\nu)}) &= \mathbb{Z}_2^{\V} = \mathbb{Z}_2^{\Q}
,\\
\rho_{\bf g} = \Q^{\rr({\bf g}) },
\label{eq:Isingrho}
\end{align}
for $\rr \in C^1(G,\mathbb{Z}_2)$.
Since $[\Q]$ is nontrivial, the fermionic symmetry action has $\Q$-projective factors $\phi = \cbd \rr$.
The action fixes the $U$-symbols on the trivial sector to be
\begin{align}
\label{eq:IsingUsym}
U_{\bf g}(\sigma_{\vv{1}},\psi_{\vv{0}}; \sigma_{\vv{1}} )  &= U_{\bf g}(\psi_{\vv{0}},\sigma_{\vv{1}}; \sigma_{\vv{1}} ) \notag \\
&=U_{\bf g}(\sigma_{\vv{1}},\sigma_{\vv{1}};\psi_{\vv{0}}  )=(-1)^{\rr({\bf g})}.
\end{align}
Symmetry fractionalization was also analyzed in Sec.~\ref{sec:Ex_IFMTCs}, where we saw that $[\coho{O}]$ vanishes, $[\coho{O}^{\central}]$ fixes
\begin{align}
\central &= \cbd \rr
,
\end{align}
and the $\eta$-symbols are given by
\begin{align}\label{eq:Isingeta}
\eta_{a_{\vv{x}}}({\bf g,h}) &= (-1)^{\vv{x} \cdot \mathsf{p}({\bf g,h})  +\cbd \rr({\bf g,h}) \cdot \delta_{a_{\vv{x}},\psi_{\vv{0}}}}
.
\end{align}

In order to incorporate symmetry defects, we can start by constructing the theories with $\rho = \openone$ and $\central = \vv{0}$, and then generate the rest by stacking with FSPT phases.
Since $\ifo^{(\nu)}$ has trivial quasiparticle sector, all $\mathcal{G}^{\eff}$-crossed extensions with fixed $\central$ are torsorially related by stacking with FSPT phases.
In particular, we define a simple base theory by taking the direct product
\begin{align}
\ifo^{(\nu) \times}_{\mathcal{G}^{\eff} , [0,0,0]_{\vv{0}}} &= \ifo^{(\nu)} \boxtimes \spt_G^{[1]}
.
\end{align}
The topological charges of the $(\vv{x}, {\bf g})$ sectors are given by
\begin{align}
\fMTC_{\vv{0},{\bf g}} &= \{ \I_{\vv{0},{\bf g}}, \psi_{\vv{0},{\bf g}} \}\notag \\
\fMTC_{\vv{1},{\bf g}} &= \{\sigma_{\vv{1},{\bf g}} \}\notag
\end{align}
The $F$- and $R$-symbols can be determined by the product structure, and the $U$- and $\eta$-symbols are trivial (corresponding to trivial symmetry action and fractionalization).
Then the remaining theories with $\central = \vv{0}$ are obtained by stacking with FSPT phases as
\begin{align}
\ifo^{(\nu) \times}_{\mathcal{G}^{\eff}, [\xdot,\mathsf{p},\pi]_{\vv{0}}} &=
\ifo^{(0) \times}_{\mathcal{G}^{\eff}, [\xdot,\mathsf{p},\pi]_{\vv{0}}} \underset{\mathcal{G}^{\eff}}{\ftimes}  \ifo^{(\nu) \times}_{\mathcal{G}^{\eff} , [0,0,0]_{\vv{0}}}
.
\end{align}

While this is a natural way to generate the theories from the perspective of classification, there is an easier method to generate the explicit data of these theories than directly using stacking.
In particular, we can utilize the torsor functor and the $\vviso$-isomorphism of Sec.~\ref{sec:centralisomorphism}.
In this way, we generate the theories with trivial symmetry action and $\central =\vv{0}$ by applying the torsor functor to the base theory
\begin{align}
\ifo^{(\nu) \times}_{\mathcal{G}^{\eff}, [\xdot,\mathsf{p},0]_{\vv{0}}} &=
\mathcal{F}_{\psi_{\vv{0}}^{\mathsf{p}} ,{\Xdot}} \left(\ifo^{(\nu) \times}_{\mathcal{G}^{\eff} , [0,0,0]_{\vv{0}}} \right).
\end{align}
The symmetry fractionalization is completely determined by
\begin{align}
[\coho{p}] \in H^2(G,\mathbb{Z}_2^{\psi}).
\end{align}
Letting $\coho{p} \in[\coho{p}]$ be a representative of the cohomology class and using Ref.~\onlinecite{Aasen21}, the relative defectification obstruction valued in $H^4(G,\text{U}(1))$ is given by
\begin{align}
\defectO_{r}({\bf g,h,k,l}) = R^{\cohosub{p}({\bf k,l}) , \cohosub{p}({\bf g,h})}.
\end{align}
As the symmetry action on $\psi_{\vv{0}}$ is trivial and $\coho{p}({\bf g,h})\in \mathbb{Z}_2^\psi$, the relative obstruction can be succinctly written as
\begin{align}
\defectO_{r} = (-1)^{\mathsf{p} \cup \mathsf{p}}.
\end{align}
Since the base theory is unobstructed, the defectification obstruction is in fact equal to the relative obstruction.
If the theory is unobstructed, meaning that there exists an $\Xdot \in C^3(G,\text{U}(1))$ such that $\cbd \Xdot =  (-1)^{\mathsf{p} \cup \mathsf{p}}$, then one can read off all topological data of the post-torsor functor theory using the expressions in Ref.~\onlinecite{Aasen21}.
This generates all possible $\mathbb{Z}_2^{\eff} \times G$-crossed extensions of $\ifo^{(\nu )}$ with trivial symmetry action.

Next, we can generate the theories with nontrivial symmetry action $\rho_{\bf g} = \Q^{\rr({\bf g})}$ using the $\vviso$-isomorphism described in Sec.~\ref{sec:centralisomorphism}, with $\vviso = \rr$.
This isomorphism also modifies the topological data as described in Sec.~\ref{sec:centralisomorphism}, such that
\begin{align}
\fMTC_{\rr({\bf g}),{\bf g}} &=
\{ \I_{\rr({\bf g}),{\bf g}}, \psi_{\rr({\bf g}),{\bf g}}\} ,
\\
\fMTC_{\vv{1} -\rr({\bf g}),{\bf g}} & = \{ \sigma_{\vv{1} -\rr({\bf g}),{\bf g}} \}
.
\end{align}
As pointed out in Sec.~\ref{sec:centralisomorphism}, we see that a $\vviso$-isomorphism changes $\central \to \central + \cbd \vviso$.
Our initial theory had $\central = 0$, so the resulting theory will have $\central = \cbd \rr$.
However, if we choose $\rr = \pi \in H^1(G,\mathbb{Z}_2^{\Q})$, then $\cbd \rr = 0$ and $\central$ would be unchanged.
Thus, we obtain the remaining $\central = \vv{0}$ theories by applying $\vviso$-isomorphisms with $\rr = \pi \in H^1(G,\mathbb{Z}_2^{\Q})$ to the previously obtained theories
\begin{align}
\ifo^{(\nu) \times}_{\mathcal{G}^{\eff}, [\xdot,\mathsf{p},\pi]_{\vv{0}}} &=
\left( \ifo^{(\nu) \times}_{\mathcal{G}^{\eff}, [\xdot,\mathsf{p},0]_{\vv{0}}} \right)_{\vviso = \pi}
.
\end{align}

On the other hand, we also obtain all of the $\central = \cbd \rr$ theories by applying a $\vviso$-isomorphism with $\vviso = \rr$ to the $\central = \vv{0}$ theories.
We see that the set of $\mathcal{G}^{\eff}$-crossed extensions of $\ifo^{(\nu)}$ for $\nu$ odd split into subsets which are invariant under stacking with FSPT phases.
As noted at the end of Sec.~\ref{sec:Ex_IFMTCs}, these subsets are labeled by $C^1(G,\mathbb{Z}_2)/Z^1(G,\mathbb{Z}_2) \cong B^2(G,\mathbb{Z}_2)$, corresponding to the possible choices of $\central = \cbd \rr$.
There is an arbitrary choice in defining the base theory when $\central \neq 0$, as $\vviso' = \rr'$ with $\cbd \rr' = \cbd \rr$ provide an equally valid way of obtaining the same $\central$, so the classification is torsorial.

\subsection{Invertible Fermionic Topological Phases \texorpdfstring{$\ifo^{(\nu)}$}{Knuodd} for \texorpdfstring{$\nu$}{nu} even}
\label{sec:invfSET}

In this section, we consider the $\mathcal{G}^{\eff}$-crossed extensions of the invertible fermionic topological phases ${\bf K}^{(\nu)}$ for $\nu$ even with general $\mathcal{G}^{\eff} = \mathbb{Z}_2^{\eff} \times_{\central} G$.
The symmetry action and fractionalization for ${\bf K}^{(\nu)}$ was analyzed in detail in Sec.~\ref{sec:Ex_IFMTCs}.
Here, we generate the complete data of all the defect theories by first applying the defect zesting functor developed in Sec.~\ref{sec:defect_torsor_functors} to a FSPT phase with $\central =0$, which changes ${\bf K}^{(0)}$ to ${\bf K}^{(\nu)}$ for $\nu$ even.
Then, we apply the $G$-crossed torsor functor method of Ref.~\onlinecite{Aasen21} to these theories with $\central =\vv{0}$ and general $\nu$ in order to generate the theories with nontrivial $\central$ (while leaving $\nu$ fixed).
This class of examples serves as a special case of the example considered in Sec.~\ref{sec:ex_FMTCSPT}.
For $\central =\vv{0}$, there is a bijection between the FSPT (i.e. $\nu=0$) phases and $\nu$ even invertible fermionic topological phases with $\mathcal{G}^{\eff} = \mathbb{Z}_2^{\eff} \times G$, given by the defect zesting functor.
In other words, the obstructions are the same for all $\nu$ even with $\central = \vv{0}$.
When $\central$ is nontrivial, we encounter additional obstructions for $\nu \neq 0$ that are not present for the FSPT phases.
The nontrivial $\central$ case of this example was revised to provide the obstruction and complete data after Ref.~\onlinecite{Bark2021inv} appeared on the arXiv, which computed the defectification obstruction and showed that it reduces to a universal form.
Our method provides a natural choice of gauge that allows us to address all $\nu$ $\operatorname{even}$ theories simultaneously.
This also allows us to derive the addition rule of symmetry fractionalization classes under fermionic stacking, proving a conjecture made in Ref.~\onlinecite{Bark2021inv}.

When $\central =\vv{0}$, we generate the defect theories for $\nu \neq 0$ by applying the defect zesting torsor of Sec.~\ref{sec:defect_torsor_functors} to the corresponding FSPT phases.
In this way, we define
\begin{align}
\ifo^{(\nu) \times} _{\mathbb{Z}_2^{\eff} \times G, [\xdot,\mathsf{p}, \maj]_{\vv{0}}} &=
\mathcal{Z}^{(\nu)} \left(\ifo^{(0) \times}_{\mathbb{Z}_2^{\eff} \times G, [\xdot,\mathsf{p}, \maj]_{\vv{0}}} \right)
.
\end{align}
The complete data is obtained by plugging the FSPT phase data from Sec.~\ref{sec:fSPT} into Eqs.~\eqref{eq:zest_N_explicit}-\eqref{eq:zest_eta_explicit}.
There is no relative obstruction for applying the defect zesting functor, so the obstructions are the same for all $\nu$ even.
In this section, it will be very convenient to use the data obtained from applying the defect zesting functor without the additional vertex basis gauge transformation that returns the data fully into the canonical gauge, as described in Sec.~\ref{sec:defect_torsor_functors}.
While the most important canonical gauge choices are still preserved in this way, the data can be transformed to possess all the canonical gauge choices by a vertex basis gauge transformation, when desired, and the resulting properties will be isomorphic (the characterizing 3-cochains will differ by a 3-coboundary).

The data of the ${\bf g } = {\bf 0}$ sector of $\ifo^{(\nu) \times} _{\mathbb{Z}_2^{\eff} \times G, [\xdot,\mathsf{p}, \maj]_{\vv{0}}}$ will be useful for computing the relative obstruction when we apply the torsor functor, so we will write it out more explicitly.
Restricting to the ${\bf g } = {\bf 0}$ sector, we have the fusion rules
\begin{align}
a_{\vv{x}} \tortilde{\otimes} b_{\vv{y}} =  [\psi^{\frac{\nu}{2} \cdot \vv{x} \cdot \vv{y}}ab]_{\vv{x+y}}
,
\end{align}
and the following topological data
\begin{align}
\label{eq:tortildeF}
\tortilde{F}^{a_{\vv{x} }b_{\vv{y}} c_{\vv{z}}} &= i^{\frac{\nu}{2} \cdot \vv{x} \cdot \vv{y} \cdot \vv{z}}(-1)^{\frac{\nu}{2} \cdot \mathsf{a} \cdot \vv{y} \cdot \vv{z}},
\\
\label{eq:tortildeR}
\tortilde{R}^{a_{\vv{x}} b_{\vv{y}} } &= e^{i\frac{\pi}{8} \nu \cdot \vv{x} \cdot \vv{y}} (-1)^{(\mathsf{a} + \vv{x} )\cdot \mathsf{b}},
\\
\label{eq:tortildeU}
\tortilde{U}_{\bf g}(a_{\vv{x}},b_{\vv{y}} ; [\psi^{\frac{\nu}{2}\cdot \vv{x} \cdot \vv{y}}ab]_{\vv{x+y}})  &= (-1)^{(\mathsf{a} +\vv{x})\cdot\vv{y} \cdot \pi({\bf g})},
\\
\label{eq:tortildeeta}
\tortilde{\;\eta}_{a_{\vv{x}}}({\bf g,h}) & = i^{\vv{x} \cdot \pi({\bf g})\cdot \pi({\bf h})} (-1)^{\vv{x} \cdot \mathsf{p}}
.
\end{align}

Next, we generate all the nontrivial $\central$ theories and their complete data by applying the $G$-crossed torsor functor method of Ref.~\onlinecite{Aasen21} to $\ifo^{(\nu) \times} _{\mathbb{Z}_2^{\eff} \times G, [\xdot,\mathsf{p}, \maj]_{\vv{0}}} $.
A convenient way of organizing this approach is to first select an unobstructed base theory for each $\pi$, which we take to be $\ifo^{(\nu) \times} _{\mathbb{Z}_2^{\eff} \times G, [0,0, \maj]_{\vv{0}}}$, and then apply all the unobstructed torsor functors.
In this way, we define
\begin{align}
\ifo^{(\nu) \times }_{\mathcal{G}^{\eff},[\xdot,\mathsf{p}, \maj]_{\central} }&=
\mathcal{F}_{\psi_{\vv{0}}^{\mathsf{p}} \otimes \I_{\vv{1}}^{\central},{\Xdot}} \left(\ifo^{(\nu) \times} _{\mathbb{Z}_2^{\eff} \times G, [0,0, \maj]_{\vv{0}}} \right)
,
\end{align}
where $\mathsf{p} \in C^2(G,\mathbb{Z}_2^{\psi})$ and $\xdot \in C^3(G,\mathbb{R}/\mathbb{Z})$.
(Recall that $\Xdot = e^{i 2\pi \xdot}$.)
From Sec.~\ref{sec:Ex_IFMTCs}, we saw that $\mathsf{p}$ must satisfy
\begin{align}
\label{eq:nueven2cochain}
\cbd \mathsf{p} = \maj \cup \central + \frac{\nu}{2} \central \cup_1 \central
\end{align}
in order for fermionic symmetry fractionalization to be unobstructed.
Assuming such an unobstructed $\mathsf{p}$, we will now show that $\xdot$ must satisfy
\begin{align}
\label{eq:nueven3cochain}
\cbd \xdot =& -\frac{1}{4} \widetilde{\maj} \cup  \widetilde{\maj} \cup  \widetilde{\central} \notag \\
&+\frac{1}{2}(\mathsf{p} \cup \mathsf{p} + \mathsf{p} \cup \central + \maj \cup(\mathsf{p} \cup_1 \central) + (\maj \cup_1 \mathsf{p}) \cup \central ) \notag\\
&-\frac{\nu}{16}( \widetilde{\central} \cup  \widetilde{\central }+2  \widetilde{\central }\cup_1 ( \widetilde{\central} \cup_1  \widetilde{\central}) + 4 \mathsf{p}\cup_1(\central \cup_1 \central))
\end{align}
in order for defectification to be unobstructed for the given $\nu$, $\pi$, $\central$, and $\mathsf{p}$.

As reviewed in Appendix~\ref{app:relativeO}, the relative defectification obstruction $\defectO_r(\coho{t},\rho)$ signifies when the torsorial action of $\coho{t}\in Z^2_{[\rho]}(G,\mathcal{A})$ results in a theory that satisfies the $G$-crossed consistency conditions.
In this example, we have $\rho = \V^{\maj}$ and
\begin{align}
\coho{t} = \psi_{\vv{0}}^{\mathsf{p}} \otimes \I_{\vv{1}}^{\central}.
\end{align}
The relative defectification obstruction vanishes if there exists an $\Xdot$ such that $\cbd\Xdot = \left(\defectO_r^{(\nu)}(\psi^{\mathsf{p}}_{\vv{0}} \otimes \I_\vv{1}^\central, \maj) \right)^{-1}$.
Since the base theories are unobstructed, the relative obstruction is the defectification obstruction of the resulting post-torsor functor theory.

We can schematically write the relative defectification obstruction of Eq.~\eqref{eq:pwformula} as
\begin{align}
\label{eq:ifoobstruction}
\defectO_r &= \eta \frac{U }{U} R  \frac{FFF}{FFF}
.
\end{align}
By inspecting Eqs.~\eqref{eq:tortildeR}-\eqref{eq:tortildeeta}, we see that the topological data of $\ifo^{(\nu) \times} _{\mathbb{Z}_2^{\eff} \times G, [0,0, \maj]_{\vv{0}}}$ take a form nearly identical to those of the base theories $\ifo^{(0) \times} _{\mathbb{Z}_2^{\eff} \times G, [0,0, \maj]_{\vv{0}}}$, with only the $F$-symbol in Eq.~\eqref{eq:tortildeF} and $R$-symbol in Eq.~\eqref{eq:tortildeR} differing.
Correspondingly, the first cluster of terms in $\defectO_r$ evaluate to
\begin{align}
\eta \frac{U}{U} R &=
i^{ \widetilde{\pi} \cup  \widetilde{\pi} \cup  \widetilde{\central}}
(-1)^{\mathsf{p} \cup \mathsf{p}
+\mathsf{p} \cup \central
+\pi \cup(\mathsf{p} \cup_1 \central)
+ (\pi \cup_1 \mathsf{p}) \cup \central}
\notag \\
& \qquad \times e^{i \frac{\pi}{8} \nu \cdot  \widetilde{\central} \cup  \widetilde{\central}},
\end{align}
which is the same as Eq.~\eqref{eq:generaldefectoFSPT}, except for the additional $\nu$-dependent phase factor.

The $\tortilde{F}$-symbols presented in Eq.~\eqref{eq:tortildeF} result in a particularly nice form for the ratio of six $F$-symbols appearing in the remaining part of the obstruction.
Moreover, they allow us to treat all $\nu$ even simultaneously.
The $F$-symbols $\tortilde{F}^{a_{\vv{x}}b_{\vv{y}} c_{\vv{z}}}$ are symmetric in $b_{\vv{y}} $ and $c_{\vv{z}}$, implying the ratio $\frac{\tortilde{F}^{\mathfrak{t}({\bf gh},{\bf kl}) \mathfrak{t}({\bf g},{\bf h}) {}^{\bf gh}\mathfrak{t}({\bf k},{\bf l})}}
{\tortilde{F}^{\mathfrak{t}({\bf gh},{\bf kl})  {}^{\bf gh}\mathfrak{t}({\bf k},{\bf l})\mathfrak{t}({\bf g},{\bf h})}}$ drops out of the relative obstruction.
Thus, the second cluster of terms, i.e. the ratio of the six $F$-symbols, becomes
\begin{align}
&\frac{FFF}{FFF}({\bf g,h,k,l})   \\
&\quad =\frac{\tortilde{F}^{\mathfrak{t}({\bf g},{\bf hkl}) {}^{\bf g}\mathfrak{t}({\bf hk},{\bf l} ) {}^{\bf g}\mathfrak{t}({\bf h},{\bf k})}}
{\tortilde{F}^{\mathfrak{t}({\bf g},{\bf hkl}) {}^{\bf g} \mathfrak{t}({\bf h},{\bf kl}) {}^{\bf gh}\mathfrak{t}({\bf k},{\bf l}) }}
\frac{\tortilde{F}^{\mathfrak{t}({\bf ghk},{\bf l}) \mathfrak{t}({\bf gh},{\bf k} ) \mathfrak{t}({\bf g},{\bf h})}}
{\tortilde{F}^{\mathfrak{t}({\bf ghk},{\bf l}) \mathfrak{t}({\bf g},{\bf hk}) {}^{\bf g}\mathfrak{t}({\bf h},{\bf k} ) }}
\notag \\
&\quad  =
\frac{i^{\frac{\nu}{2} \cdot \central({\bf g},{\bf hkl})\cdot \central({\bf hk},{\bf l} )\cdot \central({\bf h},{\bf k})}}
{i^{\frac{\nu}{2} \cdot\central({\bf g},{\bf hkl})\cdot \central({\bf h},{\bf kl}) \cdot\central({\bf k},{\bf l}) }}
\frac{i^{\frac{\nu}{2} \cdot\central({\bf ghk},{\bf l}) \cdot\central({\bf gh},{\bf k} ) \cdot\central({\bf g},{\bf h})}}
{i^{\frac{\nu}{2} \cdot\central({\bf ghk},{\bf l}) \cdot\central({\bf g},{\bf hk}) \cdot\central({\bf h},{\bf k} ) }}
\notag \\
&\quad \times
(-1)^{\frac{\nu}{2} \cdot \mathsf{p}({\bf g},{\bf hkl})\cdot \central({\bf hk},{\bf l} )\cdot \central({\bf h},{\bf k})
- \frac{\nu}{2} \cdot\mathsf{p}({\bf g},{\bf hkl}) \cdot \central({\bf h},{\bf kl}) \cdot\central({\bf k},{\bf l}) }
\notag\\
&\quad \times (-1)^{\frac{\nu}{2} \cdot\mathsf{p}({\bf ghk},{\bf l})\cdot \central({\bf gh},{\bf k} )\cdot \central({\bf g},{\bf h})
-\frac{\nu}{2} \cdot \mathsf{p}({\bf ghk},{\bf l}) \cdot\central({\bf g},{\bf hk}) \cdot\central({\bf h},{\bf k} ) }
.
\notag
\end{align}
Using the definition of $\cup_1$ defined in Appendix~\ref{app:cup}, we can write this as
\begin{align}
\frac{FFF}{FFF} = i^{\frac{\nu}{2} \cdot \central \cup_1 (\central \cup_1 \central)} (-1)^{\frac{\nu}{2} \cdot \mathsf{p} \cup_1(\central \cup_1 \central )}.
\end{align}

Finally, observing that the defectification obstruction is equal to the relative obstruction, we drop the subscript $r$ and write Eq.~\eqref{eq:ifoobstruction} as
\begin{align}
&\defectO^{(\nu)}(\psi_{\vv{0}}^{\mathsf{p}} \otimes \I_{\vv{1}}^{\central},\pi)
= i^{\pi \cup \pi \cup \central} (-1)^{\mathsf{p} \cup \mathsf{p}
+\mathsf{p} \cup \central }
\notag \\
&\qquad \qquad \times (-1)^{\pi \cup(\mathsf{p} \cup_1 \central) + (\pi \cup_1 \mathsf{p}) \cup \central}
\notag \\
&\qquad \times e^{i \frac{\pi}{8} \nu \cdot \central \cup \central} i^{\frac{\nu}{2} \cdot \central \cup_1 (\central \cup_1 \central)} (-1)^{\frac{\nu}{2} \cdot \mathsf{p} \cup_1(\central \cup_1 \central )}
.
\end{align}
This yields Eq.~\eqref{eq:nueven3cochain}.

After finding a solution $\Xdot$ to $\cbd \Xdot  = \left( \defectO^{(\nu)}(\psi_{\vv{0}}^{\mathsf{p}} \otimes \I_{\vv{1}}^{\central}) \right)^{-1}$, we can write the complete topological data of $\ifo^{(\nu) \times }_{\mathcal{G}^{\eff},[\xdot,\mathsf{p}, \maj]_{\central} }$ by applying the torsor method described in Ref.~\onlinecite{Aasen21} to the base theory.
We use $\tor{\;\;\;}$ to denote the topological data of $\ifo^{(\nu) \times }_{\mathcal{G}^{\eff},[\xdot,\mathsf{p}, \maj]_{\central} }$ and $\tortilde{\;\;\;}$ to denote the topological data of the base theory $\ifo^{(\nu) \times} _{\mathbb{Z}_2^{\eff} \times G, [0,0, \maj]_{\vv{0}}}$ given in Eqs.~\eqref{eq:zest_N_explicit}-\eqref{eq:zest_eta_explicit} by applying the defect zesting torsor to $\ifo^{(0) \times} _{\mathbb{Z}_2^{\eff} \times G, [0,0, \maj]_{\vv{0}}}$.
The fusion rules are given by
\begin{align}
\tor{N}_{a_{\mathpzc{g}} b_{\mathpzc{h}}}^{c_{\mathpzc{gh}}} = \tortilde{N}_{a_{\mathpzc{g}} b_{\mathpzc{h}}}^{\bar{\cohosub{t}}({\bf g}, {\bf h}) \otimes c_{\mathpzc{gh}}}
.
\end{align}
Leaving uniquely determined labels implicit, the $F$-symbols are given by
\begin{widetext}
\begin{align}
 \left[\tor{F}^{a_{\mathpzc{g}} b_{\mathpzc{h}} c_{\mathpzc{k}}}_{d_{\mathpzc{ghk}}}\right]_{e_{\mathpzc{gh}}f_{\mathpzc{hk}}} =
 &
 \left[{\tortilde{F}}^{\cohosub{t}({\bf g},{\bf h})_{} e'_{\mathpzc{gh}} c_{\mathpzc{k}}}_{d'_{\mathpzc{ghk}}}\right]_{e_{\mathpzc{gh}}, d''_{\mathpzc{ghk}}}
\left[{\tortilde{F}}^{a_{\mathpzc{g}} b_{\mathpzc{h}} c_{\mathpzc{k}}} _{d''_{\mathpzc{ghk}}}\right]_{e'_{\mathpzc{gh}},f'_{\mathpzc{hk}}}
\left( {\tortilde{F}}^{\cohosub{t}({\bf gh},{\bf k} )_{} \cohosub{t}({\bf g},{\bf h})_{} d''_{\mathpzc{ghk}}}_{d_{\mathpzc{ghk}}} \right)^{-1}
{\tortilde{F}}^{\cohosub{t}({\bf g},{\bf hk})_{} {}^{\bf g}\cohosub{t}({\bf h},{\bf k} )_{} d''_{\mathpzc{ghk}}}_{d_{\mathpzc{ghk}}}
\nonumber \\
&
\left[\left( {\tortilde{F}}^{{}^{\bf g}\cohosub{t}({\bf h},{\bf k})_{} a_{\mathpzc{g}} f'_{\mathpzc{hk}}}_{d'''_{\mathpzc{ghk}}}  \right)^{-1}\right]_{d''_{\mathpzc{ghk}},a'_{\mathpzc{g}}}
\left( {\tortilde{R}}^{{}^{\bf g} \cohosub{t}({\bf h},{\bf k})_{} a_{\mathpzc{g}}}_{ a'_{\mathpzc{g}} }\right)^{-1}
\left[ {\tortilde{F}}^{a_{\mathpzc{g}} \cohosub{t}({\bf h},{\bf k})_{} f'_{\mathpzc{hk}}}_{d'''_{\mathpzc{ghk}}} \right]_{a'_{\mathpzc{g}}, f_{\mathpzc{hk}}}
{\Xdot}({\bf g},{\bf h}, {\bf k})
\end{align}
where $e'_{\mathpzc{gh}} = \bar{\coho{t}}({\bf g},{\bf h})  \otimes e_{\mathpzc{gh}}$, $f'_{\mathpzc{hk}} = \bar{\coho{t}}({\bf h},{\bf k}) \otimes f_{\mathpzc{hk}}$, $a_{\mathpzc{g}}' = \bar{\coho{t}}({\bf h},{\bf k}) \otimes a_{\mathpzc{g}} = {}^{\bf g} \bar{\coho{t}}({\bf h},{\bf k}) \otimes a_{\mathpzc{g}}$, $d'_{\mathpzc{ghk}} = \bar{\coho{t}}({\bf gh},{\bf k}) \otimes d_{\mathpzc{ghk}}$, $d''_{\mathpzc{ghk}}= \bar{\coho{t}}({\bf g},{\bf h}) \otimes \bar{\coho{t}}({\bf gh},{\bf k}) \otimes d_{\mathpzc{ghk}} = {}^{\bf g}\bar{\coho{t}}({\bf h},{\bf k}) \otimes \bar{\coho{t}}({\bf g},{\bf hk}) \otimes d_{\mathpzc{ghk}}$, and $d'''_{\mathpzc{ghk}}= \bar{\coho{t}}({\bf g},{\bf hk}) \otimes d_{\mathpzc{ghk}}$.
The $R$-symbols are given by
\begin{align}
\tor{R}^{a_{\mathpzc{g}} b_{\mathpzc{h}}}_{c_{\mathpzc{gh}}}
=
\tortilde{R}^{a'_{\mathpzc{g}} b_{\mathpzc{h}}}_{c'_{\mathpzc{gh}}}
\tortilde{F}^{\cohosub{t}({\bf g },{\bf h})\cohosub{q}({\bf g},{\bf h}) c'_{\mathpzc{gh}}}_{c_{\mathpzc{gh}}}
\left[ \left(\tortilde{F}^{\cohosub{q}({\bf g},{\bf h}) a'_{\mathpzc{g}} b_{\mathpzc{h}}}_{c''_{\mathpzc{gh}}}\right)^{-1} \right]_{c'_{\mathpzc{gh}},a_{\mathpzc{g}}}
,
\end{align}
where $\coho{q}({\bf g} , {\bf h}) = \bar{\coho{t}}({\bf g} , {\bf h}) \otimes \coho{t}({\bf h} , {\bf \bar{h}gh})$, $a'_{\mathpzc{g}} = \bar{\coho{q}}({\bf g} , {\bf h}) \otimes a_{\mathpzc{g}}$, $c'_{\mathpzc{gh}} = \bar{\coho{t}}({\bf h},{\bf \bar{h}gh}) \otimes c_{\mathpzc{gh}}$, and $c''_{\mathpzc{gh}} = \bar{\coho{t}}({\bf g},{\bf h}) \otimes c_{\mathpzc{gh}}$.
The $U$-symbols are
\begin{align}
\tor{U}_{\bf k}\left( a_{\mathpzc{g}}, b_{\mathpzc{h}}; c_{\mathpzc{gh}} \right)&=
\tortilde{U}_{\bf k}\left( a'_{\mathpzc{g}}, b'_{\mathpzc{h}}; c'''_{\mathpzc{gh}} \right)
\tortilde{U}_{\bf k}\left( {}^{\bf k}\coho{t}({\bf \bar{k}gk},{\bf \bar{k}hk}) , c'''_{\mathpzc{gh}} ; c'_{\mathpzc{gh}} \right)
\tortilde{U}_{\bf g}\left( {}^{\bf g}\coho{t}({\bf h},{\bf k}) , {}^{\bf g}\coho{q}({\bf h},{\bf k}) \right)
\left( \tortilde{R}^{{}^{\bf g} \cohosub{q}({\bf h},{\bf k}) a_{\mathpzc{g}}} \right)^{-1}
\notag \\
&  \times
\left[ \left({\tortilde{F}}^{\cohosub{q}({\bf g},{\bf k}) a'_{\mathpzc{g}} b'_{\mathpzc{h}}}_{c''''_{\mathpzc{gh}}}\right)^{-1} \right]_{c'''_{\mathpzc{gh}},a_{\mathpzc{g}} }
\left[ \left({\tortilde{F}}^{{}^{\bf g}\cohosub{q}({\bf h},{\bf k}) a_{\mathpzc{g}} b'_{\mathpzc{h}}}_{c''_{\mathpzc{gh}}}\right)^{-1} \right]_{c''''_{\mathpzc{gh}},[\cohosub{q}({\bf h},{\bf k})a_{\mathpzc{g}}]}
\left[ \tortilde{F}^{a_{\mathpzc{g}} \cohosub{q}({\bf h},{\bf k}) b'_{\mathpzc{h}}}_{c''_{\mathpzc{ gh}}} \right]_{[\cohosub{q}({\bf h},{\bf k})a_{\mathpzc{g}}],b_{\mathpzc{h}}}
\notag \\
&  \times
\frac{{\tortilde{F}}^{[\cohosub{t}({\bf gk},{\bf \bar{k}hk})  \cohosub{t}({\bf g},{\bf k})] \cohosub{q}({\bf g},{\bf k})  c'''_{\mathpzc{gh}}  }
{\tortilde{F}}^{\cohosub{t}({\bf g},{\bf hk}) {}^{\bf g}\cohosub{t}({\bf h},{\bf k}) {}^{\bf g}\cohosub{q}({\bf h},{\bf k})  }
{\tortilde{F}}^{[{}^{\bf g}\cohosub{t}({\bf h},{\bf k}) \cohosub{t}({\bf g},{\bf hk}) ] {}^{\bf g}\cohosub{q}({\bf h},{\bf k})   c''''_{\mathpzc{gh}}  }
{\tortilde{F}}^{\cohosub{t}({\bf gh},{\bf k}) \cohosub{t}({\bf g},{\bf h})  c''_{\mathpzc{gh}}  }
}
{{\tortilde{F}}^{\cohosub{t}({\bf k},{\bf \bar{k}ghk}) {}^{\bf k}\cohosub{t}({\bf \bar{k}gk},{\bf \bar{k}hk})  c'''_{\mathpzc{gh}}  }
{\tortilde{F}}^{\cohosub{t}({\bf gk},{\bf \bar{k}hk}) \cohosub{t}({\bf g},{\bf k})  \cohosub{q}({\bf g},{\bf k})  }
{\tortilde{F}}^{\cohosub{t}({\bf gh},{\bf k}) \cohosub{q}({\bf gh},{\bf k})  c'_{\bf gh}  }
}
\notag \\
&  \times
\frac{\Xdot({\bf g},{\bf k}, {\bf \bar{k}hk}) }{\Xdot({\bf g},{\bf h}, {\bf k}) \Xdot({\bf k}, {\bf \bar{k}gk}, {\bf \bar{k}hk})}
,
\end{align}
where $a'_{\mathpzc{g}} = \bar{\coho{q}}({\bf g},{\bf k}) \otimes a_{\mathpzc{g}}$, $b'_{\mathpzc{h}} = \bar{\coho{q}}({\bf h},{\bf k}) \otimes b_{\mathpzc{h}}$, $c'_{\mathpzc{gh}} = \bar{\coho{q}}({\bf gh},{\bf k}) \otimes c_{\mathpzc{gh}}$, $c''_{\mathpzc{gh}} = \bar{\coho{t}}({\bf g},{\bf h}) \otimes c_{\mathpzc{gh}}$, $c'''_{\mathpzc{gh}} = {}^{\bf k}\bar{\coho{t}}({\bf \bar{k}gk},{\bf \bar{k}hk}) \otimes c'_{\mathpzc{gh}}$, and $c''''_{\mathpzc{gh}} = \coho{q}({\bf g},{\bf k}) \otimes c'''_{\mathpzc{gh}}$.
Finally, the $\eta$-symbols are
\begin{align}
 \tor{\eta}_{x_{\mathpzc{k}}} \left( {\bf g}, {\bf h} \right) =&
\tortilde{\eta}_{x'_{\mathpzc{k}}} \left( {\bf g}, {\bf h} \right)
\frac{{\tortilde{U}}_{\bf g}\left( {}^{\bf g}\cohosub{q}({\bf \bar{g}kg},{\bf h}) , x'_{\mathpzc{k}} \right)}{
{\tortilde{U}}_{\bf g}\left( {}^{\bf g}\cohosub{t}({\bf \bar{g}kg},{\bf h}) , {}^{\bf g}\cohosub{q}({\bf \bar{g}kg},{\bf h}) \right)}
{\tortilde{R}}^{{}^{\bf k}\cohosub{t}({\bf g},{\bf h}) x_{\mathpzc{k}}}{\tortilde{R}}^{ x'_{\mathpzc{k}} \cohosub{t}({\bf g},{\bf h})}
\notag \\
& \qquad \times
\frac{
{\tortilde{F}}^{\cohosub{t}({\bf kg},{\bf h}) \cohosub{t}({\bf k},{\bf g}) \cohosub{q}({\bf k},{\bf g})  }
{\tortilde{F}}^{\cohosub{t}({\bf g},{\bf \bar{g}kgh}) {}^{\bf g}\cohosub{t}({\bf \bar{g}kg},{\bf h}) {}^{\bf g}\cohosub{q}({\bf \bar{g}kg},{\bf h})  }
{\tortilde{F}}^{\cohosub{t}({\bf gh},{\bf \bar{h}\bar{g}kgh}) \cohosub{t}({\bf g},{\bf h}) x'_{\mathpzc{k}} }
{\tortilde{F}}^{\cohosub{t}({\bf k},{\bf gh}) \cohosub{q}({\bf k},{\bf gh})  x'''_{\mathpzc{k}}  }
}
{{\tortilde{F}}^{\cohosub{t}({\bf k},{\bf gh}) {}^{\bf k}\cohosub{t}({\bf g},{\bf h}) x_{\mathpzc{k}} }
{\tortilde{F}}^{[\cohosub{t}({\bf kg},{\bf h})  \cohosub{t}({\bf k},{\bf g})] \cohosub{q}({\bf k},{\bf g})  x''_{\mathpzc{k}}  }
{\tortilde{F}}^{[\cohosub{t}({\bf g},{\bf \bar{g}kgh}) {}^{\bf g}\cohosub{t}({\bf \bar{g}kg},{\bf h}) ] {}^{\bf g}\cohosub{q}({\bf \bar{g}kg},{\bf h})   x'_{\mathpzc{k}}  }
{\tortilde{F}}^{\cohosub{q}({\bf k},{\bf gh}) x'_{\mathpzc{k}} \cohosub{t}({\bf g},{\bf h}) }
}
\notag \\
& \qquad \times
\frac{\Xdot({\bf g},{\bf \bar{g}kg}, {\bf h}) }{\Xdot({\bf g},{\bf h}, {\bf \bar{h}\bar{g}kgh}) \Xdot({\bf k}, {\bf g}, {\bf h})}
,
\end{align}
where $x'_{\mathpzc{k}} = \bar{\coho{q}}({\bf k},{\bf gh}) \otimes x_{\mathpzc{k}}$, $x''_{\mathpzc{k}} = \bar{\coho{q}}({\bf k},{\bf g}) \otimes x_{\mathpzc{k}}$, and $x'''_{\mathpzc{k}} = {\coho{t}}({\bf g},{\bf h}) \otimes x'_{\mathpzc{k}}$.

\end{widetext}

\subsubsection{Equivalence relations}
\label{sec:ifoEquivalences}

Having produced all the (unobstructed) $\nu$ even invertible FSET phases, we now determine equivalences of na\"ively distinct theories in the same way that we did for FSPT phases in Sec.~\ref{sec:equivalence_relations}.
For this, we will write the quadruple of topological data specifying an invertible FSET phase as $(\xdot,\mathsf{p},\maj ; \nu)_{\central}$.

As with all $G$-crossed theories, the torsor functor gives an equivalence between theories that differ by gluing in a bosonic SPT whose 3-cocycle is a coboundary.
In the present context, this amounts to modifying the 3-cochain $\chi$ by a 3-coboundary, yielding the equivalence relation
\begin{align}
\label{eq:3coboundaryFSPequiv_IFO}
(\xdot,\mathsf{p},\maj ; \nu)_{\central} \sim(\xdot + \cbd \bdot ,\mathsf{p},\maj ; \nu)_{\central}
,
\end{align}
where $\bdot \in C^2(G,\mathbb{R}/\mathbb{Z})$.

The relabeling equivalences work in exactly the same way as for FSPT phases, using the techniques of Ref.~\onlinecite{Aasen21}, except the composition rule of torsor functors includes an additional factor of $F$-symbols that enter the 3-cochain product rule because the general $\nu$ even $F$-symbols are not all trivial.
For the first type of relabeling
\begin{align}
a_{\vv{x},{\bf g}} \to [\coho{z}({\bf g}) a]_{\vv{x},{\bf g}}
,
\end{align}
with $\coho{z}({\bf g})  = \psi_{\mathpzc{0}}^{\mathsf{z}({\bf g})}$, the analysis is the same, except for the nontrivial contribution of $F$-symbols
\begin{widetext}
\begin{align}
\frac{F^{ [\cbd \cohosub{z}({\bf gh},{\bf k}) \otimes \cbd \cohosub{z}({\bf g},{\bf h}) ], \cohosub{p}_{\central}({\bf gh},{\bf k}) , \cohosub{p}_{\central}({\bf g},{\bf h})} }
{F^{ [\cbd \cohosub{z}({\bf g},{\bf hk}) \otimes \cbd \cohosub{z}({\bf h},{\bf k}) ], \cohosub{p}_{\central}({\bf g},{\bf hk}) , \,^{\bf g}\cohosub{p}_{\central}({\bf h},{\bf k})} }
= (-1)^{\frac{\nu}{2} [(\central \cup_1 \central) \cup_{2} \cbd \mathsf{z}] ({\bf g},{\bf h},{\bf k})}
\end{align}
to the 3-cochain composition.
The resulting equivalence of theories can be written as
\begin{align}
\label{eq:first-relab_IFO}
(\xdot,\mathsf{p},\maj ;\nu )_{\central} &\sim
\left( \xdot + \frac{1}{2}( \mathsf{p} \cup_1 \cbd \mathsf{z}  + \central \cup \mathsf{z} + \pi \cup(\central \cup_2 \cbd \mathsf{z})+ \mathsf{z} \cup \cbd \mathsf{z} + \frac{\nu}{2} (\central \cup_1 \central) \cup_{2} \cbd \mathsf{z} )  ,\mathsf{p}_{}+ \cbd \mathsf{z},\maj ;\nu \right)_{\central}
.
\end{align}
\end{widetext}
This only differs from the corresponding FSPT equivalence relation when $\central \neq 0$ and $\nu = 2 \text{ mod } 4$.

For the second type of relabeling
\begin{align}
a_{\vv{x},{\bf g}} \to  [\coho{z}'(\vv{x}) a]_{\vv{x},{\bf g}},
\end{align}
where $\coho{z}'(\mathpzc{g}) = \psi_{\vv{0}}^{\vv{x}}$, the analysis is exactly the same as for FSPT phases, because the additional factor of $F$-symbols evaluates to 1, which follows from $(\central \cup_1 \central) \cup_{2} \cbd \central = \vv{0}$.
Thus, the resulting equivalence of theories under this second type of relabeling is again
\begin{align}
\label{eq:Gfequiv_IFO}
(\xdot,\mathsf{p}_{}, \maj ;\nu)_{\central} \sim (\xdot + \frac{1}{2}(\mathsf{p} \cup_1 \central + \maj \cup \central) ,\mathsf{p}_{}+{\central}, \maj ;\nu)_{\central} .
\end{align}

Using the three equivalence relations in Eqs.~\eqref{eq:3coboundaryFSPequiv_IFO}, \eqref{eq:first-relab_IFO}, and \eqref{eq:Gfequiv_IFO}, we form equivalence classes of theories describing $\mathcal{G}^{\eff} =\mathbb{Z}_2^{\eff}\times_\central G$ invertible FSET phases with $\nu$ even.
We will use $\ifo^{(\nu) \times }_{\mathcal{G}^{\eff},[\xdot,\mathsf{p}, \maj]_{\central} }$ to denote these equivalence classes.

We note that the same equivalence relations can also be obtained from the equivalence relations of the FSPT phases when one has the addition rules for stacking $\ifo^{(\nu) \times }_{\mathcal{G}^{\eff},[\xdot,\mathsf{p}, \maj]_{\central} }$ with FSPT phases.
In particular, when $(\xdot',\mathsf{p}', \maj';0)_{\central} \sim (\xdot'',\mathsf{p}'', \maj'';0)_{\central}$ for FSPT phases, it follows that
$\ifo^{(0) \times }_{\mathcal{G}^{\eff},[\xdot',\mathsf{p}', \maj']_{\central}} \underset{\mathcal{G}^{\eff}}{\ftimes} \ifo^{(\nu) \times }_{\mathcal{G}^{\eff},[\xdot,\mathsf{p}, \maj]_{\central} }$ and $\ifo^{(0) \times }_{\mathcal{G}^{\eff}, [\xdot'',\mathsf{p}'', \maj'']_{\central}}
\underset{\mathcal{G}^{\eff}}{\ftimes} \ifo^{(\nu) \times }_{\mathcal{G}^{\eff},[\xdot,\mathsf{p}, \maj]_{\central} }$ should be equivalent theories.
This induces equivalence relations on the general $\nu$ even theories from those of $\nu =0$.
After developing the addition rules for stacking in the subsequent section, we can verify that this alternate method of deriving equivalence relations on the $\nu$ even invertible FSET phases yields the same equivalence relations found above.

\subsubsection{Addition rules for stacking \texorpdfstring{$\ifo^{(\nu) \times }_{\mathcal{G}^{\eff},[\xdot,\mathsf{p}, \maj]_{\central} }$}{ifonu} with \texorpdfstring{$\nu$}{nunu} even}\label{sec:ifoAddition}

We now discuss the addition rule for the $\nu$ even theories under fermionic stacking.
Using a condensation calculation, we prove the addition rule for symmetry fractionalization classes conjectured in Ref.~\onlinecite{Bark2021inv}.
Following a similar calculation as in Ref.~\onlinecite{Bark2021inv}, but using our conventions, gauge choices, and methods, we solve for the $3$-cochain specifying the addition rule of two $\nu$ even theories.
As in the FSPT case, the $3$-cochain we provide here is only determined up to a 3-cocycle.

We wish to compute
\begin{align}
\label{eq:group-law-even}
\ifo^{(\nu) \times }_{\mathcal{G}^{\eff},[\xdot,\mathsf{p}, \maj]_{\central} }=
\ifo^{(\nu') \times }_{\mathcal{G}^{\eff},[\xdot',\mathsf{p}', \maj']_{\central} }
\underset{\mathcal{G}^{\eff}}{\ftimes}
\ifo^{(\nu'') \times }_{\mathcal{G}^{\eff},[\xdot'',\mathsf{p}'', \maj'']_{\central} }.
\end{align}
First, we recall that under fermionic stacking, the central charge is additive and therefore,
\begin{align}
\nu = \nu' + \nu'' \text{ mod } 16.
\end{align}
For the stacked theory, we will use representatives $\cond{a}_{\vv{x},{\bf 0} }= (a_{\vv{x},{\bf 0}},\I_{\vv{x},{\bf 0}})$ for the ${\bf g = 0}$ sector.
Following an identical line of arguments that led to Eq.~\eqref{eq:maj''}, we have
\begin{align}
\maj = \maj' + \maj''
.
\end{align}

We now consider the addition of the symmetry fractionalization classes.
From Eqs.~\eqref{eq:tortildeF}-\eqref{eq:tortildeeta}, and noting that only $\eta$ changes under the torsor functor $\mathcal{F}_{\cohosub{t}, \Xdot}$, we see that the symmetry fractionalization class of a $\nu$ even theory is completely specified by
\begin{align}
 \rho_{\bf g} &= \V^{\pi({\bf g})},\\
 \label{eq:ifoU}
U_{\bf g}(a_{\vv{x}},b_{\vv{y}}) &=(-1)^{(\mathsf{a}+\vv{x})\cdot \vv{y} \cdot \pi({\bf g})},\\
\label{eq:ifoeta}
\eta_{a_{\vv{x}}}({\bf g,h}) &= i^{\vv{x}\cdot \pi({\bf g})\cdot \pi({\bf h}) }
(-1)^{\vv{x} \cdot \mathsf{p}({\bf g,h}) + \mathsf{a} \cdot \central({\bf g,h})} \notag \\
& \qquad \times e^{i \frac{2\pi}{8} \nu \cdot \vv{x} \cdot \central({\bf g,h}) }
.
\end{align}
The similarity of the data for the $\nu$ even theories and the  $\nu = 0$ theories [Eqs.~\eqref{eq:TCsymmetryaction}-\eqref{eq:etaTC}] make the calculation of the fractionalization addition nearly identical to that of the FSPT phases presented in Sec.~\ref{sec:generalFSPTphases}.
The fusion space of the condensed theory is represented by
\begin{align}
\label{eq:fusionvertexeven}
\zigvabcnu \equiv \zigvabcnurhs
,
\end{align}
where we again choose to use the representative charges $\cond{a}_{\vv{x},{\bf 0} }= (a_{\vv{x},{\bf 0}},\I_{\vv{x},{\bf 0}})$ to describe the stacked theory.
For $\nu'' = 2 \text{ mod }4$, the stacking operation does not automatically return the $F$-symbols of the resulting stacked theory into the standardized form of Eq.~\eqref{eq:tortildeF}.
In order to do so, we apply a vertex basis gauge transformation after stacking, specified by
\begin{align}
\Gamma^{a_{\vv{x}}b_{\vv{y}}}  &= (-1)^{\frac{\nu''}{2} (\delta_{a_{\vv{x}},\psi_{\vv{0}} } \cdot  \vv{y} +  \vv{x} \cdot \delta_{b_{\vv{y}},\psi_{\vv{0}} } ) }
\notag \\
&= (-1)^{\frac{\nu''}{2} \left[ \mathsf{a} \cdot (1-\vv{x}) \cdot \vv{y} +  \vv{x} \cdot \mathsf{b} \cdot (1-\vv{y})\right] }
.
\end{align}
This vertex basis gauge transformation puts the $F$-symbols of the FMTC sector in the standardized form, while leaving the $R$-, $U$-, and $\eta$-symbols unchanged.
In particular, it changes the $F$-symbols of the FMTC to $\widetilde{F}^{a_{\vv{x}} b_{\vv{y}} c_{\vv{z}}}= (-1)^{\frac{\nu''}{2} \cdot ( \mathsf{a} \cdot \vv{y} \cdot\vv{z} + \vv{x} \cdot \vv{y} \cdot \mathsf{c})} F^{a_{\vv{x}} b_{\vv{y}} c_{\vv{z}}}$.
After applying this vertex basis gauge transformation to the post-stacking theory, the $F$- and $R$-symbols are in the standardized form.
We can now define the symmetry action on the fusion space in exactly the same way as in Eq.~\eqref{eq:Uaction}.
One finds exactly the same $U$-symbols for the post-stacking theory as was found in Eq.~\eqref{eq:stackUeval}.
The resulting $U$-symbols are not in the standardized form of Eq.~\eqref{eq:ifoU}, but we can again use the symmetry action gauge transformation given in Eq.~\eqref{eq:sym_ac_g} to recover the standardized form.

Similarly, we can compute the $\eta$-symbols following an identical line of reasoning that led to Eq.~\eqref{eq:etaprimeprime}, which again produces the same result
\begin{align}
{\eta}_{\zig{a}_{\vv{x}}}({\bf g,h}) = \eta'_{a_{\vv{x}}}({\bf g,h})\eta_{\I_{\vv{x}}}''({\bf g,h})
.
\end{align}
After applying the symmetry action gauge transformation Eq.~\eqref{eq:sym_ac_g} to ${\eta}_{\zig{a}_{\vv{x}}}({\bf g,h}) $, and using Eq.~\eqref{eq:ifoeta} to evaluate $\eta'_{a_{\vv{x}}}({\bf g,h})\eta_{\I_{\vv{x}}}''({\bf g,h})$, we find
\begin{align}
\widecheck{\eta}_{\zig{a}_{\vv{x}}}({\bf g,h}) &= i^{\vv{x} \cdot \maj \cup \maj} (-1)^{\vv{x} \cdot \mathsf{p} + \mathsf{a} \cdot \central} e^{i \frac{2 \pi}{8} \nu \cdot x \cdot \central}
,
\end{align}
where
\begin{align}
\label{eq:p-add}
\mathsf{p} = \mathsf{p}' + \mathsf{p}'' + \maj' \cup \maj''.
\end{align}
This proves the symmetry fractionalization addition rule for stacking invertible fermionic topological phases.

Finally, we compute the 3-cochain addition rule
\begin{align}
\xdot = \xdot' + \xdot'' + \lambda.
\end{align}
Similar to the FSPT case, this imposes the condition,
\begin{align}
\cbd \Ydot &= \frac{\defectO^{(\nu')}(\psi^{\mathsf{p}'}_{\vv{0}}\otimes \I_{\vv{1}}^{\central},\pi')\defectO^{(\nu'')}(\psi^{\mathsf{p}''}_{\vv{0}} \otimes \I_{\vv{1}}^{\central},\pi'')}{\defectO^{(\nu' + \nu'')}(\psi^{ \mathsf{p}'+\mathsf{p}''+{\pi' \cup \pi''}}_{\vv{0}}\otimes \I_{\vv{1}}^{\central},\pi' +\pi'')},
\end{align}
where $\Ydot = e^{i 2\pi \lambda}$.
Following a similar calculation as Ref.~\onlinecite{Bark2021inv}, in App.~\ref{app:Y-details}, we verify that
\begin{align}
\label{eq:3cochnaddeven}
\lambda =& \frac{1}{4}\big( \widetilde{\pi} ' \cup  \widetilde{\pi} '' \cup  \widetilde{\pi}'' + ( \widetilde{\pi} ' \cup_1  \widetilde{\pi}'' )\cup  \widetilde{\central} \big)
\notag \\
&+\frac{1}{2}\big(
\pi' \cup(\pi' \cup_1 \pi'')\cup \pi''
+ (\pi' \cup \pi'')  \cup_1 (\mathsf{p}' + \mathsf{p}'')
\notag \\
& \qquad \qquad  + \mathsf{p}' \cup_1 \mathsf{p}''
+(\pi' \cup \central )\cup_2 \mathsf{p}''   \big)
\notag \\
&+\frac{1}{2} \frac{\nu'}{2} 
\big( (\central \cup_1 \central)\cup_2 \mathsf{p}'' +
(\central \cup_1 \central) \cup_3 (\pi'' \cup \central) \big)
\notag \\
&+ \frac{1}{2}\left( f(\mathsf{p}',\pi'; \nu') +f(\mathsf{p}'',\pi''; \nu'') - f(\mathsf{p},\pi; \nu) \right)
,
\end{align}
where
\begin{align}
\label{eq:eqnforf_IFO}
f(\mathsf{p},\pi; \nu) =& \pi \cup(\mathsf{p} \cup_2 \central) + \mathsf{p} \cup_2 (\pi \cup \central)  \notag\\
&+ \frac{\nu}{2} \big( \mathsf{p} \cup_2 (\central \cup_1 \central) + (\central \cup_1 \central) \cup_2 \mathsf{p} \big)
.
\end{align}
We note that Eq.~\eqref{eq:3cochnaddeven} depends on $\nu'$ and $\nu''$ of the stacked theories both implicitly through $\mathsf{p}$, which depends on $\nu$ due to the obstruction condition Eq.~\eqref{eq:nueven2cochain}, and explicitly through the terms on the last two lines.
Eq.~\eqref{eq:3cochnaddeven} matches the corresponding expression in Ref.~\onlinecite{Bark2021inv}.
(The last line of Eq.~\eqref{eq:3cochnaddeven}, involving the $f$ terms, does not arise in their expression; however, this difference is due to a different choice of gauge and only changes the resulting classification group structure by a group isomorphism, so it is not an important discrepancy.)

\subsection{FMTCs with Trivial Symmetry Action on the Quasiparticles}
\label{sec:ex_FMTCSPT}

In this section, we consider the defectification of a general FMTC $\fMTC$ with trivial fermionic symmetry action on the quasiparticles.
In Sec.~\ref{sec:Ex_no_perm}, we considered the fractionalization aspect of this problem.
We will again assume that $\ker (\res_{\fMTC_{\vv{0}}}) = \mathbb{Z}_2^{\V}$, but if more general examples exist, it is a straightforward generalization.
As in Sec.~\ref{sec:Ex_no_perm}, we split this class of examples into three cases: (1) $\mathcal{A}_\vv{1}\neq \varnothing$, (2) $\mathcal{A}_\vv{1}=\varnothing$ and ${\bf A}_\vv{1} \neq \varnothing$, and (3) $\mathcal{A}_\vv{1} = \varnothing$ and ${\bf A}_\vv{1} = \varnothing$.
For all cases, fractionalization is unobstructed, but not all $\central$ can be manifested.
For fixed $\central$ with vanishing $[\coho{O}^\central]$, the fermionic symmetry action is classified by $H^{1}(G,\mathbb{Z}_2^{\V})$ and the symmetry fractionalization is torsorially classified by
\begin{align}
\label{eq:fractionalziationclassif}
H^2(G,\mathcal{A}_{\vv{0}})  = H^2(G,\mathbb{Z}_2^\psi) \times H^2(G,\widehat{\mathcal{A}}_{\vv{0}})
.
\end{align}
When the defectification obstruction vanishes, the defectification is classified by $H^3(G,\text{U}(1))$.
This can be regrouped into the classification of the quasiparticle fractionalization by $H^2(G,\widehat{\mathcal{A}}_{\vv{0}})$ and the classification of $\mathcal{G}^{\eff}$-crossed extensions by $\Gr(\mathcal{G}^{\eff})$.
We now consider the different cases in more detail.

Case (2) is the simplest to deal with, so we address it first.
In this case, $[\V] = [\Q]$.
The fermionic symmetry action is thus given by $[\rho_{\bf g}] = [\Q]^{\rr({\bf g}) }$, where $\rr \in C^1(G,\mathbb{Z}_2^{\Q})$.
The obstruction $[\coho{O}^{\central}]  = (\central - \cbd \rr)$ of Eq.~\eqref{eq:trivialactionOcentraldr} requires $\central = \cbd \rr$.
These actions are all related to the completely trivial symmetry action $\rho_{\bf g} = \openone$ by applying $\vviso$-isomorphisms of Sec.~\ref{sec:centralisomorphism}, with $\vviso = \rr$.
Thus, we only need to analyze the case of trivial symmetry action $\pi =\vv{0}$, and then apply $\vviso$-isomorphisms at the end.
The complete data and defectification obstructions of the $\mathbb{Z}_2^{\eff} \times G$-crossed extensions with trivial symmetry action $\rho_{\bf g} = \openone$ can be easily obtained by applying the torsor functor of Ref.~\onlinecite{Aasen21} to the trivial $G$-crossed extension
\begin{align}
\fMTC^{\times}_{\mathbb{Z}_2^{\eff} \times G} = \mathcal{F}_{\cohosub{t}, \Xdot} \left( \fMTC \boxtimes \spt_G^{[1]} \right)
,
\end{align}
where $\coho{t} \in Z^2(G,\mathcal{A}_{\vv{0}})$ and $\Xdot \in Z^3(G,\text{U}(1))$.
(The complete data and obstruction for the case of trivial symmetry action for MTCs with no fusion multiplicities was also given in Ref.~\onlinecite{Bark2019}, using a different method and gauge choice.)
The defectification obstruction in this case is
\begin{align}
&\defectO(\coho{t})({\bf g},{\bf h},{\bf k},{\bf l}) =
R^{\cohosub{t}({\bf k},{\bf l}) \cohosub{t}({\bf g},{\bf h})}
\frac{F^{\cohosub{t}({\bf gh},{\bf kl}) \cohosub{t}({\bf g},{\bf h}) \cohosub{t}({\bf k},{\bf l})}}
{F^{\cohosub{t}({\bf gh},{\bf kl})  \cohosub{t}({\bf k},{\bf l})\cohosub{t}({\bf g},{\bf h})}}
\nonumber \\
&\qquad \times
\frac{F^{\cohosub{t}({\bf g},{\bf hkl}) \cohosub{t}({\bf hk},{\bf l} ) \cohosub{t}({\bf h},{\bf k})}}
{F^{\cohosub{t}({\bf g},{\bf hkl})  \cohosub{t}({\bf h},{\bf kl}) \cohosub{t}({\bf k},{\bf l}) }}
\frac{F^{\cohosub{t}({\bf ghk},{\bf l}) \cohosub{t}({\bf gh},{\bf k} ) \cohosub{t}({\bf g},{\bf h})}}
{F^{\cohosub{t}({\bf ghk},{\bf l}) \cohosub{t}({\bf g},{\bf hk}) \cohosub{t}({\bf h},{\bf k} ) }}
,
\label{eq:defectO_noperm}
\end{align}
which can be seen from Eq.~\eqref{eq:pwformula}.

For cases (1) and (3), we need to generate theories with $\V$ symmetry action on the vortices.
We can do this by first producing a base theory for each $\rho = \V^{\pi}$.
One way to generate such base theories is
\begin{align}
\label{eq:generalbasetheory}
\fMTC^{\times  \text{ base}}_{\mathbb{Z}_2^{\eff} \times G, \rho} = \ifo^{(0)}_{\mathbb{Z}_2^{\eff} \times G , [0,0,\maj]_{\vv{0}}} \underset{\mathbb{Z}_2^{\eff} \times G}{\ftimes} \left( \fMTC \boxtimes \spt_G^{[1]} \right),
\end{align}
while another is
\begin{align}
\label{eq:generalbasetheory2}
\fMTC^{\times  \text{ base}}_{\mathbb{Z}_2^{\eff} \times G, \rho} &= \left. \fMTC^{\times  \text{ base}}_{\mathbb{Z}_2^{\eff} \times \mathbb{Z}_{2}^{\V} } \boxtimes \spt_G^{[1]} \right|_{\mathcal{S}_{\pi}},
\\
\mathcal{S}_{\pi} &= \{ (a_{\pi({\bf g})} , \I_{\bf g} ) \}
,
\end{align}
where $\fMTC^{\times  \text{ base}}_{\mathbb{Z}_2^{\eff} \times \mathbb{Z}_{2}^{\V} }$ is generated as in Eq.~\eqref{eq:generalbasetheory} with $\rho_{\bf 1} = \V$.
Since the theories entering the right hand sides of these expressions are always unobstructed, this construction always guarantees the existence of unobstructed base theories.
Knowing that they exist, one could also produce base theories by directly solving the $G$-crossed consistency conditions.

Once we have such base theories, we can apply the torsor functor of Ref.~\onlinecite{Aasen21} to generate the complete data of all possible $\mathcal{G}^{\eff}$-crossed extensions of $\fMTC$ with trivial symmetry action on the quasiparticles, that is
\begin{align}
\fMTC^{\times}_{\mathcal{G}^{\eff},\rho } = \mathcal{F}_{\cohosub{t}, \Xdot} \left(\fMTC^{\times \text{ base}}_{\mathbb{Z}_2^{\eff} \times G,\rho} \right),
\end{align}
where $\coho{t} \in Z^2(G,\mathcal{A})$ and $\Xdot \in Z^3(G,\text{U}(1))$.
This naturally generates all the possible $\mathcal{G}^{\eff} = \mathbb{Z}_2^{\eff} \times_{\central} G$.
For case (1), the obstruction $[\coho{O}^{\central}]  = [\psi_{\vv{0}}^{\pi \cup \central} \otimes h_{\vv{0}}^{\tilde{\central} \cup_{1} \tilde{\central}}]$ of Eq.~\eqref{eq:trivialactionOw} indicates which nontrivial $\central$ can be manifested and that $\central = \vv{0}$ can always occur.
For case (3), $\mathcal{A} = \mathcal{A}_{\vv{0}}$ and the obstruction $[\coho{O}^{\central}] = \central$ of Eq.~\eqref{eq:trivialactionOcentral} indicates that only  $\central = \vv{0}$ can occur.
Since the base theories are unobstructed, the defectification obstruction for the associated symmetry action and fractionalization class is equal to the relative defectification obstruction associated with applying the torsor functor, as given in Eq.~\eqref{eq:pwformula}, that is $[\defectO(\coho{t}, \rho)]  =  [\defectO_r(\coho{t}, \rho)]$.

In order to provide more explicit details about the defectification obstruction and topological data of the resulting $\fMTC^{\times}_{\mathcal{G}^{\eff},\rho }$, we need to compute the properties of the base theories.
In principle, all the topological data of the base theories $\fMTC^{\times  \text{ base}}_{\mathbb{Z}_2^{\eff} \times G,\rho} $ can be produced using a condensation calculation; this is difficult to do in complete generality, so we will only determine partial information here.

\begin{table*}[t!]
\centering
\def\arraystretch{1}
\setlength\tabcolsep{1ex}
\begin{tabular}{c|c|c|c|}
$\zig{N}$ & $\zig{b}_{\vv{y},{\bf h}}$  & $\zig{\hat{b}}_{\vv{y},{\bf h}}$  & $\zig{b}^{s}_{\vv{y},{\bf h}}$
 \\ [0.5ex]
 \hline
&&&\\[-2.5ex]
$\zig{a}_{\vv{x},{\bf g}}$
&
$\zig{N}_{\zig{a}_{\vv{x},{\bf g}} \zig{b}_{\vv{y},{\bf h}}}^{\zig{c}_{\vv{x+y},{\bf gh}}}  = N_{a_{\vv{x}} b_{\vv{y}}}^{c_{\vv{x+y}}}$ &
\begin{tabular}{rl}
$\zig{N}_{\zig{a}_{\vv{x},{\bf g}} \zig{\hat{b}}_{\vv{y},{\bf h}}}^{\zig{\hat{c}}_{\vv{x+y},{\bf gh}}}$&$  = N_{a_{\vv{x}} b_{\vv{y}}}^{c_{\vv{x+y}}} + N_{a_{\vv{x}} b_{\vv{y}}}^{[\psi c]_{\vv{x+y}}}$
\\
\\
$\zig{N}_{\zig{a}_{\vv{x},{\bf g}} \zig{\hat{b}}_{\vv{y},{\bf h}}}^{\zig{c}^{q}_{\vv{x+y},{\bf gh}}}$&$  = N_{a_{\vv{x}} b_{\vv{y}}}^{c_{\vv{x+y}}} $  \end{tabular}
&
\begin{tabular}{rl} $\zig{N}_{\zig{a}_{\vv{x},{\bf g}} \zig{b}^{s}_{\vv{y},{\bf h}}}^{\zig{\hat{c}}_{\vv{x+y},{\bf gh}}}$& $ = N_{a_{\vv{x}} b_{\vv{y}}}^{c_{\vv{x+y}}} $
\\
\\
$\zig{N}_{\zig{a}_{\vv{x},{\bf g}} \zig{b}^{s}_{\vv{y},{\bf h}}}^{\zig{c}^{q}_{\vv{x+y},{\bf gh}}}$&$  =
N_{a_{\vv{x}} b_{\vv{y}}}^{c_{\vv{x+y}}}
\, \Xi_{\zig{a}_{\vv{x},{\bf g}} \zig{b}^{s}_{\vv{y},{\bf h}}}^{\zig{c}^{q}_{\vv{x+y},{\bf gh}}}
$
\end{tabular}
 \\ [0.5ex]
 &&&\\[-2.5ex]
 \hline
&&&\\[-2.5ex]
$\zig{\hat{a}}_{\vv{x},{\bf g}}$
&
\begin{tabular}{rl}
$\zig{N}_{\zig{\hat{a}}_{\vv{x},{\bf g}} \zig{b}_{\vv{y},{\bf h}}}^{\zig{\hat{c}}_{\vv{x+y},{\bf gh}}}  $&$= N_{a_{\vv{x}} b_{\vv{y}}}^{c_{\vv{x+y}}} + N_{a_{\vv{x}} b_{\vv{y}}}^{[\psi c]_{\vv{x+y}}}$
\\
\\
$\zig{N}_{\zig{\hat{a}}_{\vv{x},{\bf g}} \zig{b}_{\vv{y},{\bf h}}}^{\zig{c}^{q}_{\vv{x+y},{\bf gh}}}$&$  = N_{a_{\vv{x}} b_{\vv{y}}}^{c_{\vv{x+y}}} $
\end{tabular}
&
$\zig{N}_{\zig{\hat{a}}_{\vv{x},{\bf g}} \zig{\hat{b}}_{\vv{y},{\bf h}}}^{\zig{c}_{\vv{x+y},{\bf gh}}}  = N_{a_{\vv{x}} b_{\vv{y}}}^{c_{\vv{x+y}}} + N_{a_{\vv{x}} b_{\vv{y}}}^{[\psi c]_{\vv{x+y}}}$
&
$\zig{N}_{\zig{\hat{a}}_{\vv{x},{\bf g}} \zig{b}^{s}_{\vv{y},{\bf h}}}^{\zig{c}_{\vv{x+y},{\bf gh}}}  = N_{a_{\vv{x}} b_{\vv{y}}}^{c_{\vv{x+y}}} $
 \\ [0.5ex]
 &&&\\[-2.5ex]
 \hline
&&&\\[-2.5ex]
$\zig{a}^{r}_{\vv{x},{\bf g}}$ &
\begin{tabular}{rl}
$\zig{N}_{\zig{a}^{r}_{\vv{x},{\bf g}} \zig{b}_{\vv{y},{\bf h}}}^{\zig{\hat{c}}_{\vv{x+y},{\bf gh}}}$&$  = N_{a_{\vv{x}} b_{\vv{y}}}^{c_{\vv{x+y}}} $
\\
\\
$\zig{N}_{\zig{a}^{r}_{\vv{x},{\bf g}} \zig{b}_{\vv{y},{\bf h}}}^{\zig{c}^{q}_{\vv{x+y},{\bf gh}}}$&$  = N_{a_{\vv{x}} b_{\vv{y}}}^{c_{\vv{x+y}}}
\, \Xi_{\zig{a}^{r}_{\vv{x},{\bf g}} \zig{b}_{\vv{y},{\bf h}}}^{\zig{c}^{q}_{\vv{x+y},{\bf gh}}}
$
\end{tabular}
&
$\zig{N}_{\zig{a}^{r}_{\vv{x},{\bf g}} \zig{\hat{b}}_{\vv{y},{\bf h}}}^{\zig{c}_{\vv{x+y},{\bf gh}}}  = N_{a_{\vv{x}} b_{\vv{y}}}^{c_{\vv{x+y}}} $
&
$\zig{N}_{\zig{a}^{r}_{\vv{x},{\bf g}} \zig{b}^{s}_{\vv{y},{\bf h}}}^{\zig{c}_{\vv{x+y},{\bf gh}}}  = N_{a_{\vv{x}} b_{\vv{y}}}^{c_{\vv{x+y}}}
\, \Xi_{\zig{a}^{r}_{\vv{x},{\bf g}} \zig{b}^{s}_{\vv{y},{\bf h}}}^{\zig{c}_{\vv{x+y},{\bf gh}}} $
\\ [-2.5ex]
 &&&\\
 \hline
\end{tabular}
\caption{Fusion coefficients $\zig{N}$ of the base theories $\fMTC^{\times  \text{ base}}_{\mathbb{Z}_2^{\eff} \times G, \rho}$ of Eq.~\eqref{eq:generalbasetheory} given in terms of the basic data of $\fMTC$.
These are computed by determining the fusion subspaces of the stacked theory prior to condensation that are invariant under fusion with the condensate.
The three charge types listed are: $\zig{a}_{\vv{x},{\bf g}}$ for $\pi({\bf g}) =\vv{0}$ and $\vv{x}=\vv{0}$ or $\vv{1}$; $\zig{\hat{a}}_{\vv{x},{\bf g}}$ for $\pi({\bf g}) =\vv{1}$ and $\vv{x}=\vv{0}$ or $v$; and $\zig{a}^{r}_{\vv{x},{\bf g}}$ (where $r = \pm$) for $\pi({\bf g}) =\vv{1}$ and $\vv{x} = \sigma$.
The $\zig{\hat{a}}_{\vv{x},{\bf g}}$ defects are $\sigma$-type (i.e. $\psi_{\vv{0}}$-zero modes), while the $\zig{a}^{r}_{\vv{x},{\bf g}}$ defects are $v$-type.
The fusion coefficients involving two split charges $\zig{a}^{r}_{\vv{x},{\bf g}}$ include factors $\Xi$ associated with similarly splitting the fusion vector spaces due to the condensation.
These factors, which depend on the $F$- and $R$-symbols of $\fMTC$, are provided in Eqs.~\eqref{eq:Xi_1}-\eqref{eq:Xi_3} for examples with no fusion multiplicities.
}.
\label{table:noperm_defect_fusion}
\end{table*}

We first consider the topological charges of the base theory of Eq.~\eqref{eq:generalbasetheory}.
These can be determined from the topological charges of the product
$\ifo^{(0)}_{\mathbb{Z}_2^{\eff} \times G , [0,0,\maj]_{\vv{0}}} \boxtimes \left( \fMTC \boxtimes \spt_G^{[1]} \right)$ prior to the condensation step of the stacking.
We use $\zig{\phantom{a}}$ to denote quantities of the stacked theory, after the condensation step.
For $\pi({\bf g}) = \vv{0}$, we see that
\begin{align}
\left\{ (\I_{\vv{x} ,{\bf g}} , a_{\vv{x} ,{\bf g}} ) , (\psi_{\vv{x} ,{\bf g}} , [\psi a]_{\vv{x} ,{\bf g}} ) \right\}  \, \to \, \zig{a}_{\vv{x} ,{\bf g}}
,
\end{align}
which indicates that $d_{\zig{a}_{\vv{x} ,{\bf g}}} = d_{a_{\vv{x}}}$ and the number of defects in each $\mathpzc{g}$-sector with $\pi({\bf g}) = \vv{0}$ is given by
\begin{align}
|\zig{\fMTC}_{\vv{x} ,{\bf g}} | = |{\fMTC}_{\vv{x} ,{\bf 0}} |
.
\end{align}
For $\pi({\bf g}) = \vv{1}$, when $\vv{x} = \vv{0}$ or $v$ (where $v$ denotes a $v$-type vortex in $\fMTC$), i.e. $a_{\vv{x}} \neq [\psi a]_{\vv{x}}$, we have the $\sigma$-type defects in the base theory labeled as
\begin{align}
\left\{ (\sigma_{\vv{x} ,{\bf g}} , a_{\vv{x} ,{\bf g}} ) , (\sigma_{\vv{x} ,{\bf g}} , [\psi a]_{\vv{x} ,{\bf g}} ) \right\}  \, \to \, \zig{\hat{a}}_{\vv{x} ,{\bf g}}
,
\end{align}
which have $d_{\zig{\hat{a}}_{\vv{x} ,{\bf g}}} = \sqrt{2} d_{a_{\vv{x} }}$.
For $\pi({\bf g}) = \vv{1}$, when $\vv{x} = \sigma$ (where $\sigma$ denotes a $\sigma$-type vortex in $\fMTC$), i.e. $a_{\vv{x}} = [\psi a]_{\vv{x}}$, we have the $v$-type defects of the base theory labeled as
\begin{align}
(\sigma_{\vv{x} ,{\bf g}} , a_{\vv{x} ,{\bf g}} )  \, &\to \, \left\{ \zig{a}^{+}_{\vv{x} ,{\bf g}} , \zig{a}^{-}_{\vv{x} ,{\bf g}} \right\}, \\
\zig{a}^{\pm}_{\vv{x} ,{\bf g}} &= \zig{\psi}_{\mathpzc{0}} \otimes \zig{a}^{\mp}_{\vv{x} ,{\bf g}}
,
\end{align}
which have $d_{\zig{a}^{\pm}_{\vv{x} ,{\bf g}}} =  \frac{1}{\sqrt{2}} d_{a_{\vv{x} }}$.
It follows that the number of defects in each $\mathpzc{g}$-sector of the base theory with $\maj({\bf g}) = \vv{1}$ is given by
\begin{align}
|\zig{\fMTC}_{\vv{0} ,{\bf g}} | & = \frac{1}{2}|{\fMTC}_{\vv{0} ,{\bf 0}} | ,
\\
|\zig{\fMTC}_{\vv{1} ,{\bf g}} | & = \frac{1}{2}|{\fMTC}_{v ,{\bf 0}} | + 2 |{\fMTC}_{\sigma ,{\bf 0}} |
.
\end{align}

We can also compute the fusion rules of the defects of the base theory in terms of the data of $\fMTC$ in this manner.
The fusion rules of the theory after condensation are determined by the fusion spaces of the theory prior to condensation.
Once a choice of module object has been made, the fusion space can be determined using Eq.~\eqref{fusionmodob}.
For objects that do not split, it is helpful to choose representative objects from the theory prior to condensation, to define the module objects after condensation.
Eq.~\eqref{eq:representative_simple} provides a way of making this identification.
We will denote the resulting module object coming from the representative $ (\I_{\vv{x},{\bf g}} , a_{\vv{x},{\bf g}})$ as $\zig{a}_{\vv{x},{\bf g}} $, similarly for $(\sigma_{\vv{x} ,{\bf g}} , a_{\vv{x} ,{\bf g}} )$ and $\zig{\hat{a}}_{\vv{x} ,{\bf g}} $.
Said another way, $\zig{a}_{\vv{x},{\bf g}}$ is the image of $ {a}_{\vv{x},{\bf g}}$ under condensation as prescribed by Eq.~\eqref{modcond}.
When an object splits under condensation, we will denote the corresponding module object as $\zig{a}^{r}_{\vv{x} ,{\bf g}}$, which is in the image of $(\sigma_{\vv{x} ,{\bf g}} , a_{\vv{x} ,{\bf g}} )$ under condensation.
Again, these module objects take a standardized form given by Eq.~\eqref{eq:splitting}.
Fusion of objects that do not split under condensation are directly given by the fusion of representative objects from the theory prior to condensation.
The fusion rules involving the charges that do split can be found straightforwardly by solving for the $\alpha$ in Eq.~\eqref{fusionmodob}.

The resulting fusion rules for the defects of the base theory of Eq.~\eqref{eq:generalbasetheory} in terms of the data of $\fMTC$ are displayed in Table~\ref{table:noperm_defect_fusion}.
When there are no fusion multiplicities for the topological charges that split under condensation, the factors associated with these split charges are given by
\begin{align}
\Xi_{\zig{a}_{\vv{0},{\bf g}} \zig{b}^{s}_{\sigma,{\bf h}}}^{\zig{c}^{q}_{\sigma,{\bf gh}}}&  =
\frac{1 +qs F^{a_{\vv{0}} b_{\sigma} \psi_{\vv{0}}}_{c_{\sigma}}}{2}
,
\label{eq:Xi_1}
\\
\Xi_{\zig{a}^{r}_{\sigma,{\bf g}} \zig{b}_{\vv{0},{\bf h}}}^{\zig{c}^{q}_{\sigma,{\bf gh}}}  &=
\frac{ 1 +qr F^{b_{\vv{0}} a_{\sigma} \psi_{\vv{0}}}_{c_{\sigma}} }{2}
,
\label{eq:Xi_2}
\\
\Xi_{\zig{a}^{r}_{\sigma,{\bf g}} \zig{b}^{s}_{\sigma,{\bf h}}}^{\zig{c}_{\vv{0},{\bf gh}}}  &= \frac{1 + i \, rs F^{a_{\sigma} \psi_{\vv{0}} b_{\sigma}}_{c_{\vv{0}}}  R^{\psi_{\vv{0}} b_{\sigma}}}{2}
.
\label{eq:Xi_3}
\end{align}
We recall from Sec.~\ref{sec:canonical} that $\left[F^{a_{\vv{0}} b_{\sigma} \psi_{\vv{0}}}_{c_{\sigma}} \right]^2 = \left[F^{a_{\sigma} \psi_{\vv{0}} b_{\sigma} }_{c_{\vv{0}}} \right]^2 = \left[F^{\psi_{\vv{0}} a_{\sigma} b_{\vv{0}} }_{c_{\sigma}} \right]^2 = \openone$ and $R^{\psi_{\vv{0}} b_{\sigma}} = \pm i$.
When there are fusion multiplicities, Eqs.~\eqref{eq:Xi_1}-\eqref{eq:Xi_3} must be modified to account for the possibility of the fusion spaces also splitting.

We can also use condensation calculations to compute the properties of base theories needed to determine the relative defectification obstruction associated with applying the torsor functor $\mathcal{F}_{\cohosub{t}, \Xdot}$ to a base theory.
The relative defectification obstruction only depends on the $F$-, $R$-, $\rho$, $U$- and $\eta$-symbols of the quasiparticles and vortices, so we restrict our attention to these quantities for the remainder of this example, and leave the ${\bf g} = {\bf 0}$ labels on charges implicit, since we are no longer considering defects.
We show that, with appropriate gauge choices, these quantities for the base theory in Eq.~\eqref{eq:generalbasetheory} are given by
\begin{align}
\zig{N}_{\zig{a}_{\vv{x}} \zig{b}_{\vv{x}}}^{\zig{c}_{\vv{x}} } & =
N_{a_{\vv{x}} b_{\vv{x}}}^{c_{\vv{x}} }
,
\label{eq:zig-N}
\\
\zig{F}^{\zig{a}_{\vv{x}} \zig{b}_{\vv{x}} \zig{c}_{\vv{x}} }_{\zig{d}_{\vv{x+y+z}}} & =
{F}^{{a}_{\vv{x}} {b}_{\vv{x}} {c}_{\vv{x}} }_{{d}_{\vv{x+y+z}}}
,
\label{eq:zig-F}
\\
\zig{R}^{\zig{a}_{\vv{x}} \zig{b}_{\vv{y}}}_{\zig{c}_{\vv{x+y}}}
&= {R}^{{a}_{\vv{x}} {b}_{\vv{y}}}_{{c}_{\vv{x+y}}}
,
\label{eq:zig-R}
\\
\zig{\rho} &= \V^{\pi}
,
\label{eq:zig-rho}
\\
\zig{\U}_{\bf g}(\zig{a}_{\vv{x}}, \zig{b}_{\vv{y}} ; \zig{c}_{\vv{x} + \vv{y}})
&= \V(\zig{a}_{\vv{x}}, \zig{b}_{\vv{y}} ; \zig{c}_{\vv{x} + \vv{y}})^{\pi({\bf g})}
\label{eq:zig-U}
\\
\zig{\eta}_{\zig{a}_{\vv{x}}}({\bf g,h})  &=  i^{\vv{x}\cdot \pi({\bf g}) \cdot \pi({\bf h})}
,
\label{eq:zig-eta}
\end{align}
where the $N$-, $F$-, and $R$-symbols on the right hand side are those of $\fMTC$.
These $U$- and $\eta$-symbols match those of Sec.~\ref{sec:Ex_no_perm} with $\coho{w} = \I_{\vv{0}}$.
Eqs.~(\ref{eq:zig-N})-(\ref{eq:zig-eta}) specify all the topological data, including symmetry action and fractionalization, for the quasiparticle and vortex sectors of $\fMTC^{\times  \text{ base}}_{\mathbb{Z}_2^{\eff} \times G,\maj} $.
This is all that is needed to compute the relative defectification obstruction using Eq.~\eqref{eq:pwformula}.
When $\coho{t} \in Z^2(G,\mathcal{A}_{\vv{0}})$, the defectification obstruction reduces to Eq.~\eqref{eq:defectO_noperm}.

We now compute these quantities.
Focusing on the quasiparticle and vortex sectors, we take the representative topological charges $\zig{a}_{\vv{x}}  = (\I_{\vv{x}} , a_{\vv{x}})$.
With this choice of representatives, Eqs.~(\ref{eq:zig-N})-(\ref{eq:zig-R}) should be obvious.
The symmetry action on the topological charges is given by
\begin{align}
{}^{\bf g} \zig{a}_{\vv{x}} &= \zig{\rho}_{\bf g}(\zig{a}_{\vv{x}}) = (\psi_{\vv{0}}^{\, \vv{x} \cdot  \pi({\bf g})}  \otimes \I_{\vv{x}} , a_{\vv{x}})
\notag\\
&\cong (\I_{\vv{x}} , \psi_{\vv{0}}^{\, \vv{x} \cdot  \pi({\bf g})} \otimes a_{\vv{x}}) = \zig{\psi}_{\vv{0}}^{\, \vv{x} \cdot  \pi({\bf g}) } \otimes \zig{a}_{\vv{x}}
,
\end{align}
where the isomorphism is associated with fusion with the boson of the condensate.

The $U$- and $\eta$-symbols for the base theories require a bit more computation.
On the quasiparticle and vortex sectors, we find
\begin{align}
\label{eq:Uactionprime}
&\zig{\rho}_{\bf g} \left( \zigvabcprime \right) = \rhoppprime \notag \\
 &\quad \quad =\sum_{\nu}\left[\zig{U}_{\bf g}( \,^{\bf g}\zig{a}_{\bf x}, \,^{\bf g}\zig{b}_{\vv{y}}, \,^{\bf g}\zig{c}_{\vv{x+y}}) \right]_{\mu \nu} \zigvabcprimenu
,
\end{align}
which diagrammatically defines the $\zig{U}_{\bf g}$-symbols.
The $(\psi_{\vv{0}},\psi_{\vv{0}})$ strands correspond to the additional isomorphisms needed for the symmetry action to map representative objects back into the same set of representative objects.
Here, we have used the fact that the theory $\fMTC \boxtimes \spt_{G}^{[1]}$ has trivial $U$-symbols, and that the FSPT phase with $[0,0,\maj]_{\vv{0}}$ has $U_{\bf g}(\psi^{\vv{x} \cdot \pi({\bf g}) }_{\vv{0}} \otimes \I_{\vv{x}}, \psi^{ \vv{y} \cdot \pi({\bf g})}_{\vv{0}} \otimes \I_{\vv{y}}) = 1$.
It is clear that $\zig{\U}_{\bf g}(\zig{a}_{\vv{x}}, \zig{b}_{\vv{y}} ; \zig{c}_{\vv{x} + \vv{y}}) = \openone$ for $\pi({\bf g}) = \vv{0}$, while for $\pi({\bf g}) = \vv{1}$, we evaluate Eq.~\eqref{eq:Uactionprime} to find
\begin{align}
&\left[\zig{\U}_{\bf g}(\zig{a}_{\vv{x}}, \zig{b}_{\vv{y}} ; \zig{c}_{\vv{x} + \vv{y}}) \right]_{\mu \nu}
= (-1)^{\vv{x} \cdot \vv{y}} \left( R^{\psi_{\vv{0}}^{\vv{x}} [\psi^{\vv{y}}b]_{\vv{y}}} \right)^{-1}
\notag \\
& \qquad \times \sum_{\lambda}
\left[ F^{a_{\vv{x}} \psi_{\vv{0}}^{\vv{x}} [\psi^{\vv{y}} b]_{\vv{y}}}_{[\psi^{\vv{x+y}}c]_{\vv{x} + \vv{y}} } \right]_{([\psi^{\vv{x}}a]_{\vv{x}} ,\mu)([\psi^{\vv{x+y}} b]_{\vv{y}}, \lambda)}
\notag \\
& \qquad \qquad \times
\left[  F^{ a_{\vv{x} } [\psi^{\vv{x+y}}b]_{\vv{y}} \psi_{\vv{0}}^{\vv{x+y}}}_{c_{\vv{x+ y}}}  \right]_{([\psi^{\vv{x+y}}c]_{\vv{x}+ \vv{y}}, \lambda) (b_{\vv{y}} , \nu)}
.
\end{align}
From the diagrammatic definitions, we can observe that the symmetry action gauge transformation specified by
\begin{align}
\gamma_{\zig{a}_{\vv{x}}}({\bf g})  = i^{\vv{x}\cdot \pi({\bf g})} (R^{\psi_{\vv{0}}^{\vv{x}} [\psi^{\vv{x}} a]_{\vv{x}}})^{\pi({\bf g})}
\end{align}
yields
\begin{align}
\widecheck{\zig{\U}}_{\bf g}(\zig{a}_{\vv{x}}, \zig{b}_{\vv{y}} ; \zig{c}_{\vv{x} + \vv{y}})
&= \V(\zig{a}_{\vv{x}}, \zig{b}_{\vv{y}} ; \zig{c}_{\vv{x} + \vv{y}})^{\pi({\bf g})}
.
\end{align}
In other words, $\widecheck{\zig{\rho}} = \V^{\pi}$.

Using Eq.~\eqref{eq:condbraid}, we find the $\eta$-symbols for the quasiparticle and vortex sectors are
\begin{align}
\widecheck{\zig{\eta}}_{\zig{a}_{\vv{x}}}({\bf g,h})  = \zig{\eta}_{\zig{a}_{\vv{x}}}({\bf g,h})  = i^{\vv{x}\cdot \pi({\bf g}) \cdot \pi({\bf h})}
.
\end{align}
Here, we used the property $R^{\psi_{\vv{0}} a_{\vv{x}}} R^{\psi_{\vv{0}} [\psi a]_{\vv{x}}} =-1$ to find that this symmetry action gauge transformation leaves $\eta$ unchanged.

\subsection{FMTCs \texorpdfstring{$\fMTC = \ifo^{(\nu)} \boxtimes \, \widehat{\fMTC}_{\vv{0}}$}{prodthry} with Factorized Symmetry}
\label{sec:ex_KM0}

As noted in Sec.~\ref{sec:fMTC}, when a SMTC has the factorized form $\fMTC_{\vv{0}} = \mathbb{Z}_{2}^{\psi} \boxtimes \widehat{\fMTC}_{\vv{0}}$, where $\widehat{\fMTC}_{\vv{0}}$ is a MTC, the corresponding $\mathbb{Z}_{2}^{\eff}$ extensions are the FMTCs $\fMTC = \ifo^{(\nu)} \boxtimes \, \widehat{\fMTC}_{\vv{0}}$.
We note that any FMTC containing a $\ifo^{(\nu)}$ subcategory is of this form due to the factorization of MTCs~\cite{Muger2003}.
In fact, the vast majority of SMTCs and FMTCs take this form.

The fermionic topological symmetry groups associated with SMTCs and FMTCs of this form have subgroups where the topological symmetry acts independently on the two sectors, such that
\begin{align}
\on{Aut}(\widehat{\fMTC}_{\vv{0}}) &< \on{Aut}^{\eff}(\fMTC_{\vv{0}})  \\
\mathbb{Z}_2^{\V} \times \on{Aut}(\widehat{\fMTC}_{\vv{0}}) &< \on{Aut}^{\eff}(\fMTC).
\end{align}
(Recall $\on{Aut}^{\eff}(\ifo^{(\nu)}) = \mathbb{Z}_2^{\V}$.)
When the fermionic symmetry action on $\fMTC_{\vv{0}}$ takes the form $\rho_{\bf g}^{(\vv{0})} = \openone \boxtimes \hat{\rho}^{(\vv{0})}_{\bf g}$, its extensions to $\fMTC$ similarly take the factorized form $\rho_{\bf g} = \rho_{\bf g}^{\ifo^{(\nu)}} \boxtimes \hat{\rho}^{(\vv{0})}_{\bf g}$, that is
\begin{align}
[\rho]: G \to  \mathbb{Z}_{2}^{\V} \times \on{Aut}(\widehat{\fMTC}_{\vv{0}})
.
\end{align}
In this case, the symmetry enrichment is easy to analyze in terms of the two individual sectors, so we can use the analysis given for fermionic invertible phases $\ifo^{(\nu)}$ in this paper and the analysis for the bosonic MTCs given in Ref.~\onlinecite{Bark2019} (and reviewed in this paper).
The analysis for more general symmetry actions on such FMTCs depends on the particulars of the example, so we restrict our attention to the factorized form in this subsection.

When the fermionic symmetry action factorizes, the resulting fermionic symmetry fractionalization also factorizes.
The fractionalization obstruction is simply the product of the corresponding obstructions of the fermionic invertible phase $\ifo^{(\nu)}$ and the bosonic MTC $\widehat{\fMTC}_{\vv{0}}$.
In particular, whether the fermionic symmetry group $\mathcal{G}^{\eff} = \mathbb{Z}_{2}^{\eff} \times_{\central} G$ can be manifested is determined entirely by whether it can be manifested for $\ifo^{(\nu)}$, that is $\coho{O}^{\central} = \coho{O}^{\central}_{\ifo^{(\nu)}}$.
When unobstructed, the classification of symmetry fractionalization on $\fMTC$ for a given $\mathcal{G}^{\eff}$ is given torsorially by
\begin{align}
H^{2}_{[\rho^{(\vv{0})}]}(G , \mathcal{A}_{\vv 0 }) = H^{2}(G , \mathbb{Z}_{2}^{\psi}) \times  H^{2}_{[\hat{\rho}^{(\vv{0})}]}(G , \widehat{\mathcal{A}}_{\vv 0 })
.
\end{align}

The defectification obstruction is similarly given by the product of those of the fermionic invertible phase $\ifo^{(\nu)}$ and the bosonic MTC $\widehat{\fMTC}_{\vv{0}}$.
The resulting classification of the unobstructed $\mathcal{G}^{\eff}$-crossed FMTCs for a fixed fermionic symmetry action is given torsorially by $H^{2}_{[\rho^{(\vv{0})}]}(G , \mathcal{A}_{\vv 0 })$ and $H^{3}(G , \text{U}(1))$, up to relabelings.
From the fermionic stacking perspective (which cycles through fermionic symmetry actions related by $\mathbb{Z}_{2}^{\V}$), the $\mathcal{G}^{\eff}$-crossed FMTCs classification is given torsorially by $H^{2}_{[\hat{\rho}^{(\vv{0})}]}(G , \widehat{\mathcal{A}}_{\vv 0 })$ and $\Gr(\mathcal{G}^{\eff} )$.

The data for the $\mathcal{G}^{\eff}$-crossed FMTCs can be generated from the data of $\ifo^{(\nu) \times}_{\mathcal{G}^{\eff}}$ and $(\widehat{\fMTC}_{\vv{0}})_{G}^{\times}$ by gluing together theories as \begin{align}
\fMTC_{\mathcal{G}^{\eff}}^{\times} = \ifo^{(\nu) \times}_{\mathcal{G}^{\eff}} \underset{G}{\boxtimes} (\widehat{\fMTC}_{\vv{0}})_{G}^{\times}
.
\end{align}
All of the possible $\mathcal{G}^{\eff}$-crossed extensions (for $[\rho_{\bf g}^{(\vv{0})}] = [\hat{\rho}^{(\vv{0})}_{\bf g}]$) are generated in this way.

\subsection{Pfaffian States \texorpdfstring{$\fMTC = \operatorname{Ising}^{(\nu)} \boxtimes \, \mathbb{Z}_{8}^{(n+1/2)}$}{IZ8} }
\label{sec:ex_MooreRead}

In this section, we consider the FMTCs
\begin{align}
\fMTC &= \operatorname{Ising}^{(\nu)} \boxtimes \, \mathbb{Z}_{8}^{(n+1/2)},
\end{align}
where $\nu$ is an odd integer (mod 16), $n$ is an integer (mod 8), and the physical fermion corresponds to~\footnote{An alternative choice of the physical fermion as $(\empsi,0)$ would lead to less interesting theories that have product structure $\fMTC = \ifo^{(\nu)} \boxtimes \mathbb{Z}_{8}^{(n+1/2)}$ and are fully described by Sec.~\ref{sec:ex_KM0}.}
\begin{align}
\psi_{\vv{0}} \equiv (\empsi,{4})
.
\end{align}
The corresponding SMTCs $\fMTC_\vv{0}$ describe the quasiparticles of various topological phases, including the Moore-Read Pfaffian~\cite{Moore1991}, anti-Pfaffian~\cite{Lee07,Levin07}, ${\bf T}$- or ${\bf PH}$-Pfaffian~\cite{Bonderson13,Chen2014b}, etc. states, which include candidate descriptions of the fractional quantum Hall state at $5/2$ filling.
There are some redundancies when labeled this way, as there are 8 distinct FMTCs of this form for a given chiral central charge.
We can also consider the other 8 $\mathbb{Z}_{2}^{\eff}$ extensions of these SMTCs, which are related to $\fMTC$ by fermionic stacking with $\ifo^{(\nu)}$ where $\nu$ is odd.
A similar analysis applies for these 16-fold way relatives, though we have not computed the explicit data of these other FMTCs, so we cannot be as explicit in our analysis.

The topological charges of $\fMTC$ are labeled $(a,j)$ such that $a\in \{ \I, \sigma, \empsi \}$ and $j\in \{0,1,2,\ldots , 7\}$.
These satisfy the fusion rules
\begin{align}
\label{eq:MRfusion}
(a,j) \otimes (b,k) = (a \otimes b, [j+k]),
\end{align}
where the square brackets denote addition modulo $8$.
The $F$- and $R$-symbols are determined by the product structure.
The first factor of the product is the Ising theory, whose data is presented in the third panel of Table~\ref{table:sixteen}.
The $F$- and $R$-symbols of the $\mathbb{Z}_{8}^{(n+1/2)}$ theory are
\begin{align}
F^{j k l} &= e^{\frac{i \pi}{8}j(k+l-[j+k])} \frac{L^{j,k}L^{[j+k],l}}{L^{k,l }L^{j,[k+l]}}\\
 R^{j k} &= e^{\frac{2\pi i}{8}\left(n+\frac{1}{2} \right)jk} \frac{L^{k,j}}{L^{j,k}}
\end{align}
where the factors $L^{j,k}$ are vertex basis gauge transformations, which we take to be
\begin{align}
L^{j,k}  =
\begin{cases}
-1 & \text{if $(j,k) \in \{ (5,4),(7,4) \}$}\\
+1 & \text{otherwise}.
\end{cases}
\end{align}
These gauge transformations are required in order for the topological data of the product theory to satisfy the canonical gauge choices made in Sec.~\ref{sec:canonical}.

The choice of $(\empsi,{4})$ as the physical fermion determines the quasiparticle and vortex sectors:
\begin{align}
{\fMTC}_{\vv{0}} &= \{ (\I,{2j}), \, (\sigma, {2j+1}), \,  (\empsi, {2j}) :j \in \mathbb{Z}_4 \},
\\ {\fMTC}_{\vv{1}} &= \{ (\I , {2j+1}), \, (\sigma, {2j}), \, (\empsi, {2j+1}): j \in \mathbb{Z}_4 \}.
\end{align}
Notice that $R^{j,j} = e^{\frac{i \pi}{4}\left(n+\frac{1}{2}\right)j^2}$, from which we see that  $(\I,4)$ is an emergent boson while $(\empsi,0)$ is an emergent fermion.
We also note that the vortices of these FMTCs are all $v$-type, and caution the reader against potential confusion between $\sigma$-type vortices and the $\sigma$ topological charge labels of the Ising sector of these theories, i.e. $(\sigma, j)$ are not $\sigma$-type objects with respect to the physical fermion $\psi_{\vv{0}} = (\empsi,4)$.

We see that $\mathcal{A} = \mathbb{Z}_{2} \times \mathbb{Z}_{8}$, where $\mathbb{Z}_{2}$ can be generated by either $(\empsi,0)$ or $\psi_{\vv{0}} = (\empsi,4)$, and $\mathbb{Z}_{8}$ by $(\I , 1)$.
We also have $\mathcal{A}_{\vv{0}} = \mathbb{Z}_{2}^{\psi} \times \mathbb{Z}_{4}$, with $\mathbb{Z}_{4}$ generated by $(\I , 2)$.
Quotienting out the physical fermion, we have $\widehat{\bf A} = \mathbb{Z}_{8}$ and  $\widehat{\mathcal{A}}_{\vv{0}} = \mathbb{Z}_{4}$.
We can write the central extension $\widehat{\bf A} = \mathbb{Z}_{4} \times_{\hat{h}} \mathbb{Z}_{2}^{\eff}$ in terms of $\hat{e}_{\vv{1}} = \widehat{(\I , 1)}$ and $\hat{h}(1,1) = \hat{h}_{\vv{0}} = \widehat{(\I , 2)}$, and similarly $\mathcal{A} = \mathbb{Z}_{2}^{\psi} \times \mathbb{Z}_{4} \times_{h} \mathbb{Z}_{2}^{\eff}$ in terms of $e_{\vv{1}} = (\I , 1)$ and $h(1,1) = h_{\vv{0}} =(\I , 2)$.

We now consider $\Autf{{\fMTC} } $.
First, we know that $\Q$ is trivial since ${\fMTC}_{\vv{1}} $ has Abelian vortices.
Next, we know that there is a nontrivial vortex symmetry $\V$ since ${\fMTC}$ is a FMTC; the $U$-symbols of $\V$ are defined in Eq.~\eqref{eq:VsymmetryU}.
Additionally, ${\fMTC}$ has a charge conjugation symmetry ${\bf C}$, whose action on the topological charges is defined as
\begin{align}\label{eq:C-def}
{\bf C}((a,j)) = \overline{(a,j)} =  (a,[-j]).
\end{align}
The $U$-symbols are given by~\cite{Bark2019}
\begin{align}
U_{\bf C}((a,j),(b,k); (c,{[j+k]})) = (-1)^{j n_k},
\end{align}
where $n_k = 1$ if $k = 2,6$ and is zero otherwise (equivalently $n_k = k(k+1)^2/2$).
It is straightforward to see that ${\bf C}^{2} = \openone$.
Therefore, we have
\begin{align}
\Autf{{\fMTC} }  = \mathbb{Z}_2^{\bf C} \times \mathbb{Z}_2^{\V}
.
\end{align}

We can consider the general (bosonic) symmetry group $G$ and symmetry action $\rho^{(0)}_{\bf g} = {\bf C}_{\vv{0}}^{\vv{m}({\bf g})}$ on $\fMTC_{\vv{0}}$, where $\vv{m} \in H^1(G,\mathbb{Z}_2)$ is a homomorphism and ${\bf C}_{\vv{0}} = \res_{\fMTC_{\vv{0}}}({\bf C})$.
Using the section $s([{\bf C}_{\vv{0}}])=[{\bf C}]$, we see that $O^{\rho} = \openone$, so this can always be extended to a symmetry action on $\fMTC$, and the extended symmetry actions are given by
\begin{align}
\rho_{\bf g} = {\bf C}^{\vv{m}({\bf g})} \V^{\pi({\bf g})}
,
\end{align}
where $\pi \in H^1(G,\mathbb{Z}_2)$.
Since $\kappa_{{\bf C},{\bf C}} = \openone$, we can choose $\beta^{(0)}_{a_{\vv{0}}}({\bf g,h})=1$ and $\beta_{a_{\vv{x}}}({\bf g,h})= i^{\vv{x} \cdot \pi({\bf g}) \cdot \pi({\bf h})}$.

It follows the fractionalization obstructions $\coho{O}^{(\vv{0})} = \hat{\I}_{\vv{0}}$ and $\coho{O} = \I_{\vv{0}}$ vanish for all possible symmetry actions, so the symmetry can be fractionalized on both $\fMTC_{\vv{0}}$ and $\fMTC$, with projective phases given by
\begin{align}
\eta^{(0)}_{a_{\vv{0}}}({\bf g},{\bf h} ) &= M_{a_{\vv{0}} \cohosub{w}^{(0)}({\bf g},{\bf h})} ,
\\
\eta_{a_{\vv{x}}}({\bf g},{\bf h} ) &= i^{\vv{x} \cdot \pi({\bf g}) \cdot \pi({\bf h})}  M_{a_{\vv{x}} \cohosub{w}({\bf g},{\bf h})}
,
\end{align}
where $\coho{w}^{(0)} \in Z_{[\widehat{\rho}]}^2 (G, \widehat{\bf A})$ and $\coho{w} \in Z_{[\rho]}^2 (G, \mathcal{A})$.

However, symmetry fractionalization may not be realized for all possible $\mathcal{G}^{\eff} = \mathbb{Z}_{2}^{\eff} \times_{\central} G$.
For this, we compute the obstructions
\begin{align}
\coho{O}^{(0)\central} &= \cbd ( \bar{\hat{e}}_{\vv{1}}^{\central} ) = \hat{h}_{\vv{0}}^{ \tilde{\vv{m}} \cup \tilde{\central} + \tilde{\central} \cup_{1} \tilde{\central} }
, \\
\coho{O}^{\eta} &= \cbd \overline{\ell( \coho{w}^{(0)} )} = \psi_{\vv{0}}^{\pi \cup \central }
, \\
\coho{O}^{\central} &= \cbd ( \bar{e}_{\vv{1}}^{\central} ) = \psi_{\vv{0}}^{\pi \cup \central } \otimes h_{\vv{0}}^{ \tilde{\vv{m}} \cup \tilde{\central} + \tilde{\central} \cup_{1} \tilde{\central} }
,
\label{eq:MROw}
\end{align}
where we used the choice $\dot{\central} = \vv{0}$, which is possible because $\eta_{a_{\vv{x}}} = \beta_{a_{\vv{x}}}$ is a solution, and the lift $\ell( \hat{a}_{\vv{x}} ) = (\I , \hat{a}_{\vv{0}}, \vv{x}) = (\I , \hat{a}_{\vv{0}}, \vv{0}) \otimes e_{\vv{1}}^{\vv{x}}$ for which $\ell( \hat{e}_{\vv{1}} ) = e_{\vv{1}}$ and $\ell( \hat{h}_{\vv{0}} ) = h_{\vv{0}}$.
When the obstructions vanish, the symmetry fractionalization on $\fMTC_{\vv{0}}$ with fixed $\central$ is torsorially classified by $H_{[\widehat{\rho}^{(0)}]}^2 (G, \widehat{\mathcal{A}}_{\vv{0}})$, the extensions of fractionalization to vortices is torsorially classified by $H^2 (G, \mathbb{Z}_{2}^{\psi})$, and the fractionalization on $\fMTC$  with fixed $\central$ is torsorially classified by $H_{[\rho^{(0)}]}^2 (G, \mathcal{A}_{\vv{0}}) = H^2 (G, \mathbb{Z}_{2}^{\psi}) \times H_{[\widehat{\rho}^{(0)}]}^2 (G, \widehat{\mathcal{A}}_{\vv{0}})$.

In order to analyze defectification, we begin by considering $G=\mathbb{Z}_2$.
In this case, we can see that the fractionalization obstructions are trivial if $\central = \vv{0}$, and are only trivial for $\central \neq \vv{0}$ when the symmetry action is trivial, i.e. $\vv{m} = \pi = \vv{0}$.
Thus, we can have either $\mathcal{G}^{\eff} = \mathbb{Z}_{2}^{\eff} \times \mathbb{Z}_{2}$ or $\mathbb{Z}_{4}^{\eff}$ for $\rho_{\bf 1} = \openone$, but only $\mathcal{G}^{\eff} = \mathbb{Z}_{2}^{\eff} \times \mathbb{Z}_{2}$ for $\rho_{\bf 1} = {\bf C}$, $\V$, and ${\bf C} \V$.
For all the unobstructed situations, the symmetry fractionalization with fixed $\central$ is classified torsorially by $H_{[\rho^{(0)}]}^2 (\mathbb{Z}_{2}, \mathcal{A}_{\vv{0}}) = \mathbb{Z}_{2} \times \mathbb{Z}_{2}$.
Since $H^4 (\mathbb{Z}_2,\text{U}(1)) = \mathbb{Z}_{1}$, the defectification obstruction automatically vanishes, and defectification is classified torsorially by $H^3 (\mathbb{Z}_2,\text{U}(1)) = \mathbb{Z}_{2}$.

Next, we can generate $\mathbb{Z}_2^{\eff } \times  \mathbb{Z}_2^{\bf C} \times \mathbb{Z}_2^{\V}$ defect theories, by which we mean $G = \mathbb{Z}_2 \times \mathbb{Z}_2$ with $\rho_{({\bf g}_{1},{\bf g}_{2})} = {\bf C}^{{\bf g}_{1}} \V^{{\bf g}_{2}}$.
One approach to this is to first take the product of an $\fMTC_{\mathbb{Z}_2^{\eff} \times \mathbb{Z}_2^{\bf C}}^{\times}$ theory with a bosonic $\mathbb{Z}_2$-SPT theory to generate a $\mathbb{Z}_2^{\eff} \times \mathbb{Z}_2 \times \mathbb{Z}_2$ theory, and then stack this with a $\mathbb{Z}_2^{\eff} \times \mathbb{Z}_2 \times \mathbb{Z}_2$-FSPT characterized by $[0,0,\tau]_{\vv{0}}$ where $\tau(({\bf g}_{1},{\bf g}_{2})) = {\bf g}_{2}$ in order to obtain the desired symmetry action.
Choosing the $\fMTC^{\times}_{\mathbb{Z}_2^{\eff} \times \mathbb{Z}_2^{\bf C} } $ theory with $\coho{w} ({\bf 1,1})=(\I,0)$ and Frobenius-Schur indicator $\kappa =1$ for all Abelian defects to be the base theory, we can write the $\mathbb{Z}_2^{\eff } \times  \mathbb{Z}_2^{\bf C} \times \mathbb{Z}_2^{\V}$  base theory generated this way as
\begin{align}
\fMTC^{\times \text{ base}}_{\mathbb{Z}_2^{\eff} \times \mathbb{Z}_2^{\bf C} \times \mathbb{Z}_2^{\V}} =
\ifo^{(0),[0,0,\tau]_{\vv{0}}}_{\mathbb{Z}_2^{\eff} \times \mathbb{Z}_2 \times \mathbb{Z}_2} \ftimes
\left( \fMTC^{\times \text{ base}}_{\mathbb{Z}_2^{\eff} \times \mathbb{Z}_2^{\bf C} } \boxtimes \spt_{\mathbb{Z}_2}^{[1]} \right)
.
\end{align}
Since $\central = \vv{0}$, all the elements in this expression are unobstructed.

Finally, we can generate defect theories with general fermionic symmetry group $\mathcal{G}^{\eff}$ and action $\rho_{\bf g} = {\bf C}^{\vv{m}({\bf g})} \V^{\pi({\bf g})}$ by generating a $\mathbb{Z}_2^{\eff} \times G$ base theory
\begin{align}
\fMTC^{\times  \text{ base}}_{\mathbb{Z}_2^{\eff} \times G , \rho} &= \left. \fMTC^{\times  \text{ base}}_{\mathbb{Z}_2^{\eff} \times \mathbb{Z}_2^{\bf C} \times \mathbb{Z}_2^{\V}} \boxtimes \spt_{G}^{[1]} \right|_{\mathcal{S}_{\vv{m},\pi}} \\
\mathcal{S}_{\vv{m},\pi} &= \{ (a_{(\vv{x} , \vv{m}({\bf g}), \pi({\bf g}) )} , \I_{\bf g} ) \}
.
\end{align}
Since all the elements used to construct this base theory were unobstructed, this base theory is also unobstructed.
From this base theory, we can obtain all $\mathcal{G}^{\eff}$-crossed extensions using the torsor method
\begin{align}
\fMTC^{\times}_{\mathcal{G}^{\eff} , \rho} = \mathcal{F}_{\cohosub{t}, \Xdot} \left(\fMTC^{\times \text{ base}}_{\mathbb{Z}_2^{\eff} \times G , \rho} \right)
,
\end{align}
where $\coho{t} \in Z^2_{\rho}(G, \mathcal{A})$ will automatically only generate the $\central$ with vanishing obstruction $[\coho{O}^{\central}] = [\I_{\vv{0}}]$.
In particular, $\coho{t}$ with $M_{\psi_{\vv{0}} \cohosub{t}} = (-1)^{\central}$ only exists if and only if Eq.~\eqref{eq:MROw} is a coboundary.
The $\Xdot$ satisfies $\cbd \Xdot = \defectO_{r}^{-1}$, where the defectification obstruction $[\defectO] = [\defectO_{r}]$ is equal to the relative defectification obstruction given in Eq.~\eqref{eq:pwformula}.

It is useful to consider the case where $G=\mathbb{Z}_{2}$ and $\rho_{\bf 1} = {\bf C}$ in more detail.
The 2-cocycle condition simplifies to $\coho{w}({\bf 1},{\bf 1}) = \overline{\coho{w}}({\bf 1},{\bf 1})$, i.e. the 2-cocycles $Z_{[\rho^{(0)}]}^2 (\mathbb{Z}_{2}, \mathcal{A})$ correspond to charge conjugation invariant Abelian topological charges $\mathcal{A}^{\bf C} = \{(\I,0), (\empsi,0), (\I,4),(\empsi,4)\}$.
The 2-coboundaries are trivial.
Thus, $H_{[\rho^{(0)}]}^2 (\mathbb{Z}_{2}, \mathcal{A}_{\vv{0}}) = \mathcal{A}^{\bf C}$.

Since there are six charge conjugation invariant topological charges, i.e. $|{\fMTC}^{\bf C}|=6$, there are six ${\bf C}$-defects.
We can use this, together with consistency conditions to determine the fusion rules of the defects.
For this, we first note that ${\bf g}$-defects are zero modes of at least one of the charges in the fusion channels of $^{\bf g}a_{\bf 0} \otimes \bar{a}_{\bf 0}$ for all $a_{\bf 0} \in \fMTC$, that is, there exists at least one $c_{\bf 0}$ such that $N_{^{\bf g}a_{\bf 0} , \bar{a}_{\bf 0}}^{c_{\bf 0}} \neq 0$ and $N_{c_{\bf 0} b_{\bf g}}^{b_{\bf g}} = N_{b_{\bf g} \overline{b_{\bf g}}}^{c_{\bf 0}} \neq 0$ for all $a_{\bf 0}$.
This corresponds to the physical process where $a_{\bf 0}$ and $\bar{a}_{\bf 0}$ are pair-created from vacuum, $a_{\bf 0}$ is braided around the ${\bf g}$-defect, and then fused back together with $\bar{a}_{\bf 0}$ to result in the charge $c_{\bf 0}$; since the process results in an extra charge $c_{\bf 0}$ and does not change the localized charge of the defect, any ${\bf g}$-defect is a $c_{\bf 0}$ zero mode.
Using $a_{\bf 0} = (\I,j)$ and $(\empsi ,j)$, which are Abelian, we see that ${\bf C}$-defects are necessarily zero-modes of $(\I ,2j)$.
Using  $a_{\bf 0} = (\sigma,j)$, there is the possibility that ${\bf C}$-defects are also zero-modes of $(\empsi ,2j)$, but we will show that they are not.
Next, we pick one of the ${\bf C}$-defects' topological charges that has the smallest quantum dimension and label the charge $X_{(\I,0)}$.
We can label the defects obtained by fusion with charges in $\fMTC$ as
\begin{align}
X_{(a,j)} = (a,j) \otimes X_{(\I,0)}
.
\end{align}
Since the fusion outcome with an Abelian charge is a unique topological charge, $X_{(a,j)}$ is clearly a topological charge (simple object) for $(a,j)$ Abelian.
On the other hand, when $a = \sigma$, this is also true because if $(a,j) \otimes X_{(\I,0)}$ had multiple fusion channels, then the sum of their quantum dimensions would be at least $2 d_{X_{(\I,0)}}$ (since we defined $X_{(\I,0)}$ to have the minimal quantum dimension of the defects), which is strictly larger than $\sqrt{2} d_{X_{(\I,0)}}$, violating Eq.~\eqref{eq:d_relation}.
Finally, the fact that the ${\bf C}$-defects are zero-modes of $(\I ,2)$ yields the equivalence relations
\begin{align}
X_{(a,j)} = (\I,2) \otimes X_{(a,j)} =  X_{(a,[j+2])}
.
\end{align}
This yields six distinct ${\bf C}$-defect topological charges,
\begin{align}
\fMTC_{\bf C} = \left\{ X_{(\I,0)} , X_{(\I,1)}, X_{(\sigma,0)}, X_{(\sigma,1)}, X_{(\empsi,0)}, X_{(\empsi,1)} \right\}
,
\end{align}
and also demonstrates that the ${\bf C}$-defects cannot also be zero-modes of $(\empsi ,2j)$, since that would reduce the number of distinct defect charges to four.
This also allows us to determine the quantum dimensions
\begin{align}
d_{X_{(a,j)}} = 2 d_{a}
,
\end{align}
since the total quantum dimension of defects $\mathcal{D}^{2}_{\fMTC_{\bf C}} = 8 d_{X_{(a,j)}}^{2}$ must equal that of the FMTC $\mathcal{D}^{2}_{\fMTC} = 32$.

Using the above arguments together with associativity of fusion, we find the fusion rules
\begin{align}
(a,j) \otimes X_{(b,k)} &= \sum_{c} N_{a b}^{c} \, X_{(c,[j+k]\text{mod}2)} , \\
X_{(a,j)} \otimes (b,k) &= \sum_{c} N_{a b}^{c} \, X_{(c,[j+k]\text{mod}2)} , \\
\end{align}
where $j ,k \in \mathbb{Z}_{2}$ and $N_{a b}^{c}$ are the fusion coefficients of the Ising MTCs.
Since $\central =\vv{0}$, fusion preserves vorticity, that is
\begin{align}
M_{X_{(a,j)},(\empsi,4)} &= M_{(a,j),(\empsi,4)} M_{X_{(0,0)},(\empsi,4)} \notag \\
&=(-1)^{j + \delta_{a , \sigma} } M_{X_{(0,0)},(\empsi,4)}
.
\end{align}
We can choose $X_{(\I,0)}$ to have $+1$ vorticity, and thus divide the defects into their vorticity sectors as
\begin{align}
{\fMTC}_{\vv{0},{\bf 1}}  &= \{ X_{(\I, 0)}, \,  X_{(\sigma, 1)}, \, X_{(\empsi,0)}\},\\
{\fMTC}_{\vv{1},{\bf 1}}  &= \{ X_{(\I, 1)}, \,X_{(\sigma, 0)}  , \, X_{(\empsi,1)}\}.
\end{align}

In order to determine the fusion of defects with defects, we first observe that the zero-mode property gives
\begin{align}
X_{(\I,0)} \otimes \overline{X_{(\I,0)}} &= (\I,0) + (\I,2) + (\I,4) + (\I,6)
.
\end{align}
Since fusion preserves vorticity, we can see that $\overline{X_{(\I,0)}}$ can either be $X_{(\I,0)}$ or $X_{(\empsi,0)}$.
In fact, both of these are possible and correspond to theories with different fractionalization classes.
We now see that the fusion rules of defects can be written as
\begin{align}
X_{(a,j)} \otimes X_{(b,k)} = \coho{t}({\bf 1,1}) \otimes \sum_{\substack{ l \in \mathbb{Z}_{4} \\ c }} N_{ab}^{c} \, (c,[j+k+2l])
,
\end{align}
where $\coho{t}({\bf 1,1}) \in \mathcal{A}^{\bf C}$ correspond to the four possible fractionalization classes.

In order to generate the rest of the $\mathcal{G}^{\eff} = \mathbb{Z}_{2}^{\eff} \times \mathbb{Z}_{2}$ defectification data for $\rho_{\bf 1} = {\bf C}$, one can exploit the product structure to write the $\mathcal{G}^{\eff}$-crossed data in terms of the (bosonic) $\mathbb{Z}_{2}$-crossed MTC defect data extending $\operatorname{Ising}^{(\nu)}$ and $\mathbb{Z}_{8}^{(n+1/2)}$, individually.
In particular, we observe that
\begin{align}
{\fMTC}_{\mathbb{Z}_2^{\bf C}}^{\times}  = \left [\operatorname{Ising}^{(\nu)} \right]_{\mathbb{Z}_2}^{\times} \underset{\mathbb{Z}_{2}}{\boxtimes} \left[ \mathbb{Z}_{8}^{(n+1/2)} \right]_{\mathbb{Z}_2^{\bf C}}^{\times}
\end{align}
yields all of the $\mathbb{Z}_{2}^{\eff} \times \mathbb{Z}_{2}$ extensions of $\fMTC$ with $\rho_{\bf 1} = {\bf C}$, when we identify $(\empsi,{4}) \in {\fMTC}_{\vv{0},{\bf 0}}$ as the physical fermion.
(The product has $G = \mathbb{Z}_{2} \times \mathbb{Z}_{2}$, but the diagonal product restricts to group elements in $\mathbb{Z}_{2} \cong \{({\bf 0,0}),({\bf 1,1})\}$.)
The Ising sector provides the $(\empsi , 0)$ generator and the $ \mathbb{Z}_{8}$ sector provides the $(\I ,4)$ generator of the $H_{[\rho^{(0)}]}^2 (\mathbb{Z}_{2}, \mathcal{A}_{\vv{0}}) = \mathbb{Z}_{2} \times \mathbb{Z}_{2}$ fractionalization classification, and the defectification classes of either sector provides the $H^3 (\mathbb{Z}_2,\text{U}(1)) = \mathbb{Z}_{2}$ defectification classification.
The complete data for $\mathbb{Z}_2$-crossed extensions $\left [\operatorname{Ising}^{(\nu)} \right]_{\mathbb{Z}_2}^{\times}$ and a significant portion of the data for $\left[ \mathbb{Z}_{8}^{(n+1/2)} \right]_{\mathbb{Z}_2^{\bf C}}^{\times}$ were presented in Ref.~\onlinecite{Bark2019}.

\subsection{\texorpdfstring{$\mathbb{Z}_{2}^{\eff}$}{Z2f} Modular Extensions of \texorpdfstring{$\fMTC_{\vv{0}} =\operatorname{\text{SO}(3)}_{4n+2}$}{SO34np2} }
\label{sec:ex_SU2}

In this section, we consider the SMTCs $\fMTC_\vv{0} =\text{SO}(3)_{4n+2}$ and their $\mathbb{Z}_{2}^{\eff}$ modular extensions.
We will focus specifically on the modular extension $\fMTC=\text{SU}(2)_{4n+2}$, but same analysis applies for the other modular extensions with the vortex sectors modified by stacking with $\ifo^{(\nu)}$.
This example has $\widehat{\bf A}_{\vv{1}} = \varnothing$, so $\Q$ acts nontrivially on the quasiparticle sector $\fMTC_\vv{0}$.
We provide the minimal data required to classify symmetry action, fractionalization, and defectification of the theories $\fMTC_{\vv{0}}=\text{SO}(3)_{4n+2}$ and $\fMTC=\text{SU}(2)_{4n+2}$;
the complete list of topological data is given in Ref.~\onlinecite{Bond2007}.
The topological charges for $\fMTC=\text{SU}(2)_{4n+2}$ are given by
\begin{align}
\fMTC_{\vv{0}} &= \{ 0, 1, 2, \cdots , 2n+1 \}, \\
\fMTC_{\vv{1}} &= \left \{ \frac{1}{2}, \frac{3}{2}, \cdots,2n+\frac{1}{2} \right \},
\end{align}
with quantum dimensions and topological spins
\begin{align}
d_{j} &=\sin(\frac{(2j+1)\pi}{4n+4})/\sin(\frac{\pi}{4n+4} )
\\ \theta_{j} &= e^{i 2 \pi \frac{j(j+1)}{4n+4}}.
\end{align}
We see that only the vacuum charge $\I_\vv{0}=0$ and the fermion $\psi_\vv{0} =2n+1$ are Abelian.
For $\fMTC=\text{SU}(2)_{4n+2}$, $\psi_\vv{0}\otimes j = (2n+1) -j$ implies there is a single $\sigma$-type object given by the vortex $n+\frac{1}{2}$.

Our first step is to compute $\Autf{\text{SO}(3)_{4n+2}}$.
As all charges $j\in \fMTC_\vv{0}$ have distinct topological spin, we immediately see that there are no charge-permuting topological symmetries and the only potentially nontrivial automorphism is $\Q$.
While Sec.~\ref{sec:Topo_symmetry} tells us that $\widehat{{\bf A}}_{\vv{1}} =\varnothing$ implies $\Q$ is nontrivial, it is helpful to explicitly verify that $\Q$ cannot be trivialized by a fermionic natural isomorphism $\Upsilon^{\eff}$.
From
\begin{align}
\Upsilon^{\eff} \circ \Q (\ket{n,n;n}) &=\gamma_n \ket{n,n;n}, \label{eq:first1}\\
\Upsilon^{\eff} \circ \Q (\ket{n,\psi;n+1}) &=- \frac{\gamma_n}{\gamma_{n+1} } \ket{n,\psi ;n+1}, \label{eq:second2}
\end{align}
we see that in order for $\Upsilon^{\eff}$ to trivialize $\Q$, we would need to simultaneously satisfy $\gamma_n=\gamma_{n+1}=1$ and $\gamma_n=-\gamma_{n+1}$.
As these constraints are incompatible, $\Q$ must be nontrivial and
\begin{align}
\Autf{\text{SO}(3)_{4n+2}} = \mathbb{Z}_2^{\Q}.
\end{align}

We next choose a fermionic symmetry action on $\fMTC_\vv{0}$
\begin{align}
\rho^{(\vv{0})}: G \to \Autf{\text{SO}(3)_{4n+2}}.
\end{align}
All possible $\rho^{(\vv{0})}$ correspond to $\Q$-projective homomorphisms, which can be parameterized by
\begin{align}
\label{eq:SO3action}
\rho^{(0)}_{\bf g} = \Q^{\rr({\bf g}) }
\end{align}
for $\rr \in C^1(G,\mathbb{Z}_2^{\Q}) $ .
The $\Q$-projective symmetry action satisfies ${[\Q]^{\cbd \rr({\bf g,h})} \cdot [\rho^{(\vv{0})}_{\bf g}] \cdot [\rho^{(\vv{0})}_{\bf h}] = [\rho^{(\vv{0})}_{\bf gh}]}$.
It follows that $\kappa_{\bf g,h} = \Q^{\cbd \rr({\bf g,h}) }$, hence $\beta_{a}({\bf g,h}) = (-1)^{\delta_{a,\psi} \cdot \cbd \rr({\bf g,h})}$, $\kappa^{\eff}({\bf g,h}) = \openone$, and $\beta_{a}^{\eff} = 1$.
With respect to fixed $\central$, $\rho^{(\vv{0})}$ is classified torsorially by $H^1(G,\mathbb{Z}_2^{\Q})$.

To fractionalize the symmetry on $\fMTC_{\vv{0}}$ with $\rho^{(\vv{0})}$, we need to check that the obstruction $[{\coho{O}}^{(\vv{0})}] \in H^3_{[\hat{\rho}]}(G,\widehat{{\bf A}})$ vanishes [see Eq.~\eqref{eq:H3SMTC}].
This follows from $\widehat{A}=\hat{\I}$.
The obstruction $[{\coho{O}}^{(\vv{0})\central}] = \central - \cbd \rr$ [see Eq.~\eqref{eq:Gf_obstruction_general_M0_1}] vanishing requires the fermionic symmetry group to have $\central=\cbd \rr$.

Considering now the FMTC $\fMTC$, we compute the fermionic topological symmetry group $\Autf{\text{SU}(2)_{4n+2}}$.
We already saw that $\Q$ is nontrivial on the quasiparticle sector, and thus $\Q$ is nontrivial on the full FMTC.
The vortex topological spins indicate that the only potential permutation of charges is given by $j\to \psi_\vv{0}\otimes j$; this permutation is in fact nontrivial and corresponds to $\V$.
Therefore,
\begin{align}
\on{Aut}^{\eff}(\text{SU}(2)_{4n+2}) = \mathbb{Z}_2^{\V}\times \mathbb{Z}_2^{\Q}.
\end{align}
Note that $\Autpsi{\text{SU}(2)_{4n+2}} = \on{Aut}^{\eff}(\text{SU}(2)_{4n+2}) /\mathbb{Z}_2^{\Q} = \mathbb{Z}_2^{\V}$.

To check whether the symmetry action $\rho^{(\vv{0})}$ extends to the full FMTC, we must show that $[O^{\rho}]$ vanishes [see Eq.~\eqref{eq:Orho}].
We compute the obstruction by picking a section $s: \res_{\fMTC_{\vv{0}}}( \Autf{\fMTC} ) \to \Autf{\fMTC}$, which is generically parametrized as $s[\rho^{(0)}] = [\V]^{w} \circ [\Q]^{r}$, where $w\in C^1(G,\mathbb{Z}_2^{\V})$.
One finds $O^\rho = [\V]^{\cbd w}$, which is obviously a coboundary, and thus every $\Q$-projective symmetry action on $\text{SO}(3)_{4n+2}$ extends to a $\Q$-projective symmetry action on $\text{SU}(2)_{4n+2}$.

In general, a fermionic symmetry action on $\text{SU}(2)_{4n+2}$ can be specified by a normalized 1-cochain $\rr \in C^1(G,\mathbb{Z}_2^{\Q})$ and a group homomorphism $\maj \in H^1(G,\mathbb{Z}^{\V}_2)$ as
\begin{align}
\rho_{\bf g} = \Q^{\rr({\bf g}) } \circ \V^{\maj({\bf g}) }.
\end{align}
The group homomorphism $\maj$ corresponds to the torsorial classification of extensions of the symmetry action [see Eq.~\eqref{eq:H1rhotorsor}].
Similar to $\rho^{(0)}$, we have $[\Q]^{\cbd \rr ({\bf g,h}) } \cdot [\rho_{\bf g}] \cdot [\rho_{\bf h}] =  [\rho_{\bf gh}]$, reflecting the $\Q$-projective symmetry action.
One can check that $\kappa_{\bf g,h}(a_{\vv{x}}, b_{\vv{y}}; c_{\vv{x+y}})  = (-1)^{(\delta_{a_{\vv{x}}, \psi}+\delta_{b_{\vv{y}}, \psi}+\delta_{c_{\vv{x+y}}, \psi})\cdot \cbd \rr({\bf g,h}) } (-1)^{\pi({\bf g} ) \cdot \pi({\bf h}) \cdot \vv{x} \cdot \vv{y}}$.
Hence, we may take $\beta_{a_{\vv{x}} }  = (-1)^{\delta_{\psi,a_{\vv{x}} } \cdot \cbd \rr({\bf g,h}) } i^{\vv{x} \cdot \pi({\bf g}) \cdot \pi({\bf h}) }$, $\kappa_{\bf g,h}^{\eff}(a_{\vv{x}},b_{\vv{y}};c_{\vv{x+y}}) = (-1)^{\pi({\bf g})\cdot \pi({\bf h}) \cdot \vv{x} \cdot \vv{y}}$, and $\beta_{a_{\vv{x}} }^{\eff}  = i^{\pi({\bf g}) \cdot \pi({\bf h}) \cdot \vv{x}}$.

To fractionalize the symmetry on $\fMTC$ with $\rho$, we need to check whether $[\coho{O}]$ vanishes (Sec.~\ref{sec:O3invclass}).
One can check by direct computation, or using the properties of cup products, that $\Omega_{a_{\vv{x}}} = \cbd( i^{\pi \cup \pi}) = 1$, and therefore $[\coho{O}] = [\openone]$.
The associated symmetry fractionalization torsor is given by the action of $Z^2_{[\rho]}(G,\mathcal{A})/B^2_{[\rho]}(G,\mathcal{A}_\vv{0})=H^2(G,\mathbb{Z}_2^{\psi})$, since $\mathcal{A}_\vv{1}=\varnothing$ and $\mathcal{A}_\vv{0} = \mathbb{Z}_2^{\psi}$.
Symmetry fractionalization is thus specified by
\begin{align}
\eta_{a_{\vv{x}}}({\bf g,h})  = (-1)^{\vv{x} \cdot \mathsf{p}({\bf g,h}) } \beta_{a_{\vv{x}}}({\bf g,h})
,
\end{align}
where $\coho{t} = \psi_{\vv{0}}^{\mathsf{p}} \in Z^2(G,\mathbb{Z}_2^{\psi})$.
In particular, this yields
\begin{align}
\eta_{\psi_{\vv{0}}}({\bf g,h}) = \beta_{\psi_{\vv{0}}}({\bf g,h}) = (-1)^{\cbd \rr({\bf g,h}) }.
\end{align}
Similar to before, the obstruction $[\coho{O}^{\central}] = \central - \cbd\rr  \in Z^2(G,\mathbb{Z}_2^{\eff})$ [see Eq.~\eqref{eq:Gf_obstruction_general_1}] vanishing requires $\central = \cbd\rr$.

We could alternatively analyze fractionalization (after verifying $[O^\rho]$ vanishes) by computing the obstructions to quasiparticle fractionalization $[\coho{O}^{(0)}]$ and $[\coho{O}^{(0)\central}]$, along with the obstruction to extending fractionalization to the vortex sector $[\coho{O}^{\eta}]$ [see Eq.~\eqref{eq:obstructioneta}].
The first two obstructions were computed above, the final obstruction vanishes as $\widehat{{\bf A}} = \mathbb{Z}_{1}$ and a lift can always be chosen with $\ell(\widehat{\I}_{\vv{0}}) = \I_{\vv{0}}$.

Finally, we turn to defectification.
The simplest defect theory corresponds to the direct product
\begin{align}
\fMTC_{\mathbb{Z}_{2}^{\eff}\times G}^{\times \text{ base}}  &=  \text{SU}(2)_{4n+2} \boxtimes \spt_{G}^{[1]} .
\end{align}
We can generate all remaining theories from this base theory via various torsor actions, which can change the $\Q$-projective symmetry action, $\V$-symmetry action, fractionalization class, and defectification class.
We can generate the previously-considered $\Q$-projective symmetry actions by applying the $\vviso$-isomorphism from Sec.~\ref{sec:centralisomorphism} with $\vviso=\rr$.
This isomorphism sends $\central \to \central + \cbd \rr$ and $\rho_{\bf g} \to \Q^{\rr({\bf g}) } \circ \rho_{\bf g}$.
The corresponding defect sectors are
\begin{align}
\fMTC_{\rr({\bf g}),{\bf g}} &= \{ 0_{\bf g}, 1_{\bf g}, 2_{\bf g}, \cdots , (2n+1)_{\bf g} \},\\
\fMTC_{[\vv{1}+\rr({\bf g})], {\bf g}} &= \left \{ \left(\frac{1}{2}\right)_{\bf g}, \left(\frac{3}{2}\right)_{\bf g}, \cdots,\left(2n+\frac{1}{2}\right)_{\bf g} \right \}.
\end{align}
Stacking with FSPT phases generates all distinct $\V$ symmetry actions, symmetry fractionalizations, and defectification classes.
Altogether, with respect to a fixed $\central=\cbd \rr$, we have a torsorial classification of the quasiparticle symmetry action by $H^1(G,\mathbb{Z}_2^{\Q})$ and of the FSPT stacking by $\Gr(\mathbb{Z}_2^{\eff}\times_{\cbd \rr}G)$.

Alternatively, one could use the torsor method with $\coho{t} \in Z^2(G,\mathbb{Z}_2^{\psi})$ to change the symmetry fractionalization class.
The associated obstruction is $\defectO_r = (-1)^{\mathsf{p}\cup \mathsf{p}} \cbd (-1)^{r \cup \mathsf{p}}  $; when $[\defectO_r]$ vanishes the possible defectification classes are torsorially generated by gluing in bosonic SPT phases $\spt^{[\alpha]}_G$, $[\alpha]\in H^3(G,U(1))$.
In this formulation, the torsorial classification with respect to a fixed $\central=\cbd \rr$ is given by $H^1(G,\mathbb{Z}_2^{\Q})$, $H^2(G,\mathbb{Z}_2^{\psi})$, and $H^3(G,U(1))$.

\section{Discussion}
\label{sec:discussion}

In this paper, we have developed the categorical description of symmetry, fractionalization, and defects for fermionic topological phases.
Our fermionic symmetry fractionalization and $\mathcal{G}^{\eff}$-crossed defect theory has provided a classification of FSET phases.
We have highlighted the similarities and differences compared to the bosonic classification, and illustrated our results through a number of explicit examples.
We now make contact with previous classification results obtained using other formalisms, as well as discuss possible extensions of our findings.

In Sec.~\ref{sec:fSPT}, we observed that the set of FSPT phases with $\mathcal{G}^{\eff} = \mathbb{Z}_2^{\eff} \times G$ are in one-to-one correspondence with the elements of the Pontryagin dual of the spin bordism group computed in Ref.~\onlinecite{Brumfiel2018a}.
In particular, the triples of cochains $[\Xdot, \mathsf{p}, \maj]_{\central}$, which provide solutions to the pentagon and heptagon equations, satisfy the same equations needed for elements in the Pontryagin dual of the spin bordism group.
Physically, the algebraic data describing the FSPT phase is equivalent to the data needed to describe topological invariants of spin 3-manifolds.
We also mention that elements of the Pontryagin dual of the spin bordism group can be naturally interpreted as a decorated domain wall construction for lattice realizations of FSPT phases (see, for example Ref.~\onlinecite{Wang2020}).
Correspondingly, one can view the algebraic theories of the FSPT phases studied here as describing the topological data of the excitations of these lattice constructions.
This correspondence provides strong evidence that the spin bordism, decorated domain wall, and $\mathcal{G}^{\eff}$-crossed classifications of FSPT phases are all equivalent.

In Sec.~\ref{sec:symmetric-fermions}, we detailed several new obstructions associated with symmetry fractionalization.
In particular, we observed that $[O^{\rho}] \in H^{2}(G , \mathbb{Z}_2^{\V})$ and $[{\coho{O}}^{\eta}] \in H^3(G,\mathbb{Z}_{2}^{\psi})$ describe obstructions to extending the symmetry action and fractionalization, respectively, from the SMTC describing the quasiparticles to the full FMTC including the vortices, in a local manner for $(2+1)$D fermionic topological phases.
These obstructions can be naturally interpreted in terms of surface terminations of $(3+1)$D FSPT phases with known defect decoration constructions, where $[O^{\rho}]$ correspond to Majorana chain decorations of tri-junctions of $G$-foams and $[\coho{O}^{\eta}]$ correspond to fermion parity decorations of quad-junctions of $G$-foams.
The latter surface termination had been investigated in Ref.~\onlinecite{Fidkowski2018}, while the former appears to be previously unexplored.

Another novel possibility arises when the quasiparticle theory has fractionalization class $\coho{w}^{(\vv{0})}({\bf g,h}) \in \widehat{\bf A}_{\vv{1}} \neq \widehat{\mathcal{A}}_{\vv{1}}$.
In this case, the extension of the fractionalization to the full FMTC is obstructed for a strictly $(2+1)$D fermionic topological phase, since the fractionalization class would correspond to a superAbelian $\sigma$-type vortex (i.e. one with quantum dimension $\sqrt{2}$).
In terms of defects, this anomaly suggests that the fusion of a ${\bf g}$-defect with a ${\bf h}$-defect into a ${\bf gh}$ defect leaves behind a Majorana zero mode at the tri-junction of the defect branch lines, which clearly does not fit the standard $G$-crossed formalism.
It would be interesting if such an anomalous theory could be realized, possibly as a surface termination of a $(3+1)$D phase with nontrivial Majorana chain decorations of tri-junctions of $G$-foams.

In Sec.~\ref{sec:GfSMTC}, our discussion of $\mathcal{G}^{\eff}$-crossed SMTCs focused on those that correspond to $G$-crossed FMTCs.
More generally, one could consider locally consistent $\mathcal{G}^{\eff}$-crossed SMTCs that cannot be written as $G$-crossed FMTCs.
Simple examples of this include $G$-crossed SMTCs, i.e. leaving the vortex sector empty.
However, one may imagine more complicated examples, such as ones for which the vortices have nontrivial symmetry action on the $G$ symmetry defects.
Such theories are exotic, as fermion parity cannot be gauged without breaking the $G$-symmetry and mapping class group transformations on the torus may not preserve the global fermion parity.
These observations indicate that such theories would represent a new type of (mixed) anomaly.
Developing the algebraic theories and explicit lattice models for these various anomalous $(2+1)$D theories and the corresponding $(3+1)$D phases that host them as surface terminations is an interesting future direction.

In this paper, we provided a full classification of $(2+1)$D FSET phases when the symmetry is on-site and unitary.
While the $G$-crossed formalism can be adapted to some of the more general locality preserving symmetries, such as translational symmetry~\cite{Cheng2016}, the generalization to spacetime reflecting symmetries has remained elusive.
However, the symmetry fractionalization formalism can be applied to general symmetries, including spacetime reflecting symmetries, and provides a classification of fractionalization.
Since the full classification of $(2+1)$D SET and FSET phases for on-site and unitary symmetries can be expressed in terms of the classifications of quasiparticle fractionalization and the classification of SPT and FSPT phases, respectively, we speculate that a similar form may enter the classification of SET and FSET phases with more general symmetries.

Finally, we note that throughout this paper we have focused on describing the local algebraic theory of symmetric fermionic topological phases.
Placing the theory on a surface with nontrivial topology requires additional care, including explicitly tracking the spin structure, as well as accounting for additional equivalence relations and their corresponding reduction in the Hilbert space dimension (see discussion at the end of Sec.~\ref{sec:SMTC_FMTC}).
Studying the mapping class group representations of spin manifolds with $G$-bundles is an interesting application we leave to future work.

\acknowledgements
We thank M.~Barkeshli, D.~Bulmash, Y.~Chen, T.~Ellison, C.~Jones, R.~Thorngren, Z.~Wang for useful discussions.
DA, PB, and CK acknowledge the hospitality and support of the Aspen Center for Physics, which is supported by National Science Foundation grant PHY-1607611.
DA was supported in part by a postdoctoral fellowship from the Gordon and Betty Moore Foundation, under the EPiQS initiative, Grant GBMF4304.
CK was supported in part by the Institute for Quantum Information and Matter, an NSF Frontier Center funded by the Gordon and Betty Moore Foundation, as well as by the Walter Burke Institute for Theoretical Physics at Caltech.

\emph{Note added}: During the preparation of this manuscript, we learned of related unpublished work that appeared concurrently on the arXiv in Refs.~\onlinecite{Bark2021frac,Bark2021cascade,Bark2021inv}.
Ref.~\onlinecite{Bark2021frac} analyzes characterization and classification of fermionic symmetry fractionalization for SMTCs, i.e. quasiparticles only.
Ref.~\onlinecite{Bark2021cascade} analyzes the tiered structure of the anomalies/obstructions for FSET phases.
Ref.~\onlinecite{Bark2021inv} analyzes the classification of the invertible FSET phases.
Subsequently, Ref.~\onlinecite{Ning2021}, which also analyzes the classification of the invertible FSET phases, appeared on the arXiv.

Among the additions and expansions to our paper in the v2 revision, several of them are directly related to these other papers.
We have updated our discussion of $\ker(\res_{\fMTC_{\vv{0}}})$ in Appendix~\ref{app:kerres} to include the possibility of certain elements discussed in Ref.~\onlinecite{Bark2021cascade}, and ruled out such possibilities for $\widehat{\bf A}_{\vv{1}} \neq \varnothing$ for which case we have proven our conjecture that $\ker(\res_{\fMTC_{\vv{0}}}) = \mathbb{Z}_{2}^{\V}$.
We have proven relations between $[O^{\rho}]$ for different $\mathbb{Z}_{2}^{\eff}$ modular extensions of the same SMTC in Sec.~\ref{sec:symmetry_action} [Eq.~\eqref{eq:Orho_nu_relation}] and relations between $[\coho{O}^{\central}]$ and $[\coho{O}^{\eta}]$ for different $\mathbb{Z}_{2}^{\eff}$ modular extensions of the same SMTC in Sec.~\ref{sec:symmetry_frac_extension} [Eqs.~\eqref{eq:Ocentral_relation} and \eqref{eq:Oeta_relation}], resolving conjectures made in v2 of Ref.~\onlinecite{Bark2021cascade} about the independence of these quantities on the choice of modular extension.
We have updated our analysis of invertible FSET phases in the case of $\nu$ even with $\central \neq 0$ (in particular Secs.~\ref{sec:inv-class} and \ref{sec:invfSET}) to correct our mis-evaluation of the fractionalization obstruction and to include the full details of classification, which was contained in Refs.~\onlinecite{Bark2021inv} and~\onlinecite{Ning2021}.
We also provide the full topological data of the unobstructed defect theories.
We carried this out using the definitions and methods from the original version of our paper, while following a similar calculation and utilizing important formulas for 4-coboundaries from Ref.~\onlinecite{Bark2021inv}.
Our result for the addition rule of 3-cochains for stacking invertible FSET phases [Eq.~\eqref{eq:3cochnaddeven}] reproduces that of Ref.~\onlinecite{Bark2021inv}.
We also use a condensation calculation to prove the addition rule of symmetry fractionalization classes for stacking invertible FSET phases [Eq.~\eqref{eq:p-add}], which was conjectured in v2 of Ref.~\onlinecite{Bark2021inv}.

Among the changes to our paper in the v4 revision, we have corrected an evaluation error in comparing the fractionalization obstruction $\coho{O}$ for different extensions of a symmetry action from quasiparticles to vortices, refined the presentation of the comparisons of $[\coho{O}^{\eta}]$ for different modular extensions, and extended the proof of Eq.~\eqref{eq:Oeta_relation} beyond the case where $\mathcal{A}_{\vv{1}} \neq \varnothing$ to all cases, except for $\nu$ odd when $\widehat{\bf A}_{\vv{1}} = \varnothing$.
We thank M.~Barkeshli and D.~Bulmash for helping uncover our error in comparing the $\coho{O}$ obstruction for different symmetry actions and for discussions of the comparisons of $[\coho{O}^{\eta}]$ for different modular extensions that motivated us to refine our presentation of these comparisons and extend our proof, as well as to calculate the obstruction comparisons for anti-unitary symmetries.
In light of our results, they have also updated their conjectures in Ref.~\onlinecite{Bark2021cascade} regarding the independence of the obstructions on modular extensions to statements that require the quotient by certain terms, which can be understood as corresponding to the boundary anomalous FSPT states associated with $(3+1)$D FSPT phases described in Ref.~\onlinecite{Wang2020}.

\appendix

\section{Factorization of Abelian BTCs with a Transparent Fermion}
\label{app:A_factorize}

In this Appendix, we prove that any Abelian BTC $\mathcal{A}$ containing a transparent $\mathbb{Z}_{2}$ fermion $\psi$ factorizes as
$\mathcal{A} =\mathbb{Z}_{2}^{(\psi)}  \boxtimes \widehat{\mathcal{A}}$, where $\widehat{\mathcal{A}}$ is a BTC.
By definition, $\psi \otimes \psi = \I$, $\theta_\psi =-1$, and $M_{\psi a} =1$ for all ${a \in \mathcal{A}}$.

We begin by noting that the fusion rules of an Abelian BTC are given by  group multiplication of the finite Abelian group $A$.
Finite Abelian groups can be expressed as the product of finite cyclic groups $A = \prod_{j=0}^{k} \mathbb{Z}_{N_{j}}$, where $N_{j} = P_{j}^{r_{j}}$ for (not necessarily distinct) prime numbers $P_{j}$ and positive integers $r_{j}$.
Restricting to the objects in one of these $\mathbb{Z}_{N_{j}}$ yields a subcategory $\mathbb{Z}_{N_j}^{p_{j}} \lhd A$.
Identifying the objects of $\mathcal{A}$ with the group elements of $A$, we assign $j=0$ to the subcategory containing the fermion $\psi$.
The fermion's $\mathbb{Z}_2$ fusion requires $N_0$ is even, and that $\psi$ is identified with $N_0/2\in \mathbb{Z}_{N_0}^{(p_0)}$.
Since the braiding of objects identified with elements of $\mathbb{Z}_{N}^{(p)}$ is given by
\begin{align}
\theta_{a} &= e^{i \frac{2 \maj}{N} p a^2}, \\
M_{ab} &= e^{i \frac{4 \pi}{N} p ab},
\end{align}
the braiding properties of $\psi$ indicate that $N_0 = 2$ and $p_0 = 1$ (see App.~\ref{app:AbelianFactor} for a review of the BTC $\mathbb{Z}_{N}^{(p)}$).
This demonstrates that (the group describing) the fusion rules factorize as $A = \mathbb{Z}_{2} \times \widehat{A}$, where $\widehat{A} = \prod_{j=1}^{k} \mathbb{Z}_{N_{j}}$.

Showing that the BTC factorizes requires a bit more work.
In order to see that the $F$-symbols factorize, we can note that the pentagon equation for an Abelian theory is just a $3$-cocycle condition and the vertex basis gauge transformations are just factoring out $3$-coboundaries, and hence the possible FTCs are classified by $H^{3}(A,\text{U}(1))$.
There are three possible types of cocyles for such theories given in Ref.~\onlinecite{Propitius1995}: type I are diagonal in $\mathbb{Z}_{N_{j}}$ subgroups, type II are pairwise terms between two cyclic subgroups, and type III are triple terms.
The nontrivial type III $F$-symbols are not compatible with braiding at all (i.e. no solutions to the hexagon equations).
The nontrivial type II $F$-symbols are not compatible with braiding when one of the two cyclic groups involved is $\mathbb{Z}_{2}$.
(This can be seen by inspecting the hexagon equations, which yields inconsistent conditions on the braiding, e.g. for $\left(R^{\psi , {N_j}/{2} }\right)^{2} =1$ and $-1$.)
Finally, the type I $F$-symbols in the $\mathbb{Z}_{2}$ containing $\psi$ must be trivial to be compatible with the self-statistics of the fermion.
Thus, the $F$-symbols of $\mathcal{A}$ can be chosen to be completely independent of the $\mathbb{Z}_{2}^{(\psi)}$, that is
\begin{align}
F^{abc} = F^{\hat{a} \hat{b} \hat{c}}
,
\end{align}
where we write $a = (a_{\psi} , \hat{a})$, and use the shorthand $\psi = (1,0)$ and $\hat{a} = (0,\hat{a})$.
In other words, the $F$-symbols can be factorized as claimed.

Using these $F$-symbols in the hexagon equations, together with the braiding properties of $\psi$, we find that
\begin{align}
R^{[a\otimes b] \psi} &= R^{a \psi} R^{b \psi} ,\\
R^{a \psi} &= R^{\psi a} = - R^{[\psi a] \psi} = \pm 1
.
\end{align}
With these relations, one can focus on the generators of $\widehat{\mathcal{A}}$, i.e. the generators $1_j \in \mathbb{Z}_{N_{j}}$.
These braiding conditions require the generators of $\mathbb{Z}_{N_{j}}$ with $N_{j}$ odd to satisfy $R^{a_j \psi} = 1$.
For $N_{j}$ even, there is freedom here to choose the generator such that $R^{1_j \psi} = 1$; if one inadvertently chose the generator such that $R^{1_j \psi} = -1$, one could simply do a relabeling $1'_j = \psi \otimes 1_j$ to obtain the desired form.
In this way, the resulting $R$-symbols are found to take the form
\begin{align}
R^{a b} &= R^{a_\psi b_\psi} R^{\hat{a} \hat{b}}.
\end{align}
Thus, the fusion rules, $F$-symbols, and $R$-symbols have be found (in a certain choice of gauge) to factorize, yielding
\begin{align}
\mathcal{A} =\mathbb{Z}_{2}^{(\psi)}  \boxtimes \widehat{\mathcal{A}}
.
\end{align}

\section{\texorpdfstring{$\mathbb{Z}_{N}^{(p)}$}{Znp} BTCs}
\label{app:AbelianFactor}

The $\mathbb{Z}_N^{(p)}$ BTCs can have $p\in\mathbb{Z}$ for all $N$ and $p\in\mathbb{Z}+\frac{1}{2}$ for $N$ even.
Topological charges are labeled by $a=0,1,\dots, N-1$, which obey the fusion rules $a\times b=[a+b]_N$, where $[]_N$ means mod $N$.
The $F$-symbols (in a particular choice of gauge) are
\begin{equation}
[F^{abc}_{[a+b+c]_N}]_{[a+b]_N,[b+c]_N}=e^{i \frac{2\pi p}{N}a(b+c-[b+c]_N)}.
\end{equation}
The $F$-symbols are all equal to $1$ when $p\in\mathbb{Z}$.
For $p\in\mathbb{Z}+\frac{1}{2}$, some of the $F$-symbols are equal to $-1$, in which case they cannot all be set to $1$ using gauge freedom.

The $R$-symbols (in our choice of gauge) are
\begin{equation}
R^{ab}_{[a+b]_N}=e^{i \frac{2\pi }{N}p ab}.
\end{equation}
The twist factors are $\theta_a=e^{i \frac{2\pi }{N} p a^2}$.

\section{Condensation Review}\label{app:condensation}

This appendix reviews the details of topological Bose condensation necessary to verify claims in the main text.
More comprehensive expositions of anyon condensation in both the mathematics and physics literature are given in Refs.~\onlinecite{Kirillov2001,Fuchs2002,Bais2009,Bais2014,Kong14,Bischoff19}.
The diagrammatic formalism of this appendix mostly follows that of Ref.~\onlinecite{Kirillov2001}; this formalism does not appear in the main text but can be used to verify all calculations described therein.

We consider a braided tensor category $\bMTC$ that contains a boson $b$ among its simple objects.
Physically, condensing $b$ means that we identify $b$ with the trivial object $\I $ and solve for the resulting theory.
Mathematically, we define
\begin{align}\label{eq:A}
A &= \I \oplus b
\end{align}
and solve for the `condensed theory' $\bMTC/A$.
Below, we define the simple objects, fusion rules, and $F$-symbols of $\bMTC/A$.
Although most of what we say holds more generally, we restrict our attention to bosons $b$ which satisfy
\begin{align}
\label{bdef}
b \otimes b &= \I ,&R^{bb} &= 1,& F^{bbb}_b&= 1.
\end{align}

An `algebra object' refers to the pair $(A,m)$, where $A \in \bMTC$ is the direct sum of vacuum and condensing boson defined in Eq.~\eqref{eq:A}, and $m$ is a co-multiplication morphism $m:A \to A\times A$ subject to certain consistency conditions, see e.g. Ref.~\onlinecite{Kirillov2001}.
Technically, we need a commutative Frobenius Algebra object, which requires a unit, a co-unit, a multiplication, and a comultiplication.
These maps need to satisfy several consistency conditions, of which we just give a representative list here, and the interested reader can look to~\cite{Kong14} for more details.
The comultiplication is defined diagrammatically as
\begin{align}
\label{algebramultiplication}
\AlgL = \AlgRa+\AlgRb+\AlgRc+\AlgRd.
\end{align}
The dual, multiplication map $m: A \times A \to A$ is found by reflecting the diagrams.
Given that $b = \bar{b}$ and $\kappa_b = +1$, we do not draw arrows on the $b$ strands.
A crucial consistency condition on $m$ is,
\begin{align}
\label{AlgObConsistency}
\AlgAssocLeft=\AlgAssocRight.
\end{align}
Using the assumptions in Eq.~\eqref{bdef} it is easy to check the comultiplication defined in Eq.~\eqref{algebramultiplication} satisfies this condition.
The Frobenius algebra object is called commutative if it further satisfies,
\begin{align}
\AlgCommLeft =\AlgCommRight.
\end{align}
Which can be easily verified using Eqs.~\eqref{algebramultiplication} and \eqref{bdef}.

The condensed theory $\bMTC/A$ inherits the objects and tensor product of objects from the parent theory $\bMTC$.
The key difference between the theories is that some objects in $\bMTC$ (namely, those related by fusion with $b$) become isomorphic in $\bMTC/A$.
This is written mathematically by modifying the morphisms so that
\begin{align}
\operatorname{mor}_{\bMTC/A}(x \to y) \cong \operatorname{mor}_{\bMTC}(x \to y \otimes A).
\end{align}
For  $f \in \operatorname{mor}_{\bMTC/A}(x \to y)$ and ${g \in \operatorname{mor}_{\bMTC/A}(x' \to y')}$, we define the tensor product as
\begin{align}\label{eq:mor-comp}
 \mortensa\;\; \otimes\;\; \mortensb \;\;\equiv \;\;\mortensc.
\end{align}
Eqs.~(\ref{eq:A}-\ref{eq:mor-comp}) define all the basic data and operations in $\bMTC/A$ in terms of that of the parent theory $\bMTC$.
We now use this data to compute the `module objects' in $\bMTC/A$, which, in somewhat of an abuse of terminology, we will refer to as the simple objects of $\bMTC/A$.

A `module object'  $(x,\mu)$ in $\bMTC/A$ is an object $x \in \bMTC$, along with a collection of morphisms $\mu :  x \to x \times A$  defined diagrammatically by
\begin{align}
(x,\mu) \; = \; \modob
\end{align}
and subject to the consistency condition
\begin{align}
\label{modcond}
\modoba\;\; = \; \; \modobb.
\end{align}
Notice that the pair $(A,m)$ automatically satisfy this condition, and indeed represent the ``new'' vacuum; see Eq.~\eqref{AlgObConsistency}.
Typically $x$ will either be a simple object which is fixed under fusion with $b$ or a direct sum of objects related by fusion with $b$.
In the latter case, we often use a representative object to label the isomorphism class and the algebra object to write down the corresponding simple module object,
\begin{align}
\label{eq:representative_simple}
\repmodob.
\end{align}
Writing the module object in this way can often simplify calculations.

A `module morphism' $f$ between objects $(x,\mu)$ and $(y, \nu)$ is given by $f \in \operatorname{mor}_{\bMTC/A}(x\to y)$ such that $[f \otimes\operatorname{id}] \circ (x, \mu)  = (y, \nu) \circ f$.
Two module objects $(x,\mu)$ and $(y, \nu)$ are isomorphic if there exists an isomorphism $f: x \to y$ such that
\begin{align}
\label{ismorphicmodob}
[ f \otimes\operatorname{id}] \circ (x, \mu)  = (y, \nu) \circ f.
\end{align}
In practice, it is not necessary to keep track of the whole isomorphism class of module objects, but rather a list of representatives, such as in Eq.~\eqref{eq:representative_simple}.
The endomorphism algebra of a module object $(x,\mu)$ is defined as
\begin{align}
&\operatorname{End}_{\bMTC/A}(x,\mu)= \notag
\\ &\quad \quad \{f \in \operatorname{End}_{\bMTC}(x): [ f \otimes\operatorname{id}] \circ (x, \mu)  = (x, \mu) \circ f\}.
\end{align}
A `simple' module object is one whose endomorphism algebra is $\mathbb{C}$.
We say the set of simple module objects $\{ (x,\mu)\}$ is complete if, for every morphism $\varphi\in  \operatorname{mor}_{\bMTC/A}(x\to y)$, there exists an
$f_{(z,\nu)\to y} \in \operatorname{mor}_{\bMTC/A}(z \to y)$ and a $g_{x\to(z,\nu)} \in \operatorname{mor}_{\bMTC/A}(x\to z)$ such that $\varphi = \sum_{(z,\nu)} f_{(z,\nu)\to y}\circ g_{x\to(z,\nu)}$.

Once a complete collection of simple module objects has been specified, we can compute the fusion rules and splitting spaces of the condensed theory by solving the relations
\begin{align}
\label{fusionmodob}
\modobfusL =
\modobfusR.
\end{align}
Eq.~\ref{fusionmodob} can be viewed as a set of equations determining $\alpha$.
If multiple linearly independent $\alpha$ satisfy Eq.~\eqref{fusionmodob}, then the condensed theory will have fusion multiplicity, and the fusion spaces are labeled by $\alpha_i$, with $i = 1,\cdots, N_{(x,\mu),(y,\nu)}^{(z,\sigma)}$.
When the condensed theory has no fusion multiplicity $\alpha$ is unique and in such cases we will not label $\alpha$ for each fusion vertex, but rather leave it implicitly determined by the surrounding simple module objects.

When the condensed theory is multiplicity free, the associators ($F$-symbols) of the condensed theory $\bMTC/A$ are defined by the relation
\begin{widetext}
\begin{align}
\label{eq:fcondmove}
\FCondL \!\!\!\!\!\!\!\!= \!\!
\sum \limits_{(q, \delta), \mu, \nu}
\left[F^{(x,\mu)(y,\nu)(z,\sigma)}_{(w, \rho)} \right]_{[(p,\lambda),\alpha,\beta][(q,\delta),\mu,\nu]}
\FCondR.
\end{align}
\end{widetext}
Including multiplicity indices is straightforward.
The associators are completely determined by the simple module objects, the fusion vertices, and the $F$- and $R$-symbols of the parent theory $\bMTC$.
One can check that this definition of the $F$-symbols satisfies the pentagon equation.

We now compute some examples relevant to the main text.
A simple object in the parent theory $\bMTC$ will either be fixed under fusion with $b$ ($s\otimes b = s$), or not ($r\otimes b= [rb] \neq r$).
These two cases correspond to two kinds of simple module objects $(x,\mu)$ in $\bMTC/A$: (1) those for which $x=r\oplus {[rb]}$; and (2) those for which $x=s$.

For case (1), there is a one-parameter family of solutions to Eq.~\eqref{modcond}, given by
\begin{align}
(r\oplus [rb],\mu) = \; \raaa\!\! +\;  \rbbb \!\! + \; \mu \; \rccc\!\! +\; \frac{[F^{rbb}]^*}{\mu}\;\rddd.
\end{align}
In this equation, $\mu$ is any non-zero complex number.
Typically we take $\mu = +1$ as a representative for this simple module object.
We can also use Eq.~\eqref{ismorphicmodob} to check  that $(r\oplus [rb],\mu) \cong (r\oplus [rb],\nu)$ via the isomorphism $f = \mu \operatorname{id}_r + \nu \operatorname{id}_{[rb]}$.
Equivalently $f \circ (r\oplus [rb],\mu)\circ f^{-1}= (r\oplus [rb],\nu)$.

In case (2), the object $(s,\mu)$ will `split' into a pair of simple module objects, which we label by a $\pm$ sign:
\begin{align}
\label{eq:splitting}
(s,\pm) =\; \saaa \pm\;\; \tau\;\; \sbbb
\end{align}
where $\tau$ is one of the roots of $\tau^2 = [F^{sbb}]^*$.

We now compute the fusion rules in two useful cases when $\mathcal{C}$ is multiplicity free.
Assuming $F^{xbb}= F^{bbx}= 1$ (as is the case for all examples we consider), and letting $s_1$ and $s_2$ denote two objects invariant under fusion with $b$, we solve Eq.~\eqref{fusionmodob} to find~\footnote{At first glance, one might worry this product is not gauge-invariant.  This is not a concern as the gauge has been fully determined by the assumption $F^{xbb}=1$.}
\begin{align}
N_{(s_1,\alpha),(s_2,\beta)}^{(r \oplus [br],\mu)} =
\begin{cases}
N_{s_1s_2}^{r}& \text{if $F^{s_1b{s}_2}_r[R^{b{s}_2}]^* = \alpha \beta$},\\
0 & \text{otherwise}.\\
\end{cases}
\end{align}
Where $N_{s_1s_2}^{r}$ comes from $\mathcal{C}$.
Similarly, for $r$ and $w$ which are not fixed under fusion with $b$ we have,
\begin{align}
N_{(r \oplus [rb], 1) , (w \oplus [wb], 1) }^{(p \oplus [pb], 1)} = N_{rw}^p.
\end{align}
Where $ N_{rw}^p$ comes from $\mathcal{C}$.
Additional fusion rules can be inferred using the pivotal structure of the condensed theory.
For example, $N_{(s_1, \alpha), ({s}_2, \beta)}^{ (r \oplus [rb],\mu)} = N_{({s}_2, \beta),  \overline{  (r \oplus [rb],\mu) }}^{\overline{(s_1, \alpha)}}$.
The dual objects are defined by~\cite{Kirillov2001}
\begin{align}
\overline{(x, \mu)}\;\; \cong \;\; \modobdual.
\end{align}

Finally, we remark that we are only interested in the subcategory of $\bMTC/A$ which braids trivially with $b$.
All objects with nontrivial monodromy with $b$ will be confined and therefore discarded.

\section{Associativity of Stacking Fermionic Theories}\label{app:associativity}

In this appendix we prove that if $(A,m)$ and $(A,m')$ are two commutative Frobenius algebra objects in $(\mathbb{Z}_2^{\psi})^n$, then they are isomorphic.
A direct consequence is that the algebra condensed in $(\fMTC_{\mathcal{G}^{\eff}}^{(1)\times} \underset{\mathcal{G}^{\eff}}{\ftimes} \fMTC_{\mathcal{G}^{\eff}}^{(2)\times})\underset{\mathcal{G}^{\eff}}{\ftimes} \fMTC_{\mathcal{G}^{\eff}}^{(3)\times}$ is isomorphic to the algebra condensed in $\fMTC_{\mathcal{G}^{\eff}}^{(1)\times} \underset{\mathcal{G}^{\eff}}{\ftimes}( \fMTC_{\mathcal{G}^{\eff}}^{(2)\times}\underset{\mathcal{G}^{\eff}}{\ftimes} \fMTC_{\mathcal{G}^{\eff}}^{(3)\times})$, and thus that the resulting theories are isomorphic.

It is convenient to label topological charges in $(\mathbb{Z}_2^{\psi})^{ n}$ with elements of $\mathbb{Z}_2^n$.
We will write them as vectors $x = (x_1,\cdots, x_n)$ with each $x_i \in \{ 0,1\}$.
The braiding and fusion are,
\begin{align}
x \otimes y = x +y \mod2, \quad R^{xy} = (-1)^{(x, y)}, \quad F^{xyz}  = 1
\end{align}
In the exponent we have used the dot product $(x , y) = \sum_{j=1}^{n} x_j y_j$.
For future use we note that the topological spin of an object $x$ is given by $(-1)^{\eff} = (-1)^{x \cdot x}$.

We now label all commutative Frobenius algebra objects $(A,m) $ in the BTC $ (\mathbb{Z}_2^{\psi})^{ n}$.
A commutative Frobenius algebra object in $(\mathbb{Z}_2^{\psi})^n$ consists of an object $A \in (\mathbb{Z}_2^{\psi})^{ n}$ and a collection of multiplications $m$.
We can identify $A$ with a subgroup $K$ of $\mathbb{Z}_2^n$ specifying a collection of bosons which we wish to condense:
\begin{align}
A = \bigoplus_{x \in K} \psi_{1}^{x_1} \psi_2^{x_2} \cdots \psi_n^{x_n}.
\end{align}
In order for $x\in K$ to be a boson, we must have $(-1)^{(x,x)}=1$.
The associativity condition on the multiplication implies $m \in Z^2(K,\text{U}(1))$.
Requiring $A$ to be a commutative algebra object additionally requires the multiplication to satisfy \begin{align}
\label{eq:commutativity}
m_{x,y} =(-1)^{(x , y)} m_{y,x}.
\end{align}

We now prove that any two 2-cocycles $m, m' \in H^2(K,\text{U}(1))$ satisfying Eq.~\eqref{eq:commutativity} are cohomologous.
This implies that the algebra objects $(A,m)$ and $(A,m')$ are related by an isomorphism, and therefore condensing them results in isomorphic theories.
We first note that the difference of any two 2-cocycles satisfying Eq.~\eqref{eq:commutativity} is a symmetric 2-cocycle $m' m^{-1}$ valued in $\text{U}(1)$.
It is known that all symmetric 2-cocycles on an Abelian group valued in $\text{U}(1)$ are cohomologous.
To see this we note that $H^2(K,\text{U}(1))_{\on{symmetric}}$ defines an Abelian extension of $K$ by $\text{U}(1)$.
Furthermore, equivalence classes of Abelian extensions of $K$ by $\text{U}(1)$ are in one-to-one correspondence with elements of $\on{Ext}^1(K,\text{U}(1))$.
As $\text{U}(1)$ is injective, $\on{Ext}^1(K,\text{U}(1)) = 0$, and there is only one extension of $K$ by $\text{U}(1)$.
Thus $H^2(K,\text{U}(1))_{\on{symmetric}}  = \mathbb{Z}_1$, meaning all 2-cocycles on $K$ satisfying Eq.~\eqref{eq:commutativity} are isomorphic.
That is, there exists an isomorphism $f: A \to A$ such that $m' = m \cbd f$ for any 2-cocycles $m$ and $m'$ satisfying Eq.~\eqref{eq:commutativity}.

Thus we have shown that there is only one multiplication, up to isomorphism, on a commutative Frobenius algebra object $(A,m) \in (\mathbb{Z}_2^{\psi})^n$.
As such, when fermionic stacking, the algebra objects we condense will always be isomorphic, meaning that the condensates will be the same, up to a relabeling of topological charges and gauge transformations.

When we stack and condense fermionic theories in different orders, the algebra objects we condense are necessarily isomorphic, and so the condensates will be the same up to a relabeling of topological charges and gauge transformations.
Therefore, fermionic stacking is associative.
We denote the one Frobenius algebra object generated by $n$-fermions and its multiplication morphism (up to isomorphism) by $A[\psi_1,\psi_2,\cdots, \psi_n]$.

\section{Group Cohomology}
\label{app:cohomology}

This appendix offers a brief review of the group cohomology used in the main text.
A more detailed review can be found in Ref.~\onlinecite{Brown1982}.

Let $G$ be a finite group and $M$ an Abelian group with group action $\rho: G \times M \to M$.
In particular, $\rho$ satisfies
\begin{align}
\rho_{\bf g}(\rho_{\bf h} (a)) &= \rho_{\bf gh}(b)\\
\rho_{\bf g}(a+b) &= \rho_{\bf g}(a) + \rho_{\bf g}(b)
\end{align}
where $a,b \in M$ and we have used additive notation.

Let $\omega({\bf g}_1, {\bf g}_2, \cdots, {\bf g}_n)$ be a function from $n$ group elements valued in $M$.
Such a function is called an `$n$-cochain', and the set of all $M$-valued $n$-cochains is denoted $C^{n}(G,M)$.
The set of $n$-cochains also forms an Abelian group with multiplication given by
\begin{align}
[\omega + \omega']({\bf g}_1, \cdots, {\bf g}_n)  =
\omega({\bf g}_1, \cdots, {\bf g}_n)+
\omega'({\bf g}_1, \cdots, {\bf g}_n).
\end{align}
The inverse of $\omega$ is $-\omega$ and the identity element is $\omega({\bf g}_1, \cdots, {\bf g}_n) = 0 \in M$.

The coboundary operator $\cbd: C^n(G,M) \to C^{n+1}(G,M)$ is defined by
\begin{align}
&[\cbd \omega]({\bf g}_1, \cdots, {\bf g}_{n+1}) = \rho_{{\bf g}_1}[\omega({\bf g}_2, \cdots, {\bf g}_n)]\nonumber \\
&\quad + \sum_{j=1}^n (-1)^j \omega({\bf g}_1, \cdots, {\bf g}_{j-1}, {\bf g}_j {\bf g}_{j+1}, {\bf g}_{j+2}, \cdots,{\bf g}_{n+1}) \nonumber \\
&\quad +(-1)^{n+1}\omega({\bf g}_1, \cdots, {\bf g}_{n}).
\end{align}
One can check that $\cbd\cbd \omega = 0 \in M$ for any $n$-cochain.

We can now define $n$-cocycles using the coboundary operator $\cbd$.
An $n$-coycle is given by an $n$-cochain $\omega$ such that $\cbd \omega = 0$.
The set of $n$-cocycles is given by
\begin{align}
Z^n_\rho(G,M) = \{ \omega \in C^n(G,M) : \cbd \omega = 0 \}.
\end{align}
If an $n$-cocycle $\omega = \cbd \mu$, then we say $\omega$ is an $n$-coboundary.
We denote the set of $n$-coboundaries as,
\begin{align}
B^n_\rho(G,M) = \{ \cbd \mu \in Z^n_\rho(G,M) : \mu \in C^{n-1}(G,M) \}
\end{align}
One can check that $B^n_\rho(G,M) $ is a normal subgroup of $Z^n_\rho(G,M) $.
The $n$-th cohomology group is defined as the quotient of $Z^n_\rho(G,M) $ by $B^n_\rho(G,M) $:
\begin{align}
H^n(G,M) = \frac{Z^n_\rho(G,M)}{B^n_\rho(G,M)}.
\end{align}

It is helpful to write down a few examples.
A 2-cochain $\omega \in C^2(G,M)$ is a 2-cocyle if it satisfies,
\begin{align}
\rho_{\bf g}[\omega({\bf h},{\bf k})] -\omega({\bf gh,k}) +\omega({\bf g,hk})-\omega({\bf g,h}) = 0.
\end{align}
A 2-cochain $\omega \in C^2(G,M)$ is a 2-cobundary if there exists a $\mu \in C^1(G,M)$ such that,
\begin{align}
\omega({\bf g,h})  = [\cbd\mu]({\bf g,h}) = \rho_{\bf g}[\mu({\bf h})] -\mu({\bf gh}) + \mu({\bf g}).
\end{align}

\subsection{Short Exact Sequences}\label{app:sequences}

Given three groups $A$, $B$, and $C$, and two homomorphisms $\alpha: A \to B$, and $\beta: B \to C$, we say the sequence $A \xrightarrow{\alpha} B \xrightarrow{\beta} C$ is exact if $\ker \beta = \im \alpha$.
A short exact sequence of Abelian groups $A$, $B$, and $C$ is written as
\begin{align}
\label{eq:exctseq}
0\to A \xrightarrow{i} B \xrightarrow{\pi} C\to 0
\end{align}
where each homomorphism is exact.
We have that $C \cong B/A$.
When the image of $i$ is in the center of $B$ the sequence defines a central extension.
In the main text we are primarily concerned with the case where $A$, $B$, and $C$ are Abelian groups, in which case the extension is always a central extension.
To fully specify the group $B$ in terms of $A$ and $C$, we need to specify the extension class $[\phi] \in H^2(C, A)$.
Letting $\phi \in [\phi] $ be a representative 2-cocycle, we can specify the group group elements of $B$ by pairs $(a,c) \in A \times C$ with group operation,
\begin{align}
(a,c) \times (a',c')  = (a + a' + \phi(c,c'), cc').
\end{align}
We have assumed $A$ is Abelian and $i(A)$ is in the center of $B$.
For later use, we also note that a section of the map $f: B \to C$ is a function $s: C \to B$ such that $s \circ f = \text{id}_C$.

We first encounter short exact sequences in the context of the full symmetry group $\mathcal{G}^{\eff}$:
\begin{align}
\mathbb{Z}_2^{\eff} \to \mathcal{G}^{\eff} \to G,
\end{align}
specified by the $2$-cocycle $\central\in H^2(G,\mathbb{Z}_2)$.

\subsection{Cup Product}\label{app:cup}

The cup product is well defined when $M$ is a ring.
Recall that a ring is an Abelian group with an additional multiplication operation that is associative and distributive.
We will denote $\cdot$ for the additional multiplication operation, and reserve $+$ for the multiplication of the Abelian group.
Our typical example will be the ring given by integers $\mathbb{Z}/2\mathbb{Z}$ with the usual addition (+) and multiplication ($\cdot$) laws.
We will denote it $\mathbb{F}_2$.
We use the definitions given in Ref.~\onlinecite{Steenrod1947} in the conventions used in this paper.
The cup product is a map of cochains:
\begin{align}
\cup: C^p(G, M) \times C^q(G,M) \to C^{p+q}(G,M)
\end{align}
defined by,
\begin{align}
&[a_p \cup b_q]({\bf g}_1, \cdots, {\bf g}_{p+q}) \\
&\quad\quad = a_p({\bf g}_1, \cdots, {\bf g}_p) \cdot b_q({\bf g}_{p+1}, \cdots, {\bf g}_{p+q}),\notag
\end{align}
for $p$- and $q$-cochains $\omega_{p}$ and $b_q$.
It satisfies,
\begin{align}
\cbd[a_p \cup b_q] = (\cbd a_p) \cup b_q +(-1)^p a_p \cup \cbd b_q
\end{align}
The higher cup products introduced by Steenrod~\cite{Steenrod1947} are
\begin{align}
\cup_i: C^p(G, M) \times C^q(G,M) \to C^{p+q-i}(G,M).
\end{align}

The explicit formulae for all cup products is known; here we only give those used in the main text.
The cup product $\cup_1$ for a $p$-cochain $a_p$ and $q$-cochain $b_q$ is
\begin{widetext}
\begin{align}
\label{cuponenh}
[ a_p \cup_1 b_q]({\bf g}_1, \cdots {\bf g}_{p+q-1}) =& (-1)^{p(q+1)}
a_p({\bf g}_1 {\bf g}_2 \cdots {\bf g}_q, {\bf g}_{q+1}, \cdots, {\bf g}_{p+q-1})\cdot
b_q({\bf g}_1, \cdots, {\bf g}_q) \notag \\
&+ (-1)^{(p-1)(q+1)}a_p({\bf g}_1, {\bf g}_2{\bf g}_3 \cdots {\bf g}_{q+1}, {\bf g}_{q+2}, \cdots, {\bf g}_{p+q-1})
\cdot b_q({\bf g}_2, \cdots, {\bf g}_{q+1}) \notag \\
&\quad\quad\quad\quad \vdots \notag  \\
&+(-1)^{(q+1)}a_p({\bf g}_1, {\bf g}_2, \cdots {\bf g}_{p-1}, {\bf g}_{p}{\bf g}_{p+1}\cdots {\bf g}_{p+q-1})
\cdot b_q({\bf g}_{p+1}, \cdots, {\bf g}_{q+p-1}).
\end{align}
A few instances of $\cup_1$ appearing in the main text are,
\begin{align}
[a_1 \cup_1 b_1]({\bf g}) & = a_1({\bf g}) \cdot b_1({\bf g}), \\
[a_1 \cup_1 b_2]({\bf g},{\bf h})  &= -a_1({\bf g h}) \cdot b_2({\bf g},{\bf h}), \\
[a_2 \cup_1 b_2]({\bf g},{\bf h},{\bf k})  &= a_2({\bf g h}, {\bf k}) \cdot b_2({\bf g},{\bf h}) -a_2({\bf  g}, {\bf hk}) \cdot b_2({\bf h},{\bf k}).
\end{align}
When $a_1, b_1, c_1 = a_1 + b_1$, and $a_2$ are $\mathbb{F}_2$ valued 1- and 2-cocycles we have,
\begin{align}
\cbd \widetilde{a}_{1} &= 2 \widetilde{a}_1 \cup \widetilde{a}_1,\\
\cbd ( \widetilde{a}_1 \cup_1 \widetilde{b}_1) & =
- \widetilde{a}_1 \cup \widetilde{b}_1 -  \widetilde{b}_1 \cup \widetilde{a}_1
+ 2(\widetilde{a}_1 \cup_1 \widetilde{b}_1)  \cup(\widetilde{a}_1 + \widetilde{b}_1)
+ 2(\widetilde{a}_1 + \widetilde{b}_1) \cup(\widetilde{a}_1 \cup_1 \widetilde{b}_1)
-4 (\widetilde{a}_1 \cup_1 \widetilde{b}_1) \cup (\widetilde{a}_1 \cup_1 \widetilde{b}_1),\\
& = \widetilde{a}_1 \cup \widetilde{a}_1+\widetilde{b}_1 \cup \widetilde{b}_1 - \widetilde{c}_1 \cup \widetilde{c}_1\\
\cbd \widetilde{a}_2 & = -2 \widetilde{a}_2 \cup_1 \widetilde{a}_2.
\end{align}
Where, the $\widetilde{\;\;}$ denotes integer addition.
These expressions can be verified directly using the definitions given above.

As the name suggests, there are $\cup_i$ products for any $i\geq 0$.
We only use $i \geq 2$ cup products when $M = \mathbb{F}_2$.
Relevant formulae for the higher cup products appearing in the main text are,
\begin{align}
[a_2 \cup_2 b_2]({\bf g_1,g_2})
&= a_2({\bf g_1,g_2}) \cdot b_2({\bf g_1,g_2})\\
[a_3 \cup_2 b_2]({\bf  g_1,g_2,g_3})
& =
a_3({\bf  g_1,g_2,g_3}) \cdot b_2({\bf g_1,g_2 g_3})
+a_3({\bf  g_1,g_2,g_3}) \cdot b_2({\bf g_2, g_3})\\
[a_3 \cup_2 b_3]({\bf  g_1,g_2,g_3,g_4})
& =
a_3({\bf g_1,g_2,g_3})\cdot b_3({\bf g_1,g_2 g_3,g_4}) +
a_3({\bf g_1g_2,g_3,g_4})\cdot b_3({\bf g_1,g_2, g_3g_4})   \\
&\quad \quad
+a_3({\bf g_1,g_2,g_3})\cdot b_3({\bf g_2 ,g_3,g_4}) +
a_3({\bf g_1,g_2g_3,g_4})\cdot b_3({\bf g_2, g_3,g_4})
\end{align}
more generally we have the relation,
\begin{align}
[a_n \cup_n b_n]({\bf g_1},\cdots,{\bf g_n})&=
  a_n({\bf g_1},\cdots,{\bf g_n}) \cdot b_n({\bf g_1},\cdots,{\bf g_n}).
\end{align}
and
\begin{align}
\cbd(a_p \cup_i b_q ) = a_p \cup_{i-1}b_q + b_q \cup_{i-1} a_p+(\cbd a_p)\cup_i b_q+ a_p\cup_i \cbd b_q.
\end{align}
\end{widetext}

\section{Characterizing \texorpdfstring{$\ker(\res_{\fMTC_{\vv{0}}})$}{kerres}}
\label{app:kerres}

In this appendix, we first prove that $\ker(\res_{\fMTC_{\vv{0}}}) = \mathbb{Z}_{2}^{\V}$ when $\widehat{\bf A}_{\vv{1}} \neq \varnothing$.
Then, lacking a similar proof for $\widehat{\bf A}_{\vv{1}} = \varnothing$, we discuss the properties $\ker(\res_{\fMTC_{\vv{0}}})$ must possess if it is not equal to $\mathbb{Z}_{2}^{\V}$.

\subsection{Proof of \texorpdfstring{$\ker(\res_{\fMTC_{\vv{0}}}) = \mathbb{Z}_{2}^{\V}$}{kerresZV} when \texorpdfstring{$\mathcal{A}_{\vv{1}} \neq \varnothing$}{A1neqvarn}}

We first consider the case where $\fMTC$ has $\mathcal{A}_{\vv{1}} \neq \varnothing$ and $\widehat{\bf A}_{\vv{1}} \neq \varnothing$.
Note that this case has only $v$-type vortices.
As explained in Sec.~\ref{sec:SMTC_FMTC}, we can pick an arbitrary Abelian vortex charge $e_{\vv{1}} \in \mathcal{A}_{\vv{1}}$, and represent all vortex charges with respect to it as $a_{\vv{1}} = a_{\vv{0}} \otimes e_{\vv{1}}$, with corresponding fusion rules
\begin{align}
a_{\vv{x}} \otimes b_{\vv{y}} = \sum_{c_{\vv{0}} } N_{a_{\vv{0}} b_{\vv{0}}}^{c_{\vv{0}}} h_{\vv{0}}^{\mathsf{x} \cdot \mathsf{y}} c_{\vv{x+y}}
.
\end{align}
It follows that the action of a topological symmetry $\varphi$ on vortex charges is determined by its action on quasiparticle charges and on $e_{\vv{1}}$, since $\varphi(a_{\vv{1}}) = \varphi(a_{\vv{0}}) \otimes \varphi(e_{\vv{1}})$.
Recall from Eq.~\eqref{eq:vortex_lifted_permutation} that the action of $\varphi$ on supersectors of topological charge is determined entirely by the action of $\varphi^{(0)}= \res_{\fMTC_{\vv{0}}}(\varphi)$ on the quasiparticle charges.
In other words, $\varphi^{(0)}$ on $\fMTC_{\vv{0}}$ determines $\varphi(e_{\vv{1}})$ up to fusion with $\psi_{\vv{0}}$.
Together, this implies the action of $\varphi$ on topological charges is determined by $\varphi^{(0)}$, up to an application of $\V$.
In particular, all extensions of $[\openone^{(0)}]$ to $\fMTC$ act on vortex charges as either $\openone$ or $\V$.

In order to prove that $\ker(\res_{\fMTC_{\vv{0}}}) = \mathbb{Z}_{2}^{\V}$, it only remains to show that any $[\varphi] \in \ker(\res_{\fMTC_{\vv{0}}})$ that leaves all topological charges fixed acts on the full topological state space as a fermionic natural isomorphism.
We now consider $\varphi$ such that
\begin{align}
\varphi (a_{\vv{x}} ) &= a_{\vv{x}} ,\\
\varphi ( \ket{a_{\vv{x}}, b_{\vv{y}} ; c_{\vv{x+y}}, \mu } ) &= \sum_{\mu'} [u^{a_{\vv{x}} b_{\vv{y}}}_{c_{\vv{x+y}}} ]_{\mu \mu'} \ket{a_{\vv{x}}, b_{\vv{y}} ; c_{\vv{x+y}}, \mu' }
,\\
u^{a_{\vv{0}} b_{\vv{0}}}_{c_{\vv{0}}} &= \openone
.
\end{align}
(The last line in this equation is a particular gauge choice for elements of $\ker(\res_{\fMTC_{\vv{0}}})$.)

Next, we apply the topological symmetry $\varphi$ to $[F^{e_{\vv{1}} a_{\vv{0}} b_{\vv{0}}}_{c_{\vv{1}}} ]_{(a_{\vv{1}},\alpha)( c_{\vv{0}},\mu )}$ to obtain
\begin{align}
[u^{a_{\vv{1}} b_{\vv{0}}}_{c_{\vv{1}}} ]_{\alpha \alpha'} &= \frac{u^{e_{\vv{1}} c_{\vv{0}}}_{c_{\vv{1}}} }{ u^{e_{\vv{1}} a_{\vv{0}}}_{a_{\vv{1}}} } \delta_{\alpha \alpha'}
.
\end{align}
This utilizes the facts that $e_{\vv{1}}$ is Abelian, the fusion channels $a_{\vv{1}}$ and $c_{\vv{0}}$ are uniquely determined, and $F$ is invertible.
Applying $\varphi$ to $R^{e_{\vv{1}} b_{\vv{0}}}_{b_{\vv{1}}}$, we obtain
\begin{align}
u^{e_{\vv{1}} b_{\vv{0}}}_{b_{\vv{1}}}  &= u^{ b_{\vv{0}} e_{\vv{1}}}_{b_{\vv{1}}}
,\\
[u^{a_{\vv{1}} b_{\vv{0}}}_{c_{\vv{1}}} ]_{\alpha \alpha'} &= [u^{ b_{\vv{0}} a_{\vv{1}}}_{c_{\vv{1}}} ]_{\alpha \alpha'}
.
\end{align}
Applying $\varphi$ to $F^{a_{\vv{0}} e_{\vv{1}} e_{\vv{1}}}_{[ha]_{\vv{0}}} $, we obtain
\begin{align}
u^{a_{\vv{1}} e_{\vv{1}}}_{[ha]_{\vv{0}}} &= u^{ e_{\vv{1}} a_{\vv{1}}}_{[ha]_{\vv{0}}} = \frac{u^{e_{\vv{1}} e_{\vv{1}}}_{h_{\vv{0}}}}{u^{a_{\vv{0}} e_{\vv{1}}}_{a_{\vv{1}}}}
.
\end{align}
Applying $\varphi$ to $[F^{a_{\vv{1}} e_{\vv{1}} b_{\vv{0}}}_{[hc]_{\vv{0}}} ]_{([ha]_{\vv{0}},\beta)( b_{\vv{1}},\nu )}$, we obtain
\begin{align}
[u^{a_{\vv{1}} b_{\vv{1}}}_{[hc]_{\vv{0}}} ]_{\nu \nu'} &= \frac{u^{a_{\vv{1}} e_{\vv{1}}}_{[ha]_{\vv{0}}} }{ u^{e_{\vv{1}} b_{\vv{0}}}_{b_{\vv{1}}} } \delta_{\nu \nu'}
=\frac{u^{e_{\vv{1}} e_{\vv{1}}}_{h_{\vv{0}}} }{ u^{e_{\vv{1}} a_{\vv{0}}}_{a_{\vv{1}}} u^{e_{\vv{1}} b_{\vv{0}}}_{b_{\vv{1}}} } \delta_{\nu \nu'}
.
\end{align}

It is now straightforward to check that $\varphi = \Upsilon$ for a fermionic natural isomorphism given by
\begin{align}
\gamma_{a_{\vv{0}}} &=1 ,\\
\gamma_{a_{\vv{1}}} &= \frac{(u^{e_{\vv{1}} e_{\vv{1}}}_{h_{\vv{0}}})^{\frac{1}{2}} }{u^{e_{\vv{1}} a_{\vv{0}}}_{a_{\vv{1}}} }
.
\end{align}
Thus, $[\varphi] = [\openone]$.

\subsection{Proof of \texorpdfstring{$\ker(\res_{\fMTC_{\vv{0}}}) = \mathbb{Z}_{2}^{\V}$}{kreresm0} when \texorpdfstring{$\widehat{\bf A}_{\vv{1}} \neq \varnothing$}{wA1}}

We next consider the case where $\fMTC$ has $\mathcal{A}_{\vv{1}} = \varnothing$ and $\widehat{\bf A}_{\vv{1}} \neq \varnothing$.
The proof is similar to the $\mathcal{A}_{\vv{1}} \neq \varnothing$ case, but modified to account for $\fMTC$ having only $\sigma$-type vortices, rather than only $v$-type vortices.
Note that $[\V] = [\Q]$ in this case.
We can pick an arbitrary superAbelian vortex charge $e_{\vv{1}} \in \widehat{\bf A}_{\vv{1}} $, and represent all vortex charges with respect to it as $a_{\vv{1}} = a_{\vv{0}} \otimes e_{\vv{1}} = [\psi a]_{\vv{0}} \otimes e_{\vv{1}} $, with corresponding fusion rules
\begin{align}
a_{\vv{0}} \otimes b_{\vv{1}} &= \sum_{c_{\vv{0}} } N_{a_{\vv{0}} b_{\vv{0}}}^{c_{\vv{0}}} c_{\vv{1}} ,\\
a_{\vv{1}} \otimes b_{\vv{1}} &= \sum_{c_{\vv{0}} } N_{a_{\vv{0}} b_{\vv{0}}}^{c_{\vv{0}}} (h_{\vv{0}} \oplus [\psi h]_{\vv{0}}) \otimes c_{\vv{0}}
.
\end{align}
It follows that the action of a topological symmetry $\varphi$ on vortex charges is determined by its action on quasiparticle charges and on $e_{\vv{1}}$, since $\varphi(a_{\vv{1}}) = \varphi(a_{\vv{0}}) \otimes \varphi(e_{\vv{1}})$.
Since there are only $\sigma$-type vortices, each vortex charge forms its own supersector.
Thus, Eq.~\eqref{eq:vortex_lifted_permutation} implies that $\varphi^{(0)}$ on $\fMTC_{\vv{0}}$ uniquely determines $\varphi(e_{\vv{1}})$, and hence also $\varphi(a_{\vv{1}}) $ for all vortices.
In particular, all extensions of $[\openone^{(0)}]$ to $\fMTC$ act on vortex charges as $\openone$.

In order to prove that $\ker(\res_{\fMTC_{\vv{0}}}) = \mathbb{Z}_{2}^{\V}= \mathbb{Z}_{2}^{\Q}$, it only remains to show that any $[\varphi] \in \ker(\res_{\fMTC_{\vv{0}}})$ that leaves all topological charges fixed acts on the full topological state space as a natural isomorphism; when treated as fermionic natural isomorphism, these divide into the two classes $[\openone]$ and $[\Q]$.
We now consider $\varphi$ such that
\begin{align}
\varphi (a_{\vv{x}} ) &= a_{\vv{x}} ,\\
\varphi ( \ket{a_{\vv{x}}, b_{\vv{y}} ; c_{\vv{x+y}}, \mu } ) &= \sum_{\mu'} [u^{a_{\vv{x}} b_{\vv{y}}}_{c_{\vv{x+y}}} ]_{\mu \mu'} \ket{a_{\vv{x}}, b_{\vv{y}} ; c_{\vv{x+y}}, \mu' } ,\\
u^{a_{\vv{0}} b_{\vv{0}}}_{c_{\vv{0}}} &= \openone
.
\end{align}

Next, we apply the topological symmetry $\varphi$ to $[F^{e_{\vv{1}} a_{\vv{0}} b_{\vv{0}}}_{c_{\vv{1}}} ]_{(a_{\vv{1}},\alpha)( c_{\vv{0}},\mu )}$ to obtain
\begin{align}
[u^{a_{\vv{1}} b_{\vv{0}}}_{c_{\vv{1}}} ]_{\alpha \alpha'} &= \frac{u^{e_{\vv{1}} c_{\vv{0}}}_{c_{\vv{1}}} }{ u^{e_{\vv{1}} a_{\vv{0}}}_{a_{\vv{1}}} } \delta_{\alpha \alpha'}
.
\end{align}
This utilizes the facts that $e_{\vv{1}}$ is superAbelian, the fusion channels $a_{\vv{1}}$ and $c_{\vv{0}}$ are uniquely determined, and $F$ is invertible.
In this case, in order to see that the fusion channel $c_{\vv{0}}$ is uniquely determined, we use the fact that $\widehat{\bf A}_{\vv{1}} \neq \varnothing$ implies that $N_{a_{\vv{0}} b_{\vv{0}}}^{c_{\vv{0}}} N_{a_{\vv{0}} b_{\vv{0}}}^{[\psi c]_{\vv{0}}} =0$ for all $a_{\vv{0}}, b_{\vv{0}} , c_{\vv{0}} \in \fMTC_{\vv{0}}$ (see Sec.~\ref{sec:Topo_symmetry}).
Applying $\varphi$ to $R^{e_{\vv{1}} b_{\vv{0}}}_{b_{\vv{1}}}$, we obtain
\begin{align}
u^{e_{\vv{1}} b_{\vv{0}}}_{b_{\vv{1}}}  &= u^{ b_{\vv{0}} e_{\vv{1}}}_{b_{\vv{1}}}
,\\
[u^{a_{\vv{1}} b_{\vv{0}}}_{c_{\vv{1}}} ]_{\alpha \alpha'} &= [u^{ b_{\vv{0}} a_{\vv{1}}}_{c_{\vv{1}}} ]_{\alpha \alpha'}
.
\end{align}

Applying $\varphi$ to $F^{a_{\vv{0}} e_{\vv{1}} e_{\vv{1}}}_{[ha]_{\vv{0}}} $ and $F^{a_{\vv{0}} e_{\vv{1}} e_{\vv{1}}}_{[\psi ha]_{\vv{0}}} $, we obtain
\begin{align}
u^{a_{\vv{1}} e_{\vv{1}}}_{[ha]_{\vv{0}}} &= u^{ e_{\vv{1}} a_{\vv{1}}}_{[ha]_{\vv{0}}} = \frac{u^{e_{\vv{1}} e_{\vv{1}}}_{h_{\vv{0}}}}{u^{a_{\vv{0}} e_{\vv{1}}}_{a_{\vv{1}}}}
,\\
u^{a_{\vv{1}} e_{\vv{1}}}_{[\psi ha]_{\vv{0}}} &= u^{ e_{\vv{1}} a_{\vv{1}}}_{[\psi ha]_{\vv{0}}} = \frac{u^{e_{\vv{1}} e_{\vv{1}}}_{[\psi h]_{\vv{0}}}}{u^{a_{\vv{0}} e_{\vv{1}}}_{a_{\vv{1}}}}.
\end{align}
Applying $\varphi$ to $[F^{a_{\vv{1}} e_{\vv{1}} b_{\vv{0}}}_{[hc]_{\vv{0}}} ]_{([ha]_{\vv{0}},\beta)( b_{\vv{1}},\nu )}$ and $[F^{a_{\vv{1}} e_{\vv{1}} b_{\vv{0}}}_{[\psi hc]_{\vv{0}}} ]_{([\psi ha]_{\vv{0}},\beta)( b_{\vv{1}},\nu )}$, we obtain
\begin{align}
[u^{a_{\vv{1}} b_{\vv{1}}}_{[hc]_{\vv{0}}} ]_{\nu \nu'} &= \frac{u^{a_{\vv{1}} e_{\vv{1}}}_{[ha]_{\vv{0}}} }{ u^{e_{\vv{1}} b_{\vv{0}}}_{b_{\vv{1}}} } \delta_{\nu \nu'}
=\frac{u^{e_{\vv{1}} e_{\vv{1}}}_{h_{\vv{0}}} }{ u^{e_{\vv{1}} a_{\vv{0}}}_{a_{\vv{1}}} u^{e_{\vv{1}} b_{\vv{0}}}_{b_{\vv{1}}} } \delta_{\nu \nu'}
,\\
[u^{a_{\vv{1}} b_{\vv{1}}}_{[\psi hc]_{\vv{0}}} ]_{\nu \nu'} &= \frac{u^{a_{\vv{1}} e_{\vv{1}}}_{[\psi ha]_{\vv{0}}} }{ u^{e_{\vv{1}} b_{\vv{0}}}_{b_{\vv{1}}} } \delta_{\nu \nu'}
=\frac{u^{e_{\vv{1}} e_{\vv{1}}}_{[\psi h]_{\vv{0}}} }{ u^{e_{\vv{1}} a_{\vv{0}}}_{a_{\vv{1}}} u^{e_{\vv{1}} b_{\vv{0}}}_{b_{\vv{1}}} } \delta_{\nu \nu'}.
\end{align}
Applying $\varphi$ to $F^{\psi_{\vv{0}} e_{\vv{1}} a_{\vv{0}}}_{a_{\vv{1}}} $, we obtain
\begin{align}
u^{\psi_{\vv{0}} a_{\vv{1}} }_{a_{\vv{1}}} &= u^{ a_{\vv{1}} \psi_{\vv{0}}}_{a_{\vv{1}}} = u^{\psi_{\vv{0}} e_{\vv{1}} }_{e_{\vv{1}}}
.
\end{align}

It is now straightforward to check that $\varphi = \Upsilon$ for a natural isomorphism given by
\begin{align}
\gamma_{a_{\vv{0}}} &= M_{a_{\vv{0}} \hat{z}} ,\\
\gamma_{a_{\vv{1}}} &= \frac{\gamma_{a_{\vv{0}}} ( \gamma_{h_{\vv{0}}} u^{e_{\vv{1}} e_{\vv{1}}}_{h_{\vv{0}}})^{\frac{1}{2}} }{u^{e_{\vv{1}} a_{\vv{0}}}_{a_{\vv{1}}} }
,
\end{align}
where $\hat{z} \in \widehat{\bf {A}}$.
Since this natural isomorphism must have
\begin{align}
\gamma_{\psi_{\vv{0}}} &= u^{\psi_{\vv{0}} e_{\vv{1}} }_{e_{\vv{1}}}
,
\end{align}
and canonical gauge requires that $u^{\psi_{\vv{0}} e_{\vv{1}} }_{e_{\vv{1}}} =\pm 1$, we cannot always set $\gamma_{a_{\vv{0}}}$ to 1 for all $a_{\vv{0}}$.
Indeed, we must choose $\hat{z} \in \widehat{\bf {A}}_{\vv{0}}$ when $u^{\psi_{\vv{0}} e_{\vv{1}} }_{e_{\vv{1}}} = 1$, and $\hat{z} \in \widehat{\bf {A}}_{\vv{1}}$ when $u^{\psi_{\vv{0}} e_{\vv{1}} }_{e_{\vv{1}}} = -1$.
Thus, under equivalences by fermionic natural isomorphism, we find $[\varphi] =[\openone]$ when $u^{\psi_{\vv{0}} e_{\vv{1}} }_{e_{\vv{1}}} = 1$, and $[\varphi] =[\Q]$ when $u^{\psi_{\vv{0}} e_{\vv{1}} }_{e_{\vv{1}}} = -1$.

\subsection{Properties of \texorpdfstring{$\ker(\res_{\fMTC_{\vv{0}}})$}{kerresM0} when \texorpdfstring{$\widehat{\bf A}_{\vv{1}} = \varnothing$}{A1varn}}

In the case where $\fMTC$ has $\widehat{\bf A}_{\vv{1}} = \varnothing$, we lack a proof that $\ker(\res_{\fMTC_{\vv{0}}})$ is $\mathbb{Z}_{2}^{\V}$, so we will consider the general properties it must satisfy.
In this case, $[\Q]$ is nontrivial on $\fMTC_{\vv{0}}$ and $\fMTC$ may contain both $v$-type and $\sigma$-type vortices.
From Eq.~\eqref{eq:vortex_lifted_permutation}, we know that any extension $\varphi$ of $\openone^{(0)}$ to $\fMTC$ must leave the vortex supersectors of charge fixed.
Thus, the elements of $\ker(\res_{\fMTC_{\vv{0}}})$ can only have $\varphi(a_{v}) = a_{v}$ or $[\psi a]_{v}$, and $\varphi(a_{\sigma}) = a_{\sigma}$.

The possibility of different $v$-type vortices having uncorrelated permutations of charge within the supersector, e.g. $\varphi(a_{v}) = a_{v}$ and $\varphi(b_{v}) = [\psi b]_{v}$, was previously raised by Ref.~\onlinecite{Bark2021cascade}, through consideration of the modular $S$-matrix of $\fMTC$.
Their argument is based on the observation that the permutations of $a_{v}$ and $b_{v}$ must be correlated when $S_{a_{v} b_{v}} \neq 0$.
With this in mind, they partition the $v$-type vortices into subsets $\fMTC_{v_1}, ... , \fMTC_{v_n}$, where $S_{a_{v} b_{v}} =0$ for all $a_{v} \in \fMTC_{v_j}$ and $b_{v} \in \fMTC_{v_k}$ with $j\neq k$, and all vortices within a subset $\fMTC_{v_j}$ can be connected by a chain of nonzero $S$-matrix elements.
Each subset must have correlated permutation of charge within their supersectors, i.e. either $\varphi(a_{v}) = a_{v}$ for all $a_{v} \in \fMTC_{v_j}$ or $\varphi(a_{v}) =[\psi a]_{v}$ for all $a_{v} \in \fMTC_{v_j}$.
Charges from different subsets might be allowed to have uncorrelated permutations.
We can equivalently make the same argument using the fusion rules.
In particular, the permutations of $a_{v}$ and $b_{v}$ must be correlated when $N_{a_{v} b_{v}}^{c_0} \neq N_{a_{v} b_{v}}^{[\psi c]_0}$ for any $c_{\vv{0}}$.
Thus, the condition $S_{a_{v} b_{v}} =0$ can be replaced by the condition that $N_{a_{v} b_{v}}^{c_0} = N_{a_{v} b_{v}}^{[\psi c]_0}$ for all $c_{\vv{0}} \in \fMTC_{\vv{0}}$.
These $S$-matrix and fusion rule arguments do not establish that there are elements of $\ker(\res_{\fMTC_{\vv{0}}})$ with such uncorrelated permutations, but rather that they cannot be immediately ruled out without inspecting the full topological data ($F$- and $R$-symbols).
Whether examples exist or it can be proven to never occur remains to be determined.

We can make a similar argument for $\sigma$-type vortices, where permutation within supersectors of charges is replaced by the action on the fusion vector spaces, e.g. $u^{\psi_{\vv{0}} a_{\sigma} }_{a_{\sigma}} = \pm 1$.
In this case, when $S^{(\psi_\vv{0})}_{a_{\sigma} b_{\sigma}} \neq 0$, we must have correlated coefficients $u^{\psi_{\vv{0}} a_{\sigma} }_{a_{\sigma}} = u^{\psi_{\vv{0}} b_{\sigma} }_{b_{\sigma}}$.
We partition the $\sigma$-type vortices into subsets $\fMTC_{\sigma_1}, ... , \fMTC_{\sigma_m}$ according to whether the $\psi_\vv{0}$-punctured $S^{(\psi_\vv{0})}$-matrix is zero for all charges in different subsets, i.e. $S^{(\psi_\vv{0})}_{a_{\sigma} b_{\sigma}} =0$.
We can similarly re-expressed this condition in terms of fusion rules.
In particular, we must have $u^{\psi_{\vv{0}} a_{\sigma} }_{a_{\sigma}} = u^{\psi_{\vv{0}} b_{\sigma} }_{b_{\sigma}}$ when $N_{a_{\sigma} b_{\sigma}}^{c_0}$ is odd for any $c_{\vv{0}}$; the possibility of uncorrelated actions requires $N_{a_{\sigma} b_{\sigma}}^{c_0}$ to be even for all $c_{\vv{0}}$.

Additionally, the potentially uncorrelated actions on $\sigma$-type vortices may also be uncorrelated with the actions on $v$-type vortices.
Thus, at this na\"ive level of consideration, there are potentially $n+m$ independent generators in $\ker(\res_{\fMTC_{\vv{0}}})$.
The order of each of these potential generators must be even, since the inspected quantities display action of order 2.

Inspecting the full topological data might rule out some or all of these, except for the fully correlated $\V$, which always exists.
On the other hand, such an inspection might reveal additional elements that are not captured by the above considerations, i.e. that leave all topological charges fixed and $u^{\psi_{\vv{0}} a_{\sigma} }_{a_{\sigma}} = 1$ for all $a_{\sigma}$, but that are nonetheless nontrivial extensions of $\openone^{(0)}$.
It is easy to see that when $\fMTC$ has no fusion multiplicities (i.e. $N_{ab}^{c} \in \{0,1 \}$), the topological symmetries that do not permute any topological charges form an Abelian subgroup, since the $u^{ab}_{c}$-symbols would simply be phases.
We conjecture that $\ker(\res_{\fMTC_{\vv{0}}})$ always forms an Abelian group.

\section{Extension of Fermionic Symmetry Action from \texorpdfstring{$\fMTC_{\vv{0}}$}{fMTCvv0} to \texorpdfstring{$\fMTC$}{fMTCff} for \texorpdfstring{$\ker (\res_{\fMTC_{\vv{0}}} ) \neq \mathbb{Z}_{2}^{\V}$}{kerres}}
\label{app:rho_extension_general}

In this appendix, we consider the extension of fermionic symmetry action from $\fMTC_{\vv{0}}$ to $\fMTC$ in the case where $\ker (\res_{\fMTC_{\vv{0}}} ) \neq \mathbb{Z}_{2}^{\V}$.
For this problem, it is more convenient to initially work in terms of ordinary actions (homomorphisms), extending $[\rho^{(\vv{0})}] : G \to \res_{\fMTC_{\vv{0}}}( \Autpsi{\fMTC} )$ to $[\rho] : G \to \Autpsi{\fMTC}$, and then impose $\gamma_{\psi_{\vv{0}}}({\bf g})=+1$ at the end.
To characterize the obstruction in this way, we first define an arbitrary section on the image of the restriction $s: \res_{\fMTC_{\vv{0}}}( \Autpsi{\fMTC} ) \to \Autpsi{\fMTC}$, such that $\res_{\fMTC_{\vv{0}}} \circ s ([\varphi^{(\vv{0})}]) = [\varphi^{(\vv{0})}]$.
Then we define
\begin{align}
\label{eq:Orho_gen}
O^{\rho}({\bf g,h}) &= s[ \rho^{(\vv{0})}_{\bf gh} ] \cdot s[ \rho^{(\vv{0})}_{\bf h} ]^{-1} \cdot s[ \rho^{(\vv{0})}_{\bf g} ]^{-1}
.
\end{align}

Since $[\rho^{(\vv{0})}]$ is a symmetry action and the restriction map is a homomorphism, it follows that $\res_{\fMTC_{\vv{0}}}( O^{\rho}({\bf g,h}) )= [\openone^{(\vv{0})}]$, and hence $O^{\rho}({\bf g,h}) \in \ker (\res_{\fMTC_{\vv{0}}} )$.
We can also see that
\begin{align}
\cbd O^{\rho}({\bf g,h,k}) &= s[ \rho^{(\vv{0})}_{\bf g} ] \cdot O^{\rho}({\bf h,k}) \cdot s[ \rho^{(\vv{0})}_{\bf g} ]^{-1} \cdot O^{\rho}({\bf gh,k})^{-1} \notag \\
 & \qquad \cdot O^{\rho}({\bf g,hk}) \cdot O^{\rho}({\bf g,h})^{-1}
\notag \\
&=  O^{\rho}({\bf g,h})^{-1} \cdot O^{\rho}({\bf gh,k})^{-1} \notag \\
& \quad \cdot O^{\rho}({\bf g,hk}) \cdot s[ \rho^{(\vv{0})}_{\bf g} ] \cdot O^{\rho}({\bf h,k}) \cdot s[ \rho^{(\vv{0})}_{\bf g} ]^{-1}
\notag \\
&= [\openone]
\end{align}
where the coboundary operator includes a group action through conjugation by $s[ \rho^{(\vv{0})}]$.
The second equivalence is obtained by noting that $\ker (\res_{\fMTC_{\vv{0}}})$ is a normal subgroup, which is assumed to be Abelian.
(As previously discussed, $\ker (\res_{\fMTC_{\vv{0}}})$ is known to be Abelian for most cases of interest, e.g. when there are no fusion multiplicities, and is conjectured to generally be true.)
The third equivalence follows from the definition of $O^{\rho}$.

If we had made a different arbitrary choice of section, $\grave{s}: \res_{\fMTC_{\vv{0}}}( \Autpsi{\fMTC} ) \to \Autpsi{\fMTC}$, we would have $\grave{s}[ \rho^{(\vv{0})}_{\bf g} ] = \varsigma({\bf g})^{-1} \cdot s[ \rho^{(\vv{0})}_{\bf g} ]$ for some $\varsigma({\bf g}) \in C^{1}(G , \ker(\res_{\fMTC_{\vv{0}}}))$.
This different choice of section would yield
\begin{align}
\grave{O}^{\rho}({\bf g,h}) &= \grave{s}[ \rho^{(\vv{0})}_{\bf gh} ] \cdot \grave{s}[ \rho^{(\vv{0})}_{\bf h} ]^{-1} \cdot \grave{s}[ \rho^{(\vv{0})}_{\bf g} ]^{-1}
\notag \\
&= s[ \rho^{(\vv{0})}_{\bf gh} ] \cdot s[ \rho^{(\vv{0})}_{\bf h} ]^{-1} \cdot s[ \rho^{(\vv{0})}_{\bf g} ]^{-1} \notag \\
& \qquad \cdot s[ \rho^{(\vv{0})}_{\bf g} ] \cdot \varsigma({\bf h}) \cdot s[ \rho^{(\vv{0})}_{\bf g} ]^{-1} \cdot \varsigma({\bf gh})^{-1} \cdot \varsigma({\bf g})
\notag \\
&= {O}^{\rho}({\bf g,h}) \cdot \cbd \varsigma({\bf g,h})
.
\end{align}
We note that the effect of the group action included in the coboundary operator here is also independent of the choice of section, since all the pertinent quantities are in $\ker(\res_{\fMTC_{\vv{0}}})$.
Since the arbitrary choice of section should not matter in the definition of an obstruction, we treat these as equivalent definitions, and define the obstruction to be the equivalence class under multiplication with 2-coboundaries $B^{2}_{s[ \rho^{(\vv{0})}]}(G , \ker(\res_{\fMTC_{\vv{0}}}))$.
Thus, we have
\begin{align}
\label{eq:O2obs_gen}
[O^{\rho}] \in H^{2}_{s[ \rho^{(\vv{0})}]}(G , \ker(\res_{\fMTC_{\vv{0}}}))
,
\end{align}
which is independent of the choice of section.

In order to see that $[O^{\rho}]$ represents an obstruction to extending the fermionic symmetry action, we note that if $[\rho_{\bf g}]$ is an extension of $[\rho^{(\vv{0})}_{\bf g}]$, then it can be written as $[\rho_{\bf g}] = \xi({\bf g}) \cdot  s[ \rho^{(\vv{0})}_{\bf g} ]$ for some $\xi({\bf g}) \in C^1(G,\ker (\res_{\fMTC_{\vv{0}}}))$.
Then we see that the condition that $[\rho]:G \to \Autpsi{\fMTC}$ is a homomorphism translates into the condition
\begin{align}
O^{\rho}({\bf g,h}) &= s[ \rho^{(\vv{0})}_{\bf g} ] \cdot \xi({\bf h}) \cdot s[ \rho^{(\vv{0})}_{\bf g} ]^{-1} \cdot  \xi^{-1}({\bf gh}) \cdot \xi({\bf g})
\notag \\
&= \cbd \xi ({\bf g, h})
.
\end{align}
This shows $[O^{\rho}]$ is indeed an obstruction to extending the symmetry action $[\rho^{(\vv{0})}]$, as it is impossible to satisfy this equation unless $O^{\rho} \in B^{2}_{s[ \rho^{(\vv{0})}]}(G , \ker(\res_{\fMTC_{\vv{0}}}))$ is a 2-coboundary, that is unless $[O^{\rho}] = [\openone]$.

It also follows from this analysis that, for any $[\rho]$ that is an extension of $[\rho^{(\vv{0})}]$, we can define another valid extension by $[\rho'_{\bf g}] = [\varphi _{\bf g}] \cdot [\rho_{\bf g}]$, where $[\varphi _{\bf g}] \in H^{1}_{s[ \rho^{(\vv{0})}]}(G, \ker (\res_{\fMTC_{\vv{0}}}))$.
Thus, the set of extensions of $[\rho^{(\vv{0})}] $ to $\psi$-fixed topological symmetries on $\fMTC$ is a $H^{1}_{s[ \rho^{(\vv{0})}]}(G, \ker (\res_{\fMTC_{\vv{0}}}))$ torsor.

Finally, in order to lift this analysis to apply to the extension of fermionic symmetry actions, we replace $\Autpsi{\fMTC}$ with $\Autf{\fMTC}$ by imposing $\gamma_{\psi_{\vv{0}}}({\bf g})=1$ and allowing $\Q$-projective actions, when $[\Q]$ is nontrivial.

When $[\Q] = [\openone]$ (i.e. $\mathcal{A}_{\vv{1}} \neq \varnothing$) or when $\res_{\fMTC_{\vv{0}}} [\Q] \in \Autf{\fMTC_{\vv{0}}}$ is nontrivial (i.e. $\widehat{\bf A}_{\vv{1}} = \mathcal{A}_{\vv{1}} = \varnothing$), the results for $\Autpsi{\fMTC}$ lift without modification.
When $[\Q] \in \Autf{\fMTC}$ is nontrivial and $\res_{\fMTC_{\vv{0}}} [\Q] = [\openone] \in \Autf{\fMTC_{\vv{0}}}$, i.e. when $\mathcal{A}_{\vv{1}} = \varnothing$ and $\widehat{\bf A}_{\vv{1}} \neq \varnothing$, this allows the freedom to change the extension $[\rho_{\bf g}]$ by $[\Q]^{r({\bf g})}$, where $r({\bf g}) \in C^1 (G, \mathbb{Z}_{2}^{\Q})$.
In this case, there is an extra $C^1 (G, \mathbb{Z}_{2}^{\Q})$ freedom in the classification corresponding to a choice of $\Q$-projectiveness of the extension.

\begin{widetext}
\section{Verifying \texorpdfstring{$\mathscr{Y}$}{ydot} for Nontrivial Symmetry Action}
\label{app:Y-details}

In this appendix we show that the 3-cochain appearing in the addition of FSPTs and more generally $\nu$ $ \text{even}$ invertible fermionic phases satisfy the necessary condition:
\begin{align}
\label{eq:ydoteq}
\cbd \Ydot &= \frac{\defectO^{(\nu')}(\psi^{\mathsf{p}'}_{\vv{0}}\otimes \I_{\vv{1}}^{\central},\pi')\defectO^{(\nu'')}(\psi^{\mathsf{p}''}_{\vv{0}} \otimes \I_{\vv{1}}^{\central},\pi'')}{\defectO^{(\nu' + \nu'')}(\psi^{ \mathsf{p}'+\mathsf{p}''+{\pi' \cup \pi''}}_{\vv{0}}\otimes \I_{\vv{1}}^{\central},\pi' +\pi'')}
,
\end{align}
where
\begin{align}
&\defectO^{(\nu)}(\psi^{\mathsf{p}}_{\vv{0}} \otimes \I_{\vv{1}}^\central,\pi)
 = i^{\pi \cup \pi \cup \central} (-1)^{
\mathsf{p} \cup \mathsf{p}
+\mathsf{p} \cup \central
+\pi \cup(\mathsf{p} \cup_1 \central)
+ (\pi \cup_1 \mathsf{p}) \cup \central
}e^{i \frac{\pi}{8} \nu ( \central \cup \central + 2 \central \cup_1 (\central \cup_1 \central) + 4 \mathsf{p} \cup_1 (\central \cup_1 \central))}
,
\end{align}
and $\cbd \mathsf{p} = \pi \cup \central + \frac{\nu}{2} \central \cup_1 \central$.

Ref.~\onlinecite{Brumfiel2018a} gave the general solution for Eq.~\eqref{eq:ydoteq} when $\central = \vv{0}$.
Ref.~\onlinecite{Bark2021inv} extended this solution to the case of $\central$ nontrivial.
Here, we follow a similar calculation as the one in Ref.~\onlinecite{Bark2021inv} and reproduce their (corrected) result.
After writing $\Ydot = e^{i 2 \pi \ydot}$, one finds
\begin{align}
\cbd \lambda  =& \frac{1}{4} ( \widetilde{\pi}' \cup \widetilde{\pi}'  +  \widetilde{\pi}'' \cup \widetilde{\pi}''  -  \widetilde{\pi} \cup \widetilde{\pi}) \cup \central \notag \\ \notag
&+ \frac{1}{2}(\mathsf{p}' \cup \mathsf{p}'' + \mathsf{p}''\cup \mathsf{p}' + (\mathsf{p}' +\mathsf{p}'')\cup \pi' \cup \pi'' + \pi ' \cup \pi'' \cup (\mathsf{p}'+\mathsf{p}'') + \pi ' \cup \pi'' \cup \central + \pi' \cup \pi'' \cup \pi' \cup \pi'') \\ \notag
&+\frac{1}{2}( \pi'  \cup(\mathsf{p}'' \cup_1 \central) + \pi ' \cup ((\pi ' \cup \pi'') \cup_1 \central) + (\pi' \cup_1 \mathsf{p}'') \cup \central + (\pi' \cup_1 (\pi '\cup \pi'')) \cup \central)\\ \notag
&+ \frac{1}{2} (
\pi''  \cup(\mathsf{p}' \cup_1 \central) + \pi '' \cup ((\pi ' \cup \pi'') \cup_1 \central) + (\pi'' \cup_1 \mathsf{p}') \cup \central + (\pi'' \cup_1 (\pi ' \cup \pi'')) \cup \central)
\notag \\
&+ \frac{1}{4}(\nu' \mathsf{p}'' + \nu'' \mathsf{p}' ) \cup_1(\central \cup_1 \central) + \frac{1}{4}(\nu' + \nu'') (\pi ' \cup \pi'')\cup_1(\central \cup_1 \central)
\\
 = &
\underbracket{-\frac{1}{4} ( \widetilde{\pi} ' \cup \widetilde{\pi} '' + \widetilde{\pi} '' \cup \widetilde{\pi}') \cup \central+ \frac{1}{2}
( \pi \cup(\pi' \cup_1 \pi'') + (\pi' \cup_1 \pi'')\cup \pi) \cup \central }_{1}
\notag  \\ \notag
&+ \frac{1}{2}(\underbracket{\mathsf{p}' \cup \mathsf{p}'' + \mathsf{p}''\cup \mathsf{p}'}_{2} + \underbracket{(\mathsf{p}' +\mathsf{p}'')\cup \pi' \cup \pi'' + \pi ' \cup \pi'' \cup (\mathsf{p}'+\mathsf{p}'')}_{3} + \pi ' \cup \pi'' \cup \central + \underbracket{\pi' \cup \pi'' \cup \pi' \cup \pi''}_{4})
\\ \notag
&+\frac{1}{2}(
\underbracket{\pi'  \cup(\mathsf{p}'' \cup_1 \central) }_{5}
+ \underbracket{\pi ' \cup ((\pi ' \cup \pi'') \cup_1 \central) }_{6}
+\underbracket{\cbd \mathsf{p}' \cup_1 \mathsf{p}''}_{7} + \underbracket{\pi'\cup( \central \cup_1 \mathsf{p}'')}_{8}
\\ \notag
&+ \frac{1}{2} (
\underbracket{\pi''  \cup(\mathsf{p}' \cup_1 \central)}_{9}
+ \underbracket{\pi '' \cup ((\pi ' \cup \pi'') \cup_1 \central) }_{10}
+\underbracket{\cbd \mathsf{p}'' \cup_1 \mathsf{p}' }_{11}+ \underbracket{\pi''\cup( \central \cup_1 \mathsf{p}')}_{12}
\\ \notag
&+\frac{1}{2}(\underbracket{\cbd \mathsf{p}' \cup_1 (\pi' \cup \pi'') }_{13}+
\underbracket{\pi' \cup (\central \cup_1 (\pi' \cup \pi''))}_{14})\\ \notag
&+\frac{1}{2}(\underbracket{\cbd \mathsf{p}'' \cup_1 (\pi' \cup \pi'')}_{15}
+\underbracket{ \pi'' \cup (\central \cup_1 (\pi' \cup \pi''))}_{16}) \notag\\
&+\underbracket{ \frac{1}{4}(\nu' \mathsf{p}'' + \nu'' \mathsf{p}' ) \cup_1(\central \cup_1 \central) +  \frac{1}{4} (\central \cup_1 \central) \cup_1 (\nu' \mathsf{p}'' + \nu'' \mathsf{p}' )}_{23}
\notag \\
&+
\underbracket{\frac{1}{4}(\nu' + \nu'') (\pi ' \cup \pi'')\cup_1(\central \cup_1 \central)
+ \frac{1}{4}(\nu' + \nu'')(\central \cup_1 \central) \cup_1 (\pi' \cup \pi'')}_{24}
\end{align}
As a guide to the eye, we have used an underbracket along with an integer to identify specific terms, the utility of this will become clear below.
In the above, we have used
\begin{align}
\frac{1}{4}(\widetilde{\pi}' \cup \widetilde{\pi}'  +  \widetilde{\pi}'' \cup \widetilde{\pi}''  -  \widetilde{\pi} \cup \widetilde{\pi})
&=-\frac{1}{4}(\widetilde{\pi}' \cup \widetilde{\pi}''+\widetilde{\pi}'' \cup \widetilde{\pi}') + \frac{1}{2}({\pi \cup(\pi' \cup_1 \pi'')+(\pi' \cup_1 \pi'')\cup \pi})
\end{align}
to simplify the first line, and the Hirsch identity with $\mathbb{F}_2$ coefficients~\cite{Hirsch1955} ,
\begin{align}
(a \cup b)\cup_1 c= a \cup(b \cup_1 c) + (a \cup_1 c) \cup b
\end{align}
to expand the last two lines, as well as the fact that $\cbd \mathsf{p} = \maj \cup \central  + \frac{\nu}{2} \central \cup_1 \central$.

We now compute several coboundaries which appear in $\cbd \lambda$, we again use the underbracket along with an integer to indicate which terms cancel after summing the coboundaries below.
\begin{align}
\label{eq:firstcob}
\cbd \left( \frac{1}{4} \pi ' \cup \pi '' \cup \pi'' \right)
& = \frac{1}{2}\underbracket{ \pi' \cup \pi' \cup \pi'' \cup \pi''}_{17}\\
\cbd\left( \frac{1}{2} \pi' \cup(\pi' \cup_1 \pi'')\cup \pi'' \right)
&=\frac{1}{2} \underbracket{\pi ' \cup \pi' \cup \pi'' \cup \pi''}_{17} + \frac{1}{2} \underbracket{\pi ' \cup \pi'' \cup \pi' \cup \pi''}_{4}\\
\cbd \left( \frac{1}{4} (\pi ' \cup_1 \pi'' )\cup \central  \right)
&= \underbracket{-\frac{1}{4}(\widetilde{\pi}' \cup \widetilde{\pi}'' + \widetilde{\pi}'' \cup \widetilde{\pi}') \cup \central + \frac{1}{2}((\pi'\cup_1 \pi'') \cup \pi + \pi \cup(\pi' \cup_1 \pi'') )\cup \central}_{1} \notag \\
&+ \frac{1}{2}(\pi' \cup_1 \pi'') \cup(\central \cup_1 \central)
\\
\cbd \left( \frac{1}{2} (\pi' \cup \pi'')  \cup_1 (\mathsf{p}' + \mathsf{p}'') \right) &=
\underbracket{\pi' \cup \pi''\cup(\mathsf{p}'+\mathsf{p}'') +(\mathsf{p}'+\mathsf{p}'')  \cup  \pi' \cup \pi''}_{3} + \underbracket{(\pi' \cup \pi'') \cup_1(\cbd \mathsf{p}' + \cbd \mathsf{p}'')}_{18}\\
\cbd \left( \frac{1}{2} \mathsf{p}' \cup_1 \mathsf{p}'' \right) &=
\frac{1}{2}( \underbracket{\mathsf{p}' \cup \mathsf{p}''+\mathsf{p}'' \cup \mathsf{p}'}_{2} +
\underbracket{(\cbd \mathsf{p}')  \cup_1 \mathsf{p}''}_{19} +
\underbracket{\mathsf{p}' \cup_1 \cbd \mathsf{p}'' }_{20})\\
\cbd \left( \frac{1}{2}(\cbd \mathsf{p}' )\cup_2 \mathsf{p}'' \right) &=
\frac{1}{2}( \underbracket{(\cbd \mathsf{p}' )\cup_1 \mathsf{p}''}_{7}
+\underbracket{\mathsf{p}'' \cup_1 \cbd \mathsf{p}' }_{21}
 + \underbracket{(\cbd \mathsf{p}') \cup_2 \cbd \mathsf{p}''}_{22} )\\
 \cbd \left( \frac{1}{2} \mathsf{p}' \cup_2 \cbd \mathsf{p}'' \right) &=
\frac{1}{2}(
\underbracket{\mathsf{p}' \cup_1 \cbd \mathsf{p}''}_{20}+
\underbracket{(\cbd \mathsf{p}'' )\cup_1 \mathsf{p}' }_{11}
 + \underbracket{(\cbd \mathsf{p}') \cup_2 \cbd \mathsf{p}''}_{22} )\\
 \cbd \left( \frac{1}{2} (\pi' +\pi'') \cup((\pi' \cup\pi'')\cup_2 \central) \right)
 &= \frac{1}{2}\pi' \cup(\underbracket{(\pi'\cup \pi'') \cup_1 \central}_{6} +\underbracket{ \central \cup_1 (\pi' \cup \pi'')}_{14}) \notag \\
  &+ \frac{1}{2}\pi'' \cup(\underbracket{(\pi'\cup \pi'') \cup_1 \central}_{10} + \underbracket{\central \cup_1 (\pi' \cup \pi'')}_{16}) \\
 \cbd \left( \frac{1}{2} (\pi'' \cup(\mathsf{p}' \cup_2 \central) + \pi' \cup(\mathsf{p}'' \cup_2 \central)) \right) &=
\frac{1}{2}( \pi'' \cup( \underbracket{\mathsf{p}' \cup_1 \central}_{9} + \underbracket{\central \cup_1 \mathsf{p}'}_{12})
+ \pi'' \cup((\cbd \mathsf{p}') \cup_2 \central)) \notag \\
 &+\frac{1}{2}(  \pi' \cup( \underbracket{\mathsf{p}'' \cup_1 \central}_{5} + \underbracket{\central \cup_1 \mathsf{p}''}_{8})
 + \pi' \cup((\cbd \mathsf{p}'') \cup_2 \central))\\
 \cbd \left( \frac{1}{2}(\pi' \cup \pi'') \cup_2 (\cbd \mathsf{p}'+\cbd \mathsf{p}'') \right)
 &=
 \frac{1}{2}\underbracket{(\pi' \cup \pi'') \cup_1(\cbd \mathsf{p}'+\cbd \mathsf{p}'')}_{18}  +\frac{1}{2} (\underbracket{\cbd \mathsf{p}'}_{13}+\underbracket{\cbd \mathsf{p}''}_{15}) \cup_1(\pi' \cup \pi'')\\
  \cbd \left(\frac{1}{2} \mathsf{p}'' \cup_2 \cbd \mathsf{p}' \right) & =
 \frac{1}{2}( \underbracket{\mathsf{p}'' \cup_1 \cbd \mathsf{p}' }_{21}+
 \underbracket{(\cbd \mathsf{p}') \cup_1 \mathsf{p}''}_{19} +
 (\cbd \mathsf{p}'') \cup_2 \cbd \mathsf{p}')\label{eq:lastcob} \\
 \cbd \left( \frac{1}{4}(\central \cup_1 \central) \cup_2 (\nu' \mathsf{p}'' + \nu'' \mathsf{p}') \right) &=
\underbracket{\frac{1}{4}(\nu' \mathsf{p}'' + \nu'' \mathsf{p}' ) \cup_1(\central \cup_1 \central) +  \frac{1}{4} (\central \cup_1 \central) \cup_1 (\nu' \mathsf{p}'' + \nu'' \mathsf{p}' )}_{23}\notag \\
&+ \frac{1}{4}(\central \cup_1 \central ) \cup_2 (\underbracket{\nu' \cbd \mathsf{p}''}_{25} + \nu'' \cbd \mathsf{p}') \\
\cbd \left( \frac{1}{4} (\nu' + \nu'') (\central \cup_1 \central) \cup_2  (\pi' \cup \pi'')\right)& = \underbracket{\frac{1}{4}(\nu' + \nu'') (\pi ' \cup \pi'')\cup_1(\central \cup_1 \central)
+ \frac{1}{4}(\nu' + \nu'')(\central \cup_1 \central) \cup_1 (\pi' \cup \pi'')}_{24}\\
\cbd\left( \frac{\nu'}{4} (\central \cup_1 \central) \cup_3 (\pi'' \cup \central)\right) & =
\frac{\nu'}{4} \left( \underbracket{(\central \cup_1 \central ) \cup_2(\pi'' \cup \central) }_{25} + (\pi'' \cup \central) \cup_2(\central \cup_1 \central ) \right)
\end{align}
Not all terms cancel directly.
Indeed, one needs to use that that,
\begin{align}
\label{eq:uncancelled}
\frac{1}{2}\pi ' \cup \pi'' \cup \central & =
\frac{1}{2} \left(
(\pi' \cup_1 \pi'') \cup(\central \cup_1 \central)
+ \pi'' \cup((\maj ' \cup \central)  \cup_2 \central)+ \pi' \cup((\maj '' \cup \central) \cup_2 \central) + ((\maj '' \cup \central)) \cup_2 (\maj ' \cup \central)\right).
\end{align}
and,
\begin{align}
\label{eq:identitycentral}
0 & = \frac{1}{2}(\central \cup_1 \central) \cup_2 \central,\\
\label{eq:identitycentral2}
0 & =\frac{1}{2} (\central \cup_1 \central) \cup_2 (\central \cup_1 \central) .
\end{align}
Using the explicit cochain definitions for $\cup_1$ and $\cup_2$ in Appendix~\ref{app:cup}, the fact that $\central$ is a 2-cocycle, $\pi'$, $\pi''$ are 1-cocycles, and that $\cbd \mathsf{p}' = \pi \cup \central$, and $\cbd \mathsf{p}'' = \pi'' \cup \central$, one can directly verify Eqns.~\eqref{eq:uncancelled}-\eqref{eq:identitycentral2} are satisfied.
Thus,
\begin{align}
\lambda &= \frac14(\pi ' \cup \pi '' \cup \pi'' + (\pi ' \cup_1 \pi'' )\cup \central ) \notag \\
&+\frac{1}{2}(
 \pi' \cup(\pi' \cup_1 \pi'')\cup \pi''
+ (\pi' \cup \pi'')  \cup_1 (\mathsf{p}' + \mathsf{p}'')
+ \mathsf{p}' \cup_1 \mathsf{p}''
+(\cbd \mathsf{p}' )\cup_2 \mathsf{p}''
+ \mathsf{p}' \cup_2 \cbd \mathsf{p}''
\notag \\
&+ (\pi' +\pi'') \cup((\pi' \cup\pi'')\cup_2 \central)
+ \pi'' \cup(\mathsf{p}' \cup_2 \central) + \pi' \cup(\mathsf{p}'' \cup_2 \central)
+(\pi' \cup \pi'') \cup_2 (\cbd \mathsf{p}'+\cbd \mathsf{p}'')
\notag \\
&
+ \mathsf{p}'' \cup_2 \cbd \mathsf{p}'
+\frac{1}{4}(\central \cup_1 \central) \cup_2 (\nu' \mathsf{p}'' + \nu'' \mathsf{p}')
+\frac{1}{4} (\nu' + \nu'') (\central \cup_1 \central) \cup_2  (\pi' \cup \pi'')
+\frac{\nu'}{4} (\central \cup_1 \central) \cup_3 (\pi'' \cup \central)
\end{align}
The expression can be rewritten as
\begin{align}
\label{eq:lambdaeq}
\lambda =& \frac{1}{4}\big(\pi ' \cup \pi '' \cup \pi'' + (\pi ' \cup_1 \pi'' )\cup \central \big)
\notag \\
&+ \frac{1}{2}  \big( \pi' \cup(\pi' \cup_1 \pi'')\cup \pi''
+ (\pi' \cup \pi'')  \cup_1 (\mathsf{p}' + \mathsf{p}'')
+ \mathsf{p}' \cup_1 \mathsf{p}''
+(\pi' \cup \central )\cup_2 \mathsf{p}'' \big)
\notag \\
&+\frac{\nu'}{4} \big((\central \cup_1 \central )\cup_2 \mathsf{p}'' +  (\central \cup_1 \central) \cup_3 (\pi'' \cup \central) \big)
+ \frac{1}{2}\big( f(\mathsf{p}',\pi'; \nu') +f(\mathsf{p}'',\pi''; \nu'') - f(\mathsf{p},\pi; \nu) \big),
\end{align}
where we have defined
\begin{align}
f(\mathsf{p},\pi; \nu) =&\pi \cup(\mathsf{p} \cup_2 \central) + (\pi \cup \central) \cup_2 \mathsf{p} + \mathsf{p} \cup_2 \cbd \mathsf{p} + (\cbd \mathsf{p}) \cup_2 \mathsf{p}  \notag\\
=& \pi \cup(\mathsf{p} \cup_2 \central) + \mathsf{p} \cup_2 (\pi \cup \central) + \frac{\nu}{2} \big( \mathsf{p} \cup_2 (\central \cup_1 \central) + (\central \cup_1 \central) \cup_2 \mathsf{p} \big)
.
\end{align}
If we redefine $\xdot \to \xdot + \frac{1}{2}f(\mathsf{p},\pi; \nu)$, the three additional pieces $\frac{1}{2}(f(\mathsf{p}',\pi'; \nu') +f(\mathsf{p}'',\pi''; \nu'') - f(\mathsf{p},\pi; \nu) )$ in Eq.~\eqref{eq:lambdaeq} cancel.
The resulting expression is identical to the one that was found in Ref.~\onlinecite{Bark2021inv} [see Eq. (17) in Table~I of Ref.~\onlinecite{Bark2021inv}] when restricted to FSPTs.

\end{widetext}
\section{Examples of the Group Law of FSPT Phases from Cochain Triples}\label{app:fSPTclassificationExamples}

We revisit the $\mathbb{Z}_8$ and $\mathbb{Z}_1$ classifications of FSPT phases with $\mathcal{G}^{\eff} = \mathbb{Z}_2^{\eff} \times \mathbb{Z}_2$ and $\mathbb{Z}_4^{\eff}$ using the technology presented in Sec.~\ref{sec:fSPTClassification}.

\subsection{\texorpdfstring{$\mathcal{G}^{\eff} = \mathbb{Z}_2^{\eff} \times \mathbb{Z}_2$}{GfZ2effZ2} }

We can check that the results of Sec.~\ref{sec:fSPTClassification} are compatible with the $\mathbb{Z}_8$ classification of FSPT phases with  $\mathcal{G}^{\eff}=\mathbb{Z}_2^{\eff}\times \mathbb{Z}$ found in Section~\ref{sec:Z8}.
Let $\maj = \text{id}: \mathbb{Z}_2 \to \mathbb{Z}_2$ be the nontrivial group homomorphism.
We will show that
\begin{align}
\label{eq:root}
[0,0,\maj]
\end{align}
is order $8$ and generates all FSPT phases with $\mathcal{G}^{\eff} = \mathbb{Z}_2^{\eff} \times \mathbb{Z}_2$.
For this triple, the defectification obstruction given by $\defectO(\I_{\vv{0}}) = 0$ vanishes, and so the triple given in Eq.~\eqref{eq:root} is admissible.
It follows from Eq.~\eqref{eq:lambdaeqmaiantext} that
\begin{align}
[0,0,\maj]^2 = [ \frac{1}{2}{ \maj \cup(\maj \cup_1 \maj) \cup \maj
+\frac{1}{4}\maj \cup \maj \cup \maj},\maj \cup \maj ].
\end{align}

Stacking this theory with itself and applying Eq.~\eqref{eq:lambdaeqmaiantext} again, we find
\begin{align}\label{eq:stack-fourth}
[0,0,\maj]^4 =  [\frac{1}{2}{ \maj \cup \maj \cup \maj},0,0 ],
\end{align}
which we recognize as the nontrivial bosonic SPT phase with $G = \mathbb{Z}_2$.
To arrive at Eq.~\eqref{eq:stack-fourth} we used
\begin{align}
(\maj \cup \maj) \cup_1(\maj \cup \maj) &= 0 \mod 2,
\end{align}
which can be seen from direct computation.

Finally, stacking $[0,0,\maj]^4$ with itself again, we find
\begin{align}
[0,0,\maj]^8 = [0,0,0].
\end{align}

The data of all $\mathcal{G}^{\eff}= \mathbb{Z}_2^{\eff}\times \mathbb{Z}_2$-graded extensions with nontrivial symmetry action can be tabulated in a relatively efficient way using the product rule to evaluate $[\maj,0,0]^n$:
\begin{align}
n&=0 && [ 0,0,0]\\
n&=1 && [0,0,\maj]\\
n&=2 && [{ ({3}/{4})\maj \cup \maj \cup \maj},\maj \cup \maj,0 ]\\
n&=3 && [{({3}/{4})\maj \cup \maj \cup \maj},\maj \cup \maj,\maj ]\\
n&=4 && [({1}/{2}){ \maj \cup \maj \cup \maj},0,0]\\
n&=5 && [({1}/{2}){ \maj \cup \maj \cup \maj},0,\maj]\\
n&=6 && [{({1}/{4})\maj \cup \maj \cup \maj},\maj \cup \maj,0 ]\\
n&=7 && [{({1}/{4})\maj \cup \maj \cup \maj} ,\maj \cup \maj,\maj].
\end{align}
One can check that the equivalence relations Eqs.~\eqref{eq:equiv1}, \eqref{eq:equiv2} and \eqref{eq:equiv3} do not relate any of these phases.
In the above list, we used $\maj \cup(\maj \cup_1 \maj) \cup \maj = \maj \cup \maj \cup \maj$.

We can also compute all the groups presented in Table~\ref{table:fsptgroups}
\begin{align}
\Gr&= \mathbb{Z}_8\\
\GSPT& = \mathbb{Z}_2\\
\Gtriv&=\mathbb{Z}_4 \\
\Grt& =\Gtriv/\GSPT =\mathbb{Z}_2 \\
\Gro&= \Gr/\Grto  = \Grto/\Grt = \mathbb{Z}_2\\
\Grto& = \Gr/\GSPT = \mathbb{Z}_4
\end{align}

\subsection{\texorpdfstring{$\mathcal{G}^{\eff} = \mathbb{Z}_4^{\eff}$}{GfZ4eff}}

We now consider a nontrivial extension of ${G = \mathbb{Z}_2}$ by $\mathbb{Z}_2^{\eff}$ with
\begin{align}
\central = \maj \cup \maj
\end{align}
where $\maj({1}) = {1}$ and $\maj(0) = 0$.
This is the same scenario encountered in Sec.~\ref{sec:z4fclassification}, and so we should anticipate $G(\mathbb{Z}_4^{\eff}) = \mathbb{Z}_1$.
We have $H^2(\mathbb{Z}_2,\mathbb{Z}_2^{\psi}) = \mathbb{Z}_2$ and can define the nontrivial 2-cocycle by $\coho{p} =\psi_{\vv{0}}^{ \maj \cup \maj}$.
One can check that
\begin{align}
\defectO_r(\I_{ \maj \cup \maj}) = \defectO_r(\psi^{ \maj \cup \maj}_{\vv{1}}\otimes \I_{ \maj \cup \maj})  = 1.
\end{align}
Hence we have four triples satisfying given by:
\begin{align}
\label{Z41}
&(0,0,0)_{ \maj \cup \maj}\\
\label{Z42}
&((1/2){ \maj\cup\maj\cup\maj},{ \maj \cup \maj},0)_{ \maj \cup \maj}\\
\label{Z43}
&((1/2){ \maj\cup\maj\cup\maj},0,0)_{ \maj \cup \maj}\\
\label{Z44}
&(0,{\maj\cup\maj} ,0)_{ \maj \cup \maj}.
\end{align}

We now quotient by the equivalence relations given by Eqs.~\eqref{eq:equiv2} and ~\eqref{eq:equiv3}.
If we set $\coho{z}({\bf g}) = \psi^{\pi({\bf g}) }$ in Eq.~\eqref{eq:equiv2}, we find
\begin{align}
[(1/2){\maj\cup\maj\cup\maj},0,0]_{\pi \cup \pi} &=[0,0,0]_{\pi \cup \pi},\notag\\
[(1/2){ \maj\cup\maj\cup\maj},{\pi \cup \pi},0]_{\pi \cup \pi} &=[0,{\pi \cup \pi},0]_{\pi \cup \pi}
\end{align}
which identifies Eq.~\eqref{Z41} with Eq.~\eqref{Z43}, and Eq.~\eqref{Z42} with Eq.~\eqref{Z44}.
Next we consider Eq.~\eqref{eq:equiv3}, and see that
\begin{align}
[0,{\pi \cup \pi},0]_{\pi \cup \pi} = [(1/2){\pi \cup \pi \cup \pi},0,0]_{\pi \cup \pi},
\end{align}
thereby identifying Eq.~\eqref{Z43} with Eq.~\eqref{Z44}.
Combining the above confirms that there is only one FSPT with $\mathcal{G}^{\eff} = \mathbb{Z}_4^{\eff}$.

We can compute all the groups presented in Table~\ref{table:fsptgroups}
\begin{align}
\Gr&= \mathbb{Z}_1\\
\GSPT& = \mathbb{Z}_1\\
\Gtriv&=\mathbb{Z}_1 \\
\Grt& =\Gtriv/\GSPT =\mathbb{Z}_1 \\
\Gro&= \Gr/\Grto  = \Grto/\Grt = \mathbb{Z}_1\\
\Grto& = \Gr/\GSPT = \mathbb{Z}_1
\end{align}


\begin{widetext}
\section{Relative Defectification Obstruction}
\label{app:relativeO}

We review the formula for the relative defectification obstruction.
Given a consistent $G$-crossed BTC, the relative obstruction signifies when the torsorial action of $\coho{t} \in Z^2_{[\rho]}(G,\mathcal{A})$ results in a consistent theory.
When the obstruction class $[\defectO_r(\coho{t})] \in Z^4(G,U(1)) $ is a coboundary, we say the relative obstruction vanishes.
This relative defectification obstruction was first described in the math literature Ref.~\onlinecite{ENO}, and later explicitly computed in Ref.~\onlinecite{Bark2019b}, where an explicit parametrization of the post-torsor functor associators was provided.
In Ref.~\onlinecite{Aasen21}, the torsorial action was extended to the full $G$-crossed BTC.
Here, we assume we have a consistent $G$-crossed BTC with all topological data known.
Let $\coho{t} \in Z^2_{[\rho]}(G,\mathcal{A})$ be a 2-cocycle; we can apply a torsor fuctor to the $G$-crossed BTC by $\coho{t}$ as described in Ref.~\onlinecite{Aasen21}, so long as the relative obstruction
\begin{align}
\label{eq:pwformula}
\defectO_{r}(\coho{t},\rho)({\bf g},{\bf h},{\bf k},{\bf l})&=
\eta_{{}^{\bf gh}\cohosub{t}({\bf k},{\bf l})}({\bf g},{\bf h})
\frac{U_{\bf g}({}^{\bf g} \coho{t}({\bf h},{\bf kl}),{}^{\bf gh}\coho{t}({\bf k},{\bf l})) }
{U_{\bf g}({}^{\bf g}\coho{t}({\bf hk},{\bf l}),{}^{\bf g}\coho{t}({\bf h},{\bf k}))}
R^{{}^{\bf gh} \cohosub{t}({\bf k},{\bf l}) \cohosub{t}({\bf g},{\bf h})}
\nonumber \\
&\qquad \qquad \times
\frac{F^{\cohosub{t}({\bf gh},{\bf kl}) \cohosub{t}({\bf g},{\bf h}) {}^{\bf gh}\cohosub{t}({\bf k},{\bf l})}}
{F^{\cohosub{t}({\bf gh},{\bf kl})  {}^{\bf gh}\cohosub{t}({\bf k},{\bf l})\cohosub{t}({\bf g},{\bf h})}}
\frac{F^{\cohosub{t}({\bf g},{\bf hkl}) {}^{\bf g}\cohosub{t}({\bf hk},{\bf l} ) {}^{\bf g}\cohosub{t}({\bf h},{\bf k})}}
{F^{\cohosub{t}({\bf g},{\bf hkl}) {}^{\bf g} \cohosub{t}({\bf h},{\bf kl}) {}^{\bf gh}\cohosub{t}({\bf k},{\bf l}) }}
\frac{F^{\cohosub{t}({\bf ghk},{\bf l}) \cohosub{t}({\bf gh},{\bf k} ) \cohosub{t}({\bf g},{\bf h})}}
{F^{\cohosub{t}({\bf ghk},{\bf l}) \cohosub{t}({\bf g},{\bf hk}) {}^{\bf g}\cohosub{t}({\bf h},{\bf k} ) }}
\end{align}
vanishes.
We have included the dependance of $\defectO_r$ on $\rho$ explicitly.
When the pre-torsor functor theory is unobstructed, the defectification obstruction is equal to the relative defectification obstruction, $\defectO(\coho{t},\rho)=\defectO_r (\coho{t},\rho)$.

\end{widetext}

\hbadness=10000	
\bibliography{references}

\begin{thebibliography}{84}%
\makeatletter
\providecommand \@ifxundefined [1]{%
 \@ifx{#1\undefined}
}%
\providecommand \@ifnum [1]{%
 \ifnum #1\expandafter \@firstoftwo
 \else \expandafter \@secondoftwo
 \fi
}%
\providecommand \@ifx [1]{%
 \ifx #1\expandafter \@firstoftwo
 \else \expandafter \@secondoftwo
 \fi
}%
\providecommand \natexlab [1]{#1}%
\providecommand \enquote  [1]{``#1''}%
\providecommand \bibnamefont  [1]{#1}%
\providecommand \bibfnamefont [1]{#1}%
\providecommand \citenamefont [1]{#1}%
\providecommand \href@noop [0]{\@secondoftwo}%
\providecommand \href [0]{\begingroup \@sanitize@url \@href}%
\providecommand \@href[1]{\@@startlink{#1}\@@href}%
\providecommand \@@href[1]{\endgroup#1\@@endlink}%
\providecommand \@sanitize@url [0]{\catcode `\\12\catcode `\$12\catcode
  `\&12\catcode `\#12\catcode `\^12\catcode `\_12\catcode `\%12\relax}%
\providecommand \@@startlink[1]{}%
\providecommand \@@endlink[0]{}%
\providecommand \url  [0]{\begingroup\@sanitize@url \@url }%
\providecommand \@url [1]{\endgroup\@href {#1}{\urlprefix }}%
\providecommand \urlprefix  [0]{URL }%
\providecommand \Eprint [0]{\href }%
\providecommand \doibase [0]{http://dx.doi.org/}%
\providecommand \selectlanguage [0]{\@gobble}%
\providecommand \bibinfo  [0]{\@secondoftwo}%
\providecommand \bibfield  [0]{\@secondoftwo}%
\providecommand \translation [1]{[#1]}%
\providecommand \BibitemOpen [0]{}%
\providecommand \bibitemStop [0]{}%
\providecommand \bibitemNoStop [0]{.\EOS\space}%
\providecommand \EOS [0]{\spacefactor3000\relax}%
\providecommand \BibitemShut  [1]{\csname bibitem#1\endcsname}%
\let\auto@bib@innerbib\@empty
\bibitem [{\citenamefont {Wen}(2004)}]{Wen04}%
  \BibitemOpen
  \bibfield  {author} {\bibinfo {author} {\bibfnamefont {Xiao-Gang}\
  \bibnamefont {Wen}},\ }\href@noop {} {\emph {\bibinfo {title} {Quantum field
  theory of many-body systems: from the origin of sound to an origin of light
  and electrons}}}\ (\bibinfo  {publisher} {Oxford University Press on
  Demand},\ \bibinfo {year} {2004})\BibitemShut {NoStop}%
\bibitem [{\citenamefont {Nayak}\ \emph {et~al.}(2008)\citenamefont {Nayak},
  \citenamefont {Simon}, \citenamefont {Stern}, \citenamefont {Freedman},\ and\
  \citenamefont {Das~Sarma}}]{Nayak08}%
  \BibitemOpen
  \bibfield  {author} {\bibinfo {author} {\bibfnamefont {Chetan}\ \bibnamefont
  {Nayak}}, \bibinfo {author} {\bibfnamefont {Steven~H.}\ \bibnamefont
  {Simon}}, \bibinfo {author} {\bibfnamefont {Ady}\ \bibnamefont {Stern}},
  \bibinfo {author} {\bibfnamefont {Michael}\ \bibnamefont {Freedman}}, \ and\
  \bibinfo {author} {\bibfnamefont {Sankar}\ \bibnamefont {Das~Sarma}},\
  }\bibfield  {title} {\enquote {\bibinfo {title} {Non-abelian anyons and
  topological quantum computation},}\ }\href {\doibase
  10.1103/RevModPhys.80.1083} {\bibfield  {journal} {\bibinfo  {journal} {Rev.
  Mod. Phys.}\ }\textbf {\bibinfo {volume} {80}},\ \bibinfo {pages}
  {1083--1159} (\bibinfo {year} {2008})},\ \Eprint
  {http://arxiv.org/abs/arXiv:0707.1889} {arXiv:0707.1889} \BibitemShut
  {NoStop}%
\bibitem [{\citenamefont {Senthil}(2015)}]{Senthil15}%
  \BibitemOpen
  \bibfield  {author} {\bibinfo {author} {\bibfnamefont {T.}~\bibnamefont
  {Senthil}},\ }\bibfield  {title} {\enquote {\bibinfo {title}
  {Symmetry-protected topological phases of quantum matter},}\ }\href {\doibase
  10.1146/annurev-conmatphys-031214-014740} {\bibfield  {journal} {\bibinfo
  {journal} {Annual Review of Condensed Matter Physics}\ }\textbf {\bibinfo
  {volume} {6}},\ \bibinfo {pages} {299--324} (\bibinfo {year}
  {2015})}\BibitemShut {NoStop}%
\bibitem [{\citenamefont {Chen}\ \emph {et~al.}(2013)\citenamefont {Chen},
  \citenamefont {Gu}, \citenamefont {Liu},\ and\ \citenamefont {Wen}}]{Chen13}%
  \BibitemOpen
  \bibfield  {author} {\bibinfo {author} {\bibfnamefont {Xie}\ \bibnamefont
  {Chen}}, \bibinfo {author} {\bibfnamefont {Zheng-Cheng}\ \bibnamefont {Gu}},
  \bibinfo {author} {\bibfnamefont {Zheng-Xin}\ \bibnamefont {Liu}}, \ and\
  \bibinfo {author} {\bibfnamefont {Xiao-Gang}\ \bibnamefont {Wen}},\
  }\bibfield  {title} {\enquote {\bibinfo {title} {Symmetry protected
  topological orders and the group cohomology of their symmetry group},}\
  }\href {\doibase 10.1103/physrevb.87.155114} {\bibfield  {journal} {\bibinfo
  {journal} {Physical Review B}\ }\textbf {\bibinfo {volume} {87}} (\bibinfo
  {year} {2013}),\ 10.1103/physrevb.87.155114},\ \Eprint
  {http://arxiv.org/abs/arXiv:1106.4772} {arXiv:1106.4772} \BibitemShut
  {NoStop}%
\bibitem [{\citenamefont {Barkeshli}\ \emph {et~al.}(2019)\citenamefont
  {Barkeshli}, \citenamefont {Bonderson}, \citenamefont {Cheng},\ and\
  \citenamefont {Wang}}]{Bark2019}%
  \BibitemOpen
  \bibfield  {author} {\bibinfo {author} {\bibfnamefont {Maissam}\ \bibnamefont
  {Barkeshli}}, \bibinfo {author} {\bibfnamefont {Parsa}\ \bibnamefont
  {Bonderson}}, \bibinfo {author} {\bibfnamefont {Meng}\ \bibnamefont {Cheng}},
  \ and\ \bibinfo {author} {\bibfnamefont {Zhenghan}\ \bibnamefont {Wang}},\
  }\bibfield  {title} {\enquote {\bibinfo {title} {Symmetry fractionalization,
  defects, and gauging of topological phases},}\ }\href {\doibase
  10.1103/PhysRevB.100.115147} {\bibfield  {journal} {\bibinfo  {journal}
  {Phys. Rev. B}\ }\textbf {\bibinfo {volume} {100}},\ \bibinfo {pages}
  {115147} (\bibinfo {year} {2019})},\ \Eprint
  {http://arxiv.org/abs/arXiv:1410.4540} {arXiv:1410.4540} \BibitemShut
  {NoStop}%
\bibitem [{\citenamefont {Bhardwaj}\ \emph {et~al.}(2017)\citenamefont
  {Bhardwaj}, \citenamefont {Gaiotto},\ and\ \citenamefont
  {Kapustin}}]{Bhardwaj2017}%
  \BibitemOpen
  \bibfield  {author} {\bibinfo {author} {\bibfnamefont {Lakshya}\ \bibnamefont
  {Bhardwaj}}, \bibinfo {author} {\bibfnamefont {Davide}\ \bibnamefont
  {Gaiotto}}, \ and\ \bibinfo {author} {\bibfnamefont {Anton}\ \bibnamefont
  {Kapustin}},\ }\bibfield  {title} {\enquote {\bibinfo {title} {{State sum
  constructions of spin-TFTs and string net constructions of fermionic phases
  of matter}},}\ }\href {\doibase 10.1007/jhep04(2017)096} {\bibfield
  {journal} {\bibinfo  {journal} {Journal of High Energy Physics}\ }\textbf
  {\bibinfo {volume} {2017}} (\bibinfo {year} {2017}),\
  10.1007/jhep04(2017)096},\ \Eprint {http://arxiv.org/abs/arXiv:1605.01640}
  {arXiv:1605.01640} \BibitemShut {NoStop}%
\bibitem [{\citenamefont {Wang}\ and\ \citenamefont {Gu}(2020)}]{Wang2020}%
  \BibitemOpen
  \bibfield  {author} {\bibinfo {author} {\bibfnamefont {Qing-Rui}\
  \bibnamefont {Wang}}\ and\ \bibinfo {author} {\bibfnamefont {Zheng-Cheng}\
  \bibnamefont {Gu}},\ }\bibfield  {title} {\enquote {\bibinfo {title}
  {Construction and classification of symmetry-protected topological phases in
  interacting fermion systems},}\ }\href {\doibase 10.1103/PhysRevX.10.031055}
  {\bibfield  {journal} {\bibinfo  {journal} {Phys. Rev. X}\ }\textbf {\bibinfo
  {volume} {10}},\ \bibinfo {pages} {031055} (\bibinfo {year} {2020})},\
  \Eprint {http://arxiv.org/abs/arXiv:1811.00536} {arXiv:1811.00536}
  \BibitemShut {NoStop}%
\bibitem [{\citenamefont {Kapustin}(2014)}]{Kapustin2014}%
  \BibitemOpen
  \bibfield  {author} {\bibinfo {author} {\bibfnamefont {Anton}\ \bibnamefont
  {Kapustin}},\ }\href@noop {} {\enquote {\bibinfo {title} {Symmetry protected
  topological phases, anomalies, and cobordisms: Beyond group cohomology},}\ }
  (\bibinfo {year} {2014}),\ \Eprint {http://arxiv.org/abs/1403.1467}
  {arXiv:1403.1467} \BibitemShut {NoStop}%
\bibitem [{\citenamefont {Gaiotto}\ and\ \citenamefont
  {Kapustin}(2016)}]{Gaiotto2016}%
  \BibitemOpen
  \bibfield  {author} {\bibinfo {author} {\bibfnamefont {Davide}\ \bibnamefont
  {Gaiotto}}\ and\ \bibinfo {author} {\bibfnamefont {Anton}\ \bibnamefont
  {Kapustin}},\ }\bibfield  {title} {\enquote {\bibinfo {title} {{Spin TQFTs
  and fermionic phases of matter}},}\ }\href {\doibase
  10.1142/S0217751X16450445} {\bibfield  {journal} {\bibinfo  {journal}
  {International Journal of Modern Physics A}\ }\textbf {\bibinfo {volume}
  {31}},\ \bibinfo {pages} {1645044} (\bibinfo {year} {2016})},\ \Eprint
  {http://arxiv.org/abs/arXiv:1505.05856} {arXiv:1505.05856} \BibitemShut
  {NoStop}%
\bibitem [{\citenamefont {Lu}\ and\ \citenamefont {Vishwanath}(2012)}]{Lu12}%
  \BibitemOpen
  \bibfield  {author} {\bibinfo {author} {\bibfnamefont {Yuan-Ming}\
  \bibnamefont {Lu}}\ and\ \bibinfo {author} {\bibfnamefont {Ashvin}\
  \bibnamefont {Vishwanath}},\ }\bibfield  {title} {\enquote {\bibinfo {title}
  {{Theory and classification of interacting integer topological phases in two
  dimensions: A Chern-Simons approach}},}\ }\href {\doibase
  10.1103/physrevb.86.125119} {\bibfield  {journal} {\bibinfo  {journal}
  {Physical Review B}\ }\textbf {\bibinfo {volume} {86}} (\bibinfo {year}
  {2012}),\ 10.1103/physrevb.86.125119},\ \Eprint
  {http://arxiv.org/abs/arXiv:1205.3156} {arXiv:1205.3156} \BibitemShut
  {NoStop}%
\bibitem [{\citenamefont {Wen}(2002)}]{Wen02}%
  \BibitemOpen
  \bibfield  {author} {\bibinfo {author} {\bibfnamefont {Xiao-Gang}\
  \bibnamefont {Wen}},\ }\bibfield  {title} {\enquote {\bibinfo {title}
  {Quantum orders and symmetric spin liquids},}\ }\href {\doibase
  10.1103/PhysRevB.65.165113} {\bibfield  {journal} {\bibinfo  {journal} {Phys.
  Rev. B}\ }\textbf {\bibinfo {volume} {65}},\ \bibinfo {pages} {165113}
  (\bibinfo {year} {2002})}\BibitemShut {NoStop}%
\bibitem [{\citenamefont {Levin}\ and\ \citenamefont {Stern}(2012)}]{Levin12}%
  \BibitemOpen
  \bibfield  {author} {\bibinfo {author} {\bibfnamefont {Michael}\ \bibnamefont
  {Levin}}\ and\ \bibinfo {author} {\bibfnamefont {Ady}\ \bibnamefont
  {Stern}},\ }\bibfield  {title} {\enquote {\bibinfo {title} {Classification
  and analysis of two-dimensional abelian fractional topological insulators},}\
  }\href {\doibase 10.1103/physrevb.86.115131} {\bibfield  {journal} {\bibinfo
  {journal} {Physical Review B}\ }\textbf {\bibinfo {volume} {86}} (\bibinfo
  {year} {2012}),\ 10.1103/physrevb.86.115131}\BibitemShut {NoStop}%
\bibitem [{\citenamefont {Essin}\ and\ \citenamefont
  {Hermele}(2013)}]{Essin13}%
  \BibitemOpen
  \bibfield  {author} {\bibinfo {author} {\bibfnamefont {Andrew~M.}\
  \bibnamefont {Essin}}\ and\ \bibinfo {author} {\bibfnamefont {Michael}\
  \bibnamefont {Hermele}},\ }\bibfield  {title} {\enquote {\bibinfo {title}
  {Classifying fractionalization: Symmetry classification of gapped
  $\mathbb{Z}_2$ spin liquids in two dimensions},}\ }\href {\doibase
  10.1103/physrevb.87.104406} {\bibfield  {journal} {\bibinfo  {journal}
  {Physical Review B}\ }\textbf {\bibinfo {volume} {87}} (\bibinfo {year}
  {2013}),\ 10.1103/physrevb.87.104406},\ \Eprint
  {http://arxiv.org/abs/arXiv:1212.0593} {arXiv:1212.0593} \BibitemShut
  {NoStop}%
\bibitem [{\citenamefont {Lu}\ and\ \citenamefont {Vishwanath}(2016)}]{Lu16}%
  \BibitemOpen
  \bibfield  {author} {\bibinfo {author} {\bibfnamefont {Yuan-Ming}\
  \bibnamefont {Lu}}\ and\ \bibinfo {author} {\bibfnamefont {Ashvin}\
  \bibnamefont {Vishwanath}},\ }\bibfield  {title} {\enquote {\bibinfo {title}
  {Classification and properties of symmetry-enriched topological phases:
  Chern-simons approach with applications to $\mathbb{Z}_2$ spin liquids},}\
  }\href {\doibase 10.1103/physrevb.93.155121} {\bibfield  {journal} {\bibinfo
  {journal} {Physical Review B}\ }\textbf {\bibinfo {volume} {93}} (\bibinfo
  {year} {2016}),\ 10.1103/physrevb.93.155121},\ \Eprint
  {http://arxiv.org/abs/arXiv:1302.2634} {arXiv:1302.2634} \BibitemShut
  {NoStop}%
\bibitem [{\citenamefont {Turaev}(2000)}]{turaev2000}%
  \BibitemOpen
  \bibfield  {author} {\bibinfo {author} {\bibfnamefont {Vladimir}\
  \bibnamefont {Turaev}},\ }\href@noop {} {\enquote {\bibinfo {title} {Homotopy
  field theory in dimension 3 and crossed group-categories},}\ } (\bibinfo
  {year} {2000}),\ \Eprint {http://arxiv.org/abs/math/0005291}
  {arXiv:math/0005291} \BibitemShut {NoStop}%
\bibitem [{\citenamefont {Turaev}(2008)}]{turaev2008}%
  \BibitemOpen
  \bibfield  {author} {\bibinfo {author} {\bibfnamefont {Vladimir}\
  \bibnamefont {Turaev}},\ }\bibfield  {title} {\enquote {\bibinfo {title}
  {Crossed group-categories},}\ }\href@noop {} {\bibfield  {journal} {\bibinfo
  {journal} {Arabian Journal for Science and Engineering}\ }\textbf {\bibinfo
  {volume} {33}},\ \bibinfo {pages} {483--504} (\bibinfo {year}
  {2008})}\BibitemShut {NoStop}%
\bibitem [{\citenamefont {Etingof}\ \emph {et~al.}(2009)\citenamefont
  {Etingof}, \citenamefont {Nikshych}, \citenamefont {Ostrik},\ and\
  \citenamefont {with an appendix~by Ehud~Meir}}]{ENO}%
  \BibitemOpen
  \bibfield  {author} {\bibinfo {author} {\bibfnamefont {Pavel}\ \bibnamefont
  {Etingof}}, \bibinfo {author} {\bibfnamefont {Dmitri}\ \bibnamefont
  {Nikshych}}, \bibinfo {author} {\bibfnamefont {Victor}\ \bibnamefont
  {Ostrik}}, \ and\ \bibinfo {author} {\bibnamefont {with an appendix~by
  Ehud~Meir}},\ }\href@noop {} {\enquote {\bibinfo {title} {Fusion categories
  and homotopy theory},}\ } (\bibinfo {year} {2009}),\ \Eprint
  {http://arxiv.org/abs/0909.3140} {arXiv:0909.3140} \BibitemShut {NoStop}%
\bibitem [{\citenamefont {Gu}\ and\ \citenamefont {Levin}(2014)}]{Gu14}%
  \BibitemOpen
  \bibfield  {author} {\bibinfo {author} {\bibfnamefont {Zheng-Cheng}\
  \bibnamefont {Gu}}\ and\ \bibinfo {author} {\bibfnamefont {Michael}\
  \bibnamefont {Levin}},\ }\bibfield  {title} {\enquote {\bibinfo {title}
  {Effect of interactions on two-dimensional fermionic symmetry-protected
  topological phases with $\mathbb{Z}_2$ symmetry},}\ }\href {\doibase
  10.1103/physrevb.89.201113} {\bibfield  {journal} {\bibinfo  {journal}
  {Physical Review B}\ }\textbf {\bibinfo {volume} {89}} (\bibinfo {year}
  {2014}),\ 10.1103/physrevb.89.201113},\ \Eprint
  {http://arxiv.org/abs/arXiv:1304.4569} {arXiv:1304.4569} \BibitemShut
  {NoStop}%
\bibitem [{\citenamefont {{David Aasen and Parsa Bonderson and Christina
  Knapp}}(2021)}]{Aasen21}%
  \BibitemOpen
  \bibfield  {author} {\bibinfo {author} {\bibnamefont {{David Aasen and Parsa
  Bonderson and Christina Knapp}}},\ }\href@noop {} {\enquote {\bibinfo {title}
  {{Torsorial actions on G-crossed braided tensor categories}},}\ } (\bibinfo
  {year} {2021}),\ \Eprint {http://arxiv.org/abs/2107.10270} {arXiv:2107.10270}
  \BibitemShut {NoStop}%
\bibitem [{\citenamefont {Barkeshli}\ and\ \citenamefont
  {Cheng}(2020)}]{Bark2019b}%
  \BibitemOpen
  \bibfield  {author} {\bibinfo {author} {\bibfnamefont {Maissam}\ \bibnamefont
  {Barkeshli}}\ and\ \bibinfo {author} {\bibfnamefont {Meng}\ \bibnamefont
  {Cheng}},\ }\bibfield  {title} {\enquote {\bibinfo {title} {{Relative
  Anomalies in (2+1)D Symmetry Enriched Topological States}},}\ }\href
  {\doibase 10.21468/SciPostPhys.8.2.028} {\bibfield  {journal} {\bibinfo
  {journal} {SciPost Phys.}\ }\textbf {\bibinfo {volume} {8}},\ \bibinfo
  {pages} {28} (\bibinfo {year} {2020})},\ \Eprint
  {http://arxiv.org/abs/arXiv:1906.10691} {arXiv:1906.10691} \BibitemShut
  {NoStop}%
\bibitem [{\citenamefont {Freed}\ and\ \citenamefont
  {Hopkins}(2019)}]{Freed19}%
  \BibitemOpen
  \bibfield  {author} {\bibinfo {author} {\bibfnamefont {Daniel~S.}\
  \bibnamefont {Freed}}\ and\ \bibinfo {author} {\bibfnamefont {Michael~J.}\
  \bibnamefont {Hopkins}},\ }\href@noop {} {\enquote {\bibinfo {title}
  {Reflection positivity and invertible topological phases},}\ } (\bibinfo
  {year} {2019}),\ \Eprint {http://arxiv.org/abs/1604.06527} {arXiv:1604.06527}
  \BibitemShut {NoStop}%
\bibitem [{\citenamefont {Kapustin}\ \emph {et~al.}(2015)\citenamefont
  {Kapustin}, \citenamefont {Thorngren}, \citenamefont {Turzillo},\ and\
  \citenamefont {Wang}}]{Kapustin15}%
  \BibitemOpen
  \bibfield  {author} {\bibinfo {author} {\bibfnamefont {Anton}\ \bibnamefont
  {Kapustin}}, \bibinfo {author} {\bibfnamefont {Ryan}\ \bibnamefont
  {Thorngren}}, \bibinfo {author} {\bibfnamefont {Alex}\ \bibnamefont
  {Turzillo}}, \ and\ \bibinfo {author} {\bibfnamefont {Zitao}\ \bibnamefont
  {Wang}},\ }\bibfield  {title} {\enquote {\bibinfo {title} {Fermionic symmetry
  protected topological phases and cobordisms},}\ }\href {\doibase
  10.1007/jhep12(2015)052} {\bibfield  {journal} {\bibinfo  {journal} {Journal
  of High Energy Physics}\ }\textbf {\bibinfo {volume} {2015}},\ \bibinfo
  {pages} {1--21} (\bibinfo {year} {2015})},\ \Eprint
  {http://arxiv.org/abs/arXiv:1406.7329} {arXiv:1406.7329} \BibitemShut
  {NoStop}%
\bibitem [{\citenamefont {Kapustin}\ \emph {et~al.}(2018)\citenamefont
  {Kapustin}, \citenamefont {Turzillo},\ and\ \citenamefont
  {You}}]{Kapustin18}%
  \BibitemOpen
  \bibfield  {author} {\bibinfo {author} {\bibfnamefont {Anton}\ \bibnamefont
  {Kapustin}}, \bibinfo {author} {\bibfnamefont {Alex}\ \bibnamefont
  {Turzillo}}, \ and\ \bibinfo {author} {\bibfnamefont {Minyoung}\ \bibnamefont
  {You}},\ }\bibfield  {title} {\enquote {\bibinfo {title} {Spin topological
  field theory and fermionic matrix product states},}\ }\href {\doibase
  10.1103/physrevb.98.125101} {\bibfield  {journal} {\bibinfo  {journal}
  {Physical Review B}\ }\textbf {\bibinfo {volume} {98}} (\bibinfo {year}
  {2018}),\ 10.1103/physrevb.98.125101}\BibitemShut {NoStop}%
\bibitem [{\citenamefont {Wang}\ \emph {et~al.}(2016)\citenamefont {Wang},
  \citenamefont {Lin},\ and\ \citenamefont {Levin}}]{Wang2016}%
  \BibitemOpen
  \bibfield  {author} {\bibinfo {author} {\bibfnamefont {Chenjie}\ \bibnamefont
  {Wang}}, \bibinfo {author} {\bibfnamefont {Chien-Hung}\ \bibnamefont {Lin}},
  \ and\ \bibinfo {author} {\bibfnamefont {Michael}\ \bibnamefont {Levin}},\
  }\bibfield  {title} {\enquote {\bibinfo {title} {Bulk-boundary correspondence
  for three-dimensional symmetry-protected topological phases},}\ }\href
  {\doibase 10.1103/PhysRevX.6.021015} {\bibfield  {journal} {\bibinfo
  {journal} {Phys. Rev. X}\ }\textbf {\bibinfo {volume} {6}},\ \bibinfo {pages}
  {021015} (\bibinfo {year} {2016})}\BibitemShut {NoStop}%
\bibitem [{\citenamefont {Tarantino}\ \emph {et~al.}(2016)\citenamefont
  {Tarantino}, \citenamefont {Lindner},\ and\ \citenamefont
  {Fidkowski}}]{Tarantino2016}%
  \BibitemOpen
  \bibfield  {author} {\bibinfo {author} {\bibfnamefont {Nicolas}\ \bibnamefont
  {Tarantino}}, \bibinfo {author} {\bibfnamefont {Netanel~H}\ \bibnamefont
  {Lindner}}, \ and\ \bibinfo {author} {\bibfnamefont {Lukasz}\ \bibnamefont
  {Fidkowski}},\ }\bibfield  {title} {\enquote {\bibinfo {title} {Symmetry
  fractionalization and twist defects},}\ }\href {\doibase
  10.1088/1367-2630/18/3/035006} {\bibfield  {journal} {\bibinfo  {journal}
  {New Journal of Physics}\ }\textbf {\bibinfo {volume} {18}},\ \bibinfo
  {pages} {035006} (\bibinfo {year} {2016})}\BibitemShut {NoStop}%
\bibitem [{\citenamefont {Wang}\ and\ \citenamefont {Gu}(2018)}]{Wang2018}%
  \BibitemOpen
  \bibfield  {author} {\bibinfo {author} {\bibfnamefont {Qing-Rui}\
  \bibnamefont {Wang}}\ and\ \bibinfo {author} {\bibfnamefont {Zheng-Cheng}\
  \bibnamefont {Gu}},\ }\bibfield  {title} {\enquote {\bibinfo {title} {Towards
  a complete classification of symmetry-protected topological phases for
  interacting fermions in three dimensions and a general group supercohomology
  theory},}\ }\href {\doibase 10.1103/PhysRevX.8.011055} {\bibfield  {journal}
  {\bibinfo  {journal} {Phys. Rev. X}\ }\textbf {\bibinfo {volume} {8}},\
  \bibinfo {pages} {011055} (\bibinfo {year} {2018})},\ \Eprint
  {http://arxiv.org/abs/arXiv:1703.10937} {arXiv:1703.10937} \BibitemShut
  {NoStop}%
\bibitem [{\citenamefont {Cheng}\ \emph {et~al.}(2018)\citenamefont {Cheng},
  \citenamefont {Bi}, \citenamefont {You},\ and\ \citenamefont
  {Gu}}]{Cheng2018}%
  \BibitemOpen
  \bibfield  {author} {\bibinfo {author} {\bibfnamefont {Meng}\ \bibnamefont
  {Cheng}}, \bibinfo {author} {\bibfnamefont {Zhen}\ \bibnamefont {Bi}},
  \bibinfo {author} {\bibfnamefont {Yi-Zhuang}\ \bibnamefont {You}}, \ and\
  \bibinfo {author} {\bibfnamefont {Zheng-Cheng}\ \bibnamefont {Gu}},\
  }\bibfield  {title} {\enquote {\bibinfo {title} {Classification of
  symmetry-protected phases for interacting fermions in two dimensions},}\
  }\href {\doibase 10.1103/PhysRevB.97.205109} {\bibfield  {journal} {\bibinfo
  {journal} {Phys. Rev. B}\ }\textbf {\bibinfo {volume} {97}},\ \bibinfo
  {pages} {205109} (\bibinfo {year} {2018})},\ \Eprint
  {http://arxiv.org/abs/arXiv:1501.01313} {arXiv:1501.01313} \BibitemShut
  {NoStop}%
\bibitem [{\citenamefont {Chen}\ \emph {et~al.}(2019)\citenamefont {Chen},
  \citenamefont {Kapustin}, \citenamefont {Turzillo},\ and\ \citenamefont
  {You}}]{Turzillo2019}%
  \BibitemOpen
  \bibfield  {author} {\bibinfo {author} {\bibfnamefont {Yu-An}\ \bibnamefont
  {Chen}}, \bibinfo {author} {\bibfnamefont {Anton}\ \bibnamefont {Kapustin}},
  \bibinfo {author} {\bibfnamefont {Alex}\ \bibnamefont {Turzillo}}, \ and\
  \bibinfo {author} {\bibfnamefont {Minyoung}\ \bibnamefont {You}},\ }\bibfield
   {title} {\enquote {\bibinfo {title} {Free and interacting short-range
  entangled phases of fermions: Beyond the tenfold way},}\ }\href {\doibase
  10.1103/PhysRevB.100.195128} {\bibfield  {journal} {\bibinfo  {journal}
  {Phys. Rev. B}\ }\textbf {\bibinfo {volume} {100}},\ \bibinfo {pages}
  {195128} (\bibinfo {year} {2019})},\ \Eprint
  {http://arxiv.org/abs/arXiv:1904.11550} {arXiv:1904.11550} \BibitemShut
  {NoStop}%
\bibitem [{\citenamefont {Brumfiel}\ and\ \citenamefont
  {Morgan}(2018)}]{Brumfiel2018a}%
  \BibitemOpen
  \bibfield  {author} {\bibinfo {author} {\bibfnamefont {Greg}\ \bibnamefont
  {Brumfiel}}\ and\ \bibinfo {author} {\bibfnamefont {John}\ \bibnamefont
  {Morgan}},\ }\href@noop {} {\enquote {\bibinfo {title} {The pontrjagin dual
  of 3-dimensional spin bordism},}\ } (\bibinfo {year} {2018}),\ \Eprint
  {http://arxiv.org/abs/1612.02860} {arXiv:1612.02860} \BibitemShut {NoStop}%
\bibitem [{\citenamefont {Lan}\ \emph {et~al.}(2017)\citenamefont {Lan},
  \citenamefont {Kong},\ and\ \citenamefont {Wen}}]{Lan2017}%
  \BibitemOpen
  \bibfield  {author} {\bibinfo {author} {\bibfnamefont {Tian}\ \bibnamefont
  {Lan}}, \bibinfo {author} {\bibfnamefont {Liang}\ \bibnamefont {Kong}}, \
  and\ \bibinfo {author} {\bibfnamefont {Xiao-Gang}\ \bibnamefont {Wen}},\
  }\bibfield  {title} {\enquote {\bibinfo {title} {Classification of
  (2+1)-dimensional topological order and symmetry-protected topological order
  for bosonic and fermionic systems with on-site symmetries},}\ }\href
  {\doibase 10.1103/PhysRevB.95.235140} {\bibfield  {journal} {\bibinfo
  {journal} {Phys. Rev. B}\ }\textbf {\bibinfo {volume} {95}},\ \bibinfo
  {pages} {235140} (\bibinfo {year} {2017})},\ \Eprint
  {http://arxiv.org/abs/arXiv:1602.05946} {arXiv:1602.05946} \BibitemShut
  {NoStop}%
\bibitem [{\citenamefont {Fidkowski}\ \emph {et~al.}(2018)\citenamefont
  {Fidkowski}, \citenamefont {Vishwanath},\ and\ \citenamefont
  {Metlitski}}]{Fidkowski2018}%
  \BibitemOpen
  \bibfield  {author} {\bibinfo {author} {\bibfnamefont {Lukasz}\ \bibnamefont
  {Fidkowski}}, \bibinfo {author} {\bibfnamefont {Ashvin}\ \bibnamefont
  {Vishwanath}}, \ and\ \bibinfo {author} {\bibfnamefont {Max~A.}\ \bibnamefont
  {Metlitski}},\ }\href@noop {} {\enquote {\bibinfo {title} {Surface
  topological order and a new 't hooft anomaly of interaction enabled 3+1d
  fermion spts},}\ } (\bibinfo {year} {2018}),\ \Eprint
  {http://arxiv.org/abs/1804.08628} {arXiv:1804.08628} \BibitemShut {NoStop}%
\bibitem [{\citenamefont {Thorngren}(2020)}]{Thorngren2020}%
  \BibitemOpen
  \bibfield  {author} {\bibinfo {author} {\bibfnamefont {Ryan}\ \bibnamefont
  {Thorngren}},\ }\bibfield  {title} {\enquote {\bibinfo {title} {Anomalies and
  bosonization},}\ }\href@noop {} {\bibfield  {journal} {\bibinfo  {journal}
  {Communications in Mathematical Physics}\ }\textbf {\bibinfo {volume}
  {378}},\ \bibinfo {pages} {1775--1816} (\bibinfo {year} {2020})},\ \Eprint
  {http://arxiv.org/abs/arXiv:1810.04414} {arXiv:1810.04414} \BibitemShut
  {NoStop}%
\bibitem [{\citenamefont {Tata}\ \emph {et~al.}(2021)\citenamefont {Tata},
  \citenamefont {Kobayashi}, \citenamefont {Bulmash},\ and\ \citenamefont
  {Barkeshli}}]{Tata2021}%
  \BibitemOpen
  \bibfield  {author} {\bibinfo {author} {\bibfnamefont {Srivatsa}\
  \bibnamefont {Tata}}, \bibinfo {author} {\bibfnamefont {Ryohei}\ \bibnamefont
  {Kobayashi}}, \bibinfo {author} {\bibfnamefont {Daniel}\ \bibnamefont
  {Bulmash}}, \ and\ \bibinfo {author} {\bibfnamefont {Maissam}\ \bibnamefont
  {Barkeshli}},\ }\bibfield  {title} {\enquote {\bibinfo {title} {{Anomalies in
  (2+ 1) D fermionic topological phases and (3+ 1) D path integral state sums
  for fermionic SPTs}},}\ }\href@noop {} {\  (\bibinfo {year}
  {2021})}\BibitemShut {NoStop}%
\bibitem [{\citenamefont {Kitaev}(2006)}]{Kitaev2006}%
  \BibitemOpen
  \bibfield  {author} {\bibinfo {author} {\bibfnamefont {Alexei}\ \bibnamefont
  {Kitaev}},\ }\bibfield  {title} {\enquote {\bibinfo {title} {Anyons in an
  exactly solved model and beyond},}\ }\href {\doibase
  10.1016/j.aop.2005.10.005} {\bibfield  {journal} {\bibinfo  {journal} {Annals
  of Physics}\ }\textbf {\bibinfo {volume} {321}},\ \bibinfo {pages} {2 -- 111}
  (\bibinfo {year} {2006})},\ \Eprint
  {http://arxiv.org/abs/arXiv:cond-mat/0506438} {arXiv:cond-mat/0506438}
  \BibitemShut {NoStop}%
\bibitem [{\citenamefont {Kitaev}(2003)}]{Kitaev2003}%
  \BibitemOpen
  \bibfield  {author} {\bibinfo {author} {\bibfnamefont {A.Yu.}\ \bibnamefont
  {Kitaev}},\ }\bibfield  {title} {\enquote {\bibinfo {title} {Fault-tolerant
  quantum computation by anyons},}\ }\href {\doibase
  https://doi.org/10.1016/S0003-4916(02)00018-0} {\bibfield  {journal}
  {\bibinfo  {journal} {Annals of Physics}\ }\textbf {\bibinfo {volume}
  {303}},\ \bibinfo {pages} {2 -- 30} (\bibinfo {year} {2003})},\ \Eprint
  {http://arxiv.org/abs/arXiv:quant-ph/9707021} {arXiv:quant-ph/9707021}
  \BibitemShut {NoStop}%
\bibitem [{\citenamefont {Aasen}\ \emph {et~al.}(2019)\citenamefont {Aasen},
  \citenamefont {Lake},\ and\ \citenamefont {Walker}}]{Aasen19}%
  \BibitemOpen
  \bibfield  {author} {\bibinfo {author} {\bibfnamefont {David}\ \bibnamefont
  {Aasen}}, \bibinfo {author} {\bibfnamefont {Ethan}\ \bibnamefont {Lake}}, \
  and\ \bibinfo {author} {\bibfnamefont {Kevin}\ \bibnamefont {Walker}},\
  }\bibfield  {title} {\enquote {\bibinfo {title} {Fermion condensation and
  super pivotal categories},}\ }\href@noop {} {\bibfield  {journal} {\bibinfo
  {journal} {Journal of Mathematical Physics}\ }\textbf {\bibinfo {volume}
  {60}},\ \bibinfo {pages} {121901} (\bibinfo {year} {2019})},\ \Eprint
  {http://arxiv.org/abs/arXiv:1709.01941} {arXiv:1709.01941} \BibitemShut
  {NoStop}%
\bibitem [{\citenamefont {Walker}(2017)}]{WalkerSimmons}%
  \BibitemOpen
  \bibfield  {author} {\bibinfo {author} {\bibfnamefont {Kevin}\ \bibnamefont
  {Walker}},\ }\href
  {http://scgp.stonybrook.edu/video_portal/video.php?id=3238} {\enquote
  {\bibinfo {title} {{L}ow-dimensional {G}-bordism and {G}-modular {TQFT}s},}\
  }\bibinfo {howpublished} {presented at the {S}imons {C}enter workshop:
  {S}trongly {C}orrelated {T}opological {P}hases of {M}atter} (\bibinfo {year}
  {2017})\BibitemShut {NoStop}%
\bibitem [{\citenamefont {Bonderson}\ \emph {et~al.}()\citenamefont
  {Bonderson}, \citenamefont {Cheng}, \citenamefont {Mong},\ and\ \citenamefont
  {Tran}}]{Bonderson20}%
  \BibitemOpen
  \bibfield  {author} {\bibinfo {author} {\bibfnamefont {Parsa}\ \bibnamefont
  {Bonderson}}, \bibinfo {author} {\bibfnamefont {Meng}\ \bibnamefont {Cheng}},
  \bibinfo {author} {\bibfnamefont {Roger S.~K.}\ \bibnamefont {Mong}}, \ and\
  \bibinfo {author} {\bibfnamefont {Alan~D.}\ \bibnamefont {Tran}},\
  }\href@noop {} {\enquote {\bibinfo {title} {Fermionic topological phases and
  modular transformations},}\ }\BibitemShut {NoStop}%
\bibitem [{\citenamefont {Johnson-Freyd}\ and\ \citenamefont
  {Reutter}(2021)}]{JohnsonFreyd2021}%
  \BibitemOpen
  \bibfield  {author} {\bibinfo {author} {\bibfnamefont {Theo}\ \bibnamefont
  {Johnson-Freyd}}\ and\ \bibinfo {author} {\bibfnamefont {David}\ \bibnamefont
  {Reutter}},\ }\href@noop {} {\enquote {\bibinfo {title} {Minimal
  nondegenerate extensions},}\ } (\bibinfo {year} {2021}),\ \Eprint
  {http://arxiv.org/abs/arXiv:2105.15167} {arXiv:2105.15167} \BibitemShut
  {NoStop}%
\bibitem [{\citenamefont {Bruillard}\ \emph {et~al.}(2017)\citenamefont
  {Bruillard}, \citenamefont {Galindo}, \citenamefont {Hagge}, \citenamefont
  {Ng}, \citenamefont {Plavnik}, \citenamefont {Rowell},\ and\ \citenamefont
  {Wang}}]{Bruillard17}%
  \BibitemOpen
  \bibfield  {author} {\bibinfo {author} {\bibfnamefont {Paul}\ \bibnamefont
  {Bruillard}}, \bibinfo {author} {\bibfnamefont {C{\'e}sar}\ \bibnamefont
  {Galindo}}, \bibinfo {author} {\bibfnamefont {Tobias}\ \bibnamefont {Hagge}},
  \bibinfo {author} {\bibfnamefont {Siu-Hung}\ \bibnamefont {Ng}}, \bibinfo
  {author} {\bibfnamefont {Julia~Yael}\ \bibnamefont {Plavnik}}, \bibinfo
  {author} {\bibfnamefont {Eric~C.}\ \bibnamefont {Rowell}}, \ and\ \bibinfo
  {author} {\bibfnamefont {Zhenghan}\ \bibnamefont {Wang}},\ }\bibfield
  {title} {\enquote {\bibinfo {title} {Fermionic modular categories and the
  16-fold way},}\ }\href {\doibase 10.1063/1.4982048} {\bibfield  {journal}
  {\bibinfo  {journal} {Journal of Mathematical Physics}\ }\textbf {\bibinfo
  {volume} {58}},\ \bibinfo {pages} {041704} (\bibinfo {year} {2017})},\
  \Eprint {http://arxiv.org/abs/arXiv:1603.09294} {arXiv:1603.09294}
  \BibitemShut {NoStop}%
\bibitem [{\citenamefont {Hansson}\ \emph {et~al.}(2004)\citenamefont
  {Hansson}, \citenamefont {Oganesyan},\ and\ \citenamefont
  {Sondhi}}]{Hansson04}%
  \BibitemOpen
  \bibfield  {author} {\bibinfo {author} {\bibfnamefont {T.H.}\ \bibnamefont
  {Hansson}}, \bibinfo {author} {\bibfnamefont {Vadim}\ \bibnamefont
  {Oganesyan}}, \ and\ \bibinfo {author} {\bibfnamefont {S.L.}\ \bibnamefont
  {Sondhi}},\ }\bibfield  {title} {\enquote {\bibinfo {title} {Superconductors
  are topologically ordered},}\ }\href {\doibase 10.1016/j.aop.2004.05.006}
  {\bibfield  {journal} {\bibinfo  {journal} {Annals of Physics}\ }\textbf
  {\bibinfo {volume} {313}},\ \bibinfo {pages} {497--538} (\bibinfo {year}
  {2004})},\ \Eprint {http://arxiv.org/abs/arXiv:cond-mat/0404327v1}
  {arXiv:cond-mat/0404327v1} \BibitemShut {NoStop}%
\bibitem [{\citenamefont {Bais}\ and\ \citenamefont
  {Slingerland}(2009)}]{Bais2009}%
  \BibitemOpen
  \bibfield  {author} {\bibinfo {author} {\bibfnamefont {F.~A.}\ \bibnamefont
  {Bais}}\ and\ \bibinfo {author} {\bibfnamefont {J.~K.}\ \bibnamefont
  {Slingerland}},\ }\bibfield  {title} {\enquote {\bibinfo {title}
  {Condensate-induced transitions between topologically ordered phases},}\
  }\href {\doibase 10.1103/PhysRevB.79.045316} {\bibfield  {journal} {\bibinfo
  {journal} {Phys. Rev. B}\ }\textbf {\bibinfo {volume} {79}},\ \bibinfo
  {pages} {045316} (\bibinfo {year} {2009})},\ \Eprint
  {http://arxiv.org/abs/arXiv:0808.0627} {arXiv:0808.0627} \BibitemShut
  {NoStop}%
\bibitem [{\citenamefont {Galindo}\ and\ \citenamefont
  {Venegas-Ram{\'i}rez}(2017)}]{Galindo2017}%
  \BibitemOpen
  \bibfield  {author} {\bibinfo {author} {\bibfnamefont {C{\'e}sar}\
  \bibnamefont {Galindo}}\ and\ \bibinfo {author} {\bibfnamefont
  {C{\'e}sar~F.}\ \bibnamefont {Venegas-Ram{\'i}rez}},\ }\href@noop {}
  {\enquote {\bibinfo {title} {Categorical fermionic actions and minimal
  modular extensions},}\ } (\bibinfo {year} {2017}),\ \Eprint
  {http://arxiv.org/abs/1712.07097} {arXiv:1712.07097} \BibitemShut {NoStop}%
\bibitem [{\citenamefont {Vishwanath}\ and\ \citenamefont
  {Senthil}(2013)}]{Vishwanath2013}%
  \BibitemOpen
  \bibfield  {author} {\bibinfo {author} {\bibfnamefont {Ashvin}\ \bibnamefont
  {Vishwanath}}\ and\ \bibinfo {author} {\bibfnamefont {T.}~\bibnamefont
  {Senthil}},\ }\bibfield  {title} {\enquote {\bibinfo {title} {Physics of
  three-dimensional bosonic topological insulators: Surface-deconfined
  criticality and quantized magnetoelectric effect},}\ }\href {\doibase
  10.1103/PhysRevX.3.011016} {\bibfield  {journal} {\bibinfo  {journal} {Phys.
  Rev. X}\ }\textbf {\bibinfo {volume} {3}},\ \bibinfo {pages} {011016}
  (\bibinfo {year} {2013})}\BibitemShut {NoStop}%
\bibitem [{\citenamefont {Fidkowski}\ \emph {et~al.}(2013)\citenamefont
  {Fidkowski}, \citenamefont {Chen},\ and\ \citenamefont
  {Vishwanath}}]{Fidkowski2013}%
  \BibitemOpen
  \bibfield  {author} {\bibinfo {author} {\bibfnamefont {Lukasz}\ \bibnamefont
  {Fidkowski}}, \bibinfo {author} {\bibfnamefont {Xie}\ \bibnamefont {Chen}}, \
  and\ \bibinfo {author} {\bibfnamefont {Ashvin}\ \bibnamefont {Vishwanath}},\
  }\bibfield  {title} {\enquote {\bibinfo {title} {Non-abelian topological
  order on the surface of a 3d topological superconductor from an exactly
  solved model},}\ }\href {\doibase 10.1103/PhysRevX.3.041016} {\bibfield
  {journal} {\bibinfo  {journal} {Phys. Rev. X}\ }\textbf {\bibinfo {volume}
  {3}},\ \bibinfo {pages} {041016} (\bibinfo {year} {2013})},\ \Eprint
  {http://arxiv.org/abs/arXiv:1305.5851} {arXiv:1305.5851} \BibitemShut
  {NoStop}%
\bibitem [{Note1()}]{Note1}%
  \BibitemOpen
  \bibinfo {note} {If generalizing to a system with multiple physical fermions
  $\psi _\protect \mathrm {0}^{(j)}$, all must satisfy $\gamma _{\psi _\protect
  \mathrm {0}^{(j)}}=1$.}\BibitemShut {Stop}%
\bibitem [{Note2()}]{Note2}%
  \BibitemOpen
  \bibinfo {note} {We do not know an example where $\protect \mathsf
  {res}_{\protect \mathcal {M}_{\protect \mathrm {0}}}( \protect \text
  {Aut}^{\psi / \protect \operatorname {f}}({\protect \mathcal {M}} ) )
  \protect \neq \protect \text {Aut}^{\psi / \protect \operatorname
  {f}}({\protect \mathcal {M}_{\protect \mathrm {0}}} )$ when topological
  symmetries are not space or time reflecting. However, they are ubiquitous
  when considering symmetries that reflect space or time. For example, when
  $\protect \mathcal {M}$ has a space or time reflecting topological symmetry,
  there are no nontrivial space or time reflecting topological symmetries of
  $\protect \mathcal {M}\boxtimes \protect \operatorname {{\protect \bf
  K}}^{(\nu )}/A[\psi ^{(1)}, \psi ^{(2)}]$ for $\nu \protect \neq 0$ or $8$,
  since such topological symmetries imply $c_{-} = 0 \protect \text { mod } 4$
  for a MTC.}\BibitemShut {Stop}%
\bibitem [{Note3()}]{Note3}%
  \BibitemOpen
  \bibinfo {note} {The notion of a ``projective homomorphisms,'' generalizing
  the notion of projective representations, can be defined to be a map $p: G
  \to K$ between two groups such that $p ({\protect \bf g}_1 \cdot {\protect
  \bf g}_2) = q({\protect \bf g}_1,{\protect \bf g}_2) p({\protect \bf g}_1)
  p({\protect \bf g}_2)$, where $q({\protect \bf g}_1,{\protect \bf g}_2) \in
  C$ is in a central subgroup $C \lhd Z(K)$. In order to be well-defined, we
  are required to have $q \in Z^2(G,C)$. Introducing an equivalence relation of
  projective homomorphisms $h' \sim k \cdot h$ translates into an equivalence
  $q' = \protect \text {d}k \cdot q$. This allows us to take the quotient by
  $B^2(G,C)$ to classify projective homomorphisms by cohomology classes $[q]
  \in H^2(G,C)$. Similar to projective representations, one can lift a
  projective homomorphism $p$ to an ordinary homomorphism $h: \protect \mathcal
  {G} \to K$ from a central extension $\protect \mathcal {G} = C \times
  _{\varepsilon } G$ with $\varepsilon \in Z^2(G,C)$ when there is a
  homomorphism $r$ such that $r \circ \varepsilon =q$.}\BibitemShut {Stop}%
\bibitem [{\citenamefont {Bulmash}\ and\ \citenamefont
  {Barkeshli}(2021{\natexlab{a}})}]{Bark2021cascade}%
  \BibitemOpen
  \bibfield  {author} {\bibinfo {author} {\bibfnamefont {Daniel}\ \bibnamefont
  {Bulmash}}\ and\ \bibinfo {author} {\bibfnamefont {Maissam}\ \bibnamefont
  {Barkeshli}},\ }\href@noop {} {\enquote {\bibinfo {title} {Anomaly cascade in
  (2+1)d fermionic topological phases},}\ } (\bibinfo {year}
  {2021}{\natexlab{a}}),\ \Eprint {http://arxiv.org/abs/2109.10922}
  {arXiv:2109.10922} \BibitemShut {NoStop}%
\bibitem [{Note4()}]{Note4}%
  \BibitemOpen
  \bibinfo {note} {In v2 of Ref.~\protect \rev@citealp {Bark2021cascade}, it
  was conjectured that the obstruction class $[\protect \textswab {O}^{\eta }]$
  is the same for all modular extensions of $\protect \mathcal {M}_{\protect
  \mathrm {0}}$ for which there is a valid extension of the symmetry action
  $[\rho ^{(0)}]$. In v2 of our paper, we computed the dependence of $[\protect
  \textswab {O}^{\eta }]$ on different extensions and showed that this
  conjecture is generally false. In light of this dependence, the authors of
  Ref.~\protect \rev@citealp {Bark2021cascade} have revised their statements
  regarding the independence of the obstructions on modular extensions to
  involve quotients by terms corresponding to the boundary anomalous FSPT
  states associated with $(3+1)$D FSPT phases described in Ref.~\protect
  \rev@citealp {Wang2020}. These terms that they quotient by precisely match
  the $\nu $ and $\pi $ dependence presented in v4 of our paper, which proves
  their statements.}\BibitemShut {Stop}%
\bibitem [{\citenamefont {Bonderson}\ \emph {et~al.}(2013)\citenamefont
  {Bonderson}, \citenamefont {Nayak},\ and\ \citenamefont {Qi}}]{Bonderson13}%
  \BibitemOpen
  \bibfield  {author} {\bibinfo {author} {\bibfnamefont {Parsa}\ \bibnamefont
  {Bonderson}}, \bibinfo {author} {\bibfnamefont {Chetan}\ \bibnamefont
  {Nayak}}, \ and\ \bibinfo {author} {\bibfnamefont {Xiao-Liang}\ \bibnamefont
  {Qi}},\ }\bibfield  {title} {\enquote {\bibinfo {title} {A time-reversal
  invariant topological phase at the surface of a 3d topological insulator},}\
  }\href {\doibase 10.1088/1742-5468/2013/09/p09016} {\bibfield  {journal}
  {\bibinfo  {journal} {Journal of Statistical Mechanics: Theory and
  Experiment}\ }\textbf {\bibinfo {volume} {2013}},\ \bibinfo {pages} {P09016}
  (\bibinfo {year} {2013})},\ \Eprint {http://arxiv.org/abs/arXiv:1306.3230}
  {arXiv:1306.3230} \BibitemShut {NoStop}%
\bibitem [{\citenamefont {{Barkeshli}}\ \emph {et~al.}(2019)\citenamefont
  {{Barkeshli}}, \citenamefont {{Bonderson}}, \citenamefont {{Cheng}},
  \citenamefont {{Jian}},\ and\ \citenamefont {{Walker}}}]{Barkeshli2019b}%
  \BibitemOpen
  \bibfield  {author} {\bibinfo {author} {\bibfnamefont {Maissam}\ \bibnamefont
  {{Barkeshli}}}, \bibinfo {author} {\bibfnamefont {Parsa}\ \bibnamefont
  {{Bonderson}}}, \bibinfo {author} {\bibfnamefont {Meng}\ \bibnamefont
  {{Cheng}}}, \bibinfo {author} {\bibfnamefont {Chao-Ming}\ \bibnamefont
  {{Jian}}}, \ and\ \bibinfo {author} {\bibfnamefont {Kevin}\ \bibnamefont
  {{Walker}}},\ }\bibfield  {title} {\enquote {\bibinfo {title} {{Reflection
  and Time Reversal Symmetry Enriched Topological Phases of Matter: Path
  Integrals, Non-orientable Manifolds, and Anomalies}},}\ }\href {\doibase
  10.1007/s00220-019-03475-8} {\bibfield  {journal} {\bibinfo  {journal}
  {Communications in Mathematical Physics}\ }\textbf {\bibinfo {volume}
  {374}},\ \bibinfo {pages} {1021--1124} (\bibinfo {year} {2019})},\ \Eprint
  {http://arxiv.org/abs/arXiv:1612.07792} {arXiv:1612.07792} \BibitemShut
  {NoStop}%
\bibitem [{\citenamefont {{Cheng}}\ \emph {et~al.}(2016)\citenamefont
  {{Cheng}}, \citenamefont {{Zaletel}}, \citenamefont {{Barkeshli}},
  \citenamefont {{Vishwanath}},\ and\ \citenamefont {{Bonderson}}}]{Cheng2016}%
  \BibitemOpen
  \bibfield  {author} {\bibinfo {author} {\bibfnamefont {Meng}\ \bibnamefont
  {{Cheng}}}, \bibinfo {author} {\bibfnamefont {Michael}\ \bibnamefont
  {{Zaletel}}}, \bibinfo {author} {\bibfnamefont {Maissam}\ \bibnamefont
  {{Barkeshli}}}, \bibinfo {author} {\bibfnamefont {Ashvin}\ \bibnamefont
  {{Vishwanath}}}, \ and\ \bibinfo {author} {\bibfnamefont {Parsa}\
  \bibnamefont {{Bonderson}}},\ }\bibfield  {title} {\enquote {\bibinfo {title}
  {{Translational Symmetry and Microscopic Constraints on Symmetry-Enriched
  Topological Phases: A View from the Surface}},}\ }\href {\doibase
  10.1103/PhysRevX.6.041068} {\bibfield  {journal} {\bibinfo  {journal}
  {Physical Review X}\ }\textbf {\bibinfo {volume} {6}},\ \bibinfo {eid}
  {041068} (\bibinfo {year} {2016})},\ \Eprint
  {http://arxiv.org/abs/arXiv:1511.02263} {arXiv:1511.02263} \BibitemShut
  {NoStop}%
\bibitem [{Note5()}]{Note5}%
  \BibitemOpen
  \bibinfo {note} {We thank the authors of Ref.~\protect \rev@citealp
  {Bark2021cascade} for sharing an early draft of their v3 revision, where they
  predicted, from anomaly matching considerations combined with our result for
  the unitary case, that the anti-unitary case should have a $q \cup \pi \cup
  \pi $ term when it has a $\pi \cup \protect \mathrm {w}$ term.}\BibitemShut
  {Stop}%
\bibitem [{\citenamefont {{Metlitski}}\ \emph {et~al.}(2014)\citenamefont
  {{Metlitski}}, \citenamefont {{Fidkowski}}, \citenamefont {{Chen}},\ and\
  \citenamefont {{Vishwanath}}}]{Metlitski2014}%
  \BibitemOpen
  \bibfield  {author} {\bibinfo {author} {\bibfnamefont {Max~A.}\ \bibnamefont
  {{Metlitski}}}, \bibinfo {author} {\bibfnamefont {Lukasz}\ \bibnamefont
  {{Fidkowski}}}, \bibinfo {author} {\bibfnamefont {Xie}\ \bibnamefont
  {{Chen}}}, \ and\ \bibinfo {author} {\bibfnamefont {Ashvin}\ \bibnamefont
  {{Vishwanath}}},\ }\bibfield  {title} {\enquote {\bibinfo {title}
  {{Interaction effects on 3D topological superconductors: surface topological
  order from vortex condensation, the 16 fold way and fermionic Kramers
  doublets}},}\ }\href@noop {} {\bibfield  {journal} {\bibinfo  {journal}
  {arXiv e-prints}\ } (\bibinfo {year} {2014})},\ \Eprint
  {http://arxiv.org/abs/arXiv:1406.3032} {arXiv:1406.3032} \BibitemShut
  {NoStop}%
\bibitem [{\citenamefont {{Wang}}\ and\ \citenamefont
  {{Levin}}(2017)}]{Wang2017}%
  \BibitemOpen
  \bibfield  {author} {\bibinfo {author} {\bibfnamefont {Chenjie}\ \bibnamefont
  {{Wang}}}\ and\ \bibinfo {author} {\bibfnamefont {Michael}\ \bibnamefont
  {{Levin}}},\ }\bibfield  {title} {\enquote {\bibinfo {title} {{Anomaly
  Indicators for Time-Reversal Symmetric Topological Orders}},}\ }\href
  {\doibase 10.1103/PhysRevLett.119.136801} {\bibfield  {journal} {\bibinfo
  {journal} {\prl}\ }\textbf {\bibinfo {volume} {119}},\ \bibinfo {eid}
  {136801} (\bibinfo {year} {2017})},\ \Eprint
  {http://arxiv.org/abs/arXiv:1610.04624} {arXiv:1610.04624} \BibitemShut
  {NoStop}%
\bibitem [{\citenamefont {Bonderson}(2007)}]{Bond2007}%
  \BibitemOpen
  \bibfield  {author} {\bibinfo {author} {\bibfnamefont {Parsa~Hassan}\
  \bibnamefont {Bonderson}},\ }\emph {\bibinfo {title} {Non-Abelian anyons and
  interferometry}},\ \href@noop {} {Ph.D. thesis},\ \bibinfo  {school}
  {California Institute of Technology} (\bibinfo {year} {2007})\BibitemShut
  {NoStop}%
\bibitem [{Note6()}]{Note6}%
  \BibitemOpen
  \bibinfo {note} {Anomalous theories that do not fully satisfy the $G$-crossed
  consistency conditions can also be generated by applying the torsor functor
  for $\protect \textswab {t}$ and $\protect \EuScript {X}$ that have
  nontrivial obstructions. These may occur, for example, as surface termination
  states at the boundaries of $(3+1)$D SET phases.}\BibitemShut {Stop}%
\bibitem [{\citenamefont {Delaney}\ \emph {et~al.}(2020)\citenamefont
  {Delaney}, \citenamefont {Galindo}, \citenamefont {Plavnik}, \citenamefont
  {Rowell},\ and\ \citenamefont {Zhang}}]{delaney2020}%
  \BibitemOpen
  \bibfield  {author} {\bibinfo {author} {\bibfnamefont {Colleen}\ \bibnamefont
  {Delaney}}, \bibinfo {author} {\bibfnamefont {C{\'e}sar}\ \bibnamefont
  {Galindo}}, \bibinfo {author} {\bibfnamefont {Julia}\ \bibnamefont
  {Plavnik}}, \bibinfo {author} {\bibfnamefont {Eric~C.}\ \bibnamefont
  {Rowell}}, \ and\ \bibinfo {author} {\bibfnamefont {Qing}\ \bibnamefont
  {Zhang}},\ }\href@noop {} {\enquote {\bibinfo {title} {Braided zesting and
  its applications},}\ } (\bibinfo {year} {2020}),\ \Eprint
  {http://arxiv.org/abs/2005.05544} {arXiv:2005.05544} \BibitemShut {NoStop}%
\bibitem [{\citenamefont {Garc\'{i}a-Etxebarria}\ and\ \citenamefont
  {Montero}(2019)}]{Garcia2019}%
  \BibitemOpen
  \bibfield  {author} {\bibinfo {author} {\bibfnamefont {I\~{n}aki}\
  \bibnamefont {Garc\'{i}a-Etxebarria}}\ and\ \bibinfo {author} {\bibfnamefont
  {Miguel}\ \bibnamefont {Montero}},\ }\bibfield  {title} {\enquote {\bibinfo
  {title} {{D}ai-{F}reed anomalies in particle physics},}\ }\href {\doibase
  10.1007/jhep08(2019)003} {\bibfield  {journal} {\bibinfo  {journal} {Journal
  of High Energy Physics}\ }\textbf {\bibinfo {volume} {2019}} (\bibinfo {year}
  {2019}),\ 10.1007/jhep08(2019)003},\ \Eprint
  {http://arxiv.org/abs/arXiv:1808.00009} {arXiv:1808.00009} \BibitemShut
  {NoStop}%
\bibitem [{\citenamefont {Cui}\ \emph {et~al.}(2016)\citenamefont {Cui},
  \citenamefont {Galindo}, \citenamefont {Plavnik},\ and\ \citenamefont
  {Wang}}]{Cui2016}%
  \BibitemOpen
  \bibfield  {author} {\bibinfo {author} {\bibfnamefont {Shawn~X}\ \bibnamefont
  {Cui}}, \bibinfo {author} {\bibfnamefont {C{\'e}sar}\ \bibnamefont
  {Galindo}}, \bibinfo {author} {\bibfnamefont {Julia~Yael}\ \bibnamefont
  {Plavnik}}, \ and\ \bibinfo {author} {\bibfnamefont {Zhenghan}\ \bibnamefont
  {Wang}},\ }\bibfield  {title} {\enquote {\bibinfo {title} {On gauging
  symmetry of modular categories},}\ }\href@noop {} {\bibfield  {journal}
  {\bibinfo  {journal} {Communications in Mathematical Physics}\ }\textbf
  {\bibinfo {volume} {348}},\ \bibinfo {pages} {1043--1064} (\bibinfo {year}
  {2016})}\BibitemShut {NoStop}%
\bibitem [{\citenamefont {Gaiotto}\ and\ \citenamefont
  {Johnson-Freyd}(2019{\natexlab{a}})}]{Gaiotto2019}%
  \BibitemOpen
  \bibfield  {author} {\bibinfo {author} {\bibfnamefont {Davide}\ \bibnamefont
  {Gaiotto}}\ and\ \bibinfo {author} {\bibfnamefont {Theo}\ \bibnamefont
  {Johnson-Freyd}},\ }\bibfield  {title} {\enquote {\bibinfo {title} {Symmetry
  protected topological phases and generalized cohomology},}\ }\href {\doibase
  10.1007/jhep05(2019)007} {\bibfield  {journal} {\bibinfo  {journal} {Journal
  of High Energy Physics}\ }\textbf {\bibinfo {volume} {2019}} (\bibinfo {year}
  {2019}{\natexlab{a}}),\ 10.1007/jhep05(2019)007},\ \Eprint
  {http://arxiv.org/abs/arXiv:1712.07950} {arXiv:1712.07950} \BibitemShut
  {NoStop}%
\bibitem [{\citenamefont {Barkeshli}\ \emph {et~al.}(2021)\citenamefont
  {Barkeshli}, \citenamefont {Chen}, \citenamefont {Hsin},\ and\ \citenamefont
  {Manjunath}}]{Bark2021inv}%
  \BibitemOpen
  \bibfield  {author} {\bibinfo {author} {\bibfnamefont {Maissam}\ \bibnamefont
  {Barkeshli}}, \bibinfo {author} {\bibfnamefont {Yu-An}\ \bibnamefont {Chen}},
  \bibinfo {author} {\bibfnamefont {Po-Shen}\ \bibnamefont {Hsin}}, \ and\
  \bibinfo {author} {\bibfnamefont {Naren}\ \bibnamefont {Manjunath}},\
  }\bibfield  {title} {\enquote {\bibinfo {title} {Classification of (2+1)d
  invertible fermionic topological phases with symmetry},}\ }\href@noop {} {\
  (\bibinfo {year} {2021})},\ \Eprint {http://arxiv.org/abs/2109.11039}
  {arXiv:2109.11039} \BibitemShut {NoStop}%
\bibitem [{Note7()}]{Note7}%
  \BibitemOpen
  \bibinfo {note} {The defectification obstruction of FSPT phases with trivial
  symmetry action is equal to the relative defectification obstruction, i.e.,
  $[\protect \EuScript {O}_r(\protect \textswab {p}_{\protect \mathrm {w}},0)]
  = [\protect \EuScript {O}(\protect \textswab {p}_{\protect \mathrm {w}}
  ,0)]$.}\BibitemShut {Stop}%
\bibitem [{\citenamefont {Gu}\ and\ \citenamefont {Wen}(2014)}]{Gu2014}%
  \BibitemOpen
  \bibfield  {author} {\bibinfo {author} {\bibfnamefont {Zheng-Cheng}\
  \bibnamefont {Gu}}\ and\ \bibinfo {author} {\bibfnamefont {Xiao-Gang}\
  \bibnamefont {Wen}},\ }\bibfield  {title} {\enquote {\bibinfo {title}
  {Symmetry-protected topological orders for interacting fermions: Fermionic
  topological nonlinear $\sigma$ models and a special group supercohomology
  theory},}\ }\href {\doibase 10.1103/physrevb.90.115141} {\bibfield  {journal}
  {\bibinfo  {journal} {Physical Review B}\ }\textbf {\bibinfo {volume} {90}}
  (\bibinfo {year} {2014}),\ 10.1103/physrevb.90.115141}\BibitemShut {NoStop}%
\bibitem [{\citenamefont {Gaiotto}\ and\ \citenamefont
  {Johnson-Freyd}(2019{\natexlab{b}})}]{Freyd2019}%
  \BibitemOpen
  \bibfield  {author} {\bibinfo {author} {\bibfnamefont {Davide}\ \bibnamefont
  {Gaiotto}}\ and\ \bibinfo {author} {\bibfnamefont {Theo}\ \bibnamefont
  {Johnson-Freyd}},\ }\bibfield  {title} {\enquote {\bibinfo {title} {Symmetry
  protected topological phases and generalized cohomology},}\ }\href {\doibase
  10.1007/jhep05(2019)007} {\bibfield  {journal} {\bibinfo  {journal} {Journal
  of High Energy Physics}\ }\textbf {\bibinfo {volume} {2019}} (\bibinfo {year}
  {2019}{\natexlab{b}}),\ 10.1007/jhep05(2019)007},\ \Eprint
  {http://arxiv.org/abs/arXiv:1712.07950} {arXiv:1712.07950} \BibitemShut
  {NoStop}%
\bibitem [{\citenamefont {M{\"u}ger}(2003)}]{Muger2003}%
  \BibitemOpen
  \bibfield  {author} {\bibinfo {author} {\bibfnamefont {Michael}\ \bibnamefont
  {M{\"u}ger}},\ }\bibfield  {title} {\enquote {\bibinfo {title} {On the
  structure of modular categories},}\ }\href@noop {} {\bibfield  {journal}
  {\bibinfo  {journal} {Proceedings of the London Mathematical Society}\
  }\textbf {\bibinfo {volume} {87}},\ \bibinfo {pages} {291--308} (\bibinfo
  {year} {2003})},\ \Eprint {http://arxiv.org/abs/arXiv:math/0201017}
  {arXiv:math/0201017} \BibitemShut {NoStop}%
\bibitem [{Note8()}]{Note8}%
  \BibitemOpen
  \bibinfo {note} {An alternative choice of the physical fermion as
  $(\varepsilon ,0)$ would lead to less interesting theories that have product
  structure $\protect \mathcal {M}= \protect \operatorname {{\protect \bf
  K}}^{(\nu )} \boxtimes \protect \mathbb {Z}_{8}^{(n+1/2)}$ and are fully
  described by Sec.~\ref {sec:ex_KM0}.}\BibitemShut {Stop}%
\bibitem [{\citenamefont {Moore}\ and\ \citenamefont {Read}(1991)}]{Moore1991}%
  \BibitemOpen
  \bibfield  {author} {\bibinfo {author} {\bibfnamefont {Gregory}\ \bibnamefont
  {Moore}}\ and\ \bibinfo {author} {\bibfnamefont {Nicholas}\ \bibnamefont
  {Read}},\ }\bibfield  {title} {\enquote {\bibinfo {title} {Nonabelions in the
  fractional quantum hall effect},}\ }\href {\doibase
  https://doi.org/10.1016/0550-3213(91)90407-O} {\bibfield  {journal} {\bibinfo
   {journal} {Nuclear Physics B}\ }\textbf {\bibinfo {volume} {360}},\ \bibinfo
  {pages} {362--396} (\bibinfo {year} {1991})}\BibitemShut {NoStop}%
\bibitem [{\citenamefont {Lee}\ \emph {et~al.}(2007)\citenamefont {Lee},
  \citenamefont {Ryu}, \citenamefont {Nayak},\ and\ \citenamefont
  {Fisher}}]{Lee07}%
  \BibitemOpen
  \bibfield  {author} {\bibinfo {author} {\bibfnamefont {Sung-Sik}\
  \bibnamefont {Lee}}, \bibinfo {author} {\bibfnamefont {Shinsei}\ \bibnamefont
  {Ryu}}, \bibinfo {author} {\bibfnamefont {Chetan}\ \bibnamefont {Nayak}}, \
  and\ \bibinfo {author} {\bibfnamefont {Matthew P.~A.}\ \bibnamefont
  {Fisher}},\ }\bibfield  {title} {\enquote {\bibinfo {title} {Particle-hole
  symmetry and the $\nu=5/2$ quantum hall state},}\ }\href@noop {} {\bibfield
  {journal} {\bibinfo  {journal} {Physical Review Letters}\ }\textbf {\bibinfo
  {volume} {99}} (\bibinfo {year} {2007})},\ \Eprint
  {http://arxiv.org/abs/arXiv:0707.0478} {arXiv:0707.0478} \BibitemShut
  {NoStop}%
\bibitem [{\citenamefont {Levin}\ \emph {et~al.}(2007)\citenamefont {Levin},
  \citenamefont {Halperin},\ and\ \citenamefont {Rosenow}}]{Levin07}%
  \BibitemOpen
  \bibfield  {author} {\bibinfo {author} {\bibfnamefont {Michael}\ \bibnamefont
  {Levin}}, \bibinfo {author} {\bibfnamefont {Bertrand~I.}\ \bibnamefont
  {Halperin}}, \ and\ \bibinfo {author} {\bibfnamefont {Bernd}\ \bibnamefont
  {Rosenow}},\ }\bibfield  {title} {\enquote {\bibinfo {title} {Particle-hole
  symmetry and the pfaffian state},}\ }\href {\doibase
  10.1103/physrevlett.99.236806} {\bibfield  {journal} {\bibinfo  {journal}
  {Physical Review Letters}\ }\textbf {\bibinfo {volume} {99}} (\bibinfo {year}
  {2007}),\ 10.1103/physrevlett.99.236806},\ \Eprint
  {http://arxiv.org/abs/arXiv:0707.0483} {arXiv:0707.0483} \BibitemShut
  {NoStop}%
\bibitem [{\citenamefont {{Chen}}\ \emph {et~al.}(2014)\citenamefont {{Chen}},
  \citenamefont {{Fidkowski}},\ and\ \citenamefont {{Vishwanath}}}]{Chen2014b}%
  \BibitemOpen
  \bibfield  {author} {\bibinfo {author} {\bibfnamefont {Xie}\ \bibnamefont
  {{Chen}}}, \bibinfo {author} {\bibfnamefont {Lukasz}\ \bibnamefont
  {{Fidkowski}}}, \ and\ \bibinfo {author} {\bibfnamefont {Ashvin}\
  \bibnamefont {{Vishwanath}}},\ }\bibfield  {title} {\enquote {\bibinfo
  {title} {{Symmetry enforced non-Abelian topological order at the surface of a
  topological insulator}},}\ }\href {\doibase 10.1103/PhysRevB.89.165132}
  {\bibfield  {journal} {\bibinfo  {journal} {\prb}\ }\textbf {\bibinfo
  {volume} {89}},\ \bibinfo {eid} {165132} (\bibinfo {year} {2014})},\ \Eprint
  {http://arxiv.org/abs/arXiv:1306.3250} {arXiv:1306.3250} \BibitemShut
  {NoStop}%
\bibitem [{\citenamefont {Bulmash}\ and\ \citenamefont
  {Barkeshli}(2021{\natexlab{b}})}]{Bark2021frac}%
  \BibitemOpen
  \bibfield  {author} {\bibinfo {author} {\bibfnamefont {Daniel}\ \bibnamefont
  {Bulmash}}\ and\ \bibinfo {author} {\bibfnamefont {Maissam}\ \bibnamefont
  {Barkeshli}},\ }\bibfield  {title} {\enquote {\bibinfo {title} {Fermionic
  symmetry fractionalization in (2+1)d},}\ }\href@noop {} {\  (\bibinfo {year}
  {2021}{\natexlab{b}})},\ \Eprint {http://arxiv.org/abs/2109.10913}
  {arXiv:2109.10913} \BibitemShut {NoStop}%
\bibitem [{\citenamefont {Ning}\ \emph {et~al.}(2021)\citenamefont {Ning},
  \citenamefont {Qi}, \citenamefont {Gu},\ and\ \citenamefont
  {Wang}}]{Ning2021}%
  \BibitemOpen
  \bibfield  {author} {\bibinfo {author} {\bibfnamefont {Shang-Qiang}\
  \bibnamefont {Ning}}, \bibinfo {author} {\bibfnamefont {Yang}\ \bibnamefont
  {Qi}}, \bibinfo {author} {\bibfnamefont {Zheng-Cheng}\ \bibnamefont {Gu}}, \
  and\ \bibinfo {author} {\bibfnamefont {Chenjie}\ \bibnamefont {Wang}},\
  }\bibfield  {title} {\enquote {\bibinfo {title} {Enforced symmetry breaking
  by invertible topological order},}\ }\href@noop {} {\  (\bibinfo {year}
  {2021})},\ \Eprint {http://arxiv.org/abs/2109.15307} {arXiv:2109.15307}
  \BibitemShut {NoStop}%
\bibitem [{\citenamefont {de~Wild~Propitius}(1995)}]{Propitius1995}%
  \BibitemOpen
  \bibfield  {author} {\bibinfo {author} {\bibfnamefont {Mark}\ \bibnamefont
  {de~Wild~Propitius}},\ }\href@noop {} {\enquote {\bibinfo {title}
  {Topological interactions in broken gauge theories},}\ } (\bibinfo {year}
  {1995}),\ \Eprint {http://arxiv.org/abs/hep-th/9511195}
  {arXiv:hep-th/9511195} \BibitemShut {NoStop}%
\bibitem [{\citenamefont {Jr}\ and\ \citenamefont
  {Ostrik}(2001)}]{Kirillov2001}%
  \BibitemOpen
  \bibfield  {author} {\bibinfo {author} {\bibfnamefont {Alexander~Kirillov}\
  \bibnamefont {Jr}}\ and\ \bibinfo {author} {\bibfnamefont {Viktor}\
  \bibnamefont {Ostrik}},\ }\href@noop {} {\enquote {\bibinfo {title} {On
  q-analog of {McKay} correspondence and {ADE} classification of {SL2}
  conformal field theories},}\ } (\bibinfo {year} {2001}),\ \Eprint
  {http://arxiv.org/abs/math/0101219} {arXiv:math/0101219} \BibitemShut
  {NoStop}%
\bibitem [{\citenamefont {Fuchs}\ \emph {et~al.}(2002)\citenamefont {Fuchs},
  \citenamefont {Runkel},\ and\ \citenamefont {Schweigert}}]{Fuchs2002}%
  \BibitemOpen
  \bibfield  {author} {\bibinfo {author} {\bibfnamefont {Jurgen}\ \bibnamefont
  {Fuchs}}, \bibinfo {author} {\bibfnamefont {Ingo}\ \bibnamefont {Runkel}}, \
  and\ \bibinfo {author} {\bibfnamefont {Christoph}\ \bibnamefont
  {Schweigert}},\ }\bibfield  {title} {\enquote {\bibinfo {title} {{TFT}
  construction of {RCFT} correlators {I}: partition functions},}\ }\href
  {\doibase 10.1016/s0550-3213(02)00744-7} {\bibfield  {journal} {\bibinfo
  {journal} {Nuclear Physics B}\ }\textbf {\bibinfo {volume} {646}},\ \bibinfo
  {pages} {353--497} (\bibinfo {year} {2002})},\ \Eprint
  {http://arxiv.org/abs/arXiv:hep-th/0204148} {arXiv:hep-th/0204148}
  \BibitemShut {NoStop}%
\bibitem [{\citenamefont {Eli\"{e}ns}\ \emph {et~al.}(2014)\citenamefont
  {Eli\"{e}ns}, \citenamefont {Romers},\ and\ \citenamefont {Bais}}]{Bais2014}%
  \BibitemOpen
  \bibfield  {author} {\bibinfo {author} {\bibfnamefont {I.~S.}\ \bibnamefont
  {Eli\"{e}ns}}, \bibinfo {author} {\bibfnamefont {J.~C.}\ \bibnamefont
  {Romers}}, \ and\ \bibinfo {author} {\bibfnamefont {F.~A.}\ \bibnamefont
  {Bais}},\ }\bibfield  {title} {\enquote {\bibinfo {title} {Diagrammatics for
  bose condensation in anyon theories},}\ }\href {\doibase
  10.1103/physrevb.90.195130} {\bibfield  {journal} {\bibinfo  {journal}
  {Physical Review B}\ }\textbf {\bibinfo {volume} {90}} (\bibinfo {year}
  {2014}),\ 10.1103/physrevb.90.195130},\ \Eprint
  {http://arxiv.org/abs/arXiv:1310.6001} {arXiv:1310.6001} \BibitemShut
  {NoStop}%
\bibitem [{\citenamefont {Kong}(2014)}]{Kong14}%
  \BibitemOpen
  \bibfield  {author} {\bibinfo {author} {\bibfnamefont {Liang}\ \bibnamefont
  {Kong}},\ }\bibfield  {title} {\enquote {\bibinfo {title} {Anyon condensation
  and tensor categories},}\ }\href {\doibase 10.1016/j.nuclphysb.2014.07.003}
  {\bibfield  {journal} {\bibinfo  {journal} {Nuclear Physics B}\ }\textbf
  {\bibinfo {volume} {886}},\ \bibinfo {pages} {436--482} (\bibinfo {year}
  {2014})},\ \Eprint {http://arxiv.org/abs/arXiv:1307.8244} {arXiv:1307.8244}
  \BibitemShut {NoStop}%
\bibitem [{\citenamefont {Bischoff}\ \emph {et~al.}(2019)\citenamefont
  {Bischoff}, \citenamefont {Jones}, \citenamefont {Lu},\ and\ \citenamefont
  {Penneys}}]{Bischoff19}%
  \BibitemOpen
  \bibfield  {author} {\bibinfo {author} {\bibfnamefont {Marcel}\ \bibnamefont
  {Bischoff}}, \bibinfo {author} {\bibfnamefont {Corey}\ \bibnamefont {Jones}},
  \bibinfo {author} {\bibfnamefont {Yuan-Ming}\ \bibnamefont {Lu}}, \ and\
  \bibinfo {author} {\bibfnamefont {David}\ \bibnamefont {Penneys}},\
  }\bibfield  {title} {\enquote {\bibinfo {title} {Spontaneous symmetry
  breaking from anyon condensation},}\ }\href {\doibase
  10.1007/jhep02(2019)062} {\bibfield  {journal} {\bibinfo  {journal} {Journal
  of High Energy Physics}\ }\textbf {\bibinfo {volume} {2019}} (\bibinfo {year}
  {2019}),\ 10.1007/jhep02(2019)062},\ \Eprint
  {http://arxiv.org/abs/arXiv:1811.00434} {arXiv:1811.00434} \BibitemShut
  {NoStop}%
\bibitem [{Note9()}]{Note9}%
  \BibitemOpen
  \bibinfo {note} {At first glance, one might worry this product is not
  gauge-invariant. This is not a concern as the gauge has been fully determined
  by the assumption $F^{xbb}=1$.}\BibitemShut {Stop}%
\bibitem [{\citenamefont {Brown}(1982)}]{Brown1982}%
  \BibitemOpen
  \bibfield  {author} {\bibinfo {author} {\bibfnamefont {K}~\bibnamefont
  {Brown}},\ }\bibfield  {title} {\enquote {\bibinfo {title} {Cohomology of
  groups},}\ }\href@noop {} {\bibfield  {journal} {\bibinfo  {journal} {New
  York}\ } (\bibinfo {year} {1982})}\BibitemShut {NoStop}%
\bibitem [{\citenamefont {Steenrod}(1947)}]{Steenrod1947}%
  \BibitemOpen
  \bibfield  {author} {\bibinfo {author} {\bibfnamefont {N.~E.}\ \bibnamefont
  {Steenrod}},\ }\bibfield  {title} {\enquote {\bibinfo {title} {Products of
  cocycles and extensions of mappings},}\ }\href
  {http://www.jstor.org/stable/1969172} {\bibfield  {journal} {\bibinfo
  {journal} {Annals of Mathematics}\ }\textbf {\bibinfo {volume} {48}},\
  \bibinfo {pages} {290--320} (\bibinfo {year} {1947})}\BibitemShut {NoStop}%
\bibitem [{\citenamefont {Hirsch}(1955)}]{Hirsch1955}%
  \BibitemOpen
  \bibfield  {author} {\bibinfo {author} {\bibfnamefont {G.}~\bibnamefont
  {Hirsch}},\ }\bibfield  {title} {\enquote {\bibinfo {title} {Quelques
  propri\'et\'es des produits de steenrod},}\ }\href@noop {} {\bibfield
  {journal} {\bibinfo  {journal} {C. R. Acad. Sci. Paris}\ }\textbf {\bibinfo
  {volume} {241}},\ \bibinfo {pages} {923--925} (\bibinfo {year}
  {1955})}\BibitemShut {NoStop}%
\end{thebibliography}%

\end{document}